  \documentclass[12pt,a4paper,twoside,openright]{report}
  \usepackage[ngerman, english]{babel}  % mehrsprachiger Textsatz
  \usepackage[latin1]{inputenc}    % Passen Sie die Option an.
  \usepackage[T1]{fontenc}         % Schriftenkodierung
  \usepackage{graphicx}            % zum Einbinden von Grafiken
  \usepackage{lmodern}             % Ersatz fuer Computer Modern-Schriften
	\usepackage[nottoc]{tocbibind} % Fuegt das Literatur- und andere Verzieichnisse in das Inhaltsverzeichnis und in die Lesezeichen der PDF-Datei; die Optionen koennen nach Bedarf angepasst werden                                 

	\usepackage{fancyhdr}   
	\usepackage{dsfont}   % option sans                      
	\usepackage[bf,small]{caption}    

\usepackage{amssymb}
\usepackage[square, comma, numbers, sort&compress]{natbib} 	
\usepackage{hypernat}  
\usepackage{multibib}
	\usepackage{hyperref}

\newcommand{\dcauthorpre}{Herrn Dipl.-Phys.} 
\newcommand{\dcauthorsurname}{Gerhold} 
\newcommand{\dcauthorname}{Philipp Frederik Clemens} 
\newcommand{\dcauthoradd}{geboren am 24.09.1979 in Den Helder, NL}

\newcommand{\dctitle}{Upper and lower Higgs boson mass bounds from a chirally invariant lattice Higgs-Yukawa model} 
\newcommand{\dcsubtitle}{~}  
\newcommand{\dcapprovala}{Prof. Dr. Michael M\"uller-Preussker} % 
\newcommand{\dcapprovalb}{Dr. habil. Karl Jansen} %  
\newcommand{\dcapprovalc}{Prof. Dr. Andreas Wipf}  % 

\newcommand{\dcdegree}{doctor rerum naturalium\\(Dr. rer. nat.)} 
\newcommand{\dcsubject}{Physik} 
\newcommand{\dcfaculty}{Mathematisch-Naturwissenschaftlichen \\Fakult\"at I}
\newcommand{\dcuniversity}{Humboldt-Universit\"at zu Berlin}
\newcommand{\dcdean}{Prof. Dr. Lutz-Helmut Sch\"on}
\newcommand{\dcpresident}{Prof. Dr. Dr. h.c. Christoph Markschies}

\newcommand{\dcdatesubmitted}{23. Juli 2009} %auch wenn nicht auf dem Titelblatt, bitte erfüllen!
\newcommand{\dcdateexam}{15. Oktober 2009} %

\newcommand{\dckeydea}{Obere Higgs-Massenschranke}
\newcommand{\dckeydeb}{Untere Higgs-Massenschranke}
\newcommand{\dckeydec}{Higgs-Yukawa Modell}
\newcommand{\dckeyded}{Higgs-Teilchen}

\newcommand{\dcpdfsubject}{Dissertation}
	\hypersetup{
		pdftitle={\dctitle},
		pdfauthor={\dcauthorname\ \dcauthorsurname},
		pdfsubject={\dcpdfsubject}, % optional
		pdfkeywords={\dckeydea, \dckeydeb, \dckeydec, \dckeyded},
		pdfpagemode=UseOutlines,
  	colorlinks,
		linkcolor=black,
		filecolor=black,
		urlcolor=black,	
		citecolor=black,
		plainpages=false,
		hypertexnames=false,
	}

\newcommand{\vs}{\vspace}
\newcommand{\hs}{\hspace}

\newcommand{\nin}{\noindent}

\newcommand{\bdm}{\begin{displaymath}}
\newcommand{\edm}{\end{displaymath}}
\newcommand{\beq}{\begin{equation}}
\newcommand{\eeq}{\end{equation}}
\newcommand{\bea}{\begin{eqnarray}}
\newcommand{\eea}{\end{eqnarray}}
\newcommand{\bit}{\begin{itemize}}
\newcommand{\eit}{\end{itemize}}
\newcommand{\bc}{\begin{center}}
\newcommand{\ec}{\end{center}}
\newcommand{\re}{\relax{\textrm{I\kern-.18em R}}}
\newcommand{\ID}{\mathds{1}}
\newcommand{\Z}{\mathds{Z}}
\newcommand{\N}{\mathds{N}}

\newcommand{\Comp}{\mathds{C}}
\newcommand{\fhs}[1]{\mbox{\hs{#1}}}
\newcommand{\ie}{\textit{i.e.}$\;$}

\newcommand{\Dgen}{{\mathcal D}}
\newcommand{\Dov}{{\mathcal D}^{(ov)}}
\newcommand{\D}{\Dov}
\newcommand{\GammaOp}{\Gamma^{(ov)}}
\newcommand{\DMom}{{\tilde{\mathcal D}}^{(ov)}}
\newcommand{\GammaOpMom}{{\tilde\Gamma}^{(ov)}}
\newcommand{\Dqcd}{{\mathcal D}^{(QCD)}}
\newcommand{\Dprime}{{\mathcal D'}^{(ov)}}
\newcommand{\DnoK}{{\breve{\mathcal D}}^{(ov)}}
\newcommand{\SD}{{\mathcal D}^{(ov)}}

\newcommand{\Dw}{{\mathcal D}^{(W)}}
\newcommand{\SDw}{{\mathcal D}^{(W)}}
\newcommand{\DDmtwoRhoInv}{{\mathcal A}}
\newcommand{\ImpSpace}{{\mathcal P}}
\newcommand{\ImpBasis}{{\mathcal Q}}
\newcommand{\sumFL}{\sum\limits_{i=1}^{N_f}}

\newcommand{\fermiMat}{{\mathcal M}}

\newcommand{\kapCrit}{\kappa_\mathrm{crit}}
\newcommand{\timeRev}[1]{#1_\mathrm{trev}}
\newcommand{\mAvg}{{\langle m \rangle}}
\newcommand{\sAvg}{{\langle s \rangle}}
\newcommand{\LagrangeMul}{\bar \omega}

\newcommand{\UVUA}{\mbox{U}(1)_V\times\mbox{U}(1)_A}
\newcommand{\ULUR}{\mbox{U}(1)_L\times\mbox{U}(1)_R}

\newcommand{\mergedBlockA}{{\mathcal A}}
\newcommand{\eq}[1]{Eq.~(\ref{#1})}
\newcommand{\eqs}[2]{Eq.~(\ref{#1}-\ref{#2})}
\newcommand{\sect}[1]{section~\ref{#1}}
\newcommand{\appen}[1]{appendix~\ref{#1}}
\newcommand{\sects}[2]{sections~\ref{#1} and \ref{#2}}
\newcommand{\chap}[1]{chapter~\ref{#1}}

\newcommand{\fig}[1]{Fig.~\ref{#1}}
\newcommand{\tab}[1]{Tab.~\ref{#1}}
\newcommand{\sign}{\mbox{sign}}
\newcommand{\diag}{\mbox{diag}}
\newcommand{\diagY}{\mbox{diag}(\hat y_t, \hat y_b)}
\newcommand{\fermiMatDouble}{\fermiMat\fermiMat^\dagger}
\newcommand{\cond}{\mbox{cond}}
\newcommand{\newfermiMat}{{\mathcal N}}

\newcommand{\precP}{Q}
\newcommand{\precR}{R}
\newcommand{\FT}{\mbox{FFT}}
\newcommand{\krylovSpace}{{\mathcal K}}
\newcommand{\latVolExp}[2]{$#1^3\times #2$}
\newcommand{\latVol}[2]{\ifnum #1=#2 $#1^4$ \else $#1^3\times #2$\fi}
\newcommand{\lattice}[2]{\ifnum #1=#2 $#1^4$-lattice \else $#1^3\times #2$-lattice\fi}
\newcommand{\latticeX}[3]{\ifnum #1=#2 $#1^4$-lattice#3 \else $#1^3\times #2$-lattice#3\fi}
\newcommand{\lattices}[2]{\ifnum #1=#2 $#1^4$-lattice \else $#1^3\times #2$-lattices\fi}
\newcommand{\intd}[1]{\mbox{d}#1\,}
\newcommand{\intdfour}[1]{\mbox{d}^4#1\,}
\newcommand{\intD}[1]{\mbox{D}#1\,}
\newcommand{\derive}[2]{\frac{\mbox{d}#1}{\mbox{d}#2}}
\newcommand{\funcInt}{\int \intD{\Phi} \intD{\psi} \intD{\bar\psi}}
\newcommand{\GEV}[1]{#1\,\mbox{GeV}}
\newcommand{\TEV}[1]{#1\,\mbox{TeV}}
\newcommand{\per}[1]{#1\,\%}
\newcommand{\Tr}{\mbox{Tr}}
\newcommand{\RE}{\mbox{Re}}
\newcommand{\IM}{\mbox{Im}}
\newcommand{\Ref}[1]{Ref.~\cite{#1}}
\newcommand{\Refs}[1]{Refs.~\cite{#1}}
\newcommand{\seeeg}[1]{see e.g. \Ref{#1}}
\newcommand{\seeegs}[1]{see e.g. \Refs{#1}}
\newcommand{\SUU}{$\mbox{SU(2)}_L\times\mbox{U(1)}_Y\mbox{ }$}
\newcommand{\proz}[1]{#1\,\%}
\newcommand{\arctanh}{\mbox{arctanh}}
\newcommand{\timeOrdering}{{\mathcal T}}
\newcommand{\OneBosonLoopContEuc}{{\mathcal J}}
\newcommand{\OneGoldstoneLoopContEuc}{{\mathcal I}}
\newcommand{\FuncIntAvg}{{\mathcal F}}
\newcommand{\IntegratorSymbol}{{\mathcal Q}}
\newcommand{\traLength}{\tau_{\mathcal Q}}
\newcommand{\intSteps}{N_{\mathcal Q}}
\newcommand{\intStepSize}{\epsilon_{\mathcal Q}}
\newcommand{\res}{\mbox{res}}
\newcommand{\MCtime}{t_{MC}}
\newcommand{\confNR}{i_{MC}}
\newcommand{\ACtime}{\tau_{MC}}
\newcommand{\LatSpace}{{\Gamma}}
\newcommand{\thermTime}{{N_{{therm}}}}
\newcommand{\confSpace}{{\mathcal V}}
\newcommand{\SFeff}{S_F^{eff}}
\newcommand{\Seff}{S^{eff}}
\newcommand{\SNzero}{S_{N_f}^{(0)}}
\newcommand{\SNone}{S_{N_f}^{(1)}}
\newcommand{\Ssigma}{S_{\sigma}}
\newcommand{\SsigmaOmega}{S_{\sigma,\omega}}
\newcommand{\SsigmaOmegaApprox}{S^{(a)}_{\sigma,\omega}}
\newcommand{\SBarOmega}{S_{\bar\omega}}
\newcommand{\Bscaled}{\underline{B}}
\newcommand{\BscaledHat}{\underline{\hat B}}
\newcommand{\detPrime}{\mbox{det}'}
\newcommand{\detPrimePrime}{\mbox{det}''}
\newcommand{\TrPrimePrime}{\mbox{Tr}''}
\newcommand{\detStar}{\mbox{det}^*}
\newcommand{\CoupStr}{\Xi}
\newcommand{\Nconf}{N_{Conf}}
\newcommand{\lowBound}{m_{H}^{low}(\Lambda)}
\newcommand{\upBound}{m_{H}^{up}(\Lambda)}
\newcommand{\fermiMatv}{\fermiMat_v}
\newcommand{\fermiMatvbreve}{\fermiMat_{\breve v}}
\newcommand{\fermiMatvMom}{{\tilde\fermiMat}_v}
\newcommand{\zetaF}{{\mathcal Z}}
\newcommand{\LanczosMatrix}{{A}}
\newcommand{\LanczosRelAcc}{{\delta_\mathrm{pres}}}
\newcommand{\UoneEM}{{U(1)_\mathrm{em}}}
\newcommand{\UoneY}{{U(1)_\mathrm{Y}}}
\newcommand{\SUthreeC}{{SU(3)_\mathrm{c}}}
\newcommand{\SUtwoTimesUoneY}{{SU(2)_\mathrm{L}\times U(1)_\mathrm{Y}}}
\newcommand{\SUtwoL}{{SU(2)_\mathrm{L}}}
\newcommand{\SUthreeCSUtwoTimesUoneY}{{SU(3)_\mathrm{c}\times SU(2)_\mathrm{L}\times U(1)_\mathrm{Y}}}
\newcommand{\SUthreeCUoneEM}{{SU(3)_\mathrm{c}\times U(1)_\mathrm{em}}}
\newcommand{\Nproc}{N_\mathrm{proc}}
\newcommand{\Ncore}{N_\mathrm{core}}
\newcommand{\Ppeak}{P_\mathrm{peak}}
\newcommand{\Ncache}{N_\mathrm{cache}}
\newcommand{\Acache}{A_\mathrm{cache}}
\newcommand{\Amem}{A_\mathrm{mem}}

\newcommand{\includeFigSingleSmall}[4]{
\bc
\begin{figure}[htb]
\centering
\includegraphics[width=0.48\textwidth]{#1}
\caption[#4]{#3}
\label{#2}
\vs{-2mm}
\end{figure}
\ec
\vs{-6mm}
}

\newcommand{\includeFigSingleMedium}[4]{
\bc
\begin{figure}[htb]
\centering
\includegraphics[width=0.75\textwidth]{#1}
\caption[#4]{#3}
\label{#2}
\vs{-2mm}
\end{figure}
\ec
\vs{-6mm}
}

\newcommand{\includeFigSingleLarge}[4]{
\bc
\begin{figure}[htb]
\centering
\includegraphics[width=1.0\textwidth]{#1}
\caption[#4]{#3}
\label{#2}
\vs{-2mm}
\end{figure}
\ec
\vs{-6mm}
}

\newcommand{\includeFigDouble}[5]{
\bc
\begin{figure}[htb]
\centering
\begin{tabular}{cc}
\includegraphics[width=0.48\textwidth]{#1}
&
\includegraphics[width=0.48\textwidth]{#2}
\\
\hs{4mm}(a) & \hs{8mm}(b)  \\
\end{tabular}
\caption[#5]{#4}
\label{#3}
\vs{-2mm}
\end{figure}
\ec
\vs{-6mm}
}

\newcommand{\includeFigDoubleDoubleHere}[7]{
\bc
\begin{figure}[ht!]
\centering
\begin{tabular}{cc}
\includegraphics[width=0.48\textwidth]{#1}
&
\includegraphics[width=0.48\textwidth]{#2}
\\
\hs{4mm}(a) & \hs{8mm}(b)  \\
\includegraphics[width=0.48\textwidth]{#3}
&
\includegraphics[width=0.48\textwidth]{#4}
\\
\hs{4mm}(c) & \hs{8mm}(d)  \\
\end{tabular}
\caption[#7]{#6}
\label{#5}
\vs{-2mm}
\end{figure}
\ec
\vs{-6mm}
}

\newcommand{\includeFigTrippleDouble}[9]{
\bc
\begin{figure}[htb]
\centering
\begin{tabular}{ccc}
\includegraphics[width=0.32\textwidth]{#1}\hs{-3mm}
&
\includegraphics[width=0.32\textwidth]{#2}\hs{-3mm}
&
\includegraphics[width=0.32\textwidth]{#3}\hs{-3mm}

\\
\hs{4mm}(a) & \hs{8mm}(b) & \hs{8mm}(c) \\
\includegraphics[width=0.32\textwidth]{#4}\hs{-3mm}
&
\includegraphics[width=0.32\textwidth]{#5}\hs{-3mm}
&
\includegraphics[width=0.32\textwidth]{#6}\hs{-3mm}
\\
\hs{4mm}(d) & \hs{8mm}(e)  & \hs{8mm}(f) \\
\end{tabular}
\caption[#9]{#8}
\label{#7}
\vs{-2mm}
\end{figure}
\ec
\vs{-6mm}
}

\newcommand{\includeFigTriple}[6]{
\bc
\begin{figure}[htb]
\centering
\begin{tabular}{ccc}
\includegraphics[width=0.32\textwidth]{#1} \hs{-3mm}
&
\includegraphics[width=0.32\textwidth]{#2} \hs{-3mm}
&
\includegraphics[width=0.32\textwidth]{#3} \hs{-3mm}
\\
\hs{4mm}(a) & \hs{8mm}(b) & \hs{8mm}(c) \\
\end{tabular}
\caption[#6]{#5}
\label{#4}
\vs{-2mm}
\end{figure}
\ec
\vs{-6mm}
}

\newcommand{\includeTab}[5]{
\begin{table}[htb]
\centering
\begin{tabular}{#1}
\hline
#2
\hline
\end{tabular}
\caption[#5]{#4}
\label{#3}
\end{table}
}

\newcommand{\includeTabNoHLines}[5]{
\begin{table}[htb]
\centering
\begin{tabular}{#1}
#2
\end{tabular}
\caption[#5]{#4}
\label{#3}
\end{table}
}

\newcommand{\includeTabHERE}[5]{
\begin{table}[h!]
\centering
\begin{tabular}{#1}
\hline
#2
\hline
\end{tabular}
\caption[#5]{#4}
\label{#3}
\end{table}
}

\graphicspath{{images/}}

  \fancypagestyle{plain}{%
    \fancyhf{}                          % clear all header and footer fields
    \fancyhead[EL]{\bfseries\thepage}
    \fancyhead[OR]{\bfseries\thepage}
  }

\begin{document}

% Da es immer zitiert werden koennen muss, benutzt man am besten
% roemische Nummerierung im Vorspann und arabische fuer den Rest.
  \pagenumbering{roman}

% Titelblatt %%%%%%%%%%%%%%%%%%%%%%%%%%%%%%%%%%%%%%%%%%%%%%%%%%%%%%%%%%%%%%%%%%%

% An dem Titelblatt sollen keine Aenderungen vorgenommen werden.
% Es sei denn, Ihre Pruefungsordnung verlangt anderes Aussehen.
  %----------Generierung der Titelseite-----bitte nicht verändern!--------------------

\author{von \\ \dcauthorpre\ \dcauthorname\ \dcauthorsurname\ \\ \dcauthoradd}

%----------
\title{ \vspace{-5cm}\dctitle \\ 
\vspace{0.5cm}
\large{\dcsubtitle} \\ 
\vspace{0.5cm} {\Large{DISSERTATION}}\\ 
\vspace{0.5cm} \large{zur Erlangung des akademischen Grades \\ 
\dcdegree\\ im Fach \dcsubject \\ 
\vspace{0.5cm} eingereicht an der \\ 
\dcfaculty \\ 
\dcuniversity \\}}
%-----------------
\date{\vspace{0.5cm}
\raggedright{
Pr\"asident der Humboldt-Universit\"at zu Berlin:\\
\dcpresident \vspace{-0.3cm}
}\vspace{0.5cm}\\
\raggedright{
Dekan der \dcfaculty:\\
\dcdean \vspace{-0.3cm}
}\vspace{0.5cm}\\
\raggedright{
Gutachter:
\begin{enumerate} 
\item{\dcapprovala} \vspace{-0.3cm}
\item{\dcapprovalb} \vspace{-0.3cm}
\item{\dcapprovalc} \vspace{-0.3cm}
\end{enumerate}} \vspace{0.5cm}
%-----------------
\raggedright{
\begin{tabular}{lll}
eingereicht am: &  &\dcdatesubmitted\\ % wenn nicht in der Prüfungsordnung, die Zeile bitte auskommentieren
Tag der m\"undlichen Pr\"ufung: & & \dcdateexam
\end{tabular}}\\ 
}
%-------------------------------------
  \maketitle

% Zusammenfassung / Abstract* %%%%%%%%%%%%%%%%%%%%%%%%%%%%%%%%%%%%%%%%%%%%%%%%%%

  %-englische-Zusammenfassung---------------------------------------

\selectlanguage{english}

\begin{abstract}
\setcounter{page}{2} % Nach Bedarf anpassen!

Motivated by the advent of the Large Hadron Collider (LHC) the aim of the present work is the non-perturbative 
determination of the cutoff-dependent upper and lower mass bounds 
of the Standard Model Higgs boson based on first principle calculations, in particular not relying on additional information 
such as the triviality property of the Higgs-Yukawa sector or indirect arguments like vacuum stability considerations.
For that purpose the lattice approach is employed to allow for a non-perturbative investigation of a chirally invariant 
lattice Higgs-Yukawa model, serving here as a reasonable simplification of the full Standard Model, containing 
only those fields and interactions which are most essential for the intended Higgs boson mass determination. 
These are the complex Higgs doublet as well as the top and bottom quark fields and their mutual interactions.
To maintain the chiral character of the Standard Model Higgs-fermion coupling also on the lattice, the latter model is 
constructed on the basis of the Neuberger overlap operator, obeying then an exact global lattice chiral symmetry. 

Respecting the fermionic degrees of freedom in a fully dynamical manner by virtue 
of a PHMC algorithm appropriately adapted to the here intended lattice calculations, such mass bounds can indeed be 
established with the aforementioned approach. Supported by analytical calculations performed in the framework of the
constraint effective potential, the lower bound is found to be approximately $\lowBound=\GEV{80}$ at a cutoff of 
$\Lambda=\GEV{1000}$. The emergence of a lower Higgs boson mass bound is thus a manifest property of the pure
Higgs-Yukawa sector that evolves directly from the Higgs-fermion interaction for a given set of Yukawa coupling constants.
Its quantitative size, however, turns out to be non-universal in the sense, that it depends on the specific form, 
for instance, of the Higgs boson self-interaction. 

The upper Higgs boson mass bound is then established in the strong coupling regime of the considered Higgs-Yukawa model.
After the infinite volume extrapolation it is found to be approximately $\upBound=\GEV{630}$ at a cutoff of $\Lambda=\GEV{1500}$.
The cutoff-dependence of this upper bound can also be resolved and it is found to be in agreement with the expected
logarithmic decline with increasing cutoff $\Lambda$ according to the triviality picture of the Higgs-Yukawa
sector. Moreover, a direct comparison of the latter cutoff-dependent mass bound with corresponding results obtained
in the pure $\Phi^4$-theory, neglecting all fermion contributions, is considered with the intention to identify the fermion 
contribution to the upper mass bound. However, higher statistics is required to draw a clear conclusion in this matter.

Finally, a first account on the applicability of L\"uscher's method for the determination of the Higgs boson decay width in 
the framework of the here considered Higgs-Yukawa model is presented. At small renormalized quartic coupling constant $\lambda_r$
the obtained decay width $\Gamma_H$ is in good agreement with corresponding perturbative results, the validity of which
is trustworthy according to $\lambda_r$ being small. The good agreement thus encourages to further pursue this approach
in some follow-up investigations to eventually establish an upper bound also for the decay width of the Standard Model Higgs 
boson.

\end{abstract}

%-deutsche Zusammenfassung----------------------------------------

\selectlanguage{ngerman}

\begin{abstract}
\setcounter{page}{3} % Nach Bedarf anpassen!

Motiviert durch die Inbetriebnahme des Large Hadron Colliders (LHC) liegt die Zielsetzung der vorliegenden Arbeit 
in der nicht-perturbativen Bestimmung der Cutoff abhängigen oberen und
unteren Massenschranken des Higgs-Teilchens im Standardmodell auf der Basis von Rechnungen, die nur auf grundlegenden 
Prinzipien beruhen und sich insbesondere nicht auf zusätzliche Informationen stützen, wie zum Beispiel die 
Trivialitätseigenschaften des Higgs-Yukawa Sektors oder etwa Stabilitätsbetrachtungen des Vakuumszustandes. 
In dieser Arbeit wird der Gitteransatz verwendet, um eine nicht-perturbative Untersuchung eines chiral invarianten
Higgs-Yukawa Modells zu ermöglichen, welches hier als eine sinnvolle Vereinfachung des vollen
Standardmodells dient, und daher nur diejenigen Felder und Wechselwirkungen beinhaltet, die die Generierung 
der größten Beiträge zu den gesuchten Massenschranken erwarten lassen. Dies sind hier das komplexwertige Higgs-Dublett sowie
die Top- und Bottom-Quarkfelder und ihre gegenseitigen Wechselwirkungen. Um die chiralen Eigenschaften der Higgs-Fermion
Kopplung des Standardmodells auch auf dem Gitter zu bewahren, ist das betrachtete Higgs-Yukawa Modell auf der Grundlage
des Neuberger Dirac Operators konstruiert, so dass es eine globale chirale Symmetrie, bzw. deren Gitterpendant,
besitzt.

In der vorliegenden Arbeit werden die fermionischen Freiheitsgrade vollständig dynamisch berücksichtigt. Dies wird 
durch die Verwendung eines für das Higgs-Yukawa Modell entsprechend angepassten PHMC Algorithmus ermöglicht. 
Die vorgenannten Massenschranken können dann durch eine numerische Auswertung des betrachteten Modells bestimmt
werden. Unterstützt durch analytische Rechnungen auf der Grundlage des effektiven Potentials findet man eine untere 
Schranke von näherungsweise $\lowBound=\GEV{80}$ bei einem Cutoff von $\Lambda=\GEV{1000}$. Das Auftreten einer unteren 
Massenschranke des Higgs Bosons ist also eine grundlegende Eigenschaft des reinen Higgs-Yukawa Sektors, welche für einen 
gegebenen Satz von Yukawa Kopplungskonstanten unmittelbar durch die Higgs-Fermion Wechselwirkung generiert wird. Es stellt
sich jedoch heraus, dass die beobachtete quantitative Größe dieser Schranke in dem Sinne nicht universell ist, als 
sie von der speziellen Form der zugrunde liegenden Wechselwirkungen, zum Beispiel von der gewählten Form der bosonischen
Selbstwechselwirkung, abhängt. 

Die obere Schranke wird dann durch die Untersuchung des Modells im starken Kopplungsbereich bestimmt. Nach der 
Extrapolation in das unendliche Volumen findet man als obere Schranke näherungsweise $\upBound=\GEV{630}$
bei einem Cutoff von $\Lambda=\GEV{1500}$. Die Cutoff-Abhängigkeit der vorgenannten oberen Schranke kann ebenfalls
aufgelöst werden. Entsprechend dem Trivialitätsbild zeigt sich diese in voller Übereinstimmung mit dem erwarteten
logarithmischen Abfall bei steigenden Werten des Cutoff Parameters. Darüber hinaus werden die hier bestimmten
Massenschranken auch mit entsprechenden Resultaten aus der reinen $\Phi^4$-Theorie verglichen mit der Zielsetzung,
den fermionischen Beitrag zur oberen Massenschranke zu identifizieren. Es ist jedoch eine höhere Statistik notwendig, 
um ein zuverlässiges Ergebnis in dieser Fragestellung zu erhalten.

Schlie\ss{}lich wird außerdem ein erster Ausblick auf die Anwendbarkeit von Lüscher's Methode zur Bestimmung der 
Higgs Boson Zerfallsbreite im hier betrachteten Higgs-Yukawa Modell gegeben. Es zeigt sich, dass die vorgenannte
Zerfallsbreite bei kleinen Werten der quartischen Kopplungskonstante $\lambda_r$ in guter Übereinstimmung mit 
entsprechenden perturbativen Ergebnissen steht, welche aufgrund der klein gewählten Werte für $\lambda_r$ 
vertrauenswürdig sind. Diese gute Übereinstimmung motiviert, den betrachteten Ansatz in zukünftigen Studien
zu verwenden, um dann auch obere Grenzen für die Zerfallsbreite des Higgs Bosons bestimmen zu können. 
\end{abstract}

% Beachten Sie die Sprache des Textes und korrekte Seitenzahlen
% im PDF-Dokument:

 \selectlanguage{english}
% \setcounter{page}{4}

% Widmung* %%%%%%%%%%%%%%%%%%%%%%%%%%%%%%%%%%%%%%%%%%%%%%%%%%%%%%%%%%%%%%%%%%%%%

  %\chapter*{Widmung}

\begin{flushright}
To Cornelia,\\
my love.
\end{flushright}

% Inhaltsverzeichnis %%%%%%%%%%%%%%%%%%%%%%%%%%%%%%%%%%%%%%%%%%%%%%%%%%%%%%%%%%%

  \tableofcontents

% Hauptteil %%%%%%%%%%%%%%%%%%%%%%%%%%%%%%%%%%%%%%%%%%%%%%%%%%%%%%%%%%%%%%%%%%%%

%  \pagenumbering{arabic}
%  \pagestyle{myheadings}          % bzw. ist fancyhdr zu benutzten
  
% Kapitel %%%%%%%%%%%%%%%%%%%%%%%%%%%%%%%%%%%%%%%%%%%%%%%%%%%%%%%%%%%%%%%%%%%%%%

% part ist optional, bitte ggf. loeschen
% \part{Teil1}      

  \chapter{Introduction}
\label{chap:Introduction}

\pagenumbering{arabic}

With their postulation of the quark model, categorizing all so far discovered hadronic matter as bound 
states of the so-called up-, down-, and strange-quarks, Gell-Mann and Zweig laid one of the corner stones for our 
current understanding of Nature's most fundamental processes~\cite{GellMann:1964nj,Zweig:1981pd}. Equally driven by theoretical
considerations as well as accelerator experiments with ever increasing accessible energy scales and precision 
this initial picture of the quark model was gradually extended over the subsequent decades. In today's 
perspective on elementary particle physics all matter is assumed to be composed of leptons and quarks coming in 
three generations, each of which consisting of an up-type and a down-type quark as well as an electron-type and 
a neutrino-type lepton~\cite{Peskin:2008hj}, as listed in \tab{tab:ParticleContent}. These elementary constituents 
mutually interact through four fundamental forces, which are the gravitation, the strong interaction, the 
electromagnetic force, and the weak interaction. 

While most macroscopic physical processes in our universe are dominated by gravitation, which can be 
well described by Einstein's classical General Relativity Theory~\cite{Einstein:1916vd} at sufficiently 
large distances, this situation changes completely at the microscopic scale of elementary processes. In that
regime gravitation is supposed to be negligible as compared to the other three fundamental forces,
unless the energy scale of the considered process becomes of the order of the Planck scale $M_P\approx \GEV{10^{19}}$.
It is only this latter situation where an otherwise hidden, potential quantum structure of gravitation is
assumed to become visible and to dominate physical processes again. Despite a strenuously pursued and ongoing search
for a consistent quantum theory of gravitation based, for instance, on the competing approaches of String 
Theory~\cite{Becker:2006zu} and Loop-Quantum Gravity~\cite{Thiemann:2008zu}, such a theory could not 
be established yet. The resulting inherent necessity of neglecting gravity in any present theory of 
elementary particles, however, is generally considered as acceptable at energy scales far below the Planck scale
for the aforementioned reasons.

Contrary to the current situation with respect to understanding the intrinsic quantum nature of gravitation, consistent 
quantum field theoretical formulations of the other three fundamental forces have successfully been devised on the basis of a common
concept, which is the principle of local gauge invariance~\cite{book:Cheng}. This idea initially emerged during the development of
a quantum theory of electromagnetism and finally led to the evolution of quantum electrodynamics (QED), the foundation
of which being a local gauge symmetry based on the gauge group $\UoneEM$ with the latter subscript indicating the
electric charge to be its generating operator~\cite{book:Feynman}. In this framework the electromagnetic force,
interacting between all particles carrying an electric charge, is modeled through the exchange of a single type of gauge boson,
which is the photon in the case of QED. Strong evidence for the aforementioned gauge principle to be indeed
a fundamental concept of Nature's most elementary processes is manifest in the impressively accurate agreement between 
the theoretical predictions arising from QED and corresponding experimental measurements, culminating in the coincidence 
of the perturbatively calculated anomalous magnetic moment of the electron with its phenomenological value up to a precision 
of at least nine significant digits~\cite{Aoyama:2007mn,Amsler:2008zzb}. 

\includeTab{|c|c|c|c|}{
& 1. Generation & 2. Generation & 3. Generation \\
\hline
&&&\\
Leptons & 
$\left( 
\begin{array}{*{1}{c}}
\nu_e   \\
e
\end{array}  \right)_L\,$ $e_R\,$ $[\nu_{e,R}]$ &
$\left( 
\begin{array}{*{1}{c}}
\nu_\mu   \\
\mu
\end{array}  \right)_L\,$ $\mu_R\,$ $[\nu_{\mu,R}]$  &
$\left( 
\begin{array}{*{1}{c}}
\nu_\tau   \\
\tau
\end{array}  \right)_L\,$ $\tau_R\,$ $[\nu_{\tau,R}]$  \\
&&&\\
Quarks & 
$\left( 
\begin{array}{*{1}{c}}
u^\alpha   \\
d^\alpha
\end{array}  \right)_L\,$ $u^\alpha_R\,$ $d^\alpha_{R}$ &
$\left( 
\begin{array}{*{1}{c}}
c^\alpha   \\
s^\alpha
\end{array}  \right)_L\,$ $c^\alpha_R\,$ $s^\alpha_{R}$ &
$\left( 
\begin{array}{*{1}{c}}
t^\alpha   \\
b^\alpha
\end{array}  \right)_L\,$ $t^\alpha_R\,$ $b^\alpha_{R}$ \\
&&&\\
}
{tab:ParticleContent}
{The fundamental particle content of the Standard Model, consisting of three generations of quarks
and leptons, is shown. These particles are labeled as up (u), down (d), strange (s),
charm (c), bottom (b), and top (t) quark as well as electron (e), muon ($\mu$), and tau ($\tau$)
lepton together with their associated neutrinos ($\nu_e, \nu_\mu, \nu_\tau$).
The bracketed doublets represent here the left-handed doublets 
of the $\SUtwoL$ gauge group, while the right-handed fields are $\SUtwoL$ singlets.
The right-handed neutrinos are listed here in square brackets to indicate that these fields
do not interact with the fundamental forces according to present knowledge, 
which would be equivalent to their non-existence. The index $\alpha$ denotes
the colour index.}
{Fundamental particle content of the Standard Model.}

Driven by the great success of QED, the concept of local gauge invariance has also been applied to the description
of the strong interactions, which finally led to the development of quantum chromodynamics (QCD) based on the gauge
group $\SUthreeC$ with the latter subscript indicating the respective group elements to act on the so-called colour index
assigned to all quark fields~\cite{book:Greiner}. The introduction of the colour index has previously been found to be inevitable
to overcome Pauli's exclusion principle, which otherwise would contradict the observation, for instance, of the $\Delta^{++}$-baryon 
consisting of three identical quarks with equal spin orientations~\cite{Han:1965pf}. As listed in \tab{tab:ParticleContent} the 
leptonic fields, on the other hand, do not carry this additional index in agreement with their non-participation in the strong 
interaction. In the framework of QCD these forces among the quarks are considered to be mediated by the exchange of so-called 
gluons, being the gauge bosons of the $\SUthreeC$ colour group. The probably most striking feature of QCD is its ability to 
explain the asymptotic freedom~\cite{Politzer:1973fx,Gross:1973id} of quarks at small spatial separations, manifest through 
the phenomenological observation of Bjorken scaling~\cite{Bjorken:1968dy} in deep inelastic scattering experiments~\cite{Miller:1971qb}, 
while accounting also for the phenomenological non-observation of free quarks according to the concept of quark 
confinement~\cite{Wilson:1974sk}. Though not rigorously proven yet, the latter picture
of quark confinement can presumably be explained in the framework of QCD by the emergence of a separation independent force 
acting between remote quarks at large spatial separations. To overcome the aforementioned uncertainty ongoing effort is 
currently invested in the elaborate investigation of the confinement picture, \seeeg{book:Greensite} and the references therein.

In contrast to the electromagnetic and the strong forces, the weak interaction, the effects of which were first observed in 
the $\beta$-decay of nuclei, acts differently on left- and right-handed particles. The latter left-right asymmetry becomes 
most apparently reflected by our current understanding that right-handed neutrinos do not interact with any of the fundamental 
forces, or equivalently, that right handed-neutrinos do not exist at all as indicated in \tab{tab:ParticleContent}.

A reasonably good understanding of the weak interaction at low energies could be provided by the (V-A)-Fermi 
theory~\cite{Marshak:1958zu,Feynman:1958zu}, accounting for the chiral nature of that interaction by introducing 
different coupling structures for the vector and axial-vector currents. This approach, however, turned out to be 
non-renormalizable due to a fermionic four-point coupling term explicitly appearing in the underlying Lagrangian, 
thus rendering the (V-A)-Fermi theory to be at most an effective model that cannot be assumed to remain valid 
in the limit of large energy scales. Instead, it eventually has to to go over into some more complete, renormalizable 
theory, a candidate for which has later been devised by Glashow, Salam, and Weinberg~\cite{Glashow:1961tr,Salam:1968zu,Weinberg:1967tq}. 
The key idea finally leading to the renormalizability~\cite{'tHooft:1971fh,'tHooft:1971rn,'tHooft:1972fi} of 
this so-called GSW-theory is to apply the concept of local gauge invariance also to the weak interaction by modeling 
the latter forces as being mediated through the exchange of gauge bosons.

The gauge group underlying the GSW-theory is the direct product $\SUtwoTimesUoneY$, where the subscripts indicate that the 
$\SUtwoL$ symmetry only refers to the left-handed doublets listed in \tab{tab:ParticleContent}, and that the underlying generating 
operator of the gauge group $\UoneY$ is actually given by the so-called hypercharge $Y$, which differs for left- and
right-handed particles in agreement with the chiral character of the weak interaction. With an adequate reparametrization the
four gauge bosons of the aforementioned gauge group $\SUtwoTimesUoneY$ can then be identified as the three bosons 
$W_\pm$, $Z$ associated to the weak interaction as well as the photon $\gamma$ mediating the electromagnetic force, thus
yielding an unified description of the so-called electroweak interaction incorporating both of the two latter forces. 
It is this observation which finally allowed to establish an unified formulation of the strong, the weak, and the
electromagnetic interactions based on a common principle, which is the concept of local gauge invariance with respect to 
the group
\beq
\label{eq:SMGaugeGroup}
\SUthreeCSUtwoTimesUoneY.
\eeq
This local gauge symmetry constitutes the foundation of the so-called Standard Model of elementary particles~\cite{Rosner:2001zy}, 
being one of the most thoroughly and successfully tested models~\cite{Altarelli:2004fq} that have so far been conceived in 
elementary particle physics.

The outstanding success of the underlying GSW-theory became first heralded through the theoretical prediction~\cite{Weinberg:1967tq} of the 
only later experimentally confirmed~\cite{Arnison:1983rp,Banner:1983jy,Arnison:1983mk} masses of the $W_\pm$ and $Z$ bosons, which have to be massive 
to explain the typically very short interaction range of the weak forces. Remarkably, the aforementioned mass generation of the $W_\pm$ 
and $Z$ gauge bosons is describable in the framework of the GSW-theory without the inclusion of any explicit mass terms 
into the underlying Lagrangian, which otherwise would have destroyed the renormalizability of that theory. Instead
a gauge invariant and renormalizable~\cite{'tHooft:1971fh,'tHooft:1971rn,'tHooft:1972fi} mechanism of mass generation could be established within the 
GSW-theory, known as the celebrated Higgs-mechanism~\cite{Higgs:1964zu, Englert:1964zu, Guralnik:1964zu}, which explains the emergence of all 
fundamental particle masses and those of the massive $W_\pm$ and $Z$ gauge bosons, as the consequence of spontaneous symmetry 
breaking~\cite{Nambu:1961zy,Nambu:1961zu2,Goldstone:1961eq}. This mechanism is 
based on the postulated existence of an additional scalar field $\varphi$, interacting with the fermions by means of the so-called 
Yukawa couplings~\cite{Yukawa:1935xg}, the strength of which is controlled by the respective Yukawa coupling constants, and interacting also
with the latter gauge bosons $W_\pm$ and $Z$ through the appearance of its covariant derivative in the underlying Lagrangian. 
The aforementioned concept of spontaneous symmetry breaking then refers to Nature's selection of a particular ground state 
$\Omega_0$ out of a manifold of degenerate vacua, such that expectation values with respect to the particular vacuum 
$\Omega_0$ can break the intrinsic symmetries of the underlying Lagrangian. In the framework of the Standard Model the original 
gauge symmetry based on the gauge group in \eq{eq:SMGaugeGroup} is spontaneously broken according to
\beq
\SUthreeCSUtwoTimesUoneY \fhs{2mm} \rightarrow \fhs{2mm} \SUthreeCUoneEM
\eeq
thus maintaining only the colour and the electromagnetic local gauge symmetry. The consequences of this spontaneous symmetry breaking
are then most apparently reflected by the vacuum expectation value $|\langle \Omega|\varphi|\Omega \rangle|$ of the introduced scalar 
field $\varphi$ acquiring a non-zero value $v\neq 0$, which will be labeled as 'vev' in the following. 

In the broken phase the boson masses $m_{W_\pm}$ and $m_Z$ as well as the fermion masses are then generated through the implicit 
generation of corresponding mass terms in the Lagrangian arising from the aforementioned coupling structures, while the underlying
theory still remains renormalizable~\cite{'tHooft:1971rn}. The scalar field $\varphi$ itself, which is actually defined as a complex doublet 
valued field, can then be decomposed by virtue of the Goldstone theorem~\cite{Goldstone:1961eq,Goldstone:1962es} into a massive mode $h$, 
the so-called Higgs mode, as well as three massless modes $g^\alpha$, $\alpha=1,2,3$, 
denoted as Goldstone modes. The latter Goldstone modes, however, do not refer to additional, physically observable particles 
themselves but instead become identified as the longitudinal polarization modes of the three -- now massive -- gauge bosons 
$W_\pm$ and $Z$, which is often expressed by the sloppy formulation that 'the Goldstone modes are eaten by the massive gauge 
bosons'. 

\begin{figure}[htb]
\centering
\begin{tabular}{cc}
\includegraphics[width=0.4675\textwidth]{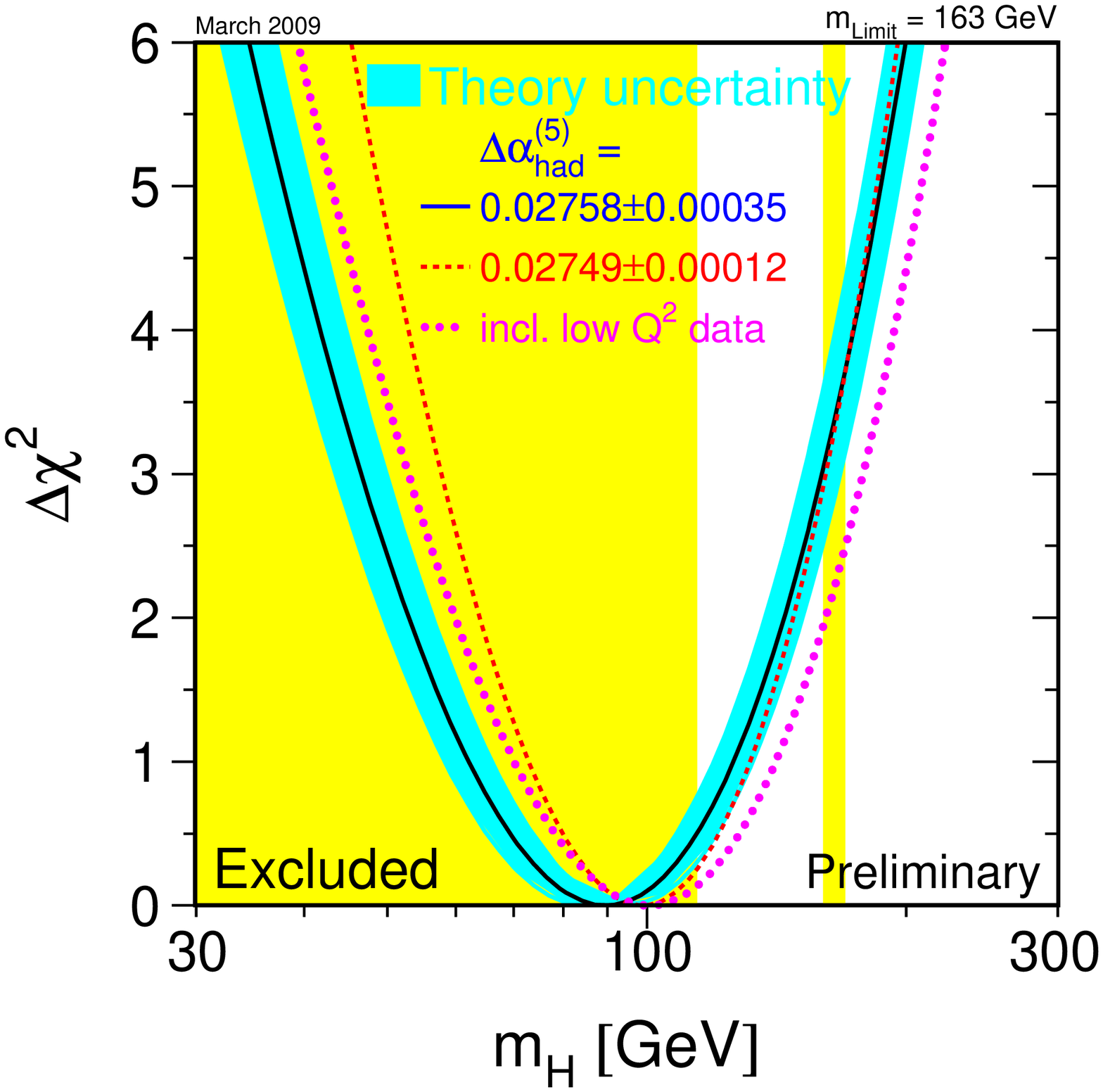}
&
\includegraphics[angle=90, width=0.48\textwidth]{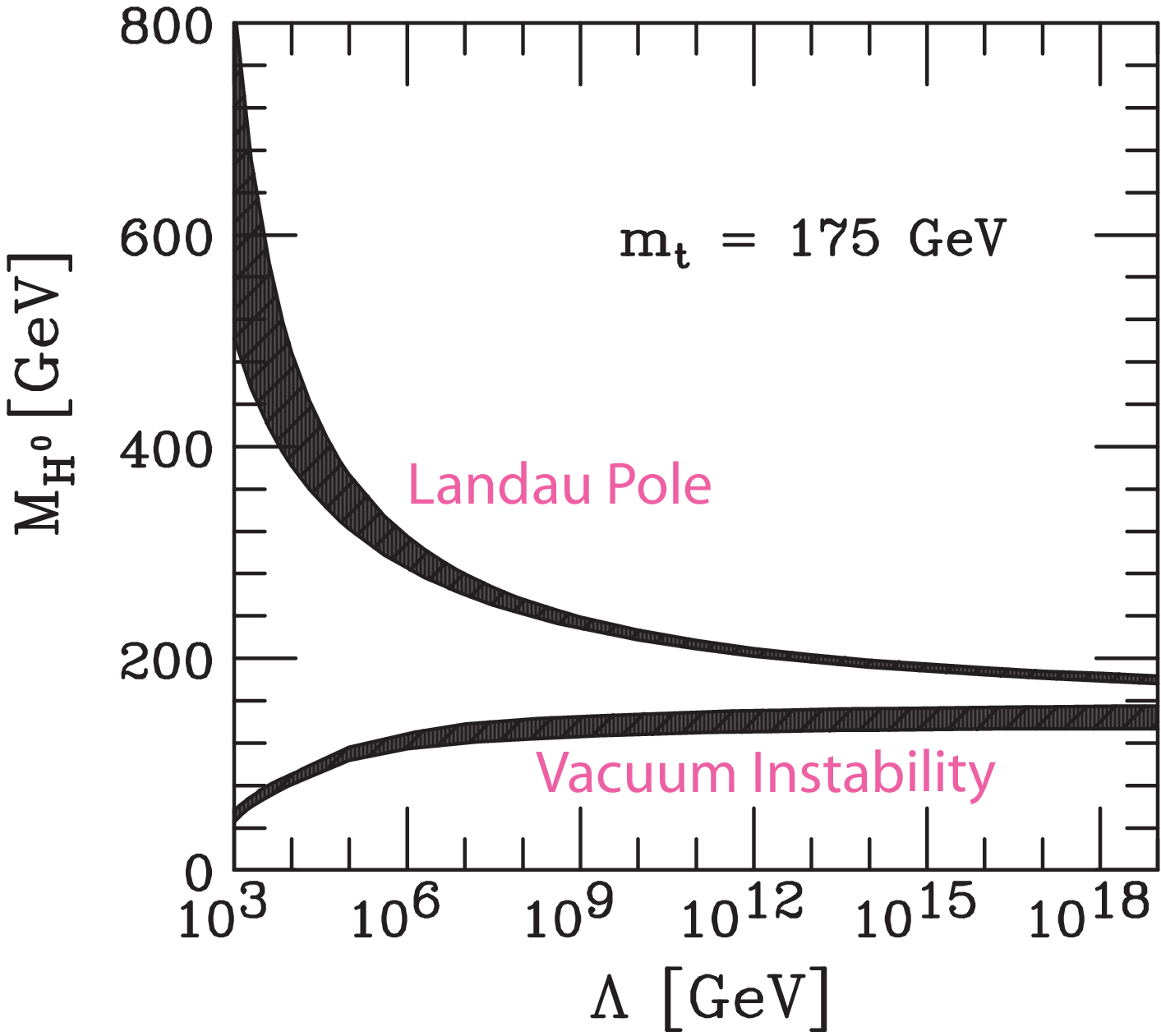}
\\
\hs{4mm}(a) & \hs{8mm}(b)  \\
\end{tabular}
\caption[Experimentally excluded Higgs boson masses and perturbatively obtained upper and lower Higgs boson mass bounds.]
{Panel (a) shows the probability distribution of the expected value for the Higgs boson mass as derived from fitting
high precision electroweak measurements with corresponding theoretical predictions arising from the Standard Model
under the assumption of a specific value of the Higgs boson mass. This probability density is given here in terms of the value
$\Delta\chi^2\equiv \chi^2-\chi^2_{min}$ with the latter $\chi^2$-value and its minimum $\chi^2_{min}$ obtained from
the aforementioned fit procedures. The solid line and its associated blue error band depict the main fit result. The additional
dotted curves refer to different sets of considered measurements and varying model parameters. The highlighted yellow areas 
mark the Higgs boson masses excluded by direct searches at a $\proz{95}$ confidence level. For more details the reader is referred 
to \Ref{LEPEWWGHomePage}, which is also the source of this plot. Panel (b) shows the cutoff-dependent upper and lower Higgs boson 
mass bounds that have been derived from the consideration of the Landau-pole as well as the vacuum instability at a top quark mass of 
$\GEV{175}$ as reviewed in \Ref{Hagiwara:2002fs}, which is also the source of this plot. The error bands indicate here the 
uncertainties of the theoretical calculations. }
\label{fig:PerturbativeHiggsMassBounds}
\vs{-8mm}
\end{figure}

Associated to the emerging massive Higgs mode is a corresponding massive physical particle denoted as the Higgs boson that
should in principle be observable in accelerator experiments. Though not discovered yet, its existence is widely considered
as very likely according to the aforementioned great success of the Standard Model in general and the GSW-theory
in particular. The ongoing search for the Higgs boson, being one of the main driving forces behind the construction of 
the Large Hadron Collider (LHC)~\cite{LHCMachine:2008zu}, has so far excluded the existence of a Higgs particle with a mass $m_H$
below $\GEV{114.4}$ at the $\proz{95}$ confidence level~\cite{Amsler:2008zzb,LEPEWWGHomePage} leaving open, however, the possibility 
of an even heavier Higgs boson. Apart from these exclusion limits arising from direct Higgs searches, one can also derive information
on the probability distribution of the expected value for the Higgs boson mass by comparing high precision electroweak measurements
with corresponding theoretical predictions arising from the Standard Model under the assumption of a particular Higgs boson mass. 
The aforementioned direct exclusion limit and an example of the latter probability distribution are presented together in 
\fig{fig:PerturbativeHiggsMassBounds}a which has been taken from \Ref{LEPEWWGHomePage}. From these data one can infer that a 
rather light Higgs boson mass between $\GEV{114.4}$ and $\GEV{200}$ is to be expected. 

In contrast to the situation of the gauge bosons $W_\pm$ and $Z$, the masses of which could successfully be predicted by 
virtue of the GSW-theory on the basis of some at that time available low energy quantities, the Higgs boson mass $m_H$ itself
cannot directly be estimated from that theory due to its immediate dependence on the unknown strength of the Higgs 
self-interaction. In the Standard Model the latter self-interaction is assumed to be of the form $\lambda|\varphi|^4$ 
with the bare quartic coupling constant $\lambda$ being apriori unconstrained. At tree-level one finds the Higgs boson mass
to be given through $m_H^2 \propto \lambda_rv_r^2$, where the renormalized vacuum expectation value $v_r$ is phenomenologically
known. The renormalized quartic self-coupling constant $\lambda_r$, however, is so far undetermined and could in principle
take any value due to the unconstrained interval of eligible bare coupling constants $\lambda$, leaving thus apriori open 
the possibility of an arbitrarily large Higgs boson mass.
 
A careful analysis of the pure Higgs sector of the Standard Model, however, reveals the underlying Higgs self-coupling structure
to be trivial~\cite{Aizenman:1981du,Frohlich:1982tw,Luscher:1988uq, Hasenfratz:1987eh, Kuti:1987nr, Hasenfratz:1988kr,Gockeler:1992zj} 
in four space-time dimensions. The notion of triviality refers here to the behaviour of the renormalized quartic coupling constant
$\lambda_r$ in dependence on the cutoff parameter $\Lambda$. The latter cutoff has to be introduced to regularize the considered theory,
yielding only then a mathematically well defined model with all otherwise divergent loop integrals, appearing for instance in the
perturbative evaluation of the theory, rendered finite. In a renormalizable theory this auxiliary parameter can later be removed again 
in the sense of sending it to infinity while holding all physical observables constant, making the physical predictions arising from such 
a theory eventually independent from the previously introduced auxiliary parameter $\Lambda$, as desired. This is generally referred to
as the renormalization procedure. In a trivial theory, however, all renormalized coupling constants vanish as functions of $\Lambda$ 
in the limit $\Lambda\rightarrow\infty$, leading thus to a free, non-interacting theory when trying to remove the cutoff parameter 
$\Lambda$ by sending it to infinity. 

At least two consequences arise from the latter observation. The first is that the Higgs sector of the Standard Model can only be 
considered as an effective theory inherently connected with a non-removable cutoff parameter $\Lambda$, which can be interpreted as 
the maximal energy scale up to which the underlying effective theory can be valid at most. The description of physical processes with 
even larger energy scales would thus require an extension of the considered theory beyond the latter scale $\Lambda$.
As a second consequence the renormalized quartic coupling constant at a given cutoff parameter $\Lambda$ is bounded from above 
according to $\lambda_r(\Lambda)\le \lambda_r^{up}(\Lambda)$ directly translating also into corresponding upper bounds $\upBound$ 
on the Higgs boson mass itself.

The remarkable conclusion that can be drawn when combining the latter two observations is the following: Once the Higgs boson
has actually been discovered, for instance at the LHC, and its physical mass $m_H$ is eventually known, it would be possible 
to infer from the comparison of $m_H$ with its cutoff-dependent upper bound $m_H^{up}(\Lambda)$ up to which energy scale the 
Standard Model can be valid at most. Such a consideration is of fundamental interest, since it provides essential information 
concerning the development of future extensions to the Standard Model which, despite its tremendous success, leaves a 
couple of urging questions unanswered. Among them are, for instance, the so far lacking explanation for dark matter and dark 
energy~\cite{Kamionkowski:2007wv}, the observed neutrino oscillations~\cite{Akhmedov:1999uz}, the baryon asymmetry~\cite{Cline:2006ts},
and the hierarchy problem~\cite{Quigg:2009vq}. It is moreover unclear, how gravitation can be incorporated into the Standard Model, 
whether there are more than the so far observed three quark and lepton generations, and whether the rather large number of parameters 
in the Standard Model can be reduced by some further unification of the strong and the electroweak interactions. Promising candidates 
for solving at least some of the aforementioned problems in part are, for instance, the Minimal Supersymmetric Standard 
Model~\cite{Nilles:1984zu}, String Theory~\cite{Becker:2006zu}, Loop Quantum Gravity~\cite{Thiemann:2008zu}, Grand Unified 
Theories~\cite{Georgi:1974sy}, Little Higgs Models~\cite{Schmaltz:2005ky}, and Extra Dimensions~\cite{PerezLorenzana:2005iv}. 
Having the phenomenological value of the Higgs boson mass at hand would then allow to specify an energy scale from which on an 
extension of the Standard Model, potentially according to one of the aforementioned approaches, becomes definitely necessary.

This latter energy scale can further be narrowed, since a cutoff-dependent lower Higgs boson mass bound $\lowBound$
can also be established, however not on the basis of the triviality property of the considered model. 
In renormalized perturbation theory such cutoff-dependent lower bounds have been derived from vacuum stability considerations, 
\seeeg{Cabibbo:1979ay, Linde:1975sw,Weinberg:1976pe,Linde:1977mm, Sher:1988mj, Lindner:1988ww, Altarelli:1994rb,Casas:1994qy, Casas:1996aq},
while corresponding upper bounds have been established based on the criterion of the Landau pole being situated
beyond the cutoff of the theory, \seeeg{Cabibbo:1979ay, Dashen:1983ts, Lindner:1985uk}, as well as from unitarity 
requirements, \seeeg{Dicus:1992vj, Lee:1977eg, Marciano:1989ns}. A review on these various calculations of the 
sought-after cutoff-dependent upper and lower Higgs boson mass bounds can be found in 
\Ref{Hagiwara:2002fs}, the summarizing plot of which is presented in \fig{fig:PerturbativeHiggsMassBounds}b.
From these already established results one can, for instance, infer that an extension of the Standard Model would become 
inevitable at energy scales beyond $\Lambda=\GEV{10^6}$, in case the Higgs boson mass would be found to be lighter than 
$\GEV{120}$.

\bc
\vs{-7mm}
\ec  

However, the results relying on the aforementioned vacuum instability arguments have to be considered with care. In the underlying
calculations the effective potential has been determined in terms of the renormalized coupling constants by means of 
perturbation theory. It is then usually found that the renormalized expression for the effective potential exhibits an 
instability, meaning that the effective potential becomes unbounded from below, if the Higgs boson mass becomes too small
for a given setting of the renormalized fermion masses, thus inducing a lower bound on the Higgs boson mass, as desired. 
It is, however, also known that the effective potential is convex for all values of the bare coupling constants, where the
underlying theory is well-defined, and can therefore never be unstable~\cite{O'Raifeartaigh:1986hi}, which is apparently in 
contrast to the aforementioned observation of an unstable effective potential at small values of the Higgs boson mass. Additionally, 
it has recently been argued that the appearance of a vacuum instability in the renormalized calculation is actually caused by the 
incorrect removal of the cutoff $\Lambda$ from the underlying trivial theory~\cite{Holland:2003jr,Holland:2004sd}. 
  
This apparent paradox can be resolved~\cite{Lin:1990ue}, if one keeps in mind that the definition domain of any renormalized quantity 
might be constrained, though its bare counterpart is not. This is particularly obvious in the framework of a trivial theory, 
such as the Higgs-Yukawa sector of the Standard Model. In that scenario the bare quartic coupling constant $\lambda$ can take any positive 
value, but its renormalized counterpart is constrained to $\lambda_r\in [\lambda_r^{low}(\Lambda),\lambda_r^{up}(\Lambda)]$, 
where the latter bounds are apriori undetermined. For clarification it is pointed out again that the cutoff $\Lambda$ always has to 
be kept finite due to the triviality of the model, since otherwise one would inevitably end up with $\lambda_r\equiv 0$ being the 
only eligible choice for the renormalized quartic coupling constant. As long as one chooses the renormalized coupling constants 
underlying the perturbative calculation of the effective potential according to their respective definition domains, the arising 
result on the latter potential will actually refer to a certain configuration of bare model parameters at the considered cutoff, for 
which the effective potential is known to be convex. If, however, one chooses the aforementioned renormalized coupling constants
not to be contained within their respective definition domains, then one cannot establish a connection of the arising
renormalized result on the effective potential, which might predict an instability, to any configuration of bare parameters. 
In case of $\lambda_r<\lambda_r^{low}(\Lambda)$ this can be expressed by saying that the associated bare quartic coupling
constant is negative. For $\lambda_r\not\in [\lambda_r^{low}(\Lambda),\lambda_r^{up}(\Lambda)]$ one therefore actually studies the 
analytical continuation of the renormalized expression for the effective potential in a regime of the renormalized parameters, 
where the originally underlying model, which is given as the regularized model in terms of bare parameters and a cutoff parameter, 
no longer exists, thus resolving the above paradox.

In the present work, however, the question of the validity of the vacuum instability arguments shall not be addressed in detail. 
It is instead the aim to investigate whether the lower mass constraint on the Higgs boson mass can also be established 
without relying on the aforementioned vacuum stability considerations. Additionally, the validity of the perturbatively 
obtained upper Higgs boson mass bound is also not fully clear, since the underlying perturbative calculation of the 
aforementioned Landau pole has to be performed at rather large values of the renormalized quartic coupling constant. The 
latter two remarks make the Higgs boson mass bound determination an interesting subject for a non-perturbative investigation
relying only on first principles. In the present work the lattice approach~\cite{book:Rothe} will therefore be employed to 
study the upper and lower Higgs boson mass bounds. 

For the purpose of performing such lattice investigations an appropriate lattice formulation of the full Standard Model would 
be needed which, however, has not completely been devised yet. While a very successful lattice formulation of QCD is well established since 
long~\cite{Wilson:1974sk}, a consistent construction of lattice theories with dynamical fermions obeying the local 
$\SUtwoTimesUoneY$ chiral gauge symmetry is much more complicated due to the conceptual obstacles reviewed in \Ref{Shamir:1995zx}.
In fact, it has even been a long standing problem to establish a global chiral symmetry on the lattice while decoupling
the fermion doubler modes from the physical spectrum, which has, however, recently been solved~\cite{Luscher:1998pq} on the 
basis of the Ginsparg-Wilson relation~\cite{Ginsparg:1981bj}. 
Concerning the extension of this latter approach also to the case of the local gauge symmetry, one of the major difficulties is to 
guarantee the local chiral gauge symmetry to be exactly preserved on the lattice also beyond a tree-level consideration. In
case of the aforementioned approach based on the Ginsparg-Wilson relation the problem arises from the fact that the gauge fields
explicitly appear in the definition of the left- and right-handed projection operators~\cite{Niedermayer:1998bi} leading then to 
a gauge field dependent phase ambiguity~\cite{Luscher:1998du,Luscher:2000hn} when decomposing the fermionic integration measure into 
its left- and right-handed components. Fixing this ambiguous phase such that local gauge symmetry is exactly preserved on the lattice 
poses a severe problem for the construction of chiral gauge theories on the lattice. However, remarkable progress has recently 
been achieved by proving that chirally invariant left- and right-handed fermion integration measures are in fact constructable at least
for the case of Abelian chiral gauge theories~\cite{Luscher:1998du} as well as $\SUtwoTimesUoneY$ chiral gauge theories in 
topological sectors with vanishing $U(1)$ magnetic flux~\cite{Kadoh:2007xb}. Though ongoing research continuously pushes towards 
the practical evaluation of these theories on the lattice~\cite{Kadoh:2003ii,Kadoh:2007wz}, there are still many difficulties 
hindering such practical calculations, for instance the problem that the fermion determinant is not necessarily real in these 
models~\cite{Luscher:1998du,Kadoh:2007xb}.

Any intended lattice calculation in the framework of the Standard Model therefore has to approximate the full theory by 
some less comprehensive working model, still reflecting, however, its essential features being most relevant for 
the considered physical processes. 

For the above reasons, the question for the upper and lower Higgs boson mass bounds $\lowBound$ and $\upBound$ has already been 
investigated in numerous lattice studies mainly concentrating on the pure Higgs sector of the Standard Model, describing only the Higgs 
dynamics and its self-interaction~\cite{Hasenfratz:1987eh,Kuti:1987nr,Hasenfratz:1988kr,Bhanot:1990ai}, as well as the pure 
Higgs-Yukawa sector of the Standard Model, additionally respecting also the Higgs-fermion 
coupling~\cite{Holland:2003jr,Holland:2004sd,Lin:1993hp,Bock:1992gt,Lee:1989xq}. The latter investigations, however, suffered from 
their inability to establish the aforementioned global chiral symmetry, which could also not be restored in the continuum limit 
while lifting the unwanted fermion doublers at the same time. The pure Higgs models on the other hand could obviously not 
investigate the fermionic influence on the upper Higgs boson mass bounds, while a lower bound could not be established at 
all in these models due to the lacking fermion dynamics. The latter remarks also apply to the investigations of the pure Higgs-gauge 
sector, which has been studied on the lattice, for instance, in \Refs{Montvay:1984wy,Evertz:1985fc,Hasenfratz:1987uc,Evertz:1989hb}.

There are two main developments that warrant the reconsideration of these questions. First, with the advent of the LHC, 
we can expect that the properties of the Standard Model Higgs boson, such as its mass and decay width, will be revealed 
experimentally. Second, there is, in contrast to the situation of earlier investigations of lattice 
Higgs-Yukawa  models~\cite{Smit:1989tz,Shigemitsu:1991tc,Golterman:1990nx,book:Jersak,book:Montvay,Golterman:1992ye,Jansen:1994ym}, 
a consistent formulation of a Higgs-Yukawa model with an exact lattice global chiral symmetry~\cite{Luscher:1998pq} based on the 
Ginsparg-Wilson relation~\cite{Ginsparg:1981bj}. This new development allows to maintain the chiral character of the 
Higgs-fermion coupling structure on the lattice while simultaneously lifting the fermion doublers, thus 
eliminating manifestly the main objection to the earlier investigations. The interest in lattice Higgs-Yukawa models has therefore 
recently been renewed~\cite{Bhattacharya:2006dc,Gerhold:2007yb,Giedt:2007qg,Gerhold:2007gx,Poppitz:2007tu,Fodor:2007fn,Gerhold:2009ub}.

In the present work the question for the Higgs boson mass bounds shall be revisited in the framework of the latter chirally invariant 
lattice Higgs-Yukawa model. Conceptually, this will then allow for a non-perturbative investigation of the fermionic contribution to the upper 
Higgs boson mass bound as well as the determination of the lower Higgs boson mass bound, while relying on first principles only. The 
main target of this study are the cutoff-dependent mass bounds, since they will eventually lead to conclusions on the energy scale up to which
the Standard Model can be valid at most, provided that the Higgs boson and its mass will experimentally be discovered. 

For clarification it is pointed out that the fermions included within the considered lattice Higgs-Yukawa model will be treated in a fully 
dynamical manner. It is further remarked that the notion of chiral symmetry will always refer to the global chiral symmetry in this work. 
In fact, no gauge fields are contained at all within the here considered pure Higgs-Yukawa model, and its underlying field content 
only consists of the scalar field $\varphi$ as well as the top-bottom quark doublet interacting with each other 
through respective Yukawa couplings. 

For the purpose of establishing the announced non-perturbative upper and lower Higgs boson mass bounds the following steps will
be taken in the subsequent chapters. First, a brief and general introduction to the theoretical background of Higgs-Yukawa models
on the lattice will be given in \chap{chap:GeneralBackground}, followed by the introduction of the actually considered chirally 
invariant lattice Higgs-Yukawa model presented in \chap{chap:TheModel}. The phase structure of that model is then studied in detail 
in \chap{chap:PhaseDiagram}, which will later allow to identify those regions in bare parameter space where the eventual lattice 
calculations of phenomenological interest have to be performed. A technical prerequisite for the intended lattice calculations 
certainly is the availability of a suitable numerical algorithm. Such an algorithm has been implemented here on the basis of the 
well-known PHMC-method~\cite{Frezzotti:1997ym} adapted, however, to the particular scenario of the considered Higgs-Yukawa 
model. The various algorithmic improvements underlying this implementation, which eventually lead to a substantial performance gain 
crucial for the successful processing of the targeted numerical calculations, are detailed in \chap{chap:SimAlgo}. The actual 
determination of the cutoff-dependent lower Higgs boson mass bound is then presented in \chap{chap:ResOnLowerBound} supported 
by a couple of analytical calculations based on the constraint effective potential and lattice perturbation theory. In 
particular, the dependence of the lower bound on the fermion masses will be studied, as well as the question for its universality 
with respect to the actual form of the Higgs self-interaction. This is an interesting question, since the latter self-interaction is 
not constrained to have the form $\lambda|\varphi|^4$ by the usual renormalization arguments due to the Higgs-Yukawa sector being 
only an effective theory. The upper Higgs boson mass bound arising in the considered Higgs-Yukawa model is then established in 
\chap{chap:ResOnUpperBound}. In this case it is of particular interest to compare the obtained results to the bounds observed 
in the pure Higgs sector, thus allowing to estimate the strength of the fermionic contribution. This will be done by means of 
a direct comparison with the lattice $\Phi^4$-theory. Finally, a first and brief outlook towards a future 
determination of the decay width of the Higgs boson on the basis of L\"uscher's method~\cite{Luscher:1990ux,Luscher:1991cf} 
shall be given in \chap{chap:ResOnDecayWidth} before summarizing the conclusions of the present work in \chap{chap:Conclusions}.

  \chapter{General background of the Higgs-Yukawa sector on the lattice}
\label{chap:GeneralBackground}

The aim of the present chapter is to give a brief introduction into the general theoretical background
that underlies the non-perturbative investigation of the pure Higgs-Yukawa sector of the Standard Model
on a finite, Euclidean space-time lattice. As a first step the functional integral formulation of the Higgs-Yukawa
sector and its extension to Euclidean space-time will be introduced in \sect{sec:FuncFormInEucTime}.
Continuing with the Euclidean formulation of the Higgs-Yukawa model the qualitative propagator pole structure of an 
unstable Higgs particle is discussed in \sect{sec:UnstableSignature} by means of a perturbative one-loop 
calculation performed in the pure $\Phi^4$-theory. The resulting knowledge about the pole structure will become
of great importance for the later definition of observables aiming at the determination of the Higgs boson
mass as well as its decay properties in \sect{sec:SimStratAndObs}.

Finally, the basic ideas and conceptual problems of discretizing the Euclidean Higgs-Yukawa sector on a finite
space-time lattice shall be discussed in \sect{sec:NaiveDiscret}. This is also where the discretization 
induced breakdown of chiral symmetry, that obstructed the investigation of the Higgs-Yukawa sector on the lattice in the
past, will be addressed. The chapter then ends with a discussion of the continuum limit in general and the question
of how to adopt these concepts to the case of the Higgs-Yukawa sector with its property of triviality, in particular.

%--------------------------------------------------------------------------------------------------------------------------
\section{The Higgs-Yukawa model in the functional integral formulation}
\label{sec:FuncFormInEucTime}

In the operator formalism the Higgs-Yukawa sector of the Standard Model describes the dynamics and the mutual
interactions of the bosonic field operator $\hat \varphi$ and the quark field operators $\hat t, \hat b, \hat c, \hat s, \hat u, \hat d$
of the top, bottom, charm, strange, up, and down quarks, respectively.
For the sake of brevity, however, we will restrict the presentation to the bosonic, the top, and the
bottom quark operator fields only. To avoid confusion it is remarked again that all couplings to gauge fields 
are explicitly excluded in the pure Higgs-Yukawa sector.

In this section the formulation of the pure Higgs-Yukawa sector in terms of the functional integration approach shall be 
briefly discussed. As a starting point we begin here with the introduction of the functional integral formalism in Minkowski 
space-time. It is well known that the vacuum expectation value of a time-ordered product of field operators in a quantum field
theory can be expressed in terms of a functional integral performed over all fields underlying the 
considered theory~\cite{Feynman:1948ur,Feynman:1950ir,Matthews:1955zi,Berezin:1966zu,book:Montvay, book:Rothe}. 
To be more precise the integration variables in this functional integration approach, 
\ie the field variables at all space-time points, are $\Comp$-numbers or anti-commuting Grassmann variables~\cite{Candlin:1956zu, Berezin:1966zu}, 
respectively, instead of operators. Up to a normalization constant the vacuum expectation value of the considered time-ordered product of 
field operators is then given as the functional integral of the respective product of field variables weighted
by a factor $\exp(iS)$ with $S\equiv S[\varphi, t, \bar t, b, \bar b]$ denoting the action associated to the set of integration variables.
For the pure Higgs-Yukawa sector of the Standard Model, this finding can be expressed as
\beq
\label{eq:IntroOfFuncIntMink}
\langle \Omega| \timeOrdering\, O(\hat\varphi, \hat t, \hat{\bar t}, \hat b, \hat{\bar b})  |\Omega\rangle
= \frac{1}{Z} \FuncIntAvg[O(\varphi, t, \bar t, b, \bar b)]
\equiv \langle O(\varphi, t, \bar t, b, \bar b) \rangle, \quad Z=\FuncIntAvg[1]
\eeq
where the aforementioned normalization constant $Z$ is the so-called partition function and the functional integral 
$\FuncIntAvg[O(\varphi, t, \bar t, b, \bar b)]$ is defined as
\bea
\label{eq:IntroOfFuncIntMink2}
\FuncIntAvg[O(\varphi, t, \bar t, b, \bar b)] = \int \intD{\varphi}\intD{\varphi^\dagger} \intD{t} \intD{\bar t}\intD{b} \intD{\bar b} \,\,
O(\varphi, t, \bar t, b, \bar b)  \cdot e^{iS[\varphi, t, \bar t, b, \bar b]}.
\eea
 
Here, the state $\Omega$ denotes the normalized vacuum state of the interacting theory, which was assumed to be non-degenerate in
the derivation of \eq{eq:IntroOfFuncIntMink}. The subtleties induced by a degenerate vacuum will be addressed at the end of this section.
The time ordering operation $\timeOrdering$ sorts the field operators according to their time indices such that field operators with larger 
time index appear on the left hand side of those with smaller time index. 
The integral in \eq{eq:IntroOfFuncIntMink2} is then a functional integral over the complex, doublet field
$\varphi$ and the quark fields $t,\bar t, b, \bar b$, which are represented by anti-commuting Grassmann numbers.
With some shorthand notation the observable $O(\hat\varphi, \hat t, \hat{\bar t}, \hat b, \hat{\bar b})$ is actually meant to be some
polynomial of the field operators at a finite number of space-time points, given by a set labeled here as $M$, according to
\beq
O(\hat\varphi, \hat t, \hat{\bar t}, \hat b, \hat{\bar b}) \equiv 
O(\{\hat\varphi_x, \hat t_x, \hat{\bar t}_x, \hat b_x, \hat{\bar b}_x\,\,:\quad x\in M\}).
\eeq
The expression $O(\varphi, t, \bar t, b, \bar b)$ on the right hand side of \eq{eq:IntroOfFuncIntMink} denotes the same function but with 
its arguments being the corresponding integration variables. 

It is remarked that the integration measure in \eq{eq:IntroOfFuncIntMink2} is lacking a thorough mathematical definition in continuous 
space-time. Only for a discrete and finite set of space-time points $\Gamma$ is such a definition 
possible according to 
\beq
\label{eq:DefOfFuncIntMeasure}
\intD{\varphi}\intD{\varphi^\dagger} \intD{t} \intD{\bar t}\intD{b} \intD{\bar b} = 
\prod_{x\in \Gamma} \intd{\varphi_x}\intd{\varphi^\dagger_x} \intd{t_x} \intd{\bar t_x}\intd{b_x} \intd{\bar b_x},
\eeq
where at each space-time point $x\in \Gamma$ the measure $\intd{\varphi_x}\intd{\varphi^\dagger_x}$ is given as the Lebesgue measure in $\Comp^2$ 
and $\intd{t_x}, \intd{\bar t_x}, \intd{b_x}, \intd{\bar b_x}$ denote the Grassmann integration measures~\cite{Berezin:1966zu} of the independent 
Grassmann variables $t_x, \bar t_x, b_x, \bar b_x$. For the rest of this section we continue, however, with the formal expression in \eq{eq:IntroOfFuncIntMink} 
based on continuous space-time, and postpone the discussion of its actual mathematical meaning to \sect{sec:NaiveDiscret}.

For the considered pure Higgs-Yukawa sector, the total action $S$ appearing in the given functional integral is given 
by its bosonic and fermionic contributions according to
\bea
S[\varphi, t, \bar t, b, \bar b] &\fhs{-2mm}=\fhs{-2mm}& S_\varphi[\varphi] + S_F[\varphi, t, \bar t, b, \bar b], \\
S_\varphi[\varphi] &\fhs{-2mm}=\fhs{-2mm}& \int \intd{^4x}\, \frac{1}{2} \partial_\mu\varphi_x^\dagger\partial^\mu\varphi_x - \frac{1}{2} m_0^2 \varphi^\dagger_x\varphi_x
-\lambda\left(\varphi_x^\dagger \varphi_x \right)^2, \\
S_F[\varphi, t, \bar t, b, \bar b] &\fhs{-2mm}=\fhs{-2mm}& \int \intd{^4x}\, \bar t_x i\gamma^\mu \partial_\mu t_x + \bar b_x i\gamma^\mu \partial_\mu b_x
- y_b \left(\bar t_x, \bar b_x \right)_L \varphi_x b_{R,x} 
- y_t \left(\bar t_x, \bar b_x \right)_L \tilde\varphi_x t_{R,x} \quad\nonumber\\
&\fhs{-2mm}+\fhs{-2mm}& c.c.,
\eea
which is obtained by restricting the Lagrangian of the full Standard Model, \seeeg{Peskin:2008hj}, to the pure Higgs-Yukawa sector and 
replacing all field operators by their corresponding counterparts in terms of integration variables. It is remarked that we will
consistently include a factor of $1/2$ in front of the bosonic kinetic and mass terms throughout this work, which deviates a bit from
the standard notation. Moreover, we have used $\tilde \varphi = i\tau_2\varphi^*$ in the above equation with $\tau_1$, $\tau_2$, and $\tau_3$ 
denoting the three Pauli-matrices. 

A major downside of the given functional integral expression in \eq{eq:IntroOfFuncIntMink} is that it is very ill-suited for a numerical evaluation.
The obvious reason is that a numerical evaluation, by its nature, would have to approximate the functional integral by a sum over a finite
set of field configurations. This finite sum, however, converges very slowly, since the exponential factor appearing in the integrand has 
a purely imaginary argument, thus inducing a heavily oscillating complex weight factor of constant norm. This implies that all contributions of the 
field configurations included in such a finite sum, be it almost classical solutions to the equations of motion with minimal action or completely randomly 
picked field configurations, are of the same order of magnitude. The expected dominance of the classical solutions in the functional integral is therefore 
not established through some direct suppression of the physically less relevant field configurations, but instead by a mutual cancellation of the 
contributions arising from the latter configurations due to the complex factor $\exp(iS)$. The concept of importance sampling is thus not applicable to 
the approximation of \eq{eq:IntroOfFuncIntMink} by a finite sum, which makes the given functional integral highly unfavourable for a direct numerical evaluation.

The described numerical difficulties can be cured, if one considers the underlying theory in Euclidean space-time 
instead of the physical Minkowski space-time. The transition to Euclidean space-time is achieved by the following 
two steps. First, one allows for complex numbers of the time coordinates in Minkowski space, \ie 
$x_0=t-i\tau\in \Comp$. The newly introduced imaginary part of $x_0$ will be referred to as Euclidean time $\tau$, 
whereas the real part of $x_0$ will be denoted as the Minkowski time $t$. For purely imaginary $x_0$ the space-time 
point $x=(-i\tau,\vec x)$ can then be parametrized in terms of the real Euclidean coordinates $x_E=(\tau,\vec x)$. 
Furthermore, the field operators associated to the newly introduced complex times are given as the 
analytical continuation of the field operators in Minkowski time according to
\beq
\hat \varphi_{x_0,\vec x} = e^{i\hat H x_0}\, \hat \varphi_{0,\vec x}\, e^{-i\hat H x_0}
\eeq
for the case of the bosonic field and analogously for the quark fields, where $x_0$ is now a complex 
number and $\hat H$ is still the same Hamiltonian that is associated to the underlying theory in the 
original Minkowski space. 

The second step is to perform a Wick rotation~\cite{Wick:1954eu} to Euclidean time by substituting $x_0\rightarrow-i x_0$, which will
be expressed in the following by the symbolic operation $W[\ldots]$ replacing all space-time coordinates of its argument by
Wick-rotated coordinates. In this notation the Wick-rotation of the considered time-ordered observable then reads
\bea
W[ \timeOrdering\, O(\hat\varphi, \hat t, \hat{\bar t}, \hat b, \hat{\bar b}) ] 
&=&  \timeOrdering_E\, O_E(\hat\varphi, \hat t, \hat{\bar t}, \hat b, \hat{\bar b}) 
\eea
where the Euclidean time ordering operation $\timeOrdering_E$ sorts the field operators 
according to the Euclidean time and the shorthand expression $O_E(\hat\varphi, \hat t, \hat{\bar t}, \hat b, \hat{\bar b})$ 
is given as
\bea
O_E(\hat\varphi, \hat t, \hat{\bar t}, \hat b, \hat{\bar b}) &=& 
O(\{\hat\varphi_x, \hat t_x, \hat{\bar t}_x, \hat b_x, \hat{\bar b}_x\,\, : \quad x\in W[M] \}).
\eea

The fundamental assumption underlying the idea of the Wick-rotation is that the vacuum expectation value 
$\langle \Omega| \timeOrdering\, O(\hat\varphi, \hat t, \hat{\bar t}, \hat b, \hat{\bar b})  |\Omega\rangle$ 
is analytical as a function of the space-time coordinates $x\in M$ of the involved field operators. 
The basic idea of the Wick-rotation approach is then to obtain the desired vacuum expectation value 
$\langle \Omega| \timeOrdering\, O(\hat\varphi, \hat t, \hat{\bar t}, \hat b, \hat{\bar b})  |\Omega\rangle$
by analytically continuing the expectation value
$\langle \Omega| W[\timeOrdering\, O(\hat\varphi, \hat t, \hat{\bar t}, \hat b, \hat{\bar b})]  |\Omega\rangle$
back to Minkowski space-time~\cite{Schwinger:1958zu,Schwinger:1959zz}. 
The question whether Euclidean Greens functions can be analytically continued to Minkowski space-time at all
was answered by Osterwalder and Schrader in their famous work~\cite{Osterwalder:1973dx, Osterwalder:1974tc}, where 
they provided necessary and sufficient conditions under which such an analytical continuation is possible. 
Based on these conditions the required analytical continuity has explicitly been proven for the pure $\Phi^4$-theory~\cite{Glimm:1974ki}.
It will therefore also be assumed for the considered Higgs-Yukawa theory in the following as a working hypothesis.

Following the same arguments that led to the derivation of the functional integral in \eq{eq:IntroOfFuncIntMink}
one finds that the vacuum expectation value of the considered Wick-rotated observable
can again be given in terms of a functional integral~\cite{book:Montvay, book:Rothe} according to
\beq
\label{eq:IntroOfFuncIntEuc}
\langle \Omega| W[\timeOrdering\, O(\hat\varphi, \hat t, \hat{\bar t}, \hat b, \hat{\bar b})]  |\Omega\rangle
= \frac{1}{Z_E} \FuncIntAvg_E[O_E(\varphi, t, \bar t, b, \bar b)]
\equiv \langle O_E(\varphi, t, \bar t, b, \bar b) \rangle_E, 
\eeq
with $Z_E=\FuncIntAvg_E[1]$ denoting the Euclidean partition function. The Euclidean functional integral itself
is given as
\bea
\label{eq:IntroOfFuncIntEucSpaceTime}
\FuncIntAvg_E[O_E(\varphi, t, \bar t, b, \bar b)] &=& \int \intD{\varphi}\intD{\varphi^\dagger} \intD{t} \intD{\bar t}\intD{b} \intD{\bar b} 
\, O_E(\varphi, t, \bar t, b, \bar b) \cdot
 e^{-S_E[\varphi, t, \bar t, b, \bar b]},\quad
\eea
where the integration is now performed over all field variables at purely Euclidean space-time points defined 
through real Euclidean coordinates $x_E$. The significant enhancement from a numerical perspective results from 
the transition of the complex weight $\exp(iS)$ in \eq{eq:IntroOfFuncIntMink2} to the positive and real factor 
$\exp(-S_E)$ appearing in the Euclidean functional integral. The latter exponential depends on the real 
Euclidean action $S_E$ which can be parametrized in terms of Euclidean coordinates according to
\bea
\label{eq:DefOfEuclideanAction1}
S_E[\varphi, t, \bar t, b, \bar b] &\fhs{-2mm}=\fhs{-2mm}& S_{E,\varphi}[\varphi] + S_{E,F}[\varphi, t, \bar t, b, \bar b], \\
\label{eq:DefOfEuclideanAction2}
S_{E,\varphi}[\varphi] &\fhs{-2mm}=\fhs{-2mm}& \int \intd{^4x_E}\, \frac{1}{2} \partial^E_\mu\varphi_{x_E}^\dagger\partial^E_\mu\varphi_{x_E} 
+ \frac{1}{2} m_0^2 \varphi^\dagger_{x_E}\varphi_{x_E}
+\lambda\left(\varphi_{x_E}^\dagger \varphi_{x_E} \right)^2, \\
\label{eq:DefOfEuclideanAction3}
S_{E,F}[\varphi, t, \bar t, b, \bar b] &\fhs{-2mm}=\fhs{-2mm}& \int \intd{^4x_E}\, \bar t_{x_E}\, \gamma^E_\mu \partial^E_\mu t_{x_E} + \bar b_{x_E} \gamma^E_\mu\partial^E_\mu b_{x_E}\\
&+& \int \intd{^4x_E}\, y_b \left(\bar t_{x_E}, \bar b_{x_E} \right)_L \varphi_{x_E} b_{R,x_E} 
+ y_t \left(\bar t_{x_E}, \bar b_{x_E} \right)_L \tilde\varphi_{x_E} t_{R,x_E} 
+ c.c.,\quad\nonumber
\eea
where $\partial^E_\mu$ denotes the derivative with respect to the Euclidean coordinates and $\gamma^E_\mu$ are the Euclidean gamma matrices 
being related to their Minkowski counterparts $\gamma^\mu$ through
\bea
\gamma^E_0 = \gamma^0,\quad &
\gamma^E_j = -i\gamma^j,\,\, j=1,2,3, \quad &
\gamma^E_5 = \gamma^E_0\gamma^E_1\gamma^E_2\gamma^E_3.
\eea

Moreover, one also learns from \eqs{eq:DefOfEuclideanAction1}{eq:DefOfEuclideanAction3} that $S_E$ is bounded from below, if
the bare parameters are appropriately chosen. For vanishing Yukawa coupling constants this is manifestly guaranteed by the quartic 
coupling term, provided that the bare coupling constant $\lambda$ is positive. If $\lambda$ is zero, a lower bound of $S_E$ still 
exists provided that $m_0^2$ is positive. Allowing for non-vanishing Yukawa coupling constants, on the other hand, does not affect 
the question for the existence of a lower bound. This can be seen by integrating out the fermionic degrees of freedom leading then to 
an effective action in terms of the remaining field variables, \ie the bosonic field variables $\varphi_x$. With the rules of 
Grassmann integration~\cite{Berezin:1966zu} one easily finds that the contributions to the effective action resulting from this integration 
of the fermionic fields are only logarithmic in $\varphi_x$. The question for a lower bound of the effective action, and thus the question 
for the stability of the considered Euclidean theory, is therefore dictated by the algebraic terms in the effective action, which arise only 
from the purely bosonic contribution $S_{E,\varphi}$. This observation will later allow to study the system also at vanishing bare 
quartic coupling constant and non-zero Yukawa coupling constants, provided that the bare scalar mass $m_0^2$ is positive.

The consequence arising from the weight factor $\exp(-S_E)$ being real and bounded from above is that the Euclidean functional integral in 
\eq{eq:IntroOfFuncIntEuc} unlike its Minkowski counterpart becomes efficiently accessible by means of numerical techniques such as Monte-Carlo 
integration with importance sampling.  An example of such a numerical algorithm for the evaluation of the Euclidean functional integral restricted
to a finite space-time lattice will be presented in \sect{sec:HMCAlgorithm}.

The remaining disadvantage of the Euclidean approach is, that all computed observables are calculated in Euclidean space-time, \ie for complex 
Minkowski times. To derive the corresponding result in physical space-time, the obtained results have to be analytically continued to real 
Minkowski time, \ie to complex Euclidean times. This continuation is trivial in the case of observables that depend only on one point in space-time, 
such as the vacuum expectation value $v$ of the scalar field operator $\hat\varphi_x$. For other observables like n-point functions, for instance, 
this continuation is highly non-trivial. It is, however, also possible to derive certain observables, such as the energy spectrum of the underlying 
Hamiltonian $\hat H$, and thus the masses and other properties of the particles described by the theory, without having to rely on the analytical 
continuation of the Euclidean observables. This is possible, since the eigenvalues of the Hamiltonian are also encoded in the Euclidean n-point functions. 
The Higgs boson mass, for instance, can directly be derived from the temporal exponential decay of the Euclidean 2-point function of the Higgs field at 
zero spatial momentum as will be discussed in \sect{sec:SimStratAndObs}.

Finally, the  question of how a degenerate vacuum affects the results in \eq{sec:FuncFormInEucTime} and \eq{eq:IntroOfFuncIntEuc} shall be addressed.
This question is of particular interest when studying the situation of spontaneous symmetry breaking, because in this case there exists a set 
$\bar \Omega=\{\Omega\}$ of vacua $\Omega$, which are degenerate with respect to the Hamiltonian $\hat H$. In the situation of spontaneous symmetry breaking
one of these degenerate states is selected by Nature as the true ground state $\Omega_0\in\bar\Omega$.
Physical observables are then given as vacuum expectation values with respect to this true vacuum state $\Omega_0$. The functional
integral approaches in \eq{sec:FuncFormInEucTime} and \eq{eq:IntroOfFuncIntEuc}, however, compute actually the sum of all vacuum expectation values 
over all vacua $\Omega\in\bar\Omega$. This is because a gap in the energy spectrum of the Hamiltonian $\hat H$ has been assumed for the derivation of 
\eq{sec:FuncFormInEucTime} and \eq{eq:IntroOfFuncIntEuc} separating a single ground state from the rest of the states. If the ground state is degenerate, 
one actually obtains\footnote{In case of a continuous symmetry the sums have to be understood as corresponding integrals.}
\bea
\label{eq:IntroOfFuncIntEucDegenVacua}
\langle O_E(\varphi, t, \bar t, b, \bar b) \rangle_E
&=&\sum\limits_{\Omega_1, \Omega_2\in\bar\Omega}
\langle \Omega_1| \timeOrdering_E\, O_E(\hat\varphi, \hat t, \hat{\bar t}, \hat b, \hat{\bar b})  |\Omega_2\rangle
\cdot \left[ \sum\limits_{\Omega\in\bar\Omega} \langle \Omega|\Omega\rangle  \right]^{-1},
\eea
in the Euclidean case and the equivalent result for Minkowski space-time. 

Restricting the discussion to the Euclidean case one would in principle have to impose boundary conditions on the integration variables in the functional 
integral at $\tau=\pm\infty$ to get the correct vacuum expectation value with respect to $\Omega_0$. These boundary conditions on the
field variables are given by the vacuum expectation values of the corresponding field operators at $\tau=\pm\infty$ with respect to the selected vacuum.
In case of the bosonic field, for instance, one would have to set 
$\varphi_{\tau=\pm\infty,\vec x}:=\langle\Omega_0| \hat\varphi_{\tau=\pm\infty,\vec x}|\Omega_0\rangle$, 
at each spatial coordinate $\vec x$, which means to constrain the field variables at $\tau=\pm\infty$ to the vev $v$. This approach is, however, 
not very practical in actual numerical calculations. Instead one usually applies other strategies, which are
equivalent in infinite volume. One such approach is to break the degeneracy of the vacua explicitly by introducing an external, non-vanishing current $J$ 
according to
\beq
S_{E,J} = S_E + \sum\limits_{x}J^\dagger \varphi_x,
\eeq
which is given here as a complex, two component, space-time independent vector $J$ according to an assumed translational invariance
of the ground state $\Omega_0$.

Certainly, the dynamics induced by $S_{E,J}$ is not the same as the one arising from $S_E$. The essential 
observation~\cite{Gockeler:1989wp,Gockeler:1990zn,Gockeler:1991ty}, however, is that in infinite volume 
the degeneracy of the vacua is broken for any non-zero value of $J$, while the targeted dynamics of the 
system described by $S_E$ is restored in the limit $J\rightarrow 0$. If $J$ is chosen such, that $\Omega_0$ becomes 
the state with lowest energy eigenvalue of the Hamiltonian $\hat H_J$ associated to $S_{E,J}$ one finally arrives at
\bea
\label{eq:IntroOfFuncIntEucDegenVacuaWithCurrent}
\langle \Omega_0| \timeOrdering_E\, O_E(\hat\varphi, \hat t, \hat{\bar t}, \hat b, \hat{\bar b})  |\Omega_0\rangle &=&
\lim\limits_{J\rightarrow 0} \langle O_E(\varphi, t, \bar t, b, \bar b)  \rangle_{E,J}, \\
\langle O_E(\varphi, t, \bar t, b, \bar b)  \rangle_{E,J} &=&
\frac{1}{Z_{E,J}} \FuncIntAvg_{E,J}[O_E(\varphi, t, \bar t, b, \bar b)], \quad Z_{E,J} = \FuncIntAvg_{E,J}[1], \quad
\eea
where the definition of the functional integral $\FuncIntAvg_{E,J}[O_E(\varphi, t, \bar t, b, \bar b)]$ in the presence of a source $J$ is analogously obtained 
from \eq{eq:IntroOfFuncIntEucSpaceTime} by replacing $S_E$ with $S_{E,J}$.

%--------------------------------------------------------------------------------------------------------------------------
\section{Signature of an unstable Higgs boson in Euclidean space-time}
\label{sec:UnstableSignature}

In the full formulation of the Standard Model one would expect the Higgs boson to be an unstable particle. The 
Standard Model Higgs boson decays, for instance, into pairs of fermions $f\bar f$ or pairs of weak interaction gauge bosons
$ZZ$, $W_\pm W_\mp$. In the pure Higgs-Yukawa sector this feature is inherited and the Higgs boson can decay into fermion
pairs, but the decay into gauge boson pairs does not exist, since no gauge bosons are included within the pure Higgs-Yukawa
sector. Instead, the Higgs particle can go over into a pair of Goldstone bosons, which is the equivalent of the decay 
into the weak interaction gauge boson pairs according to the Goldstone equivalence theorem~\cite{Lee:1977eg, Chanowitz:1985hj}. 

In this section the signature of the unstable Higgs boson decaying into pairs of Goldstone bosons shall be briefly discussed in the 
framework of continuous Euclidean space-time. These considerations will play an important role for the determination of the
Higgs boson mass in the later lattice calculations. 

As a starting point we begin with the consideration of the pure $\Phi^4$-theory with only one real component in continuous Euclidean 
space-time. The corresponding action is given as
\beq
S_E = \int \intd{^4x_E} \,\,  \frac{1}{2}\partial^E_\mu\Phi_{x_E}\partial^E_\mu\Phi_{x_E} + \frac{1}{2} m_0^2 \Phi_{x_E}\Phi_{x_E} 
+ \lambda \Phi^4_{x_E}.
\eeq
We will consider this theory in the broken phase with a non-zero vacuum expectation value of the scalar field $\Phi$ denoted as the vev $v$.
The bare Higgs field $h$ is then defined as the fluctuation around $v$ according to
\beq
h_{x_E} = \Phi_{x_E}-v.
\eeq
The corresponding propagator is given as the connected two-point function, which is equal to its disconnected part in this case.
In momentum space the Higgs propagator reads
\beq
\tilde G_H(p_E) = \lim\limits_{J\rightarrow 0} \langle \tilde h_{p_E} \tilde h_{-p_E}   \rangle_{E,J}
\eeq
where $\langle\ldots\rangle_{E,J}$ denotes the expectation value defined through the functional integral in Euclidean space-time as introduced
in the previous section, however, appropriately restricted to the pure $\Phi^4$-theory, and $\tilde h_{p_E}$ is the four-dimensional Fourier
transformation of $h_{x_E}$ given as
\beq
\tilde h_{p_E} = \frac{1}{\sqrt{(2\pi)^4}}\int \intd{^4x_E}\, h_{x_E} \cdot e^{-ip_Ex_E} .
\eeq
This propagator can be calculated by means of perturbation theory in Euclidean space-time. In the broken phase with non-zero values of $v$ 
one easily finds that the inverse propagator can be written as
\beq
\tilde G^{-1}_H(p_E) = p_E^2 + m_0^2 + 12\lambda v^2- \tilde\Sigma_H(p_E)
\eeq
where the Higgs boson self-energy $\tilde \Sigma_H(p_E)$ is given as the sum over all amputated, connected, one-particle irreducible diagrams with two external
Higgs field legs (prior to their amputation), both carrying the momentum $p_E$. It is remarked, that the coupling vertices of the Higgs
field $h$ in the broken phase are different from those of the scalar field $\Phi$. In particular, the Higgs field couples to itself also through
3-point vertices unlike the field $\Phi$. All this will be worked out in more detail for the case of lattice perturbation theory in 
\sect{sec:PredFromPertTheory}. Here, however, we just continue with the calculation of the self-energy $\tilde \Sigma_H(p_E)$. At one-loop order there are only two 
diagrams constituting the self-energy $\tilde \Sigma_H(p_E)$ as sketched in \fig{fig:FeynmanDiagramsForPurePhi4HiggsPropagator}. Their respective 
contributions will be labeled as $\tilde\Sigma_{H1}(p_E)$ and $\tilde\Sigma_{H2}(p_E)$.

%\includeFigSingleSmall{LPTDiagramsForPurePhi4HiggsPropagator}
\includeFigSingleSmall{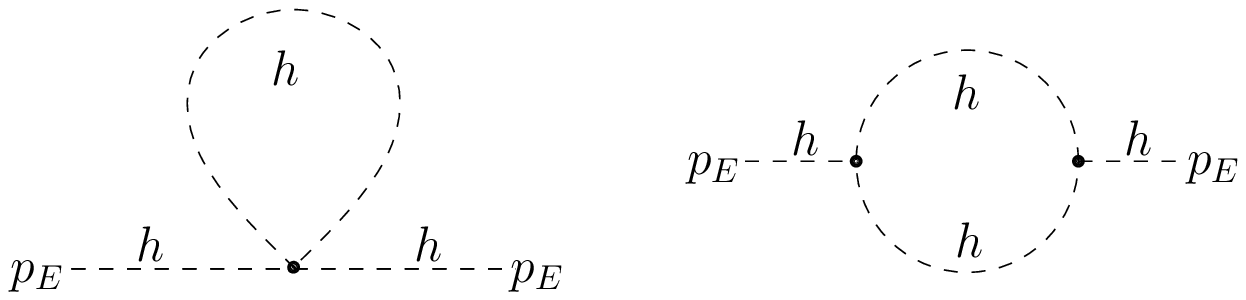}
{fig:FeynmanDiagramsForPurePhi4HiggsPropagator}
{The diagrams contributing to the Higgs propagator $\tilde G_H(p_E)$ at the one-loop level in the real, 
one-component $\Phi^4$-theory. Their contributions to the self-energy $\tilde\Sigma_H(p_E)$ are labeled as $\tilde\Sigma_{H1}(p_E)$ and 
$\tilde\Sigma_{H2}(p_E)$, respectively.}
{Diagrams contributing to the Higgs propagator in the pure one-component $\Phi^4$-theory at the one-loop level.}

The first diagram in \fig{fig:FeynmanDiagramsForPurePhi4HiggsPropagator} is of order $\lambda$ but the associated loop integral does not depend on 
the external momentum $p_E$. It will therefore only generate a constant, momentum-independent contribution $\tilde\Sigma_{H1}(p_E)\equiv\tilde\Sigma_{H1}$ to the
self-energy and is thus not explicitly computed in the following. In a renormalization procedure
it would be absorbed into the renormalization of the Higgs boson mass. The second diagram is of
order $\lambda^2$ and will give rise to a momentum dependence of $\tilde\Sigma_H(p_E)$. Its contribution $\tilde\Sigma_{H2}(p_E)$ is given as
\bea
\label{eq:GuessOfPropFormBosonicLoop1}
\tilde\Sigma_{H2}(p_E) &=& \frac{288\lambda^2v^2}{(2\pi)^4}\int\limits_{k_E=0}^{|k_E|=\Lambda} \fhs{-2mm} \intdfour{k_E}\, 
\frac{1}{k_E^2+\bar m^2} \cdot \frac{1}{(p_E-k_E)^2+\bar m^2}
\eea
where the given integral is regularized here by a hard momentum cutoff $\Lambda$, the given factor of 288 arises from the 
multiplicity of the considered diagram, and $\bar m^2 \equiv m_0^2+12\lambda v^2$ is just a convenient abbreviation. Using 
the Feynman parameter trick this expression can be transformed into
\beq
\tilde\Sigma_{H2}(p_E) = \frac{288\lambda^2v^2}{(2\pi)^4} \int\limits_{k_E=0}^{|k_E|=\Lambda}\fhs{-2mm} \intdfour{k_E} \int\limits_{0}^1 \intd{x}\,   
\Big([k_E^2+\bar m^2]x + [(p_E-k_E)^2+\bar m^2]\cdot (1-x)\Big)^{-2}   
\eeq
which further simplifies to 
\bea
\label{eq:GuessOfPropFormTransToOneDimInt}
\tilde\Sigma_{H2}(p_E) &=& \frac{288 \lambda^2 v^2}{(2\pi)^4} \int\limits_{0}^{\Lambda} \intd{\tilde k} \int\limits_{0}^1 \intd{x}   
\frac{ 2\pi^2 \tilde k^3}{\Big(\tilde k^2+\bar m^2 + p_E^2\cdot[x-x^2]\Big)^{2}},
\eea
if one applies the substitution $\tilde k_E = k_E + (x-1)p_E$ and $\tilde k = [\tilde k_E^2]^{1/2}$. It is remarked that the shift in 
the integration bounds has been neglected in \eq{eq:GuessOfPropFormTransToOneDimInt} which is reasonable for sufficiently large values 
of the cutoff $\Lambda$. This latter expression can now be evaluated by some numerical analysis software~\cite{Mathematica:2007yu} yielding
\bea
\label{eq:GuessOfPropFormMathematicaResult}
\frac{8\pi^2}{288\lambda^2 v^2}\tilde\Sigma_{H2}(p_E) &=& - \sqrt{\frac{4\bar m^2+p_E^2}{p_E^2}} \cdot \arctanh\left[ \sqrt{\frac{p_E^2}{4\bar m^2 + p_E^2}}\, \right] 
+ \frac{1}{2}\log\left( 1+\frac{\Lambda^2}{\bar m^2} \right) \quad \quad\nonumber \\
&+& \frac{1}{\sqrt{p_E^2}} \frac{4\bar m^2+2\Lambda^2+p_E^2}{\sqrt{4\bar m^2+4\Lambda^2+p_E^2}} \cdot 
\arctanh\left[ \sqrt{\frac{p_E^2}{4\bar m^2+4\Lambda^2+p_E^2}}\, \right].
\eea
For the inverse propagator one obtains
\bea
\label{eq:StructureOfPropInPhi4Theory}
\tilde G^{-1}_H(p_E) &=& p_E^2 + m_0^2 +C(\Lambda,m_0,\lambda,v) 
+ 36\pi^{-2}\lambda^2 v^2 \OneBosonLoopContEuc(p_E^2, \bar m^2)  
\eea
where the constant $C(\Lambda,m_0,\lambda,v)$ is independent of the momentum $p_E$ for $|p_E|\ll \Lambda$, which is 
assumed to be the case and the one-loop contribution $\OneBosonLoopContEuc(p_E^2, \bar m^2)$ is given as
\beq
\label{eq:DefOfContEuc1LoopBosContrib}
\OneBosonLoopContEuc(p_E^2, \bar m^2) = \frac{\arctanh\left( q \right)}{q}, \quad q = \sqrt{\frac{p_E^2}{4 \bar m^2 + p_E^2}}.
\eeq
In terms of renormalized quantities the inverse propagator in this one-loop calculation can be expressed as
\bea
\label{eq:StructureOfPropInPhi4TheoryRenormalized}
\tilde G^{-1}_H(p_E) &=& p_E^2 + m_H^2 
+ 36\pi^{-2}\lambda_r^2 v_r^2 \Big( \OneBosonLoopContEuc(p_E^2, m_H^2) - C_{H0}\Big), \\
C_{H0} &=& \OneBosonLoopContEuc(p_E^2, m_H^2)\Bigg|_{p_E^2=-m_H^2} = \frac{\pi}{2\sqrt{3}},
\eea
where the renormalized mass $m_H$ is defined by the pole $\bar p_E$ of the propagator through 
$\bar p_E^2 = -m_H^2$ with the extra minus sign appearing due to the Wick rotation. For the sake of brevity a specific definition of the renormalized 
quartic coupling constant $\lambda_r$ and the renormalized vev $v_r$ is postponed to later sections. Here, it is sufficient 
to assume that these renormalized quantities reproduce their bare counterparts at lowest order, \ie $\lambda_r=\lambda+O(\lambda_r^2)$ and 
$v_r=v+O(\lambda_r)$, to guarantee the deviation between \eq{eq:StructureOfPropInPhi4Theory} and \eq{eq:StructureOfPropInPhi4TheoryRenormalized}
to be of order $O(\lambda_r^3)$.

The main conclusions that can be drawn from this calculation are as follows. The propagator $\tilde G_H(p_E)$ has a pole at $p_E^2=-m_H^2$ by construction.
In addition to that, it has a branch cut starting at $p_E^2=-4m_H^2$. This branch cut will play a major role in the rest of this section. It is
induced by the function $\OneBosonLoopContEuc(p_E^2, m_H^2)$ appearing in \eq{eq:StructureOfPropInPhi4TheoryRenormalized}. Here the branch cut of the
square root in \eq{eq:DefOfContEuc1LoopBosContrib} was chosen to be located along the negative real axis. The resulting pole and branch cut 
structure of the Higgs propagator $\tilde G_H(p_0,0,0,0)$ at zero spatial momentum is presented in \fig{fig:HiggsPropagatorBranchCutIllu}a. Across the 
branch cut the real part of $\tilde G_H(p_E)$ is continuous but the imaginary part has a discontinuity. It is remarked that the branch points 
of $\tilde G_H(p_0,0,0,0)$, which are given as $p_0=\pm 2 im_H$, are not divergent singularities but only gaps in the definition domain of the 
propagator $\tilde G_H(p_E)$.

Since the physical decay width $\Gamma$ of a particle in Euclidean space-time and its mass are given through the real and imaginary part
of the pole on the second Riemann sheet of the analytically continued propagator according to
\beq
\label{eq:DefOfDecayWidth}
\tilde G^{-1}_{H, II}(im_H+\Gamma_H/2,0,0,0) = 0
\eeq
for the case of the Higgs boson~\cite{Luscher:1988uq}, the above calculation shows that the Higgs boson has zero decay width 
in the considered one-component $\Phi^4$-theory, and is thus a stable particle. 

For clarification it is remarked that the first Riemann sheet of the propagator is given by \eq{eq:StructureOfPropInPhi4TheoryRenormalized}.
This function can be analytically continued across the branch cuts, where the propagator in \eq{eq:StructureOfPropInPhi4TheoryRenormalized} is 
discontinuous. The analytically continued function is then denoted as the propagator on the second Riemann sheet. In fact the analytical 
continuation depends on the side from which the branch cut is approached. These subtleties shall, however, not be discussed here. Due to 
the symmetries of the considered one-loop expression for the propagator in \eq{eq:StructureOfPropInPhi4TheoryRenormalized}
they would result here only in adequate changes of the sign of the pole's real and imaginary part in \eq{eq:DefOfDecayWidth}.

%\includeFigDouble{BranchCutOfHiggsProp}{BranchCutOfHiggsPropNComp}
\includeFigDouble{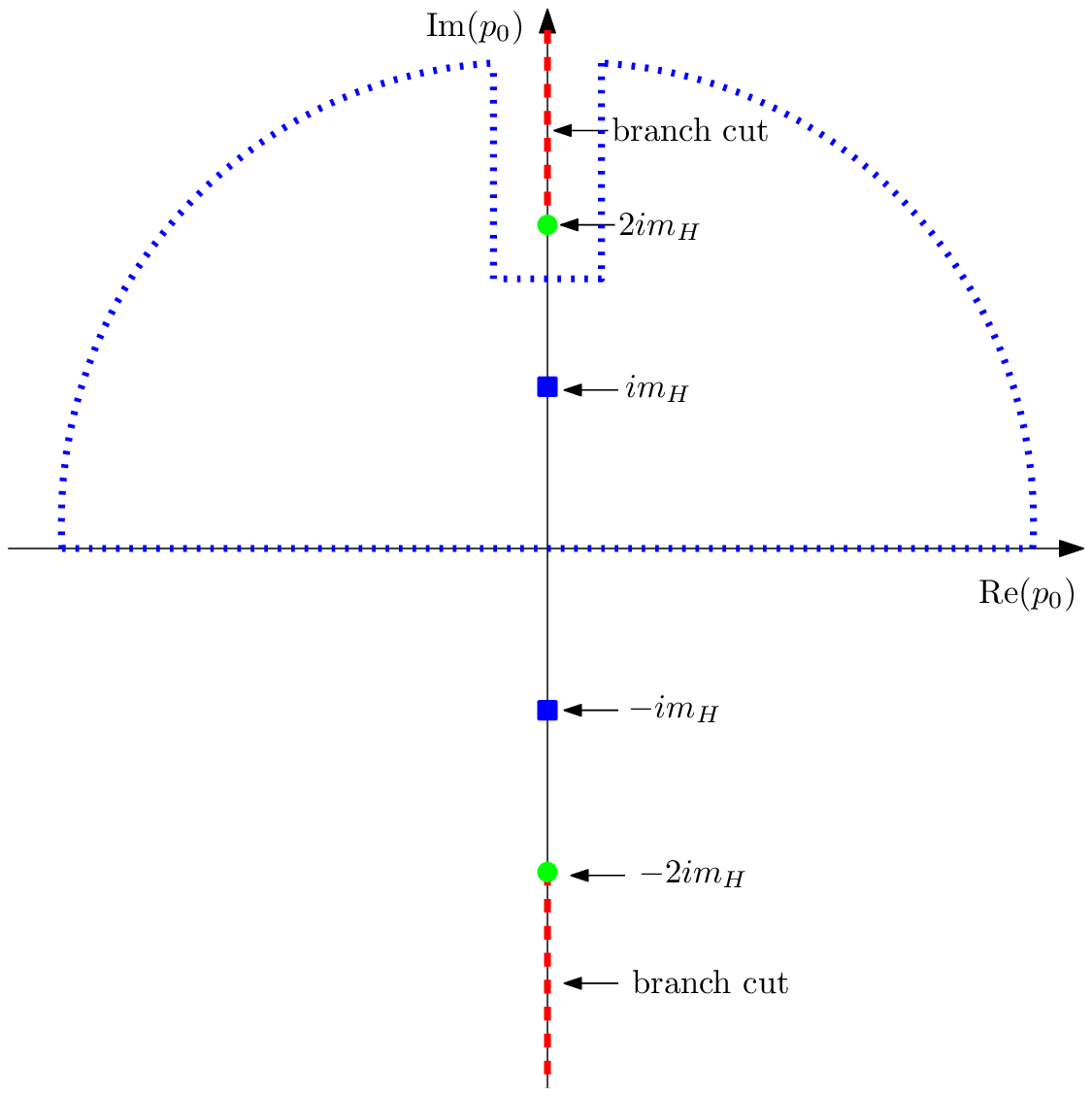}{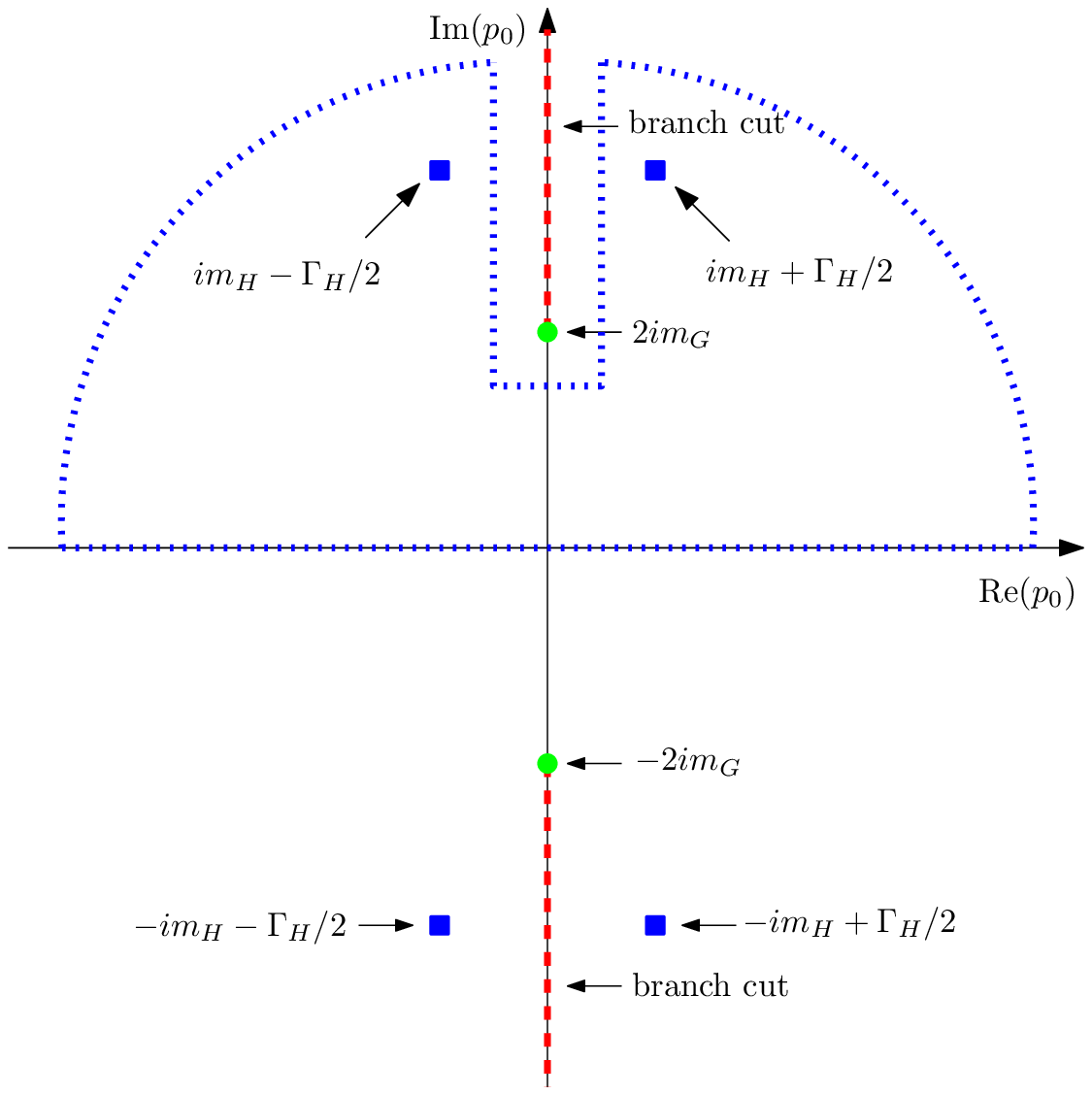}
{fig:HiggsPropagatorBranchCutIllu}
{The pole and branch cut structure of the Higgs propagator $\tilde G_H(p_0,0,0,0)$ in the complex plane spanned by 
the real and imaginary part of $p_0$. The poles are depicted by the blue square symbols, while the green circular symbols represent 
branch points. The red dashed lines indicate the branch cuts, where the propagator has a discontinuity in its
imaginary part. The blue dotted curves indicate the path used for the contour integration as discussed in the
main text. Panel (a) shows the result for the one-component $\Phi^4$-theory while panel (b) presents the situation
in the n-component $\Phi^4$-theory. The poles in panel (b) are actually located on second Riemann sheets.}
{Pole and branch cut structure of the Higgs propagator in the pure $\Phi^4$-theory.}

The preceding finding that the Higgs boson is stable in the one-component $\Phi^4$-theory is an obvious result, since there are 
no other particles present the Higgs boson could decay into. However, this situation changes completely if one considers the $\Phi^4$-theory 
with $n$ real components instead. In this extended theory one has one massive Higgs mode and $n-1$ massless Goldstone modes. However, the 
renormalized mass $m_G$ of the Goldstone modes is considered here as a free parameter to obtain a more general result for later use. Following 
the same steps as above, the Higgs propagator at one-loop order in the n-component $\Phi^4$-theory can be computed by calculating the one-Higgs-loop 
contribution and the $n-1$ one-Goldstone-loop contributions yielding
\bea
\label{eq:StructureOfPropInNCompPhi4TheoryRenormalized}
\tilde G^{-1}_H(p_E) &=& p_E^2 - \left(im_H + \Gamma_H/2\right)^2
+ 36\pi^{-2}\lambda_r^2 v_r^2 \Big( \OneBosonLoopContEuc(p_E^2, m_H^2) - C_{H0}\Big) \\
&+& 4\pi^{-2}(n-1)\lambda_r^2 v_r^2 \Big( \OneBosonLoopContEuc(p_E^2, m_G^2) - C_{G0}\Big), \nonumber\\
\label{eq:DefOfRenHiggsPropConstantCH0}
C_{H0} &=& \OneBosonLoopContEuc_{II}(p_E^2, m_H^2)\Big|_{p_E^2=\left(im_H^2+\Gamma_H/2\right)^2}, \\
\label{eq:DefOfRenHiggsPropConstantCG0}
C_{G0} &=& \OneBosonLoopContEuc_{II}(p_E^2, m_G^2)\Big|_{p_E^2=\left(im_H^2+\Gamma_H/2\right)^2},
\eea
where the factor $32(n-1)$ comes again from the multiplicity of the considered diagrams and $\OneBosonLoopContEuc_{II}$ denotes the analytical
continuation of $\OneBosonLoopContEuc$ onto the second Riemann sheet. 

It is remarked that the renormalized quantities in \eq{eq:StructureOfPropInNCompPhi4TheoryRenormalized}  
cannot be chosen independently of each other. This is clear from a physical perspective, since the latter quantities are connected to each 
other by emerging from the same set of bare parameters. In fact, the underlying theory is specified by only three bare parameters, namely
$\lambda$, $m_0^2$, and $\Lambda$, while the propagator is parametrized here in terms of the five renormalized constants $m_H$, $\Gamma_H$, 
$\lambda_r$, $v_r$, and $m_G$ which makes apparent the fact that the latter numbers, as emerging from the same underlying bare 
theory, must be related to each other, thus obeying additional constraints. Otherwise the renormalized result given in \eq{eq:StructureOfPropInNCompPhi4TheoryRenormalized}
would have more degrees of freedom than the underlying bare theory. Such constraints are parametrized in terms of the 
renormalization conditions which, however, have not explicitly been given yet. One constraint can, for instance, be inferred 
from the requirement that $\tilde G_H(p_E^2)$ be real for purely positive values of $p_E^2$, which is not automatically guaranteed 
by the general expression in \eq{eq:StructureOfPropInNCompPhi4TheoryRenormalized}, if all renormalized quantities could be 
chosen arbitrarily. Together with the later presented definition of the renormalized quartic coupling constant in \eq{eq:DefOfRenQuartCoupling}
these constraints reduce the independent renormalized model parameters, for instance, to $m_H$, $m_G$, and $\lambda_r$, where
$m_G$ would actually be zero in infinite volume. It is only kept as a free parameter here to obtain a more general result for later use.
Conversely, the aforementioned constraints allow, for instance, for the perturbative calculation of the Higgs boson decay width, 
when the aforementioned independent quantities $m_H$, $m_G$, and $\lambda_r$ are known, as will be discussed in \chap{chap:ResOnDecayWidth}.

The branch and pole structure of the Higgs propagator $\tilde G_H(p_0,0,0,0)$ in the considered n-component $\Phi^4$-theory
at zero spatial momentum is presented in \fig{fig:HiggsPropagatorBranchCutIllu}b. The main observation is, that the 
function $\OneBosonLoopContEuc(p_E^2, m_G^2)$ induces a branch cut starting already at $p_0=\pm 2im_G$. This implies that the pole
$\bar p_E=(\bar p_0,0,0,0)$ of the Higgs propagator $\tilde G_H(p_E)$ can not be situated at purely imaginary values $\bar p_0=im_H$ 
any longer, if the renormalized Goldstone mass $m_G$ is smaller than $m_H/2$. Instead the pole is shifted away from the imaginary axis
onto the second Riemann sheet of the analytically continued Higgs propagator. This is illustrated in \fig{fig:HiggsPropagatorBranchCutIllu}b.
In fact, there is a deep physical reason why the pole has to be shifted onto the second Riemann sheet and can not appear on the first sheet
provided that the underlying Hamiltonian $\hat H$ is hermitian. This consideration will be given at the end of this section.

The shifted pole thus leads to a non-zero decay width $\Gamma_H$ according to \eq{eq:DefOfDecayWidth}. This is the mathematical formulation 
of the expected result that the Higgs particle becomes unstable and can decay into pairs of Goldstone bosons provided that the quartic coupling 
constant is non-zero and that the Higgs boson mass is larger than the mass of the Goldstone pair, which is zero in infinite volume.

The signature of an unstable Higgs particle in Euclidean space-time is thus that the pole of the Higgs propagator has non-zero real and imaginary 
parts.This will play a role for the approaches to determine the Higgs boson mass in the later numerical calculations. Assuming the validity of the 
one-loop calculation the decay width can then be calculated from \eq{eq:StructureOfPropInNCompPhi4TheoryRenormalized}. However, the 
further discussion of the Higgs boson decay properties is postponed to \chap{chap:ResOnDecayWidth}. 

Moreover, it is remarked that extending this analysis to the Higgs-Yukawa model, including also the coupling to the fermion fields, yields
very similar results. At one-loop order the fermion loop coupling to the Higgs boson induces again a branch cut starting at 
$p_E^2 = -4m^2_F$, where $m_F$ denotes here the renormalized mass of the considered fermion. This branch cut would again shift the
pole of the Higgs propagator off the imaginary axis, provided that the renormalized Higgs boson mass is larger than $2m_F$.
The obvious interpretation is that the Higgs particle is unstable and decays into pairs of the considered fermions, if $m_H>2 m_F$, as expected.

Finally, the Euclidean time-correlation function $C_H(\Delta \tau)$ of the Higgs field shall be discussed. This function will play
an important role in the determination of the Higgs boson mass. It is defined as
\bea
\label{eq:HiggsTimeCorrelatorInEucTime1}
C_H(\Delta \tau) &=& \lim\limits_{J\rightarrow 0}\,
\frac{1}{\sqrt{(2\pi)^{7}}}
\int \intd{\tau}\intd{^3\vec x} \intd{^3\vec y}\,\, \langle h_{\tau+\Delta \tau,\vec x} h_{\tau,\vec y} \rangle_{E,J} \\
\label{eq:HiggsTimeCorrelatorInEucTime2}
&=& \frac{1}{\sqrt{2\pi}}\int \intd{p_0}\,\, e^{ip_0\Delta\tau} \cdot \tilde G_H(p_0,0,0,0)
\eea
and can be calculated by closing the contour of the integral in \eq{eq:HiggsTimeCorrelatorInEucTime2} in the complex plane as illustrated in
\fig{fig:HiggsPropagatorBranchCutIllu}. For the case of the one-component $\Phi^4$-theory one then obtains
\beq
C_H(\Delta \tau) = \left[ \frac{\sqrt{2\pi}}{2m_H} + O(\lambda_r) \right] \cdot e^{-m_H \Delta\tau} + \int_{2m_H}^\infty\intd{E}\, \rho(E) e^{-E\Delta\tau}
\eeq
where the first term is induced by the residuum 
\beq
\res(\tilde G_H(p_0,0,0,0))\Big|_{p_0=im_H} = \frac{1}{2im_H} + O(\lambda_r)
\eeq
of the Higgs propagator at its pole and the integral over the spectral function $\rho(E)$ is given by the discontinuity of the Higgs propagator
in \eq{eq:StructureOfPropInPhi4TheoryRenormalized} along the branch cut according to
\beq
\label{eq:DefOfSpectralFunction}
\rho(E) = \lim\limits_{\epsilon\rightarrow 0}\,
\frac{1}{\sqrt{2\pi}}\IM\left[\tilde G_H(iE-\epsilon,0,0,0) -  \tilde G_H(iE+\epsilon,0,0,0) \right],
\eeq
since its real part is continuous. The spectral function $\rho(E)$ is
thus real. From the result derived for the Higgs propagator in \eq{eq:StructureOfPropInPhi4TheoryRenormalized} one finds that $\rho(E)$
is positive as illustrated in \fig{fig:SpectralFunctionsFromPT}a for some specific choice of the renormalized parameters.
The main conclusion for the one-component $\Phi^4$-theory is that the time correlation function $C_H(\Delta\tau)$
is a sum of decaying exponential functions with purely positive weights. The argument of the slowest decaying exponential is given by
the Higgs boson mass $m_H$, which is clearly separated from the continuous spectrum described by the spectral function $\rho(E)$ starting
only at $E=2m_H$. In this scenario one can thus determine the Higgs mass $m_H$ by matching the time correlation function $C_H(\Delta \tau)$
at large Euclidean time separations $\Delta\tau$ to a single exponential decay behaviour according to $C_H(\Delta \tau)\propto \exp(-m_H\Delta\tau)$
for $\Delta\tau\gg 1$.
 
This situation changes dramatically for the n-component $\Phi^4$-theory. Following the same steps as above one now finds for the time correlation
function
\beq
\label{eq:DefOfTimeCorrFunction2}
C_H(\Delta \tau) = \int_{2m_G}^\infty\intd{E}\, \rho(E) e^{-E\Delta\tau},
\eeq
where the spectral function $\rho(E)$ is again given by \eq{eq:DefOfSpectralFunction} but now based on the discontinuity of the Higgs propagator 
in \eq{eq:StructureOfPropInNCompPhi4TheoryRenormalized}. It is explicitly remarked that the poles of the Higgs propagator do not contribute
to the contour integration, since they are only situated on the second Riemann sheet. They are therefore not within the integration contour,
which is completely located on the first Riemann sheet. The main conclusion at this point is that the time correlation function $C_H(\Delta\tau)$
is again given by a sum of decaying exponentials with positive weights, but in this case the spectral function $\rho(E)$ of the exponential decay
rates is completely continuously distributed as illustrated in \fig{fig:SpectralFunctionsFromPT}b,c. In fact, there is no separated decay rate of the
Higgs correlator $C_H(\Delta \tau)$ in contrast to the previously discussed case of the one-component $\Phi^4$-theory.

%\includeFigTriple{SpectralFunctionN1}{SpectralFunctionN4Weak}{SpectralFunctionN4Strong}
\includeFigTriple{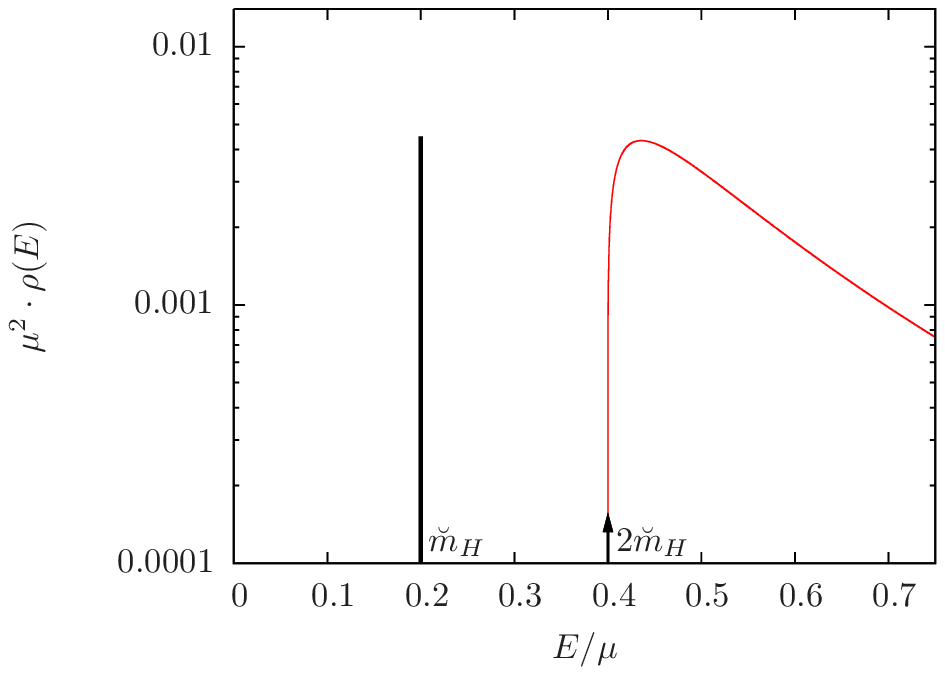}{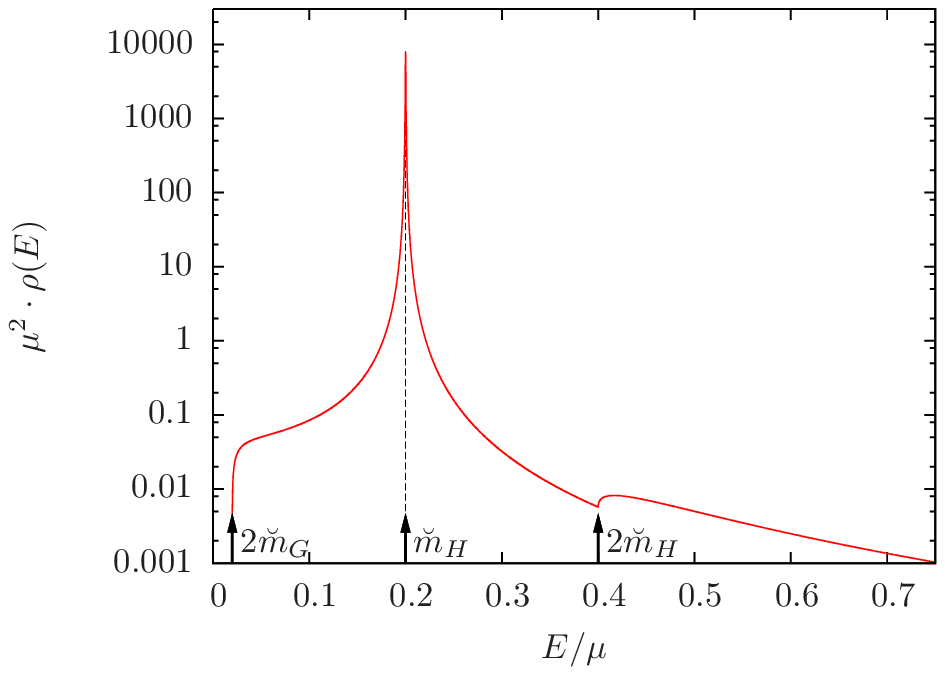}{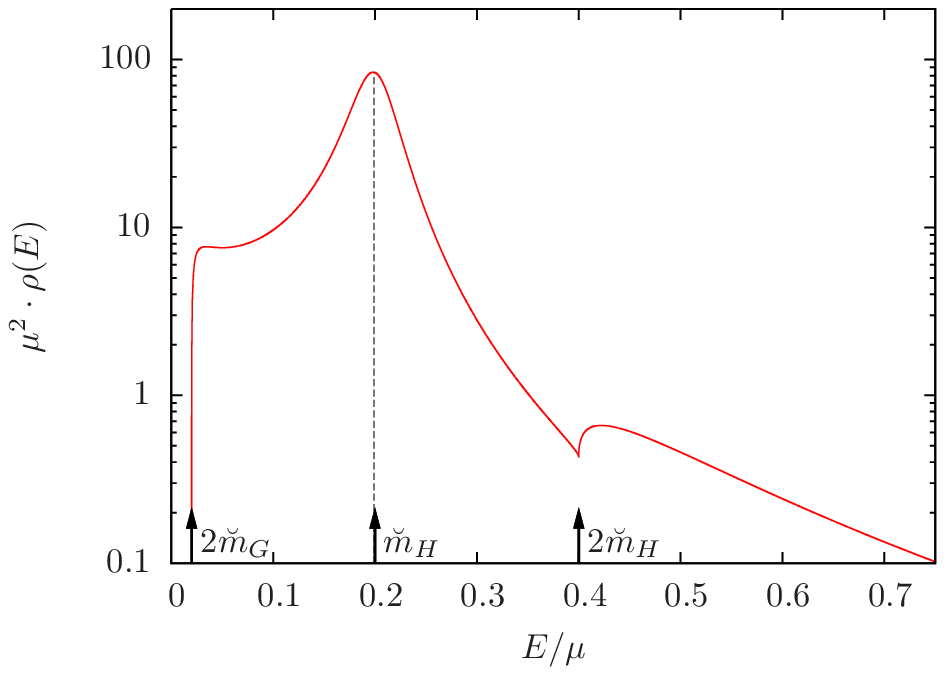}
{fig:SpectralFunctionsFromPT}
{The red curve depicts the spectral function $\rho(E)$ as defined in \eq{eq:DefOfSpectralFunction} for some specific selections of the 
renormalized parameters, given here in units of a scale $\mu$. In panels (b), (c) the Goldstone mass is $\breve m_G = m_G/\mu=0.01$, while 
the Higgs boson mass is $\breve m_H=m_H/\mu=0.2$ in all three cases. The parameters $v_r$ and $\Gamma_H$ were fixed by the only later given 
renormalization condition in \eq{eq:DefOfRenQuartCoupling} and requiring $G_H(p_E^2)$ to be real at purely positive values of $p^2_E$ as 
discussed in the main text. (a) One-component $\Phi^4$-theory with $n=1$ and $\lambda_r=0.01$. The thick vertical black line symbolizes the 
contribution of the Higgs pole to $C_H(\Delta\tau)$ in terms of a $\delta$-contribution at $E=m_H$ which, however, is actually not part of the 
definition in \eq{eq:DefOfSpectralFunction}. (b) The n-component $\Phi^4$-theory with $n=4$ and $\lambda_r=0.01$. (c) The n-component $\Phi^4$-theory 
with $n=4$ and $\lambda_r=1.0$. In all panels the arrows and dashed lines are only meant to guide the eye.}
{Spectral function $\rho(E)$ in the pure $\Phi^4$-theory for different values of the renormalized coupling constants.}

Conceptually, one can thus not apply
the same procedure of matching $C_H(\Delta \tau)$ to a single exponential behaviour at large Euclidean times to determine $m_H$. 
However, as demonstrated in \fig{fig:SpectralFunctionsFromPT}b for some specific choice of the renormalized parameters with a rather small 
value of $\lambda_r$ the spectral function has a peak at $E\approx m_H$ and is completely dominated by that contribution in the presented setup. For 
sufficiently small values of the renormalized coupling constant $\lambda_r$ the above procedure of determining $m_H$ thus remains applicable at least 
from a practical point of view. For larger values of $\lambda_r$ on the other hand the peak at $E\approx m_H$ becomes less prominent. Therefore, 
the crucial observation is that analyzing the time correlation function in terms of a single exponential fit even at $\Delta\tau\gg 1$ cannot be 
supposed to work well in the strong coupling regime. This will become an issue for the determination of the Higgs boson mass in the later 
numerical calculations.

It is remarked here that the above results are completely consistent with the equivalent picture in the operator formalism. In this approach the 
time correlation function of a given operator $\hat O(\tau)$, depending only on field operators at the Euclidean time $\tau$ with 
$\hat O(0)$ being hermitian, is 
\bea
\label{eq:DefOfOpCorrFunctionInOpForm1}
\langle \Omega_0 | \hat O(\tau) \hat O(0) | \Omega_0\rangle &=&
\langle \Omega_0 | e^{+\hat H \tau}\hat O(0) e^{-\hat H \tau} \hat O(0) | \Omega_0\rangle \\
&=&
\sum\limits_n e^{-\tau (E_n-E_0)} \cdot |\langle \Omega_0| \hat O(0)| E_n \rangle|^2, \nonumber
\eea
where $E_n$ and $|E_n\rangle$ denote the eigenvalues and the eigenvectors of the Hamiltonian $\hat H$. Provided that the Hamiltonian $\hat H$
is hermitian, all eigenvalues are real and the time correlation function is a sum of decaying exponentials with positive weights. This is the 
reason why the pole of the Higgs propagator can not be situated on the first Riemann sheet if the Hamiltonian is hermitian. Otherwise there 
would be contributions of the corresponding residua to the time correlation function in \eq{eq:HiggsTimeCorrelatorInEucTime1} resulting in an 
oscillatory behaviour of $C_H(\Delta \tau)$ which would be a contradiction to the result in \eq{eq:DefOfOpCorrFunctionInOpForm1}.

%--------------------------------------------------------------------------------------------------------------------------
\section{Naive discretization of the Higgs-Yukawa sector on a space-time lattice}
\label{sec:NaiveDiscret}

It was already noted in \sect{sec:FuncFormInEucTime} that the functional integrals in \eq{eq:IntroOfFuncIntMink} and \eq{eq:IntroOfFuncIntEuc}
are lacking a thorough mathematical definition in continuous space-time and that a sound definition of the integration measure can 
only be given for a discrete, finite set of space-time points $x\in\Gamma$. The given functional integrals therefore have actually to be understood 
as the limit of a high-, but finite-dimensional integration over the field variables 
$\varphi_x,\varphi^\dagger_x,t_x, \bar t_x, b_x, \bar b_x$ with $x\in\Gamma$ according to the integration measure given in \eq{eq:DefOfFuncIntMeasure} in the 
limit where the discrete points $x\in\Gamma$ become infinitely densely packed and $\Gamma$ extends over the 
whole space-time. Moreover, discretized versions $S_{\Gamma}$ and $S_{E,\Gamma}$ of the actions $S$ and $S_E$ 
depending only on the field variables at the discrete points $x\in\Gamma$ are required for this limit procedure and will be discussed later in this section.

Here, we continue with giving a precise meaning to the aforementioned limit procedure by specifying a particular form of the set $\Gamma$ of 
space-time points according to
\beq
\Gamma\equiv \Gamma_{L_\mu,a} = \Big\{(a\cdot n_0,a\cdot n_1,a\cdot n_2,a\cdot n_3): \quad n_\mu\in\N_0, n_\mu<L_\mu,\,\mu=0,\ldots,3    \Big\},
\eeq
which will be used in the following\footnote{The restriction that the given set $\Gamma_{L_\mu,a}$ only covers
space-time points with non-negative coordinates would be trivially resolvable by adequate shifts of the specified coordinates.
For the sake of brevity, however, we stay with the given expression.}. This construction arranges the points $x\in \Gamma$ in a hypercubic 
order and will thus be referred to as a lattice with side lengths $L_0,\ldots,L_3$ and lattice spacing $a$. Restricting ourselves 
to the consideration of the Euclidean functional integral in the following a well-defined meaning is then given to \eq{eq:IntroOfFuncIntEuc} by
\bea
\label{eq:DefOfFuncIntOnLattice}
\langle O_E(\varphi, t, \bar t, b, \bar b)\rangle_E &=& \lim\limits_{a\rightarrow 0} \,\lim\limits_{L_\mu\rightarrow\infty} \,
\langle O_E(\varphi, t, \bar t, b, \bar b)  \rangle_{E,\Gamma}, \\
\label{eq:DefOfFuncIntOnLattice2}
\langle O_E(\varphi, t, \bar t, b, \bar b)  \rangle_{E,\Gamma} &=&
\frac{1}{Z_{E,\Gamma}} \FuncIntAvg_{E,\Gamma}[O_E(\varphi, t, \bar t, b, \bar b)], \quad Z_{E,\Gamma} = \FuncIntAvg_{E,\Gamma}[1],
\eea
where the finite-dimensional (quasi) functional integral $\FuncIntAvg_{E,\Gamma}[O_E(\varphi, t, \bar t, b, \bar b)]$ associated to the
lattice $\Gamma_{L_\mu,a}$ is defined as
\beq
\label{eq:DefOfFuncIntOnLattice3}
\FuncIntAvg_{E,\Gamma}[O_E(\varphi, t, \bar t, b, \bar b)] = 
\int \fhs{-2mm}\prod_{x\in \Gamma_{L_\mu,a}} \fhs{-2mm} \intd{\varphi_x}\intd{\varphi^\dagger_x} \intd{t_x} \intd{\bar t_x}\intd{b_x} \intd{\bar b_x} \,\,
O_E(\varphi, t, \bar t, b, \bar b)\cdot e^{-S_{E,\Gamma}}.
\eeq
The inner limit in \eq{eq:DefOfFuncIntOnLattice} will be referred to in the following as the infinite volume limit, while the outer limit will be 
called the continuum limit $a\rightarrow 0$. It is, however, already remarked at this point that the notion of the 'continuum limit' 
does actually not simply refer to sending the lattice spacing $a$ to zero. In fact, the bare coupling parameters have to be varied simultaneously in order
to obtain a convergent result. This will be discussed in more detail in \sect{sec:LatDiscOfContinuumLimit}. For the time being we continue
here with using the notion 'continuum limit' simply as a synonym for sending $a$ to zero. 

It is exactly this definition of the functional integral $\FuncIntAvg_{E,\Gamma}[O_E(\varphi, t, \bar t, b, \bar b)]$ that builds the basis of the
so-called lattice approach. The lattice approach aims at calculating $\langle O_E(\varphi, t, \bar t, b, \bar b)\rangle_E$ by the direct numerical 
evaluation of the right-hand side of \eq{eq:DefOfFuncIntOnLattice}. This is done by calculating $\langle O_E(\varphi, t, \bar t, b, \bar b)  \rangle_{E,\Gamma}$
for a series of different lattices $\Gamma_{L_\mu,a}$ with finite extent and non-zero lattice spacing $a$. The obtained results have then to be 
extrapolated to the continuum and infinite volume limit in order to yield the correct result for $\langle O_E(\varphi, t, \bar t, b, \bar b)\rangle_E$. 

A numerical method suitable for the evaluation of \eq{eq:DefOfFuncIntOnLattice} on a finite space-time lattice will be discussed in \sect{sec:HMCAlgorithm}. 
In the following subsections we continue with the discussion of the discretized action $S_{E,\Gamma}$, the choice of which is not
unique and has a strong impact on the convergence properties of the limit procedure in \eq{eq:DefOfFuncIntOnLattice}. Besides that different
lattice actions can also differ strongly with respect to the symmetries they obey. The latter observation will play a major role for the 
construction of the chirally invariant lattice Higgs-Yukawa model in \sect{sec:modelDefinition}.

%--------------------------------------------------------------------------------------------------------------------------
\subsection{Discretization of the purely bosonic part of the action}
\label{sec:LatDiscOfBosPart}

In this section the discretization of the purely bosonic part of the Euclidean action given in
\eq{eq:DefOfEuclideanAction2} shall be discussed. Discretizing a continuous action is, however, not
a unique procedure. The only strict requirement is that for any given set of smooth fields the 
discretized form of the action converges to its continuum counterpart in the limit $a\rightarrow 0$. 
This rather loose condition leaves open much freedom for the construction of lattice actions, which
can be used to improve the convergence properties of the lattice action in the aforementioned limit.

An obvious way to obtain a discretized version of the continuum action, that obeys the requirement
to converge to its continuum counterpart, is given by replacing all differential operators with corresponding
difference quotients of field variables at neighbouring lattice sites. Analogously, all space-time integrals
would be represented as finite Riemann sums in this approach. Omitting the subscript $E$ indicating
the Euclidean space-time from now on, this procedure applied to the purely bosonic part of the Euclidean 
action in \eq{eq:DefOfEuclideanAction2} yields
\bea
\label{eq:BosonicLatticeHiggsActionContNot}
S_{\Gamma,\varphi}[\varphi] &=& a^4\sum\limits_{x,\mu} 
\frac{1}{2} \nabla^f_\mu \varphi^\dagger_x
\nabla^f_\mu \varphi_x
+ a^4\sum\limits_{x} \frac{1}{2} m_0^2 \varphi^\dagger_x\varphi_x
+a^4\sum\limits_{x} \lambda\left(\varphi_x^\dagger \varphi_x \right)^2,
\eea
where the unspecified summation over the space-time points $x$ is always meant to be performed over the 
set of lattice points $\Gamma_{L_\mu,a}$ in the following and the summation over $\mu=a\cdot \hat e_\mu$ 
runs over the four unit basis vectors $\hat e_\mu$ along the four Euclidean space-time 
dimensions. The applied lattice forward derivative operator $\nabla^f_\mu$ in direction $\mu$ is defined as
\bea
\nabla^f_\mu \varphi_x = \frac{\varphi_{x+\mu} - \varphi_x}{a},
&\mbox{ and } &
\nabla^b_\mu \varphi_x = \frac{\varphi_{x} - \varphi_{x-\mu}}{a},
\eea
where $\nabla^b_\mu$ denotes the corresponding backward derivative. At the boundary of $\Gamma_{L_\mu,a}$ 
the space-time point $x+\mu$ may leave the set $\Gamma_{L_\mu,a}$ and the field variable $\varphi_{x+\mu}$ would be
undefined. In that case an appropriate modulo operation according to $\varphi_{x+\mu}\equiv \varphi_{x+\mu-aL_\mu\hat e_\mu}$
is implicitly assumed here. The lattice is then said to have periodic boundary conditions.

In the given lattice action the nearest neighbour forward difference quotient has been chosen to substitute 
the differential operator. The discretization error of this lattice operator is known to be of order $O(a)$. However, 
the total error term of the given lattice action only starts at $O(a^2)$. This is because the lattice sum over the
given product of difference quotients can be rewritten in terms of the symmetric, second derivative difference
quotient expression according to
\beq
\sum\limits_{x} 
\nabla^f_\mu \varphi^\dagger_x \nabla^f_\mu \varphi_x
=
-\sum\limits_{x} 
\varphi^\dagger_x \nabla^b_\mu  \nabla^f_\mu \varphi_x
=
-\sum\limits_{x} 
\frac{\varphi_{x}^\dagger \cdot \left(\varphi_{x+\mu}+\varphi_{x-\mu} -2\varphi_{x}\right) }{a^2},
\eeq
the discretization error of which is of order $O(a^2)$. Furthermore, due to the aforementioned periodicity 
of the lattice the Riemann sum in \eq{eq:BosonicLatticeHiggsActionContNot} corresponds to a trapezoidal 
rule~\cite{Press:2007zu} approximating the targeted integral in continuous space-time with an error term starting 
at $O(a^2)$. The total discretization error of the given lattice action is thus of order $O(a^2)$.

However, the order of the discretization error of the discretized action alone does not specify the order of 
the discretization error that one will encounter in the evaluation of a specific observable. The reason for that
is twofold. First of all the considered discretized observable itself may contain deviations from its continuum 
counterpart with different order in $a$ than the discretization error of the action. This applies at least to all
observables containing differential operators, such as the topological charge observable in lattice QCD, since the
aforementioned differential operators contained within such observables have to be discretized as well. Secondly, and 
more interestingly, the integration over all field configurations in the functional integral in \eq{eq:DefOfFuncIntOnLattice3}
generates additional contributions to the expectation value of the considered observable that would
not be present in the continuum, and these disturbing contributions can be of lower order in the lattice spacing $a$ 
than the lowest order discretization error observed in the lattice action or in the discretization of the considered observable
itself. In general, all possible terms that have the same dimension as the considered 
observable and that are not prohibited by the symmetries simultaneously obeyed by the theory and the considered observable
can be generated in this way, \ie by quantum fluctuations. This was first discussed in the famous work of Symanzik 
introducing the Symanzik improvement program~\cite{Symanzik:1983dc,Symanzik:1983gh}.

The main idea of the Symanzik improvement program is to identify all terms that can potentially be generated in the
functional integral, \ie all terms that are compatible with the symmetries of the model, and to add corresponding
counterterms with appropriately chosen coefficients to the underlying action, such that all unwanted terms of a certain 
order in the lattice spacing $a$ are exactly canceled. 

The analysis of the purely bosonic lattice theory specified through \eq{eq:BosonicLatticeHiggsActionContNot} 
shows that parity and time reflection symmetry greatly constrain the terms that can be generated by quantum fluctuations, 
provided that the considered observable also obeys these symmetries. In that case, one finds that discretization errors can 
only have even powers in the lattice spacing~\cite{Symanzik:1983dc,Symanzik:1983gh}. The lowest order of a possible 
discretization error is thus $O(a^2)$.  

With the apparatus of the Symanzik improvement program at hand it is possible to construct improved lattice actions 
such that the observables of interest have even better convergence orders than $O(a^2)$. 
However, since the fermionic part of the lattice action that will actually be used
in the later calculations is of order $O(a^2)$, we will stay with the expression given
in \eq{eq:BosonicLatticeHiggsActionContNot}.

Finally, the eigenvectors and eigenvalues of the lattice derivative operator $-\nabla^b_\mu\nabla^f_\mu$ 
shall be introduced for later use. As in the continuum the eigenvectors are 
given by plane waves, the allowed four-component momenta $p$ of which, however, are restrained to the 
set of lattice momenta
\beq
\label{eq:DefOfAllowedLatticeMomenta}
p \in \ImpSpace \equiv \ImpSpace_{\Gamma} = \left\{ \left(\frac{2\pi}{aL_0}n_0,\frac{2\pi}{aL_1}n_1,\frac{2\pi}{aL_2}n_2,\frac{2\pi}{aL_3}n_3\right):\,
n_\mu\in\N_0,\, n_\mu<L_\mu,\, \mu=0,\ldots,3 \right\}
\eeq
associated to the set of lattice points $\LatSpace_{L_\mu,a}$.
The corresponding eigenvalues are then given by the squared lattice momenta $\hat p^2$ according to
\beq
-\sum\limits_\mu \nabla^b_\mu \nabla^f_\mu e^{ip\cdot x} = \hat p^2 \cdot e^{ip\cdot x} 
\eeq
with
\beq
\label{eq:DefOfSquaredLatticeMomHatP}
\hat p^2 \equiv \frac{1}{a^2}\sum\limits_\mu  4\sin^2(ap_\mu/2) = p^2 - \frac{a^2}{12} \sum\limits_\mu p_\mu^4 +  O(a^4).
\eeq
In the non-interacting case the lattice propagator of the bosonic field $\varphi$ in momentum space
\beq
\tilde G_\varphi(p) = \langle \tilde\varphi_p \tilde\varphi_{-p} \rangle_{\Gamma}
\eeq
with $p\in \ImpSpace_{\Gamma}$ can then be written in terms of the aforementioned eigenvalues according to
\beq
\label{eq:FreeBosonicInvPropagator}
\tilde G^{-1}_\varphi(p) \equiv \tilde G^{-1}_\varphi(\hat p^2) = \hat p^2 + m_0^2 = p^2 + m_0^2 - \frac{a^2}{12} \sum\limits_\mu p_\mu^4 +  O(a^4),
\eeq
where the field variables $\tilde \varphi_p$ in momentum space are given as
\beq
\label{eq:DefOfScalarFieldInMomSpace}
\tilde \varphi_p = \frac{1}{\sqrt{V}} \sum\limits_x e^{-ipx} \varphi_x
\eeq
with $V=L_0\cdot\ldots\cdot L_3$ denoting the lattice volume.

The concluding remark at this point is that the obtained lattice propagator and in particular the location of its pole deviate
from their continuum results by discretization errors starting at order $O(a^2)$, as expected. One also learns from the appearance
of the term $p_\mu^4$ in \eq{eq:FreeBosonicInvPropagator} that Lorentz-symmetry, which becomes an $O(4)$-symmetry in Euclidean 
space-time is explicitly broken by the lattice, as expected. However, the latter symmetry is finally restored in the continuum 
limit $a\rightarrow 0$ and the symmetry breaking terms disappear rather quickly at order $O(a^2)$.

%--------------------------------------------------------------------------------------------------------------------------
\subsection{Naive discretization of the fermionic part of the action}
\label{sec:LatDiscOfFermiPart}

Following the same straightforward ansatz, which worked satisfactorily for the case of the purely bosonic 
part of the action, a first approach to a lattice formulation of the Euclidean fermion action given in 
\eq{eq:DefOfEuclideanAction3} would be
\bea
\label{eq:DefOfEuclideanLatticeFermionAction}
S_{\Gamma,F}[\varphi, t, \bar t, b, \bar b] &=& a^4 \sum_{x, \mu}
\bar t_{x}\, \gamma_\mu \nabla^f_\mu t_{x} 
+ \bar b_{x} \gamma_\mu \nabla^f_\mu b_{x}\\
&+& a^4 \sum\limits_x\, y_b 
\left(\bar t_{x}, \bar b_{x} \right)_L \varphi_x b_{R,x} 
+ y_t \left(\bar t_{x}, \bar b_{x} \right)_L \tilde\varphi_x t_{R,x} + c.c.\,.\quad\nonumber
\eea
An obvious disadvantage of this discretization is its rather bad convergence behaviour in the limit
$a\rightarrow 0$ due to the discretization error of the given action being of order $O(a)$ as induced 
by the non-symmetric lattice derivative operator $\nabla^f_\mu$. The most severe problem with this approach, 
however, is more subtle. The application of the non-symmetric lattice derivative operator $\nabla^f_\mu$ breaks 
explicitly reflection positivity such that the hermiticity of the Hamiltonian $\hat H_\Gamma$ associated to the 
discrete system, and thus the positivity of the so-called transfer matrix being the matrix representation of
the transfer operator $\exp(-a\hat H_\Gamma)$, is no longer guaranteed~\cite{Sadooghi:1996ip, book:Montvay}.
Moreover, it could be generally shown within the framework of QED that any application of 
a non-symmetric form of a lattice derivative operator for the discretization of the fermionic action
leads to the emergence of non-covariant terms in the fermion self-energy and other observables in  
the continuum limit~\cite{Sadooghi:1996ip}. This is a severe defect, since the continuum physics would not
be correctly reproduced in the continuum limit $a\rightarrow 0$.

This severe problem can be cured by replacing the non-symmetric lattice derivative operator $\nabla^f_\mu$ 
with the symmetrized, anti-hermitian operator $\nabla^s_\mu$ defined as
\beq
\nabla^s_\mu = \frac{1}{2} \left[ \nabla^f_\mu +  \nabla^b_\mu \right] = -\left[\nabla^s_\mu\right]^\dagger,
\eeq
which does not break reflection positivity. Moreover, the discretization error of the symmetric operator
only starts at order $O(a^2)$. At tree-level the lattice artefacts of the resulting fermion action itself 
are thus of order $O(a^2)$. As discussed in the previous section, it is not guaranteed that this apparent 
tree-level $O(a)$-improvement also holds when quantum fluctuations are considered. In fact, a detailed analysis 
for the case of QCD following the ideas of the Symanzik improvement program~\cite{Symanzik:1983dc,Symanzik:1983gh}
shows that the apparent $O(a)$-improvement is indeed broken at the quantum level. If quantum 
corrections are respected in this analysis, one finds that the discretization errors actually start
at order $O(a)$ in the interacting case~\cite{Sheikholeslami:1985ij}. $O(a)$-improvement can then only be achieved by a 
modification of the fermion lattice action. It was found that at least on-shell 
$O(a)$-improvement can be established through the extension of the action by the so-called Sheikholeslami-Wohlert 
term~\cite{Sheikholeslami:1985ij}. 

Here, however, we will no further address the question of the convergence properties of the introduced 
fermion action in the continuum limit, since it will be seen shortly, that it actually suffers from a more severe defect. 
This defect becomes most obvious when considering the fermion propagator in momentum
space, which is given as
\beq
\label{eq:FermionPropagatorFreeTheory1}
\tilde G_f(p) = \langle \tilde f_p \tilde {\bar f}_{p} \rangle
\eeq
with
\bea
\tilde f_p = \frac{1}{\sqrt{V}} \sum\limits_x e^{-ipx} f_x &\mbox{ and }&
\tilde {\bar f}_p = \frac{1}{\sqrt{V}} \sum\limits_x e^{ipx} \bar f_x 
\eea
where $f\in\{t,b \}$ selects the quark flavour, $\tilde f$ and $\tilde {\bar f}$ denote the Fourier transforms
of the respective quark fields, $\tilde G_f(p)$ is a $4\times 4$ matrix, 
and $p\in \ImpSpace_\Gamma$ is a four-component lattice momentum. For the massless case in the non-interacting theory
the fermion propagator is easily obtained by the rules of Grassmann integration yielding
\beq
\tilde G_f(p) = \frac{-i \gamma_\mu \tilde p_\mu }{ \tilde p^2},
\eeq
where $i\tilde p_\mu$ denotes the eigenvalue of the symmetrized lattice derivative $\nabla^s_\mu$ associated 
to the eigenvector $\exp(ipx)$ according to
\bea
\label{eq:DefOfLatticeMomTildeP}
\nabla^s_\mu e^{ip\cdot x} = i\tilde p_\mu \cdot e^{ip\cdot x} &\mbox{ with }& 
\tilde p_\mu \equiv \frac{1}{a}\sin(ap_\mu) = p_\mu + O(a^2).
\eea

The main observation is that the free lattice propagator $G_f(p)$ matches its continuum counterpart
for small momenta, since one has $\tilde p_\mu \rightarrow p_\mu$ in the continuum limit $a\rightarrow 0$.
In particular, the propagator in \eq{eq:FermionPropagatorFreeTheory1} has a pole at $p=0$ in agreement
with the considered fermion being massless. However, the additional zeros of the sine at the edges of 
the Brillouin zone induce 15 additional poles in the propagator that give contributions
to any computed observable as if there were 15 additional fermions present in the theory. This phenomenon
is called 'fermion doubling' and is directly related to the application of the symmetrized derivative
operator, which had to be chosen due to the aforementioned constraints.

A widely spread approach aiming at the suppression of the contributions of the unwanted fermion doublers
in the continuum limit is the introduction of the so-called Wilson term into the fermion lattice action.
Adopting this strategy to the Higgs-Yukawa model the fermion lattice action becomes
\bea
\label{eq:DefOfEuclideanLatticeFermionActionWithWilsonTerm}
S_{\Gamma,F}[\varphi, t, \bar t, b, \bar b] &=& a^4 \sum_{x, \mu}
\bar t_{x}\, \Dw t_{x} 
+ \bar b_{x} \Dw b_{x}\\
&+& a^4 \sum\limits_x\, y_b 
\left(\bar t_{x}, \bar b_{x} \right)_L \varphi_x b_{R,x} 
+ y_t \left(\bar t_{x}, \bar b_{x} \right)_L \tilde\varphi_x t_{R,x} + c.c.,\quad\nonumber
\eea
with the pioneering Wilson Dirac operator $\Dw$ given as
\beq
\label{eq:DefOfWilsonOperator}
\Dw = \sum\limits_\mu \gamma_\mu \nabla^s_\mu - a\frac{r}{2} \nabla^b_\mu\nabla^f_\mu,
\eeq
where the so-called Wilson parameter $r$ is usually chosen as $r=1$, which is a reasonable choice as discussed
in \Ref{Sheikholeslami:1985ij}.

The effect of the Wilson term can again most obviously be studied by considering the free fermion
propagator, which now becomes
\bea
\tilde G_f(p) &=& \frac{-i\gamma_\mu \tilde p_\mu +M(p)}{\tilde p^2 + \left[M(p)\right]^2}, \\
M(p) &=& \frac{2r}{a} \sum\limits_\mu \sin^2(ap_\mu/2).
\eea
The free propagator still has 15 additional, unphysical poles. The corresponding quasi particles, however, 
have now acquired a mass given by a constant times $2r/a$. In the continuum limit the unwanted fermion 
doublers thus become infinitely massive and their contribution to any observable vanishes.

The inclusion of the Wilson term thus solves the fermion doubling problem. The price one pays when adopting
this approach is twofold. Most obviously, the inclusion of the Wilson term induces $O(a)$ lattice artefacts
already at tree-level. More interesting from a conceptual point of view is, however, the loss of chiral symmetry, 
meaning that the standard chiral symmetry relation
\beq
\label{eq:StandardChiralSymRel}
\Dgen \gamma_5 + \gamma_5 \Dgen = 0,
\eeq
where $\Dgen$ is some Dirac operator does not hold for the Wilson Dirac operator $\Dw$,
since the Wilson-term does not anti-commute with $\gamma_5$. 

In fact, the celebrated Nielsen-Ninomiya-theorem~\cite{Nielsen:1980rz,Nielsen:1981xu,Nielsen:1981hk,Friedan:1982nk}
excludes the possibility of constructing a real, bilinear lattice fermion action that is local and translational
invariant while obeying the chiral symmetry relation in \eq{eq:StandardChiralSymRel} and lifting the unwanted 
fermion doublers at the same time. Any approach of investigating chiral symmetry on the lattice thus has to 
abandon at least one of the aforementioned requisites. This will be discussed in more detail in 
\sect{sec:NeubergeOperator}.

%--------------------------------------------------------------------------------------------------------------------------
\subsection{Continuum limit}
\label{sec:LatDiscOfContinuumLimit}

As already pointed out the notion 'continuum limit' does not simply refer to sending
the lattice spacing $a$ to zero. In fact, physical observables in interacting theories do not even
converge to finite results in general, when the lattice spacing $a$ is sent to zero, while the bare 
parameters are held constant. This observation is closely connected to the general appearance of 
infinities in bare perturbation theory of unregulated, interacting theories, since the introduction
of a finite spaced lattice is nothing else but a regularization of the considered theory and the limit
$a\rightarrow 0$ directly corresponds to its removal. This lattice regularization restricts the set of
respected momenta to the  allowed lattice momenta $p\in\ImpSpace$ given in \eq{eq:DefOfAllowedLatticeMomenta}.
The lattice spacing $a$ is thus manifestly related to a momentum cutoff parameter $\Lambda$.

To avoid confusion it is remarked that the lattice regularization does not exactly correspond to a hard momentum
cutoff. While the discrete lattice momenta $p$ themselves are nominally cut off at the edges of the Brillouin zone
according to $|p_\mu|<\pi/a$, their respective contribution to physical observables already deviates from 
the targeted continuum contribution at significantly smaller momenta, which becomes apparent, for instance,
in the analytical expressions for the bare fermion and bosonic propagators considered in the preceding sections.
In this respect the nominal threshold value $\pi/a$ can actually not be considered as a hard momentum
cutoff and a physically meaningful cutoff parameter can in fact not uniquely be defined. Instead, only its
scale is physically meaningful. Here and in the following we will choose the definition of the cutoff parameter 
to be $\Lambda=1/a$. The underlying ambiguity, however, will later become an issue when trying to compare the 
eventually derived cutoff-dependent mass bounds with results obtained in other regularizations.

To establish a meaningful continuum limit the bare parameters thus have to be varied along with sending 
the lattice spacing $a$ to zero. Following the Wilsonian picture of renormalization~\cite{Wilson:1973jj},
this can be achieved by choosing a number of linearly independent physical observables $O_1,\ldots,O_N$ equal to the 
number $N$ of bare parameters available in the considered theory, the expectation values of which are kept
constant by suitable adjustments of the bare parameters, while sending the lattice spacing $a$ to zero. 
Through this procedure one defines trajectories of
constant physics in terms of the selected physical observables $O_1,\ldots,O_N$ in bare parameter space spanned by 
the bare parameters and the lattice spacing $a$. 

While the tuned expectation values $\langle O_1\rangle,\ldots,\langle O_N\rangle$ converge to finite
values with $a\rightarrow 0$ by construction, provided that the aforementioned trajectories exist at all and can be extended down to 
arbitrarily small lattice spacings $a>0$, such a convergent behaviour is not automatically guaranteed
for all other observables. In fact, this is a property of the considered theory. For the purpose of categorization,
the notion of 'non-perturbative renormalizability' refers to all those theories, in which for any point in bare 
parameter space a trajectory of constant physics can be constructed by a suitable choice of $O_1,\ldots, O_N$ 
and extended down to arbitrarily small lattice spacings $a>0$, such that the expectation values of all physical 
observables - not only of the tuned ones - converge to finite values when sending the lattice spacing $a$ to 
zero along the constructed trajectory. In such a non-perturbatively renormalizable theory a meaningful continuum limit
can thus be established by adjusting the bare parameters to a trajectory of constant physics while sending $a\rightarrow 0$.

In order to extract actual continuum physics as observed in Nature one would moreover have to select that specific
trajectory of constant physics that reproduces phenomenology. This is done in practice by tuning the expectation values
$\langle O_1\rangle,\ldots,\langle O_N\rangle$ of the selected observables to their respective phenomenologically known
values. 

In the case of the pure Higgs-Yukawa sector one has four free bare parameters at hand for a given lattice spacing $a$, if one respects 
only the heaviest fermion doublet, \ie the top-bottom doublet. These are the top and bottom Yukawa coupling constants $y_t$, $y_b$, the  
scalar mass $m_0^2$ and the quartic self-coupling constant $\lambda$. The phenomenologically relevant trajectory of 
constant physics can then, for instance, be constructed by tuning these bare parameters to reproduce the phenomenological top and 
bottom quark masses, the phenomenological value of the vev, and a constant Higgs boson mass. Since the Higgs boson mass
is not determined yet, a one-dimensional freedom in the fixation of the phenomenologically relevant trajectory is left
open, which can, for instance, be parametrized in terms of the quartic self-coupling parameter $\lambda$ at a given lattice 
spacing $a$. This is exactly the approach that will be applied in the present work.

For clarification it is remarked that the pure Higgs-Yukawa sector of the Standard Model is actually not considered
as a non-perturbatively renormalizable theory due to the triviality of the Higgs sector, meaning that the trajectories of
constant physics can not be extended to arbitrary small $a$. The lattice spacing $a$, and thus the cutoff $\Lambda$, can therefore
not be removed from the theory and any physical results thus have to depend on this scale. This is the mathematical 
manifestation of the Higgs-Yukawa sector being an effective theory only valid up to some cutoff $\Lambda$. The strategy for
the extraction of the cutoff dependent results is nevertheless the same, which is to construct trajectories of constant physics 
compatible with phenomenology and to extend them as far as possible. The cutoff dependent result is then taken from the constructed 
trajectory at the specified value of the cutoff $\Lambda$, \ie at the specified lattice spacing $a$.

In a practical implementation, however, it is useful to exploit the fact that all appearances
of the lattice spacing $a$ can be completely removed from the evaluation of the functional integral in \eq{eq:DefOfFuncIntOnLattice3}
through a rescaling of the field variables and bare parameters according to
\bea
\label{eq:DefOfDimLessParameters}
\breve m_0 = a \cdot m_0,\quad&
\breve y_{t,b} = y_{t,b},\quad&
\breve \lambda = \lambda, \\
\label{eq:DefOfDimLessParameters2}
\breve\varphi_x = a\cdot \varphi_x, \quad&
\breve t, \breve{\bar t} = a^{3/2} t, {\bar t},\quad&
\breve b, \breve{\bar b} = a^{3/2} b, {\bar b}.
\eea
The expectation value of the dimensionless observable $O(\breve\varphi,\breve t,\breve {\bar t},\breve b,\breve{\bar b})$, from 
which the actually targeted expectation value $\langle O(\varphi,t,{\bar t},b,{\bar b})\rangle$ can directly be obtained
by a trivial rescaling, is then given by the dimensionless expression
\bea
\label{eq:DefOfFuncIntOnLatticeDimensionless}
\langle O(\breve\varphi,\breve t,\breve {\bar t},\breve b,\breve {\bar b})\rangle_{\Gamma,0} &=& 
\frac{1}{Z_{\Gamma,0}} \FuncIntAvg_{\Gamma,0}[O(\breve\varphi,\breve t,\breve {\bar t},\breve b,\breve {\bar b})], \quad 
Z_{\Gamma,0} = \FuncIntAvg_{\Gamma,0}[1],
\eea
where the functional integral $\FuncIntAvg_{\Gamma,0}$ is defined as
\bea
\FuncIntAvg_{\Gamma,0}[O(\breve\varphi,\breve t,\breve {\bar t},\breve b,\breve {\bar b})] &\fhs{-4mm}=\fhs{-4mm}& 
\int \prod_{x} \intd{\breve \varphi_x}\intd{\breve \varphi^\dagger_x} 
\intd{\breve t_x} \intd{\breve {\bar t_x}}\intd{\breve b_x} \intd{\breve {\bar b_x}}  \,\,
O(\breve \varphi, \breve t, \breve {\bar t}, \breve b, \breve {\bar b}))\cdot e^{-\breve S_{\Gamma}[\breve \varphi, \breve t, \breve {\bar t}, \breve b,
\breve {\bar b}]}.\quad\quad\quad
\eea
In the above equation the total action $\breve S_{\Gamma} = \breve S_{\Gamma,\varphi}+\breve S_{\Gamma,F}$  is expressed in terms of the 
rescaled quantities according to
\bea
\label{eq:BosonicLatticeHiggsActionContNotDimensionless}
\breve S_{\Gamma,\varphi}[\breve \varphi] &=& \sum\limits_{x} 
\frac{1}{2} \breve\nabla^f_\mu \breve \varphi^\dagger_x
\breve\nabla^f_\mu \breve \varphi_x
+ \frac{1}{2} \breve m_0^2 \breve \varphi^\dagger_x\breve \varphi_x
+\breve\lambda\left(\breve \varphi_x^\dagger \breve \varphi_x \right)^2, \\
\label{eq:DefOfEuclideanLatticeFermionActionWithWilsonTermDimensionless}
\breve S_{\Gamma,F}[\breve \varphi, \breve t, \breve {\bar t}, \breve b, \breve {\bar b}] &=& \sum_{x, \mu}
\breve {\bar t}_{x}\, \breve \Dw \breve t_{x} 
+ \breve {\bar b}_{x} \breve \Dw \breve b_{x}\\
&+&  \sum_x  \breve y_b 
\left(\breve {\bar t}_{x}, \breve {\bar b}_{x} \right)_L \breve \varphi_x \breve b_{R,x} 
+ \breve y_t \left(\breve {\bar t}_{x},\breve {\bar b}_{x} \right)_L \breve {\tilde\varphi}_x\breve  t_{R,x} + c.c.,\quad\nonumber
\eea
where the dimensionless lattice operators are given as
\bea
\label{eq:DefOfLatDiffOperators}
\breve \Dw = a\Dw, \quad&
\breve\nabla^f_\mu = a\nabla^f_\mu, \quad&
\breve\nabla^b_\mu = a\nabla^b_\mu.
\eea

The technical advantage that arises from the fact that the lattice spacing $a$ can be removed by introducing a set of rescaled fields
and bare parameters is that the required tuning of the $N$ bare parameters, as described above, becomes simplified. This is because
the lattice spacing can directly be reintroduced into the dimensionless system in \eq{eq:DefOfFuncIntOnLatticeDimensionless} by demanding 
that one expectation value of the selected observables, for instance $\langle O_1\rangle$, exactly matches its phenomenological value. 
Practically speaking, one then only needs to fine-tune the bare parameters with respect to the reduced set of the $N-1$ remaining 
conditions, matching $\langle O_2\rangle, \ldots, \langle O_N\rangle$ to their phenomenological values. In the case of the considered 
Higgs-Yukawa sector one can, for instance, reintroduce the lattice spacing $a$ by demanding the vev $v$ to match its phenomenologically known 
value of $\GEV{246}$. The remaining task would then be to fine-tune the Yukawa coupling constants to reproduce the physical quark 
masses, while the bare quartic coupling constant would stay a free parameter, as discussed above. This procedure will be presented 
in more detail in \sect{sec:SimStratAndObs}.

For practical lattice calculations it is therefore very convenient to consider the dimensionless expression
given in \eq{eq:DefOfFuncIntOnLatticeDimensionless}, which will be the actual starting point of the later
lattice computations. The construction of the above described continuum limit then requires the existence 
of a critical point in the dimensionless system where all correlation lengths diverge. Such a critical point
is necessary to allow for holding the physical observables constant in the limit $a\rightarrow 0$ according to 
the underlying scaling relations in \eqs{eq:DefOfDimLessParameters}{eq:DefOfDimLessParameters2}. The same reasoning moreover 
requires the associated phase transition to be of second order. Knowing the phase structure of the considered model 
is therefore an important prerequisite for the construction of the continuum limit. For the later introduced lattice
Higgs-Yukawa model the phase structure will therefore be investigated in detail in \chap{chap:PhaseDiagram}.

Finally, it is remarked here that for the sake of brevity the notation will be shortened by 
discarding the hat-symbols on top of all field variables, operators, and parameters in 
\eqs{eq:DefOfDimLessParameters}{eq:DefOfLatDiffOperators} 
as well as the subscripts $\Gamma$, and $0$ from now on.

  \chapter{The chirally invariant \texorpdfstring{$\mbox{SU(2)}_L\times\mbox{U(1)}_Y$}{SU(2) x U(1)} lattice Higgs-Yukawa model}
\label{chap:TheModel}

As pointed out in the introduction the earlier non-perturbative investigations of lattice
Higgs-Yukawa models suffered from their inability to restore chiral symmetry in the
continuum limit while lifting the unwanted fermion doublers at the same time.  
Since then, however, several new approaches for the construction of chirally invariant lattice
fermions have been developed~\cite{Kaplan:1992bt,Shamir:1993zy,Neuberger:1997fp,Neuberger:1998wv,Luscher:1998pq}. 
These new fermion formulations allow for a consistent formulation 
of the chiral Higgs-fermion coupling structure of the Standard Model on the lattice while simultaneously 
lifting the fermion doublers, thus eliminating manifestly the main objection to the earlier investigations. 

The lattice Higgs-Yukawa model that is investigated in the present work is based on the Neuberger overlap 
operator~\cite{Neuberger:1997fp,Neuberger:1998wv} which is a solution of the Ginsparg-Wilson relation~\cite{Ginsparg:1981bj}. 
The properties of this operator will be briefly discussed in \sect{sec:NeubergeOperator}. The first chirally 
invariant lattice Higgs-Yukawa model, obeying an $\UVUA$ lattice chiral symmetry, was originally introduced by 
L\"uscher~\cite{Luscher:1998pq}. It will be presented in \sect{sec:modelDefinitionU1VU1A} in some reformulation being 
intuitively better accessible. Based on these considerations it is a small step to establish the Higgs-Yukawa 
model that will actually be studied in this work obeying a lattice \SUU chiral symmetry. This will be done 
in \sect{sec:modelDefinition}.

The main strategy for approaching the eventual aim of the present work, which is the non-perturbative
determination of upper and lower Higgs boson mass bounds, will be discussed in \sect{sec:SimStratAndObs}.
Most of the observables investigated in the rest of this thesis are also explicitly defined there.

A first account on the lattice techniques used to evaluate the considered Higgs-Yukawa model numerically 
is then given in \sect{sec:HMCAlgorithm}. Serving as a starting point for the numerous algorithmic enhancements,
which will be presented in detail in \chap{chap:SimAlgo}, a standard implementation of an HMC-algorithm~\cite{Duane:1987de,Gottlieb:1987mq} is 
given at that point, which will also extensively be used in \chap{chap:PhaseDiagram} for the numerical evaluation of 
the phase diagram of the considered Higgs-Yukawa model. 

For the later numerical treatment of the model the properties of the fermion determinant play a vital role,
in particular the question, whether and, if yes, how much its complex phase fluctuates during the Markov process
of generating field configurations. The complex phase of the fermion determinant associated to the considered 
Higgs-Yukawa model is therefore discussed in \sect{sec:ComplexPhaseOfFermionDet}.

%--------------------------------------------------------------------------------------------------------------------------
\section{The chirally invariant Neuberger Dirac operator}
\label{sec:NeubergeOperator}

It is well known that the Nielsen-Ninomiya-theorem~\cite{Nielsen:1980rz,Nielsen:1981xu,Nielsen:1981hk,Friedan:1982nk}
excludes the possibility of constructing a real, bilinear lattice fermion action that is local and translational
invariant while obeying the standard chiral symmetry relation given in \eq{eq:StandardChiralSymRel} and lifting 
the unwanted fermion doublers at the same time. Any approach of investigating chiral symmetry on the lattice thus 
has to abandon at least one of the aforementioned requisites.

A nowadays very commonly used ansatz for studying chiral properties on the lattice is based on the 
observation~\cite{Ginsparg:1981bj} that the Nielsen-Ninomiya-theorem can be circumvented by replacing the continuum chiral 
symmetry in \eq{eq:StandardChiralSymRel} with a lattice modified version thereof, that recovers the actual
symmetry only in the continuum limit. The construction of lattice theories exactly obeying this modified 
chiral symmetry is then not excluded by the Nielsen-Ninomiya-theorem. The great advantage of this approach lies 
in the existence of an exactly preserved symmetry established at any lattice spacing even far away from the 
continuum limit. It is the form of this exact symmetry itself that converges to the continuum chiral symmetry 
in the continuum limit.

More precisely, it was found that any operator $\Dgen$ that satisfies the Ginsparg-Wilson relation~\cite{Ginsparg:1981bj}
\beq
\label{eq:GinspargWilsonRel}
\Dgen\gamma_5 + \gamma_5 \Dgen = a  \Dgen\gamma_5 R \Dgen,
\eeq
where $a$ denotes the lattice spacing and $R$ is some hermitian, positive definite operator, being local in position space
and proportional to the identity in Dirac space, then obeys the lattice modified chiral symmetry relation
\beq
\gamma_5\Dgen + \Dgen\hat\gamma_5 = 0
\eeq
with the modified $\gamma_5$-matrix
\beq
\label{eq:DefOfGammaHat}
\hat \gamma_5 = \gamma_5 \left( 1-a R\Dgen \right).
\eeq
It can directly be seen from the appearance of the lattice spacing $a$ on the right hand side of \eq{eq:DefOfGammaHat}
that $\hat\gamma_5$ converges to $\gamma_5$ in the continuum limit, thus recovering the continuum chiral
symmetry for $a\rightarrow 0$ and one easily verifies that any operator of the form
\beq
\Dgen = \frac{1}{a}R^{-1} \left(\ID - V\right)
\eeq
solves the Ginsparg-Wilson relation in \eq{eq:GinspargWilsonRel} provided that the operator $V$ satisfies
\bea
V^\dagger V = \ID &\mbox{ and }& V^\dagger = \gamma_5 V \gamma_5.
\eea
This, however, does not guarantee the resulting Dirac operator $\Dgen$ to be free of fermion doublers. 

A specific solution of the Ginsparg-Wilson relation for which it could explicitly be shown that it lifts the unwanted 
fermion doublers was presented in \Ref{Neuberger:1998wv}. This so-called Neuberger overlap operator is given as
\bea
\label{eq:DefOfNeuberDiracOp}
\SD &=& \frac{\rho}{a}\left\{1+\frac{ A}{\sqrt{ A^\dagger  A}}   \right\},
\quad A = \SDw - \frac{\rho}{a}, \quad 0 < \rho < 2r
\eea
where $\rho$ is a free, dimensionless parameter within the specified constraints that is related to the operator $R$ in 
\eq{eq:GinspargWilsonRel} through $R= \rho^{-1}\ID$ and $\SDw$ denotes here the Wilson operator as defined in \eq{eq:DefOfWilsonOperator}.
Furthermore, the given operator was proven to be local in a field theoretical sense also in the presence of QCD gauge fields
which, however, are not considered here, at least if the latter fields underlying the considered Dirac operator obey certain smoothness 
conditions~\cite{Hernandez:1998et}. The locality properties were then found to depend on the parameter $\rho$ and the strength
of the gauge coupling constant~\cite{Hernandez:1998et}. At vanishing gauge coupling the most local operator was shown to be 
obtained at $\rho=1$. Here, the notion 'most local' has to be understood in the sense of the most 
rapid exponential decrease with the distance $|x-y|$ of the coupling strength induced by the matrix elements $\SD_{x,y}$ 
between the field variables at two remote space-time points $x$ and $y$. For that reason the setting $\rho=1$
will be adopted for the rest of this work.

The locality properties, the successful lifting of the unwanted fermion doublers and the existence of an exact,
though modified, chiral symmetry make the Neuberger operator an excellent candidate for the 
construction of chiral theories on the lattice. The downside of this operator, however, are the tremendous numerical
costs required for its application on a given vector, which mainly result from the appearance of the inverse square root 
of the hermitian operator $A^\dagger A$ in \eq{eq:DefOfNeuberDiracOp}. Much effort has therefore been spent on developing 
more efficient algorithms for performing lattice calculations with dynamical fermions based on the Neuberger overlap 
operator~\cite{vandenEshof:2002ms, Arnold:2003sx,Cundy:2004pza,Cundy:2005pi}.

In the present work aiming at the investigation of the pure Higgs-Yukawa sector, however, the situation is much different, 
since no gauge fields are included in this sector of the Standard Model.
It is the absence of gauge fields that makes the eigenvectors and eigenvalues of the Neuberger 
operator analytically available. In momentum space with the allowed four-component momenta $p\in\ImpSpace$
the eigenvectors of the doublet Dirac operator $\D_{8\times 8}=\diag(\SD,\SD)$ that will actually be used in the 
construction of the Higgs-Yukawa model in the following are given as
\beq
\label{eq:DefOfEigenvectorsOfD1}
\Psi_x^{p,\zeta\epsilon k} = e^{i p \cdot x} \cdot u^{\zeta\epsilon k}(p),\quad
u^{\zeta\epsilon k}(p) = \sqrt{\frac{1}{2}}
\left(
\begin{array}{*{1}{c}}
u^{\epsilon k}(p) \\
\zeta u^{\epsilon k}(p) \\
\end{array}
\right), \quad
\zeta=\pm 1, \,
\epsilon=\pm 1, \,
k \in \{1,2\}
\eeq
with $u^{\epsilon k}(p)$ denoting the usual four-component spinor structure 
\beq
\label{eq:DefOfEigenvectorsOfD2}
u^{\epsilon k}(p) = 
\sqrt{\frac{1}{2}}
\left(
\begin{array}{*{1}{c}}
\xi_k \\
\epsilon\frac{\tilde p \bar\theta}{\sqrt{\tilde p^2}} \xi_k 
\end{array}
\right)
\,
\mbox{for }\tilde p\neq 0
\quad
\mbox{and}
\quad
u^{\epsilon k}(p) = 
\sqrt{\frac{1}{2}}
\left(
\begin{array}{*{1}{c}}
\xi_k \\
\epsilon \xi_k \\
\end{array}
\right)
\,
\mbox{for }\tilde p= 0,
\eeq
where the lattice momenta $\tilde p$ have been defined in \eq{eq:DefOfLatticeMomTildeP}. 
Here $\xi_k\in\Comp^2$ are two orthonormal vectors and the four component 
quaternionic vectors $\theta$, $\bar\theta$ are defined as $\theta = (\ID, -i\vec\tau)$ and 
$\bar\theta = (\ID, +i\vec\tau) = \theta^\dagger$ with $\vec\tau$ denoting the vector of Pauli matrices. 
The corresponding eigenvalues are then given as 
\bea
\label{eq:eigenValOfFreeND}
\nu^\epsilon(p)&=& \frac{\rho}{a} + \frac{\rho}{a}\cdot\frac{\epsilon i\sqrt{\tilde p^2} + 
a\frac{r}{2}\hat p^2 - \frac{\rho}{a}}{\sqrt{\tilde p^2 + (a\frac{r}{2}\hat p^2 -
\frac{\rho}{a})^2}},
\eea
where the squared lattice momenta $\hat p^2$ have been given in \eq{eq:DefOfSquaredLatticeMomHatP}.
Here $r$ denotes the coefficient of the Wilson term in the underlying Wilson Dirac operator and
$\epsilon=\pm 1$ equals the sign of the imaginary part of the complex eigenvalues according to 
$\mbox{Im}[\nu^\pm(p)] \gtrless 0$. The eigenvalues $\nu^{\epsilon}(p)$ thus form a circle in the 
complex plane, the radius of which is given by the parameter $\rho/a$. 

It is this analytic availability of the eigenvalues and eigenvectors that allows for an extremely efficient 
construction of the Neuberger overlap operator in momentum space as will be discussed in \sect{sec:HMCAlgorithm}.

%--------------------------------------------------------------------------------------------------------------------------
\section{A first \texorpdfstring{$\UVUA$}{U(1) x U(1)} lattice Higgs-Yukawa model}
\label{sec:modelDefinitionU1VU1A}

The first formulation of a chirally invariant lattice Higgs-Yukawa model has been given by L\"uscher~\cite{Luscher:1998pq}. It was shown in that 
work that a chirally invariant lattice theory can be constructed starting from any Dirac operator that satisfies the Ginsparg-Wilson relation 
in \eq{eq:GinspargWilsonRel} and that the resulting model is free of fermion doublers at least for sufficiently small coupling constants provided 
that the underlying Dirac operator possesses that property. This first formulation of an $\UVUA$ chirally invariant lattice Higgs-Yukawa model was
given as\footnote{With some abuse of notation the specification of the partition function is taken here and in the following as a shorthand
notation for defining a lattice model. More precisely one would have to specify the functional integral $\FuncIntAvg[O]$
in terms of which the expectation value of an observable is given according to $\langle O \rangle = \FuncIntAvg[O]/Z$, $Z=\FuncIntAvg[1]$.}
\bea
\label{eq:LuescherPartitionFunc}
Z &=& \int \intD{\phi}\intD{\phi^\dagger} \intD{\psi} \intD{\bar\psi} \intD{\chi} \intD{\bar\chi} e^{ -S_\phi - S_Y  -S_F^{kin} } \\
\label{eq:DefLuescherYukawaCouplingTerm}
S_Y &=& a^4\sum\limits_{x} \hat y (\bar\psi_x+\bar\chi_x) \left[P_-\phi_x + P_+\phi_x^\dagger  \right] (\psi_x+\chi_x) \\
S_F^{kin} &=& a^4\sum\limits_{x,y}\bar\psi_x \Dgen_{x,y} \psi_y - \frac{2}{a}\bar\chi_x R_{x,y}^{-1}\chi_y
\eea
where $\phi$ denotes a one-component complex scalar field, $\hat y$ is the Yukawa coupling constant, and $\psi$, $\chi$ are Dirac fields. 
The left- and right-handed projection operators $P_\pm$ are moreover given in the usual manner according to
\bea
P_\pm &=& \frac{1\pm \gamma_5}{2}.
\eea
The Higgs action $S_\phi$ is not further specified, except for the requirements that it solely depends on the field $\phi$ and that 
it is invariant under global $U(1)$ transformations of the scalar field $\phi$. The lattice $\Phi^4$-action would thus be an eligible
candidate for $S_\phi$. The particular choice of the Dirac operator $\Dgen$ is also left open with the only requirement of being a solution 
to the Ginsparg-Wilson relation. 

For clarification it is remarked that the field $\chi$ does not propagate and that it does not have any
physical interpretation at all. It was only introduced in \Ref{Luscher:1998pq} as an auxiliary tool facilitating the
construction of the chirally invariant coupling structure among the fermionic and bosonic fields. This auxiliary
field is introduced here in such a way that the sum $\psi+\chi$ transforms according to the continuum chiral symmetry.

One can then easily verify that this model obeys a global $U(1)_V$ symmetry
as well as an exact, but lattice modified, $U(1)_A$ chiral symmetry according to
\bea
\label{eq:LatticeChiralTrans1}
\delta\psi=i\epsilon\left[\gamma_5\left(1-\frac{a}{2}R\Dgen \right)\psi + \gamma_5\chi\right], & 
\delta\chi = i\epsilon\gamma_5\frac{a}{2}R\Dgen\psi,& 
\delta\phi =2i\epsilon\phi,    \\
\label{eq:LatticeChiralTrans2}
\delta\bar\psi=i\epsilon\left[\bar\psi\left(1-\frac{a}{2}\Dgen R\right)\gamma_5 + \bar\chi\gamma_5\right], & 
\delta\bar\chi = i\epsilon\bar\psi\frac{a}{2}\Dgen R\gamma_5,& 
\delta\phi^\dagger =-2i\epsilon\phi^\dagger,    
\eea
where $\epsilon\in \re$ is some infinitesimally small parameter. The given symmetry then recovers the actual continuum 
symmetry in the limit $a \rightarrow 0$.

A more enlightening representation of the partition function given in \eq{eq:LuescherPartitionFunc} can be obtained by integrating
out the fermion fields $\psi$ and $\chi$ leading to 
\beq
Z = \int \intD{\phi}\intD{\phi^\dagger} \det\left(\Dgen + \hat y \left(P_-\phi+P_+\phi^\dagger\right)\left[\ID - \frac{a}{2}R \Dgen \right]  \right) \cdot e^{ -S_\phi} 
\eeq
up to a constant factor. Here the expressions $P_-\phi$ and $P_+\phi^\dagger$ have actually to be understood as the space-time-diagonal matrix 
$\diag(P_-\phi_x)$ and $\diag(P_+\phi_x^\dagger)$. This type of shorthand notation will extensively be used in the following.
Starting at the given representation one can make the left- and right-handed coupling structure of the theory
more explicit and thus intuitively better understandable by defining the modified left- and right-handed projection operators 
$\hat P_{\pm}$ as well as the modified $\gamma_5$-matrix $\hat\gamma_5$ according to
\bea
\label{eq:GeneralDefOfModifedProjectors}
\hat P_\pm = (1 \pm \hat \gamma_5)/2, \quad & &
\hat\gamma_5 = \gamma_5 \left(\ID - a R \Dgen \right).
\eea
Employing the Ginsparg-Wilson relation these modified operators can be shown to satisfy the relations
\beq
\hat P_{\pm} + \hat P_{\mp} = \ID, \quad 
\left[ \hat P_\pm \right]^2 = \hat P_\pm,  \quad
\hat P_\pm \hat P_\mp = 0, \quad 
\left[\hat \gamma_5\right]^2 =\ID, \quad 
\hat \gamma_5 \hat P_\pm = \pm \hat P_\pm, 
\eeq
and 
\bea
\label{eq:ProjectorCombination}
P_\pm \Dgen = \Dgen \hat P_\mp,
&&
P_\pm\hat P_\pm = P_\pm \left( \ID-\frac{a}{2} R \Dgen \right),
\eea
allowing to rewrite the partition function after reintroducing the fermion fields $\psi$, $\bar\psi$ in the 
equivalent\footnote{The notion of two partition functions $Z_1$ and $Z_2$ being equivalent expresses that the 
expectation values of any observable $O$ determined by the functional integrals $\FuncIntAvg_1$, $\FuncIntAvg_2$
associated to the given partition functions according to $\langle O \rangle_1 = \FuncIntAvg_1[O]/Z_1$
and $\langle O \rangle_2 = \FuncIntAvg_2[O]/Z_2$ are identical. Here the partition function in \eq{eq:LuescherPartitionFuncReformulated}
is equivalent to the one in \eq{eq:LuescherPartitionFunc}, if one restricts the consideration to physical observables,
\ie observables not depending on $\chi$. Analogously, this notion of equivalence is also used for the underlying actions $S_1$ and $S_2$.}
but physically more enlightening form
\bea
\label{eq:LuescherPartitionFuncReformulated}
Z &=& \int \intD{\phi}\intD{\phi^\dagger} \intD{\psi} \intD{\bar\psi}\,\, e^{ -S_\phi -S_F }, \\
\label{eq:LuescherActionReformulated}
S_F &=&  a^4\sum\limits_{x,y} \bar\psi_x \left(\Dgen_{x,y} + \hat y\left[P_-\phi \hat P_- + P_+\phi^\dagger \hat P_+ \right]_{x,y}  \right)\psi_y,
\eea
which no longer contains the auxiliary field $\chi$. The obvious physical interpretation of this model is that 
the left- and right-handed modes of the field $\psi$ are given here by the expressions $\hat P_-\psi$ and $\hat P_+\psi$
based on the modified projection operators, while the left- and right-handed modes of $\bar\psi$ are still defined
via the original projection operators $P_\pm$~\cite{Niedermayer:1998bi,Luscher:1998du}. 

Finally, the chiral symmetry of the model can also be written in a more enlightening form exploiting the definition of the
above projection operators. One finds that the model is invariant under the $\ULUR$ transformation
\bea
\label{eq:ChiralSymmetryTrafoU1VU1A1}
\psi\rightarrow  U_R \hat P_+ \psi + U_L \hat P_- \psi,
&\quad&
\bar\psi\rightarrow  \bar\psi P_+ U_L^\dagger + \bar\psi P_- U^\dagger_R, \\
\label{eq:ChiralSymmetryTrafoU1VU1A2}
\phi \rightarrow  U_R  \phi U_L^\dagger,
&\quad&
\phi^\dagger \rightarrow U_L \phi^\dagger U_R^\dagger,
\eea
where $U_{L,R}\in \mbox{U}(1)$ are arbitrarily chosen.

%--------------------------------------------------------------------------------------------------------------------------
\section{The definition of the \texorpdfstring{\SUU}{SU(2) x U(1)} lattice Higgs-Yukawa model}
\label{sec:modelDefinition}

The model that will actually be considered in the following, is a four-dimensional, chirally invariant 
$SU(2)_L \times U(1)_Y$ lattice Higgs-Yukawa model based on the Neuberger overlap operator, aiming at the 
implementation of the chiral Higgs-fermion coupling structure of the pure Higgs-Yukawa 
sector of the Standard Model reading
\beq
\label{eq:StandardModelYuakwaCouplingStructure}
L_Y = y_b \left(\bar t, \bar b \right)_L \varphi b_R 
+y_t \left(\bar t, \bar b \right)_L \tilde\varphi t_R  + c.c.,
\eeq
with $\tilde \varphi = i\tau_2\varphi^*$ and $y_{t,b}$ denoting the bare top and bottom Yukawa coupling constants.
In this model the consideration is restricted to the top-bottom doublet $(t,b)$ interacting with the complex Higgs 
doublet $\varphi$, which is a reasonable simplification, since the Higgs dynamics is dominated by the 
coupling to the heaviest fermions. 

The fields considered in this model are one four-component, real scalar field $\Phi$, being equivalent to the
complex doublet $\varphi$ of the Standard Model, and $N_f$ top-bottom
doublets represented by eight-component spinors $\psi^{(i)}\equiv (t^{(i)}, b^{(i)})$, $i=1,...,N_f$.
In this work, the chiral character of the targeted coupling structure in \eq{eq:StandardModelYuakwaCouplingStructure}
will be preserved on the lattice by constructing the fermionic action $S_F$ from the (doublet) Neuberger overlap 
operator $\D_{8\times 8}=\diag(\D,\D)$ which will, however, be denoted as $\D$ in the following for the sake of 
brevity. Starting directly from the formulation in \eq{eq:LuescherActionReformulated}
one can write down a chirally invariant \SUU lattice Higgs-Yukawa model according to
\bea
\label{eq:DefOfModelPartitionFunctionOriginal}
Z &=& \funcInt e^{-S_\Phi[\Phi] - S_F[\Phi,\psi,\bar\psi]}  \quad \mbox{with} \\
\label{eq:DefYukawaCouplingTerm}
S_F[\Phi,\psi,\bar\psi] &=& \sumFL\,
\bar\psi^{(i)} \left[\D + P_+ \phi^\dagger \fhs{1mm}\mbox{diag}\left(\hat y_t,\hat y_b\right) \hat P_+
+ P_- \fhs{1mm}\mbox{diag}\left(\hat y_t,\hat y_b\right) \phi \hat P_-    \right]  \psi^{(i)}, \quad\quad\quad
\eea
where the four-component scalar field $\Phi_x$, defined on the Euclidean site indices $x=(t,\vec x)$ of a $L_s^3\times L_t$-lattice,
was rewritten as a quaternionic, $2 \times 2$ matrix $\phi_x = \Phi_x^\mu \theta_\mu,\, \theta_0=\ID,\, \theta_j =  -i\tau_j$  
with $\vec\tau$ denoting the vector of Pauli matrices, acting on the flavour index of the fermion doublets.
Here and in the following the lattice spacing is assumed to be $a=1$. The lattice spacing can later be reintroduced as discussed 
in \sect{sec:LatDiscOfContinuumLimit}. According to the more general definition in \eq{eq:GeneralDefOfModifedProjectors}
the modified left- and right-handed projection operators $\hat P_{\pm}$ associated to the Neuberger Dirac operator
are given as
\bea
\hat P_\pm = \frac{1 \pm \hat \gamma_5}{2},  & \quad &
\hat\gamma_5 = \gamma_5 \left(\ID - \frac{1}{\rho} \D \right),
\eea
where the free parameter $\rho$ of the overlap operator will be set to one in the actual implementation of the model
as discussed in \sect{sec:NeubergeOperator}. This action now obeys an exact global $\mbox{SU}(2)_L\times \mbox{U}(1)_Y$ 
lattice chiral symmetry. For $\Omega_L\in \mbox{SU}(2)$ and $\epsilon\in \re$ the action is invariant under the transformation
\bea
\label{eq:ChiralSymmetryTrafo1}
\psi\rightarrow  U_R \hat P_+ \psi + U_L\Omega_L \hat P_- \psi,
&\quad&
\bar\psi\rightarrow  \bar\psi P_+ \Omega_L^\dagger U_{L}^\dagger + \bar\psi P_- U^\dagger_{R}, \\
\label{eq:ChiralSymmetryTrafo2}
\phi \rightarrow  U_\phi  \phi \Omega_L^\dagger,
&\quad&
\phi^\dagger \rightarrow \Omega_L \phi^\dagger U_\phi^\dagger
\eea
with $U_{L,R,\phi} \equiv \exp(i\epsilon Y)$ denoting the respective representations of the global $U(1)_Y$ 
symmetry group. Employing the explicit form of the hypercharge being related to the isospin component $I_3$ and the 
electric charge $Q$ according to $Y = Q-I_3$, the above $U(1)_Y$ matrices can explicitly be parametrized as 
\bea
U_L = 
\left(
\begin{array}{*{2}{c}}
e^{+i\epsilon/6}&\\
&e^{+i\epsilon/6}\\
\end{array}
\right),
&
U_R = 
\left(
\begin{array}{*{2}{c}}
e^{2i\epsilon/3}&\\
&e^{-i\epsilon/3}\\
\end{array}
\right),
&
U_{\phi} = 
\left(
\begin{array}{*{2}{c}}
e^{+i\epsilon/2}&\\
&e^{-i\epsilon/2}\\
\end{array}
\right),\quad \quad
\eea
for the case of the considered top-bottom doublet. For clarification it is remarked that the right-handed fields are isospin
singlets and have only been written here in form of doublets for the sake of a shorter notation. Note also that in the mass-degenerate 
case, \ie $\hat y_t=\hat y_b$, the above global symmetry is extended to $\mbox{SU}(2)_L\times \mbox{SU}(2)_R$.
In the continuum limit the symmetry in \eqs{eq:ChiralSymmetryTrafo1}{eq:ChiralSymmetryTrafo2} recovers the continuum 
$\mbox{SU}(2)_L\times \mbox{U}_Y(1)$ global chiral symmetry and the lattice Higgs-Yukawa coupling becomes 
equivalent to \eq{eq:StandardModelYuakwaCouplingStructure} when identifying 
\bea
\varphi_x = 
C\cdot \left(
\begin{array}{*{1}{c}}
\Phi_x^2 + i\Phi_x^1\\
\Phi_x^0-i\Phi_x^3\\
\end{array}
\right),\quad
&
\tilde\varphi_x = i\tau_2\varphi^*_x = 
C\cdot \left(
\begin{array}{*{1}{c}}
\Phi_x^0 + i\Phi_x^3\\
-\Phi_x^2+i\Phi_x^1\\
\end{array}
\right), \quad
&
y_{t,b} = \frac{\hat y_{t,b}}{C} \quad\quad\quad
\eea
for some real, non-zero constant $C$.
 
The formulation in \eqs{eq:DefOfModelPartitionFunctionOriginal}{eq:DefYukawaCouplingTerm} is helpful for 
identifying the chiral symmetries of the model. For its further treatment, however, the fermion degrees of freedom
have to be integrated out. This can explicitly be done by the rules of Grassmann integration. The expectation value
of some observable $O[\Phi, t, \bar t, b, \bar b]$ is then obtained as
\beq
\label{eq:DefOfLatticeObsWithoutPsi}
\langle O[\Phi, t, \bar t, b, \bar b] \rangle = \frac{1}{Z_\Phi}\FuncIntAvg_\Phi[O'[\Phi]], 
\quad Z_\Phi = \FuncIntAvg_\Phi[1],
\eeq
where the remaining functional integral over the field $\Phi$ is given as
\bea
\label{eq:DefOfLatticeFuncIntWithoutPsi}
\FuncIntAvg_\Phi[O'[\Phi]] &=& \int \intD{\Phi}\, O'[\Phi] \cdot \left[\det(\fermiMat[\Phi])\right]^{N_f} \cdot e^{-S_\Phi[\Phi]} \\
\label{eq:DefOfLatticeFuncIntWithoutPsi2}
&=& \int \intD{\Phi}\, O'[\Phi]     \cdot e^{-\Seff[\Phi]},
\eea
with the effective action $\Seff[\Phi]$ being
\bea
\label{eq:DefOfEffAction}
\Seff[\Phi] &=& S_\Phi[\Phi] + \SFeff[\Phi], \\
\label{eq:DefOfEffFermionAction}
\SFeff[\Phi] &=& -N_f\cdot \log\det\left(\fermiMat[\Phi]\right).
\eea
With the help of \eq{eq:ProjectorCombination} the fermion matrix $\fermiMat[\Phi]$ appearing in the above equations 
can then be written as 
\bea
\label{eq:DefOfFermiMatM}
\fermiMat[\Phi] &=& \D + B[\Phi]\GammaOp, \\
\GammaOp &=& \ID - \frac{1}{2\rho}\D, \\
\label{eq:DefOfMatB}
B[\Phi] &=& \diag\left( \hat B(\Phi_x)   \right), \\
\label{eq:DefOfBHat}
\hat B(\Phi_x) &=& \Phi_x^\mu \hat B_\mu, \\
\hat B_\mu &=& P_+\theta_\mu^\dagger \mbox{diag}\left(\hat y_t,\hat y_b\right) + P_- \mbox{diag}\left(\hat y_t,\hat y_b\right)
\theta_\mu,  
\eea
where this rather fragmentary presentation was chosen here to establish the notation that will be used in the forthcoming
analytical considerations. It is remarked that the observable $O'[\Phi]$ 
is different from $O[\Phi, t, \bar t, b, \bar b]$ in general and depends only on the scalar field $\Phi$. Its specific form 
has to be worked out by the rules of Grassmann integration, leading then to the potential appearance of matrix elements of the inverse 
fermion matrix $\fermiMat^{-1}$ in the respective expression for $O'[\Phi]$. It is also remarked, that the symmetry with respect
to the transformation of the scalar field given in \eqs{eq:ChiralSymmetryTrafo1}{eq:ChiralSymmetryTrafo2} is directly inherited 
to the determinant $\det(\fermiMat[\Phi])$. Further properties of the latter determinant, in particular the behaviour of its 
potentially complex phase will be addressed in \sect{sec:ComplexPhaseOfFermionDet}.
 
The so far unspecified purely bosonic action $S_{\Phi}$ is chosen here to be the lattice $\Phi^4$-action
parametrized by the hopping parameter $\kappa$ and the lattice quartic coupling constant $\hat\lambda$ according to
\beq
\label{eq:LatticePhiAction}
S_\Phi = -\kappa\sum_{x,\mu} \Phi_x^{\dagger} \left[\Phi_{x+\mu} + \Phi_{x-\mu}\right]
+ \sum_{x} \Phi^{\dagger}_x\Phi_x + \hat\lambda \sum_{x} \left(\Phi^{\dagger}_x\Phi_x - N_f \right)^2,
\eeq
which is a convenient parametrization for the actual numerical computations. However, this form of the lattice action 
is fully equivalent to the lattice action in continuum notation given in \eq{eq:BosonicLatticeHiggsActionContNot}
written in terms of the bare mass $m_0$, the bare quartic coupling constant $\lambda$, and the lattice derivative
operator $\nabla^f_\mu$. The connection is established through a rescaling of the scalar field $\Phi$ and the 
involved coupling constants according to
\beq
\label{eq:RelationBetweenHiggsActions}
\varphi_x = \sqrt{2\kappa}
\left(
\begin{array}{*{1}{c}}
\Phi_x^2 + i\Phi_x^1\\
\Phi_x^0-i\Phi_x^3\\
\end{array}
\right),
\quad
\lambda=\frac{\hat\lambda}{4\kappa^2}, \quad
m_0^2 = \frac{1 - 2N_f\hat\lambda-8\kappa}{\kappa}, \quad
y_{t,b} = \frac{\hat y_{t,b}}{\sqrt{2\kappa}}.
\eeq

In the given model the expectation value $\langle \varphi \rangle$ would always be identical to zero due to the 
symmetries in \eqs{eq:ChiralSymmetryTrafo1}{eq:ChiralSymmetryTrafo2}. To study 
the mechanism of spontaneous symmetry breaking, one usually introduces an external current $J$, which has 
to be removed after taking the thermodynamic limit, leading then to the existence of symmetric and broken 
phases with respect to the order parameter $\langle \varphi\rangle$ as discussed in \sect{sec:FuncFormInEucTime}. 
In this scenario the bare vev $v$ is directly given by $\langle \varphi\rangle_J$ in the limit $J\rightarrow 0$. 
An alternative approach, which was shown to be equivalent in the thermodynamic 
limit~\cite{Hasenfratz:1989ux,Hasenfratz:1990fu,Gockeler:1991ty}, is to define the vev $v$ as the expectation 
value of the {\textit{rotated}} field $\varphi^{rot}$ given by a global transformation of the original field $\varphi$ 
according to
\beq
\label{eq:GaugeRotation}
\varphi^{rot}_x = U[\varphi] \varphi_x
\eeq
with the $\mbox{SU}(2)$ matrix $U[\varphi]$ selected for each configuration of field variables $\{\varphi_x:\, x\in \Gamma\}$
such that 
\beq
\label{eq:GaugeRotationRequirement}
\sum\limits_x \varphi_x^{rot} = 
\left(
\begin{array}{*{1}{c}}
0\\
\left|\sum\limits_x \varphi_x \right|\\
\end{array}
\right).
\eeq
Here we use this second approach. According to the notation in \eq{eq:BosonicLatticeHiggsActionContNot}, which already 
includes a factor $1/2$, the relation between the vev $v$ and the expectation value of $\varphi^{rot}$
is then given as 
\beq
\label{eq:DefOfVEV}
\langle \varphi^{rot} \rangle = 
\left(
\begin{array}{*{1}{c}}
0\\
v\\
\end{array}
\right).
\eeq
In this setup the unrenormalized Higgs mode $h_x$ and the Goldstone modes $g^1_x,g^2_x,g^3_x,$ can 
directly be read out of the rotated field according to
\beq
\label{eq:DefOfHiggsAndGoldstoneModes}
\varphi_x^{rot}  = 
\left(
\begin{array}{*{1}{c}}
g_x^2 + ig_x^1\\
v + h_x - i g_x^3\\
\end{array}
\right).
\eeq
The great advantage of this approach is that no limit procedure $J\rightarrow 0$ has to be performed, which
simplifies the numerical evaluation of the model tremendously. It is finally remarked that from now on we can
assume the degeneracy of the vacua to be broken whenever we consider analytical calculations due
to the equivalence of this technique with the explicit introduction of an external current $J$.

%--------------------------------------------------------------------------------------------------------------------------
\section{Simulation strategy and considered observables}
\label{sec:SimStratAndObs}

The eventual aim of this work is the non-perturbative determination of the cutoff dependent upper and lower
bounds of the Higgs boson mass as defined in \eq{eq:DefOfDecayWidth}. The general strategy that will be applied for 
that purpose is the non-perturbative evaluation of the whole range of Higgs boson masses that are attainable within the 
pure Higgs-Yukawa sector at a fixed value of the cutoff in consistency with phenomenology. This will be done by
numerically evaluating the finite lattice model of the Higgs-Yukawa sector introduced in the preceding section
and extrapolating the obtained results to the infinite volume limit. 

As discussed in \sect{sec:LatDiscOfContinuumLimit} the requirement of reproducing phenomenology restricts the freedom in the choice 
of the bare parameters $m_0^2, y_{t,b}, \lambda$ due to the phenomenological knowledge of the renormalized quark masses
and the renormalized vacuum expectation value of the scalar field. Throughout this work $m_{t}/a = \GEV{175}$, $m_{b}/a = \GEV{4.2}$, 
and $v_r/a = \GEV{246}$ will be assumed\footnote{In its current listing~\cite{Amsler:2008zzb} the Particle Data Group specifies 
$m_b=\GEV{4.20+0.17-0.07}$ in the $\overline{MS}$ scheme, $m_t=\GEV{171.2\pm 2.1}$ as the perturbative top pole mass, and 
$G_F/(\hbar c)^3=1.166367(5)\cdot 10^{-5}\, \mbox{GeV}^{-2}$, 
where the Fermi coupling constant $G_F$ is related to the renormalized vev through $v_r=1/\sqrt{\sqrt{2}\cdot G_F}$. 
However, due to the ambiguity of the latter quark masses with respect to the chosen renormalization scheme, and since 
statistical and systematic errors will dominate the uncertainty of the finally obtained results in this work,
it is not so meaningful to adopt the most recent results on the aforementioned quantities for the intended lattice calculations.
Instead, the settings given in the main text were taken over from prior Higgs-Yukawa model investigations~\cite{Holland:2003jr}.}. 
Here $m_t$, $m_b$, and $v_r$ denote the renormalized top and bottom quark masses as well as the renormalized vev in dimensionless 
lattice units. However, these conditions leave open an one-dimensional freedom in the bare parameters, which will be addressed at 
the end of this section.

Furthermore, the model has to be evaluated in the broken phase to respect the observation of spontaneous symmetry breaking, 
however close to a second order phase transition to a symmetric phase as discussed in \sect{sec:LatDiscOfContinuumLimit}. 

The physical scale, \ie the inverse lattice spacing $a^{-1}$, can then be reintroduced by comparing the 
renormalized vev $v_r=v/\sqrt{Z_G}$ measured on the lattice with its phenomenologically known value according to
\bea
\label{eq:FixationOfPhysScale}
246\, \mbox{GeV} &=& \frac{v_r}{a} \equiv \frac{v}{\sqrt{Z_G}\cdot a},
\eea
where $Z_G$ denotes the Goldstone renormalization constant. Associated to the lattice spacing is a cutoff parameter $\Lambda$
of the underlying lattice regularization, which is defined here as
\beq
\label{eq:DefOfCutoffLambda}
\Lambda = a^{-1}.
\eeq

As already pointed out in \sect{sec:LatDiscOfContinuumLimit}, this definition is not unique, and other authors use different definitions, 
for instance $\Lambda=\pi/a$ motivated by the value of the momenta at the edge of the Brillouin zone. However, since the quantities that
actually enter any lattice calculation are rather the lattice momenta $\tilde p$ instead of the momenta $p$, which are connected through
the application of a sine function according to \eq{eq:DefOfLatticeMomTildeP}, it seems natural to choose the definition of the 
cutoff $\Lambda$ given in \eq{eq:DefOfCutoffLambda}. 

In the Euclidean continuum the aforementioned Goldstone and Higgs renormalization constants, more precisely their inverse
values $Z^{-1}_G$ and $Z^{-1}_H$, are usually defined as the real part of the derivative of the inverse Goldstone and Higgs 
propagators in momentum space with respect to the continuous squared momentum $p_c^2$ at some scale $p_c^2=-\mu_G^2$ and 
$p_c^2=-\mu_H^2$, respectively. The restriction to the real part is introduced to make this definition applicable also in 
the case of an unstable Higgs boson, due to the branch cut structure of the propagators at negative values of $p_c^2$ 
induced by the massless Goldstone modes as discussed in \sect{sec:UnstableSignature}.
This is the targeted definition that shall be adopted to the later lattice calculations. 

On the lattice, however, the propagators are only defined at the discrete lattice momenta $p\in \ImpSpace$
according to
\bea
\tilde G_H(p) &=& \langle \tilde h_p \tilde h_{-p}\rangle, \\
\tilde G_G(p) &=& \frac{1}{3}\sum\limits_{\alpha=1}^3 \langle \tilde g^\alpha_p \tilde g^\alpha_{-p}\rangle, 
\eea
where the Higgs and Goldstone fields in momentum representation read
\bea
\tilde h_p = \frac{1}{\sqrt{V}}\sum\limits_x e^{-ipx} h_x &\mbox{ and }&
\tilde g^\alpha_p = \frac{1}{\sqrt{V}}\sum\limits_x e^{-ipx} g^\alpha_x
\eea
with $V=L_s^3\cdot L_t$ denoting the lattice volume.

Computing the derivative of the lattice propagators is thus not a well-defined operation. Moreover, the lattice
propagators are not even functions of $p^2$, since rotational invariance is explicitly broken by the discrete
lattice structure. To adopt the above described concept to the lattice nevertheless, some lattice scheme has to be
introduced here that converges to the continuum definitions of $Z_G$ and $Z_H$ in the limit $a\rightarrow 0$.
The idea is to use some analytical fit formulas $f_{G}(p_c)$, $f_{H}(p_c)$ derived from renormalized 
perturbation theory in the Euclidean continuum to approximate the measured lattice propagators $\tilde G_G(p)$ and 
$\tilde G_H(p)$ at small momenta $\hat p^2<\gamma$ for some appropriate value of $\gamma$ such that the discretization 
errors are acceptable. The details of this fit procedure are discussed in \sect{sec:ParticleMassDetDetailsLowerBound} 
and \sect{sec:CorrVsPropMass}. One can then define the analytically continued lattice propagators as
\bea
\tilde G_G^{c}(p_c) = f_{G}(p_c) &\mbox{ and }&
\tilde G_H^{c}(p_c) = f_{H}(p_c).
\eea
The targeted Goldstone and Higgs renormalization constants $Z_G(\mu_G^2)$ and $Z_H(\mu_H^2)$ can then be 
defined\footnote{Exploiting the $O(4)$-symmetry of the continuous propagators one has $\tilde G^c_{G,H}(p_c)\equiv \tilde G^c_{G,H}(p_c^2)$
with some abuse of notation, where an underlying mapping $p_c\leftrightarrow p_c^2$ is implicitly assumed.} as 
\bea
Z^{-1}_G(\mu^2_G) &=& \derive{}{p_c^2} \RE\left(\left[\tilde G^{c}_G(p_c^2)\right]^{-1}\right)\Big|_{p_c^2 = -\mu^2_G},\\
Z^{-1}_H(\mu^2_H) &=& \derive{}{p_c^2} \RE\left(\left[\tilde G^{c}_H(p_c^2)\right]^{-1}\right)\Big|_{p_c^2 = -\mu^2_H}.
\eea

A natural choice for the fixation of the scales $\mu^2_G$, $\mu^2_H$ is the so-called on-shell scheme, where one 
sets the scales $\mu^2_G$, $\mu^2_H$ to the squared masses $m^2_{G}$, $m^2_{H}$ of the Higgs and Goldstone bosons.
As already discussed in \sect{sec:UnstableSignature} the physical masses are given by the 
poles of the respective propagators on the second Riemann sheet. To adopt this definition to the introduced 
lattice scheme we define the Higgs boson mass $m_H$, its decay width $\Gamma_H$, and the mass $m_G$ of the stable
Goldstone bosons through
\bea
\label{eq:DefOfHiggsAndGoldstoneMassByPole}
\left[\tilde G^{c}_{H,II}(im_H+\Gamma_H/2,0,0,0)\right]^{-1} = 0, &\mbox{ and }&
\left[\tilde G^{c}_G(im_G,0,0,0)\right]^{-1} = 0,
\eea
where $\tilde G^{c}_{H,II}(p)$ denotes the analytical continuation of $\tilde G^{c}_{H}(p)$ onto the second Riemann sheet.

Extracting the Higgs mass $m_H$ and its decay width $\Gamma_H$ according to this definition would, however, 
require a very accurate analytical continuation of the Higgs propagator onto the second Riemann sheet. While 
such a continuation can in principle be obtained with the help of a decent fit formula $f_{H}(p)$ as will 
be discussed in more detail in \sect{sec:ParticleMassDetDetailsLowerBound} and \sect{sec:CorrVsPropMass}, 
reaching the necessary level of accuracy to extract the mass and the decay width reliably from 
\eq{eq:DefOfHiggsAndGoldstoneMassByPole} is a highly non-trivial task. Such an analytical continuation 
will nevertheless later be tried.

Owing to these difficulties the Goldstone and Higgs renormalization factors are rather determined at the 
scales $\mu_{G}^2 = m^2_{Gp}$ and $\mu_{H}^2 = m^2_{Hp}$ given by the masses $m_{Hp}$ and $m_{Gp}$, which 
will be referred to in the following as propagator masses in contrast to the pole masses $m_H$ and $m_G$. 
We thus define
\bea
\label{eq:DefOfRenormalFactors}
Z_G \equiv Z_G(m^2_{Gp})  &\mbox{ and }&
Z_H \equiv Z_H(m^2_{Hp}),
\eea
where the propagator masses $m_{Hp}$, $m_{Gp}$ are defined through a vanishing real-part of the inverse 
propagators according to
\bea
\label{eq:DefOfPropagatorMinkMass}
\RE\left(\left[G_G^{c}(p_c^2)\right]^{-1}\right)\Big|_{p_c^2 = -m^2_{Gp}} = 0 &\mbox{ and }& 
\RE\left(\left[G_H^{c}(p_c^2)\right]^{-1}\right)\Big|_{p_c^2 = -m^2_{Hp}} = 0.
\eea
The reasoning for selecting these latter definitions for the Higgs and Goldstone masses is, that the required 
analytical continuation in the case of the Higgs propagator is much more robust, since it only needs to extend 
the measured lattice propagator to purely negative values of $p_c^2$ in contrast to the situation resulting from 
the definition in \eq{eq:DefOfHiggsAndGoldstoneMassByPole}. It is remarked here that the Goldstone propagator
mass $m_{Gp}$ was only introduced for the sake of an uniform notation, since $m_{G}$ is identical to $m_{Gp}$, 
due to the Goldstone bosons being stable.

As for the unstable Higgs boson, however, one finds that the discrepancy between the pole mass $m_H$ and the
propagator mass $m_{Hp}$ is directly related to the size of the decay width $\Gamma_H$. In the weak coupling 
regime of the theory the two mass definitions $m_H$ and $m_{Hp}$ can thus be considered to coincide up to
small perturbative corrections, due to a vanishing decay width in that limit. For the pure $\Phi^4$-theory
the deviation between $m_{Hp}$ and $m_H$ has explicitly been worked out in renormalized perturbation theory~\cite{Luscher:1988uq}. 
In infinite volume the finding is
\beq
\label{eq:ConnectionMhMhp}
m_H = m_{Hp} \cdot \left(1+\frac{\pi^2}{288} (n-1)^2 \left[\frac{4!\cdot\lambda_r}{16\pi^2}\right]^2  + O(\lambda_r^3)  \right),
\eeq
where $\lambda_r$ denotes the renormalized quartic self-coupling constant and $n$ is the number of components 
of the scalar field $\Phi$, \ie $n=4$ for the here considered case. This calculation was performed in 
the pure $\Phi^4$-theory, thus neglecting any fermionic degrees of freedom, and for exactly massless 
Goldstone particles. However, one learns from this result that the definition of $m_{Hp}$ in \eq{eq:DefOfPropagatorMinkMass} 
as the Higgs boson mass is very reasonable at least for sufficiently small values of the renormalized coupling constants.

The definition of the renormalized quartic self-coupling constant $\lambda_r$ that was used in the derivation of \eq{eq:ConnectionMhMhp}
is
\beq
\label{eq:DefOfRenQuartCoupling}
\lambda_r = \frac{m^2_{Hp}-m^2_{Gp}}{8v_r^2},
\eeq
which shall also be taken over to the considered Higgs-Yukawa model. 
It is remarked, that it is in principle possible to determine a renormalized quartic coupling constant $\lambda_r$ through
the evaluation of the amputated, connected, one-particle-irreducible four-point function at a specified momentum configuration 
as it is usually done in perturbation theory. However, the signal to noise ratio of the corresponding lattice observable is suppressed by the
lattice volume. It is thus extremely hard to measure the renormalized quartic coupling constant in lattice calculations by means
of the direct evaluation of such four-point functions. Instead, the alternative definition of $\lambda_r$ given in \eq{eq:DefOfRenQuartCoupling}
will be adopted here. It is further remarked that this definition was shown~\cite{Luscher:1988uq} to coincide with the bare coupling 
parameter $\lambda$ to lowest order in the pure $\Phi^4$-theory. 

Besides the already given definitions it is also possible to determine the mass spectrum of the considered theory through the numerical 
evaluation of the lowest lying energy eigenvalues of the Hamiltonian $\hat H$ underlying the theory in continuous space-time. This can,
for instance, be done by studying the two-point function of an hermitian operator $\hat O(0)$ with $\hat O(t)$ depending only on field 
operators at the Euclidean time $t$. As discussed in \sect{sec:UnstableSignature} the connected two-point function is directly related 
to the energy eigenvalues $E_n$ of the Hamiltonian $\hat H$ through
\bea
\label{eq:CorrMassRelation}
\langle \Omega_0 | \hat O(t) \hat O(0) | \Omega_0\rangle -
|\langle \Omega_0 | \hat O(0) | \Omega_0\rangle |^2 &=&
\sum\limits_{n\neq 0} e^{-t (E_n-E_0)} \cdot \left| \langle \Omega_0| \hat O(0)| E_n \rangle \right|^2,
\eea
where $|E_0\rangle$ equals the vacuum state. Choosing an operator $\hat O(t)$ with zero spatial momentum
and certain quantum numbers thus allows to determine the mass at rest of the lightest particle possessing the specified 
quantum numbers. This can be done by exploiting the fact that the exponential decay of the above time-correlation function is 
dominated at large Euclidean times $t\gg 1$ by the lowest lying energy eigenvalue, the associated eigenstate of which 
having a non-zero overlap with the selected operator.

Taking over this idea to the discrete lattice one can define the so-called correlator 
Higgs boson mass $m_{Hc}$ through the exponential decay of the lattice Higgs time-slice correlation 
function specified here as 
\bea
\label{eq:DefOfHiggsTimeSliceCorr}
C_H(\Delta t) &=& \frac{1}{L_t} \sum\limits_{t=0}^{L_t-1} \left\langle O_H(t+\Delta t)\cdot O_H(t) \right\rangle
\eea
based on the observable
\bea
\label{eq:HiggsFieldTSoperator}
O_H(t) = \frac{1}{L_s^3} \sum\limits_{\vec x} h_{t,\vec x} \quad 
\eea
at large Euclidean time separations $\Delta t$ according to
\bea
\label{eq:DefOfHiggsMassFromTimeSliceCorr}
C_H(L_t/2\ge\Delta t \gg 1) &\propto& \mbox{cosh}\Big[m_{Hc}\cdot (L_t/2-\Delta t)\Big],
\eea
where the details of the actual determination of $m_{Hc}$ are postponed to \sect{sec:HiggsTSCanalysisLowerBound}.

The advantage of this approach is that it does not suffer from any unknown systematic uncertainty induced by the analytical continuation 
of the Higgs propagator. As discussed in \sect{sec:UnstableSignature} for the case of the pure $\Phi^4$-theory in continuous space-time 
the correlator mass $m_{Hc}$ coincides with $m_H$ in the weak coupling regime, at least from a practical point of view, if one can
consider the spectral function $\rho(E)$ defined in \eq{eq:DefOfSpectralFunction} to be dominated by its peak at $E\approx m_H$. This 
situation changes in the strongly coupling regime of the model, where contributions from the continuously distributed, low lying 
two-Goldstone states to the correlation function $C_H(\Delta \tau)$ can not be neglected.

However, as a helpful side effect of the finiteness of the lattice the spectral function $\rho(E)$ on the lattice is discrete in 
contrast to its continuous counterpart. One can thus try to improve the extraction of the correlator mass $m_{Hc}$ in the strong 
coupling regime by respecting a small number of discrete two-Goldstone contributions in the analysis of the correlation function 
$C_H(\Delta \tau)$ at large time separations. The details of this procedure are discussed in 
\sect{sec:HiggsTSCanalysisLowerBound} and \sect{sec:HiggsTSCanalysisUpperBound}.

In addition to the uncertainties arising from the influence of low lying two-Goldstone states there are further sources
which give rise to deviations between $m_H$ and $m_{Hc}$. These deviations are induced, for instance, by finite volume effects, 
discretization effects, as well as contributions to $m_{Hc}$ arising from excited states according to the finiteness of $\Delta\tau$
in a practical calculation. Some of these effects will explicitly be observed in the later numerical investigation of the model.

It is remarked here that the Hamiltonian $\hat H_\Gamma$ describing the discretized system is not identical to the 
Hamiltonian $\hat H$ in continuous space-time. For the definition in \eq{eq:DefOfHiggsMassFromTimeSliceCorr} to make sense
on a finite lattice it is not sufficient to demand that $\hat H_\Gamma$ converges to $\hat H$ in the
continuum limit. In addition one has to demand that $\hat H_\Gamma$ is hermitian for
any finite lattice $\Gamma$ in order to guarantee a monotonic behaviour of the time-slice correlator for $0\le \Delta t \le L_t/2$, 
which is needed for extracting the correlator mass $m_{Hc}$ as specified in \eq{eq:DefOfHiggsMassFromTimeSliceCorr}. 
The condition of $\hat H_\Gamma$ being hermitian then leads to the actually required positivity of the so-called transfer matrix 
$T_{\alpha,\beta}$, given as $T_{\alpha,\beta}=\langle\alpha|\exp(-a\hat H_\Gamma)|\beta\rangle$ with $|\alpha\rangle$ and 
$|\beta\rangle$ denoting two states of the underlying Hilbert space. The positivity of the transfer matrix in the pure lattice 
$\Phi^4$-theory can explicitly be proven~\cite{book:Montvay}. While the positivity of the transfer matrix associated to lattice 
QCD based on the Wilson operator could also be demonstrated~\cite{Luscher:1976ms}, the situation is less clear in case of the Neuberger
overlap operator, which could, however, be proven to lead to a positive transfer matrix at least in the non-interacting 
case~\cite{Luscher:2000hn}. As a working hypothesis we will therefore assume the underlying lattice Hamiltonian to be hermitian
for the rest of this work.

Correspondingly, one can also extract the physical top and bottom quark masses $m_t, m_b$ by studying the fermionic time correlation
functions $C_f(\Delta t)$ at large Euclidean time separations $\Delta t$, where $f=t,b$ denotes the quark flavour here. 
On the lattice the fermionic time correlation functions can be defined as
\bea
\label{eq:DefOfFermionTimeSliceCorr}
C_f(\Delta t) &=& \frac{1}{L_t\cdot L_s^6} \sum\limits_{t=0}^{L_t-1} \sum\limits_{\vec x, \vec y}
\Big\langle 2\,\RE\,\Tr\,\left(f_{L,t+\Delta t, \vec x}\cdot \bar f_{R,t,\vec y}\right) \Big\rangle,
\eea
where the left- and right-handed spinors are given through the projection operators according to
\bea
\label{eq:DefOfLeftHandedSpinors}
\left(
\begin{array}{*{1}{c}}
t\\
b\\
\end{array}
\right)_L
= \hat P_- 
\left(
\begin{array}{*{1}{c}}
t\\
b\\
\end{array}
\right)
 &\mbox{ and }&  (\bar t, \bar b)_R  = (\bar t, \bar b) P_-.
\eea
The above restriction to only the left-handed quark field correlator, which reduces the numerical costs of computing $C_f(\Delta t)$, 
is motivated by the relation $\Tr \langle \Omega_0|\hat f_L\hat {\bar f_R} + \hat f_R\hat {\bar f_L}|\Omega_0\rangle = 2\,\RE\,\Tr \langle \Omega_0|\hat f_L\hat {\bar f_R}|\Omega_0\rangle$
which holds in the continuous operator formalism at $\Delta t=0$. It is remarked that the given fermionic correlation function would 
be identical to zero due to the exact lattice chiral symmetry obeyed by the considered Higgs-Yukawa model, if one would not
rotate the scalar field $\varphi$ as discussed in \sect{sec:modelDefinition}. This rotation is implicitly assumed in the
following. The quark masses $m_{t,b}$ can then be extracted according to
\bea
\label{eq:DefOfFermionMassFromTimeSliceCorr}
C_f(L_t/2\ge\Delta t \gg 1) &\propto& \mbox{cosh}\Big[m_{f}\cdot (L_t/2-\Delta t)\Big],
\eea
where the details of the determination of $m_{t,b}$ are postponed to \sect{sec:FermionTSCanalysisLowerBound}.
It is further remarked that the full {\textit{all-to-all}} correlator as defined in \eq{eq:DefOfFermionTimeSliceCorr}) can be 
trivially computed by using sources which have a constant value of one on a whole time slice for a selected
spinor index and a value of zero everywhere else.
This all-to-all correlator yields very clear signals for the top and bottom quark mass determination.

The lacking definition of the renormalized Yukawa coupling constants can now be provided according to
\bea
\label{eq:DefOfRenYukawaConst}
y_{t,r} = \frac{m_t}{v_r} &\mbox{ and }&
y_{b,r} = \frac{m_b}{v_r},
\eea
reproducing the bare Yukawa coupling constants $y_{t,b}$ at lowest order.

Finally, some comments on the practical approach to evaluate the Higgs boson mass bounds shall be given. 
For a given cutoff $\Lambda$ the aforementioned requirements of reproducing the vev and the quark masses still leave open 
an one-dimensional freedom in the bare model parameters, which can be parametrized in terms of the bare quartic self-coupling 
constant $\lambda$. However, aiming at lower and upper Higgs boson mass bounds, this remaining freedom can be fixed, since it 
is expected from perturbation theory that the lightest Higgs boson masses are obtained at vanishing self-coupling constant 
$\lambda=0$, while the heaviest masses are attained at infinite coupling constant $\lambda=\infty$, respectively, according to 
the qualitative one-loop perturbation theory result for the Higgs boson mass shift
\beq
\label{eq:perturbTheroyResult}
\delta m_H^2/a^2 \equiv (m_H^2 - m_0^2)/a^2 \propto  \left(\lambda - y_t^2 - y_b^2 \right) \cdot \Lambda^2.
\eeq
From this relation one may conjecture that the lower Higgs boson mass bound will be obtained by evaluating the model at vanishing
bare quartic coupling constant $\lambda$, while the upper Higgs boson mass bound should be observed at $\lambda=\infty$. 
This argument is, however, not complete, since the bare mass also changes with varying coupling constants when holding
the cutoff $\Lambda$ constant. For clarification, this latter expectation will be discussed in detail in 
\sect{sec:DepOfHiggsMassOnLambda} and \sect{sec:DepOfHiggsMassonLargeLam}, where it will also be confronted with
the results of direct lattice calculations. 

Furthermore, the Yukawa coupling parameters are intended to be tuned such that the phenomenologically known values of the fermion
masses will be reproduced. As an initial guess for the adjustment of the bare Yukawa coupling parameters  in the later lattice 
calculations the tree-level relation 
\beq
\label{eq:treeLevelTopMass}
y_{t,b} = \frac{m_{t,b}}{v_r}
\eeq
will be used in this work. The actual physical fermion masses generated in the respective lattice calculations are then explicitly 
calculated by means of the correlation function given in \eq{eq:DefOfFermionTimeSliceCorr} and can be used in follow-up Monte-Carlo 
studies to fine-tune the bare Yukawa coupling constants to reproduce the phenomenological expectations with greater accuracy.

%--------------------------------------------------------------------------------------------------------------------------
\section{A hybrid Monte-Carlo algorithm for the evaluation of the model}
\label{sec:HMCAlgorithm}

For the numerical evaluation of the model an algorithm is needed 
that can compute the extremely high dimensional functional integral in 
\eq{eq:DefOfLatticeObsWithoutPsi}. Besides the very high dimension of that 
integral, being of order $O(V)$, a further difficulty arises from the 
appearance of the fermion determinant $\det(\fermiMat)$ in its integrand.
Due to the complexity class of calculating that determinant being of order $O(V^3)$, 
a direct numerical evaluation of the fermion determinant is therefore not feasible 
for lattice volumes of physical interest. 

There are, however, a couple of algorithms allowing to compute the expectation value of observables
in the framework of the considered functional integral by means of converging stochastic series, 
which are - in practice - truncated after a finite, sufficiently large number of samples. 
Among them are the HMC algorithm~\cite{Duane:1987de,Gottlieb:1987mq}, the PHMC 
algorithm~\cite{Frezzotti:1997ym,Frezzotti:1998eu,Frezzotti:1998yp} and the RHMC 
algorithm~\cite{Kennedy:1998cu,Clark:2003na}, each of which capable of including the full fermion 
dynamics into the calculation, \ie fully respecting the contribution of the fermion determinant 
$\det(\fermiMat)$ to the functional integral. 

The basic observation underlying all of the aforementioned approaches is that the determinant of an
hermitian, positive definite matrix $A$ can be written in terms of a Gaussian integration over some
complex vector $\omega$ according to
\beq
\label{eq:GaussIntegralOfDetA}
\det(A) = \int \intD{\omega} \intD{\omega^\dagger} e^{-\frac{1}{2}\omega^\dagger A^{-1} \omega}
\eeq
up to some constant factor independent of the matrix $A$.

Applying this relation to the determinant of the fermion matrix appearing in the integrand
of \eq{eq:DefOfLatticeFuncIntWithoutPsi}, would allow to respect the fermion dynamics
in the Monte-Carlo simulation in an efficient manner, \ie without having to actually compute the 
determinant itself. However, the fermion matrix $\fermiMat$ is neither positive nor hermitian, 
and thus one can not directly apply this relation to $\det(\fermiMat)$. 

To get around this issue one can instead apply \eq{eq:GaussIntegralOfDetA} to the squared operator $\fermiMatDouble$
leading then to an equivalent expression for the partition function of the considered Higgs-Yukawa model given as
\bea
\label{eq:PartFuncWithPhaseOfDet}
Z &=& \int \intD{\Phi} \intD{\pi} \intD\omega \intD{\omega^\dagger}
\,\, \left[ \arg\det(\fermiMat) \right]^{N_f} \cdot
e^{-S[\Phi, \pi, \omega]}, \\
\label{eq:IntroOfConjPhiMomentaInAction}
S[\Phi, \pi, \omega] &=& S_\Phi[\Phi] + f(\pi) + S_F[\Phi,\omega] \quad \mbox{with} \quad\\
S_F[\Phi,\omega] &=& \frac{1}{2} \sum\limits_{i=1}^{N_f/2}\omega_i^\dagger \left[\fermiMat\fermiMat^\dagger  \right]^{-1}
\omega_i,
\eea
where $N_f$ is assumed to be even and the complex phase $\arg\det(\fermiMat)$ of the fermion determinant in 
\eq{eq:PartFuncWithPhaseOfDet} corrects for considering the squared fermion matrix $\fermiMatDouble$ instead of $\fermiMat$.
Here, the expression $\omega\equiv(\omega_1,\ldots,\omega_{N_f/2})$ actually denotes a set of $N_f/2$ 
complex vectors denoted as pseudo-fermion fields. By means of the latter assumption the HMC algorithm introduced in this 
section is restricted to even values of $N_f$. The treatment of odd $N_f$ is postponed to \sect{sec:basicConceptsOfPHMC}.

An important feature of the given action in \eq{eq:IntroOfConjPhiMomentaInAction} is that the inverse operator $[\fermiMatDouble]^{-1}$
is not required to be computed completely. Only its application on a given vector $\omega_i$ needs to be calculated, which
can efficiently be done by means of a CG-algorithm~\cite{Press:2007zu}, for instance. Moreover, only the ability to 
apply the matrices $\fermiMat$ and $\fermiMat^\dagger$ on a given vector $\omega_i$, but not their complete representations 
in terms of their respective matrix elements, is necessary for the latter algorithm to work.
Due to the diagonal structure of the Dirac operator $\D$ in momentum space and the diagonal form
of the operator $B[\Phi]$ in position space, the aforementioned matrix applications $\fermiMat \omega_i$ and $\fermiMat^\dagger \omega_i$
can efficiently be computed by means of a Fast Fourier Transform (FFT), switching back and forth between position and
momentum space. With the help of the eigenvalues and eigenvectors of $\D$ given in \eqs{eq:DefOfEigenvectorsOfD1}{eq:eigenValOfFreeND}
the Dirac operator can then very efficiently be constructed in momentum space, while the operator $B[\Phi]$ is
trivially applicable in position space. To compute the underlying Fast Fourier Transforms the standard numeric software library
FFTW~\cite{FFTW05} is used as well as an improved, self-written implementation especially tuned for the case of four dimensions, 
as discussed in \appen{sec:xFFT}.

An additional integration over the conjugate momenta $\pi$ associated to the bosonic field $\Phi$ has been 
introduced in \eq{eq:PartFuncWithPhaseOfDet} for later use. For clarification it is remarked at this point that the introduction 
of this additional integration and the extension of the action in \eq{eq:IntroOfConjPhiMomentaInAction} by the positive, 
real function $0 < f(\pi)\in\re$ of the conjugate momenta has no effect on the expectation value of any observable $O[\Phi]$ 
depending only on the field $\Phi$. By virtue of Grassmann integration this is the only type of observables we are interested in,
as discussed in \sect{sec:modelDefinition}. However, for later use in \sect{sec:ExactReweighing} we extend the consideration
here to the more general class of observables $O[\Phi,\omega]$, depending also on the pseudo-fermion fields $\omega$.

The expectation value of any observable $O[\Phi, \omega]$ can then be computed by evaluating the functional integral
\bea
\label{eq:DefOfExpValueWithoutPsi}
\langle O[\Phi,\omega] \rangle &=& \frac{1}{Z}  \int \intD{\Phi} \intD{\pi} \intD{\omega} \intD{\omega^\dagger}
\,\, O[\Phi,\omega] \cdot e^{i N_f \arg\det(\fermiMat)} \cdot
e^{-S[\Phi, \pi, \omega]} \quad\quad
\eea
by means of a Monte-Carlo integration. The underlying idea of the Monte-Carlo integration technique is to generate an 
infinite, random series of sampling points $\xi_n\equiv(\Phi, \pi, \omega)_n\in \confSpace,\, n\in \N$ in terms of field configurations of the
field $\Phi$, its conjugate momenta $\pi$, and the complex fields $\omega\equiv(\omega_1,\ldots,\omega_{N_f/2})$ distributed 
according to the probability distribution $\bar p(\xi)=\exp(-S[\Phi,\pi,\omega])/Z$. Here, the space $\confSpace$ is the set of all field configuration
tuples $(\Phi,\pi,\omega)$. The desired expectation value $\langle O[\Phi,\omega] \rangle$ 
is then given as
\bea
\label{eq:DefOfExpValueInMarrrkovChain}
\langle O[\Phi,\omega] \rangle &=& \lim\limits_{N\rightarrow \infty} \bar O_{N},\\
\label{eq:DefOfAvgValueInMarrrkovChain}
\bar O_{N} &=& \frac{1}{N} \sum\limits_{n=1}^N O_n \cdot e^{i N_f \arg\det(\fermiMat)},\\
\label{eq:DefOfSampleValueInMarrrkovChain}
O_n &=& O[(\Phi, \omega)_n]\equiv O[\xi_n].
\eea

In practice, however, this infinite series is truncated after a finite number of $N$ configurations. As a consequence
one only obtains statistical estimates $\bar O_N$ for the true expectation value $\langle O \rangle$ from \eq{eq:DefOfExpValueInMarrrkovChain}.
These estimates are afflicted with statistical errors that decrease proportional to $1/\sqrt{N}$. The associated proportionality 
constant depends on the considered observable as well as on the degree of statistical (in)dependence of the
generated field configurations $(\Phi, \omega)_n$ as will be further discussed in \sect{sec:FACC}.

The vital question at this point is how the series of field configurations $(\xi_n)_{n\in\N}$ with the 
aforementioned properties can be constructed in an efficient manner. This can be achieved by means of a Markov 
process~\cite{Bremaud:2001zu}. Starting from a given configuration $\xi_1$ 
a Markov process constructs the so-called Markov chain $(\xi_n)_{n\in\N}$ by iteratively enlengthening the finite series
$(\xi_n)_{n=1,\ldots,N}$ with the N+1-th element $\xi_{N+1}$ which is randomly sampled according to the conditional 
probability distribution $P(\xi_{N+1} | \xi_N)$, also called transition probability density, which only depends 
on the preceding element $\xi_N$. Each such update will be referred to as a Markov step. Associated to this iterative procedure 
is a series of probability distributions $p_n(\xi)$, which give the probability density that the n-th element of the Markov 
chain is $\xi$. These probability densities evolve during the Markov process according to
\beq
\label{eq:MarkovFixPointEquation}
p_{n+1}(\xi) = \int \intD{\tilde \xi} P(\xi | \tilde \xi) p_{n}(\tilde\xi).
\eeq
In the theory of Markov chains one can show that this iterative relation corresponds to a contraction in the space
of probability distributions provided that the underlying transition probabilities are ergodic~\cite{Kennedy:2006ax}, 
\ie that the probability for the Markov chain to reach the configuration $\xi$ starting from $\tilde \xi$ is non-zero 
for any pair of $\xi$, $\tilde\xi$. The sequence of probability distributions $p_n(\xi)$ then converges to a fix point 
denoted as $p_{{fix}}(\xi)$. From a practical point of view the elements $\xi_n$
of the constructed Markov chain can thus be considered as being sampled according to $p_{{fix}}(\xi)$ for $n\ge \thermTime$,
where the so-called thermalization time $\thermTime$ has to be chosen sufficiently large. For $n<\thermTime$ the Markov process 
is said to be in its thermalization phase and the corresponding configurations $\xi_n$ have to be neglected in the 
computation of \eqs{eq:DefOfExpValueInMarrrkovChain}{eq:DefOfSampleValueInMarrrkovChain}. An appropriate relabeling of
the configurations is implicitly assumed here. Moreover, the quantity $\Nconf$ denotes the total number
of available field configurations with $n\ge \thermTime$ in the following, while the configuration index $\confNR=1,\ldots,\Nconf$ 
labels these available configurations.

The aim is thus to define a transition probability density $P(\xi | \tilde\xi)$ that fulfills the aforementioned
prerequisite of ergodicity of the Markov process, while leading to the probability distribution $\bar p(\xi)\equiv\exp(-S[\xi])/Z$ 
as the fix point $p_{fix}(\xi)$ of \eq{eq:MarkovFixPointEquation}. The latter requirement can trivially be met by the 
sufficient, but not necessary condition of so-called 'detailed balance'~\cite{Metropolis:1953am} according to
\beq
\label{eq:DefOfDetailedBalance}
P(\xi| \tilde \xi)\cdot \bar p(\tilde \xi) = P(R(\tilde \xi) | R(\xi)) \cdot \bar p(R(\xi)),
\eeq
where $R:\confSpace\rightarrow\confSpace$ shall be some invertible, differentiable mapping with $|\det(\partial R / \partial \xi)|=1$, $RR=\ID$,
and $\bar pR=\bar p$ which is introduced here for later use\footnote{$RR$ and $\bar pR$ are shorthand notations for the consecutive 
applications of the respective mappings.}. Obviously, any transition probability distribution $P(\xi|\tilde \xi)$ obeying this condition of 
detailed balance leads to $p_{fix}(\xi)=\bar p(\xi)$ as the fix point of \eq{eq:MarkovFixPointEquation}. One therefore usually takes 
\eq{eq:DefOfDetailedBalance} as the starting point for the construction of the desired transition probabilities.

An additional, very helpful construction tool results from the observation that for any two transition probability distributions $P_1(\xi|\tilde\xi)$
and $P_2(\xi|\tilde\xi)$, which share the common fix point $p_{fix}(\xi)$ of \eq{eq:MarkovFixPointEquation}, 
one finds that the consecutive application of their associated Markov steps leads again to the same fix point $p_{fix}(\xi)$. This follows
directly when inserting the transition probability distribution of the aforementioned consecutive application, which is given by
\beq
\left(P_1 \circ P_2  \right)(\xi|\tilde \xi) = \int \intD{\hat\xi} P_1(\xi|\hat\xi)\cdot P_2(\hat\xi|\tilde \xi),
\eeq
into \eq{eq:MarkovFixPointEquation}. This will be exploited here by splitting the targeted transition probability distribution $P(\xi|\tilde \xi)$
into three parts according to 
\beq
P = P_\Phi\circ P_\omega\circ P_\pi,
\eeq
where now only the overall product $P(\xi|\tilde\xi)$ but not each single contributions $P_\pi$, $P_\omega$, $P_\phi$ alone has to be ergodic.
The conjugate momenta $\pi$ can then be sampled directly according to the obvious solution 
\beq
\label{eq:ProbDisDesityOfConjMom}
P_\pi(\xi|\tilde\xi) = \delta(\Phi-\tilde\Phi) \cdot \delta(\omega-\tilde\omega)\cdot  e^{-f(\pi)} \cdot\left[\int\intD{\hat\pi} e^{-f(\hat\pi)}\right]^{-1}
\eeq
of the detailed balance condition in \eq{eq:DefOfDetailedBalance} for $R=\ID$, provided that the function $f(\pi)$ has a sufficiently simple form, 
such as a Gaussian distribution. We will therefore set $f(\pi) = \pi^\dagger\pi/2$ for the rest of this section. However, it is already remarked 
at this point that the choice of the function $f(\pi)$ has a great impact on the properties of the associated Markov process, such as the statistical 
(in)dependence of the generated field configurations $\xi_n$ as will be further discussed in \sect{sec:FACC}.

The complex fields $\omega$ can also be sampled directly according to the probability distribution $P_\omega(\xi|\tilde\xi)$ given as
\beq
\label{eq:DirectSamplingOfOmegaDistributionHMC}
P_\omega(\xi|\tilde\xi) = \delta(\Phi-\tilde\Phi) \cdot \delta(\pi-\tilde\pi)\cdot e^{ -\frac{1}{2}\fhs{-1mm}\sum\limits_{i=1}^{N_f/2}\fhs{-1mm} 
\omega_i^\dagger \left[\fermiMat\fermiMat^\dagger  \right]^{-1}  \omega_i  } 
\cdot
\left[\int\intD{\hat \omega} \intD{\hat \omega^\dagger}  e^{  - \frac{1}{2}\fhs{-1mm}\sum\limits_{i=1}^{N_f/2}\fhs{-1mm}
 \hat\omega_i^\dagger \left[\fermiMat\fermiMat^\dagger  \right]^{-1}  \hat\omega_i  }\right]^{-1}
\eeq
which again is a solution to the detailed balance condition with $R=\ID$, thus leading to the desired fix point in \eq{eq:MarkovFixPointEquation}.
The practical problem of sampling the fields $\omega\equiv(\omega_1,\ldots, \omega_{N_f/2})$ according to \eq{eq:DirectSamplingOfOmegaDistributionHMC}
can be solved by the simple substitution $\eta_i= \fermiMat^{-1}\omega_i$, $i=1,\ldots,N_f/2$. The fields $\eta_i$ are then Gaussian distributed
according to 
\beq
P_\eta(\eta_i) =  e^{-\frac{1}{2}\eta_i^\dagger\eta_i} \cdot
\left[\int\intD{\hat\eta_i}\intD{\hat\eta_i^\dagger} e^{-\frac{1}{2}\hat\eta_i^\dagger\hat\eta_i}\right]^{-1}
\eeq
and the fields $\omega_i$ are obtained as
\beq
\label{eq:HMCOmegaEtaRelation}
\omega_i = \fermiMat \eta_i.
\eeq

So far, the consecutive application of the already introduced Markov steps, resulting in the transition probability distribution $P_\pi\circ P_\omega$,
obviously does not yield an ergodic Markov process, since the bosonic field $\Phi$ was not updated. For that purpose the third transition probability
density $P_\Phi(\xi|\tilde\xi)$ will be introduced in the following. For the later practical numerical computations it will turn out convenient to split 
up $P_\Phi(\xi|\tilde\xi)$ into a so-called proposition probability density $P_\Phi^{(p)}(\xi|\tilde \xi)$ and an associated acceptance probability 
$P_\Phi^{(a)}(\xi|\tilde \xi)$ according to 
\beq
P_\Phi(\xi|\tilde \xi) = 
P_\Phi^{(p)}(\xi|\tilde \xi)\cdot P_\Phi^{(a)}(\xi|\tilde \xi)
+ \delta(\xi|\tilde\xi)\cdot \int \intD{\hat \xi} P_\Phi^{(p)}(\hat\xi | \tilde \xi) \cdot (1- P_\Phi^{(a)}(\hat\xi|\tilde \xi))
\eeq
where the the second summand guarantees that $P_\Phi(\xi|\tilde \xi)$ is normalized to one. The interpretation of this ansatz
is that a new field configuration is proposed sampled according to the proposition probability density $P_\Phi^{(p)}(\xi|\tilde \xi)$. In a second step, the
so-called Metropolis step~\cite{Metropolis:1953am}, the proposed configuration is only accepted as the next element in the Markov chain with a probability 
of $P_\Phi^{(a)}(\xi|\tilde \xi)$. In case of rejection, the original field configuration is taken as the next element in the Markov chain. In this setup one 
directly finds that the condition of detailed balance is always fulfilled for any proposition probability density $P_\Phi^{(p)}(\xi|\tilde\xi)$, if one chooses 
the acceptance probability to be
\beq
\label{eq:DefOfAcceptanceProb}
P_\Phi^{(a)}(\xi|\tilde\xi) = \min\left(1, \frac{\bar p(R(\xi))}{\bar p(\tilde\xi)} \cdot 
\frac{P_\phi^{(p)}(R(\tilde\xi)|R(\xi))}{P_\phi^{(p)}(\xi|\tilde\xi)} 
\right)
\eeq
which again is not a necessary, but a sufficient choice.

For the construction of the aspired Markov process one thus only needs to specify a proposition probability density $P_\Phi^{(p)}(\xi|\tilde \xi)$ 
that provides the missing link in establishing the ergodicity of $P=P_\Phi\circ P_\omega\circ P_\pi$, since the condition that $\bar p(\xi)$ is 
the fix point associated to $P(\xi|\tilde\xi)$ is already guaranteed by construction. For that purpose the invertible and differentiable 
transformation $\IntegratorSymbol:\confSpace\rightarrow\confSpace$ is introduced here. The idea is that the configuration $\IntegratorSymbol(\xi_N)$ 
will be proposed as the subsequent element in the Markov chain following the element $\xi_N$. The corresponding proposition probability density 
is thus given as
\beq
P_\Phi^{(p)}(\xi|\tilde \xi) = \delta(\xi, \IntegratorSymbol(\tilde\xi)),
\eeq
and the associated acceptance probability becomes
\beq
\label{eq:DefOfAcceptProbForQ}
P_\Phi^{(a)}(\xi|\tilde\xi) = \min\left(1, \frac{\bar p(\xi)}{\bar p(\tilde\xi)} \cdot 
\left|\det\left(\partial Q/\partial \tilde \xi \right) \right|
\right),
\eeq
where $\bar pR=\bar p$ was used and an additional requirement, the so-called reversibility of $\IntegratorSymbol$ according to
\beq
\label{eq:DefOfReversCond}
Q^{-1} = R\IntegratorSymbol R
\eeq
has been imposed and exploited. 

For a practical, numerical implementation it is useful to choose the mapping $\IntegratorSymbol:\confSpace\rightarrow\confSpace$ to be 
a symplectic integrator~\cite{Ruth:1983ad,Channell:1990zu}, since symplectic integrators have the intrinsic property that their Jacobian
determinant is one, \ie $\det\left(\partial Q/\partial \xi \right)=1$. Moreover, they can be chosen such that they obey the reversibility
condition in \eq{eq:DefOfReversCond} for some adequate choice of $R$, and that they almost preserve the probability density according to
$\bar p(\xi)\approx \bar p(\IntegratorSymbol (\xi))$. The latter property is desirable, since it guarantees the acceptance probability 
$P_\Phi^{(a)}(\xi|\tilde\xi)$ in \eq{eq:DefOfAcceptProbForQ} to be close to one, which otherwise would block the considered approach 
in practice.

Actually, a symplectic integrator is an algorithm for the numerical integration of ordinary differential equations. The idea to exploit these techniques here
is to introduce a new parameter, the so-called Monte-Carlo time $\MCtime\in [0,\traLength]$, and to consider trajectories $\xi(\MCtime)\in\confSpace$ in 
configuration space, where $\traLength$ is the so-called trajectory length. In the given setup one can now easily specify 'equations of motion' that leave 
the probability density $\bar p(\xi)$ invariant. Such a set of ordinary differential equations is, for instance, given by
\bea
\label{eq:HMCEqOfMotion1}
\frac{\intd{\Phi}}{\intd{\MCtime}}  &=& \frac{\intd{}}{\intd{\pi}} S[\Phi,\pi,\omega] = \frac{\intd{}}{\intd{\pi}} f(\pi) = \pi,\\
\label{eq:HMCEqOfMotion2}
\frac{\intd{\pi}}{\intd{\MCtime}}  &=& -F[\Phi,\omega] \quad \mbox{with}\quad F[\Phi,\omega] = \frac{\intd{}}{\intd{\Phi}} S[\Phi,\pi,\omega],\\
\label{eq:HMCEqOfMotion3}
\frac{\intd{\omega}}{\intd{\MCtime}}  &=& 0,
\eea
which is not a unique choice but the one that we will consider in the following and $F[\Phi,\omega]$ will later be referred to as molecular
dynamics force. It is furthermore remarked, that one could also have introduced
non-trivial dynamics for the field $\omega$ together with corresponding conjugate momenta. However, this is not necessary, since the fields $\omega$
have already been sampled directly by means of \eqs{eq:DirectSamplingOfOmegaDistributionHMC}{eq:HMCOmegaEtaRelation}. Moreover, allowing also for
such non-trivial dynamics for the fields $\omega$ usually leads to a greater statistical dependence of the generated field configurations, 
which could also be observed in the case of the considered Higgs-Yukawa model. For that reason, non-trivial dynamics will only be assigned to the 
bosonic field $\Phi$ and its conjugate momenta $\pi$.

The action of $\IntegratorSymbol$ on the given configuration $\tilde\xi$ shall here be defined as the result of the numerical integration of the equations 
of motion in \eqs{eq:HMCEqOfMotion1}{eq:HMCEqOfMotion3} over the specified interval $[0,\traLength]$ starting with the initial condition $\xi(0)=\tilde \xi$
according to
\beq
\IntegratorSymbol(\tilde\xi) = \xi(\traLength) \quad \mbox{with}\quad \xi(0)=\tilde \xi.
\eeq

As a practical example for a symplectic integration scheme the so-called leap-frog integrator shall be briefly discussed. 
For that purpose the linear operator $L(H)$, the so-called Liouville operator, is introduced here through its action on 
a given differentiable function $h:\confSpace\rightarrow X$, defined on the configuration space $\confSpace$, according to
\bea
\label{eq:DefOfLiouvilleOp}
L(H) h(\xi)&=& \sum\limits_x \sum\limits_{j=0}^3 \frac{\partial H(\xi)}{\partial \pi_x^j}\cdot \frac{\partial h(\xi)}{\partial \Phi_x^j} 
- \frac{\partial H(\xi)}{\partial \Phi_x^j}\cdot \frac{\partial h(\xi)}{\partial \pi_x^j}
\eea
where the differentiable mapping $H:\confSpace\rightarrow \re$ and the underlying space $X$ are not further specified 
at this point. For clarification the summation over the indices of the fields $\Phi$ and $\pi$ is explicitly given here.
For a given trajectory $\xi(\MCtime)$ obeying the equations of motion in \eqs{eq:HMCEqOfMotion1}{eq:HMCEqOfMotion3} one 
then has
\bea
\derive{}{\MCtime} h(\xi(\MCtime)) &=& L(T+V) h(\xi(\MCtime)),
\eea
leading to
\bea
\label{eq:LiouvilleApp1}
h(\xi(\traLength)) &=& e^{\traLength\cdot L(T+V)} h(\xi(0))
\eea
where $\exp(\traLength \cdot L(T+V))$ will be referred to in the following as the trajectory propagation operator and
the expressions $T$ and $V$ decompose the total action according to $S[\xi] = T[\xi]+V[\xi]$. Here, they are explicitly
given as 
\bea
T[\xi] &=& f(\pi) = \pi^\dagger\pi/2, \quad \mbox{and} \\
V[\xi] &=& S_\Phi[\Phi] + S_F[\Phi,\omega].
\eea
The main idea underlying the considered numerical integrator is then to uniformly subdivide the trajectory length $\traLength$ 
into $\intSteps$ steps, each of which with step size $\intStepSize = \traLength/\intSteps$. Exploiting the linearity of the Liouville
operator together with the Baker-Campbell-Hausdorff formula according to
\bea
\label{eq:BCHformulaSimple}
e^{\intStepSize \cdot L(T+V) + O(\intStepSize^3)} &=& e^{(\intStepSize/2) \cdot L(V)} e^{\intStepSize \cdot L(T)} e^{(\intStepSize/2) \cdot L(V)}
\eea
one arrives at the representation 
\bea
\label{eq:LeapFrogJointBCH}
e^{\traLength\cdot L(T+V)} &=& \left(1+O(\intStepSize^2)  \right) \cdot \prod\limits_{j=1}^{\intSteps}
e^{(\intStepSize/2) \cdot L(V)} e^{\intStepSize \cdot L(T)} e^{(\intStepSize/2) \cdot L(V)}
\eea
of the trajectory propagation operator. Considering \eq{eq:LiouvilleApp1} and \eq{eq:LeapFrogJointBCH} for 
the special case $h:\confSpace \rightarrow \confSpace$ with 
$h(\xi)=\xi$ one directly arrives at an iterative integration scheme, by the help of which one
can compute $\xi(\traLength)$ up to an error of order $O(\intStepSize^2)$ for a given trajectory 
start point $\xi(0)$ according to
\bea
\label{eq:LeapFrogScheme1}
\pi(\intStepSize/2) &\fhs{-3.75mm}=\fhs{-3.75mm}& -[\intStepSize/2]\cdot F[\xi(0)] + \pi(0), \\
\label{eq:LeapFrogScheme2}
\Phi([n+1]\intStepSize) &\fhs{-3.75mm}=\fhs{-3.75mm}& \intStepSize\cdot \pi(n\intStepSize+\intStepSize/2) + \Phi(n\intStepSize),\quad n=0,\ldots \intSteps-1,\quad\quad\\
\label{eq:LeapFrogScheme3}
\pi([n+1]\intStepSize + \intStepSize/2) &\fhs{-3.75mm}=\fhs{-3.75mm}& -\intStepSize\cdot F[\xi([n+1]\intStepSize)] + \pi(n\intStepSize+\intStepSize/2), 
\quad n=0,\ldots \intSteps-2, \quad\quad\quad\\
\label{eq:LeapFrogScheme4}
\pi(\traLength) &\fhs{-3.75mm}=\fhs{-3.75mm}& -[\intStepSize/2]\cdot F[\xi(\intSteps\intStepSize)] + \pi([\intSteps-1]\intStepSize+\intStepSize/2).
\eea
For the derivation of this so-called leap-frog scheme the relations 
\beq
\label{eq:LeapFrogDiscreteIntSteps1}
e^{\intStepSize\cdot L(T)}\xi = \lim\limits_{N\rightarrow\infty} \left( 1+\frac{\intStepSize}{N} L(T) \right)^N \xi 
\quad \mbox{and} \quad
e^{\intStepSize\cdot L(V)}\xi = \lim\limits_{N\rightarrow\infty} \left( 1+\frac{\intStepSize}{N} L(V) \right)^N \xi 
\eeq
with 
\beq
\label{eq:LeapFrogDiscreteIntSteps2}
\left( 1+\frac{\intStepSize}{N} L(T) \right)^N \xi
=
\left(
\begin{array}{*{3}{c}}
1 & \frac{\intStepSize}{N} \frac{\partial f}{\partial \pi} \pi^{-1} & 0 \\
0 & 1 & 0 \\
0 & 0 & 1 \\
\end{array}
\right)^N
\left(
\begin{array}{*{1}{c}}
\Phi \\
\pi \\
\omega \\
\end{array}
\right)
=
\left(
\begin{array}{*{1}{c}}
\Phi + \intStepSize \pi \\
\pi \\
\omega \\
\end{array}
\right)
\eeq
and 
\beq
\label{eq:LeapFrogDiscreteIntSteps3}
\left( 1+\frac{\intStepSize}{N} L(V) \right)^N \xi
=
\left(
\begin{array}{*{3}{c}}
1 & 0 & 0 \\
-\frac{\intStepSize}{N} \frac{\partial S}{\partial \Phi} \Phi^{-1} & 1 & 0 \\
0 & 0 & 1 \\
\end{array}
\right)^N
\left(
\begin{array}{*{1}{c}}
\Phi \\
\pi \\
\omega \\
\end{array}
\right)
=
\left(
\begin{array}{*{1}{c}}
\Phi  \\
\pi - \intStepSize F\\
\omega \\
\end{array}
\right)
\eeq
have been exploited, where $N\in \N$ and $\Phi^{-1}$, $\pi^{-1}$ each denote the inverse of the diagonal matrix given 
by the respective field variables of $\Phi$ and $\pi$.
 
According to \eq{eq:LeapFrogJointBCH} the discretization error of this leap-frog procedure is of order $O(\intStepSize^2)$. Furthermore, 
the Jacobian determinant $\det(\partial Q/\partial \xi)$ associated to this discrete integrator is one, as desired, since the underlying
discrete integration steps in \eq{eq:LeapFrogDiscreteIntSteps1} each have a determinant of one according to 
\eqs{eq:LeapFrogDiscreteIntSteps2}{eq:LeapFrogDiscreteIntSteps3}. The acceptance probability in \eq{eq:DefOfAcceptProbForQ} 
thus becomes 
\beq
\label{eq:DefOfAcceptProbForQSimplectic}
P_\Phi^{(a)}(\xi|\tilde\xi) = \min\left(1, \frac{\bar p(\xi)}{\bar p(\tilde\xi)} \right),
\eeq
which can easily be evaluated numerically. Moreover, the given leap-frog integration scheme is reversible in the sense of \eq{eq:DefOfReversCond}
when defining the mapping $R$ as
\beq
R(\xi) = (\Phi,-\pi,\omega),
\eeq
which fulfills all aforementioned requirements for the selected choice of $f(\pi)$.

For later use the special case of $\hat \lambda=\infty$ shall also be briefly considered. In that scenario the field variables $\Phi_x$
are constraint to a sphere, \ie $\Phi^\dagger_x\Phi_x=N_f$ at every space time point $x$, according to \eq{eq:LatticePhiAction}.
The differential equations in \eqs{eq:HMCEqOfMotion1}{eq:HMCEqOfMotion3}, however, do not explicitly respect that constraint and 
are thus not applicable in a practical approach. A solution is to consider some alternative equations of motion given here as
\bea
\label{eq:FACCLamInfinityPhiEqOfMotionInfiniteLambda}
\frac{d(\Phi_x\theta)}{d\MCtime} &=& \pi_x\theta \cdot \left(\Phi_x\theta\right),  \\
\label{eq:FACCLamInfinityPEqOfMotionInfiniteLambda}
\frac{d(\pi_x\theta)}{d\MCtime} &=& -\left(F_x[\Phi,\omega]\theta\right) \cdot \left(\Phi_x\theta \right)^\dagger,
\eea
obeying exactly the aforementioned constraint according to 
\bea
\derive{}{\MCtime} \left( \Phi^\dagger_x\Phi_x \right) &=& 0
\eea
where the conjugate momenta $\pi$ here only have an inner index running from 1 to 3, \ie $\pi_x\equiv (\pi^1_x,\pi^2_x,\pi^3_x)$,
and an appropriate projection is implicit in \eq{eq:FACCLamInfinityPEqOfMotionInfiniteLambda} annihilating the would-be
contribution to $\pi_x^0$. In this notation the sampling function for the conjugate momenta can again be written as $f(\pi)=\pi^\dagger\pi/2$.
The Liouville operator $L(H)$ associated to this new set of differential equations is different from the form given in 
\eq{eq:DefOfLiouvilleOp} but can analogously be derived. Decomposing the corresponding total action $S[\Phi,\pi,\omega]$ again
into the contributions $T[\xi]$ and $V[\xi]$ as defined before, the finite step size integration steps arising from the exactly integrable
expressions $\exp(\intStepSize L(T))\xi$ and $\exp(\intStepSize L(V))\xi$ then become
\bea
\label{eq:InfiniteLambdaFiniteIntStep1}
\left[\exp(\intStepSize L(T))\xi\right]_x &=& 
\left(
\begin{array}{*{1}{c}}
e^{\intStepSize \pi_x\theta} (\Phi_x\theta) \\
\pi_x \\
\omega_x \\
\end{array}
\right) \quad \quad\mbox{and}\\
\label{eq:InfiniteLambdaFiniteIntStep2}
\left[\exp(\intStepSize L(V))\xi\right]_x &=&  
\left(
\begin{array}{*{1}{c}}
\Phi_x  \\
\pi_x - \intStepSize F_x\\
\omega_x \\
\end{array}
\right).
\eea
With the help of \eqs{eq:InfiniteLambdaFiniteIntStep1}{eq:InfiniteLambdaFiniteIntStep2} the above presented leap frog scheme can
then analogously be adapted to obey the constraint $\Phi_x^\dagger\Phi_x=N_f$ exactly. 

Finally, it is remarked that there are other symplectic integration schemes with higher convergence order 
available~\cite{Yoshida:1990zz,Omelyan:2003zu}, the application of which can reduce the numerical costs of the 
integration procedure. This will be discussed in more detail in \sect{sec:MultiPolynoms}.
Moreover, it is summarized that with the made choices the transition probability density $P_\Phi$ provides the missing link to render 
$P=P_\Phi\circ P_\omega\circ P_\pi $ ergodic. The HMC-code written for studying the considered Higgs-Yukawa model implements exactly 
this transition probability distribution, besides some additional technical enhancements that will be described in detail in 
\chap{chap:SimAlgo} in the broader context of a PHMC-algorithm.

%-----------------------------------------------------------------------------------------------------
\section{The complex phase of the fermion determinant}
\label{sec:ComplexPhaseOfFermionDet}

As discussed in the previous section, the complex phase of the fermion determinant $\det(\fermiMat)$
appears in the functional integration given in \eq{eq:DefOfExpValueWithoutPsi}. For even values of $N_f$ 
this contribution gives just a factor of one provided that the fermion determinant is real. If this is not 
the case the aforementioned phase may yield an important contribution 
to the functional integral which would then need to be respected. For the later introduced PHMC-algorithm aiming at 
evaluating the model for odd values of $N_f$ such unwanted contributions can even arise if the determinant 
is real but not positive. The properties of the fermion matrix $\fermiMat$ with respect to the phase of its 
determinant are therefore studied in the following.

%-----------------------------------------------------------------------------------------------------
\subsection{Degenerate case \texorpdfstring{$y_t=y_b$}{yt=yb}}
\label{eq:ComplexPhaseOfFermionDetDegCase}

For the case of equal top and bottom Yukawa coupling parameters, which we want to start with, one can explicitly
show the fermion determinant to be real. In QCD this is usually demonstrated by exploiting the $\gamma_5$-hermiticity
of the Dirac operator. In the case of the considered Higgs-Yukawa model the fermionic matrix is, however, neither
hermitian nor $\gamma_5$-hermitian. A different approach is therefore needed here. For that purpose the idea is
to exploit the action of the operator $C=\gamma_0\gamma_2$ as well as the action of the matrix $\gamma_5$ 
and the Pauli matrix $\tau_2$ on the Yukawa coupling operator $B$ as well as the Wilson Dirac operator $\Dw$, underlying the 
here actually relevant Neuberger overlap operator $\D$ according to \eq{eq:DefOfNeuberDiracOp}. For degenerate Yukawa 
coupling constants one finds
\bea
\label{eq:GammaTransformOfDB1}
\gamma_0 \gamma_2 \Dw \gamma_2 \gamma_0 = \left[\Dw\right]^T, 
&\quad&
\gamma_0 \gamma_2 B \gamma_2 \gamma_0 = B,\\
\label{eq:GammaTransformOfDB2}
\gamma_5 \Dw \gamma_5 = \left[\Dw\right]^\dagger, 
&\quad&
\gamma_5 B \gamma_5 = B, \\
\label{eq:GammaTransformOfDB3}
\tau_2 \Dw \tau_2 = \Dw, 
&\quad&
\tau_2 B \tau_2 = B^*,
\eea
where the superscripts $T$ and $*$ denote the transposition and the complex conjugation, respectively. 
Exploiting the fact, that the Neuberger overlap operator can be written as a power series in $\Dw$
and $[\Dw]^\dagger$ according to \eq{eq:DefOfNeuberDiracOp}, the relations
\bea
\gamma_0 \gamma_2 \gamma_5 \D \gamma_5 \gamma_2 \gamma_0 = \left[\D\right]^* &\mbox{ and } &
\tau_2 \D \tau_2 = \D
\eea
are directly inherited from \eqs{eq:GammaTransformOfDB1}{eq:GammaTransformOfDB3}. Combining these observations one derives
\beq
V\fermiMat V^\dagger = \fermiMat^*, \quad \mbox{with} \quad V = \gamma_0\gamma_2\gamma_5\tau_2
\eeq
which holds in the mass degenerate case, \ie for $y_t = y_b$. 

With this finding one can easily show, that the eigenvalue spectrum of the fermion matrix $\fermiMat$
comes in complex conjugate pairs. Let $\lambda$ be an eigenvalue of $\fermiMat$
then $\lambda^*$ will also be an eigenvalue according to
\bea
&&
0 = \det\left(\fermiMat - \lambda\ID \right) = \det\left(V\left[\fermiMat - \lambda\ID\right]V^\dagger \right)
= \det\left( \fermiMat^* - \lambda\ID\right)\\
\Rightarrow &&
0 = \det\left( \fermiMat - \lambda^*\ID\right).
\eea
This directly implies that the fermion determinant is real. It does, however, not guarantee the sign of
$\det(\fermiMat)$ to be positive, since eigenvalues with zero imaginary part may appear
which would be their own complex conjugate counterpart. In that case it is possible that there is an odd number of eigenvalues
with exactly zero imaginary part and negative real part which would generate a negative sign of the fermion determinant.
That this scenario can indeed occur in practice is demonstrated in \fig{fig:ExactSpectrumOfM}, where the full spectrum of $\fermiMat$ for
a randomly chosen field configuration $\Phi$, leading to the described situation, is presented. This example is given for the rather 
small lattice volume $4^3\times 8$ such that the explicit calculation of all eigenvalues was still feasible with the help of some
numeric software libraries~\cite{LAPACK:1999zu}.
 
The consequence of the latter observations is that the phase of the fermion determinant does not
need to be considered in the HMC-algorithm presented in \sect{sec:HMCAlgorithm} provided that $N_f$ is even. 
For the case of odd $N_f$, however, which will be treated by means of the PHMC-algorithm introduced
in \sect{sec:basicConceptsOfPHMC}, the potentially alternating sign of $\det(\fermiMat)$ has to be respected 
to guarantee the correctness of the algorithm. This is referred to as 'sign-problem' in the literature. For 
practical calculations the difficulty is to determine the sign without computing all eigenvalues exactly, which 
would be excluded by the complexity of this task growing proportional to $V^3$.

%\includeFigDouble{spectrumNeubergerXiEigenValuesL4x4x4x8Split10Large}{spectrumNeubergerXiEigenValuesL4x4x4x8Split10Small}
\includeFigDouble{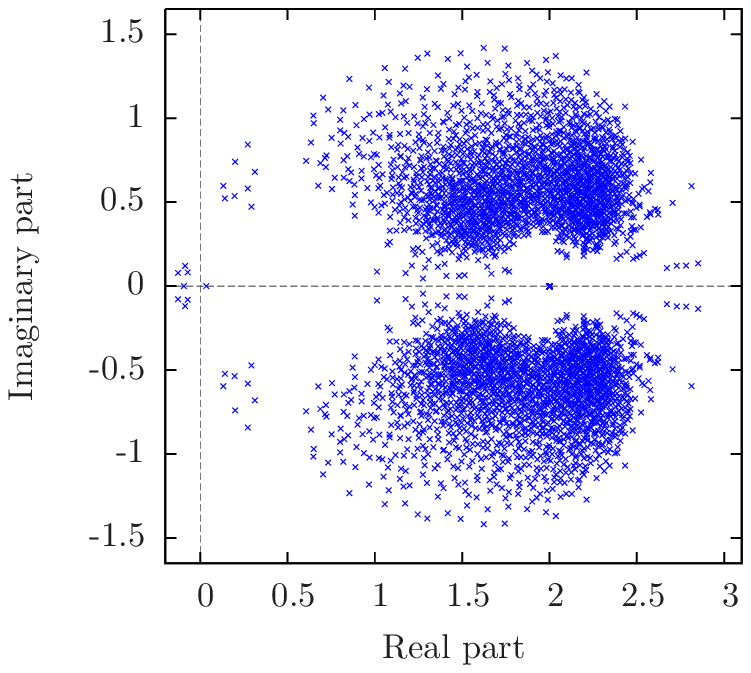}{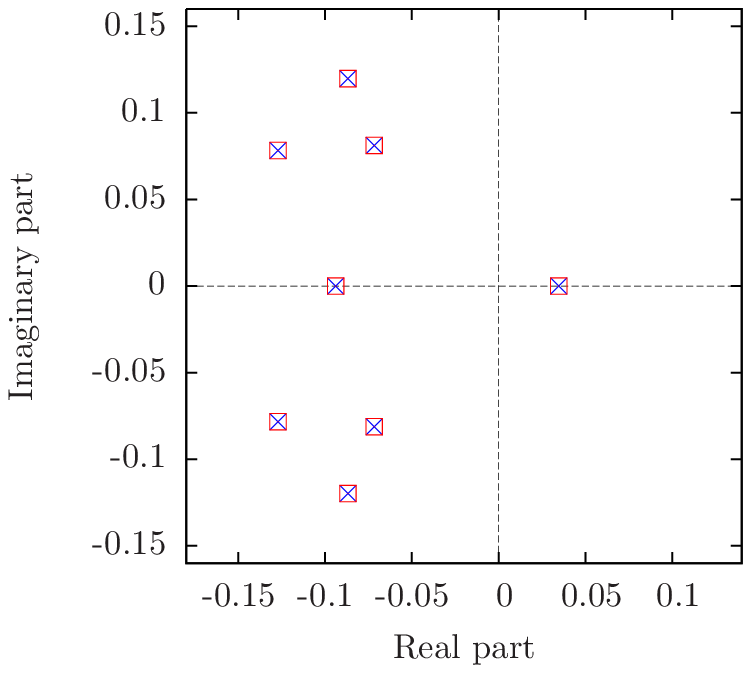}
{fig:ExactSpectrumOfM}
{The complex eigenvalue spectrum of the fermion matrix $\fermiMat[\Phi]$ for a single field configuration $\Phi$ 
on a \lattice{4}{8} is depicted by the blue crosses. The underlying field configuration $\Phi$ has been selected from a 
series of Gaussian sampled configurations, such that the fermion determinant is negative. The degenerate Yukawa coupling 
constants are $\hat y_t =\hat y_b=1.5$. The dashed lines are only meant to guide the
eye. Panel (b) is a magnification of panel (a). The red rectangles in panel (b) indicate the complex conjugate spectrum,
while the blue crosses depict the original eigenvalues, to make apparent the fact that the spectrum of $\fermiMat$ is 
complex conjugate to itself in the degenerate case with equal top and bottom Yukawa coupling constants.}
{Eigenvalue spectrum of the fermion matrix $\fermiMat$ in the degenerate case.}

Exploiting the complex conjugate structure of the eigenvalue spectrum, however, an efficient algorithm
for the determination of the sign of the fermion determinant can be constructed. The underlying observation is
that the sign of $\det(\fermiMat)$ is directly given by the number $k$ of eigenvalues with zero imaginary part 
being situated in the half plane with negative real part according to
\beq
\label{eq:RelationOfFetMwithK}
\sign\det(\fermiMat) = (-1)^k.
\eeq
These eigenvalues can actually be computed in practice with the help of an implementation of the Arnoldi-algorithm~\cite{ARPACK:1998zu},
which allows to find approximations for a specified number of eigenvalues of a non-hermitian matrix, such as the fermion matrix 
$\fermiMat$, with a specified accuracy. For our purpose, the most important feature of the latter algorithm is that the aforementioned
eigenvalues can be calculated ordered by their real part, for instance, in ascending order. By virtue of the Arnoldi-algorithm it 
is therefore feasible in practice to compute only those eigenvalues of the fermion matrix $\fermiMat$ that exhibit a negative real part,
allowing to determine the value of $k$, and thus the sign of $\det(\fermiMat)$ according to \eq{eq:RelationOfFetMwithK}, without having 
to evaluate the full eigenvalue spectrum. 

For clarification it is remarked that the described approach relies on the number of eigenvalues exhibiting a negative
real part to be sufficiently small. Otherwise the determination of these eigenvalues would again become unfeasible from 
a practical point of view. For such cases, there also exist some more advanced
techniques~\cite{Neff:2001xc,Farchioni:2008na}, which derive the sign of $\det(\fermiMat)$ from a transformed operator 
$P(\fermiMat)$ for some polynomial $P$, keeping the relevant eigenvalues, \ie those with zero imaginary part, invariant 
but rotating those with non-zero imaginary part into the positive half-plane. The idea of that approach would be to end 
up with a reduced number of eigenvalues with negative real part, thus lowering the numerical effort for their determination. 
However, it was found that only a small fraction of the eigenvalues being of order $O(10)$ has a negative real-part such 
that the simpler approach already solves the problem satisfactorily in our case. It will therefore be the method of 
choice in the following.

In the analysis of the generated field configurations the sign of the determinant has originally been monitored and included 
into the calculation of all observables as required to guarantee the correctness of the applied algorithm. However, even after 
the analysis of thousands of configurations produced in actual Monte-Carlo calculations not a single negative sign has
been observed for those parameter settings which are of physical interest in this study. At some point the further analysis of 
the sign of the fermion determinant has therefore been stopped in order to save computational resources for more promising 
calculations.

%-----------------------------------------------------------------------------------------------------
\subsection{General case \texorpdfstring{$y_t\neq y_b$}{with non-degenerate Yukawa coupling constants}}
\label{eq:ComplexPhaseOfFermionDetGenCase}

For unequal Yukawa coupling constants, \ie $y_t\neq y_b$, the relation for
the operator $B$ in \eq{eq:GammaTransformOfDB3} is violated, thus invalidating the proof
of the complex conjugate eigenvalue spectrum of $\fermiMat$ given in the preceding  
section. In fact, the spectrum of the fermion matrix is not complex conjugate for unequal 
Yukawa coupling constants as demonstrated in \fig{fig:ExactSpectrumOfMForGeneralCase}, where an 
example of the full eigenvalue spectrum is presented for a randomly chosen field $\Phi$.
It was also numerically confirmed that the corresponding fermion determinant $\det(\fermiMat)$ is 
indeed not real. 

However, in \fig{fig:ExactSpectrumOfMForGeneralCase} one can still observe a complex conjugate main
structure induced by the circular and complex conjugate spectrum of the free overlap operator. The eigenvalues
deviate around this base structure according to the strength of the Yukawa coupling constants. For sufficiently
small coupling parameters $y_t$, $y_b$ one would therefore expect the distribution of the phase $\arg\det(\fermiMat)$ 
to be dominated by a peak around its center value at zero. In the limit $y_t,y_b \rightarrow 0$ one would moreover 
assume the width of this distribution to vanish.
An example of the distribution of $\arg\det(\fermiMat)$ is presented in \fig{fig:DistributionOfDetMPhase}a.
Here, the phase of the fermion determinant has been determined numerically for several thousand randomly
chosen field configurations. The choice of the Yukawa coupling parameters corresponds to the setting that will
later be used for the physically motivated evaluations of the model. One sees that even for this not very small coupling
strength the phase of $\det(\fermiMat)$ does not wildly oscillate over the full interval $[-\pi,\pi]$ but
is already clearly restrained.

Besides the complex conjugate base structure of the fermion matrix there is an additional observation that helps
to explain the small fluctuation of the phase $\arg\det(\fermiMat)$. In fact the spectrum of $\fermiMat$ is still
complex conjugate provided that the underlying field $\Phi$ is identical to its time-reversal $\timeRev{\Phi}$. For 
clarification let $T$ be the time-reversal operator, acting on the space of pseudo fermion fields, according to
\beq
[T\omega]_{t,\vec x} = \omega_{-t,\vec x}.
\eeq
One then finds
\beq
\label{eq:ActionOfTimeRevOnDiracOp}
\gamma_0 T \D T \gamma_0 = \left[\D\right]^\dagger \quad \mbox{and} \quad
\gamma_0 T P_\pm T \gamma_0 = P_\mp,
\eeq
which can be used to establish the announced result. For that purpose let $\lambda$ be an eigenvalue of $\fermiMat$.
Employing the above result one can show that the complex conjugate value $\lambda^*$ is also an eigenvalue of $\fermiMat$
according to
\bea
\label{eq:SpectrumGenCsseTimeRev}
0 &=& \det\left( \fermiMat - \lambda\ID  \right) = \det\left(\gamma_0 T\left[ \fermiMat - \lambda\ID \right] T\gamma_0 \right)  \nonumber\\
&=& \det\left( [\D]^\dagger +  
\left[ \timeRev{\phi}^\dagger \diagY P_- + \diagY \timeRev{\phi} P_+ \right]  \left[\GammaOp \right]^\dagger - \lambda\ID \right)  \nonumber\\
&=& \det\left( \fermiMat^\dagger - \lambda\ID  \right) = \det\left( \fermiMat - \lambda^*\ID  \right)
\eea
where the time-reversed field $\timeRev{\phi}$ is given as
\beq
\label{eq:TimeRevOfPhi}
\left[\timeRev{\phi}\right]_{t,\vec x} = \phi_{-t, \vec x}
\eeq
and $\timeRev{\phi}=\phi$ has been assumed.
The spectrum of the fermion matrix is therefore complex conjugate also in the general case with $y_t \neq y_b$ 
provided that the underlying field configuration $\Phi$ is invariant under time reflection, \ie $\timeRev{\phi} = \phi$,
or equivalently, $\timeRev{\Phi} = \Phi$.

%\includeFigDouble{spectrumNeubergerXiEigenValuesL4x4x4x8Split0024Large}{spectrumNeubergerXiEigenValuesL4x4x4x8Split0024Small}
\includeFigDouble{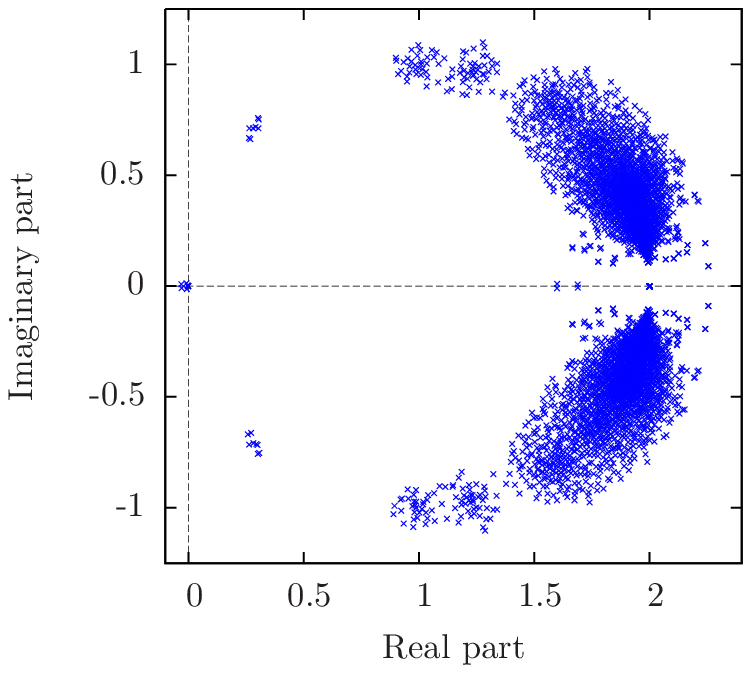}{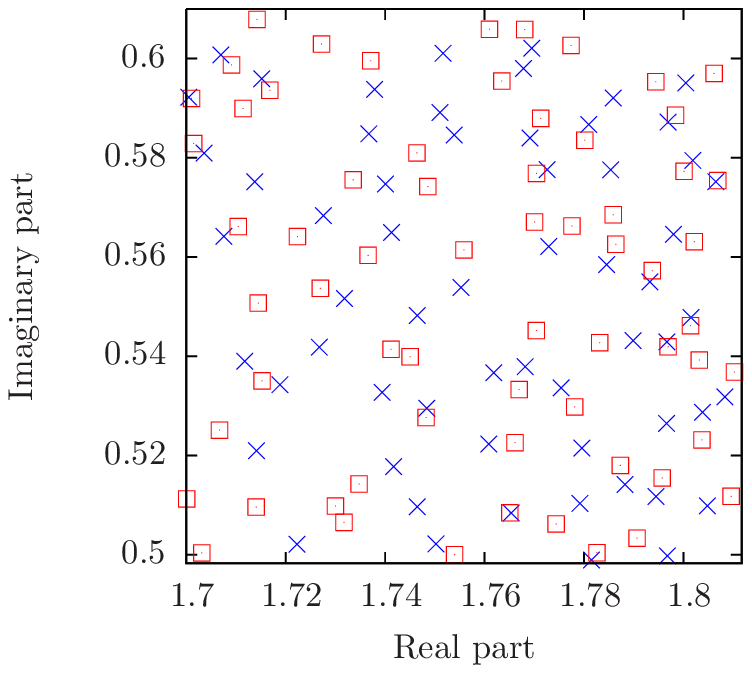}
{fig:ExactSpectrumOfMForGeneralCase}
{The complex eigenvalue spectrum of the fermion matrix $\fermiMat[\Phi]$ for a single field configuration $\Phi$ 
on a \lattice{4}{8} is depicted by the blue crosses. The underlying field configuration $\Phi$ has been randomly sampled
according to a standard Gauss distribution. The non-degenerate Yukawa coupling constants are $\hat y_t=1.0$ and $\hat y_b=0.024$. 
The dashed lines are only meant to guide the eye. Panel (b) is a magnification of panel (a). 
The red rectangles in panel (b) indicate the complex conjugate spectrum, while the blue crosses depict the original eigenvalues,
to make apparent the fact that the spectrum of $\fermiMat$ is not complex conjugate to itself in the general case.
}
{Eigenvalue spectrum of the fermion matrix $\fermiMat$ in the non-degenerate case.}

For clarification it is remarked that the correct derivation of the result in \eq{eq:SpectrumGenCsseTimeRev} actually 
requires to pay special attention to the zero modes of the operator $\GammaOp$. This can be worked in detail
by employing the projection operator in \eq{eq:DefOfProjector} and exploiting the relation in \eq{eq:DetRelationWithProjector}, 
which will be introduced and applied to the considered Higgs-Yukawa model in \sect{sec:LargeYukawaCouplings}. The result 
in \eq{eq:SpectrumGenCsseTimeRev} is, however, unaltered by this more careful calculation.

From the aforementioned result one may conjecture that the constraint on the phase $\arg\det(\fermiMat)$ observed in 
\fig{fig:DistributionOfDetMPhase}a should become even more pronounced in the background of the physically relevant field 
configurations, which exhibit a non-vanishing vacuum expectation value $v$. The reasoning is that in this case the 
amplitude of the zero momentum mode of the scalar field $\Phi$, which is invariant under time reflection, is the most 
dominant mode. An example of this scenario is shown in \fig{fig:DistributionOfDetMPhase}b, where the distribution of 
$\arg\det(\fermiMat)$ is presented as observed in a Monte-Carlo calculation performed in the broken phase with a model 
parameter setup which is typical for the later lattice investigations. One clearly observes the complex phase of the 
respective fermion matrices to be much stronger constraint than for the case of the previously discussed random configurations. 

%\includeFigTriple{PhaseOfFermiDetL4x4x4x4_Random}{PhaseOfFermiDetL4x4x4x4_MC}{PhaseOfFermiDetL6x6x6x6_MC}
\includeFigTriple{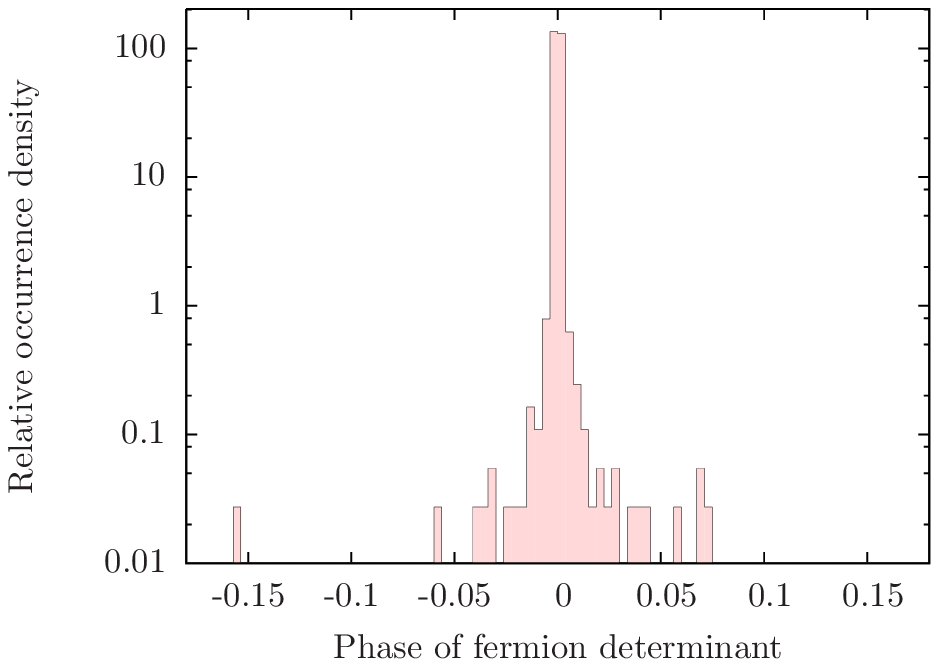}{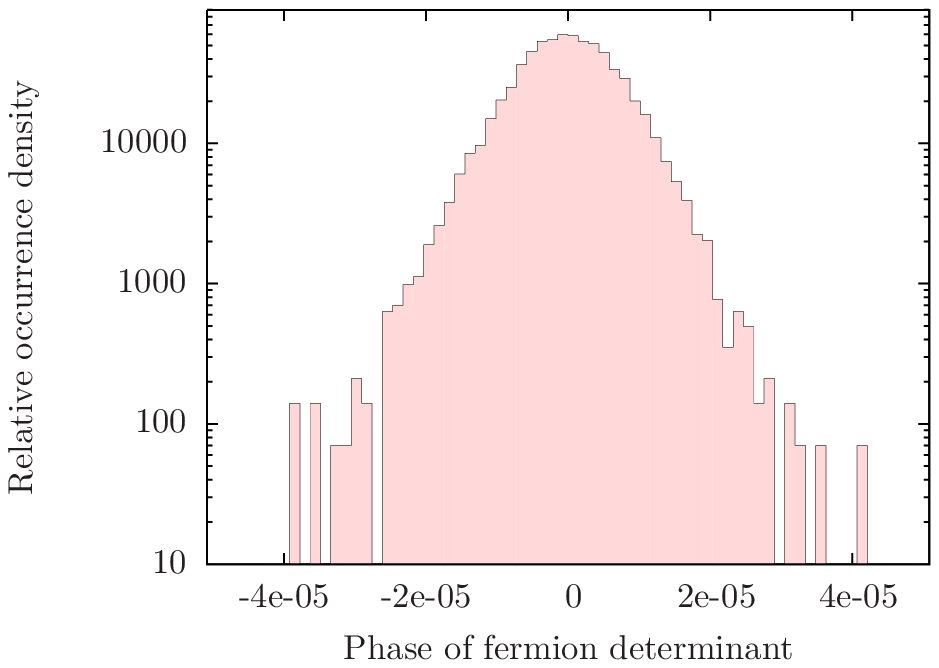}{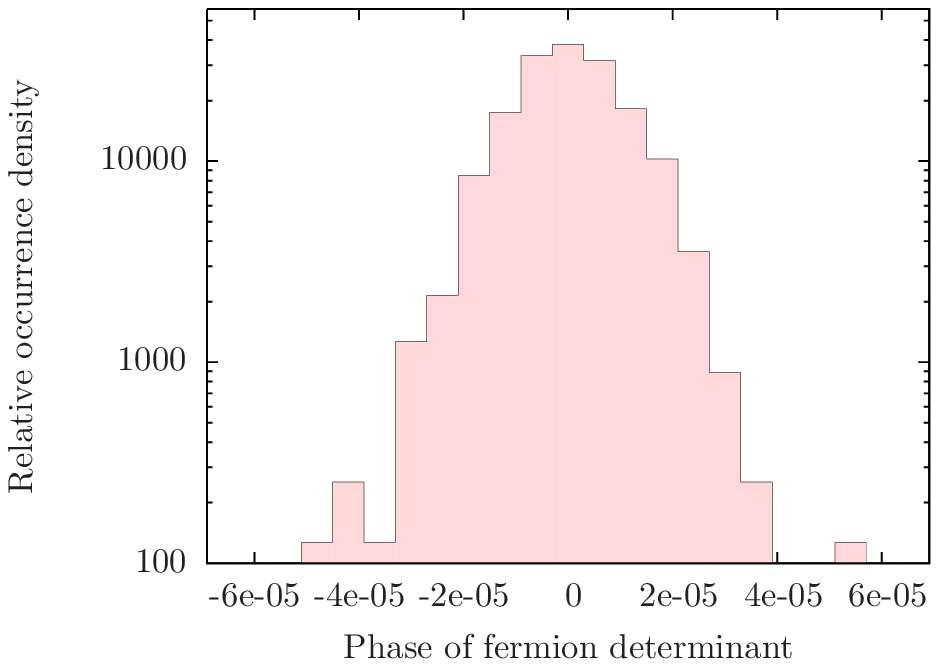}
{fig:DistributionOfDetMPhase}
{The relative occurrence density of the complex phase of the fermionic determinant as determined by the complete
calculation of all eigenvalues of the fermion matrix $\fermiMat[\Phi]$. The underlying field configurations
$\Phi$ have been randomly chosen by means of a standard Gauss distribution in panel (a), while in panels
(b) and (c) the configurations $\Phi$ have been obtained by Monte-Carlo calculations based on the PHMC 
algorithm introduced in \sect{sec:basicConceptsOfPHMC}. The computations were performed on a \lattice{4}{4}in 
panels (a), (b) and on a \lattice{6}{6} in panel (c). The Yukawa coupling constants $y_t$, $y_b$ are
non-degenerate here and were chosen according to the tree level relation in \eq{eq:treeLevelTopMass}
aiming at the reproduction of the physical top and bottom quark mass.
The quartic coupling constant was set to zero and the hopping parameter was tuned to generate an almost 
constant cutoff $\Lambda\approx \GEV{475}$. In panel (a) a single exceptional configuration with $|\arg\det(\fermiMat)|>1.5$
was cut off for a better overview. Note in particular the different scales on the horizontal axes.
}
{Probability distribution of the complex phase of the fermion determinant.}
 
The presented results on the distribution of $\arg\det(\fermiMat)$ as displayed in \fig{fig:DistributionOfDetMPhase}a 
and \fig{fig:DistributionOfDetMPhase}b have been obtained on very small \latticeX{4}{4}{s,} where the direct computations 
of all eigenvalues of $\fermiMat$ were easily feasible. The vital question is how the behaviour of the complex phase
$\arg\det(\fermiMat)$ changes on larger lattice volumes. To obtain an indication on how the lattice volume might influence
the distribution of $\arg\det(\fermiMat)$ an additional Monte-Carlo calculation has been performed on a larger
\latticeX{6}{6}{,} however with identical Yukawa coupling 
constants and tuned hopping parameter aiming at the reproduction of the previously observed lattice result on the 
vacuum expectation value $v$. The distribution of the complex phase of the fermion determinant resulting
from this larger volume computation is presented in \fig{fig:DistributionOfDetMPhase}c. The main observation is that
the increased lattice volume, which was amplified here by a factor of approximately $5$ compared to the \latticeX{4}{4}{,} does
not seem to have a pronounced impact on the distribution of $\arg\det(\fermiMat)$, which is a rather encouraging finding.

However, it should be stressed at this point that there is no obvious way of how the phase of the fermion determinant
can be accounted for in a practical numerical calculation, once the lattice volume becomes too large for
the direct computation of the full determinant. The phase of $\det(\fermiMat)$ will therefore be neglected
in our simulation algorithm in the case of {\textit{unequal}} Yukawa coupling constants.

Consequently, all results obtained for $y_t\neq y_b$ are effected by an unknown systematic uncertainty given by the 
constraint of $\arg\det(\fermiMat)$. The major analysis of the considered model will therefore be performed in the degenerate
case with $y_t=y_b$ where the simulation algorithm is exact. From the presented indications one may, however, 
conjecture that the aforementioned systematic uncertainty of the results obtained for unequal Yukawa coupling constants 
in the {\textit{broken}} phase should be on an acceptable level compared to the statistical uncertainties also on larger 
lattice volumes.

  \chapter{The phase diagram in the large \texorpdfstring{$N_f$-limit}{Nf-limit}}
\label{chap:PhaseDiagram}
 
In this chapter the phase structure of the given lattice Higgs-Yukawa model will be discussed. 
The order parameters being considered in the following to distinguish the
different phases are the vacuum expectation value $v$ of the rotated field $\varphi^{rot}$ as
defined in \eq{eq:DefOfVEV} and its staggered counterpart, the staggered vev $v_s$, which is specified below.
For the purpose of this chapter, however, an equivalent definition for the vev $v$ is used, which expresses  
$v$ and the staggered vev $v_s$ in terms of the so-called magnetization $m$ and the staggered
magnetization $s$ according to
\bea
\label{eq:MagVeVRel}
v = \sqrt{2\kappa} \cdot \langle m \rangle &\mbox{ and }&
v_s = \sqrt{2\kappa} \cdot \langle s \rangle,
\eea
where the latter observables are defined as
\bea
\label{eq:DefOfMagnetizations}
m = \left\langle \left|\frac{1}{V} \sum\limits_x \Phi_x    \right|\right\rangle &\quad \mbox{and} \quad&
s = \left\langle \left|\frac{1}{V} \sum\limits_x e^{-ip_s\cdot x}\cdot \Phi_x    \right|\right\rangle,
\eea
with $p_s$ denoting the so-called staggered momentum $p_s=(\pi,\pi,\pi,\pi)$. 

For clarification it is remarked that there actually is no direct physical interpretation assigned to the staggered
vev $v_s$ in the context of this work, though the latter quantity can indeed be physically relevant in the framework
of other studies, for instance for the investigation of anti-ferromagnetism in solid state physics. 
In the lattice approach employed here to examine the pure Higgs-Yukawa sector of the Standard Model in the spontaneously 
broken phase, however, the staggered vev has to vanish in physical units\footnote{This is a 
stronger requirement than, for instance, for the vev $v$, which only has to vanish in dimensionless lattice units in the
continuum limit.} in any attempt of constructing a physically meaningful continuum limit.
The consideration of both observables nevertheless has an 
interesting motivation beyond the obvious fact that it yields a more complete understanding of the investigated model. 
It directly marks all those regions in parameter space as physically uninteresting for the purpose of this study, 
where the model exhibits a non-vanishing staggered vev $v_s$. Furthermore, it serves as a 
probe for the structure of the ground state of the model, yielding additional
information needed for the determination whether a phase, that exhibits a vanishing vev $v$, 
actually is a symmetric phase, which would require the respective ground state to be $\Phi\equiv 0$.
For instance, a staggered-broken phase ($v=0$, $v_s\neq 0$) would mistakenly be identified as a
symmetric phase, if the staggered vev was not taken into account. 

For the eventual aim of determining the Higgs boson mass bounds the identification of the second order phase  
transition surfaces separating the broken phases with $v\neq 0$, $v_s=0$ and the symmetric phases is an important prerequisite, 
since these regions directly constitute the areas in bare parameter space where the eventual calculations of physical 
interest have to be performed, as discussed in \sect{sec:LatDiscOfContinuumLimit}. In all generality, the phase structure of 
the investigated model cannot be given in analytical form. It can, however, be investigated in certain limits
of the model. 

Following the ideas of \Ref{Hasenfratz:1992xs} different approaches will be presented in the subsequent sections 
that allow for the determination of the phase structure in the limit of an infinite number of fermion generations 
$N_f$. These calculations are performed in the mass degenerate case with equal top and bottom Yukawa coupling 
parameters $\hat y_t=\hat y_b$. The reasoning is that almost all later lattice calculations will be performed in that 
scenario due to the uncertainties\footnote{Additional concerns related to the non-degenerate case will be discussed in 
\sect{sec:SubLowerHiggsMassboundsGenCase}.} otherwise arising from the complex phase of the fermion determinant as discussed 
in \sect{eq:ComplexPhaseOfFermionDetGenCase}.

At the end of each section the analytical results in the respective approximation are compared
to corresponding direct Monte-Carlo calculations. These numerical data have been 
produced by the HMC-algorithm introduced in \sect{sec:HMCAlgorithm}, which by construction is applicable 
for even values of $N_f$ only.

\section{Small Yukawa and quartic coupling constants}
\label{sec:SmalYukCoup}

In this section the phase structure of the considered lattice Higgs-Yukawa model will be 
investigated in the large $N_f$-limit for small values of the quartic coupling parameter 
$\hat \lambda$ and the degenerate Yukawa coupling constant $\hat y \equiv \hat y_t = \hat y_b$. 
More precisely, the limit $N_f\rightarrow \infty$ is studied with the coupling constants 
scaling according to
\beq
\label{eq:LargeNBehaviourOfCouplings1}
\hat y = \frac{\tilde y_N}{\sqrt{N_f}}\,,\quad  
\hat \lambda = \frac{\tilde \lambda_N}{N_f}\,,\quad
\kappa = \tilde \kappa_N\,,
\eeq
where the parameters $\tilde y_N$, $\tilde \lambda_N$, and $\tilde \kappa_N$ are held constant
in that limit procedure.

\subsection{Analytical calculations}
\label{sec:SmallYukawaCouplings}

The starting point for the subsequent analytical calculation of the phase structure is the observation that 
in the considered limit a factor of $N_f$ can be factorized out of the action $\Seff[\Phi]$ given
in \eq{eq:DefOfEffAction} according to
\beq
\Seff[\Phi] = N_f \cdot \breve \Seff[\breve \Phi], \quad \breve \Seff[\breve \Phi] = \breve S_\Phi[\breve \Phi] 
+ \breve \SFeff[\Phi],
\quad \breve\Phi = N_f^{-1/2} \cdot \Phi,
\eeq
where the rescaled bosonic and fermionic actions $\breve S_\Phi[\breve \Phi]$ and $\breve \SFeff[\Phi]$
are given as
\bea
\breve S_\Phi[\breve \Phi] &=& -\kappa\sum_{x,\mu} \breve\Phi_x^{\dagger} \left[\breve\Phi_{x+\mu} + \breve\Phi_{x-\mu}\right]
+ \sum_{x} \breve\Phi^{\dagger}_x\breve\Phi_x + \tilde\lambda_N \sum_{x} \left(\breve\Phi^{\dagger}_x\breve\Phi_x - 1\right)^2,
\eea
and\footnote{According to $\hat y\cdot\Phi = \tilde y_N \cdot \breve \Phi$ the fermion determinant can also be
written entirely in terms of the rescaled parameters and fields, which will later explicitly be done.}
\bea
\breve \SFeff[\Phi] &=& -\log\det\left( \fermiMat[\Phi] \right),
\eea
respectively. 
In the limit $N_f\rightarrow \infty$ the functional integration over all field configurations in \eq{eq:DefOfLatticeFuncIntWithoutPsi2} 
then reduces to a sum over all absolute minima\footnote{With some abuse of notation the notion 'minimum' is used here and in the 
following also as a synonym for the more precise formulation 'location of the minimum'.}
$\breve \Phi'$ of $\breve \Seff[\breve\Phi]$. The expectation values
of the rescaled order parameters $\breve v = N_f^{-1/2}\cdot v$ and $\breve v_s = N_f^{-1/2}\cdot v_s$ 
are then directly obtainable from the knowledge of these ground states. 

To make a determination of these states feasible in practice we consider here the restricted ansatz
\beq
\label{eq:staggeredAnsatz}
\Phi'_x = \sqrt{N_f} \cdot \left(\breve m_\Phi \hat\Phi_1 + \breve s_\Phi e^{ip_sx} \hat\Phi_2    \right)
\eeq
for the ground states of $\Seff[\Phi]$, taking only a constant and a staggered mode into account, where the real numbers
$\breve m_\Phi$, $\breve s_\Phi$ and the unit vectors $\hat \Phi_1$, $\hat \Phi_2$ specify the amplitudes and orientations, 
respectively, of the constant and staggered momentum modes. This restriction is 
reasonable, since we are eventually interested in locating those phases, in which the ground states are translational
invariant. The staggered mode is also included within this ansatz to allow for the determination of $v_s$, which
is desirable for the aforementioned reasons. For a given ground state $\Phi'$ parametrized in terms of 
\eq{eq:staggeredAnsatz} the corresponding contribution to the rescaled order parameters $\breve v$ and $\breve v_s$
would then be given as $\sqrt{2\kappa}\breve m_\Phi$ and $\sqrt{2\kappa} \breve s_\Phi$, respectively.

It is, however, not guaranteed that the true ground states of the action $\Seff[\Phi]$ can actually be parametrized in terms
of \eq{eq:staggeredAnsatz}. In fact, we will later observe phases derived by the given ansatz, that simultaneously exhibit a 
non-vanishing vev $v$ and a non-vanishing staggered vev $v_s$, which can be understood as an indication for the true ground state
structure being more complex than assumed in this simple approach. 
The restriction of the consideration to the ansatz in \eq{eq:staggeredAnsatz} has thus to be understood as a working hypothesis
only. The obtained analytical results on the phase structure are therefore later explicitly confronted with corresponding 
numerical data obtained in direct Monte-Carlo calculations. 

So far, the given conceptual background of the large $N_f$-analysis is rather crude. A deeper understanding can be 
achieved by considering the constraint effective potential~\cite{Fukuda:1974ey,O'Raifeartaigh:1986hi} which is defined as
\bea
\label{eq:DefOfContraintEffPotForPhaseStruc}
VU[\underline{\tilde \Phi}_0, \underline{\tilde \Phi}_{p_s}] &=& 
-\log\left(\int \intD{\psi} \intD{\bar \psi} \left[\prod\limits_{0\neq k \neq p_s}\intd{\tilde \Phi_k}  \right]\,\,
e^{-S_\Phi[\Phi] - S_F[\Phi,\psi,\bar\psi]} 
\Bigg|_{\tilde\Phi_0=\underline{\tilde \Phi}_0, \tilde\Phi_{p_s}=\underline{\tilde \Phi}_{p_s}}     
\right)\quad\quad
\eea
up to some constant independent of the arguments $\underline{\tilde \Phi}_0$ and
$ \underline{\tilde \Phi}_{p_s}$. Here the original definition in \Ref{Fukuda:1974ey,O'Raifeartaigh:1986hi}
has been extended to depend not only on the amplitudes of the constant momentum modes
but also on the amplitudes of the staggered modes. In the above definition the expression $\tilde \Phi$ 
denotes the scalar field $\Phi$ in momentum space according to
\beq
\label{eq:FTofPhi}
\tilde \Phi_p = \frac{1}{\sqrt{V}}\, \sum\limits_x e^{-ipx}\cdot \Phi_x.
\eeq
With the help of this constraint effective potential the expectation value of any observable 
$O[{\tilde \Phi}_0, {\tilde \Phi}_{p_s}]$ depending solely on 
the constant and the staggered modes of the scalar field $\Phi$, such as the magnetizations
$m$ and $s$ defined in \eq{eq:DefOfMagnetizations}, can then be written as
\bea
\label{eq:DefOfEffIntegralOfObs}
\langle O[\tilde \Phi_0, \tilde \Phi_{p_s}] \rangle = \frac{1}{Z} \FuncIntAvg_E[O[\tilde \Phi_0, \tilde
\Phi_{p_s}]], \quad Z = \FuncIntAvg_E[1],
\eea
where the latter effective integral is given as 
\beq
\label{eq:EffectivePotnetialFuncIntegral}
\FuncIntAvg_E[O[\tilde \Phi_0, \tilde \Phi_{p_s}]] = \int \intd{\tilde \Phi_0} \intd{\tilde \Phi_{p_s}}
O[\tilde \Phi_0, \tilde \Phi_{p_s}] \cdot e^{-VU[\tilde \Phi_0, \tilde \Phi_{p_s}]}.
\eeq
The constraint effective potential thus has a manifest interpretation.
It is directly connected to the probability distribution $p(\tilde \Phi_0, \tilde \Phi_{p_s})$
of the momentum mode amplitudes $\tilde \Phi_0$ and $\tilde \Phi_{p_s}$ according to 
$p(\tilde \Phi_0, \tilde \Phi_{p_s}) = \exp(-VU[{\tilde \Phi}_0, {\tilde \Phi}_{p_s}])/Z$
as observed, for instance, in a corresponding Markov process. For sufficiently large lattice 
volumes $V$ the expectation value of the considered observable $O[{\tilde \Phi}_0, {\tilde \Phi}_{p_s}]$
can thus directly be obtained from the minima of the constraint effective potential.
This is clear from the integral representation of the latter quantity in \eq{eq:DefOfEffIntegralOfObs}, 
which is obviously dominated by the minima of $U[{\tilde \Phi}_0, {\tilde \Phi}_{p_s}]$ 
for sufficiently large lattice volumes due to the explicit appearance of the volume\footnote{The volume
appears explicitly in \eq{eq:EffectivePotnetialFuncIntegral} according to the actions $S_\Phi$ and $S_F$ 
underlying the definition of $VU[{\tilde \Phi}_0, {\tilde \Phi}_{p_s}]$ being extensive quantities, 
scaling proportional to the lattice volume $V$.} $V$ in the weight factor 
$\exp(-VU[{\tilde \Phi}_0, {\tilde \Phi}_{p_s}])$ in \eq{eq:EffectivePotnetialFuncIntegral}.

It is further remarked that the {\textit{finite}} volume constraint effective potential is not a convex function
in general~\cite{O'Raifeartaigh:1986hi}. In particular, it can, for instance, exhibit the
famous Mexican hat structure with a non-zero minimum, being unique apart from the trivial degeneracy
induced by the symmetries of the model, signaling then the situation of spontaneous symmetry breaking.
This is illustrated in \fig{fig:EffPotIllustration}a, sketching a finite volume example of the less general
potential $U[{\tilde \Phi}_0]$, defined analogously to \eq{eq:DefOfContraintEffPotForPhaseStruc} including, however,
also the staggered mode in the bosonic integration, as originally defined in \Ref{O'Raifeartaigh:1986hi}.

This is in contrast to the situation of the {\textit{infinite}} volume, where the constraint effective potential 
$U[{\tilde \Phi}_0]$ can be proven to be always convex~\cite{O'Raifeartaigh:1986hi}. In infinite volume the latter
constraint effective potential can therefore never assume the aforementioned Mexican hat structure, or any other 
non-convex form. In a model parameter setup, where spontaneous symmetry breaking occurs, the infinite volume constraint 
effective potential would then take the form of a convex hull, for instance, of the Mexican hat potential
which, however, does not possess a unique minimum even if one factors out the aforementioned degeneracy induced 
by the symmetries of the model. This observation can be generalized to the statement that the infinite volume 
constraint effective potential, which has to be convex, becomes exactly flat at its minima and possesses 
a manifold of degenerate minima in the latter sense, if the model exhibits spontaneous symmetry breaking. A corresponding 
example is sketched in \fig{fig:EffPotIllustration}c.

For the case of a {\textit{finite}} volume we shall always assume the constraint effective potential to possess a single, unique
minimum in the following, where the notion of a 'single' minimum is used here with some abuse of notation to distinguish the 
case illustrated in \fig{fig:EffPotIllustration}a from the one in \fig{fig:EffPotIllustration}c. In fact, the minimum in the 
former case is also degenerate according to the symmetries of the model. This shall, however, implicitly be understood in the 
following when using the terms 'single minimum' or 'unique minimum' versus 'degenerate minimum'. This is a reasonable notation, 
since we will only be interested in observables being invariant under the latter symmetries.

Concerning the {\textit{infinite}} volume, however, the alert reader might wonder, whether the statement, that the integral 
in \eq{eq:EffectivePotnetialFuncIntegral} is dominated by the minimum of the constraint effective potential, does actually 
make sense in the scenario of spontaneous symmetry breaking, since the infinite volume constraint effective potential then 
possesses a manifold of degenerate minima as illustrated in \fig{fig:EffPotIllustration}c. The key observation in this 
respect is the following. For any finite volume, the constraint effective potential does not need to be convex. In particular, 
it can then exhibit, for instance, the Mexican hat structure with a single, non-zero minimum. As the volume increases the 
constraint effective potential finally becomes convex. In the sketched example of the Mexican hat potential, the bump in 
the center vanishes in the thermodynamic limit, as illustrated in \fig{fig:EffPotIllustration}b. The crucial question is, 
how fast it decreases. Explicit lattice calculations~\cite{O'Raifeartaigh:1986hi} of the constraint effective potential 
performed in the four-dimensional pure $\Phi^4$-theory find that the latter decline is much slower than $O(1/V)$. The weight 
factor $\exp(-VU[{\tilde \Phi}_0, {\tilde \Phi}_{p_s}])$ appearing in \eq{eq:EffectivePotnetialFuncIntegral}, however,
explicitly contains a factor of the volume $V$, such that the effect of the integral in \eq{eq:EffectivePotnetialFuncIntegral}
being dominated by the non-degenerate minimum of the finite volume constraint effective potential becomes in fact more 
pronounced as the finite volume is increased, although the constraint effective potential itself finally becomes convex. 
Conceptually, it is therefore not the infinite volume limit of the constraint effective potential itself, that is actually 
of interest, but instead the infinite volume limit of the position of its finite volume minimum,  when studying the thermodynamic 
limit of the underlying model.

%\includeFigTriple{ContraintEffectivePotentialIllustrationFiniteVol}{ContraintEffectivePotentialIllustrationFiniteLargerVol}{ContraintEffectivePotentialIllustrationInfiniteVol}
\includeFigTriple{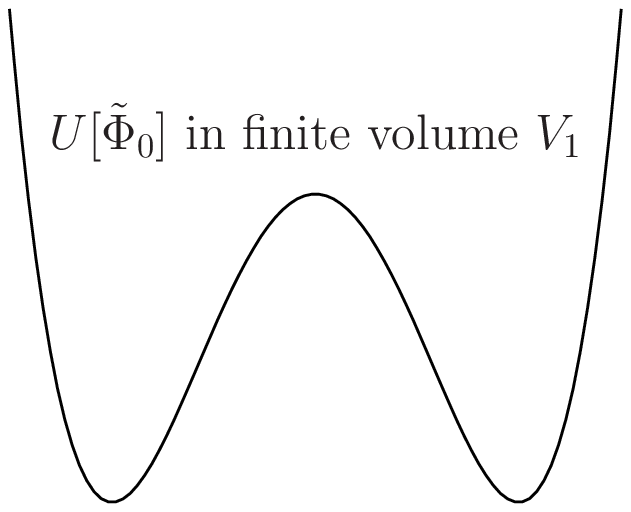}{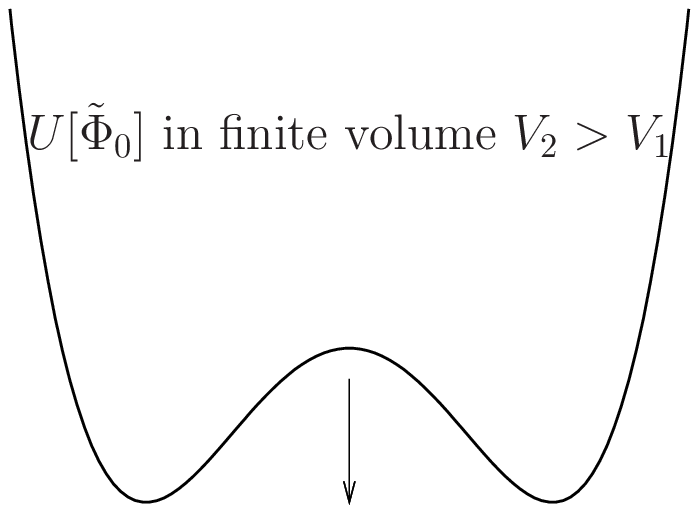}{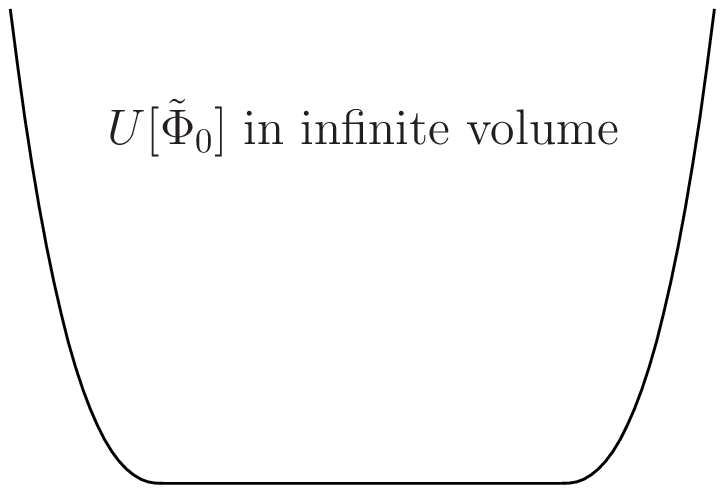}
{fig:EffPotIllustration}
{Some qualitative examples of the constraint effective potential $U[{\tilde \Phi}_0]$ in the scenario of spontaneous symmetry 
breaking are illustrated. Panels (a) and (b) represent the situation of some finite volumes $V_1$ and $V_2$, respectively, with
$V_1< V_2$, while panel (c) refers to the infinite volume. For practical reasons only the dependence of $U[{\tilde \Phi}_0]$
(vertical axis) on one component of the four-component vector $\tilde \Phi_0$ (horizontal axis) is depicted. 
These sketches only serve the purpose of illustration with the intention to express that the non-convex finite volume
constraint effective potential becomes convex in the infinite volume limit, leading then -- in the case of spontaneous
symmetry breaking -- to the emergence of degenerate minima, even if one factors out the obvious degeneracy induced 
by the symmetries of the model.
}
{Illustration of the constraint effective potential in finite and infinite volume.}
 
It is remarked that the above considerations referred to the exact constraint potential. In the following, however, we will calculate 
the constraint effective potential only to a finite order in perturbation theory, in fact to tree-level in this chapter and to one-loop 
order, at least with respect to the purely bosonic contributions, in \sect{sec:EffPot}. The computed results on the constraint effective 
potential will therefore not be convex even in the infinite volume limit. From a practical point of view the above subtleties of taking 
the infinite volume limit of the potential itself or rather of its minimum will therefore not play a role in the following. In this 
respect we will occasionally speak about 'the minimum' of the constraint effective in infinite volume, having then always in mind what 
has been said here.

It is further remarked that the constraint effective potential $U[{\tilde \Phi}_0]$ in finite volume is not the same object as 
the effective potential~\cite{Goldstone:1962es,JonaLasinio:1964cw} $\Gamma[{\tilde \Phi}_0]$, being defined by 
means of a Legendre transformation of the generating functional $W[J]=\log(Z[J])/V$. It is only the thermodynamic limit 
that drives~\cite{O'Raifeartaigh:1986hi} the constraint effective potential $U[{\tilde \Phi}_0]$ to converge to
the effective potential $\Gamma[{\tilde \Phi}_0]$. In contrast to $U[{\tilde \Phi}_0]$, the effective potential 
$\Gamma[{\tilde \Phi}_0]$ is always a convex function, also in finite volume. In fact, one can show~\cite{O'Raifeartaigh:1986hi} 
that $\Gamma[{\tilde \Phi}_0]$ is the convex hull of $U[{\tilde \Phi}_0]$. For clarification it is, however, pointed out that 
the effective potential $\Gamma[{\tilde \Phi}_0]$ will not be considered in the present work.

For the here intended investigation of the phase structure of the underlying Higgs-Yukawa model, the basic idea is to determine 
the absolute minimum of the constraint effective potential $U[{\tilde \Phi}_0, {\tilde \Phi}_{p_s}]$ 
with respect to ${\tilde \Phi}_0$ and ${\tilde \Phi}_{p_s}$, which would then directly yield the considered 
order parameters $v$ and $v_s$ at least for sufficiently large lattice volumes $V$. This can be done, for instance, 
in the large $N_f$-limit, where the analytical calculation of the constraint effective potential can greatly be simplified.
For that purpose we equivalently rewrite the definition of the constraint effective potential in \eq{eq:DefOfContraintEffPotForPhaseStruc}
in terms of a purely Gaussian contribution $S_0[\Phi',\Phi,\psi,\bar\psi]$, a solely bosonic tree-level part $S_\Phi[\Phi']$, and the
remaining, interacting contribution $S_I[\Phi',\Phi,\psi,\bar\psi]$ according to
\bea
\label{eq:DefOfEffPotInMomSpace}
e^{-VU[\Phi']} &\fhs{-3mm}=\fhs{-3mm}& e^{-S_\Phi[\Phi']} \fhs{-1mm}
\int\fhs{-1mm} \intD{\psi} \intD{\bar \psi}\fhs{-1mm} \left[\prod\limits_{0\neq k \neq p_s}\intd{\tilde \Phi_k}  \right]
e^{-S_I[\Phi',\Phi,\psi,\bar\psi]} \cdot e^{-S_0[\Phi',\Phi,\psi,\bar\psi]}
\Bigg|_{{\tilde\Phi_0= \breve m_\Phi\cdot \sqrt{VN_f}\hat\Phi_1,}\atop{ \tilde\Phi_{p_s}=\breve s_\Phi\cdot \sqrt{VN_f}\hat\Phi_2}} \quad\quad\quad
\eea
where $\Phi'$ has been defined in \eq{eq:staggeredAnsatz} and the expressions $U[\tilde\Phi_0, \tilde \Phi_{p_s}]$ and $U[\Phi']$ 
have been treated synonymously, implicitly assuming the obvious mappings 
$\Phi' \leftrightarrow (m_\Phi,s_\Phi,\hat\Phi_1,\hat\Phi_2)\leftrightarrow (\tilde\Phi_0,\tilde\Phi_{p_s})$ 
here and in the following.

For the sake of simplicity, we restrict the subsequent considerations to the case of equal orientations, \ie 
$\hat \Phi \equiv \hat\Phi_1 = \hat\Phi_2$. In that scenario the so far unspecified expressions for 
the purely Gaussian contribution $S_0[\Phi',\Phi,\psi,\bar\psi]$ and the interacting contribution
$S_I[\Phi',\Phi,\psi,\bar\psi]$ can be given as
\beq
\label{eq:DefOfGaussActionLargeNf}
S_0[\Phi',\Phi,\psi,\bar\psi] = \sum\limits_{i=1}^{N_f} \bar\psi^{(i)} \fermiMat[\Phi'] \psi^{(i)} 
+ \frac{1}{2} \sum\limits_{0\neq k \neq p_s} \tilde\Phi^\dagger_{k}               
\left[2-4\tilde\lambda_N-\sum_\mu 4\kappa \cos(k_\mu)  \right] 
\tilde\Phi_{k}  
\eeq
and\footnote{Appropriate modulo operations are implicit in 
\eq{eq:DefOfInterActionLargeNf} to guarantee all sums and differences of lattice momenta 
to be contained in $\ImpSpace$.} 
\bea
\label{eq:DefOfInterActionLargeNf}
S_I[\Phi',\Phi,\psi,\bar\psi] &=& \sum\limits_{i=1}^{N_f} \bar\psi^{(i)} B[\Phi-\Phi']\GammaOp \psi^{(i)}
+ \frac{\hat\lambda}{V} \fhs{-1mm}\sum\limits_{k_1,\ldots,k_4}\fhs{-1mm} \delta_{k_1+k_3,k_2+k_4} 
\tilde\Phi^\dagger_{k_1}\tilde\Phi_{k_2}\tilde\Phi^\dagger_{k_3}\tilde\Phi_{k_4} \quad\quad \\
&-& \frac{\hat\lambda}{V}  \left( \tilde\Phi_0^\dagger \tilde\Phi_0\tilde\Phi_0^\dagger \tilde\Phi_0
+  \tilde\Phi_{p_s}^\dagger \tilde\Phi_{p_s}\tilde\Phi_{p_s}^\dagger \tilde\Phi_{p_s}
+6 \tilde\Phi_0^\dagger \tilde\Phi_0 \tilde\Phi_{p_s}^\dagger \tilde\Phi_{p_s} \right) . \nonumber
\eea

In the given setup the constraint effective potential $U[\Phi']$ can now formally be expanded in terms of 
a diagrammatic series, each diagram of which having assigned a certain inverse power of $N_f$~\cite{Coleman:1988zu}. 
This is achieved by expanding the interacting contribution $\exp(-S_I)$ into a power series and performing the
resulting Gaussian integrations over the bosonic modes $\tilde\Phi_{k}$, $k\in\ImpSpace/\{0,p_s\}$ and  
over the fermion fields $\psi$, $\bar\psi$. Examples for such diagrams are presented in \fig{fig:EffPotExampleDiagrams} together
with their respective weight in inverse powers of $N_f$. Basically, the idea is that due to the considered scaling
given in \eq{eq:LargeNBehaviourOfCouplings1} each $\hat\lambda$-vertex carries a factor of $N_f^{-1}$, while every 
$\hat y$-vertex has a factor $N_f^{-1/2}$ assigned. Furthermore, the amplitudes of the momentum modes with $k=0$ or $k=p_s$ 
scale proportional to $N_f^{1/2}$ according to the assumed ground state structure as discussed at the beginning of this 
section, while the propagators of all other momenta are of order $O(1)$, which formally follows from the given form of the 
purely Gaussian contribution $S_0$. A further remark on the latter statement follows below.

%\includeFigSingleMedium{LPTDiagramsForEffectivePotential}
\includeFigSingleMedium{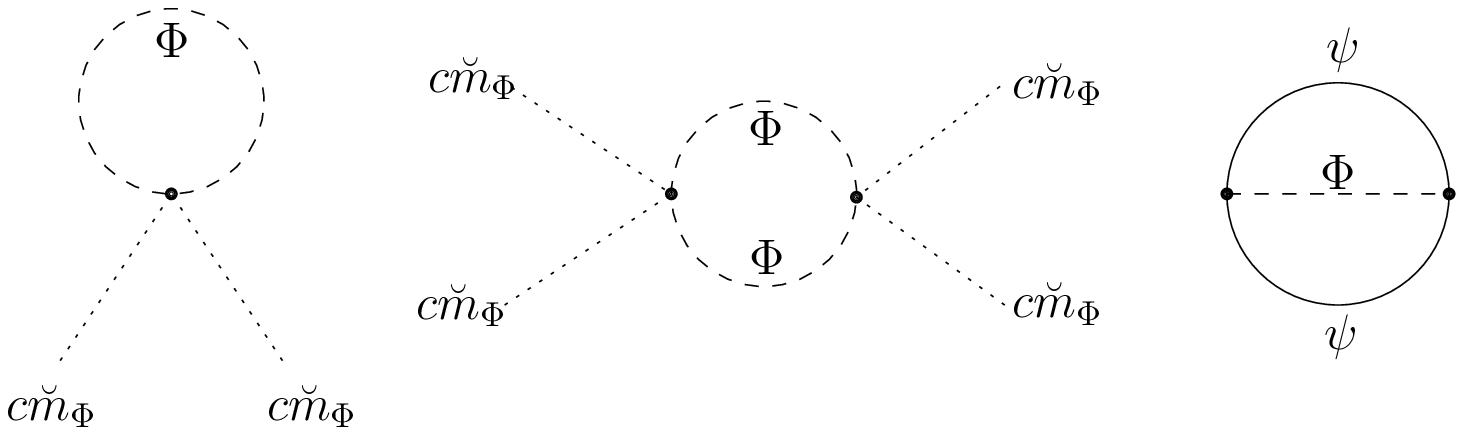}
{fig:EffPotExampleDiagrams}
{An illustration of some diagrams arising from the expansion of $\exp(-S_I)$ with $S_I$ defined in \eq{eq:DefOfInterActionLargeNf} is given. 
While all momenta of the fermion propagators enter the perturbative expansion, the momenta of the bosonic propagators appearing in the
respective loop summations are restricted to $k\in\ImpSpace/\{0,p_s\}$ excluding the constant and the staggered mode due to the 
definition of the constraint effective potential in \eq{eq:DefOfEffPotInMomSpace}.
The dotted lines actually do not indicate propagators but only a multiplication with the outer amplitudes $\tilde \Phi_{0}=c\breve m_\Phi\hat\Phi$ and
$\tilde \Phi_{p_s}=c\breve s_\Phi\hat\Phi$, respectively, where the underlying vector structure has already been multiplied out in 
the presented sketch and the constant $c=\sqrt{VN_f}$ has been introduced for the sake of a shorter notation. Diagrams with staggered outer 
momentum modes exist equally well but are not presented here. The contributions of the given 
diagrams to $U[\Phi']/N_f$ are thus of order $O(N_f^{-1})$ for the left two diagrams and $O(N_f^{-2})$ for the right diagram, which also 
depends on the outer amplitudes through their explicit appearance in the bare fermion propagator.
}
{Examples of diagrams contributing to the constraint effective potential.}
 
In the large $N_f$-limit with the coupling constants scaling according to \eq{eq:LargeNBehaviourOfCouplings1} 
one then directly finds
\bea
\frac{1}{N_f}U\left[\Phi'\right] 
&\stackrel{N_f\rightarrow\infty}\longrightarrow&
\tilde U(\breve m_\Phi, \breve s_\Phi) \cdot \left[1+O(N_f^{-1})\right], \quad\quad\\
\label{eq:DefOfScaledEffPot}
VN_f\cdot \tilde U(\breve m_\Phi, \breve s_\Phi) &=& S_\Phi[\Phi'] - N_f\cdot \log\det\left(\fermiMat[\Phi']\right) 
\eea
up to some constant factor\footnote{Here and in the following 'constant' means independent of $\breve m_\Phi$, $\breve s_\Phi$, and $\hat\Phi$.}
\footnote{$\tilde U(\breve m_\Phi, \breve s_\Phi)$ does not depend on the orientation $\hat \Phi$ as will later be seen.}
provided that at least one of the two variables $\breve m_\Phi, \breve s_\Phi\in\re$ is non-zero and that the exponents
of the Gaussian factors in the aforementioned Gaussian integrations over the bosonic modes $\tilde\Phi_{k}$, $k\in\ImpSpace/\{0,p_s\}$  
are positive, which is guaranteed for all values of $\breve m_\Phi$ and $\breve s_\Phi$, provided that 
\beq
\label{eq:SmallYAnaPrecondition}
1-2 \tilde\lambda_N - \sum\limits_\mu 2\kappa\cos(k_\mu)>0 \quad \mbox{ for all } k \in\ImpSpace/\{0,\,p_s\}.
\eeq 
For the considered purpose of determining the phase structure of the Higgs-Yukawa model in the limit $N_f\rightarrow\infty$
one is therefore left with the problem of finding the absolute minimum of $\tilde U(\breve m_\Phi, \breve s_\Phi)$ with 
respect to $\breve m_\Phi, \breve s_\Phi$, thus recovering the identical approach already derived from the simple consideration 
at the beginning of this section.

It is, however, remarked here that the given expansion is only valid if the true ground state of the system actually has the form
of \eq{eq:staggeredAnsatz}, which has to be assumed. The reason is that one needs the propagators in the aforementioned
diagrammatic expansion to be of order $O(1)$ and not $O(N_f)$, for instance, to derive the above result. In this respect one has 
not gained additional information as compared to the very simple consideration at the beginning of this section.
Given a scenario with a ground state in the form of \eq{eq:staggeredAnsatz}, however, one has learned that the 
the correction to $\tilde U(\breve m_\Phi,\breve s_\Phi)$ in the large $N_f$-limit is of order $O(N_f^{-1})$, which follows by 
working out the combinatorically possible diagrams arising from the expansion of $\exp(-S_I)$ as systematically discussed
in \Ref{Coleman:1988zu}. In contrast to the aforementioned earlier
consideration the lastly discussed approach based on the effective potential moreover opens the possibility of determining
these corrections in powers of $N_f^{-1}$. This idea will be picked up again in \sect{sec:DepOfHiggsMassOnLambda}, where 
corrections to $\tilde U(\breve m_\Phi,\breve s_\Phi)$ will actually be calculated, though in a different context than the 
large $N_f$-analysis.
 
Here, however, we continue with the consideration of the limit $N_f\rightarrow \infty$. 
As already pointed out before we have restricted the evaluation of the effective potential 
to the case of equal orientations $\hat \Phi \equiv \hat\Phi_1 = \hat\Phi_2$. In that
setup the purely bosonic part of $\tilde U(\breve m_\Phi, \breve s_\Phi)$ becomes
\beq
\label{eq:SPhiForPhiAnsatz}
\frac{S_{\Phi}[\Phi']}{VN_f}
\fhs{+1mm}=\fhs{+1mm} -8\tilde\kappa_N \Big(\breve m_\Phi^2-\breve s_\Phi^2\Big) +  \breve m_\Phi^2+\breve s_\Phi^2 
+\tilde\lambda_N \Big( \breve m_\Phi^4 +\breve s_\Phi^4 + 6\breve m_\Phi^2 \breve s_\Phi^2 -2\left(\breve m_\Phi^2+\breve s_\Phi^2 \right) \Big).
\eeq

The remaining calculation of the determinant of the fermion matrix $\fermiMat[\Phi']$ can also be performed
analytically due to the simple structure of $\Phi'$ given in \eq{eq:staggeredAnsatz}. In this approach
the fermion matrix in momentum space\footnote{With some abuse of notation no new symbol has been introduced 
for the momentum space representation of $\fermiMat$ and other operators.} $\fermiMat(p_1,p_2)$ has a 
diagonal-plus-subdiagonal-block-structure. In terms of the spinor basis given in 
\eqs{eq:DefOfEigenvectorsOfD1}{eq:DefOfEigenvectorsOfD2} it can be represented as
\bea
\label{eq:StructureOfB}
\fermiMat(p_1,p_2) &=&
N_f^{\frac{1}{2}}\breve m_\Phi\cdot\delta_{p_1,p_2}\cdot\hat B(p_2, \hat\Phi) \cdot \GammaOp(p_2)\\ 
&+&N_f^{\frac{1}{2}}\breve s_\Phi\cdot\delta_{p_1,\wp_2}\cdot \Upsilon(p_1,p_2) \cdot \hat B(p_2, \hat\Phi)  \cdot \GammaOp(p_2) \nonumber\\
&+&\delta_{p_1,p_2}\cdot\D(p_2) \nonumber \,,
\eea
where the diagonal part is caused by the zero momentum mode of the scalar field $\Phi'$, while the sub-diagonal
contribution is created by the staggered mode. In \eq{eq:StructureOfB} this is expressed
by $\wp_2$ denoting the shifted momenta $\wp_2 = p_2 + (\pi,\pi,\pi,\pi)$, where
adequate modulo-operations are implicit to guarantee $\wp_2\in\ImpSpace$. The 
matrices $\Upsilon(p_1,p_2)$, $\D(p)$, $\GammaOp(p)$, and $\hat B(p, \hat \Phi)$ 
are $8\times 8$-matrices with the indices $\zeta_1\epsilon_1 k_1,\zeta_2\epsilon_2 k_2$ and
denote the spinor basis transformation matrix
\beq
\label{eq:DefOfSpinorBasisTransMat}
\Upsilon(p_1,p_2)_{\zeta_1\epsilon_1 k_1,\zeta_2\epsilon_2 k_2} = 
\left[u^{\zeta_1\epsilon_1 k_1}(p_1)\right]^\dagger u^{\zeta_2\epsilon_2 k_2}(p_2),
\eeq
the Dirac matrix
\beq
\label{eq:MatDiracSpinorRep}
\D(p)_{\zeta_1\epsilon_1 k_1,\zeta_2\epsilon_2 k_2} = \delta_{\epsilon_1,\epsilon_2}\cdot\delta_{k_1,k_2}\cdot
\delta_{\zeta_1,\zeta_2} \cdot \nu^{\epsilon_1}(p),
\eeq
the vertex structure matrix
\beq
\label{eq:DefOfVertexStructureMatSpinorRep}
\GammaOp(p)=\ID-\D(p)/2\rho,
\eeq
and the Yukawa coupling matrix
\bea
\label{eq:MatBinSpinorRep}
\hat B(p, \hat\Phi)_{\zeta_1\epsilon_1k_1,\zeta_2\epsilon_2k_2}&\fhs{-3mm}=\fhs{-3mm}&
\left[u^{\zeta_1\epsilon_1k_1}(p)\right]^\dagger 
\,\hat B(\hat \Phi)\,
u^{\zeta_2\epsilon_2k_2}(p) \ \\
&\fhs{-3mm}=\fhs{-3mm}& \hat y\, \delta_{k_1,k_2} \Big[\delta_{\epsilon_1,\epsilon_2}\delta_{\zeta_1,\zeta_2} \hat\Phi^0
+\delta_{\epsilon_1,-\epsilon_2}\Big\{i\zeta_2\delta_{\zeta_1,\zeta_2}\hat\Phi^1 +
\delta_{\zeta_1,-\zeta_2}\left[i\hat\Phi^3+\zeta_2\hat\Phi^2\right]\Big\}\Big], \nonumber
\eea
in the spinor basis as introduced in \eq{eq:DefOfEigenvectorsOfD1}. Due to this diagonal-subdiagonal-block-structure 
the determinant in \eq{eq:DefOfScaledEffPot} can thus be factorized by merging 
the four $8\times 8$ blocks, which correspond to the momentum indices ${(p,p)}$, ${(\wp,p)}$, ${(p,\wp)}$, 
and ${(\wp,\wp)}$. Up to some constant terms, which are independent of $\Phi'$, one can thus rewrite the 
effective potential as
\beq
\label{eq:EffActionRewrittenForSmallY}
VN_f \tilde U(\breve m_\Phi, \breve s_\Phi)=
S_\Phi[\Phi'] - N_f\cdot \log\Bigg[\prod\limits_{{p\in\ImpSpace}\atop{0\le p_3<\pi}} {\det \left(\mbox{diag}\left[\D(p),\D(\wp)\right] 
+ \mergedBlockA(p)\right)} \Bigg],
\eeq
where the restriction $0\le p_3<\pi$ has been introduced to prevent the double counting that would
occur if one would perform the product over all $p\in\ImpSpace$ after having merged the blocks.
Here $\mergedBlockA$ denotes these merged, momentum dependent $16 \times 16$ matrices given by
\beq
\mergedBlockA(p) = 
\left(
\begin{array}{*{2}{c}}
\mergedBlockA^{1,1}(p) &\mergedBlockA^{1,2}(p) \\
\mergedBlockA^{2,1}(p) &\mergedBlockA^{2,2}(p) \\
\end{array}
\right)
\eeq
with
\bea
\mergedBlockA^{1,1}(p) &= & N_f^{\frac{1}{2}}\breve m_\Phi\cdot \hat B(p, \hat\Phi) \cdot \GammaOp(p),  \\
\mergedBlockA^{1,2}(p) &= & N_f^{\frac{1}{2}}\breve s_\Phi\cdot \Upsilon(p,\wp)\cdot \hat B(\wp, \hat\Phi) \cdot \GammaOp(\wp),  \\
\mergedBlockA^{2,1}(p) &= & N_f^{\frac{1}{2}}\breve s_\Phi\cdot \Upsilon(\wp,p)\cdot \hat B(p, \hat\Phi) \cdot \GammaOp(p),  \\
\mergedBlockA^{2,2}(p) &= & N_f^{\frac{1}{2}}\breve m_\Phi\cdot \hat B(\wp, \hat\Phi) \cdot \GammaOp(\wp). 
\eea
The expression in \eq{eq:EffActionRewrittenForSmallY} can be written 
more compactly, taking the fact into account that the involved matrices
are diagonal with respect to the index $k$ due to
\eq{eq:MatDiracSpinorRep}, \eq{eq:MatBinSpinorRep} and
\beq
\Upsilon(p,\wp)_{\zeta_1\epsilon_1 k_1,\zeta_2\epsilon_2 k_2} = \delta_{\zeta_1,\zeta_2} \cdot
\delta_{\epsilon_1,-\epsilon_2} \cdot \delta_{k_1,k_2}.
\eeq
Since one easily finds that the determinant in \eq{eq:EffActionRewrittenForSmallY}
is invariant under the permutation $p\leftrightarrow\wp$, one can extend the 
product in that equation, which is performed only over one half of the whole
momentum space, again to the full momentum space $\ImpSpace$ by factorizing 
out the identity $\delta_{k_1,k_2}$. One then obtains for the effective 
potential
\beq
VN_f \tilde U(\breve m_\Phi, \breve s_\Phi)
= S_\Phi[\Phi'] - N_f\cdot \log\left[\prod\limits_{p\in\ImpSpace} {\det \left(\mbox{diag}\left[\DnoK(p),\DnoK(\wp)\right]
+\breve\mergedBlockA(p)\right)} \right],
\label{eq:EffActionRewrittenForSmallYCompactified}
\eeq
with the definitions
\beq
\D(p) = \delta_{k_1,k_2} \cdot \DnoK(p), \quad
\mergedBlockA(p) = \delta_{k_1,k_2} \cdot\breve\mergedBlockA(p), \quad \mbox{and} \quad
\mergedBlockA^{a,b}(p) = \delta_{k_1,k_2} \cdot\breve\mergedBlockA^{a,b}(p),
\eeq
and $a,b\in \{1,2\}$. Ordering the indices $\zeta\epsilon$
according to $\{++,+-,-+,--\}$ the latter four $4\times 4$ matrices are 
explicitly given by
{\footnotesize
\bea
\breve\mergedBlockA^{1,1}(p)
&=& \tilde y_N \breve m_\Phi\cdot
\left(
\begin{array}{*{4}{c}}
\hat\Phi^0\gamma^+(p)&i\hat\Phi^1\gamma^-(p)&0&(i\hat\Phi^3-\hat\Phi^2)\gamma^-(p)\\
i\hat\Phi^1\gamma^+(p)&\hat\Phi^0\gamma^-(p)&(i\hat\Phi^3-\hat\Phi^2)\gamma^+(p)&0\\
0&(i\hat\Phi^3+\hat\Phi^2)\gamma^-(p)&\hat\Phi^0\gamma^+(p)&-i\hat\Phi^1\gamma^-(p)\\
(i\hat\Phi^3+\hat\Phi^2)\gamma^+(p)&0&-i\hat\Phi^1\gamma^+(p)&\hat\Phi^0\gamma^-(p)\\
\end{array}
\right) \nonumber\\
\\
\breve\mergedBlockA^{1,2}(p) 
&=& \tilde y_N \breve s_\Phi \cdot
\left(
\begin{array}{*{4}{c}}
i\hat\Phi^1\gamma^+(\wp)&\hat\Phi^0\gamma^-(\wp)&(i\hat\Phi^3-\hat\Phi^2)\gamma^+(\wp)&0\\
\hat\Phi^0\gamma^+(\wp)&i\hat\Phi^1\gamma^-(\wp)&0&(i\hat\Phi^3-\hat\Phi^2)\gamma^-(\wp)\\
(i\hat\Phi^3+\hat\Phi^2)\gamma^+(\wp)&0&-i\hat\Phi^1\gamma^+(\wp)&\hat\Phi^0\gamma^-(\wp)\\
0&(i\hat\Phi^3+\hat\Phi^2)\gamma^-(\wp)&\hat\Phi^0\gamma^+(\wp)&-i\hat\Phi^1\gamma^-(\wp)\\
\end{array}
\right)\nonumber\\
\eea}
where the abbreviation $\gamma^\epsilon(p) = 1-\nu^\epsilon(p)/2\rho$ was used.
The remaining matrices $\breve\mergedBlockA^{2,2}(p)$ and $\breve\mergedBlockA^{2,1}(p)$
are obtained from $\breve\mergedBlockA^{1,1}(p)$, $\breve\mergedBlockA^{1,2}(p)$ by 
interchanging $p$ and $\wp$. Using some algebraic manipulation package, the  
determinant of the $8\times 8$ matrix in \eq{eq:EffActionRewrittenForSmallYCompactified}
can be computed leading to the final expression for the effective potential
\bea
\label{eq:EffActionRewrittenForSmallYFinal}
\tilde U(\breve m_\Phi, \breve s_\Phi)
&=& 
-8\tilde\kappa_N \Big(\breve m_\Phi^2-\breve s_\Phi^2\Big) +  \breve m_\Phi^2+\breve s_\Phi^2 
+\tilde\lambda_N \Big( \breve m_\Phi^4 +\breve s_\Phi^4 + 6\breve m_\Phi^2\breve s_\Phi^2 -2\left(\breve m_\Phi^2+\breve s_\Phi^2 \right) \Big)\nonumber\\
&-& \frac{1}{V}\sum\limits_{p\in\ImpSpace}
\log\Bigg[\left(\left|\nu^+(p)\right|\cdot \left|\nu^+(\wp)\right| 
+\tilde y_N^2 \left[\breve m_\Phi^2 - \breve s_\Phi^2\right] \cdot 
\left|\gamma^+(p)\right| \cdot \left|\gamma^+(\wp) \right|\right)^2\nonumber \\
&+&\breve m_\Phi^2 \tilde y_N^2   
\Big(\left|\gamma^+(p)\right|\cdot \left|\nu^+(\wp) \right| 
- \left|\gamma^+(\wp)\right| \cdot \left|\nu^+(p) \right|\Big)^2 
\Bigg]^2. 
\eea
 
The resulting phase structure in the large $N_f$-limit can now be obtained by minimizing the 
effective potential $\tilde U(\breve m_\Phi, \breve s_\Phi)$ with respect to $m_\Phi$ and $s_\Phi$. 
This is done here in the infinite volume limit, where the sum over all allowed momenta $\ImpSpace$ 
in \eq{eq:EffActionRewrittenForSmallYFinal} becomes a four-dimensional momentum integral 
over continuous momenta according to
\beq
\label{eq:IntSumRelation}
\frac{1}{V} \sum\limits_{p\in\ImpSpace}\; ... \quad \rightarrow \quad 
\int\limits_{p_\mu\in[-\pi,\pi]} \frac{d^4p}{(2\pi)^4} \;...
\eeq
which was evaluated with the aid of some numeric analysis software~\cite{Mathematica:2007yu}.
The resulting phase structure is presented in \fig{fig:PhaseDiagrams1} for the selected values 
$\tilde\lambda_N=0.1$ and $\tilde\lambda_N=0.3$. Here, four different phases do emerge
which are labeled in analogy to the terminology of magnetism in solid state physics. These are\\
\begin{tabular}{ccl}
&{\textit{(I)}} &  a symmetric phase (SYM): $\bar m_\Phi=0,\, \bar s_\Phi=0$,\\
&{\textit{(II)}} & a broken $=$ ferromagnetic phase (FM): $\bar m_\Phi\neq 0,\, \bar s_\Phi=0$,   \\
&{\textit{(III)}} & a staggered-broken $=$ anti-ferromagnetic  phase (AFM): $\bar m_\Phi=0,\, \bar s_\Phi\neq 0$,  \\
&{\textit{(IV)}} & and a broken/staggered-broken $=$ ferrimagnetic phase (FI): $\bar m_\Phi\neq 0,\, \bar s_\Phi\neq 0$,   \\
\end{tabular}\\
where $\bar m_\Phi$, $\bar s_\Phi$ denote the location of the absolute minimum of the effective potential according to
$\tilde U(\bar m_\Phi, \bar s_\Phi) = \min(\tilde U(\breve m_\Phi,\breve s_\Phi))$. 

In both cases, \ie $\tilde\lambda_N=0.1$ and $\tilde\lambda_N=0.3$,
one finds a symmetric phase approximately centered around $\tilde\kappa_N=0$ 
at sufficiently small values of the Yukawa coupling constant $\tilde y_N$.
This is what one would have expected, since the model becomes the pure $\Phi^4$-theory
in the limit $\tilde y_N\rightarrow 0$. From the same consideration the accompanying phase 
transitions at small values of the Yukawa coupling constant can also be assumed to be of 
second order. This is indeed the case here as can be verified by the results presented in 
\fig{fig:ExpectedMSPlots1}. In that plots the dependence of $\bar m_\Phi$, $\bar s_\Phi$, 
and thus of the order parameters $v$ and $v_s$, on the hopping parameter $\tilde \kappa_N$ is 
shown to be continuous at sufficiently small values of $\tilde y_N$. 

\bc
\setlength{\unitlength}{0.01mm}
\begin{figure}[htb]
\centering
\begin{tabular}{cc}
$\tilde\lambda_N=0.1$ & $\tilde\lambda_N=0.3$ \\
\begin{picture}(6600,5500)
%\put(600,500){\includegraphics[width=5cm]{PhaseDiagramLam001}}
\put(600,500){\includegraphics[width=5cm]{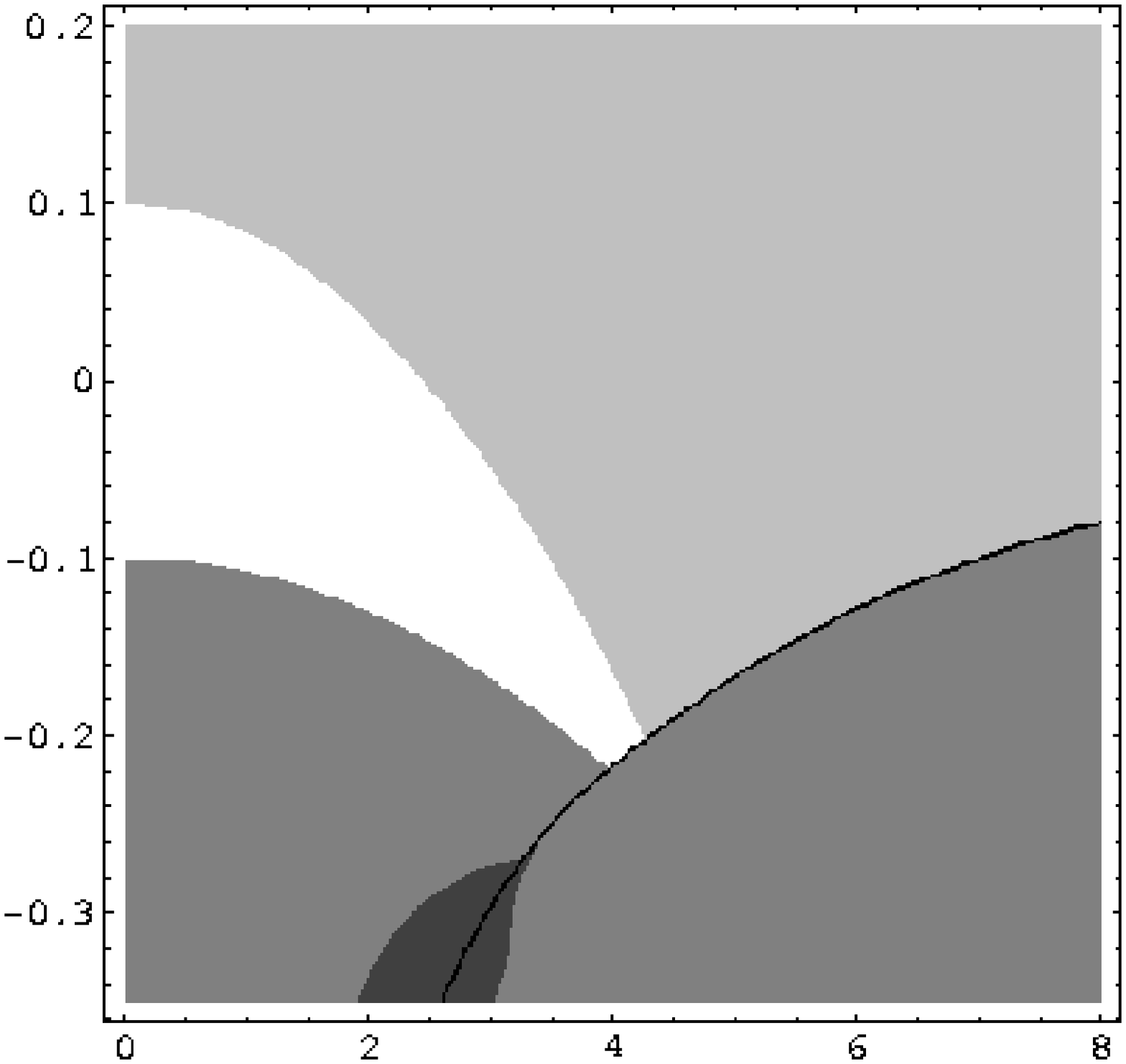}}
\put(25,3100){$\tilde \kappa_N$}
\put(3200,300){$\tilde y_N$}
\put(1300,3350){\textbf {SYM}}
\put(4500,3350){\textbf {FM}}
\put(1300,1500){\textbf {AFM}}
\put(4500,1500){\textbf {AFM}}
\put(2900,1000){$\leftarrow$\textbf {FI}}
\end{picture}
&
\begin{picture}(6600,5500)
%\put(600,500){\includegraphics[width=5cm]{PhaseDiagramLam003}}
\put(600,500){\includegraphics[width=5cm]{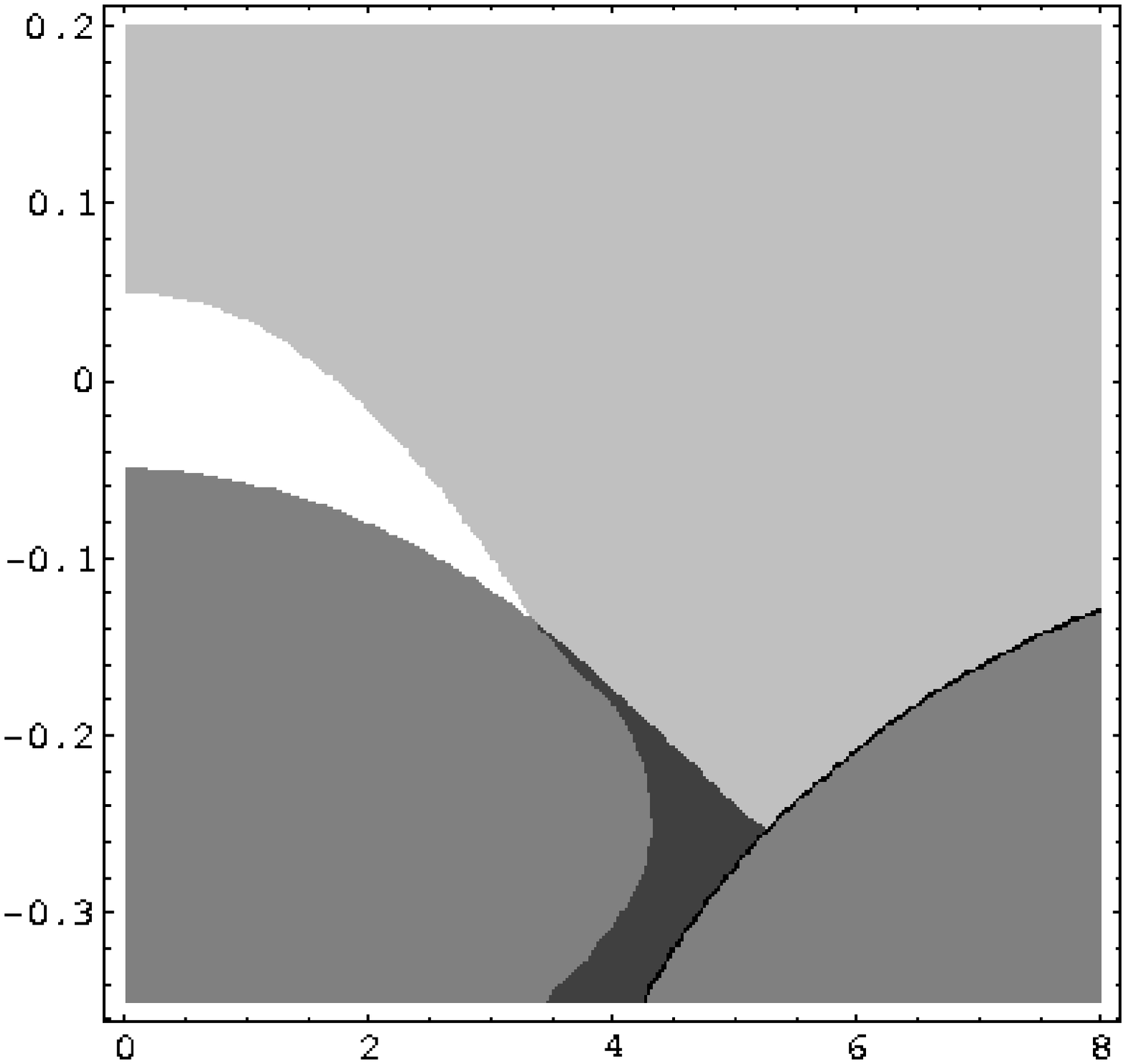}}
\put(25,3100){$\tilde \kappa_N$}
\put(3200,300){$\tilde y_N$}
\put(1300,3350){\textbf {SYM}}
\put(4500,3350){\textbf {FM}}
\put(1300,1500){\textbf {AFM}}
\put(4500,1500){\textbf {AFM}}
\put(2600,1000){\textbf {FI}$\rightarrow$}
\end{picture}
\\
\end{tabular}
\caption[Analytically obtained phase diagrams at small Yukawa coupling constants.]
{Phase diagrams with respect to the Yukawa coupling constant $\tilde y_N$ and the hopping 
parameter $\tilde \kappa_N$ for the constant quartic coupling parameters $\tilde \lambda_N=0.1$ (left)
and $\tilde \lambda_N=0.3$ (right). The black lines indicate first order phase transitions, while
all other transitions are of second order. Both phase diagrams were determined for infinite volume.
An explanation of the occurring phases is given in the main text.}
\label{fig:PhaseDiagrams1}
\end{figure}
\vs{-6mm}
\ec

With increasing values of the Yukawa coupling constant the symmetric phase bends downwards to 
negative values of the hopping parameter $\tilde \kappa_N$, unless it either hits a first order phase 
transition to an anti-ferromagnetic phase (black line in \fig{fig:PhaseDiagrams1}), 
which is the case for $\tilde\lambda_N=0.1$, or it eventually goes over into two FM-FI 
and FI-AFM second order phase transitions, as observed for the selected value $\tilde\lambda_N=0.3$. 
The order of these phase transitions can again be determined by considering the dependence
of $\bar m_\Phi$, $\bar s_\Phi$ on $\tilde \kappa_N$, which is presented in \fig{fig:ExpectedMSPlots2}.

It is remarked that the results on the $\tilde \kappa_N$-dependence of $\bar m_\Phi$, $\bar s_\Phi$ 
have only been depicted here for the setting $\tilde \lambda_N=0.3$ and not for $\tilde \lambda_N=0.1$, since the 
latter plots would not provide qualitatively new information.

However, though the obtained expression in \eq{eq:EffActionRewrittenForSmallYFinal} seems to 
be applicable for any value of $\tilde \lambda_N>0$ it actually only yields a meaningful result 
for the effective potential in the large $N_f$-limit if the precondition in \eq{eq:SmallYAnaPrecondition}
is fulfilled. This induces an absolute bound for the validity of the presented calculation of
$\tilde\lambda_N<0.5$. Moreover, the obtained phase diagrams can only be trusted for 
$|\tilde\kappa_N|<(1-2\tilde\lambda_N)/8$. Whether the obtained results beyond that threshold are still
in agreement with the true phase structure of the model has to be checked explicitly, for instance,
by direct Monte-Carlo calculations.

It is therefore an open question at this point, whether the rather surprising appearance of the ferrimagnetic 
phase (FI) observed in both presented scenarios, \ie for $\tilde\lambda_N=0.1$ and $\tilde\lambda_N=0.3$, 
even deeply inside the anti-ferromagnetic phase region, is actually present in an exact evaluation of the 
phase structure. This question will explicitly be investigated in the subsequent section.

\bc
\setlength{\unitlength}{0.0105mm}
\begin{figure}[htb]
\centering
\begin{picture}(12000,13144)
%\put(0,1000){\includegraphics[angle=0,width=0.27\textwidth]{plExpectedMSatY030Lam003}}
\put(0,1000){\includegraphics[angle=0,width=0.27\textwidth]{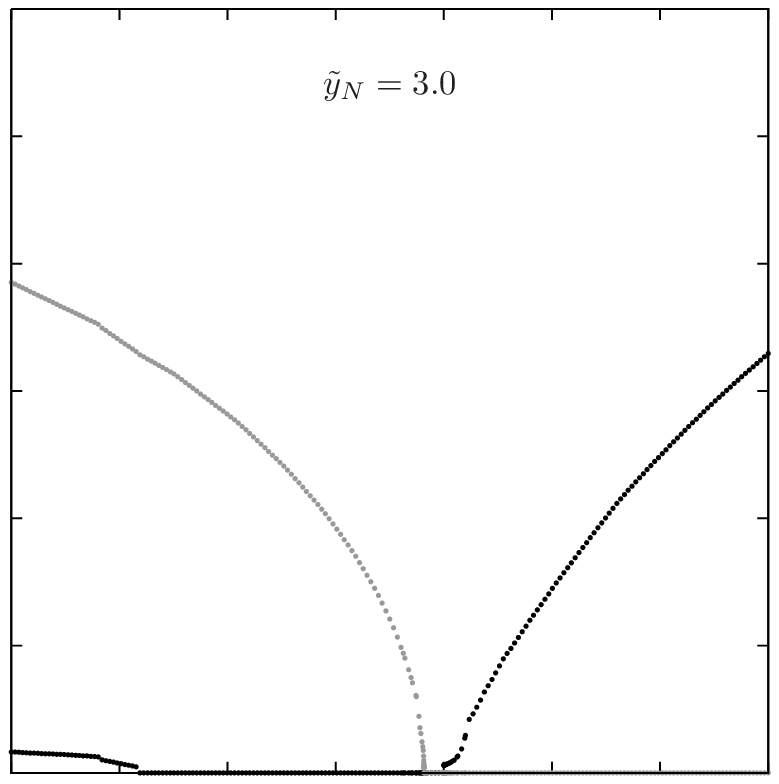}}
%\put(4000,1000){\includegraphics[angle=0,width=0.27\textwidth]{plExpectedMSatY035Lam003}}
\put(4000,1000){\includegraphics[angle=0,width=0.27\textwidth]{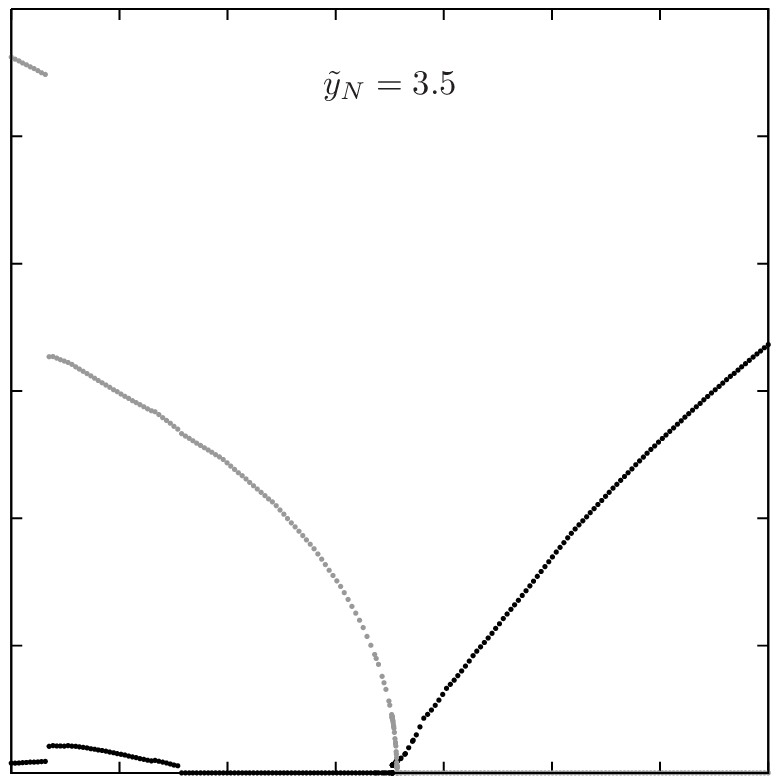}}
%\put(8000,1000){\includegraphics[angle=0,width=0.27\textwidth]{plExpectedMSatY040Lam003}}
\put(8000,1000){\includegraphics[angle=0,width=0.27\textwidth]{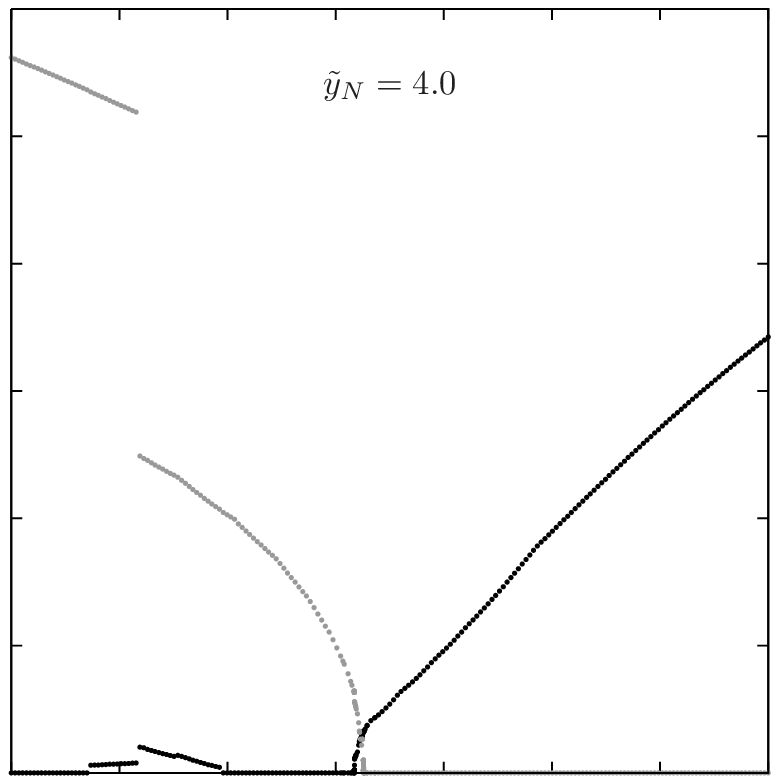}}
%\put(0,5048){\includegraphics[angle=0,width=0.27\textwidth]{plExpectedMSatY015Lam003}}
\put(0,5048){\includegraphics[angle=0,width=0.27\textwidth]{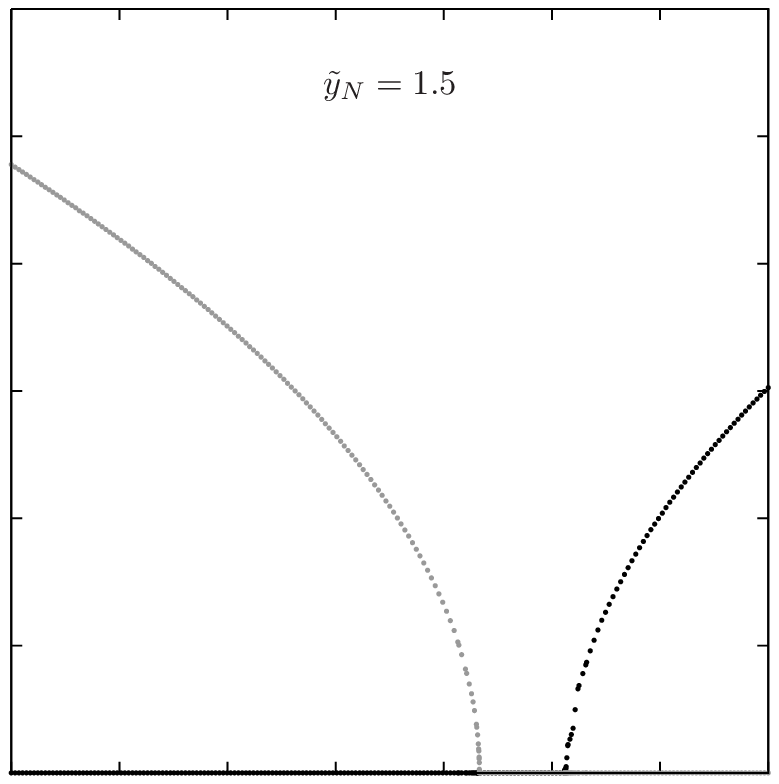}}
%\put(4000,5048){\includegraphics[angle=0,width=0.27\textwidth]{plExpectedMSatY020Lam003}}
\put(4000,5048){\includegraphics[angle=0,width=0.27\textwidth]{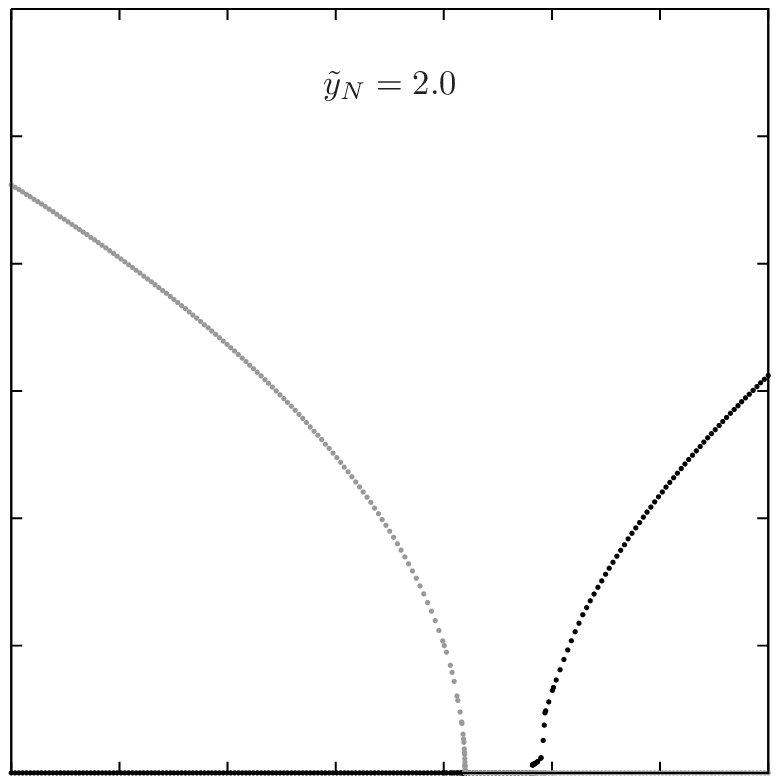}}
%\put(8000,5048){\includegraphics[angle=0,width=0.27\textwidth]{plExpectedMSatY025Lam003}}
\put(8000,5048){\includegraphics[angle=0,width=0.27\textwidth]{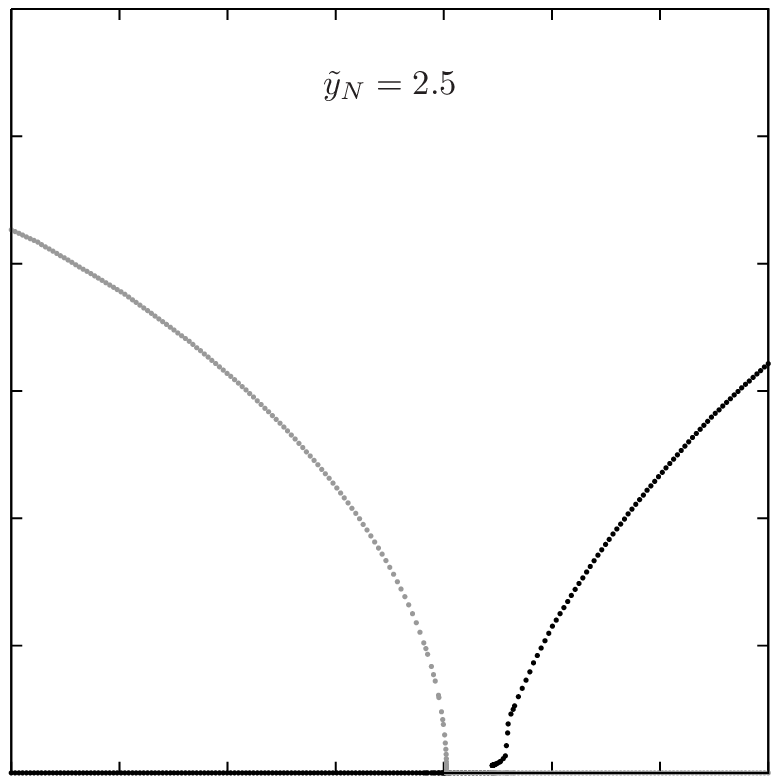}}
%\put(0,9096){\includegraphics[angle=0,width=0.27\textwidth]{plExpectedMSatY000Lam003}}
\put(0,9096){\includegraphics[angle=0,width=0.27\textwidth]{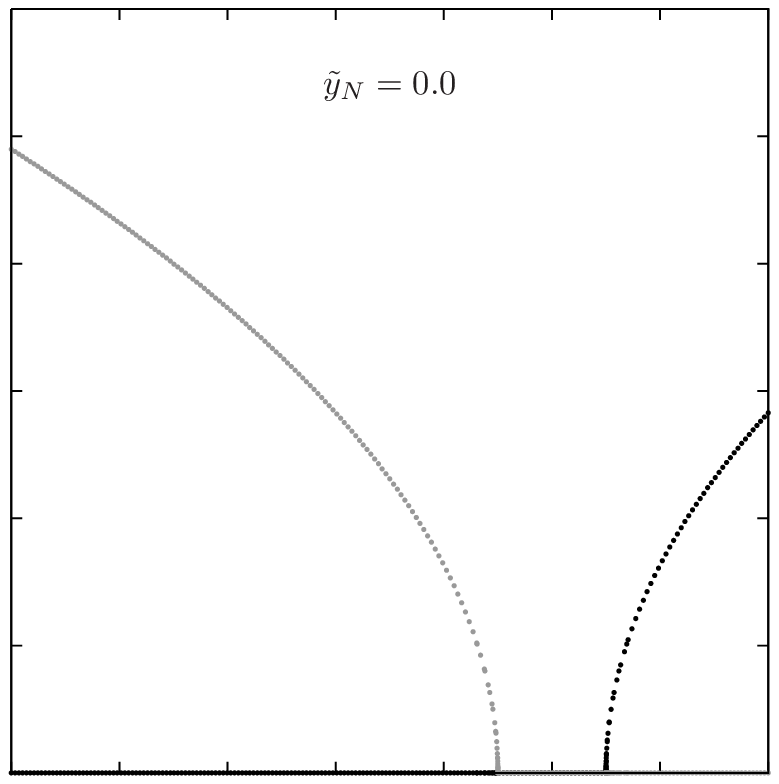}}
%\put(4000,9096){\includegraphics[angle=0,width=0.27\textwidth]{plExpectedMSatY005Lam003}}
\put(4000,9096){\includegraphics[angle=0,width=0.27\textwidth]{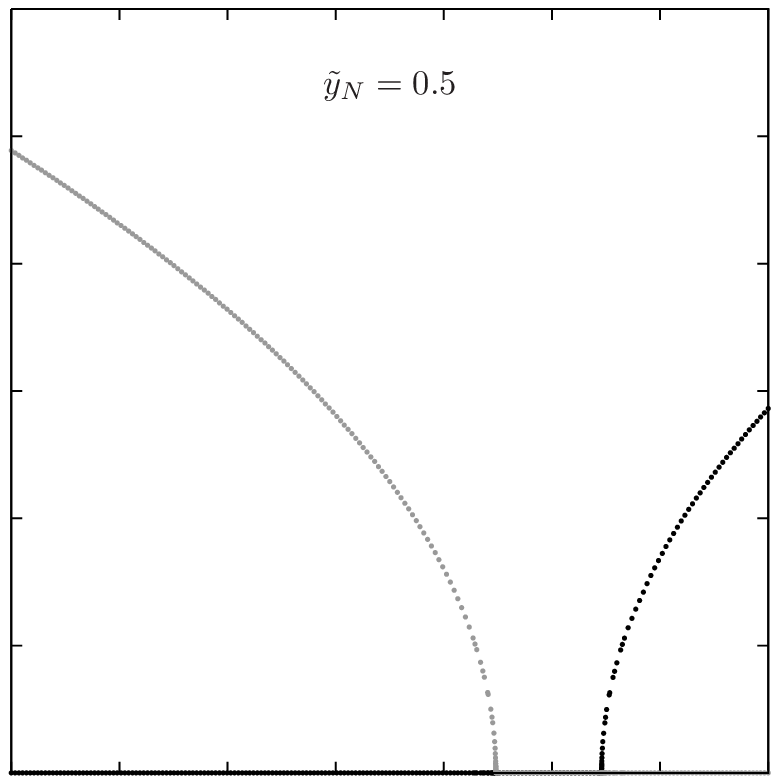}}
%\put(8000,9096){\includegraphics[angle=0,width=0.27\textwidth]{plExpectedMSatY010Lam003}}
\put(8000,9096){\includegraphics[angle=0,width=0.27\textwidth]{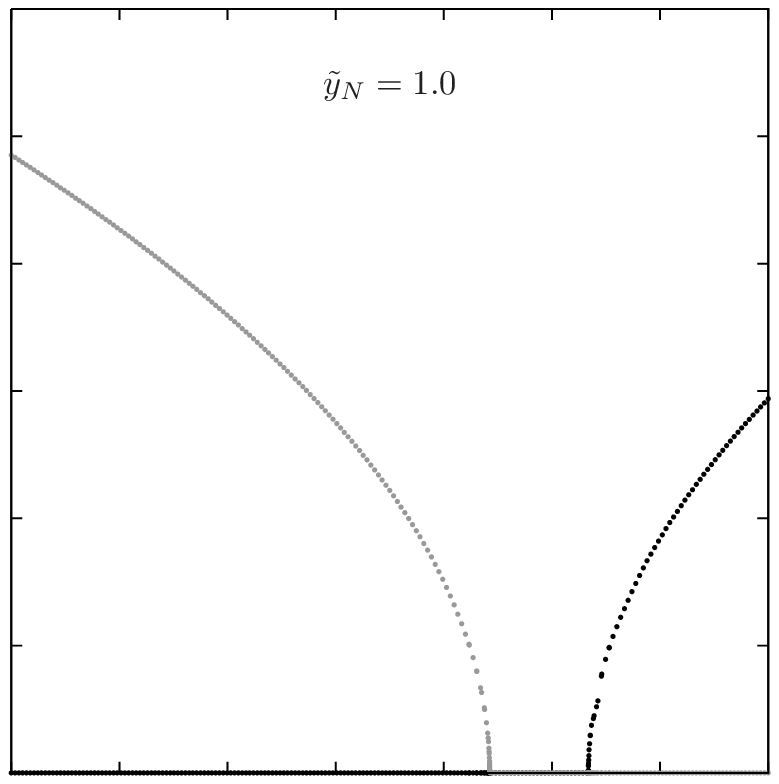}}

\put(-200,700){\tiny{-0.5}}
\put(+381,700){\tiny{-0.4}}
\put(+942,700){\tiny{-0.3}}
\put(+1513,700){\tiny{-0.2}}
\put(+2084,700){\tiny{-0.1}}
\put(+2735,700){\tiny{0.0}}
\put(+3316,700){\tiny{0.1}}

\put(3800,700){\tiny{-0.5}}
\put(4381,700){\tiny{-0.4}}
\put(4942,700){\tiny{-0.3}}
\put(5513,700){\tiny{-0.2}}
\put(6084,700){\tiny{-0.1}}
\put(6735,700){\tiny{0.0}}
\put(7316,700){\tiny{0.1}}

\put(7800,700){\tiny{-0.5}}
\put(8381,700){\tiny{-0.4}}
\put(8942,700){\tiny{-0.3}}
\put(9513,700){\tiny{-0.2}}
\put(10084,700){\tiny{-0.1}}
\put(10735,700){\tiny{0.0}}
\put(11316,700){\tiny{0.1}}
\put(11870,700){\tiny{0.2}}

\put(5800,100){$\tilde \kappa_N$}
\put(-1500,7500){$\bar m_\Phi$,}
\put(-1500,7000){$\bar s_\Phi$}

\put(-430,960){\tiny{0.0}}
\put(-430,1640){\tiny{0.5}}
\put(-430,2320){\tiny{1.0}}
\put(-430,3000){\tiny{1.5}}
\put(-430,3670){\tiny{2.0}}
\put(-430,4340){\tiny{2.5}}

\put(-430,5008){\tiny{0.0}}
\put(-430,5688){\tiny{0.5}}
\put(-430,6368){\tiny{1.0}}
\put(-430,7048){\tiny{1.5}}
\put(-430,7718){\tiny{2.0}}
\put(-430,8388){\tiny{2.5}}

\put(-430,9056){\tiny{0.0}}
\put(-430,9736){\tiny{0.5}}
\put(-430,10416){\tiny{1.0}}
\put(-430,11096){\tiny{1.5}}
\put(-430,11766){\tiny{2.0}}
\put(-430,12436){\tiny{2.5}}
\put(-430,13104){\tiny{3.0}}
\end{picture}
\caption[Dependence of $\bar m_\Phi$ and $\bar s_\Phi$ on $\tilde \kappa_N$ at small Yukawa coupling constants for $\tilde y_N\le 4.0$.]
{The dependence of $\bar m_\Phi$ and $\bar s_\Phi$ on the hopping parameter $\tilde \kappa_N$
is presented for several selected values of the Yukawa coupling constant $\tilde y_N\le 4.0$. These results
have been obtained for the constant quartic coupling parameter $\tilde \lambda_N=0.3$. The black curve depicts the dependence
of $\bar m_\Phi$ on $\tilde \kappa_N$, while the gray curve represents the corresponding results for $\bar s_\Phi$.}
\label{fig:ExpectedMSPlots1}
\end{figure}
\vs{-6mm}
\ec

\bc
\setlength{\unitlength}{0.0105mm}
\begin{figure}[htb]
\centering
\begin{picture}(12000,13144)
%\put(0,1000){\includegraphics[angle=0,width=0.27\textwidth]{plExpectedMSatY075Lam003}}
\put(0,1000){\includegraphics[angle=0,width=0.27\textwidth]{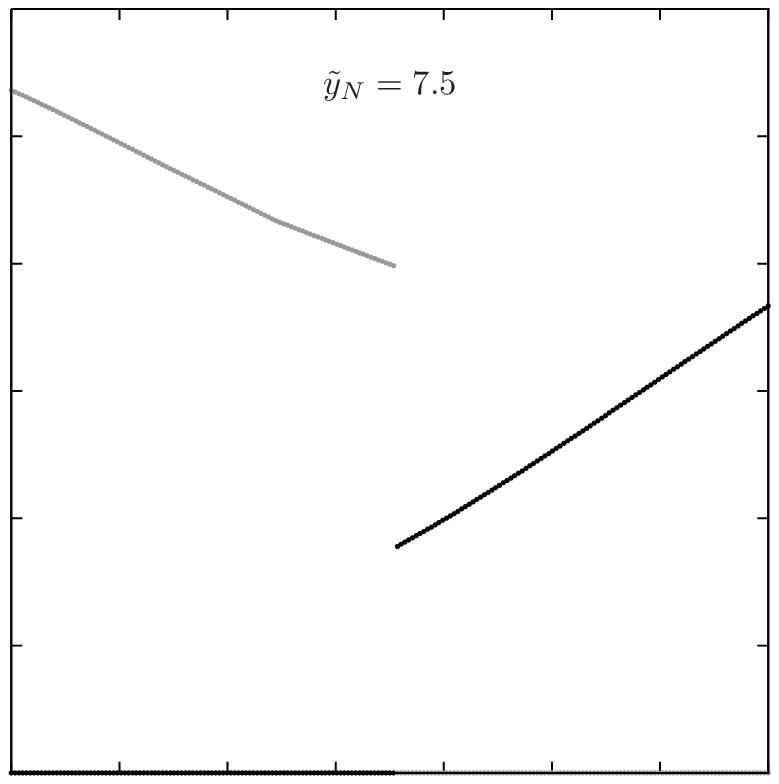}}
%\put(4000,1000){\includegraphics[angle=0,width=0.27\textwidth]{plExpectedMSatY080Lam003}}
\put(4000,1000){\includegraphics[angle=0,width=0.27\textwidth]{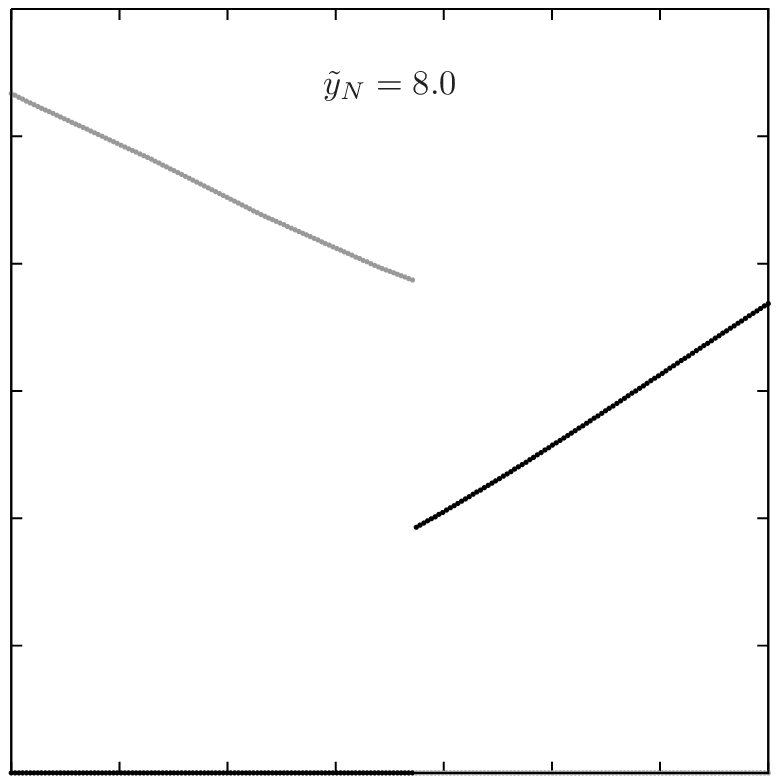}}
%\put(8000,1000){\includegraphics[angle=0,width=0.27\textwidth]{plExpectedMSatY090Lam003}}
\put(8000,1000){\includegraphics[angle=0,width=0.27\textwidth]{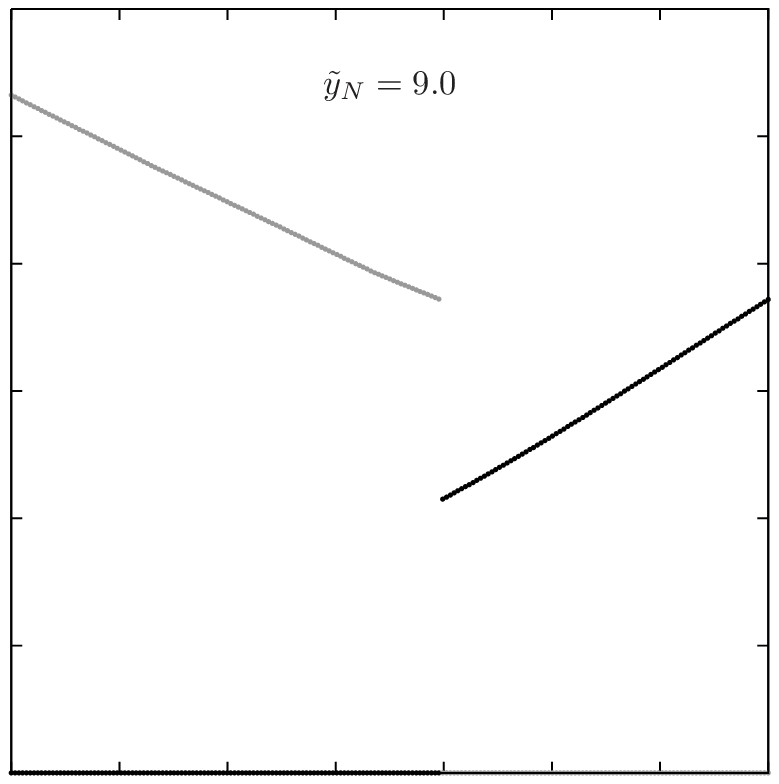}}
%\put(0,5048){\includegraphics[angle=0,width=0.27\textwidth]{plExpectedMSatY060Lam003}}
\put(0,5048){\includegraphics[angle=0,width=0.27\textwidth]{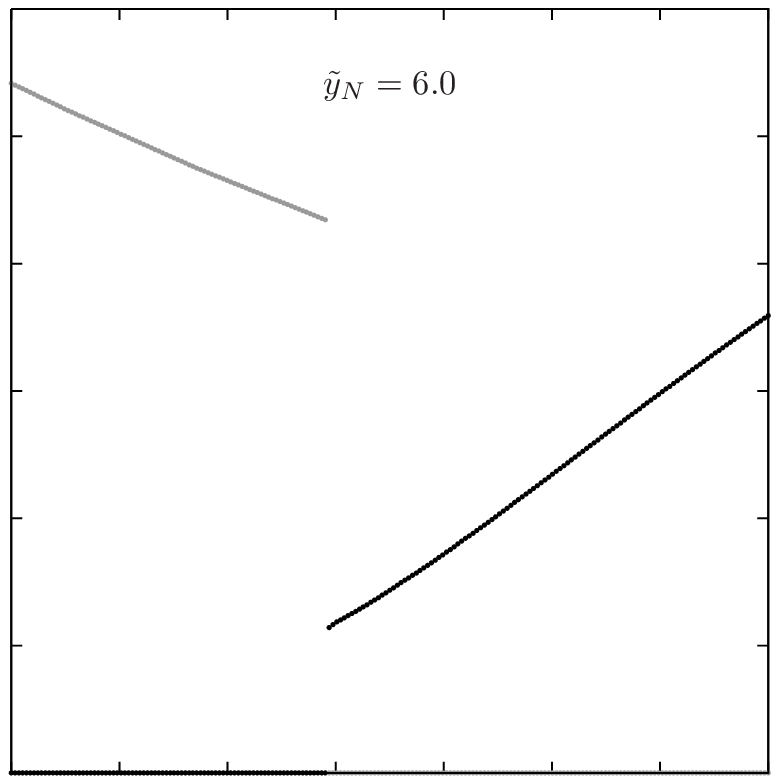}}
%\put(4000,5048){\includegraphics[angle=0,width=0.27\textwidth]{plExpectedMSatY065Lam003}}
\put(4000,5048){\includegraphics[angle=0,width=0.27\textwidth]{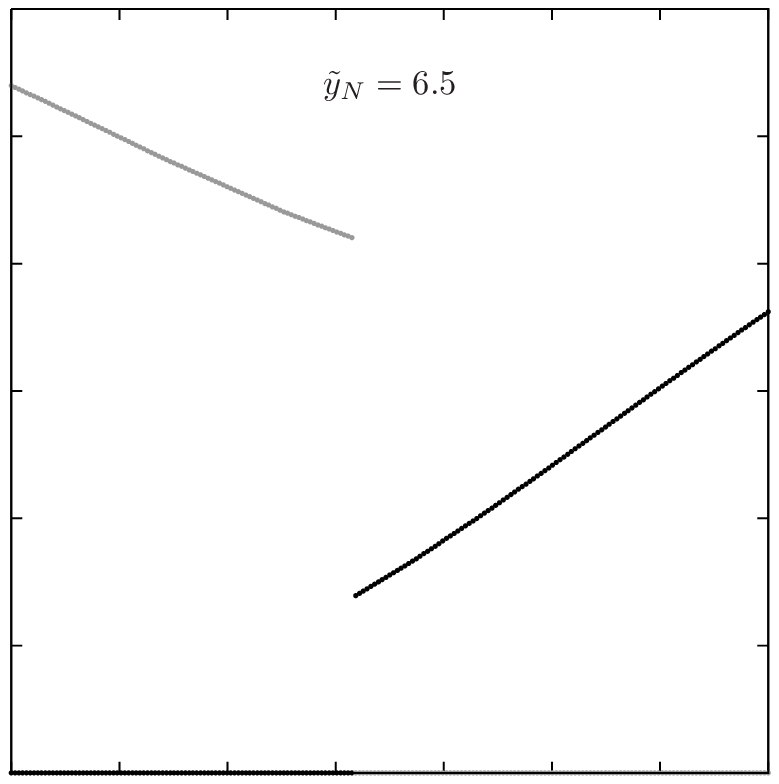}}
%\put(8000,5048){\includegraphics[angle=0,width=0.27\textwidth]{plExpectedMSatY070Lam003}}
\put(8000,5048){\includegraphics[angle=0,width=0.27\textwidth]{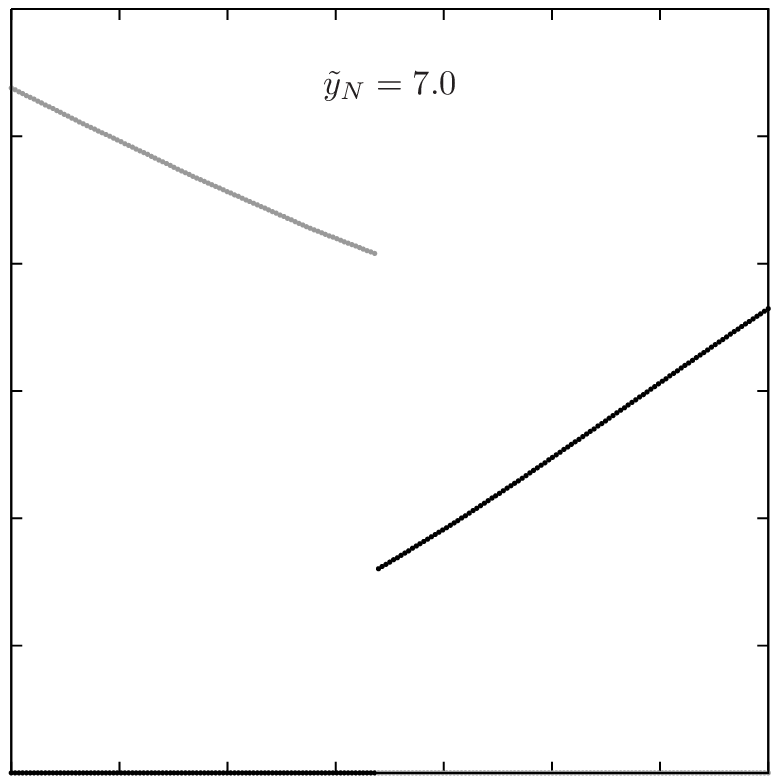}}
%\put(0,9096){\includegraphics[angle=0,width=0.27\textwidth]{plExpectedMSatY045Lam003}}
\put(0,9096){\includegraphics[angle=0,width=0.27\textwidth]{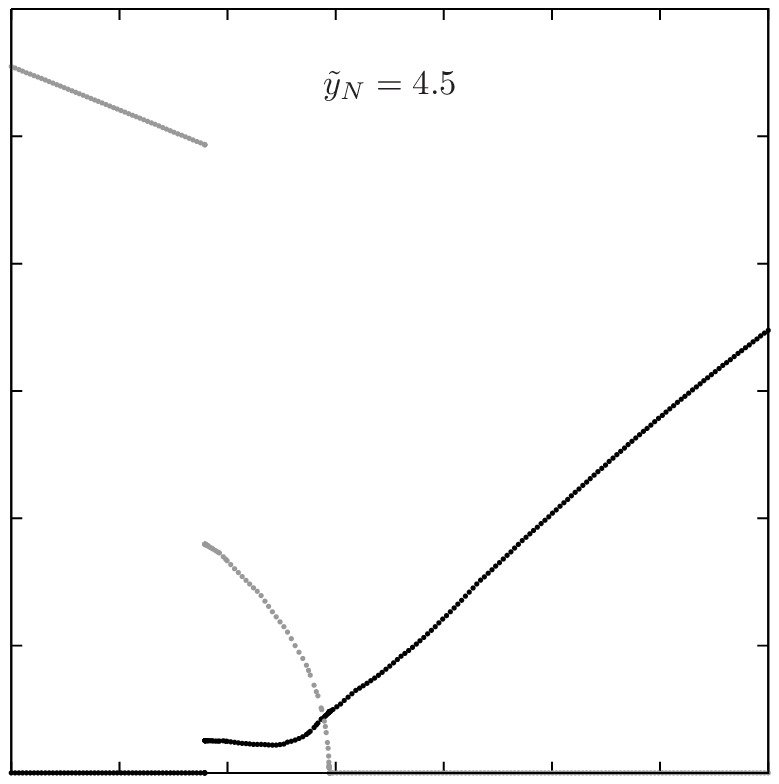}}
%\put(4000,9096){\includegraphics[angle=0,width=0.27\textwidth]{plExpectedMSatY050Lam003}}
\put(4000,9096){\includegraphics[angle=0,width=0.27\textwidth]{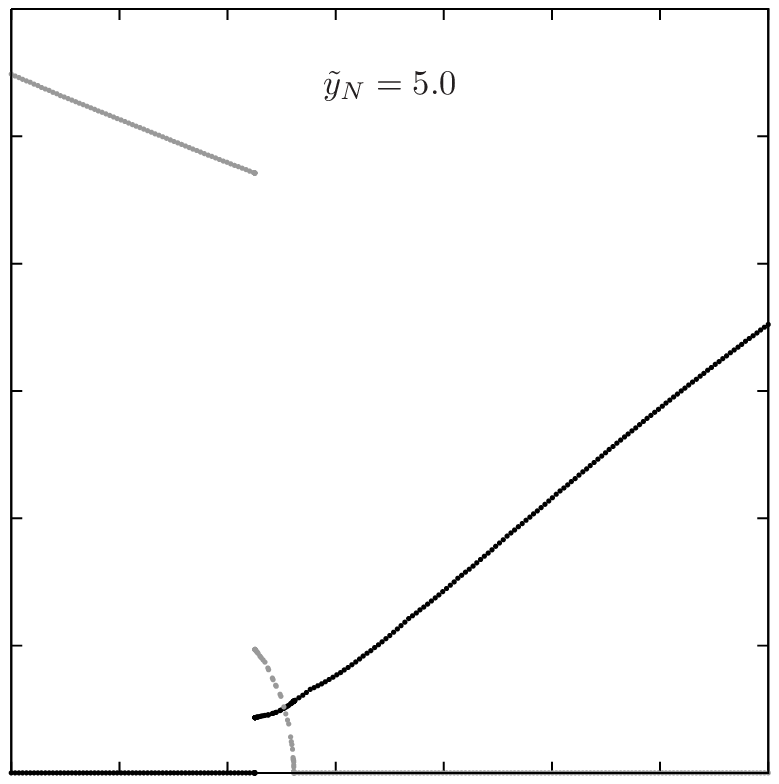}}
%\put(8000,9096){\includegraphics[angle=0,width=0.27\textwidth]{plExpectedMSatY055Lam003}}
\put(8000,9096){\includegraphics[angle=0,width=0.27\textwidth]{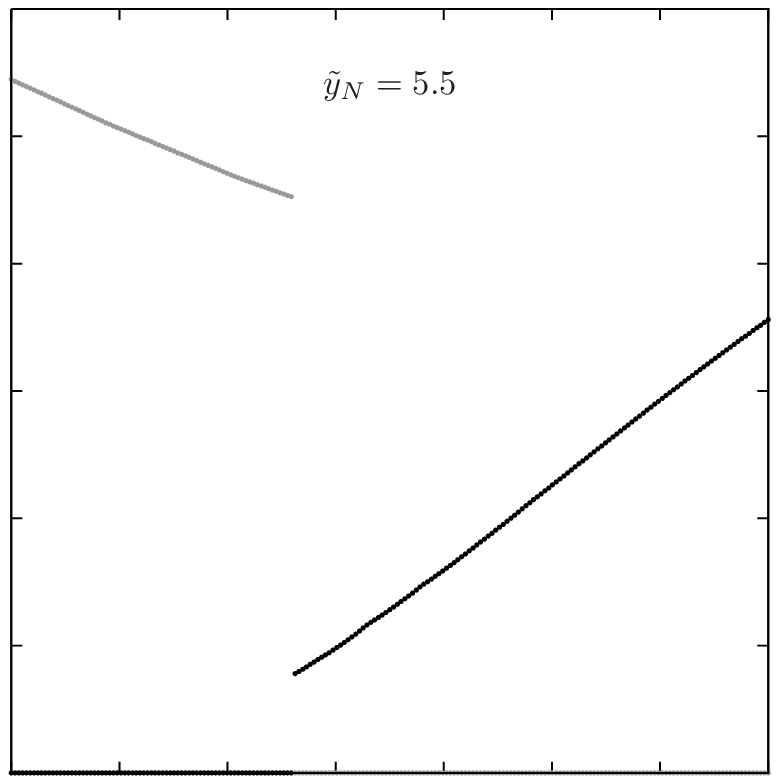}}

\put(-200,700){\tiny{-0.5}}
\put(+381,700){\tiny{-0.4}}
\put(+942,700){\tiny{-0.3}}
\put(+1513,700){\tiny{-0.2}}
\put(+2084,700){\tiny{-0.1}}
\put(+2735,700){\tiny{0.0}}
\put(+3316,700){\tiny{0.1}}

\put(3800,700){\tiny{-0.5}}
\put(4381,700){\tiny{-0.4}}
\put(4942,700){\tiny{-0.3}}
\put(5513,700){\tiny{-0.2}}
\put(6084,700){\tiny{-0.1}}
\put(6735,700){\tiny{0.0}}
\put(7316,700){\tiny{0.1}}

\put(7800,700){\tiny{-0.5}}
\put(8381,700){\tiny{-0.4}}
\put(8942,700){\tiny{-0.3}}
\put(9513,700){\tiny{-0.2}}
\put(10084,700){\tiny{-0.1}}
\put(10735,700){\tiny{0.0}}
\put(11316,700){\tiny{0.1}}
\put(11870,700){\tiny{0.2}}

\put(5800,100){$\tilde \kappa_N$}
\put(-1500,7500){$\bar m_\Phi$,}
\put(-1500,7000){$\bar s_\Phi$}

\put(-430,960){\tiny{0.0}}
\put(-430,1640){\tiny{0.5}}
\put(-430,2320){\tiny{1.0}}
\put(-430,3000){\tiny{1.5}}
\put(-430,3670){\tiny{2.0}}
\put(-430,4340){\tiny{2.5}}

\put(-430,5008){\tiny{0.0}}
\put(-430,5688){\tiny{0.5}}
\put(-430,6368){\tiny{1.0}}
\put(-430,7048){\tiny{1.5}}
\put(-430,7718){\tiny{2.0}}
\put(-430,8388){\tiny{2.5}}

\put(-430,9056){\tiny{0.0}}
\put(-430,9736){\tiny{0.5}}
\put(-430,10416){\tiny{1.0}}
\put(-430,11096){\tiny{1.5}}
\put(-430,11766){\tiny{2.0}}
\put(-430,12436){\tiny{2.5}}
\put(-430,13104){\tiny{3.0}}
\end{picture}
\caption[Dependence of $\bar m_\Phi$ and $\bar s_\Phi$ on $\tilde \kappa_N$ at small Yukawa coupling constants for $\tilde y_N\ge 4.5$.]
{The dependence of $\bar m_\Phi$ and $\bar s_\Phi$ on the hopping parameter $\tilde \kappa_N$
is presented for several selected values of the Yukawa coupling constant $\tilde y_N\ge 4.5$. These results
have been obtained for the constant quartic coupling parameter $\tilde \lambda_N=0.3$. The black curve depicts the dependence
of $\bar m_\Phi$ on $\tilde \kappa_N$, while the gray curve represents the corresponding results for $\bar s_\Phi$.}
\label{fig:ExpectedMSPlots2}
\end{figure}
\vs{-6mm}
\ec

\subsection{Comparison with direct Monte-Carlo calculations}
\label{sec:SmallY}

Now the predicted phase structure shall be confronted with the results of 
direct Monte-Carlo calculations performed by means of the HMC-algorithm
introduced in \sect{sec:HMCAlgorithm}. To determine the exact location of the phase transition,
the susceptibilities $\chi_m$, $\chi_s$ of the magnetizations 
defined as 
\begin{equation} 
\chi_m =  V\cdot \left[\langle m^2\rangle -\langle m \rangle^2 \right], 
\quad
\chi_s =  V\cdot \left[\langle s^2\rangle -\langle s \rangle^2 \right],
\label{magneticsus}
\end{equation} 
are fitted as functions of the hopping parameter $\kappa\equiv \tilde \kappa_N$ to the 
-- partly phenomenologically motivated -- ansatz~\cite{Jansen:1989gd}
\begin{equation} 
\label{finitesizesus}
\chi_{m,s} = A_1^{m,s}\cdot\left(\frac{1}{L^{-2/\nu} + A_{2,3}^{m,s}(\kappa-\kapCrit^{m,s})^2}\right)^{\gamma/2},
\end{equation} 
where $A_1^{m,s}$, $A_{2,3}^{m,s}$, and $\kapCrit^{m,s}$ are the fitting parameters 
for the magnetic susceptibility $\chi_m$ and the staggered susceptibility $\chi_s$, respectively,  
while $\nu$ and $\gamma$ denote the critical exponents\footnote{The critical exponents $\nu$, $\gamma$, 
and $\beta$ are defined here through the infinite volume behaviour of the correlation length $\xi$, the 
susceptibility $\chi$, and the magnetization $\langle m\rangle$ at the phase transition according 
to $\xi\propto |\tau|^{-\nu}$, $\chi \propto |\tau|^{-\gamma}$, and $\langle m \rangle \propto |\tau|^{\beta}$,
where $\tau=\kappa/\kapCrit -1$ is assumed to be positive in the latter case. }
of the pure $\Phi^4$-theory. Neglecting all subleading, logarithmic contributions one can infer\footnote{The 
scaling behaviour of the correlation length, \ie the inverse mass,
and the vacuum expectation value $v$ at the phase transition, which has been derived in
\Ref{Luscher:1988uq}, is explicitly given in \eq{eq:ScalingBehOfVeV} and \eq{eq:ScalingBehOfMass}.}
from \Ref{Luscher:1988uq} that the critical exponents $\nu$ and $\beta$ of the four-dimensional, 
four-component pure $\Phi^4$-theory are given as $\nu=\beta=0.5$ to lowest order. Exploiting the 
hyperscaling relations provided in \Ref{Kleinert:2001zu} the so far unspecified parameter $\gamma$ 
can then be derived from the values of $\nu$ and $\beta$ yielding the result $\gamma=1$.

For clarification it is remarked that the notation $A_{2,3}^m$ actually refers to 
two parameters, namely $A_2^m$ for $\kappa<\kapCrit^m$ and $A_3^m$ in the other case, such that 
the resulting curve is not necessarily symmetric. An analogous statement holds for $A_{2,3}^s$.
The SYM-FM and SYM-AFM phase transition points are then given by the critical hopping parameters 
$\kapCrit^m$, $\kapCrit^s$ where the magnetic and staggered susceptibilities develop their
respective maximum.

%\includeFigTriple{ExamplePhaseTransDetA}{ExamplePhaseTransDetB}{ExamplePhaseTransDetC}
\includeFigTriple{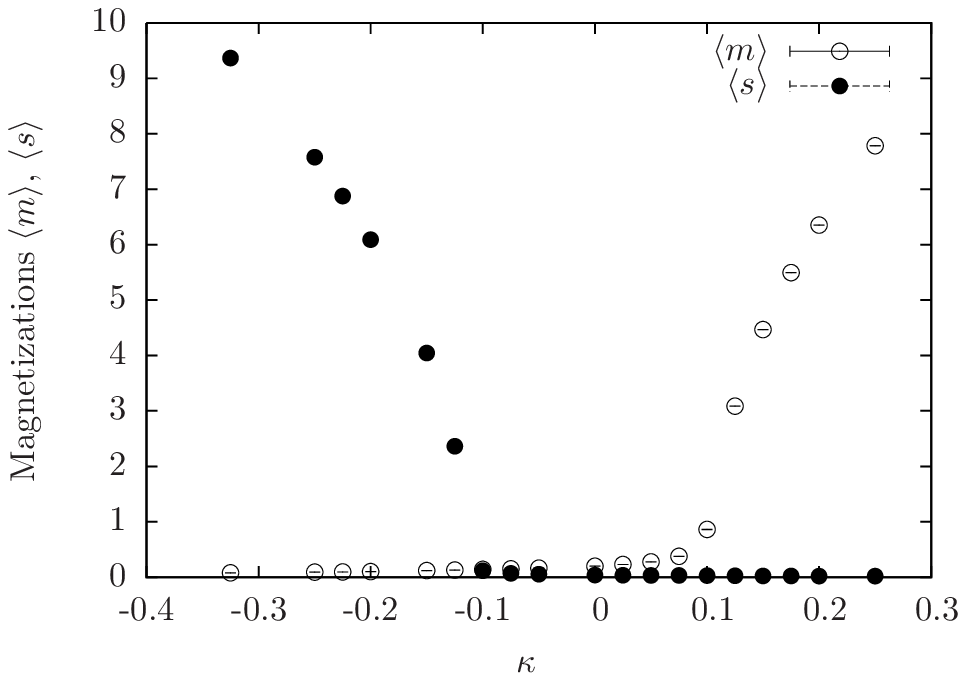}{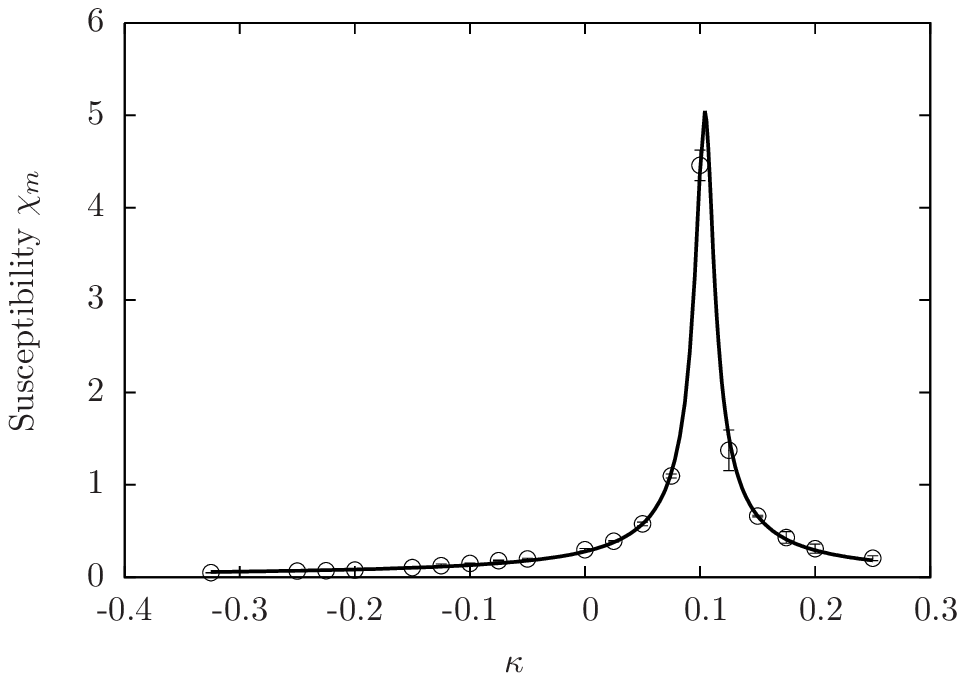}{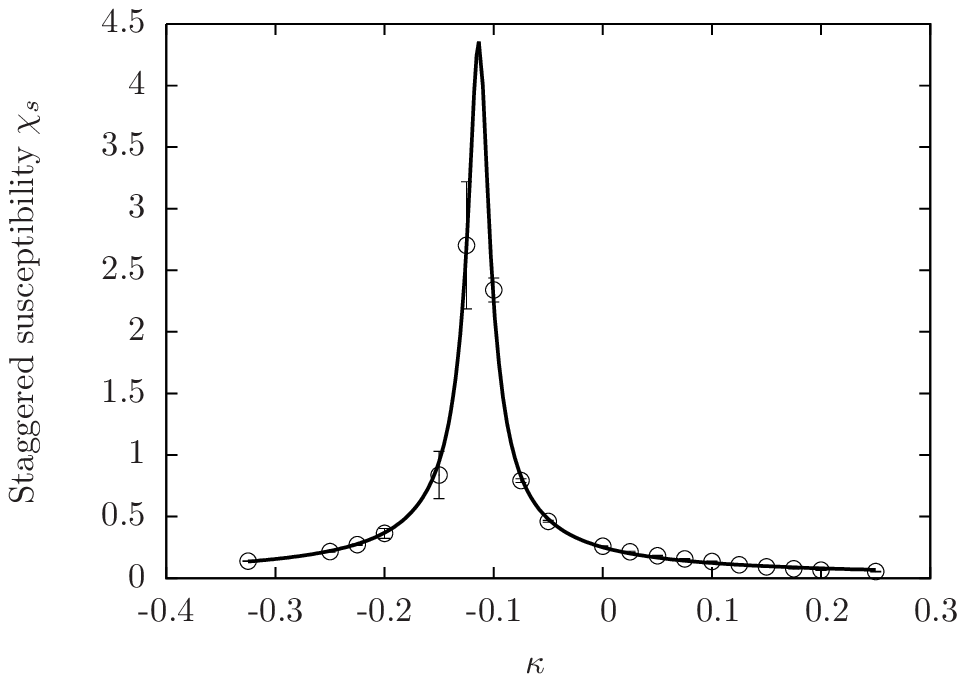}
{fig:kappascan1}
{An example for the determination of the phase transition points 
separating the ferromagnetic and the anti-ferromagnetic phase from the 
symmetric phase. The behaviour of the average magnetization $\mAvg$ and 
the staggered magnetization $\sAvg$ is shown in panel (a) as a function of $\kappa$. 
The corresponding susceptibilities are plotted in panels (b) and (c). 
The solid lines are fits with the finite size formula
in \eq{finitesizesus}. The parameters chosen are 
$\tilde y_N=0.632$, $\tilde\lambda_N=0.1$, $V=6^4$ and $N_f=10$.}
{Example for the numerical determination of the phase transition point.}

A typical example for such a determination of the phase transition points is presented 
in \fig{fig:kappascan1} for a rather small value of the Yukawa coupling 
constant. The average magnetizations $\mAvg$ and $\sAvg$ as well as the corresponding
susceptibilities are shown as a function of $\kappa$. The vanishing of the average 
magnetization and staggered magnetization is clearly observed when the 
symmetric phase is entered, except for some small finite volume effects. 
Associated to these phase transitions are peaks in the susceptibilities as expected, 
which can very well be described by the fit ansatz in \eq{finitesizesus}.
It is remarked that this ansatz, though not unique, provides a very good description 
of the numerically obtained data leading to a reliable determination of the critical 
hopping parameters $\kapCrit^{m,s}$ at least for the here considered small values
of the Yukawa and quartic coupling constants.

Using the strategy just described the values of $\kapCrit^m$ and $\kapCrit^s$ have been computed 
for various Yukawa coupling constants $\tilde y_N<5$, while holding the quartic coupling parameter
at the constant value $\tilde\lambda_N=0.1$. In \fig{fig:phasediagram1} the numerical results on 
the phase structure are presented as obtained by direct Monte-Carlo calculations on a $8^4$- and 
a $6^4$-lattice with $N_f=10$. These numerical findings are compared to the phase structure which was 
analytically determined in the infinite volume and infinite $N_f$-limit. On a qualitative level the 
analytical and the numerical results are in very good agreement. As expected, a symmetric (SYM), a 
ferromagnetic (FM) and an anti-ferromagnetic (AFM) phase can be observed in the obtained lattice data.
Moreover, the numerically determined symmetric phase bends strongly towards smaller values of the 
critical hopping parameter, when the Yukawa coupling constant is increased, as predicted. 

It is remarked that the simulations become much more demanding when entering the
anti-ferromagnetic phase, due to an increasingly bad condition number of the fermion matrix
$\fermiMat$. Within the anti-ferromagnetic phase numerical results are therefore only presented on
rather small $6^4$-lattices here.

%\includeFigSingleMedium{plPhaseTransKappaL6and8smallY}
\includeFigSingleMedium{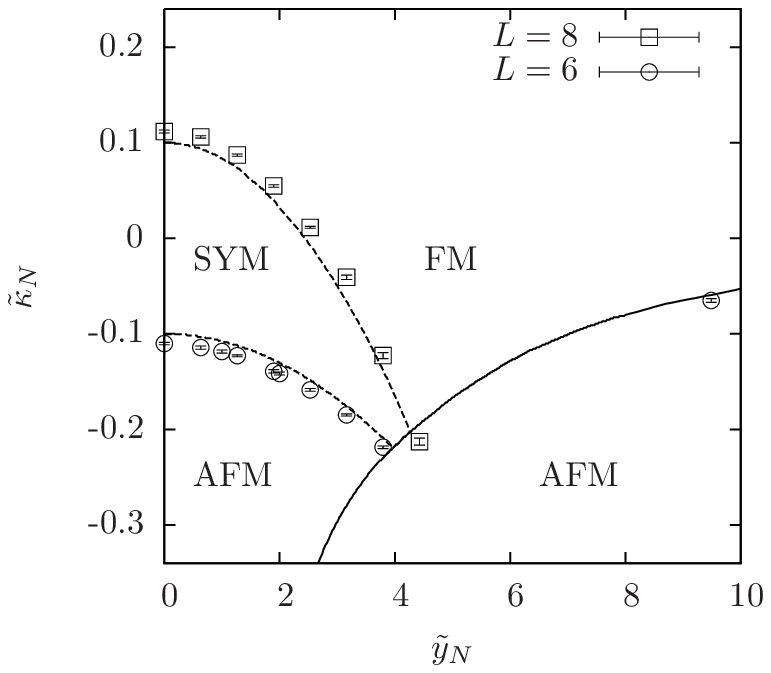}
{fig:phasediagram1}
{The phase diagram is presented at small Yukawa coupling constants as obtained by direct Monte-Carlo calculations
together with the corresponding analytical infinite volume prediction performed in the large $N_f$-limit. 
The dashed lines denote second order phase transitions while the solid line marks a first order transition.
The numerical data depicted by open squares were obtained on an \lattice{8}{8}while the ones represented 
by open circles were measured on a \latticeX{6}{6}{.} All results were obtained at
$\tilde \lambda_N=0.1$ and $N_f=10$.}
{Numerically obtained phase diagram at small Yukawa coupling constants and $N_f=10$.}

%\includeFigDouble{SignForFIphaseA}{SignForFIphaseB}
\includeFigDouble{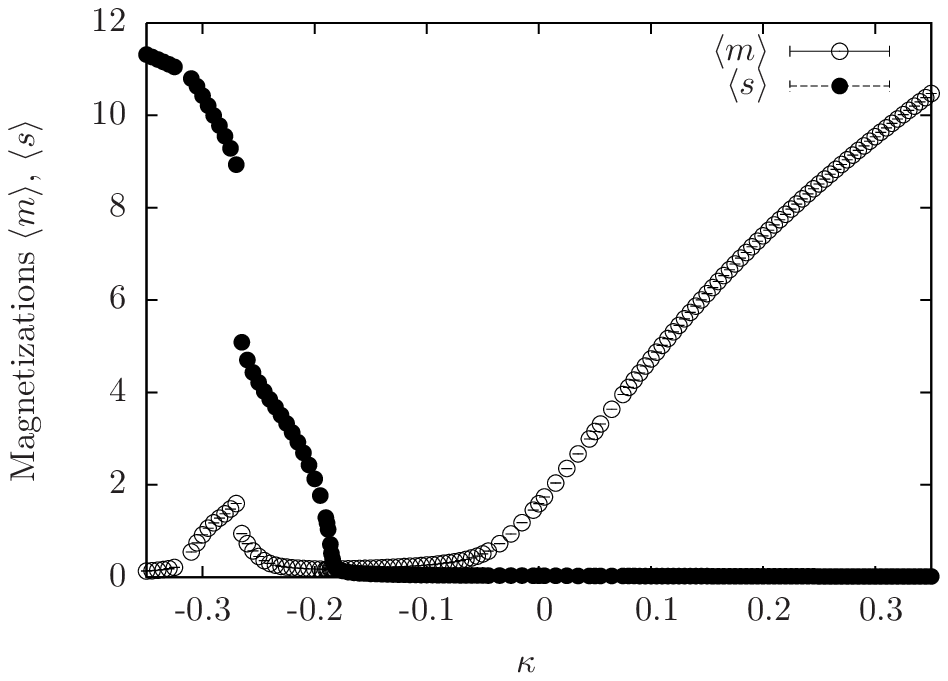}{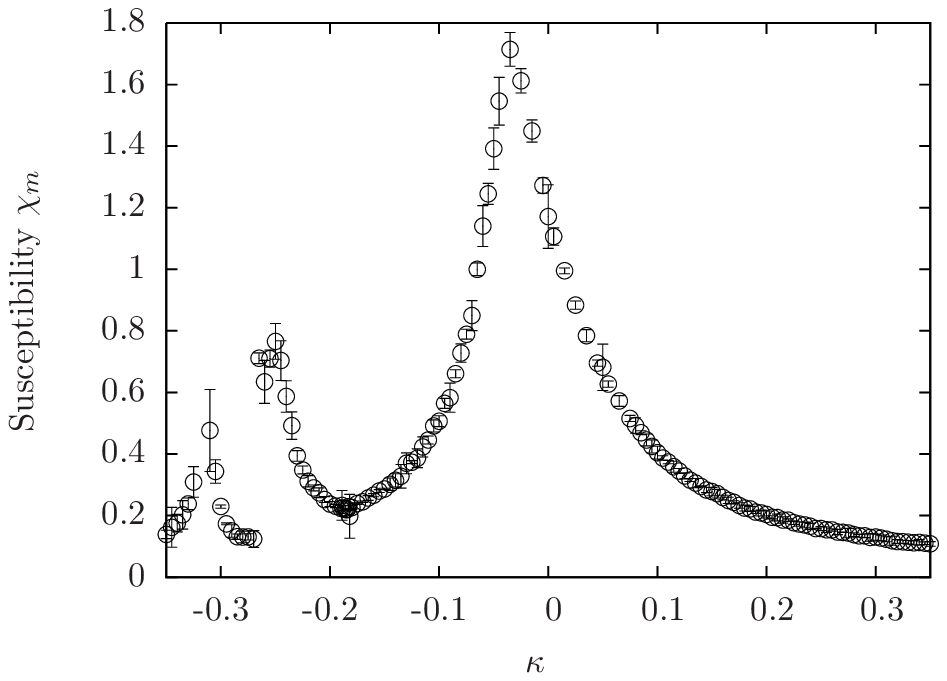}
{fig:ferrimagnetic}
{Evidence for the ferrimagnetic phase 
with $\mAvg >0$ and $\sAvg >0$ inside the anti-ferromagnetic phase.
The behaviour of the average magnetization $\mAvg$ and staggered 
magnetization $\sAvg$ is shown in panel (a) as a function of $\kappa$ for 
$\tilde y_N=3.162$, $\tilde\lambda_N=0.1$, $N_f=10$, and $V=6^4$.
The corresponding magnetic susceptibility is shown in panel (b).
From left to right its three observable peaks correspond to the phase transitions
AFM-FI, FI-AFM, and SYM-FM.
By the analytical large $N_f$, infinite volume calculation the ferrimagnetic 
phase was expected to occur approximately at $\kappa\le-0.27$, as can be inferred
from \fig{fig:PhaseDiagrams1}.}
{Evidence for the ferrimagnetic phase.}

Besides these three phases the analytical calculation in \sect{sec:SmallYukawaCouplings} has also
predicted a fourth, somewhat peculiar phase, where both, the average 
magnetization $\mAvg$ as well as its staggered counterpart $\sAvg$, are simultaneously non-zero.
This so-called ferrimagnetic (FI) phase was found to appear at 
intermediate values of the Yukawa coupling constant deeply inside the anti-ferromagnetic phase. 
In \fig{fig:ferrimagnetic} evidence for the existence of this ferrimagnetic phase is provided.
Its location within the phase diagram is in good agreement with the analytical
prediction. However, the ferrimagnetic phase is not within the prime interest of this 
study and is therefore not further investigated here.

Concerning the order of the encountered phase transitions one can conjecture that the SYM-FM 
as well as the SYM-AFM phase transition are of second order according to the 
continuous dependence of $\mAvg$ and $\sAvg$ on the hopping parameter $\kappa$
as seen for instance in \fig{fig:ferrimagnetic}a. This is in full agreement with the
corresponding predictions based on the analytical calculations in \sect{sec:SmallYukawaCouplings}. 
Contrarily, the direct FM-AFM phase transition occurring at intermediate 
values of the Yukawa coupling constant was predicted to be of first order. 
To clarify this an example of such a phase transition as obtained from 
the numerical calculations is presented in \fig{fig:kappascan2}. In panel (a) one 
can clearly observe an abrupt jump in the dependence of $\mAvg$ and $\sAvg$ on the hopping parameter
$\kappa$ indicating a discontinuous phase transition. 
Furthermore, an example of a tunneling event between two distinct ground states as observed
in a Monte-Carlo run performed close to the critical value $\kapCrit^m=\kapCrit^s$ 
is presented in subfigures (b) and (c), serving as another strong indication for the first order
nature of the considered phase transition. However, the order of this phase transition is not studied 
by more elaborate methods here, since this is not in the main interest of the present study.

Qualitatively, all presented findings are in very good agreement with the analytical 
large $N_f$ calculations in \sect{sec:SmallYukawaCouplings}. On a quantitative level, 
however, the encountered deviations in \fig{fig:phasediagram1} need 
to be further addressed. These deviations
can be ascribed to finite volume effects as well as finite $N_f$ corrections.

%\includeFigTriple{MSplotL4Nf10y2}{mTrajectoryL4Nf10y2}{sTrajectoryL4Nf10y2}
\includeFigTriple{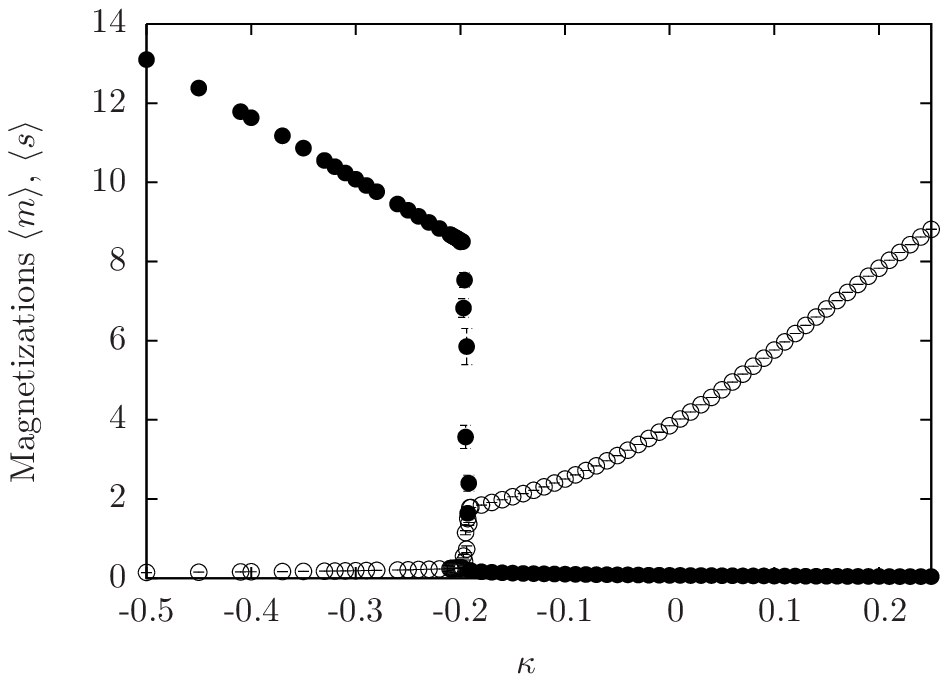}{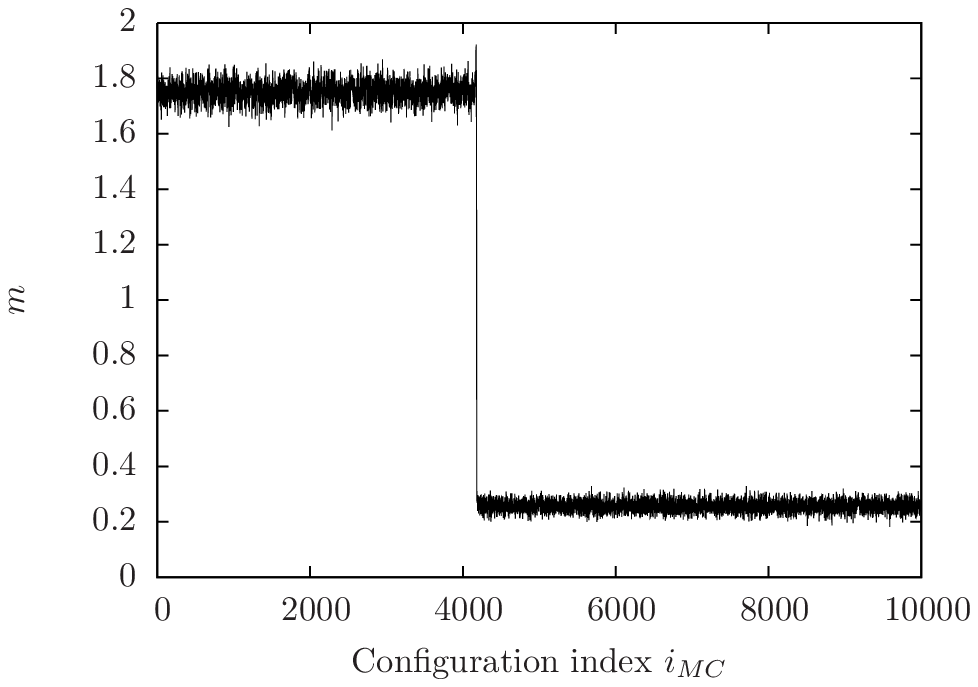}{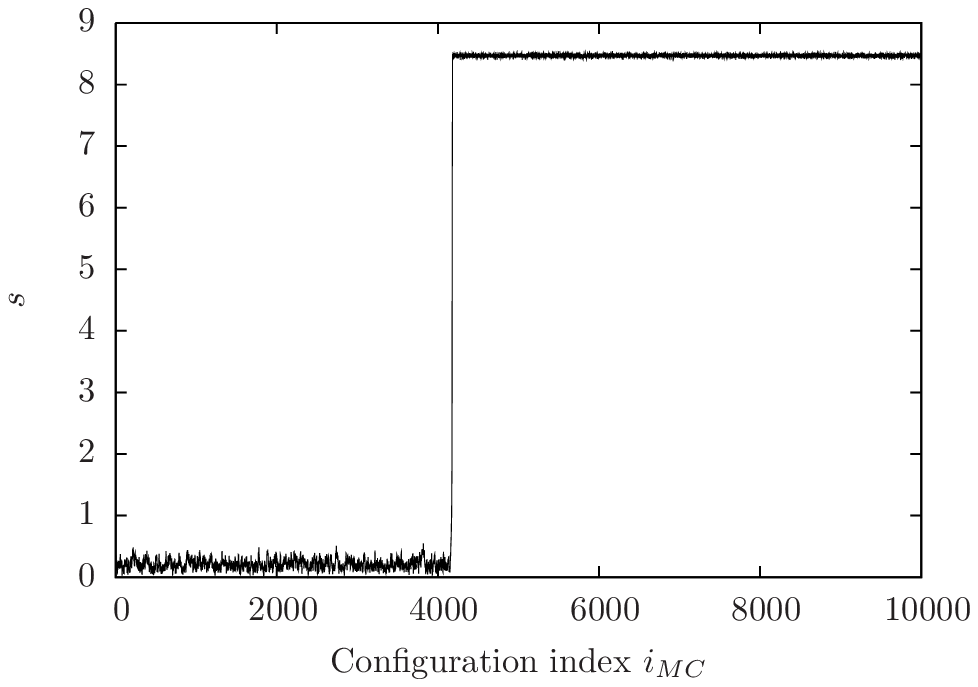}
{fig:kappascan2}
{The direct FM-AFM phase transition at intermediate values of the Yukawa
coupling constant. The behaviour of the average magnetization $\mAvg$ and staggered
magnetization $\sAvg$ is presented in panel (a) as a function of $\kappa$. 
The parameters chosen are $\tilde y_N=6.325$, $\tilde \lambda_N=0.1$, $V=4^4$, and $N_f=10$.
Panels (b) and (c) show a tunneling event between
two ground states which serves here as a strong indication for the first order nature 
of the phase transition.
Panels (b) and (c) show the magnetizations $m$ and $s$, respectively, 
versus the configuration index $\confNR$ at the hopping parameter $\kappa=-0.196$
being very close to its critical value.}
{The direct FM-AFM phase transition at intermediate values of the Yukawa coupling constant.}

%\includeFigSingleMedium{plFiniteEffectsPhaseDiagram1}
\includeFigSingleMedium{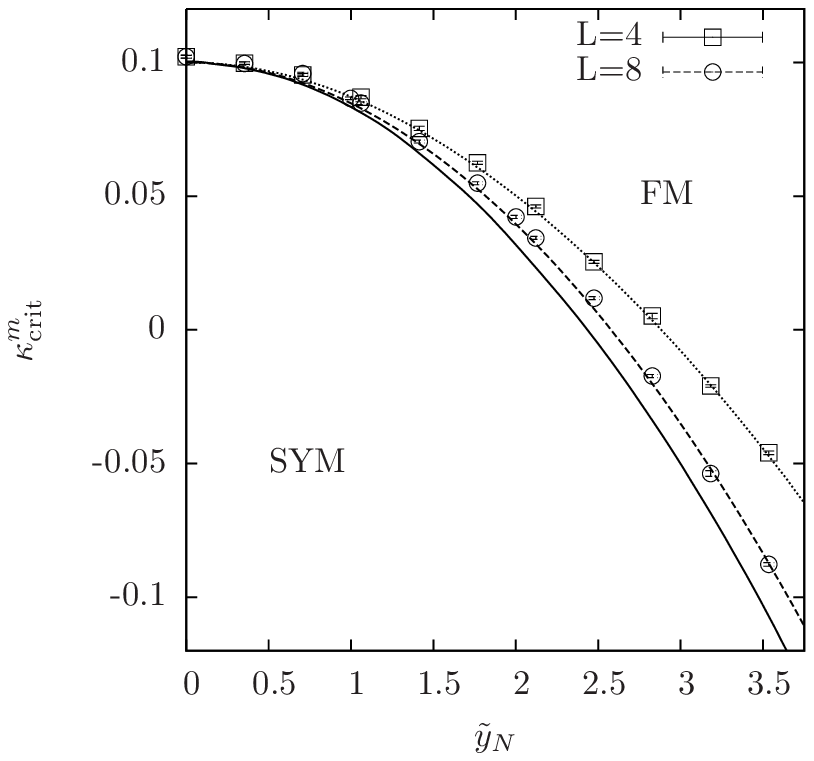}
{fig:finitesize1}
{A demonstration of finite size effects. The critical hopping parameter $\kapCrit^m$
separating the ferromagnetic and the symmetric phase is shown for several selected 
values of the Yukawa coupling constant. The open squares and open circles depict
the numerical data as obtained on a \lattice{4}{4}and on an \latticeX{8}{8}{,} respectively.
These results are compared to the analytical $V=4^4$ (dotted), $V=8^4$ (dashed), and 
$V=\infty$ (solid) phase transition lines analytically determined in the large $N_f$-limit.
The chosen parameters are $\tilde \lambda_N=0.1$ and $N_f = 50$.}
{Finite volume effects of the critical hopping parameter at small Yukawa coupling constants. }

The strength of the finite volume effects is illustrated in \fig{fig:finitesize1},
where the critical hopping parameter $\kapCrit^m$ dividing the ferromagnetic from the symmetric 
phase is presented for several selected values of the Yukawa coupling constant. These results
have been obtained from numerical calculations performed with $N_f=50$ on a \lattice{4}{4}(open squares), 
and on an \lattice{8}{8}(open circles). The rather large number $N_f=50$ has been 
chosen here in order to demonstrate the finite volume dependence isolated from the 
$N_f$-dependence. One can clearly observe that the phase transition 
line is strongly shifted towards smaller values of the hopping parameter when the lattice 
size is increased. 

This effect can also be anticipated by the analytical determination of the
phase transition line, when one imposes finite lattice sizes for the calculation 
of the effective potential in the large $N_f$ approximation. In \fig{fig:finitesize1}
the resulting analytical finite volume, large $N_f$ predictions for the phase transition lines
are depicted by the dotted, dashed, and solid curves representing the cases of the
\latticeX{4}{4}{,}the \latticeX{8}{8}{,}and the infinite volume limit, respectively. 
Comparing these results with the numerically obtained data for the critical hopping 
parameters one sees that the aforementioned curves describe the numerical findings very well. 
One can also observe the convergence of the numerically obtained critical hopping parameters
to the analytically predicted infinite volume phase transition line as the lattice size increases.

Concerning the analytical large $N_f$-calculation of the phase transition points 
on a finite lattice volume, it is remarked that the fermion determinant $det(\fermiMat[\Phi'])$ becomes 
identical to zero in finite volume, if both parameters $\breve m_\Phi$ and $\breve s_\Phi$, underlying
the considered ansatz $\Phi'$ for the sought-after ground state, vanish exactly. The analytical finite 
volume calculation will thus never predict a phase where $\breve m_\Phi$ and $\breve s_\Phi$
are simultaneously identical to zero. In case of a finite lattice volume the phase transition points to the 
symmetric phase can therefore not be determined by simply searching for that value of the hopping parameter 
$\kappa$, where the average magnetizations predicted by the analytical calculation become exactly zero. 
This difficulty does not occur in case of an infinite lattice volume, since the zero modes of $\fermiMat[\Phi']$ 
then form a set of only zero measure and the integral entering the effective potential in 
\eq{eq:EffActionRewrittenForSmallYFinal} can be shown to yield a finite value even at $\breve m_\Phi=\breve s_\Phi=0$, 
such that there actually is a symmetric phase in the sense of the ground state being $\Phi\equiv 0$ in the infinite 
volume limit. 

%\includeFigTriple{transPointNfDepLam001Y000}{transPointNfDepLam001Y010}{transPointNfDepLam001Y020}
\includeFigTriple{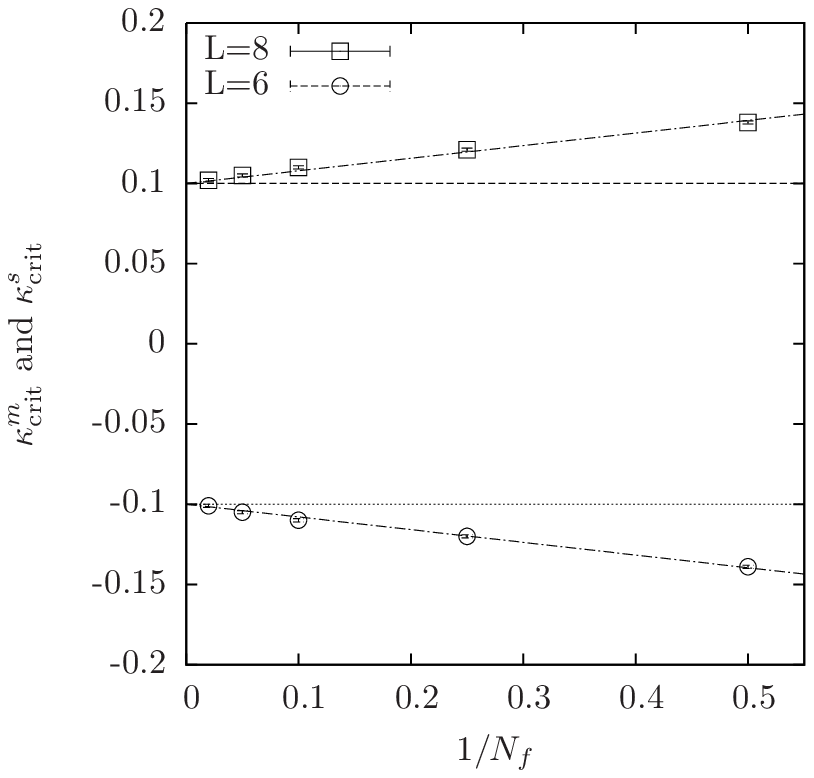}{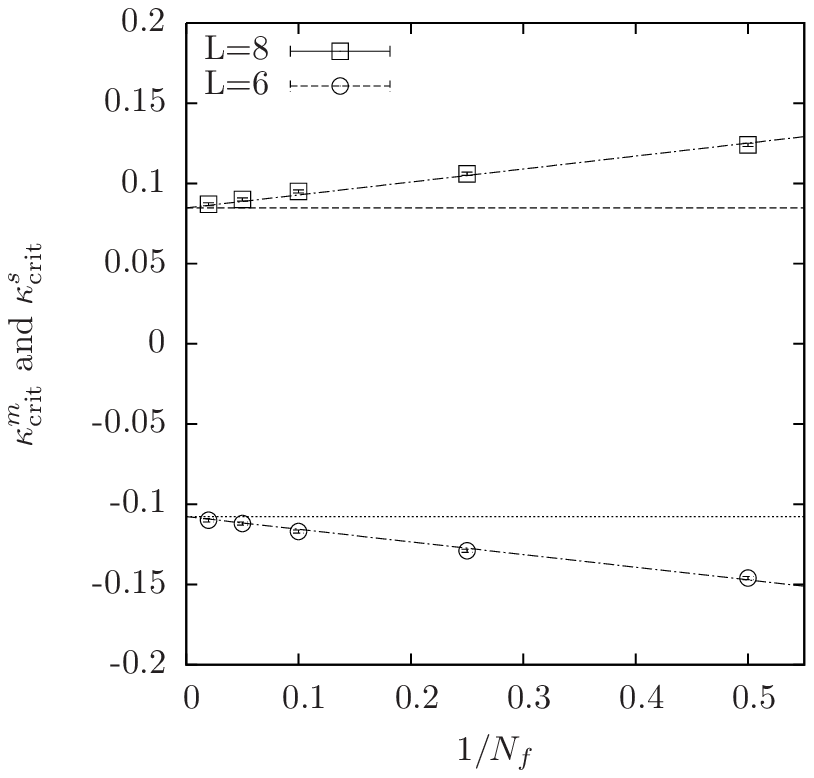}{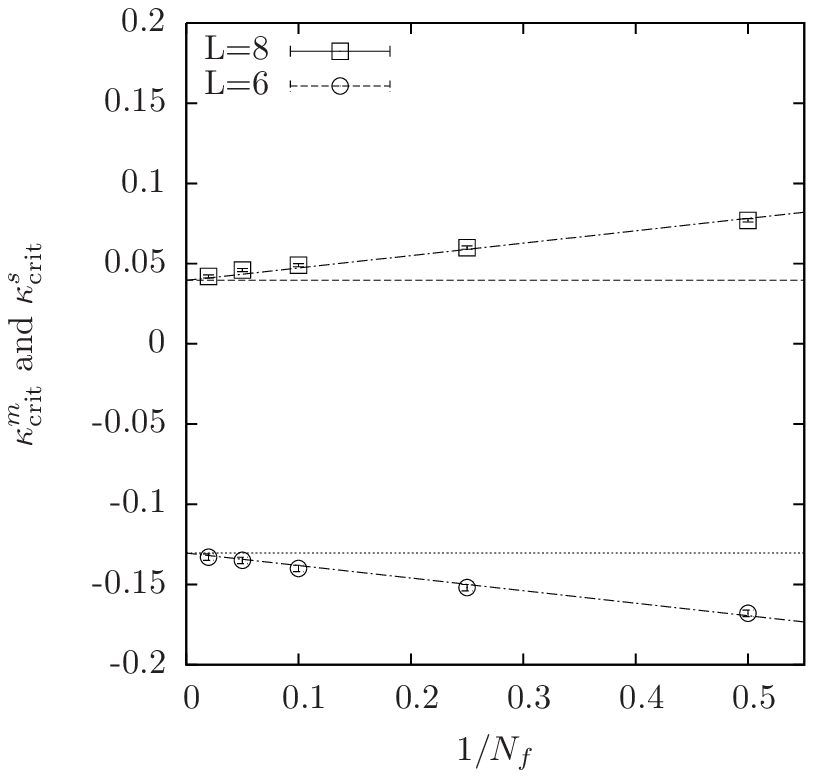}
{fig:NfDependence}
{The $N_f$-dependence of the critical hopping parameters
$\kapCrit^m$, $\kapCrit^s$ for the selected Yukawa coupling parameters
$\tilde y_N=0.0$ (a), $\tilde y_N=1.0$ (b), and $\tilde y_N=2.0$ (c).
The data with square symbols were measured on an \lattice{8}{8}while 
those represented by circular symbols were obtained on a \latticeX{6}{6}{.}The 
analytical finite volume, large $N_f$ predictions for the SYM-FM and the SYM-AFM
phase transitions are represented by the dashed and dotted horizontal lines, respectively. 
The dash-dotted lines are fits of the numerical data with the fit ansatz 
$f_{m,s}(N_f) = A_{m,s}/N_f + B_{m,s}$ where $B_{m,s}$ is set to the respective analytical 
prediction and $A_{m,s}$ is the only free fitting parameter. The results were computed with
$\tilde\lambda_N=0.1$.}
{$N_f$-dependence of the critical hopping parameter at small Yukawa coupling constants.}

For the purpose of finding the critical hopping parameters also in the finite volume scenario a different 
approach is therefore needed. Here $\kapCrit$ has been determined by finding that value of $\kappa$, 
where the absolute minimum of the effective potential $\tilde U(\breve m_\Phi, \breve s_\Phi)$ becomes flattest, 
\ie where the second derivative of the effective potential at the location of its minimum with respect 
to $\breve m_\Phi$ and $\breve s_\Phi$, respectively, becomes minimal. Since the scalar field oscillates the 
stronger around this minimum the smaller its second derivative is, this approach corresponds to finding the 
phase transition point by searching for the maximum of the susceptibility.

Finally, the $N_f$-dependence of the numerically obtained critical hopping parameters $\kapCrit^m$ 
and $\kapCrit^s$ is shown in \fig{fig:NfDependence} for several selected values of the Yukawa coupling 
constant. One clearly sees that for increasing $N_f$ the numerical results converge very well 
to the analytical finite volume, large $N_f$ predictions, as expected. It is interesting to note that
the leading order contribution of the finite $N_f$ corrections, which is of order $O(N_f^{-1})$ as 
discussed in \sect{sec:SmallYukawaCouplings}, seems to be the only practically relevant correction here, 
even at the smallest, with the HMC-algorithm accessible value $N_f=2$. This can be seen in 
\fig{fig:NfDependence} by fitting the observed deviations between the large $N_f$ predictions and the
numerical finite $N_f$ results with the fit ansatz $f_{m,s}(N_f)=A_{m,s}/N_f$ with $A_{m,s}$ being the 
only free parameter. Furthermore, one observes that the critical hopping parameter $\kapCrit^m$ is shifted
towards larger values with decreasing $N_f$ while $\kapCrit^s$ is
shifted towards smaller values.

From the presented numerical results one can conclude that the structure of the phase diagram
of the considered Higgs-Yukawa model at small values of the quartic and Yukawa coupling constant 
is very well described on a qualitative level by the results of the analytical large $N_f$ calculation
presented in \sect{sec:SmallYukawaCouplings}.
It also gives a very good understanding of the encountered finite volume effects.

%----------------------------------------------------------------------------------------------------------------------------------------------------------
\section{Large Yukawa and small quartic coupling constants}
\label{sec:LargeYukawaCoup}

In this section the phase diagram of the considered Higgs-Yukawa model 
will be studied in the regime of large values of the Yukawa coupling 
constant $\hat y$ and small values of the quartic coupling parameter
$\hat\lambda \ge 0$. This is done by considering the limit of a large
number of degenerate fermion generations $N_f$ while scaling the coupling 
parameters according to
\beq
\label{eq:DefOfLargeNScalingForStrongCoupling}
\hat y = \tilde y_N,\, \tilde y_N = \mbox{const}, \quad
\hat \lambda = \frac{\tilde \lambda_N}{N_f},\, \tilde \lambda_N = \mbox{const}, \quad
\kappa = \frac{\tilde \kappa_N}{N_f},\, \tilde \kappa_N = \mbox{const},
\eeq
where the quantities $\tilde y_N$, $\tilde \lambda_N$, and $\tilde \kappa_N$ are 
held constant in that limit procedure.

\subsection{Analytical calculations}
\label{sec:LargeYukawaCouplings}
The starting point of the analytical investigation of this large $N_f$-limit is
again the constraint effective potential given in \eq{eq:DefOfContraintEffPotForPhaseStruc}. 
Here, however, we will use the equivalent definition
\bea
\label{eq:DefOfContraintEffPotForPhaseStrucWithEffAction}
VU[\underline{\tilde \Phi}_0, \underline{\tilde \Phi}_{p_s}] &=& 
-\log\left(\int \left[\prod\limits_{0\neq k \neq p_s}\intd{\tilde \Phi_k}  \right]\,\,
e^{-S_\Phi[\Phi] - \SFeff[\Phi]} \Bigg|_{\tilde\Phi_0=\underline{\tilde \Phi}_0, \tilde\Phi_{p_s}=\underline{\tilde \Phi}_{p_s}}     \right)
\eea
based on the effective fermion action given in \eq{eq:DefOfEffFermionAction}.
 
The intended analytical calculation of the constraint effective potential 
$U[\underline{\tilde \Phi}_0, \underline{\tilde \Phi}_{p_s}]$ can then be performed 
in three steps. Firstly, the effective fermion action $\SFeff[\Phi]$ is expanded in 
powers of the inverse Yukawa coupling constant $1/\hat y$. 
Taking only the first non-vanishing contribution of this power series
into account and performing the large $N_f$-limit as specified in 
\eq{eq:DefOfLargeNScalingForStrongCoupling}, the model can then be shown to 
effectively become an $O(4)$-symmetric, non-linear sigma-model up to some finite
volume terms. Finally, the phase diagram of the latter sigma-model is 
determined by investigating the resulting expression for the effective potential
in some later specified large $N$ approximation, where $N$ denotes the number
of components of the field $\Phi$, \ie $N=4$ in our case.

For the investigation of the effective action $\SFeff[\Phi]$ in finite volume it is crucial 
to pay special attention to the modes in the unphysical corners of the Brillouin
zone given as
\beq
\ImpBasis_\pi = \left\{\Psi^{p,\zeta\epsilon k} : p_\mu\in\{0,\pi\},\, p\neq 0,\, \zeta,\epsilon=\pm 1,\, k\in\{1,2\} 
\right\},
\label{eq:DefPiModes}
\eeq
where the vectors $\Psi^{p,\zeta\epsilon k}$ have explicitly been defined in \eq{eq:DefOfEigenvectorsOfD1}.
These modes will be referred to in the following as $\pi$-modes. Given these $120$ $\pi$-modes 
one can define the corresponding projection operator
\beq
\label{eq:DefOfProjector}
P_\pi = \sum\limits_{\Psi \in \ImpBasis_\pi} \Psi \Psi^\dagger
\eeq
projecting to the sub-space $V_\pi=\mbox{span}(\ImpBasis_\pi)$ spanned by $\ImpBasis_\pi$. 
Using this notation it is straightforward to establish the very helpful relation
\bea
\label{eq:DetRelationWithProjector}
\det\left(E\left(\ID-P_\pi\right) + P_\pi F P_\pi \right)
&=&
\det\left(\left(\ID-P_\pi\right)E\left(\ID-P_\pi\right) + P_\pi F P_\pi \right)  \\
&=&
\det\left(\left(\ID-P_\pi\right)E + P_\pi F P_\pi \right) \nonumber\\
&=&
\detPrime\left( E \right) \cdot \detStar\left( F \right)\nonumber
\eea
where $E$ and $F$ are arbitrary operators defined on the same space $V_0$ as $\D$ and $B$. Here the expression
$\mbox{det}^*\left( F \right)$ denotes the determinant of $F$ with respect to the sub-space $V_\pi$ and
$\mbox{det}'\left( E \right)$ is the determinant of $E$ with respect to the complementary space 
$\bar V_\pi\equiv \mbox{span}(\ImpBasis / \ImpBasis_\pi)$, where $\ImpBasis=\{\Psi^{p,\zeta\epsilon k}:p\in\ImpSpace\}$ 
denotes the full set of all modes $\Psi^{p,\zeta\epsilon k}$. Using \eq{eq:DetRelationWithProjector} several times 
one can rewrite the effective action, neglecting constant factors independent of $\Phi$, according to
\bea
\label{eq:ReductionOfEffectiveAction}
e^{-\frac{\SFeff[\Phi]}{N_f}} 
&\fhs{-3mm}=\fhs{-3mm}&
\det\left( \D + B\left(\ID - \frac{1}{2\rho}\D\right)\right)\\
&\fhs{-3mm}=\fhs{-3mm}&
\detPrime\left( \hat y \Bscaled'\left(\Dprime-2\rho\ID'\right)-2\rho\Dprime  \right) \nonumber \\
&\fhs{-3mm}=\fhs{-3mm}&
\detPrime\left( \Dprime-2\rho\ID'\right) \cdot
\detPrime\left( \Bscaled'-\frac{2\rho}{\hat y}\Dprime\left(\Dprime-2\rho\ID'\right)^{-1}  \right) \nonumber \\
&\fhs{-3mm}=\fhs{-3mm}&
\det\left(  \Bscaled - \left(\Bscaled-\ID\right)P_\pi -\frac{2\rho}{\hat y} \DDmtwoRhoInv   \right) \nonumber \\
&\fhs{-3mm}=\fhs{-3mm}&
\det\left(\Bscaled\right) \cdot \det\left(\ID-\left(\ID-\Bscaled^{-1}\right)P_\pi\right)
\cdot \det\left(\ID - \frac{2\rho}{\hat y}\Bscaled^{-1}\DDmtwoRhoInv \left[ \ID - \left(\ID-\Bscaled^{-1}\right)P_\pi \right]^{-1} \right)
\nonumber
\eea
where the introduction of the underscore through $\Bscaled \equiv B / \hat y$ is only intended 
to make the dependence on the Yukawa coupling constant explicit for the sake of a better readability.
Here $\Dprime$, $B'$, and $\ID'$ denote the restrictions of the operators $\D$, $B$, and $\ID$ to
the sub-space $\bar V_\pi$. This restriction is introduced, since it guarantees $\Dprime-2\rho\ID'$ to be 
invertible. The operator $\DDmtwoRhoInv$ is then defined by extending the domain of the inverse of 
$\Dprime-2\rho\ID'$ again to the full space $V_0$ by inserting the projector 
$\ID-P_\pi$ according to
\beq
\label{eq:DefOfOperatorA}
\DDmtwoRhoInv = \Dprime\cdot \left[\Dprime-2\rho\ID'  \right]^{-1} \cdot 
\left(\ID - P_\pi\right),
\eeq
which is well-defined and finite over the whole space $V_0$. The last 
determinant in \eq{eq:ReductionOfEffectiveAction} can further be 
reduced by using the relation
\bea
\left[ \ID - \left(\ID-\Bscaled^{-1}\right)P_\pi \right]^{-1} &=&
\ID - P_\pi + P_\pi\left(1-P_\pi + P_\pi \Bscaled^{-1} P_\pi \right)^{-1}P_\pi \nonumber \\
&-&\left(\ID-P_\pi \right) \Bscaled^{-1} P_\pi \left(1-P_\pi + P_\pi \Bscaled^{-1} P_\pi \right)^{-1}P_\pi
\eea
and by applying again \eq{eq:DetRelationWithProjector} leading then to the
compact notation for the effective action
\bea
\label{eq:EffectiveActionFullLargeYFirst}
\SFeff[\Phi] &=& - N_f \cdot \log\det\left(\Bscaled\right) 
- N_f \cdot \log\detStar\left(\Bscaled^{-1}\right) 
- N_f \cdot \log\detStar\left(\ID + \frac{2\rho}{\hat y} F[\Phi] \right)  \nonumber\\
&-& N_f \cdot \log\det\left(\ID - \frac{2\rho}{\hat y} \DDmtwoRhoInv
\cdot \Bscaled^{-1} \right).
\eea
Here the abbreviation $F[\Phi]$ is defined as the somewhat lengthy expression
\bea
\label{eq:DefOfFiniteTermMat}
F[\Phi] &=&  \left[\ID-\frac{2\rho}{\hat y} \Bscaled^{-1}\DDmtwoRhoInv\right]^{-1}
\Bscaled^{-1} \DDmtwoRhoInv \Bscaled^{-1} P_\pi \left[\ID-P_\pi +P_\pi \Bscaled^{-1}  P_\pi   \right]^{-1}.
\eea
  
It is remarked that the latter determinants $\detStar$ only give rise to some finite volume effects, 
since these determinants are only performed over the 120-dimensional sub-space $V_\pi$. Their 
contributions to the effective action do therefore {\textit{not}} scale proportional to the volume $V$ 
as the lattice size increases in contrast to all other appearing terms.
We will come back to discussing these finite volume effects later. 
Here, however, we first continue with the evaluation of the last term in 
\eq{eq:EffectiveActionFullLargeYFirst} by rewriting the corresponding trace 
as a power series in the inverse coupling constant $1/\hat y$ according to
\bea
\label{eq:DefPowerSeries}
\mbox{Tr}\log\left(\ID - \frac{2\rho}{\hat y} \DDmtwoRhoInv \cdot \Bscaled^{-1} \right)
&=&
- \mbox{Tr}\, \sum\limits_{r=1}^{\infty} \frac{2^r}{r} \left(\frac{\rho}{\hat y}\right)^r 
\left[ \DDmtwoRhoInv \Bscaled^{-1}\right]^r 
\eea
and by eventually cutting off this power series after the first
non-vanishing term, which is well-justified for sufficiently large values of $\hat y$.
For the purpose of establishing the desired connection to a sigma-model
it is most convenient to evaluate these expressions in position space. 
Then the matrix $\Bscaled^{-1}$ is block diagonal and explicitly given by
\beq
\label{eq:InverseOfB}
\Bscaled^{-1} = \Bscaled^\dagger \cdot \left(\Bscaled \Bscaled^\dagger  \right)^{-1}, \quad
\Bscaled^{-1}_{x,y} = \delta_{x,y} \cdot \BscaledHat(\Phi_x^*/|\Phi_x|^2),
\eeq
where the notation $(\Phi^*_x)^0=\Phi^0_x$, $(\Phi^*_x)^i= -\Phi^i_x$
was used, $\hat B(\Phi_x)$ was defined in \eq{eq:DefOfBHat}, and the underscore indicates
again a simple rescaling with the factor $1/\hat y$. 
In position space the matrix $\DDmtwoRhoInv \Bscaled^{-1}$ can hence be written as
\bea
\label{eq:CalcOfABInv}
\left[\DDmtwoRhoInv \Bscaled^{-1}\right]_{x,y} &\fhs{-3mm}=& \sum\limits_{p\in\ImpSpace} \sum\limits_{\zeta\epsilon k}  \frac{e^{ipx} u^{\zeta\epsilon k}(p) \alpha^\epsilon(p)
e^{-ipy}\left[u^{\zeta\epsilon k}(p)\right]^\dagger}{|\Psi^{p,\zeta\epsilon k}|^2} \BscaledHat(\Phi^*_{y}/|\Phi_{y}|^2)  \\
&\fhs{-3mm}=& \frac{1}{V}\sum\limits_{p\in\ImpSpace} \sum\limits_{{\zeta\epsilon k}\atop{\zeta'\epsilon' k'}}  \alpha^\epsilon(p)
e^{ip(x-y)}
{u^{\zeta\epsilon k}(p)}  \left(\BscaledHat(p, \Phi^*_{y}/|\Phi_{y}|^2)\right)_{\zeta\epsilon k, \zeta'\epsilon' k'}  
{\left[u^{\zeta'\epsilon' k'}(p)\right]^\dagger} \nonumber
\eea
with $\hat B(p, \Phi_y)$ as defined in \eq{eq:MatBinSpinorRep}. The scalars $\alpha^\epsilon(p)$ 
denote the eigenvalues of the anti-hermitian operator $\DDmtwoRhoInv$ corresponding to its eigenvectors 
$\Psi^{p,\zeta\epsilon k}$ and are explicitly given by
\bea
\label{eq:defOfAlpha}
i\re\ni \alpha^\epsilon(p) &=& 
\Bigg\{
\begin{array}{*{3}{ccl}}
\frac{\nu^\epsilon(p)}{\nu^\epsilon(p) - 2\rho}  &:& p\in\ImpSpace,\,\nu^\epsilon(p)\neq 2\rho,        \\ 
0  &:& p\in\ImpSpace,\,\nu^\epsilon(p) =  2\rho.\\
\end{array}
\eea
The result for the trace of the operator $\DDmtwoRhoInv \Bscaled^{-1}$ is then directly found to be
\bea
\label{eq:TrGnPower1}
Tr\, \left[ \DDmtwoRhoInv \Bscaled^{-1} \right] 
&=& \frac{1}{V} \sum\limits_{x}\sum\limits_{p\in\ImpSpace} \mbox{Tr}_{8\times 8} \left[
|\Phi_{x}|^{-2} \DDmtwoRhoInv(p)
\BscaledHat(p, \Phi^*_{x})\right],
\eea
which can be generalized to the trace of the $r$-th power of $\DDmtwoRhoInv \Bscaled^{-1}$ yielding
\beq
\label{eq:TrGn1}
\fhs{-2mm}
\Tr\, \left[ \DDmtwoRhoInv \Bscaled^{-1} \right]^r 
\fhs{-2mm}= \fhs{-5mm}\sum\limits_{{x_1,...,x_r}\atop{p_1,...,p_r\in\ImpSpace}} \fhs{-4mm}
\Tr_{8\times 8}\, \left[  \prod\limits_{j=1}^r \frac{e^{ip_j(x_j-x_{j+1})}}{V} 
|\Phi_{x_{j+1}}|^{-2} \DDmtwoRhoInv(p_j)  \BscaledHat(p_j,\Phi^*_{x_{j+1}}) \Upsilon(p_j,p_{j+1}) \right] 
\eeq
with $p_{r+1}$ being identified with $p_1$, and $x_{r+1}$ with $x_1$. 
Here the expression $\DDmtwoRhoInv(p)$ stands for the diagonal matrix 
\beq
\DDmtwoRhoInv(p)_{\zeta_1\epsilon_1 k_1,\zeta_2\epsilon_2 k_2} = 
\delta_{\zeta_1,\zeta_2} \cdot \delta_{\epsilon_1,\epsilon_2} \cdot \delta_{k_1,k_2} \cdot
\alpha^{\epsilon_1}(p).
\eeq

The $8\times 8$ trace appearing in \eq{eq:TrGn1} can be further simplified by inserting the identity 
$\Upsilon(p_i,0)\Upsilon(0,p_i)$ at some proper places leading then to
\beq
\label{eq:TrGn2}
\fhs{-2mm}
\Tr_{8\times 8}\, \left[  \prod\limits_{j=1}^r \DDmtwoRhoInv(p_j) \BscaledHat(p_j,\Phi^*_{x_{j+1}}) \Upsilon(p_j,p_{j+1}) \right]
=
\Tr_{8\times 8}\, \left[\prod\limits_{j=1}^r \DDmtwoRhoInv(0,p_j) \BscaledHat(0,\Phi^*_{x_{j+1}}) \right]
\eeq
with $\DDmtwoRhoInv(0,p)$ given as
\bea
\DDmtwoRhoInv(0,p) &=& \Upsilon(0,p) \DDmtwoRhoInv(p) \Upsilon(p,0) \nonumber \\
&=& \frac{\alpha^+(p)}{\sqrt{\tilde p^2}} \cdot 
\diag\left[
\left(
\begin{array}{*{2}{c}}
\tilde p_0 & -\vec{\tilde p}\vec\theta\\
\vec{\tilde p}\vec\theta & -\tilde p_0\\
\end{array}
\right) 
,
\left(
\begin{array}{*{2}{c}}
\tilde p_0 & -\vec{\tilde p}\vec\theta\\
\vec{\tilde p}\vec\theta & -\tilde p_0\\
\end{array}
\right) \right].
\eea
Here, the relation $\alpha^+(p) = - \alpha^-(p)$ has been exploited. 
Due to the insertion of the spinor basis transformation matrices $\Upsilon(p_i,0)$
and $\Upsilon(0,p_i)$ the sums over the momenta in \eq{eq:TrGn1} factorize now according to
\bea
\label{eq:DefOfMatStrucT}
\Tr\,\left[\DDmtwoRhoInv \underline{B}^{-1}   \right]^r
&=& \sum\limits_{x_1,...,x_r} \Tr_{8\times 8}\, \left[  \prod\limits_{j=1}^r 
{\mathcal T}_{x_j,x_{j+1}} \right],\\
{\mathcal T}_{x,y} 
&=&
\sum\limits_{p\in\ImpSpace} \frac{e^{ip(x-y)}}{V}  \DDmtwoRhoInv(0, p) 
|\Phi_{y}|^{-2} \BscaledHat(0,\Phi^*_{y})
\eea
where the momentum sum is a four-dimensional Fourier transform of an anti-symmetric and purely imaginary 
function, hence yielding real values. With the definition 
\beq
\label{eq:DefFourTransOfAlpha}
\re\ni \CoupStr_\mu \equiv \CoupStr_\mu(\Delta x) = -\CoupStr_\mu(-\Delta x) = 
\sum\limits_{p\in\ImpSpace} \frac{e^{ip\Delta x}}{V} \alpha^+(p) \cdot \frac{\tilde p_\mu}{\sqrt{\tilde p^2}},
\quad \Delta x = x-y
\eeq
the hermitian $8\fhs{-1mm}\times\fhs{-1mm} 8$-matrix ${\mathcal T}_{x,y}$ can compactly be written
as
\beq
\label{eq:ExplicitTn}
{\mathcal T}_{x,y} = \frac{1}{|\Phi_y|^2} \cdot 
\left(
\begin{array}{*{4}{c}}
\Phi^0_y\CoupStr_0 + i\Phi^1_y\vec\CoupStr\vec\theta& -i\Phi^1_y\CoupStr_0-\Phi^0_y\vec{\CoupStr}\vec\theta& (i\Phi^3_y-\Phi^2_y)\vec{\CoupStr}\vec\theta &(-i\Phi^3_y+\Phi^2_y)\CoupStr_0 \\
\Phi^0_y\vec\CoupStr\vec\theta + i\Phi^1_y\CoupStr_0& -i\Phi^1_y\vec\CoupStr\vec\theta -
\Phi^0_y\CoupStr_0 & (i\Phi^3_y-\Phi^2_y)\CoupStr_0 &(-i\Phi^3_y+\Phi^2_y)\vec\CoupStr\vec\theta \\
(i\Phi^3_y+\Phi^2_y)\vec\CoupStr\vec\theta&-(i\Phi^3_y+\Phi^2_y)\CoupStr_0 & \Phi^0_y\CoupStr_0-i\Phi^1_y\vec\CoupStr\vec\theta &i\Phi^1_y\CoupStr_0-\Phi^0_y\vec\CoupStr\vec\theta \\
(i\Phi^3_y+\Phi^2_y)\CoupStr_0& -(i\Phi^3_y+\Phi^2_y)\vec\CoupStr\vec\theta & \Phi^0_y\vec\CoupStr\vec\theta-i\Phi^1_y\CoupStr_0 & i\Phi^1_y\vec\CoupStr\vec\theta - \Phi^0_y\CoupStr_0 \\
\end{array}
\right)
\eeq
leading directly to the result that the lowest order contribution of the power series 
in \eq{eq:DefPowerSeries} vanishes according to
\beq
\Tr\,\left[\DDmtwoRhoInv  \Bscaled^{-1}  \right] = \sum\limits_{x}\, \Tr_{8\times 8} 
\left[ {\mathcal T}_{x,x} \right] = 0.
\eeq
The first non-vanishing contribution is the second 
order term, which can be evaluated by explicitly computing the $8 \times 8$
trace, yielding
\bea
\label{eq:SecondOrderA4}
\Tr\,\left[\DDmtwoRhoInv  \Bscaled^{-1}  \right]^2 &=&  \sum\limits_{x,y} \Tr_{8\times 8}\,\left[  {\mathcal T}_{x,y} {\mathcal T}_{y,x} \right] \nonumber \\
&=&
-8 \cdot \sum\limits_\mu\sum\limits_{x,y} \frac{\Phi_{x}^\mu \Phi_{y}^\mu }{|\Phi_{x}|^2 \cdot |\Phi_{y}|^2} \cdot
\left|\CoupStr(\Delta x) \right|^2.
\eea
Cutting off the power series in \eq{eq:DefPowerSeries}
after this first non-vanishing term, which is well justified for sufficiently large values of the Yukawa
coupling constant $\hat y$, the effective fermion action can thus be written as
\bea
\label{eq:EffectiveActionFullLargeY}
\SFeff[\Phi] &=& -N_f \cdot \left(\sum\limits_{x} \log(|\Phi_x|^8) +  
\frac{(4\rho)^2}{\hat y^2}  \sum\limits_{x,y} \left|\CoupStr(\Delta x)\right|^2\frac{\Phi_{x}^\dagger
\Phi_{y}}{|\Phi_{x}|^2\cdot|\Phi_{y}|^2} \right) \\
&-& N_f \cdot \log\detStar\left(\Bscaled^{-1}\right) 
- N_f \cdot \log\detStar\left(\ID + \frac{2\rho}{\hat y} F[\Phi] \right). \nonumber
\eea

As already pointed out the determinants $\detStar$ are taken with respect to the $120$-dimensional
space $V_\pi$. In contrast to all other terms appearing in the effective action
these determinants are {\textit{not}} proportional to the volume $V$. They are therefore suppressed as 
the lattice size goes to infinity. Moreover, the very last term in \eq{eq:EffectiveActionFullLargeY}
even vanishes on finite lattice sizes when the Yukawa coupling constant $\hat y$ becomes sufficiently 
large. This is in contrast to the determinant $\detStar(\Bscaled^{-1})$ being independent of $\hat y$.
It is therefore quite instructive to evaluate the latter finite volume contribution in more detail,
which can be done at least if one restricts the consideration to the ansatz given in \eq{eq:staggeredAnsatz} 
taking only a constant and a staggered mode of the scalar field into account. The neglection of all other 
modes is not rigorously justified here, but motivated by the fact that in the broken (staggered-broken) phase the 
constant (staggered) modes yield the most dominant contributions. A crude estimate of the 
effect of $\detStar(\Bscaled^{-1})$ should thus be obtainable from this approach. Taking again only one 
orientation into account, \ie $\hat \Phi_1=\hat \Phi_2$, one finds
\beq
\label{eq:DetStarBInv}
\log\detStar \left( \underline{B}^{-1}[\Phi']\right) = 
-60\log\left( N_f \right)
+8\log\left| \tilde m_\Phi \right|
+56\log\left| \tilde m_\Phi^2 - \tilde s_\Phi^2 \right|,
\eeq
with the abbreviations
\beq
\label{eq:InverseBForstaggeredAnsatz}
\tilde m_\Phi = \frac{\breve m_\Phi}{\breve m_\Phi^2 - \breve s_\Phi^2} \quad \mbox{and} \quad
\tilde s_\Phi = \frac{\breve s_\Phi}{\breve s_\Phi^2 - \breve m_\Phi^2}.
\eeq

The resulting asymmetry in $\breve m_\Phi$ and $\breve s_\Phi$ is caused by the fact that the $8$ zero modes $\Psi^{0,\zeta\epsilon k}$ 
are not included in the sub-space $V_\pi$. The effect of the latter terms and especially the resulting asymmetry in $\mAvg$
and $\sAvg$ is clearly observed in corresponding Monte-Carlo simulations on small lattices and large values of the 
Yukawa coupling constant $\hat y$ as discussed in the subsequent section. Moreover, the result in \eq{eq:DetStarBInv}
would also hinder the expectation value $\mAvg$ from vanishing, thus obscuring the potential existence
of symmetric phases at large values of $\hat y$ on small lattices. However, as the lattice volume increases these finite 
volume effects eventually disappear relative too the contributions given in the upper row of \eq{eq:EffectiveActionFullLargeY}. 
In the following the $\detStar$-terms will therefore be neglected, which is well justified on sufficiently large lattice 
volumes.
 
To establish the announced connection to a sigma-model we now consider the
large $N_f$-limit where the coupling constants scale according to \eq{eq:DefOfLargeNScalingForStrongCoupling}.
With some implicit substitution of the integration variables the effective potential in \eq{eq:DefOfContraintEffPotForPhaseStrucWithEffAction} 
can then be written as
\beq
\label{eq:DefOfContraintEffPotInSigmaModell}
VU[\breve m_\Phi\sqrt{VN_f} \hat \Phi_1, \breve s_\Phi \sqrt{VN_f} \hat \Phi_2] = -\log\left(
\int \left[\prod\limits_{0\neq k \neq p_s}\intd{\tilde \Phi_k}  \right]\,\,
e^{- \SNzero[\Phi] - \SNone[\Phi] }  
\Bigg|_{{\tilde\Phi_0 = \breve m_\Phi\sqrt{V} \hat \Phi_1,}\atop{\tilde\Phi_{p_s} = \breve s_\Phi \sqrt{V} \hat \Phi_2}} \right)
\eeq
up to some constant terms independent of the variables $m_\Phi$, $s_\Phi$ and the orientations
$\hat \Phi_1$, $\hat \Phi_2$ where the actions $\SNzero[\Phi]$ and $\SNone[\Phi]$ are given as 
\bea
\label{eq:EffectiveActionsAfterNfSubstitution1}
\SNzero[\Phi] &=& -\tilde\kappa_N\sum\limits_{x,\mu} \Phi_x^\dagger \left[ \Phi_{x+\mu} +  \Phi_{x-\mu}        \right]
-\frac{(4\rho)^2}{\tilde y_N^2}  \sum\limits_{x,y} \left|\CoupStr(\Delta x)\right|^2\frac{\Phi_{x}^\dagger
\Phi_{y}}{|\Phi_{x}|^2\cdot|\Phi_{y}|^2}, \\
\label{eq:EffectiveActionsAfterNfSubstitution2}
\SNone[\Phi] &=& N_f \cdot\sum\limits_x \left[\Phi_x^\dagger \Phi_x 
+ \tilde\lambda_N \cdot \left(\Phi_x^\dagger \Phi_x -1 \right)^2
-4\log\left(\Phi_x^\dagger \Phi_x  \right)   \right].
\eea
From this result one finds that the effective potential is dominated by the action $\SNone[\Phi]$ in the large $N_f$-limit
which, however, is completely independent of the hopping parameter $\tilde\kappa_N$. Taking only this contribution
into account would thus not allow to study the phase transition with respect to $\tilde\kappa_N$. For that purpose 
the suppressed contribution $\SNzero[\Phi]$ also needs to be accounted for, which can be done in the large $N_f$-limit 
according to
\bea
U[\breve m_\Phi\sqrt{VN_f} \hat \Phi_1, \breve s_\Phi \sqrt{VN_f} \hat \Phi_2]
&\stackrel{N_f\rightarrow\infty}{\longrightarrow}&
U_\sigma[\breve m_\Phi\sqrt{V} \hat \Phi_1/\bar\varphi, \breve s_\Phi \sqrt{V} \hat \Phi_2/\bar\varphi]
\eea
where again constant terms have been neglected and $U_\sigma$ is given as
\bea
\label{eq:DefOfEffPotInSigmaModel}
VU_\sigma[\underline{\tilde\sigma}_0, \underline{\tilde\sigma}_{p_s}]
&=&
-\log\left(\int \left[\prod\limits_{0\neq k \neq p_s}\intd{\tilde \sigma_k}  \right]\,\,
e^{- \Ssigma[\sigma] }  \cdot \prod\limits_x \delta(\sigma_x^\dagger \sigma_x-1)     \,
\Bigg|_{{{\tilde\sigma_0=\underline{\tilde\sigma}_0,}\atop {\tilde\sigma_{p_s}=\underline{\tilde\sigma}_{p_s}}} }
\right)
\eea
according to the substitution $\tilde \sigma_k = \tilde\Phi_k/\bar\varphi$ and $\tilde \sigma$ denoting the Fourier 
transform of $\sigma$ defined analogously to \eq{eq:FTofPhi}. 
Here the dominant contribution $\SNone[\Phi]$ has enforced $\Phi_x^\dagger\Phi_x=\bar\varphi^2$ and
equivalently $\sigma_x^\dagger\sigma_x=1$ for all space-time points $x$, where the amplitude $\bar\varphi$ 
is given by the determination equation 
\bea
\label{eq:FixationOfPhiBetrag}
0 &=& -4 \cdot \frac{1}{\bar\varphi^2} + 1 + 2\tilde\lambda_N\cdot \left(\bar\varphi^2-1
\right),
\eea
while $\Ssigma[\sigma]$ reflects the sub-leading dynamics induced by $\SNzero[\Phi]$.
With this fixation of the scalar field amplitude the model in \eq{eq:EffectiveActionFullLargeY} 
effectively becomes a non-local, four-dimensional, non-linear sigma-model in
the infinite volume and large $N_f$-limit according to
\beq
\label{eq:DefOfSigmaModel}
\Ssigma[\sigma] = -\sum\limits_{x,y} \kappa^{eff}_{x,y} \cdot \sigma^\dagger_{x} \sigma_{y}, \quad \forall_x\,|\sigma_x| = 1
\eeq
with the effective, non-local coupling matrix
\beq	
\label{eq:EffectiveCouplingMatrix}
\kappa^{eff}_{x,y} =  \frac{16\rho^2}{\tilde y_N^2 \bar\varphi^2}\left| \CoupStr(\Delta x)  \right|^2 + 
\tilde\kappa_N\cdot \bar\varphi^2\cdot\sum\limits_{\mu=\pm 1}^{\pm 4} \delta_{\Delta
x, \hat e_\mu}. 
\eeq
Here the notion 'non-local' simply refers to the fact, that the field $\sigma_x$ 
at a given lattice site $x$ is coupled to $\sigma_y$ at any other site $y$ of the lattice. 
This leaves nevertheless open the possibility that the interaction is still local 
in a field theoretical sense with exponentially decaying coupling strength 
\cite{Hernandez:1998et}. This question was, however, not investigated in detail, 
since the eventual interest of this study focuses on the small Yukawa coupling regime.

It is remarked that the outcome in \eq{eq:EffectiveCouplingMatrix} reproduces the result which 
was found for a Higgs-Yukawa model based on Wilson fermions~\cite{Hasenfratz:1992xs} with the 
only difference that the coupling matrix in that case consisted solely of nearest-neighbour 
couplings.

It is further remarked that the extension of this analysis to the case, where the quartic
coupling constant is not small, \ie $\tilde\lambda_N \equiv \hat \lambda$, is straightforward. 
In that case the quartic coupling term in \eq{eq:EffectiveActionsAfterNfSubstitution2} would not scale
proportional to $N_f$ but proportional to $N_f^2$ instead, thus making the quartic coupling term the most dominant
contribution in the large $N_f$-limit. One would then still end up with a $\sigma$-model based on the
same coupling matrix as given in \eq{eq:EffectiveCouplingMatrix}, but with the difference that the 
determination equation for $\bar\varphi$ in \eq{eq:FixationOfPhiBetrag} would reduce to the trivial
condition $\bar\varphi = 1$. However, this part of the phase diagram is not in the main interest of
this study and is thus not further investigated here.

The desired phase structure of the considered Higgs-Yukawa model can now be obtained by minimizing 
the effective potential $U_\sigma[\underline{\tilde\sigma}_0, \underline{\tilde\sigma}_{p_s}]$ of the 
associated $\sigma$-model with respect to $\underline{\tilde\sigma}_0$ and $\underline{\tilde\sigma}_{p_s}$. 
However, the direct computation of $U_\sigma[\underline{\tilde\sigma}_0, \underline{\tilde\sigma}_{p_s}]$ 
is hindered by the constraint of the integration variables in \eq{eq:DefOfEffPotInSigmaModel}, which 
is a general obstacle for investigating $\sigma$-models. The crucial step towards the evaluation of 
$U_\sigma[\underline{\tilde\sigma}_0, \underline{\tilde\sigma}_{p_s}]$ is thus the removal of the
constraint $|\sigma_x|=1$. For that purpose the restriction $|\sigma_x|=1$ can be encoded as a 
$\delta$-function~\cite{Flyvbjerg:1988em, Justin:1993zu} written in terms of an integration of the term 
$\exp(i\omega^{(i)}_x(\sigma_x^\dagger\sigma_x-1))$ over the newly introduced one-component real scalar 
field $\omega^{(i)}_x$.

In this approach the constraint effective potential $U_\sigma[\underline{\tilde\sigma}_0, \underline{\tilde\sigma}_{p_s}]$ 
of the $\sigma$-model can be written as
\bea
\label{eq:DefOfEffPotInSigmaModelExtended}
VU_\sigma[\underline{\tilde\sigma}_0, \underline{\tilde\sigma}_{p_s}]
&=&
-\log\left(\int \prod\limits_{0\neq k \neq p_s}\intd{\tilde \sigma_k}  \,\int\limits_{-\infty}^{\infty}\prod\limits_{x} \intd{\omega_x^{(i)}}\,\,
e^{- \SsigmaOmega[\sigma, \omega] }       \,
\Bigg|_{{{\tilde\sigma_0=\underline{\tilde\sigma}_0,}\atop {\tilde\sigma_{p_s}=\underline{\tilde\sigma}_{p_s}}}}\right),
\eea
where constant terms have again been neglected and the action $\SsigmaOmega$ is given as
\bea
\label{eq:DefOfSigmaOmegaAction}
\SsigmaOmega[\sigma,\omega] &=& -\sum\limits_{x,y} \kappa^{eff}_{x,y} \cdot \sigma^\dagger_{x} \sigma_{y}
+\sum\limits_x \omega_x\left(\sigma^\dagger_{x} \sigma_{x} -1  \right)
\eea
with $\omega_x\equiv \omega_x^{(r)} + i\omega_x^{(i)}$. Up to this point the choice of $\omega_x^{(r)}\in \re$ is free. For later convenience
we set $\omega_x^{(r)} = \bar \omega^{(r)}$, where $\bar \omega^{(r)}\in\re$ is space-time independent. So far, the calculation 
performed with respect to the sigma model is exact. The first approximation occurs in the following step, where the action 
$\SsigmaOmega[\sigma,\omega]$ is replaced by $\SsigmaOmegaApprox[\sigma,\omega]$ defined as
\bea
\label{eq:DefOfSigmaOmegaActionApprox}
\SsigmaOmegaApprox[\sigma,\omega] &=& -\sum\limits_{x,y} \kappa^{eff}_{x,y} \cdot \sigma^\dagger_{x} \sigma_{y}
+\sum\limits_x \bar\omega\left(\sigma^\dagger_{x} \sigma_{x} -1  \right)
\eea
where $\bar\omega\equiv \bar\omega^{(r)}+i\bar\omega^{(i)}$ and $\bar\omega^{(i)}= \sum_x \omega_x^{(i)}/V$. The consideration
was here restricted to the constant mode of $\omega_x$, which was found to be justified\footnote{The approach considered in 
\Ref{Flyvbjerg:1988em} differs slightly. There the Gaussian integration over $\sigma_x^2,\ldots,\sigma_x^N$ is performed first, 
explicitly revealing then the resulting effective action to scale proportional to $N$, being thus dominated by the ground 
states of the remaining fields $\sigma_x^1$ and $\omega_x$ in the limit $N\rightarrow \infty$. The $\omega$-field content of 
these ground states was then found to be well describable by considering only the constant mode of $\omega_x$.
} in \Ref{Flyvbjerg:1988em} for sufficiently large 
values of $N$, with $N$ denoting the number of components of $\sigma_x$, \ie $N=4$ in this case. Whether this approximation
is justified here will turn out in the subsequent section, where the resulting analytical results will be compared to direct Monte-Carlo
calculations. If one accepts this approximation one can write the associated approximation of the effective potential as
\bea
\label{eq:ApproximationToEffPotential}
VU_\sigma^{(a)}[\underline{\tilde\sigma}_0, \underline{\tilde\sigma}_{p_s}] &=&
-\log\left(\int\limits_{-\infty}^{\infty} \intd{\bar\omega^{(i)}}\,\,
e^{- \SBarOmega\left[  \sqrt{\underline{\tilde\sigma}_0^\dagger \underline{\tilde\sigma}_0/V}, 
\sqrt{\underline{\tilde\sigma}_{p_s}^\dagger \underline{\tilde\sigma}_{p_s}/V}, 
\bar\omega\right] }      \right)
\eea
up to constant terms where the action $\SBarOmega$ is obtained from $\SsigmaOmega$ by integrating out the modes
$\tilde\sigma_k$, $0\in\ImpSpace/\{0,p_s\}$ according to
\bea
\label{eq:EffectiveActionReducedModel}
\SBarOmega[m_\sigma,s_\sigma,\LagrangeMul] &=& -\ln\left[ \detPrimePrime\left( -\kappa^{eff} + \LagrangeMul   \right)   \right]^{-N/2} 
+   m_\sigma^2 \cdot \left\langle0\left| -\kappa^{eff} + \LagrangeMul
\right|0\right\rangle \nonumber \\
&+& s_\sigma^2 \cdot \left\langle p_s \left| -\kappa^{eff} + \LagrangeMul \right|p_s\right\rangle
-V \LagrangeMul,
\eea
which is mathematically well-defined only if $\bar\omega^{(r)}$ was chosen sufficiently large, which will turn out to be the case,
at least for the SYM-FM phase transition, after having solved the resulting gap equations. At this point we continue with this 
formal expression and postpone the discussion of the validity of the performed Gauss integration to the end of this section.

In the latter result the shorthand notations $|0\rangle\equiv |(0,0,0,0)\rangle$ and $|p_s\rangle \equiv |(\pi,\pi,\pi,\pi)\rangle$ 
were used, denoting the constant and staggered modes according to
\beq
\label{eq:DefOfBRACKETmodes}
|k\rangle_x \; \equiv \; \sqrt{\frac{1}{V}} \; e^{ik\cdot x}
\eeq
being eigenvectors of the effective coupling matrix $\kappa^{eff}$. Furthermore, $\detPrimePrime$ is the determinant restricted to
the space $\mbox{span}(|k\rangle:\, k\in\ImpSpace/\{0,p_s\})$, thus neglecting the two latter modes. 
For convenience, this short-hand notation will also be applied in the following where it is unambiguous.

To evaluate the determinant in \eq{eq:EffectiveActionReducedModel}, the eigenvalues of the effective coupling matrix $\kappa^{eff}$ 
need to be known. The eigenvectors are simply plane waves with wave vectors $k\in\ImpSpace$ and one easily finds the corresponding 
eigenvalues according to
\bea
\label{eq:EigenValuesOfCouplingMatrixP}
\sum\limits_{y}\kappa^{eff}_{x,y} \cdot e^{iky} 
&=& 
\left(   2\tilde\kappa_N \bar \varphi^2 \sum\limits_{\mu= 1}^{4}\cos(k_\mu) 
+ \frac{16\rho^2}{\tilde y_N^2 \bar\varphi^2}\cdot q(k)\right) \cdot e^{ikx}
\eea
where $q(k)$ denotes the eigenvalues of the matrix $W_{x,y} \equiv |\CoupStr(\Delta x)|^2$ given as
\beq
\re\ni q(k) = \frac{1}{V}\sum\limits_{p\in\ImpSpace} 
\alpha^+(p)\cdot \alpha^+(\wp_k)\cdot \frac{\tilde p\cdot \tilde \wp_k}{\sqrt{\tilde p^2}\cdot
\sqrt{\tilde{\wp_k}^2}}, \quad \wp_k = k-p.
\eeq
With this notation the action in \eq{eq:EffectiveActionReducedModel} becomes
\bea
\label{eq:AlternativeApproachEffectiveActionSigmaModel}
\SBarOmega[m_\sigma,s_\sigma,\LagrangeMul] &=& \frac{N}{2}\TrPrimePrime\,\ln\left[ -\kappa^{eff} + \LagrangeMul \right] 
+ m_\sigma^2 \cdot V \cdot \left( -8\tilde\kappa_N\bar\varphi^2 - \frac{16\rho^2}{\tilde y_N^2\bar\varphi^2}q(0) + \LagrangeMul  \right) \nonumber \\
&+& s_\sigma^2 \cdot V \cdot \left( +8\tilde\kappa_N\bar\varphi^2 - \frac{16\rho^2}{\tilde y_N^2\bar\varphi^2}q(p_s) + \LagrangeMul  \right)
-V \LagrangeMul,
\eea
where the summation over the coupling matrix components has been performed 
by using \eq{eq:EigenValuesOfCouplingMatrixP} with the settings 
$k=0$ and $k=p_s$, respectively. Analogously to $\detPrimePrime$, $\TrPrimePrime$ denotes the trace neglecting 
the modes $|0\rangle$ and $|p_s\rangle$ in the aforementioned sense. In the infinite volume limit the 
considered system is dominated by the minima of the effective potential in \eq{eq:ApproximationToEffPotential}
determined through the stationary points $(\bar m_\sigma, \bar s_\sigma, \bar{\LagrangeMul})$ of $\SBarOmega$. 
These stationary points solve the following set of gap equations, which are obtained by differentiating $\SBarOmega$ 
with respect to $m_\sigma$, $s_\sigma$, and $\LagrangeMul$ leading to
\beq
\label{eq:LargeYGap1}
0=\bar m_\sigma\cdot \left[\bar{\LagrangeMul} - \left(8\tilde\kappa_N \bar\varphi^2 + \frac{16\rho^2}{\tilde y_N^2 \bar\varphi^2} \cdot q(0)\right) 
\right],
\eeq

\beq
\label{eq:LargeYGap2}
0=\bar s_\sigma\cdot \left[\bar{\LagrangeMul} - \left(-8\tilde\kappa_N \bar\varphi^2 + \frac{16\rho^2}{\tilde y_N^2 \bar\varphi^2} \cdot  
q(p_s)  \right)  \right], \quad \mbox{and}
\eeq

\beq
\label{eq:LargeYGap3}
\bar m_\sigma^2+ \bar s_\sigma^2 =  1 -  \frac{1}{V}\sum\limits_{{k\in\ImpSpace}\atop{0\neq k \neq p_s}} 
\left[-\tilde\kappa_N \bar\varphi^2 \sum\limits_{\mu= 1}^{ 4}\cos(k_\mu) - \frac{8\rho^2}{\tilde y_N^2 \bar\varphi^2} q(k) +
\frac{\bar{\LagrangeMul}}{2}\right]^{-1}
\eeq
where $N=4$ has been used.

The relation in \eq{eq:LargeYGap1} implies that $\bar m_\sigma$ or the given argument within the square brackets has to
vanish. An analogous observation can be drawn from \eq{eq:LargeYGap2}. For the investigation of the phase structure we 
now consider two different scenarios, namely a ferromagnetic phase corresponding to $\bar m_\sigma\neq 0$, $\bar s_\sigma=0$ 
and an anti-ferromagnetic phase corresponding to $\bar m_\sigma= 0$, $\bar s_\sigma\neq0$. For each of these cases one can 
then derive a self-consistency relation. For the aforementioned ferromagnetic phase~(FM) one obtains from \eq{eq:LargeYGap1}
\beq
\label{eq:LargeYGap4}
\bar{\LagrangeMul} = 8\tilde\kappa_N \bar\varphi^2 + \frac{16\rho^2}{\tilde y_N^2 \bar\varphi^2} \cdot q(0)
\eeq
and hence the self-consistency relation
\beq
\label{eq:LargeYGap5}
0< \bar m_\sigma^2 =  1 - \frac{1}{V}\sum\limits_{{k\in\ImpSpace}\atop{0\neq k \neq p_s}} 
\left[
{\mathcal W}_m(k)
\right]^{-1},
\eeq
where the quantity ${\mathcal W}_m(k)$ is defined as
\beq
{\mathcal W}_m(k)
=
\tilde\kappa_N \bar\varphi^2 \sum\limits_{\mu= 1}^{ 4}\left(1-\cos(k_\mu)\right) 
+ \frac{8\rho^2}{\tilde y_N^2 \bar\varphi^2} \left(q(0)-q(k)\right).
\eeq
For the anti-ferromagnetic phase~(AFM) on the other hand one obtains from \eq{eq:LargeYGap2}
\beq
\label{eq:LargeYGap6}
\bar{\LagrangeMul} = -8\tilde\kappa_N \bar\varphi^2 + \frac{16\rho^2}{\tilde y_N^2 \bar\varphi^2} \cdot q(p_s)
\eeq
and hence the self-consistency relation
\beq
\label{eq:LargeYGap7}
0<\bar s_\sigma^2 =  1 -  \frac{1}{V}\sum\limits_{{k\in\ImpSpace_s(\epsilon)}\atop{0\neq k \neq p_s}} 
\left[
{\mathcal W}_s(k)
\right]^{-1},
\eeq
where the quantity ${\mathcal W}_s(k)$ is defined as
\beq
{\mathcal W}_s(k)
=
-\tilde\kappa_N \bar\varphi^2 \sum\limits_{\mu= 1}^{ 4}\left(1+\cos(k_\mu)\right) 
+ \frac{8\rho^2}{\tilde y_N^2 \bar\varphi^2} \left(q(p_s)-q(k)\right).
\eeq

{\bc
\setlength{\unitlength}{0.01mm}
\begin{figure}[htb]
\centering
\begin{tabular}{cc}
$\tilde\lambda_N=0.0$ & $\tilde\lambda_N=0.1$ \\
\begin{picture}(6600,5500)
%\put(600,500){\includegraphics[width=5cm]{PhaseDiagramLargeYLam000}}
\put(600,500){\includegraphics[width=5cm]{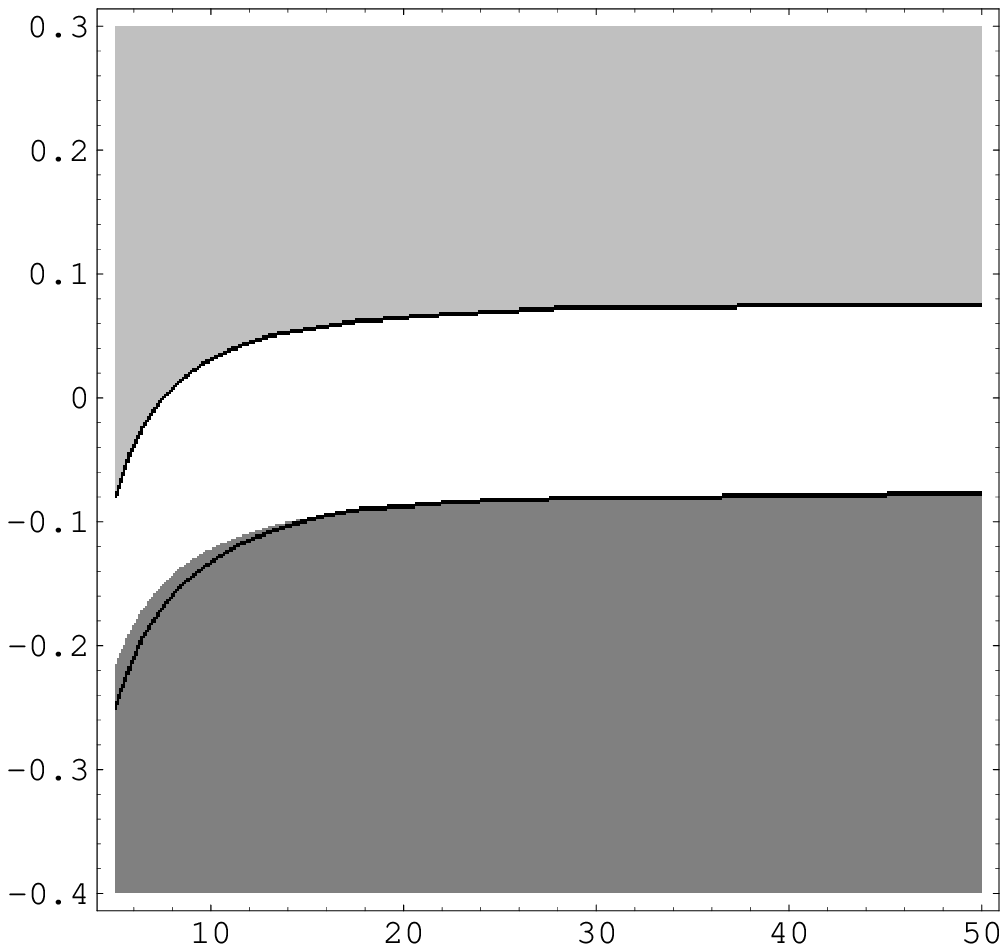}}
\put(25,3100){$\tilde \kappa_N$}
\put(3200,300){$\tilde y_N$}
\put(4500,3250){\textbf {SYM}}
\put(1500,4500){\textbf {FM}}
\put(4500,1100){\textbf {AFM}}
\end{picture}
&
\begin{picture}(6600,5500)
%\put(600,500){\includegraphics[width=5cm]{PhaseDiagramLargeYLam001}}
\put(600,500){\includegraphics[width=5cm]{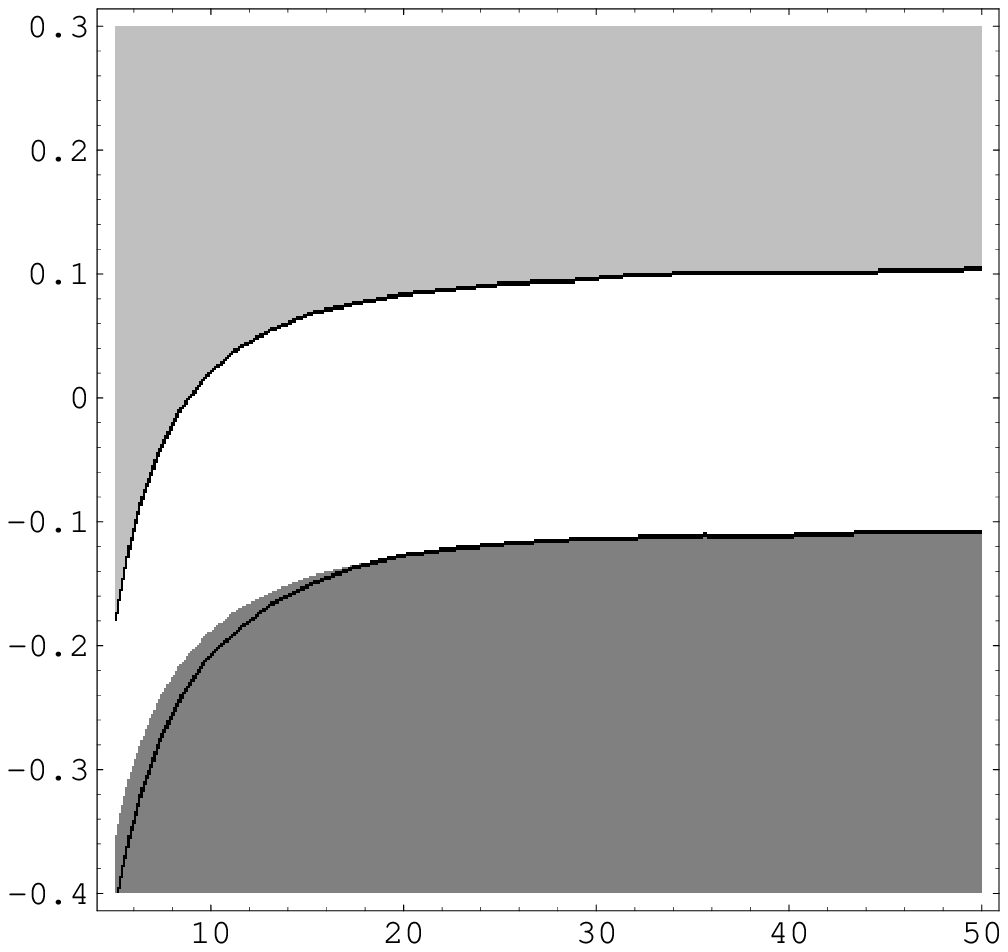}}
\put(25,3100){$\tilde \kappa_N$}
\put(3200,300){$\tilde y_N$}
\put(4500,3250){\textbf {SYM}}
\put(1500,4500){\textbf {FM}}
\put(4500,1100){\textbf {AFM}}
\end{picture}
\\
$\tilde\lambda_N=1.0$ & $\tilde\lambda_N=10.0$ \\
\begin{picture}(6600,5500)
%\put(600,500){\includegraphics[width=5cm]{PhaseDiagramLargeYLam010}}
\put(600,500){\includegraphics[width=5cm]{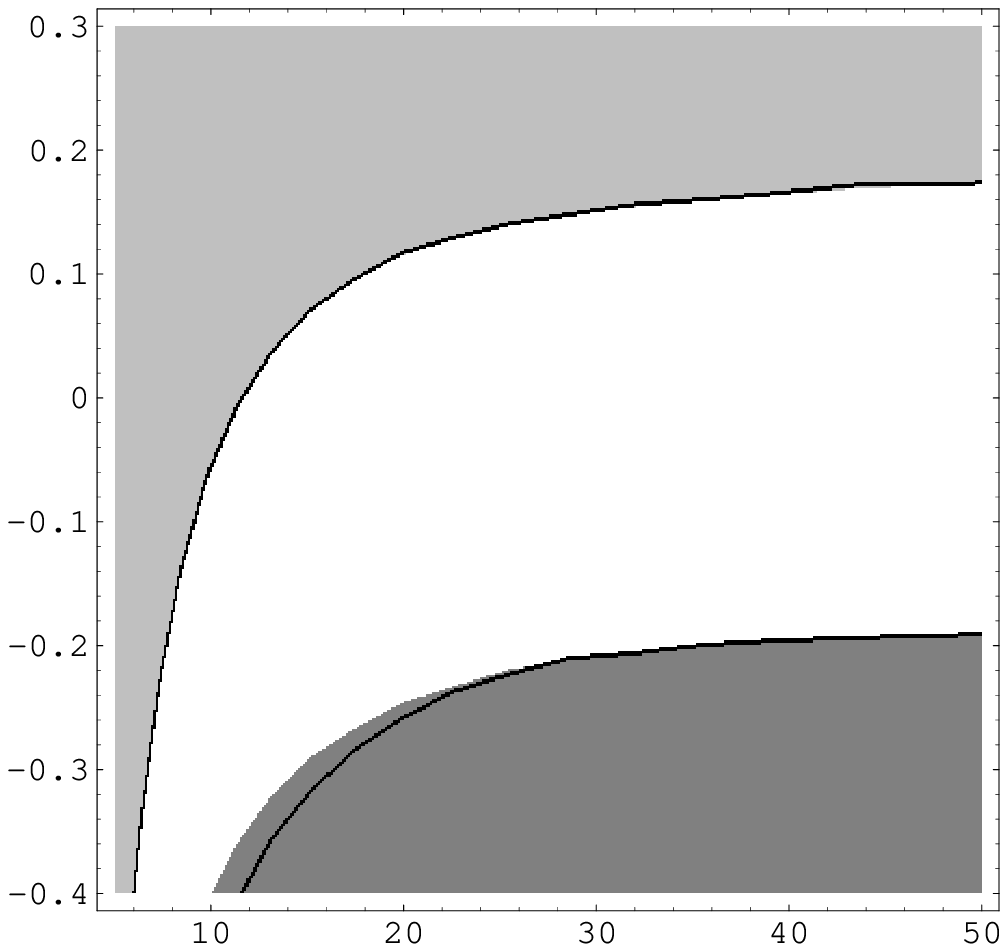}}
\put(25,3100){$\tilde \kappa_N$}
\put(3200,300){$\tilde y_N$}
\put(4500,3250){\textbf {SYM}}
\put(1500,4500){\textbf {FM}}
\put(4500,1100){\textbf {AFM}}
\end{picture}
&
\begin{picture}(6600,5500)
%\put(600,500){\includegraphics[width=5cm]{PhaseDiagramLargeYLam100}}
\put(600,500){\includegraphics[width=5cm]{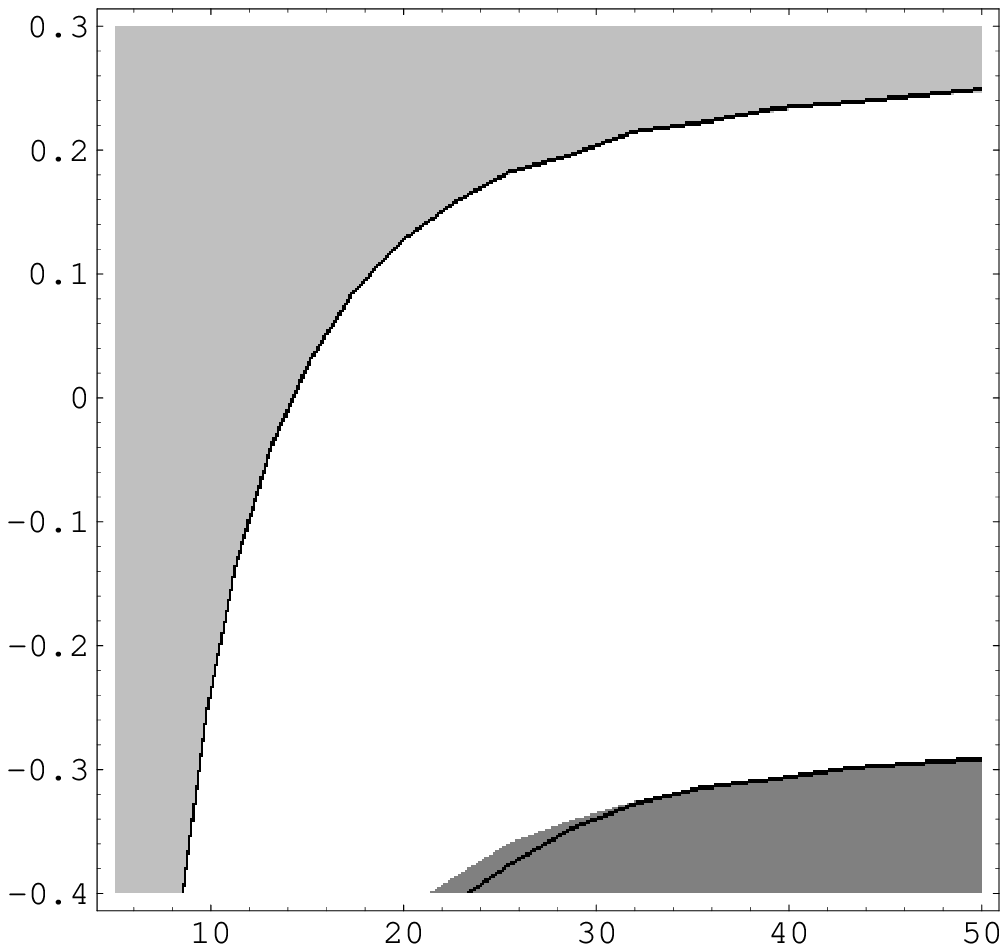}}
\put(25,3100){$\tilde \kappa_N$}
\put(3200,300){$\tilde y_N$}
\put(4500,3250){\textbf {SYM}}
\put(1500,4500){\textbf {FM}}
\put(4500,1100){\textbf {AFM}}
\end{picture}
\end{tabular}
\caption[Phase diagrams at large Yukawa coupling constant.]{Phase diagrams in the infinite volume limit with respect to the Yukawa coupling constant $\tilde y_N$ and the hopping 
parameter $\tilde \kappa_N$ for several selected values of the quartic coupling constant $\tilde \lambda_N$.
The presented phase structure was determined at $\epsilon=10^{-1}$, while the black lines show
the phase transition lines obtained at $\epsilon=10^{-3}$. For the purpose of studying the $\epsilon$-dependence the summation
in \eq{eq:LargeYGap5} has also been restricted to $k\in \ImpSpace_m(\epsilon)$ in analogy to the mandatory restriction in \eq{eq:LargeYGap7}. 
An explanation of the parameter $\epsilon$ is given in the main text.}
\label{fig:PhaseDiagrams2}
\end{figure}
\ec
\vs{-8mm}
}

The given equations \eq{eq:LargeYGap5} and \eq{eq:LargeYGap7} are denoted 
as self-consistency relations here because the assumption of a (anti-)ferromagnetic phase becomes 
inconsistent, if the resulting value for $\bar m_\sigma^2$ (or $\bar s_\sigma^2$, respectively) becomes 
non-positive. If both assumptions become inconsistent simultaneously, this corresponds to a symmetric 
phase~(SYM), while the case $\bar m_\sigma^2>0$ and $\bar s_\sigma^2>0$ would be denoted 
as a ferrimagnetic phase~(FI) in analogy to the terminology introduced in \sect{sec:SmallYukawaCouplings}.

At this point the Gauss integration leading to \eq{eq:EffectiveActionReducedModel} becomes eventually justified,
at least for the ferromagnetic phase. This is because the choice of $\LagrangeMul^{(r)}$ in this scenario according 
to \eq{eq:LargeYGap4} sufficiently shifts the eigenvalues $2{\mathcal W}_m(k)$ of the matrix $-\kappa^{eff}+\LagrangeMul$ to 
make all of their real parts positive, except for the constant mode ($k=0$) which, however, was excluded from the 
Gauss-integration.

For the anti-ferromagnetic phase, in contrast, choosing $\LagrangeMul^{(r)}$ according to \eq{eq:LargeYGap6}
does not guarantee all real parts of the eigenvalues $2{\mathcal W}_s(k)$ of $-\kappa^{eff}+\LagrangeMul$ to be positive. 
The aforementioned Gauss-integration can therefore only be performed over all those modes specified by $0\neq k\neq p_s$,
which fulfill ${\mathcal W}_s(k)\ge\epsilon$ for an arbitrary lower bound $\epsilon>0$. The sum in \eq{eq:LargeYGap7}
is therefore additionally restricted to the set $\ImpSpace_s(\epsilon)$ instead of the full set of momenta $\ImpSpace$
according to 
\bea
\label{eq:DefOfRedImpSpaceMS}
\ImpSpace_m(\epsilon) = \Big\{k\in\ImpSpace\,:\, {\mathcal W}_m(k)\ge\epsilon  \Big\} & \mbox{ and } &
\ImpSpace_s(\epsilon) = \Big\{k\in\ImpSpace\,:\, {\mathcal W}_s(k)\ge\epsilon  \Big\}.
\eea
Here the introduction of the set $\ImpSpace_m(\epsilon)$ is 
actually unnecessary due to the previous remark. A certain fraction of the momenta, depending on the choice of $\epsilon$,
has thus been ignored in \eq{eq:LargeYGap7}. The significance of this neglection on the finally resulting
phase structure has therefore to be checked explicitly, as discussed at the end of this section.

{\bc
\setlength{\unitlength}{0.0105mm}
\begin{figure}[htb]
\centering
\begin{picture}(12000,9060)
%\put(0,1000){\includegraphics[angle=0,width=0.2676\textwidth]{plExpectedMSLargeYatY010Lam001}}
\put(0,1000){\includegraphics[angle=0,width=0.2676\textwidth]{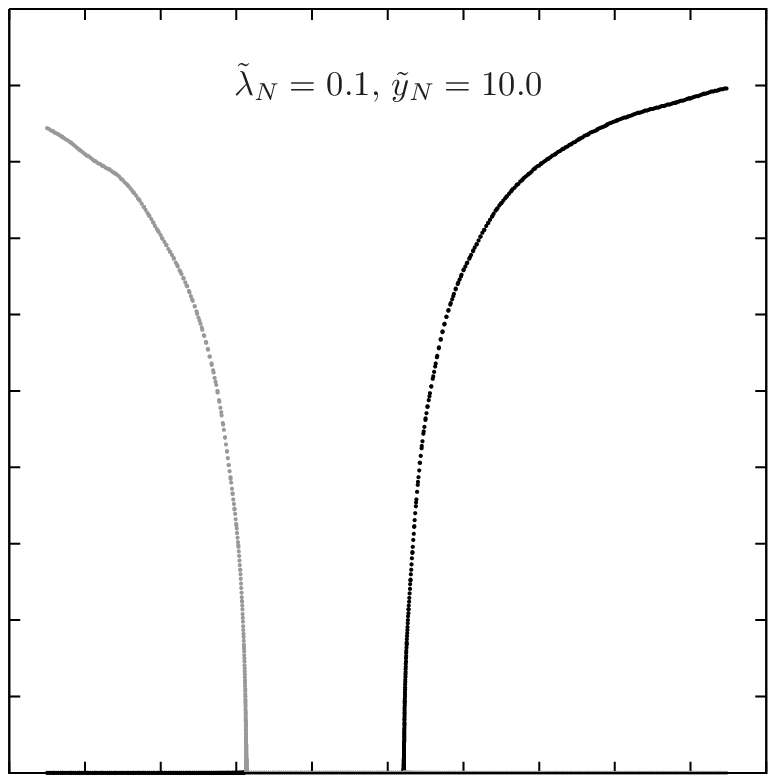}}
%\put(4000,1000){\includegraphics[angle=0,width=0.2676\textwidth]{plExpectedMSLargeYatY020Lam001}}
\put(4000,1000){\includegraphics[angle=0,width=0.2676\textwidth]{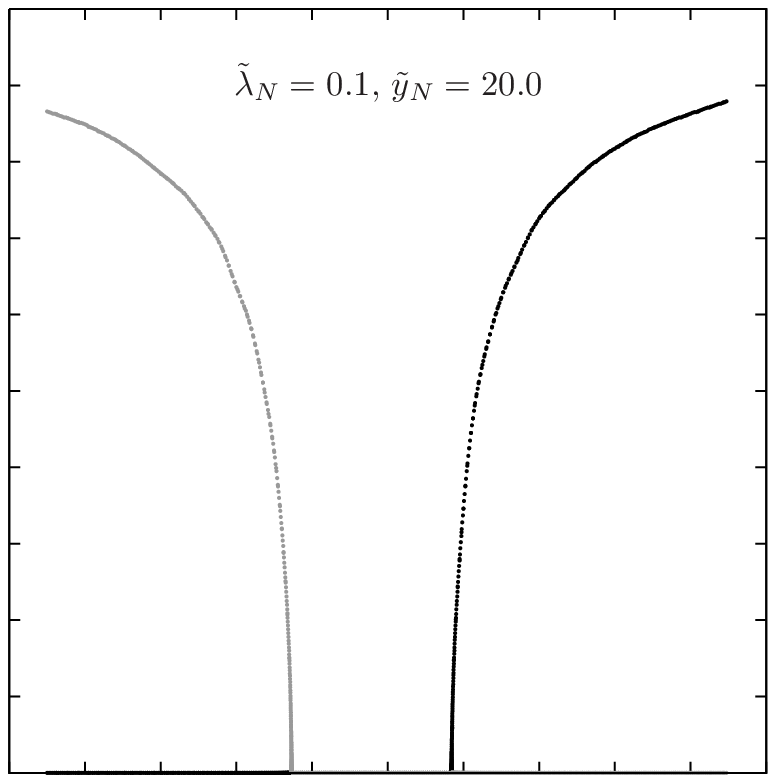}}
%\put(8000,1000){\includegraphics[angle=0,width=0.2676\textwidth]{plExpectedMSLargeYatY040Lam001}}
\put(8000,1000){\includegraphics[angle=0,width=0.2676\textwidth]{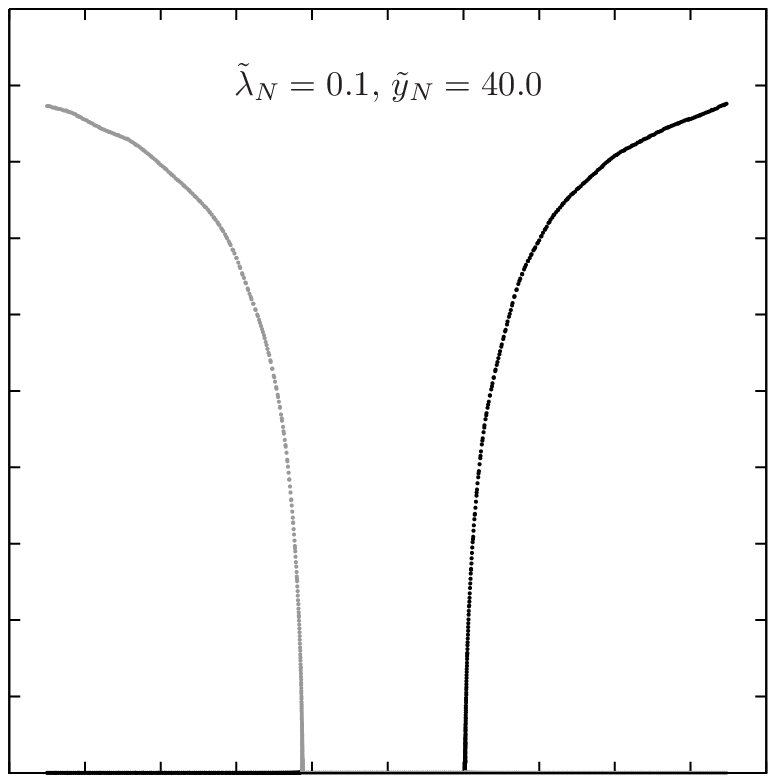}}
%\put(0,5030){\includegraphics[angle=0,width=0.2676\textwidth]{plExpectedMSLargeYatY010Lam000}}
\put(0,5030){\includegraphics[angle=0,width=0.2676\textwidth]{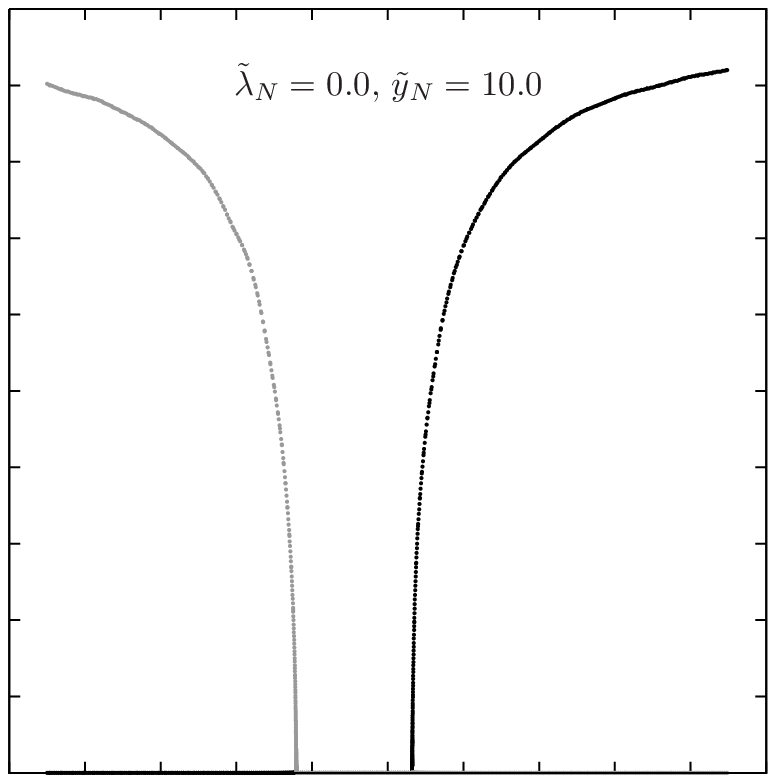}}
%\put(4000,5030){\includegraphics[angle=0,width=0.2676\textwidth]{plExpectedMSLargeYatY020Lam000}}
\put(4000,5030){\includegraphics[angle=0,width=0.2676\textwidth]{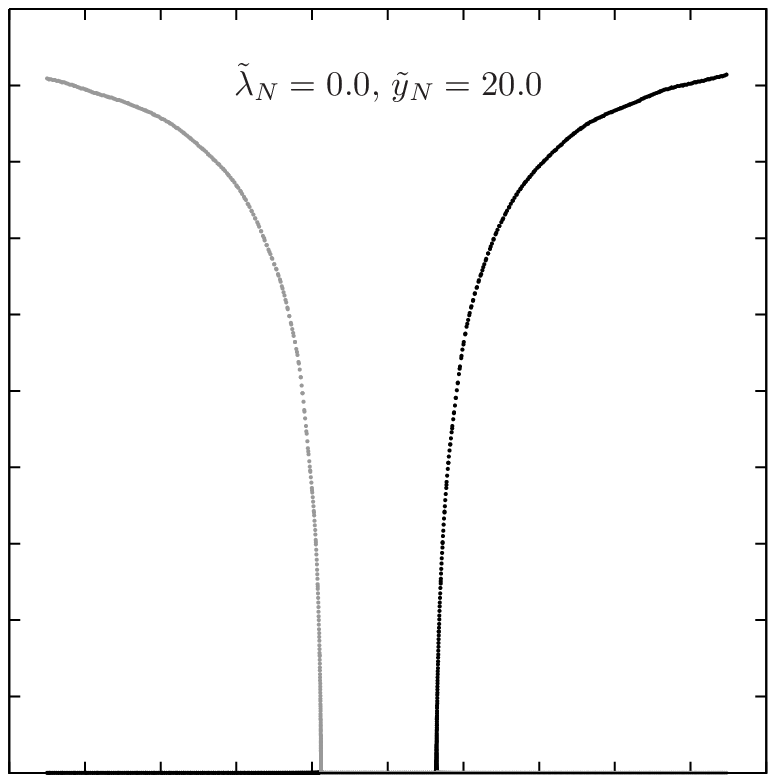}}
%\put(8000,5030){\includegraphics[angle=0,width=0.2676\textwidth]{plExpectedMSLargeYatY040Lam000}}
\put(8000,5030){\includegraphics[angle=0,width=0.2676\textwidth]{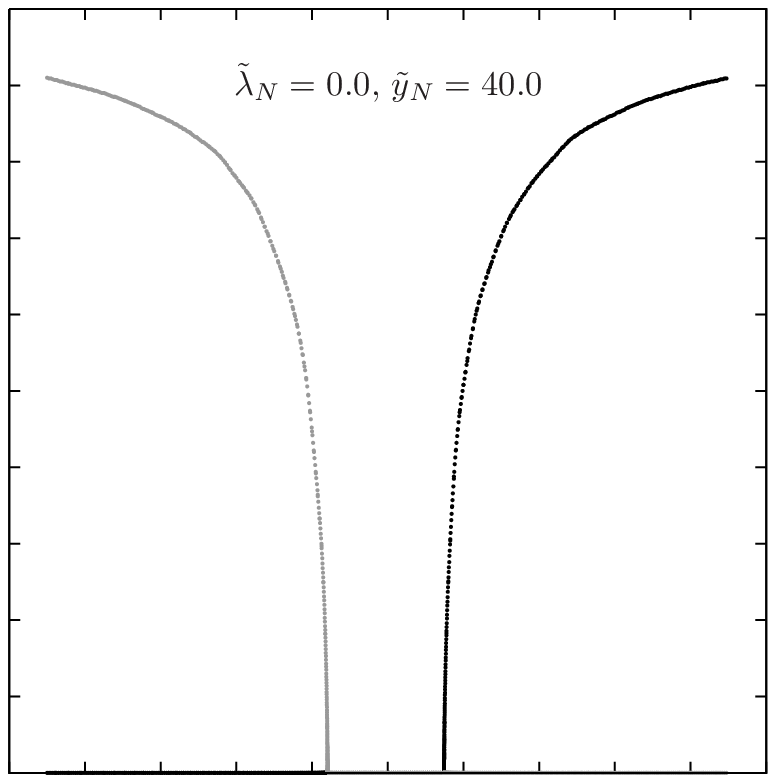}}

\put(+178,700){\tiny{-0.4}}
\put(+978,700){\tiny{-0.2}}
\put(1855,700){\tiny{0.0}}
\put(2655,700){\tiny{0.2}}
\put(3455,700){\tiny{0.4}}

\put(4178,700){\tiny{-0.4}}
\put(4978,700){\tiny{-0.2}}
\put(5855,700){\tiny{0.0}}
\put(6655,700){\tiny{0.2}}
\put(7455,700){\tiny{0.4}}

\put(8178,700){\tiny{-0.4}}
\put(8978,700){\tiny{-0.2}}
\put(9855,700){\tiny{0.0}}
\put(10655,700){\tiny{0.2}}
\put(11455,700){\tiny{0.4}}

\put(5800,100){$\tilde \kappa_N$}

\put(-430,960){\tiny{0.0}}
\put(-430,1766){\tiny{0.2}}
\put(-430,2572){\tiny{0.4}}
\put(-430,3378){\tiny{0.6}}
\put(-430,4189){\tiny{0.8}}

\put(-430,4990){\tiny{0.0}}
\put(-430,5796){\tiny{0.2}}
\put(-430,6602){\tiny{0.4}}
\put(-430,7408){\tiny{0.6}}
\put(-430,8219){\tiny{0.8}}
\put(-430,9020){\tiny{1.0}}
\end{picture}
\caption[Dependence of $\bar m_\sigma$ and $\bar s_\sigma$ on $\tilde\kappa_N$ at large Yukawa coupling constants for $\tilde \lambda_N=0.0$ and $\tilde \lambda_N=0.1$.]
{The dependence of $\bar m_\sigma$ and $\bar s_\sigma$ on the hopping parameter $\tilde\kappa_N$ is shown. 
The black curve depicts the results on $\bar m_\sigma$ while the grey curve shows the behaviour of $\bar s_\sigma$.
These results were obtained in the infinite volume limit for several selected values of the Yukawa coupling constant $\tilde y_N$ 
and the quartic coupling parameters chosen as $\tilde \lambda_N=0.0$ and $\tilde \lambda_N=0.1$.}
\label{fig:ExpectedMSPlots3}
\end{figure}
\ec
\vs{-8mm}
}

The phase structure of the considered Higgs-Yukawa model can now be obtained in the presented approximation by numerically 
evaluating \eq{eq:LargeYGap5} and \eq{eq:LargeYGap7}. For some selected values of the quartic coupling constant $\tilde \lambda_N$ 
the resulting phase diagrams with respect to the parameters $\tilde\kappa_N$ and $\tilde y_N$ are shown in \fig{fig:PhaseDiagrams2}.
All presented results were obtained in the infinite volume limit by replacing the finite sums in \eq{eq:LargeYGap5} and \eq{eq:LargeYGap7}
with corresponding integrals according to \eq{eq:IntSumRelation}.

For $\tilde y_N\rightarrow\infty$ the effective coupling matrix in \eq{eq:EffectiveCouplingMatrix} converges
to the coupling structure of a pure nearest-neighbour sigma-model. One therefore expects a symmetric phase centered
around $\tilde \kappa_N=0$ at large values of the Yukawa coupling constant $\tilde y_N$. This is indeed what one observes
in \fig{fig:PhaseDiagrams2}. Furthermore, one learns from the presented diagrams that the performed analytical calculation
predicts the symmetric phase to bend towards negative values of $\tilde \kappa_N$  when the Yukawa coupling constant $\tilde y_N$
is decreased. 

{\bc
\setlength{\unitlength}{0.0105mm}
\begin{figure}[htb]
\centering
\begin{picture}(12000,9060)
%\put(0,1000){\includegraphics[angle=0,width=0.2676\textwidth]{plExpectedMSLargeYatY010Lam100}}
\put(0,1000){\includegraphics[angle=0,width=0.2676\textwidth]{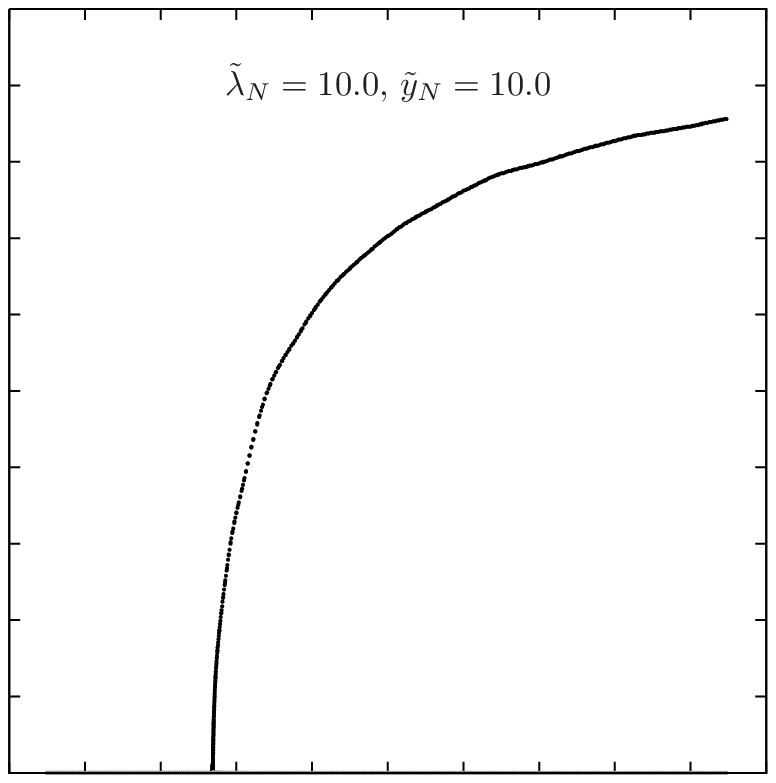}}
%\put(4000,1000){\includegraphics[angle=0,width=0.2676\textwidth]{plExpectedMSLargeYatY020Lam100}}
\put(4000,1000){\includegraphics[angle=0,width=0.2676\textwidth]{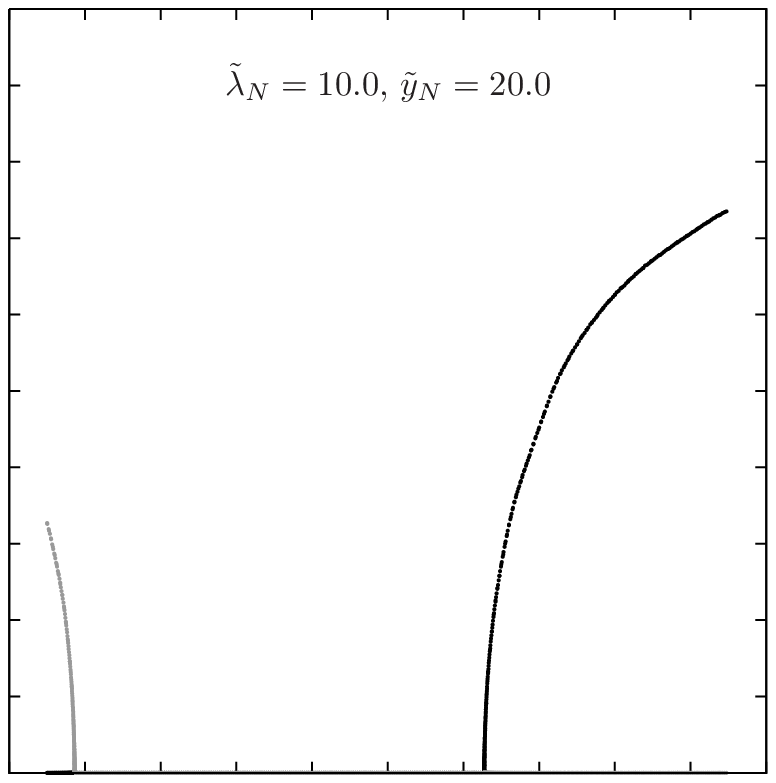}}
%\put(8000,1000){\includegraphics[angle=0,width=0.2676\textwidth]{plExpectedMSLargeYatY040Lam100}}
\put(8000,1000){\includegraphics[angle=0,width=0.2676\textwidth]{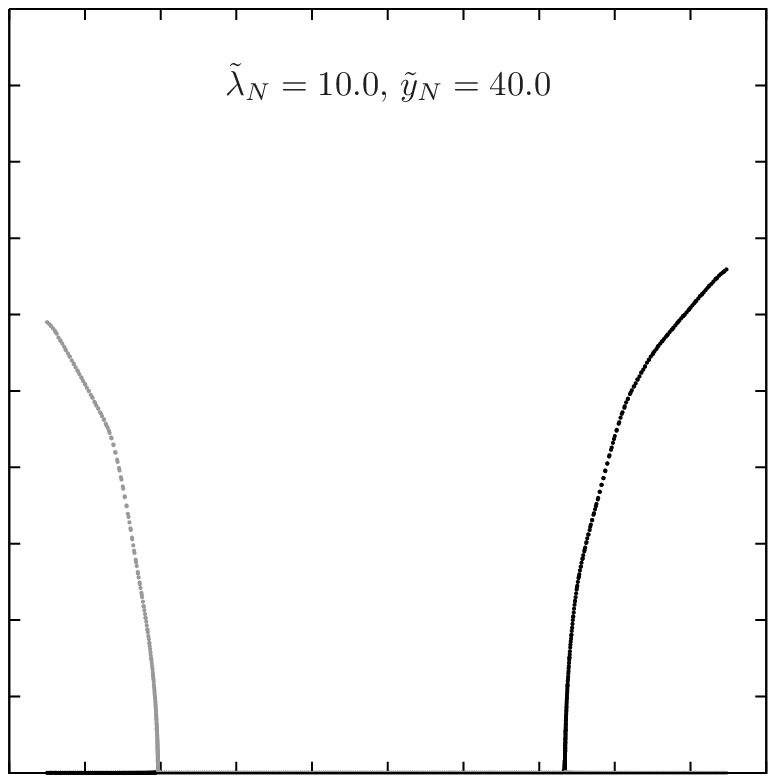}}
%\put(0,5030){\includegraphics[angle=0,width=0.2676\textwidth]{plExpectedMSLargeYatY010Lam010}}
\put(0,5030){\includegraphics[angle=0,width=0.2676\textwidth]{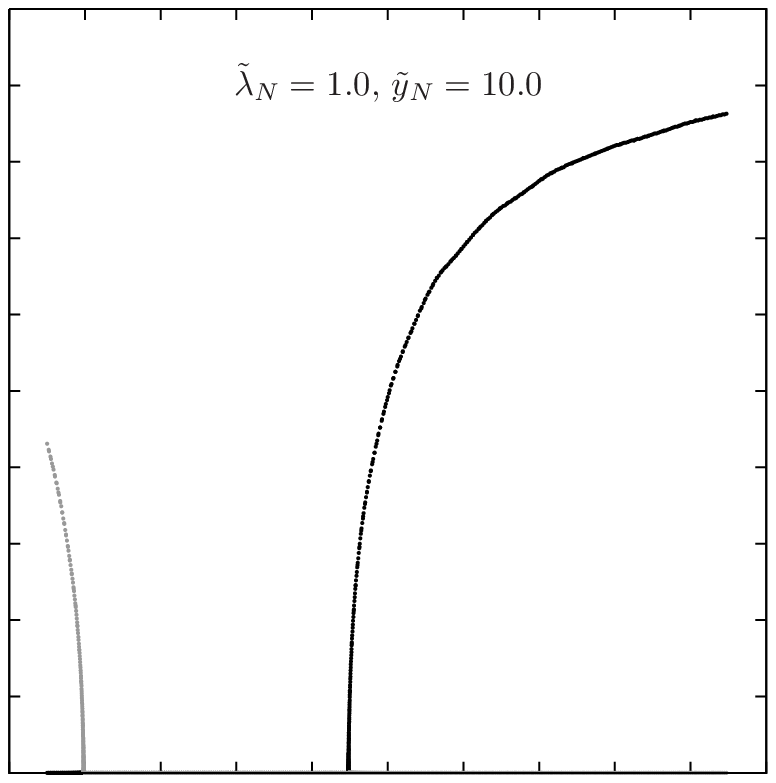}}
%\put(4000,5030){\includegraphics[angle=0,width=0.2676\textwidth]{plExpectedMSLargeYatY020Lam010}}
\put(4000,5030){\includegraphics[angle=0,width=0.2676\textwidth]{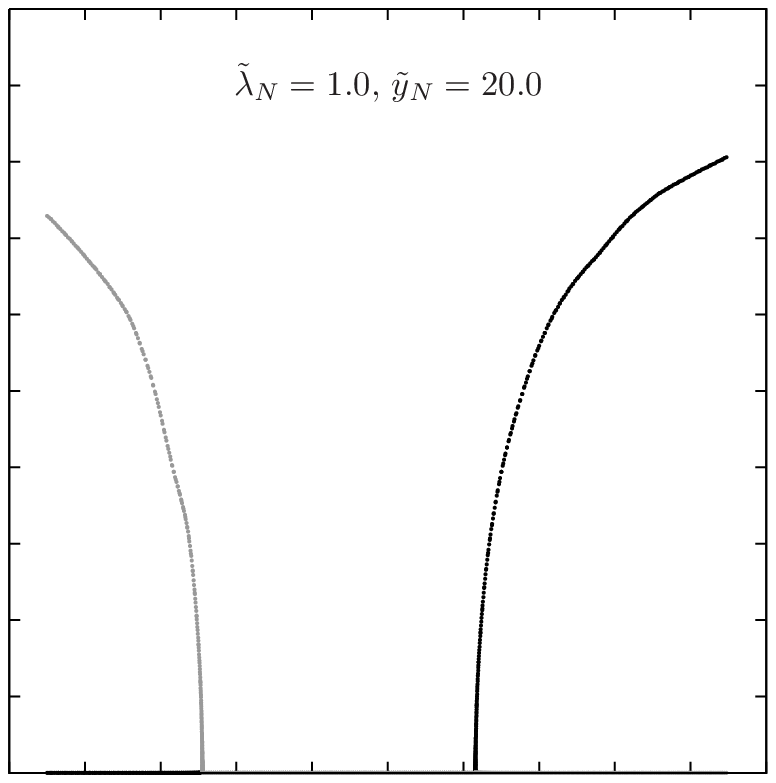}}
%\put(8000,5030){\includegraphics[angle=0,width=0.2676\textwidth]{plExpectedMSLargeYatY040Lam010}}
\put(8000,5030){\includegraphics[angle=0,width=0.2676\textwidth]{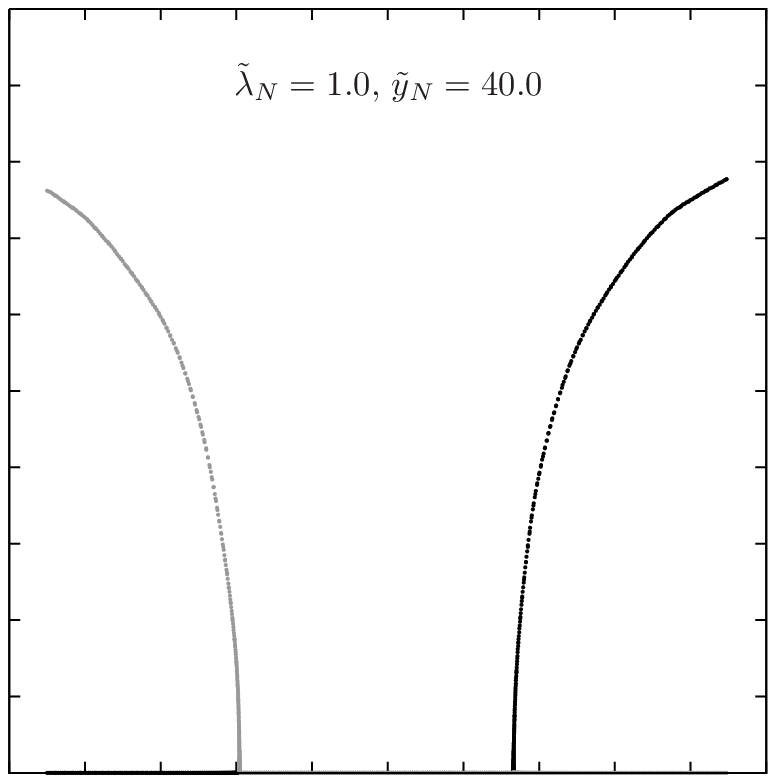}}

\put(+178,700){\tiny{-0.4}}
\put(+978,700){\tiny{-0.2}}
\put(1855,700){\tiny{0.0}}
\put(2655,700){\tiny{0.2}}
\put(3455,700){\tiny{0.4}}

\put(4178,700){\tiny{-0.4}}
\put(4978,700){\tiny{-0.2}}
\put(5855,700){\tiny{0.0}}
\put(6655,700){\tiny{0.2}}
\put(7455,700){\tiny{0.4}}

\put(8178,700){\tiny{-0.4}}
\put(8978,700){\tiny{-0.2}}
\put(9855,700){\tiny{0.0}}
\put(10655,700){\tiny{0.2}}
\put(11455,700){\tiny{0.4}}

\put(5800,100){$\tilde \kappa_N$}

\put(-430,960){\tiny{0.0}}
%\put(-430,1363){\tiny{0.1}}
\put(-430,1766){\tiny{0.2}}
%\put(-430,2169){\tiny{0.3}}
\put(-430,2572){\tiny{0.4}}
%\put(-430,2975){\tiny{0.5}}
\put(-430,3378){\tiny{0.6}}
%\put(-430,3786){\tiny{0.7}}
\put(-430,4189){\tiny{0.8}}
%\put(-430,4592){\tiny{0.9}}

\put(-430,4990){\tiny{0.0}}
%\put(-430,5393){\tiny{0.1}}
\put(-430,5796){\tiny{0.2}}
%\put(-430,6199){\tiny{0.3}}
\put(-430,6602){\tiny{0.4}}
%\put(-430,7005){\tiny{0.5}}
\put(-430,7408){\tiny{0.6}}
%\put(-430,7816){\tiny{0.7}}
\put(-430,8219){\tiny{0.8}}
%\put(-430,8622){\tiny{0.9}}
\put(-430,9020){\tiny{1.0}}
\end{picture}
\caption[Dependence of $\bar m_\sigma$ and $\bar s_\sigma$ on $\tilde\kappa_N$ at large Yukawa coupling constants for $\tilde \lambda_N=1.0$ and $\tilde \lambda_N=10.0$.]
{The dependence of $\bar m_\sigma$ and $\bar s_\sigma$ on the hopping parameter $\tilde\kappa_N$ is shown. The 
black curve depicts the results on $\bar m_\sigma$ while the grey curve shows the behaviour of $\bar s_\sigma$.
These results were obtained in the infinite volume limit for several selected values of the Yukawa coupling constant $\tilde y_N$ 
and the quartic coupling parameters chosen as $\tilde \lambda_N=1.0$ and $\tilde \lambda_N=10.0$.}
\label{fig:ExpectedMSPlots4}
\end{figure}
\ec
\vs{-8mm}
}

The order of the observed phase transitions is again determined by considering the dependence of $\bar m_\sigma$ and $\bar s_\sigma$
on the hopping parameter $\tilde\kappa_N$ as depicted in \fig{fig:ExpectedMSPlots3} and \fig{fig:ExpectedMSPlots4}. From the continuous
behaviour one can expect the occurring phase transitions to be of second order. This is what one would have expected, since the considered
Higgs-Yukawa model effectively becomes a $\sigma$-model in the limit of large Yukawa coupling constants.

Finally, it is remarked that the significance of the conceptual uncertainty for determining the SYM-AFM phase transition due to neglecting
the momenta $\ImpSpace/\ImpSpace_s(\epsilon)$ in \eq{eq:LargeYGap7} is tested in \fig{fig:PhaseDiagrams2}. Here the results for the phase 
transition lines obtained for $\epsilon=10^{-1}$ and $\epsilon=10^{-3}$ are compared to each other. 
For the SYM-AFM phase transition the respective results for the critical hopping parameter increasingly differ at decreasing values of 
$\tilde y_N$. The discrepancy between these results can serve as a crude indication down to which value of $\tilde y_N$ the approach 
in \eq{eq:LargeYGap7} can be considered as a good approximation. However, in the limit $\tilde y_N\rightarrow\infty$ the fraction of the 
neglected modes finally vanishes and the problem encountered during the Gauss-integration in 
\eq{eq:EffectiveActionReducedModel} eventually disappears. The obtained results for the SYM-AFM phase transition should thus be trustable 
for sufficiently large values of the Yukawa coupling constant.

\subsection{Comparison with direct Monte-Carlo calculations}
\label{sec:LargeY}

From the calculations in the previous section the emergence of a ferromagnetic, an anti-ferromagnetic  
and a symmetric phase is expected at large values of the Yukawa coupling constant. 
These analytical considerations also revealed that significant finite size 
effects can be present in the symmetric phase which may render its detection difficult. 
These predictions will now be confronted with the results of direct Monte-Carlo calculations
performed by the HMC-algorithm introduced in \sect{sec:HMCAlgorithm}.

In \fig{fig:kappascan3}, the numerically obtained values of 
the average magnetizations $\mAvg$ and $\sAvg$ obtained on various sized lattices are shown as 
a function of $\kappa$. These Monte-Carlo calculations have been performed with $N_f=2$, $\tilde\lambda_N=0.1$,
and the rather large value of the Yukawa coupling constant $\tilde y_N=30$. The results presented in 
\fig{fig:kappascan3} demonstrate that the expected symmetric phase indeed emerges only on 
sufficiently large lattice volumes, while on small lattices the magnetization 
does not vanish as a function of decreasing $\kappa$ even deeply within the anti-ferromagnetic
phase. Though the average magnetizations $\mAvg$ and $\sAvg$ are never expected to be exactly zero 
on a finite lattice due to stochastical fluctuations of the underlying observables $m$ and $s$ around 
their respective ground states, being only suppressed as the volume increases, one may conjecture 
from \fig{fig:kappascan3} that additional finite volume effects are present here manifest in the 
strong observed asymmetry with respect to the $\kappa$-dependence of $\mAvg$ and $\sAvg$ 
even at the considered large value of the Yukawa coupling constant being $\tilde y_N=30$. For 
clarification it is recalled that one would expect a symmetric behaviour of the latter quantities 
in infinite volume centered around $\kappa=0$ when the Yukawa coupling constant is sent to infinity, 
as discussed in the preceding section. And indeed, the observed behaviour of $\mAvg$ and $\sAvg$ is 
consistent with the analytical prediction of a finite volume effect hindering the magnetization 
$\mAvg$ but not $\sAvg$ from vanishing according to \eq{eq:DetStarBInv}.

%\includeFigTriple{msLargeY30N4Nf2Lam001}{msLargeY30N8Nf2Lam001}{msLargeY30N16Nf2Lam001}
\includeFigTriple{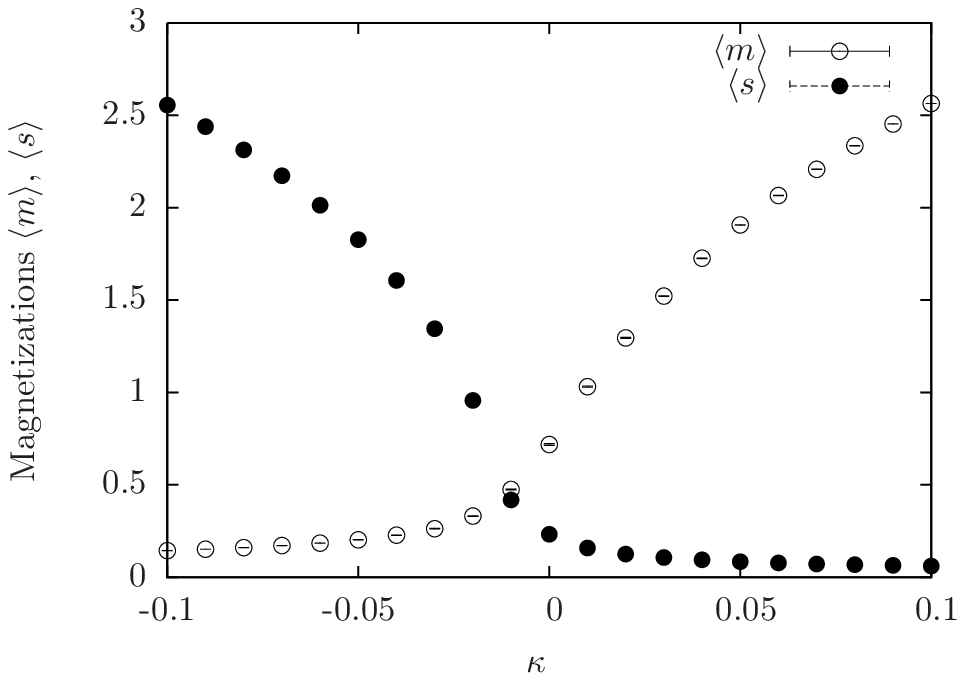}{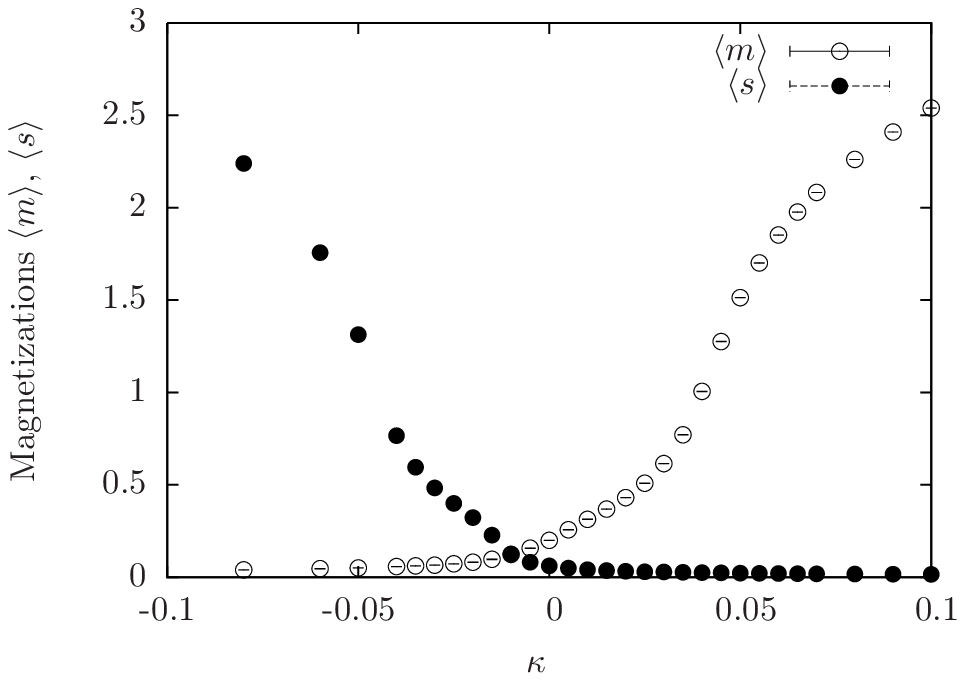}{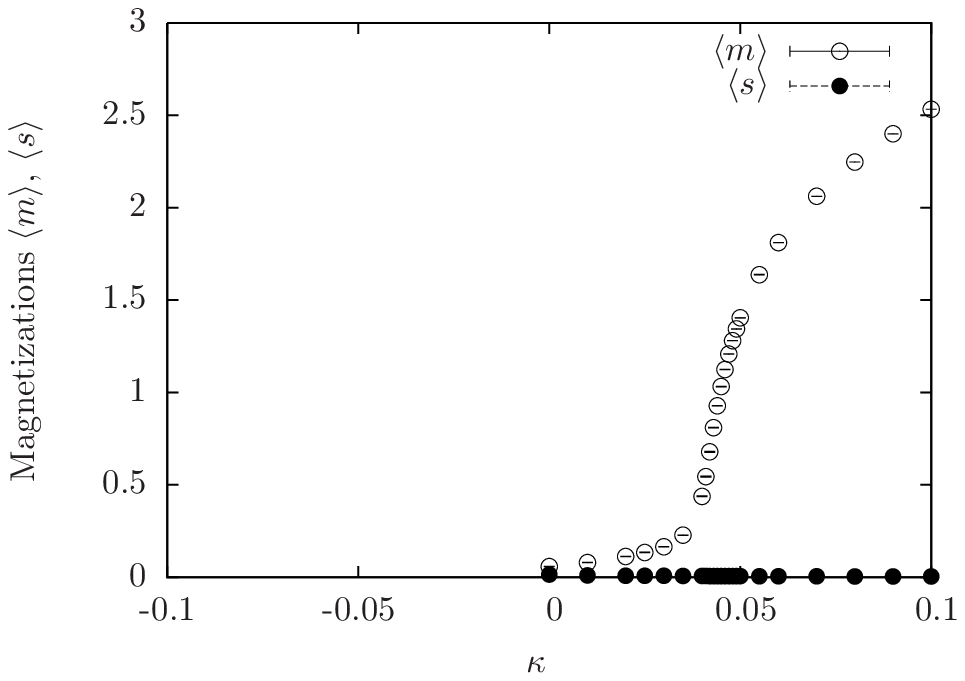}
{fig:kappascan3}
{The behaviour of the average magnetization $\mAvg$ and staggered 
magnetization $\sAvg$ as a function of $\kappa$ on a $4^4$- (a), an $8^4$- (b) 
and a $16^4$-lattice (c). In these plots the model parameters were chosen to be
$\tilde y_N=30$, $\tilde\lambda_N=0.1$, and $N_f=2$. Concerning panel (c), the numerical
computations on the \lattice{16}{16}became too demanding  at negative values of $\kappa$
for the naive implementation of the HMC-algorithm discussed in \sect{sec:HMCAlgorithm}.  }
{Dependence of $\mAvg$ and $\sAvg$ on $\kappa$ for different lattice volumes at large Yukawa coupling constants.}
 
It is quite instructive to compare the effect caused by the aforementioned finite volume contributions
to the obtained Monte-Carlo data. Meant as a quick and crude estimation the effective action $S_\Phi[\Phi]+\SFeff[\Phi]$,
with $\SFeff[\Phi]$ given in \eq{eq:EffectiveActionFullLargeY}, has therefore been evaluated at $\tilde y_N = \infty$ for 
the scalar field being restricted to the ansatz $\Phi=\Phi'$ with $\Phi'$ defined in \eq{eq:staggeredAnsatz}. 
Employing the results in \eq{eq:SPhiForPhiAnsatz} and \eq{eq:DetStarBInv} the expectation values $\mAvg$ and $\sAvg$ have 
then been estimated through the minima of the resulting expression for $S_\Phi[\Phi']+\SFeff[\Phi']$ with respect to 
$\breve m_\Phi$ and $\breve s_\Phi$. While this simple approach cannot correctly predict the phase transition of the model, 
it suffices to estimate the behaviour of $\mAvg$ and $\sAvg$ in the limit of large Yukawa coupling constants and large 
negative or positive values of the hopping parameter. This is demonstrated in \fig{fig:largeYfiniteSE}, where the average 
magnetizations numerically obtained on the \lattice{4}{4}are presented again, this time, however, together with the 
aforementioned analytical estimate of the finite volume effects, depicted by the dashed lines. One then finds in 
\fig{fig:largeYfiniteSE} that the asymptotic behaviour of $\mAvg$ as well as the aforementioned asymmetry in $\mAvg$ and 
$\sAvg$ is well explainable already with this simple approach.

%\includeFigDouble{LargeYN4Plateau}{LargeYN4PlateauZoom}
\includeFigDouble{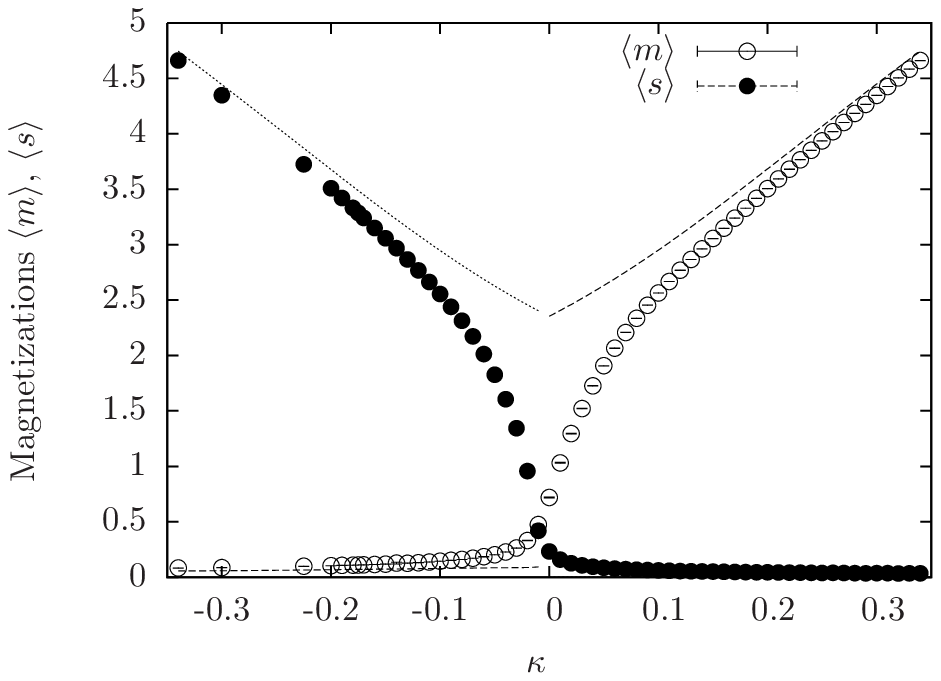}{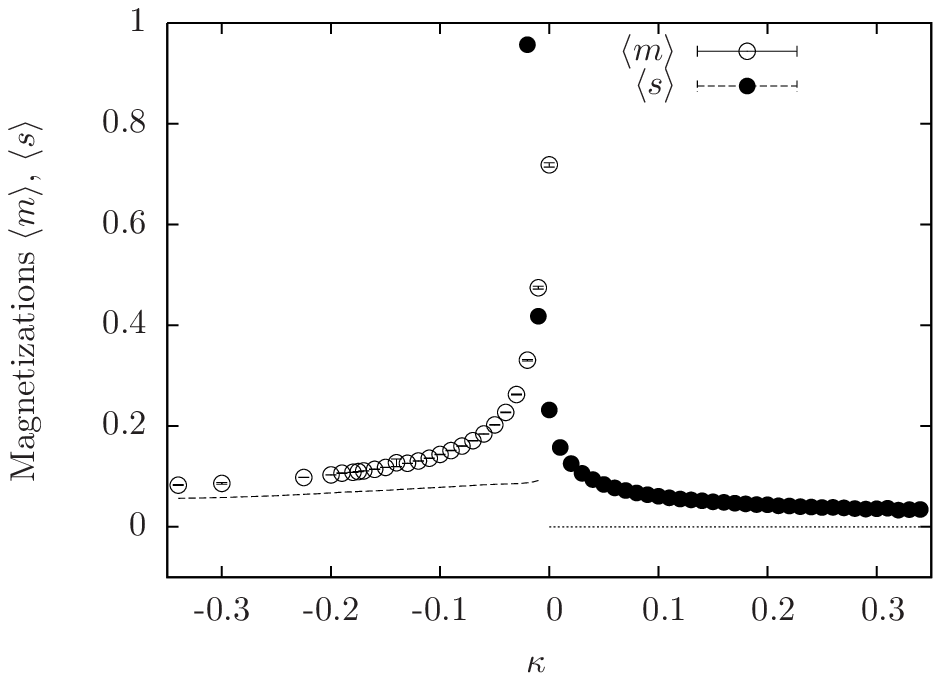}
{fig:largeYfiniteSE}
{Comparison with the analytical estimate of the finite size effects.
The dependence of the average magnetizations $\mAvg$ and $\sAvg$ of \fig{fig:kappascan3}a on $\kappa$
as obtained on the $4^4$-lattice at $\tilde y_N=30$, $\tilde\lambda_N=0.1$, and $N_f=2$ are 
compared to an analytical estimate of $\mAvg$ (dashed line)
and $\sAvg$ (dotted line) including the finite volume contributions in \eq{eq:DetStarBInv}
as discussed in the main text. Panel (b) is just a magnification of panel (a).}
{Comparison of finite size effects at large Yukawa coupling constants with analytical estimates.}

It is remarked that it is the term $\log\left|\tilde m_\Phi \right|$ appearing in \eq{eq:DetStarBInv} that hinders
$\mAvg$ from vanishing and induces the observed asymmetry. As already pointed out before this term as well as all other 
contributions arising from the $\detStar$-determinants in \eq{eq:EffectiveActionFullLargeY} are only finite volume 
contributions. As the volume increases the latter contributions loose their significance and the average magnetization
becomes increasingly less influenced by the above logarithmic term, such that a symmetric phase can finally emerge. 
This is exactly what is observed in \fig{fig:kappascan3}.
 
In \fig{fig:kappascan4} the susceptibilities $\chi_m$ corresponding 
to the magnetizations in \fig{fig:kappascan3} are presented.
On the smallest lattice, \ie the \latticeX{4}{4}{,}one observes only one peak
in the magnetic susceptibility, approximately centered at $\kappa=0$. From this result
one might draw the conclusion that the phase transition point is located at $\kappa=0$,
excluding a symmetric phase, since the staggered susceptibility reaches its
maximum at the same value of $\kappa$. However, with increasing lattice sizes
a second peak develops in the susceptibilities. This is very well observed in
\fig{fig:largeYfiniteSE}b corresponding to the larger \latticeX{8}{8}{.}It 
is actually this second peak, approximately centered around $\kappa=0.04$ in this case, 
that correctly describes the {\textit{physical}} phase transition between the ferromagnetic 
and the symmetric phase, while the first one is only caused by the aforementioned finite 
volume effects. Its height is therefore expected to be constant in contrast to the physical peak, 
which grows with increasing lattice volume. On the largest presented lattice, the 
\latticeX{16}{16}{,}the physical peak at $\kappa=0.04$ completely dominates the
scene and the former small volume peak at $\kappa=0$ has disappeared, presumably
hidden beneath the large error bars at $\kappa=0$. 
 
%\includeFigTriple{SusLargeY30N4Nf2Lam001}{SusLargeY30N8Nf2Lam001}{SusLargeY30N16Nf2Lam001}
\includeFigTriple{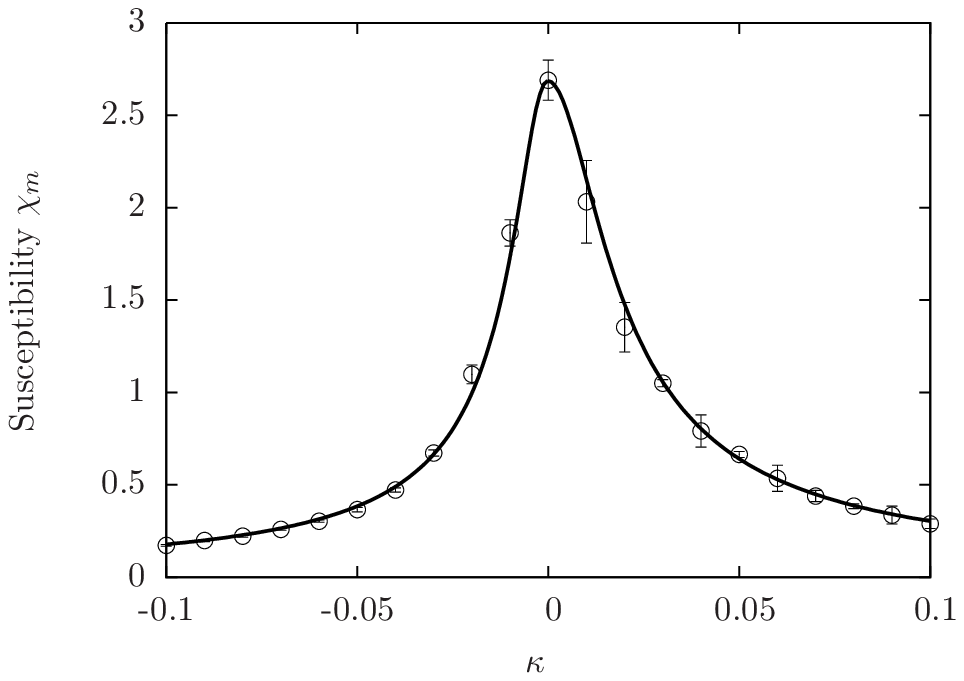}{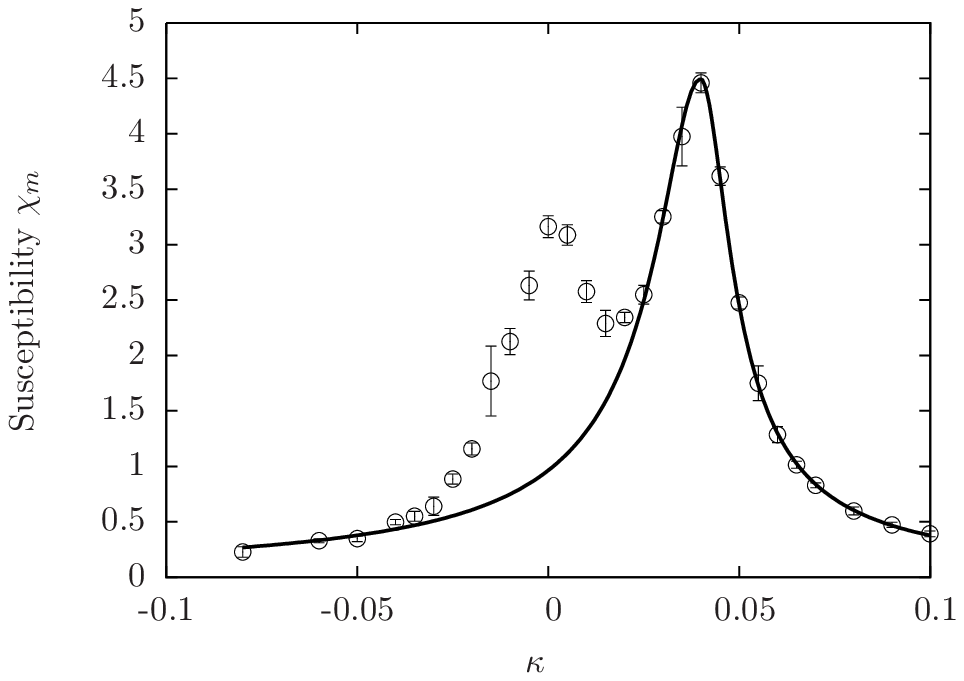}{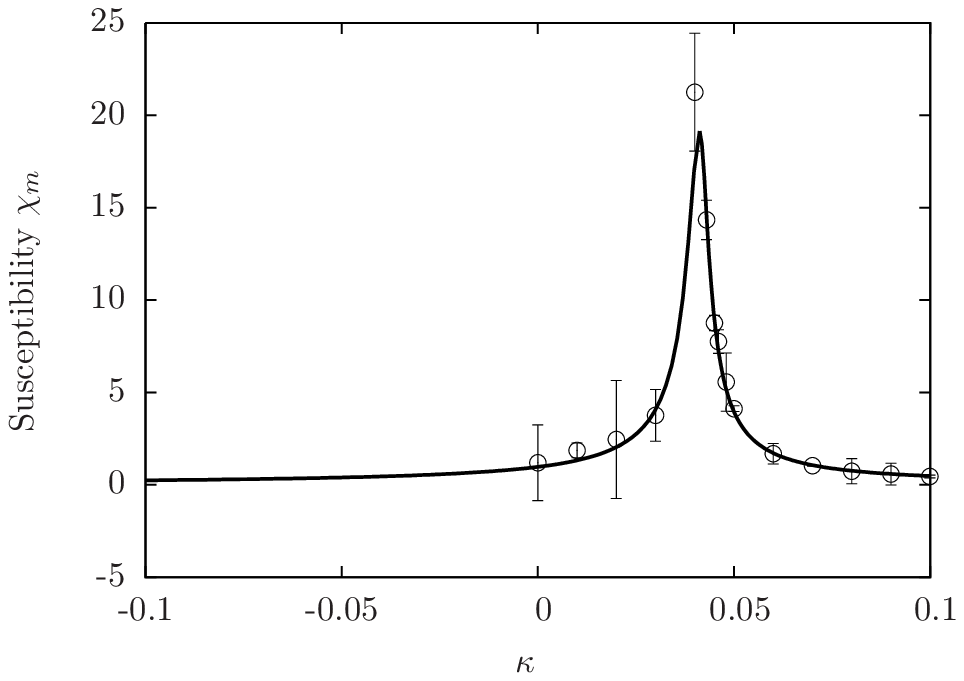}
{fig:kappascan4}
{The behaviour of the magnetic susceptibility $\chi_m$
as a function of the hopping parameter $\kappa$ on a $4^4$- (a), an $8^4$- (b) 
and a $16^4$-lattice (c). In these plots the model parameters were chosen to be 
$\tilde y_N=30$, $\tilde\lambda_N=0.1$ and $N_f=2$.
The fit in panel (b) is only applied to those points with $\kappa\ge0.025$
or $\kappa\le-0.05$ in order to reduce the influence of the unphysical peak
at $\kappa=0.0$. Note also the changing scale in the three plots.}
{Dependence of the magnetic susceptibility $\chi_m$ on $\kappa$ for large Yukawa coupling constants}
 
The SYM-FM phase transition points have then been determined by fitting the physical peaks
in the magnetic susceptibility $\chi_m$ on the intermediate \lattices{8}{8}to the finite 
volume fit ansatz in \eq{finitesizesus} by taking only the points belonging to the 
physical peak into account as demonstrated in \fig{fig:kappascan4}b. 
It is remarked that no results for the SYM-AFM phase transition are provided here,
since this phase transition points have not been reliably detectable on the \latticeX{6}{6}{,}due to the
finite volume effects discussed above, and also not on the \lattice{8}{8}owing to a much larger
numerical demand when performing the Monte-Carlo calculations in the anti-ferromagnetic phase with
the naive implementation of the HMC-algorithm introduced in \sect{sec:HMCAlgorithm}.

In \fig{fig:phasediagram3} the obtained phase transition points are presented
together with the analytical infinite volume, large $N_f$ expectation of the 
phase structure at large values of the Yukawa coupling constant taken over from
the preceding section. Qualitatively, the picture obtained from the numerical simulations 
is in full accordance with the results from the large $N_f$ approximation. There 
are second order phase transitions separating a ferromagnetic phase from a symmetric 
phase. In this symmetric phase strong finite size effects are
encountered blocking its observation on too small lattice volumes. Here the order of the 
phase transition is identified to be of second order
according to the continuous dependence of $\mAvg$ on the hopping parameter $\kappa$ as
demonstrated in \fig{fig:kappascan3}. Again a more careful analysis is omitted here, since the 
considered phase transition is not in the main interest of the present study.
Quantitatively, however, the numerical results deviate from the analytical expectation due to 
the finite settings $N_f=2$ and $V=8^4$, but are still in relatively good agreement.

As a concluding remark one can summarize that a symmetric phase at large values of the
Yukawa coupling constant does indeed exist although its existence is obscured on too small
lattice volumes by strong finite size effects, and that its location within the phase diagram is in 
acceptable agreement with the analytical large $N_f$ predictions. 
The existence of a symmetric phase at strong Yukawa couplings has also
been observed and discussed in \Ref{Giedt:2007qg}.
From these findings one can conclude that the phase structure of the considered Higgs-Yukawa model 
at large values of the Yukawa coupling constant is acceptably well described by the analytical large 
$N_f$ calculations presented in \sect{sec:LargeYukawaCouplings}.

%\includeFigSingleMedium{plLargeYPhaseDiagram1}
\includeFigSingleMedium{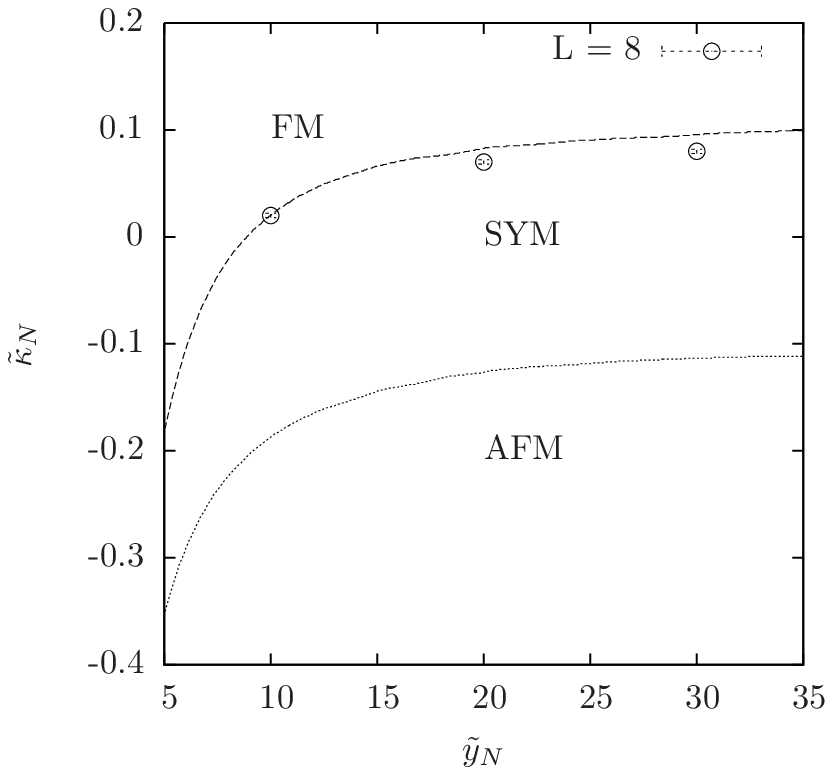}
{fig:phasediagram3}
{The phase diagram with respect to $\tilde\kappa_N$ and $\tilde y_N$ for large values of the Yukawa 
coupling constant. The numerical results on the SYM-FM phase transition points have been obtained on an 
$8^4$-lattice at $\tilde\lambda_N=0.1$ and $N_f=2$, while the SYM-AFM phase transition points could 
not reliably be determined, as discussed in the main text. The presented numerical results are compared 
to the corresponding prediction for the phase transition line obtained from the analytical infinite volume, 
$N_f=\infty$ calculation, as depicted by the dashed line. The analytically expected SYM-AFM phase transition 
is marked by the dotted curve.}
{Numerically obtained phase diagram for large values of the Yukawa coupling constant.}

  \chapter{The simulation algorithm}
\label{chap:SimAlgo}

In the preceding chapter the HMC algorithm described in \sect{sec:HMCAlgorithm}, 
which is only applicable for even values of $N_f$ by construction, was used in 
order to check the validity of the analytical calculations performed in the large $N_f$-limit. 
Since there is no need to consider odd values of $N_f$ to test the analytical large $N_f$ results, 
this easily implementable algorithm was fully sufficient for the given purpose.
The numerical evaluation of the model in the {\textit{physically relevant}} setup, however,
requires a more advanced technique, for instance a PHMC~\cite{Frezzotti:1997ym,Frezzotti:1998eu,Frezzotti:1998yp} 
or a RHMC~\cite{Kennedy:1998cu,Clark:2003na} 
algorithm, which are both capable of dealing with odd values of $N_f$. Here, a PHMC algorithm
has therefore been implemented in addition to the previously employed HMC method. The basic concepts 
of this more advanced approach will shortly be summarized in \sect{sec:basicConceptsOfPHMC}.

For the case of the considered Higgs-Yukawa model this standard algorithm could be enriched by a 
couple of technical enhancements. The most crucial improvements originate from the observation that
the condition number of the fermion matrix can be reduced by about 1-2 orders of magnitude through
the use of appropriate preconditioning techniques at least in the considered broken phase of the model, 
and that the auto-correlation times of the considered observables can be brought down from order 
$O(100)$ to order $O(1)$ by a technique known as Fourier acceleration~\cite{Batrouni:1985jn,Catterall:2001jg}.
This will be discussed in \sects{sec:CondNr}{sec:FACC}, respectively.
Furthermore, efficient techniques for an exact reweighting of the generated field configurations
as well as the direct sampling of the pseudo fermion fields are presented in \sects{sec:ExactReweighing}{sec:DirectSampling}.
The computational performance could additionally be improved by the use of a multiple time-scale integration
scheme of the molecular dynamics underlying the PHMC algorithm, as demonstrated in \sect{sec:MultiPolynoms}.
   
Moreover, a highly efficient implementation of a Fast Fourier Transformation (FFT) in four dimensions
was implemented for the purpose of switching back and forth between the momentum and the position space 
representations of the pseudo fermion fields allowing then for a very efficient application of
the Dirac operator as discussed in \sect{sec:HMCAlgorithm}. In the considered case of four dimensions
and on sufficiently large lattice volumes this implementation is significantly more efficient on the employed
hardware resources than the most widely spread standard implementation FFTW~\cite{FFTW05} as detailed 
in \appen{sec:xFFT}.

\includeTab{|cccccccc|}
{
\latVolExp{L_s}{L_t} & $N_f$ & $\kappa$  & $\hat \lambda$ & $\hat y_t$     & $\hat y_b/\hat y_t$ & $\langle m \rangle$ & $\Lambda$ \\
\hline
\latVol{16}{32}      & $1$   & $0.12301$ & $0$       & $0.35285$ & $1$       & $0.5103(10)$ & $\GEV{960.0 \pm 2.3}$\\
\latVol{16}{32}      & $1$   & $0.12313$ & $0$       & $0.35302$ & $1$       & $1.2506(5)$  & $\GEV{392.2 \pm 3.7}$\\
\latVol{16}{32}      & $1$   & $0.29929$ & $\infty$  & $0.55038$ & $1$       & $0.1183(4)$  & $\GEV{2608.1 \pm 9.4}$\\
\latVol{16}{32}      & $1$   & $0.30400$ & $\infty$  & $0.55470$ & $1$       & $0.2025(1)$  & $\GEV{1514.4 \pm 2.1}$\\
\latVol{16}{32}      & $3$   & $0.12220$ & $0$       & $0.35169$ & $0.024$   & $1.1801(8)$  & $\GEV{415.3 \pm 2.3}$\\
}
{tab:Chap5EmployedRuns}
{The model parameters of the Monte-Carlo runs constituting the testbed for the algorithmic improvements
introduced in the subsequent sections are presented together with the obtained values of the average magnetization 
$\langle m \rangle$ and the cutoff $\Lambda$ determined by \eq{eq:DefOfCutoffLambda}. The top quark Yukawa 
coupling constant has been chosen here according to the tree-level relation in \eq{eq:treeLevelTopMass} 
aiming at the reproduction of the phenomenologically known top quark mass. For the only non-degenerate setup
listed in the last line the ratio $\hat y_b/\hat y_t$ has been set to its phenomenological value.
}
{Model parameters of the Monte-Carlo runs constituting the testbed for the presented algorithmic improvements.}

The aforementioned algorithmic improvements were crucial for the successful numerical evaluation of
the model and shall therefore be presented in some detail here. In the following the applicability of 
these techniques will be discussed and tested in the framework of the physically relevant model parameter 
setups listed in \tab{tab:Chap5EmployedRuns}, which cover the whole range of cutoff values $\Lambda$ that 
will be investigated in the later chapters dealing with the actual Higgs boson mass bound determination. 

It is finally remarked, that some of the techniques discussed in the following, though only presented here for the case 
of the Higgs-Yukawa model, seem to be applicable to QCD as well.

%-----------------------------------------------------------------------------------------------------
\section{The basic PHMC algorithm}
\label{sec:basicConceptsOfPHMC}

The PHMC algorithm as well as its predecessor, the HMC algorithm, are both based on the fact, that the 
determinant of a hermitian and positive matrix can be written in terms of a Gaussian integral over some 
complex vector according to \eq{eq:GaussIntegralOfDetA}. As in the case of QCD the fermion matrix $\fermiMat$ of the considered 
Higgs-Yukawa model, however, is not positive definite. The determinant of $\fermiMat$ itself can therefore 
not be expressed in terms of such a Gaussian integration. The basic idea underlying both, the HMC as well 
as the PHMC algorithm, is to circumvent this problem by considering the positive and hermitian operator 
$\fermiMat\fermiMat^\dagger$ instead of $\fermiMat$. While this doubling of the fermion matrix is interpreted 
as a duplication of the number
of fermions in the theory making this approach applicable in case of even numbers of degenerate fermions
as discussed in \sect{sec:HMCAlgorithm}, the PHMC aims at correcting for the double counting. 

This can be achieved by taking the square root of $\fermiMat\fermiMat^\dagger$ which is well defined due
to the positivity and hermiticity of $\fermiMat\fermiMat^\dagger$. One then arrives at the expression
\bea
\label{eq:PHMCPartitionFunction}
Z_\Phi &=& \int D\Phi\,\,\, e^{iN_f\arg\det(\fermiMat)} \cdot
\left[\det(\sqrt{\fermiMat\fermiMat^\dagger})   \right]^{N_f} \cdot
e^{-S_\Phi[\Phi] } 
\eea
for the partition function of the considered Higgs-Yukawa model which is fully equivalent to the definition in 
\eq{eq:DefOfLatticeObsWithoutPsi}. The general idea of a PHMC (Polynomial Hybrid Monte-Carlo) algorithm is to make the 
square root of the operator $\fermiMat\fermiMat^\dagger$ numerically accessible by using polynomial approximations. 
More precisely, for a given polynomial $P(x)$ of degree $N_P$ approximating the function $x^{-\alpha}$, $\alpha=1/2$ 
with maximal relative deviation 
\beq
\label{eq:DefOfDeltaPOfPolynom}
\delta_P = \max\limits_{x\in [\epsilon_P,\lambda_P]} \left|\frac{P(x)-x^{-\alpha}}{x^{-\alpha}}  \right|
\eeq
in the interval $x\in [\epsilon_P,\lambda_P]$, $\lambda_P>\epsilon_P>0$ 
the determinant can be transformed into a Gaussian integration times a real weight factor $W[\Phi]$ according to 
\bea
\label{eq:PHMCPartitionFunction2}
Z_\Phi &=& \FuncIntAvg_P\left[ e^{iN_f\arg\det(\fermiMat)} \cdot W[\Phi] \right], \\
\label{eq:DefOfWeightFactor1}
W[\Phi]&=& \left[ \det\left(\sqrt{\fermiMat\fermiMat^\dagger}\cdot P(\fermiMat\fermiMat^\dagger)  \right) \right]^{N_f}, 
\eea
with the definition of the functional integral $\FuncIntAvg_P$ being\footnote{As usual constant factors have been neglected.}
\bea
\FuncIntAvg_P[O[\Phi]] &=&
\int D\Phi\,D\omega\,D\omega^\dagger\,\,  O[\Phi] \cdot e^{- S_\Phi[\Phi]- S_{PF}[\Phi,\omega]},\,\,\,\,\quad \\
\label{eq:DefOfFermionActionSPF}
S_{PF}[\Phi,\omega] &=&  \frac{1}{2} \sum\limits_{i=1}^{N_f}\omega^\dagger_i P(\fermiMat\fermiMat^\dagger)\omega_i,
\eea
which is suitable for numerical calculations as will be seen in the following. Here $\omega\equiv(\omega_1,\ldots,\omega_{N_f})$ 
denotes a set of $N_f$ complex vectors $\omega_i$ referred to as pseudo fermion fields in the following, and the calculation of 
the complex phase $\arg\det(\fermiMat)$ has already been discussed in \sect{sec:ComplexPhaseOfFermionDet}. The expectation value 
of an observable $O[\Phi]$ is then given as
\bea
\label{eq:PHMCReweighingStratGeneral}
\langle O[\Phi] \rangle &=& \frac{\langle O[\Phi] \cdot e^{iN_f\arg\det(\fermiMat)} \cdot W[\Phi] \rangle_P}
{\langle e^{iN_f\arg\det(\fermiMat)} \cdot W[\Phi] \rangle_P}
\eea
where the expectation value $\langle\ldots\rangle_P$ with respect to the modified dynamics arising from the
polynomial $P(\fermiMatDouble)$ alone, \ie neglecting the weight factors, is given as
\bea
\langle O[\Phi] \rangle_P &=& \frac{1}{Z_P} \FuncIntAvg_P\left[O[\Phi]\right], \quad Z_P = \FuncIntAvg_P\left[1\right].
\eea 
It is actually this latter functional integral that is computed in a PHMC algorithm. The multiplication of the
considered observable with the weight factors according to \eq{eq:PHMCReweighingStratGeneral}, which is known
as reweighting in the literature, then guarantees the PHMC algorithm to yield the correct result with respect to the 
full dynamics.

It is remarked at this point that the polynomial $P(x)$ with the aforementioned properties can be constructed in 
numerous ways. In this implementation a least-square minimization procedure with respect to the residual 
$r_P(x) = \left[P(x) - x^{-\alpha}  \right]$ with $\alpha=1/2$ has been used, based on the software 
code published in \Ref{Gebert:2003pk}. This method defines the scalar product of two given square-integrable, real functions 
$f(x)$, $g(x)$ as
\beq
\label{eq:DeterOfPolynomialPScalarProd}
\langle f| g\rangle_\alpha = \int\limits_{\epsilon_P}^{\lambda_P}\intd{x} x^{2\alpha}\cdot f(x) \cdot g(x) 
\eeq
leading then to the corresponding norm $|f|_\alpha = \sqrt{\langle f|f\rangle_\alpha}$. The squared norm of the residual
\beq
\label{eq:DefOfMeasureRP}
R_P^2 = |r_P|^2_\alpha = \int\limits_{\epsilon_P}^{\lambda_P}\intd{x} x^{2\alpha}\cdot \left[P(x) - x^{-\alpha}  \right]^2
\eeq
can then be minimized in the space of all polynomials with maximal degree $N_P$ by using the availability of a scalar
product to project the function $x^{-\alpha}$ to a chosen basis of that space. Here, the inclusion of the factor
$x^{2\alpha}$ in \eq{eq:DeterOfPolynomialPScalarProd} and thus in the norm of the residual implicates an optimization
with respect to the relative deviation, which is desirable for our purpose. The only difficulty is that this computation
has to be performed with very high numerical precision to get stable results also in case of high polynomial degrees. 
For that purpose the code in \Ref{Gebert:2003pk} employs some standard high precision numeric software library~\cite{cln96a}. 
Furthermore, it is remarked that this approach relies on the fact that the scalar product of the considered function $x^{-\alpha}$
and any given polynomial as defined in \eq{eq:DeterOfPolynomialPScalarProd} can be calculated analytically, since 
computing the respective integral numerically with the required accuracy is not very practicable.

For the numerical evaluation of the functional integral in \eq{eq:PHMCReweighingStratGeneral} to yield stable results 
the weight factor should in general, apart from the exceptions in the next paragraph, fluctuate as little as possible. 
In the limit $R_P\rightarrow 0$ the weight $W[\Phi]$ converges to one and the fluctuations vanish provided that 
a sufficiently large approximation interval $[\epsilon_P,\lambda_P]$ has been chosen such that all eigenvalues 
of the operator $\fermiMatDouble$ encountered in the Markov process are covered. Since decreasing $\delta_P$ or 
enlarging the approximation interval requires a higher degree $N_P$ of the underlying approximation polynomial 
there is always a trade-off between the fluctuation of $W[\Phi]$ and the computational resources needed to 
calculate $P(\fermiMatDouble)$.

While it is absolutely mandatory for the numerical stability of the algorithm that $\lambda_P$ is a true 
upper bound for all eigenvalues of $\fermiMatDouble$ encountered in the Monte-Carlo process,
the lower interval bound $\epsilon_P$ is allowed to be chosen larger than the smallest attainable
eigenvalue, which would be corrected for by the reweighting factor $W[\Phi]$. This freedom is often used to 
explicitly select a lower bound $\epsilon_P$ that is clearly larger than the lowest observed eigenvalue of
$\fermiMatDouble$. Though the resulting reduction of the required polynomial degree $N_P$ is accompanied
with a higher fluctuation of $W[\Phi]$ this technique can reduce the overall numerical
cost of the whole computation in the sense of achieved accuracy per numerical cost, in particular if there 
are only a few extremely small eigenvalues well separated from the rest of the eigenvalue spectrum~\cite{Frezzotti:1997ym}. 
It can also be advantageous when the considered observable $O[\Phi]$ itself fluctuates strongly, ideally 
anti-proportional to the smallest eigenvalue of $\fermiMat$ like, for instance, the fermion propagator~\cite{Frezzotti:1997ym}. 
In the latter case choosing a larger $\epsilon_P$ does not only
reduce the necessary polynomial degree $N_P$ but it also stabilizes the numerical calculation by compensating
the fluctuation of such an observable $O[\Phi]$ with that of $W[\Phi]$. In this work, however, a different 
reweighting approach is implemented. The details of this new reweighting technique will be given in \sect{sec:ExactReweighing}.

Concerning the practical evaluation of the matrix-valued polynomial $P(\fermiMatDouble)$ a suitable representation
for $P(x)$ has to be chosen. A representation in terms of a Chebyshev basis is numerically very stable,
but it does not directly allow for the calculation of the forces $F[\Phi,\omega]$ needed for the integration of the 
molecular dynamics equations of motion as discussed in \sect{sec:HMCAlgorithm}. A decomposition in terms of monomials 
given as
\beq
P(\fermiMatDouble) = \prod\limits_{j=1}^{N_P} \left(\fermiMatDouble -z_j\ID  \right),
\eeq
where the complex numbers $z_j$ denote the roots of the polynomial $P(x)$, provides a direct approach
for calculating the aforementioned molecular forces $F[\Phi,\omega]$ according to
\bea
F[\Phi,\omega] = \derive{}{\Phi}S_\Phi[\Phi] + \frac{1}{2}\sum\limits_{i=1}^{N_f} \omega_i^\dagger\frac{d}{d\Phi} P(\fermiMatDouble) \omega_i
\eea
with
\beq
\label{eq:SinglePHMCForce}
\omega_i^\dagger\frac{d}{d\Phi} P(\fermiMatDouble) \omega_i = \omega_i^\dagger \sum\limits_{j=1}^{N_P}
\prod\limits_{m=1}^{j-1}\left(\fermiMatDouble -z_m\ID  \right)
\cdot \left[ \frac{d}{d\Phi} \fermiMatDouble  \right] \cdot
\prod\limits_{n=j+1}^{N_P}\left(\fermiMatDouble -z_n\ID  \right) \omega_i.
\eeq
The evaluation of this latter expression requires no more than $N_P$ applications of the operator $\fermiMatDouble$
for each pseudo fermion field $\omega_i$, provided that enough computer memory is available to store a number of $N_P$ 
auxiliary vectors.

The problem with this naive representation of $P(x)$ in terms of monomials is its high numerical instability.
This disadvantage can be cured by selecting an appropriate order of the roots $z_i$ and thus of the
corresponding monomials. Different ordering schemes have been examined in the literature and the so-called
bit-reversed ordering was found to yield very stable results~\cite{Bunk:1998rm}. This is the approach that will also
be used in this implementation of the PHMC algorithm.

The Markov step for the field $\Phi$ is then constructed in the same way as for the HMC algorithm by
introducing the momenta $\pi$ conjugate to $\Phi$, integrating the equations of motions in terms of 
$\Phi(\MCtime)$ and $\pi(\MCtime)$ with $\MCtime$ denoting the Monte-Carlo time, and accepting or rejecting
the proposed configuration according to a final Metropolis step as described in \sect{sec:HMCAlgorithm}. The technique
for the direct sampling of the pseudo fermion fields $\omega_i$ also differs from the one used in the HMC
algorithm and will be discussed in \sect{sec:DirectSampling}.

Finally, it shall be remarked that instead of the introduction of $N_f$ pseudo fermion fields $\omega_i,\,
i=1,\ldots,N_f$ in \eq{eq:PHMCPartitionFunction} one could also have used only one pseudo fermion field $\omega_1$
together with a polynomial $P(x)$ approximating the function $x^{-N_f/2}$. It has, however, been discussed
in the literature~\cite{Clark:2006fx}, at least for the case of the RHMC algorithm, that this approach should be less efficient 
than the chosen method due to larger force strengths encountered in the integration of the molecular dynamics 
and -- translated into the framework of the PHMC algorithm -- the requirement for a higher degree of the 
approximation polynomial for fixed relative accuracy $\delta_P$.

%-----------------------------------------------------------------------------------------------------
\section{Reducing the fermion matrix condition number through preconditioning}
\label{sec:CondNr}

The condition number of the fermion matrix has a tremendous impact on the overall numerical cost
of the lattice calculations performed in this study. For a general, diagonalizable matrix $A$ it is 
given as the norm of the ratio of the eigenvalue $\nu_{max}$ with largest norm divided by the 
eigenvalue $\nu_{min}$ with smallest norm according to
\beq
\cond(A) = \left| \frac{\nu_{max}}{\nu_{min}}  \right|.
\eeq

This quantity strongly influences the asymptotic convergence speed of many numerical algorithms. 
For the conjugate gradient (CG) algorithm, used in the HMC approach to apply the inverse of
$\fermiMatDouble$ on a given vector, the norm of the residual goes to zero proportional to 
$\exp(-N_I/\cond(\fermiMatDouble))$ for sufficiently large iteration numbers $N_I$. In case
of the polynomial approximation used in the PHMC algorithm the condition number $\cond(\fermiMatDouble)$ 
specifies the size of the approximation interval $[\epsilon_P,\lambda_P]$, more precisely the ratio 
$\lambda_P/\epsilon_P$, needed to cover the whole eigenvalue spectrum of $\fermiMatDouble$. Thus, 
it is directly related to the numerical cost for the evaluation of the polynomial $P(\fermiMatDouble)$. 
The condition number $\cond(\fermiMat)$ also influences the convergence speed of the Arnoldi algorithm 
used to determine the sign of the fermion determinant as described in \sect{sec:ComplexPhaseOfFermionDet}. 

The main idea of this section is to reduce the numerical cost of the intended lattice calculations by
reducing the condition number of the fermion matrix. To be more precise, the fermion matrix
$\fermiMat$ will be replaced by some other matrix, which generates exactly the same physical
results but possesses a lower condition number. This is possible, at least in the broken phase of 
the model, which is the scenario that underlies all considerations in the following. In \sect{sec:PreconOfM} 
this approach will be discussed for and applied to the fermion matrix $\fermiMat$ itself in the context of reducing
the computational costs for the determination of the sign of the fermion determinant $\det(\fermiMat)$.
Since the identical approach will not work for the squared operator $\fermiMatDouble$,
a somewhat different ansatz will be presented in \sect{sec:PreconOfMDouble} capable of 
cutting down the condition number of $\fermiMatDouble$ by $1-2$ orders of magnitude, thus leading to an
outstanding performance gain in the later lattice calculations.

%-----------------------------------------------------------------------------------------------------
\subsection{Preconditioning of \texorpdfstring{$\fermiMat$}{M}}
\label{sec:PreconOfM}

Based on the observation in \sect{eq:ComplexPhaseOfFermionDetDegCase} that
the general structure of the eigenvalue spectrum of $\fermiMat$ is very much determined by the
eigenvalue spectrum of the free Dirac operator $\D$ as well as the magnetization $m$ of the 
underlying field configuration the basic idea here is to use the analytically invertible operator 
$\fermiMat[\Phi']$ to precondition the general fermion matrix $\fermiMat[\Phi]$, where the field 
$\Phi'$ has been given in \eq{eq:staggeredAnsatz} and consists solely of a constant and a staggered mode as specified 
by the amplitudes\footnote{The notation in this chapter differs from that in \chap{chap:PhaseDiagram}.
The hat symbols of $\breve m_\Phi$, $\breve s_\Phi$ underlying the definition of $\Phi'$ in \eq{eq:staggeredAnsatz}
are omitted and the factor $N_f^{1/2}$ in the definition of $\Phi'$ will always be divided out,
such that the observables $m$, $s$ evaluated on $\Phi'$ equal $m_\Phi$, $s_\Phi$.}
$m_\Phi$, $s_\Phi$ and the orientations $\hat\Phi_1$, $\hat\Phi_2$.  
For that purpose the preconditioning operator $Q[m_\Phi, s_\Phi, \hat \Phi_1, \hat \Phi_2]$ is defined as
\beq
Q[m_\Phi, s_\Phi, \hat \Phi_1, \hat \Phi_2] = \left( \fermiMat[\Phi'/N_f^{\frac{1}{2}}]  \right)^{-1}.
\eeq
The obvious expectation is that the product $\fermiMat Q$ should be close to the identity provided that the 
parameters $m_\Phi$, $s_\Phi$ of the matrix $Q$ match the magnetizations\footnote{Here and in the following it is
always assumed that at least one of the two magnetizations $m$, $s$ is non-zero.} 
$m$, $s$ of the field configuration $\Phi$ underlying the operator $\fermiMat[\Phi]$ and that the orientations
$\hat\Phi_1$, $\hat\Phi_2$ point into the directions $\sum_x\Phi_x$ and $\sum_x \exp(ip_sx)\Phi_x$, respectively. 
Consequently, the condition number $\cond(\fermiMat Q[m,s,\sum_x\Phi_x/mV, \sum_x \exp(ip_sx)\Phi_x/sV])$ is supposed to 
be close to one, at least in the broken phases. For simplicity, however, a restricted ansatz is employed 
in the following given as
\beq
Q[m_\Phi] = \left( \fermiMat[\Phi'/N_f^{\frac{1}{2}}]  \right)^{-1}\quad \mbox{with} \quad 
s_\Phi = 0, \, \hat \Phi_1^\mu = \delta_{\mu,0},
\eeq
which turns out to be still sufficiently good for the considered purpose here.

This approach seems promising for the determination of the sign of the fermion determinant $\det(\fermiMat)$. 
The reason is that the numerical cost for the computation of the eigenvalues of $\fermiMat Q[m]$ should 
be reduced due to its condition number being smaller than $\cond(\fermiMat)$. At the same time the phase, and thus 
the sign, of the determinant is unaltered by the multiplication with the operator $Q[m]$, \ie
\beq
\arg\det(\fermiMat[\Phi]) = \arg\det(\fermiMat[\Phi] Q[m])
\eeq
according to 
\beq
\arg\det(Q[m_\Phi]) = 1, \quad \forall m_\Phi\neq 0.
\eeq
The latter result is easily obtained by employing the exactly known eigenvalue spectrum of $Q[m_\Phi]$ for
some non-zero value $m_\Phi$. For the sake of completeness it is remarked that in the case of exactly zero 
magnetization, \ie $m\equiv 0$, the latter quantity can not be employed to precondition the fermion matrix 
due to $\det(Q[0])=0$. However, the set $\{m=0\}$ is a set of zero measure and will moreover practically not 
occur when investigating the broken phase as intended. Furthermore, it is straightforward to test for 
this condition in the program code to avoid the application of the preconditioning technique in that scenario.

%\includeFigTrippleDouble{spectrumOfMlam0kap012301}{spectrumOfMQlam0kap012301}{SingleMCondNrComparisonlam0kap012301}
%{spectrumOfMlamnankap029929}{spectrumOfMQlamnankap029929}{SingleMCondNrComparisonlamnankap029929}
\includeFigTrippleDouble{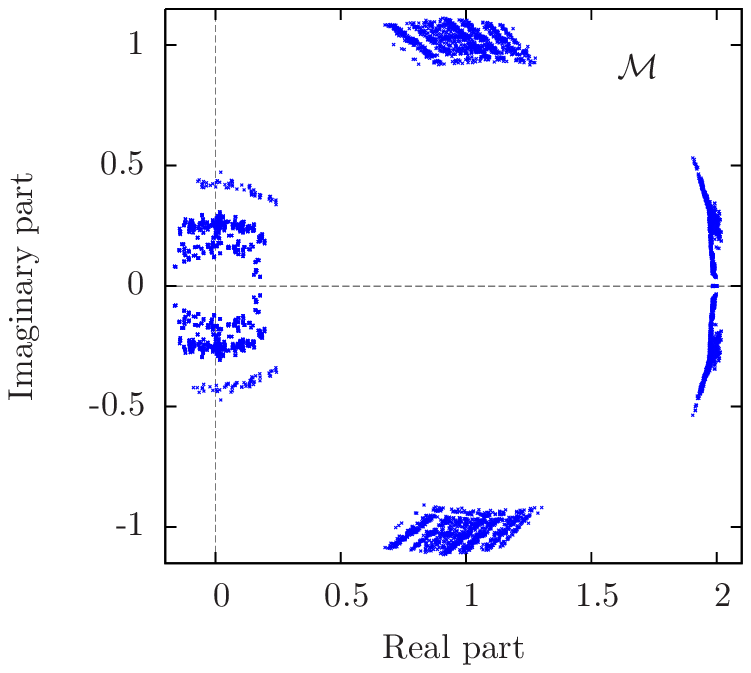}{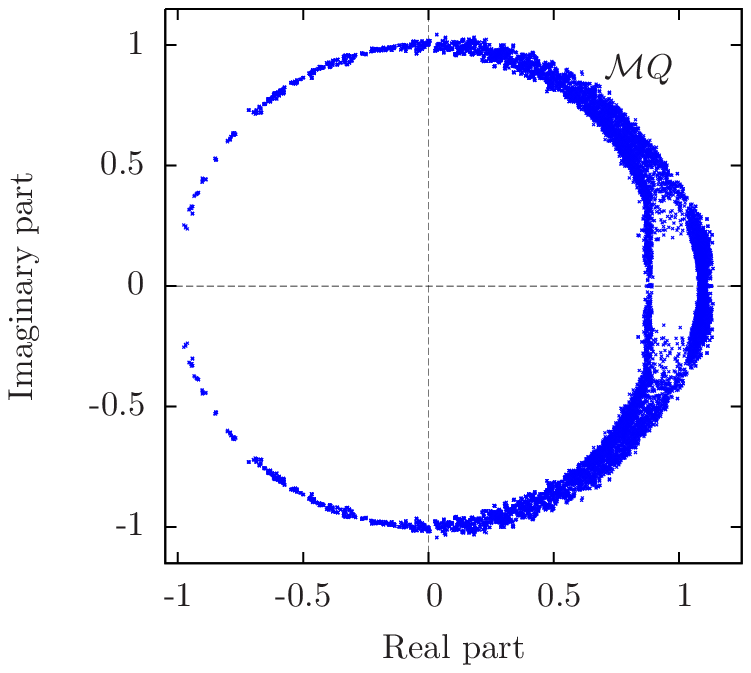}{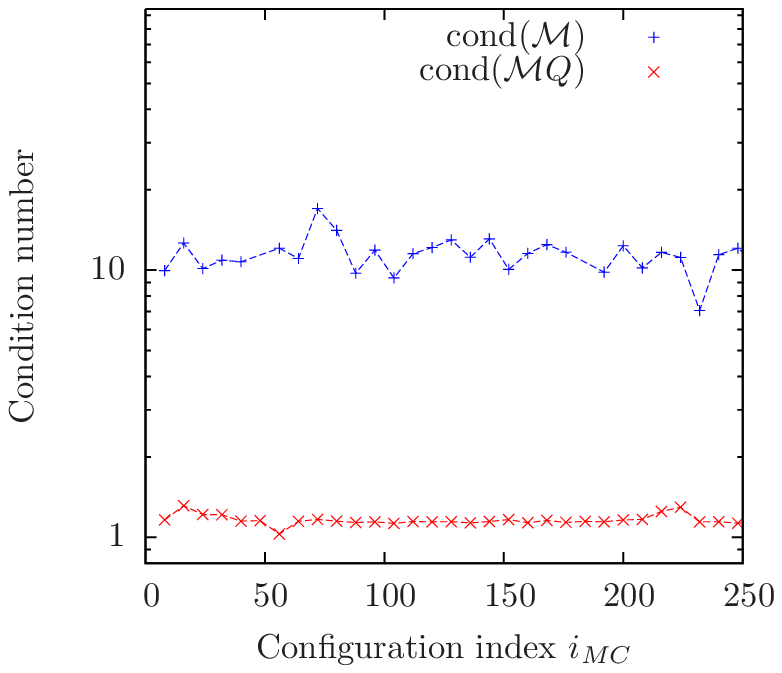}
{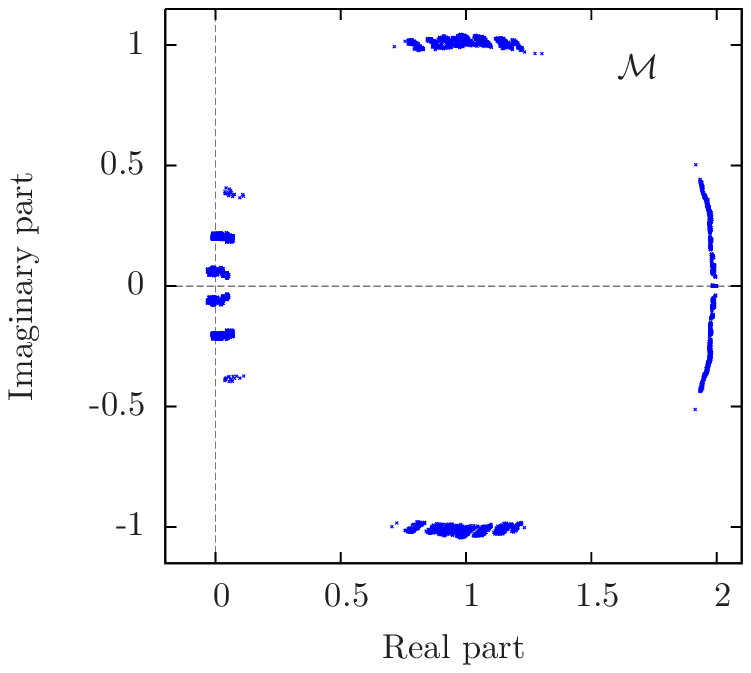}{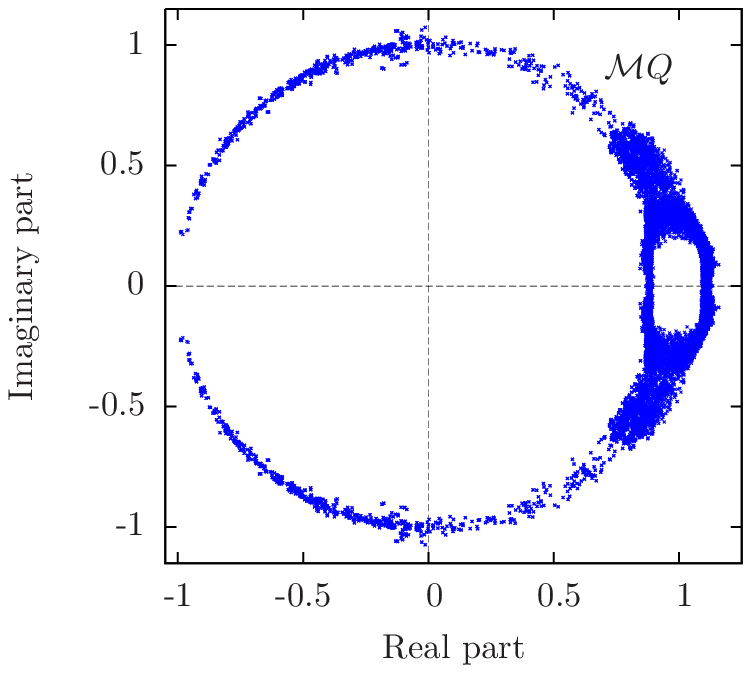}{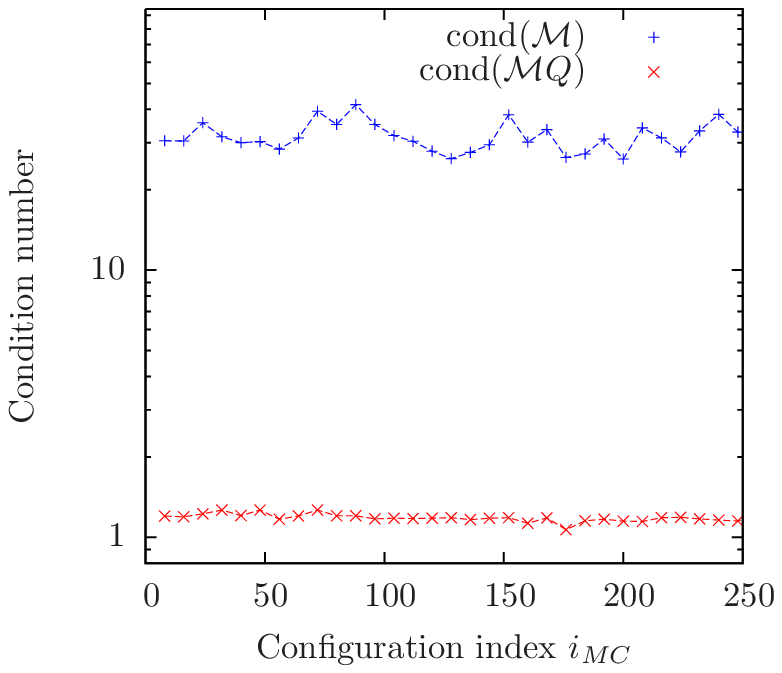}
{fig:TestOfPrecOfM}
{The structure of the eigenvalue spectra of the fermion matrices $\fermiMat[\Phi]$ and 
$\fermiMat[\Phi]Q[m]$  is illustrated. The panels in the left column each show a superposition of 
eigenvalues of the operator $\fermiMat[\Phi]$ that have been computed  
by means of the Arnoldi algorithm in the background of several field configurations $\Phi$ generated 
in the Monte-Carlo runs specified in \tab{tab:Chap5EmployedRuns}. The upper panel row refers to the listed 
run with $\kappa=0.12301$, while the lower row refers to the run with $\kappa=0.29929$.
For each field configuration $\Phi$ a total of 120 eigenvalues has been calculated, composed out of the 
20 eigenvalues exhibiting the smallest norms, the largest norms, the smallest real parts, the largest real 
parts, the smallest imaginary parts, and the largest imaginary parts, respectively. 
The corresponding results for the preconditioned matrix $\fermiMat[\Phi]Q[m]$ are presented in the middle 
panel column. Panels (c) and (f) show the respective condition numbers versus the configuration index $\confNR$,
where every eighth configuration has been evaluated, while the dashed lines are only meant to guide the eye.
}
{The eigenvalue spectra of the fermion matrices $\fermiMat$ and $\fermiMat Q$.}   

The applicability of this approach is demonstrated in \fig{fig:TestOfPrecOfM}. The results presented there
have been obtained in Monte-Carlo runs performed on \lattices{16}{32} in the broken phase, selected
to give a typical example of the improvement that has been achieved. The plots in panels (a,b,d,e) 
each show a superposition of eigenvalues determined on several field configurations. Since the full spectrum
cannot be computed for practical reasons, only 120 eigenvalues have been computed on each field configuration,
equally composed out of those eigenvalues with smallest norm, largest norm, smallest and largest real part,
as well as smallest and largest imaginary part, which could be achieved with the help of an implementation~\cite{ARPACK:1998zu} 
of the Arnoldi algorithm.
The intention of this presentation is to provide an illustration of the eigenvalue spectrum of the respectively 
considered operators. 

In panels (a,d) this has been done for the fermion operator $\fermiMat$ itself and one clearly observes the
circular base structure inherited from the free Neuberger operator $\D$. In contrast, the eigenvalue
structure of the matrix $\fermiMat Q[m]$ presented in panels (b,e) looks very much distinct. The eigenvalues no
longer lie on the Neuberger circle passing through the origin, but rather on the unit circle.
The results for the corresponding condition numbers are presented in \fig{fig:TestOfPrecOfM}c,f. The 
condition number $\cond(\fermiMat Q[m])$ is distinctly smaller than that of the original fermion
matrix $\fermiMat$ and fluctuates very close to one as intended.
 
However, the convergence speed of the Arnoldi algorithm is not solely determined by the condition number
but also by the eigenvalue density and other factors. Besides that the approach to determine the sign of 
the fermion determinant described in \sect{eq:ComplexPhaseOfFermionDetDegCase} depends very much on the 
average number of eigenvalues situated in the complex half-plane with negative real part. The actual reduction of the
numerical cost of the procedure to compute the sign of the fermion determinant $\det(\fermiMat)$ should therefore 
be tested explicitly. In \tab{tab:PrecOfMRealLifeExam} the resulting average computation times for determining 
$\sign\det(\fermiMat)$ with and without the usage of the preconditioning matrix $Q[m]$ are compared to each other. 
One sees that speed-up factors around 1-2 orders of magnitude can be achieved by this approach in practice.

\includeTab{|c|c|c|}{
Monte-Carlo run & Avg. time for $\sign\det(\fermiMat)$ & Avg. time for $\sign\det(\fermiMat Q)$ \\
\hline
$\kappa=0.12301$ & $\approx 320$ minutes & $\approx 4.6$ minutes \\
$\kappa=0.29929$ & $\approx 360$ minutes & $\approx 19$ minutes \\
}
{tab:PrecOfMRealLifeExam}
{Average times needed for the determination of the sign of $\det(\fermiMat[\Phi])$ and 
$\det(\fermiMat[\Phi]Q[m])$, respectively. For that purpose the method explained in the main
text has been applied. The underlying field configurations $\Phi$ have been generated in the Monte-Carlo
runs specified in \tab{tab:Chap5EmployedRuns}. The measurement was done by counting the number of
signs that could be determined within 48 hours on a single CPU core. The respective sets of configurations
underlying the calculation of $\sign\det(\fermiMat[\Phi])$ and $\sign\det(\fermiMat[\Phi] Q[m])$ were the same.
}
{Average times for the determination of $\sign\det(\fermiMat)$ and $\sign\det(\fermiMat Q)$.}

%-----------------------------------------------------------------------------------------------------
\subsection{Preconditioning of \texorpdfstring{$\fermiMatDouble$}{MM}}
\label{sec:PreconOfMDouble}

With the results of the previous section in mind it would be extremely desirable to exploit the
benefits of the low condition number of $\fermiMat Q[m]$ also in the actual PHMC algorithm itself
where the condition number of the squared operator $\fermiMatDouble$ determines the computational cost.
Indeed, a partition function being equivalent to \eq{eq:PHMCPartitionFunction} can be constructed
based on the assumedly improved operator $\fermiMat Q[m] Q^\dagger[m] \fermiMat^\dagger$
according to
\bea
\label{eq:PHMCPartitionFunctionImprovedWithQ}
Z_\Phi &=& \int D\Phi\,\,\, e^{iN_f\arg\det(\fermiMat Q[m])} \cdot
\left[\det(\sqrt{\fermiMat Q[m_\Phi]Q^\dagger[m_\Phi]\fermiMat^\dagger})   \right]^{N_f} \cdot
e^{-S_\Phi[\Phi] },
\eea
where the preconditioning parameter $m_\Phi$ is held constant during the whole Monte-Carlo calculation.
It should be noted here, that there is no correction factor needed for the inclusion of the
operator $Q[m_\Phi]$. This is because the parameter $m_\Phi$ is fixed, such that the determinant 
$\det(Q[m_\Phi])$ only gives a constant factor independent of the actual field configuration
$\Phi$. The idea here is to fix the parameter $m_\Phi$ at the start of the Monte-Carlo computation,
such that it matches the average magnetization, \ie $m_\Phi = \langle m \rangle$. This, of course,
involves some small amount of tuning.

From the experience of the previous section one might then expect the condition number 
$\cond(\fermiMat Q[m_\Phi] Q^\dagger[m_\Phi] \fermiMat^\dagger)$ to be of the order of the square 
of $\cond(\fermiMat Q[m_\Phi])$, \ie of order $O(1)$ according to the examples presented in 
\fig{fig:TestOfPrecOfM}c,f.

This, however, is not the case. In fact, the condition number $\cond(\fermiMat Q[m_\Phi] Q^\dagger[m_\Phi] \fermiMat^\dagger)$
can even be much larger than the one of the unpreconditioned operator $\fermiMatDouble$, as demonstrated
in \fig{fig:ExampleOfPrecMMDag1}a and \fig{fig:ExampleOfPrecMMDag1}b for identically the same Monte-Carlo
runs that served as the example setup in the preceding section. This is not a result of fixing the preconditioning
parameter $m_\Phi$ to $\langle m \rangle$. To clarify this the condition numbers of the preconditioned operator
have been determined here by setting the parameter $m_\Phi$ individually to the respective magnetization $m$ for each
evaluated configuration, \ie $\cond(\fermiMat Q[m] Q^\dagger[m] \fermiMat^\dagger)$ is presented in 
\fig{fig:ExampleOfPrecMMDag1}.

%\includeFigDouble{MMdagCondNrComparisonlam0kap012301}{MMdagCondNrComparisonlamnankap029929}
\includeFigDouble{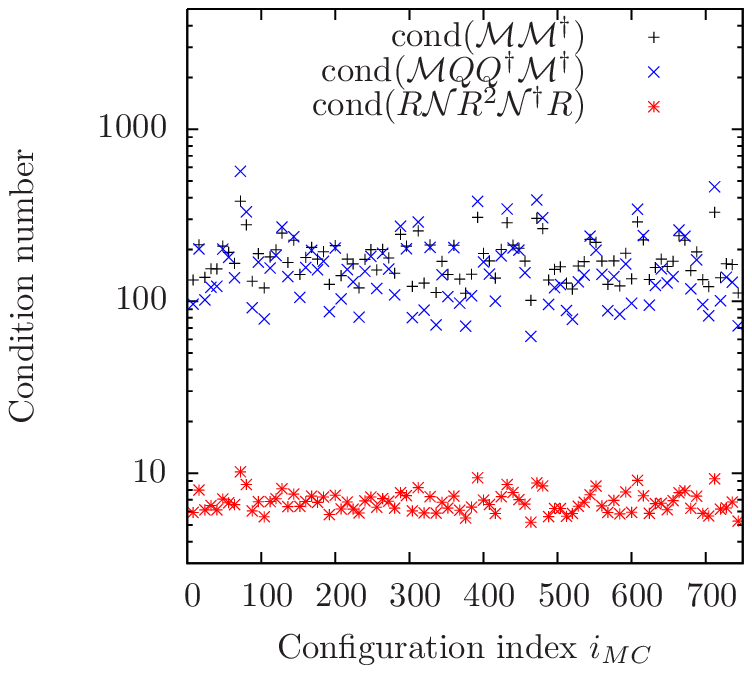}{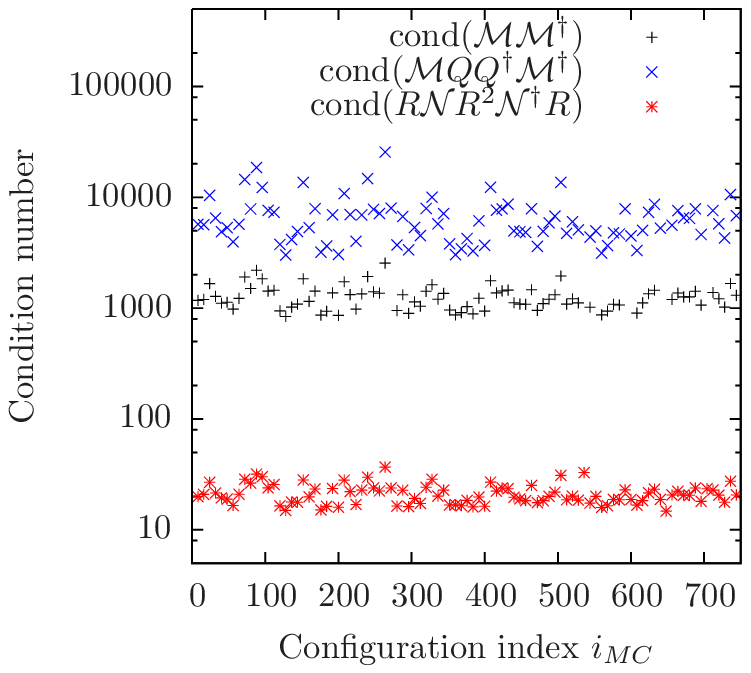}
{fig:ExampleOfPrecMMDag1}
{The condition numbers of the squared fermion matrix $\fermiMatDouble$ is compared to the corresponding
numbers of the preconditioned matrices $\fermiMat Q[m]Q^\dagger[m]\fermiMat^\dagger$ and
$R[m]\newfermiMat R^2[m]\newfermiMat^\dagger R[m]$, respectively. The underlying field configurations $\Phi$
have been generated in the Monte-Carlo runs specified in \tab{tab:Chap5EmployedRuns}. Panel (a)
refers to the listed run with $\kappa=0.12301$ and panel (b) refers to the run with $\kappa=0.29929$.
}
{Condition numbers of the squared fermion matrices $\fermiMatDouble$, $\fermiMat QQ^\dagger\fermiMat^\dagger$ ,
and $R\newfermiMat R^2\newfermiMat^\dagger R$.}

The reason for this outcome originates from the fact that the fermion matrix $\fermiMat$ is not hermitian. 
In fact, the operators $\fermiMat$ and $\fermiMat^\dagger$ do not even share a common eigenvector basis. 
This statement is also inherited to the matrices $\fermiMat Q[m]$ and $Q^\dagger[m]\fermiMat^\dagger$.
In this general setup the condition number of a product of matrices is not the product 
of their condition numbers, \ie
\beq
\cond(\fermiMat Q[m] Q^\dagger[m] \fermiMat^\dagger) \neq \cond(\fermiMat Q[m]) \cdot \cond(Q^\dagger[m]
\fermiMat^\dagger),
\eeq
which explains the observation in \fig{fig:ExampleOfPrecMMDag1}. The naive approach given in 
\eq{eq:PHMCPartitionFunctionImprovedWithQ} must therefore be considered to have failed. 

To elaborate on a potential cure to this unsatisfactory finding, we consider the following Gedanken experiment. 
We now assume to have a $\Phi$-independent, non-singular, hermitian matrix $T$ with $TT=\ID$ at hand, such that 
$\fermiMat T$ is hermitian. In case of QCD such an operator is given by $T=\gamma_5$ due to the $\gamma_5$-hermiticity
of the QCD fermion matrix. Moreover, we assume that the condition number $\beta= \cond(\fermiMat T(\breve W[m]\breve W[m])^{-1/2})$
is small, which is plausible due to the definition of the latter hermitian operator $\breve W$ given here as
\bea
\label{eq:DefOfAbstractCondMatW}
\breve W[m_\Phi] &=& \fermiMat[\Phi'/N_f^{\frac{1}{2}}] T \quad \mbox{with} \quad
s_\Phi=0,\,\, \hat\Phi_1^\mu = \delta_{\mu,0}.
\eea
Postulating that $\breve W[m]$ is non-singular, which is indeed the case according to the above definition, provided that
$m\neq 0$ as already discussed in the preceding section, one directly finds
\beq
\beta = \cond\left(\fermiMat T (\breve W[m]\breve W[m])^{-\frac{1}{2}}\right) 
= \cond\left((\breve W[m]\breve W[m])^{-\frac{1}{4}} \fermiMat T (\breve W[m]\breve W[m])^{-\frac{1}{4}}\right)
\eeq
and thus
\bea
\beta^2 &=&
\cond\left(  (\breve W[m]\breve W[m])^{-\frac{1}{4}} \fermiMat T 
(\breve W[m]\breve W[m])^{-\frac{1}{2}} \fermiMat T (\breve W[m]\breve W[m])^{-\frac{1}{4}}
\right)   
\eea
due to the hermiticity of $(\breve W[m]\breve W[m])^{-\frac{1}{4}} \fermiMat T (\breve W[m]\breve W[m])^{-\frac{1}{4}}$.
Exploiting the fact that $T$ commutes\footnote{One directly finds $T\breve W\breve W 
= T \fermiMat[\Phi'/N_f^{\frac{1}{2}}] \fermiMat^\dagger[\Phi'/N_f^{\frac{1}{2}}]TT
= T \fermiMat^\dagger[\Phi'/N_f^{\frac{1}{2}}] \fermiMat[\Phi'/N_f^{\frac{1}{2}}]TT
= \breve W\breve W T$ when exploiting 
$\fermiMat[\Phi'/N_f^{\frac{1}{2}}] \fermiMat^\dagger[\Phi'/N_f^{\frac{1}{2}}] 
= \fermiMat^\dagger[\Phi'/N_f^{\frac{1}{2}}] \fermiMat[\Phi'/N_f^{\frac{1}{2}}]$, 
which holds for the space-time independent field $\Phi'$ underlying the definition of $\breve W$ 
in \eq{eq:DefOfAbstractCondMatW}.}
with $\breve W \breve W$ according to the specific definition of $\breve W$ in 
\eq{eq:DefOfAbstractCondMatW} one can further simplify the above expression to
\bea
\beta^2 &=&
\cond\left(  \breve R[m] \fermiMat \breve R[m] \breve R[m] \fermiMat^\dagger \breve R[m],
\right)   
\eea
where the hermitian, positive matrix $\breve R[m]$ is defined as
\bea
\label{eq:CondMatRAbstractDef1}
\breve R[m_\Phi] &=& (\breve W[m]\breve W[m])^{-\frac{1}{4}} \\
\label{eq:CondMatRAbstractDef2}
&=& \left(\fermiMat[\Phi'/N_f^{\frac{1}{2}}] \fermiMat^\dagger[\Phi'/N_f^{\frac{1}{2}}] \right)^{-\frac{1}{4}}
\quad \mbox{with} \quad s_\Phi=0,\,\, \hat\Phi_1^\mu = \delta_{\mu,0}.
\eea
Moreover, one has 
\bea
\det\left( \fermiMatDouble\right) &=&
\det\left( \breve R[m_\Phi]\fermiMat \breve R^2[m_\Phi] \fermiMat^\dagger \breve R[m_\Phi] \right)
\eea
up to a $\Phi$-independent factor, provided that the constant parameter $m_\Phi$ is non-zero. Following the same arguments 
given at the beginning of this section the Monte-Carlo calculation could thus be based on the latter well conditioned operator 
having a condition number of only $\beta^2$ in the here considered hypothetical scenario.

However, when actually trying to apply this strategy to the considered Higgs-Yukawa model, one faces the problem of 
finding an operator $T$ with the aforementioned properties. Since $\fermiMat$ is not $\gamma_5$-hermitian, as discussed
in \sect{sec:ComplexPhaseOfFermionDet}, the setting $T=\gamma_5$ is not an eligible choice here. In fact, the fermion 
matrix $\fermiMat$ would not even be $\gamma_5$-hermitian, when the three components $\Phi^1_x$, $\Phi^2_x$, and $\Phi^3_x$
of a given field configuration would be identical to zero and only $\Phi^0_x$ would be non-constant over space-time. 
The origin of these bad hermiticity properties lies in the pronounced asymmetry of the fermion matrix
\beq
\fermiMat = \D + B \left(\ID-\frac{1}{2\rho}\D  \right).
\eeq

A potential cure to this problem would be the construction of a new fermion matrix $\newfermiMat[\Phi]$ having better
symmetry properties, with respect to a $\gamma_5$-hermiticity for instance, while still producing the same 
physics. The idea would then be to precondition the modified operator $\newfermiMat$ following the above strategy.

The latter requirement of producing the same physics is exactly fulfilled, if the determinant of $\fermiMat[\Phi]$ equals 
that of the new operator $\newfermiMat[\Phi]$ for all field configurations $\Phi$ up to a constant factor independent of 
$\Phi$. An example of such an operator is
\beq
\newfermiMat = P_\pi -2\rho\DDmtwoRhoInv + (\ID-P_\pi) B (\ID-P_\pi)
\eeq
where the matrix $\DDmtwoRhoInv$ has been defined in \eq{eq:DefOfOperatorA} and the projector $P_\pi$ was given in \eq{eq:DefOfProjector}.
As desired the operator $\newfermiMat$ then fulfills
\beq
\det(\fermiMat[\Phi]) = \det(\newfermiMat[\Phi])
\eeq
up to a constant factor, as can be seen by employing the relation in \eq{eq:DetRelationWithProjector}. In contrast 
to $\fermiMat$ this new operator is $\gamma_5$-hermitian provided that the three components $\Phi_x^1$, $\Phi_x^2$, and 
$\Phi_x^3$ of a given field configuration would be identical to zero and only $\Phi_x^0$ would be non-trivial. Moreover, 
in the degenerate case with equal Yukawa coupling constants, \ie $y_t=y_b$, the matrix $\newfermiMat$ is 
$\tau_2\gamma_5$-hermitian provided that only the component $\Phi_x^2$ of the underlying field configuration is identical 
to zero. From these observations one may conjecture that the operator $\newfermiMat$ is a more promising 
candidate for reducing the computational cost according to the strategy described above.

Though the newly introduced operator $\newfermiMat$ does also not -- at least not obviously -- obey an exact
$T$-hermiticity relation in the general situation, we nevertheless apply the above preconditioning strategy.
The next step is thus the definition of the preconditioning matrix for the new operator $\newfermiMat$ according to
\eq{eq:CondMatRAbstractDef2} yielding the definition
\bea
\precR[m_\Phi] &=& \left( \newfermiMat[\Phi'/N_f^{\frac{1}{2}}] \newfermiMat^\dagger[\Phi'/N_f^{\frac{1}{2}}] \right)^{-\frac{1}{4}}\quad \mbox{with} \quad
s_\Phi=0,\,\, \hat\Phi_1^\mu = \delta_{\mu,0}.
\eea
The expectation is now that the composed operator $\precR[m]\newfermiMat\precR^2[m]\newfermiMat^\dagger\precR[m]$ 
possesses a distinctly smaller condition number as compared to $\fermiMatDouble$. 

That this is indeed the case can be observed in \fig{fig:ExampleOfPrecMMDag1} where the condition number
$\cond(\precR[m]\newfermiMat\precR^2[m]\newfermiMat^\dagger\precR[m])$ is presented for exactly the same 
configurations underlying also the aforementioned results for $\cond(\fermiMat Q[m] Q^\dagger[m] \fermiMat^\dagger)$ 
and $\cond(\fermiMatDouble)$. In comparison with the operator $\fermiMatDouble$ the condition number of the improved 
operator has been reduced by a factor of around 1-2 orders of magnitude.

However, one may ask whether this technique is restricted to the case of degenerate Yukawa coupling
parameters, which have been assumed so far. Furthermore one may ask whether the success of this approach
is related to the already relatively small condition numbers of the initial operator $\fermiMatDouble$ presented
in \fig{fig:ExampleOfPrecMMDag1}. To answer these questions an additional example is given in 
\fig{fig:ExampleOfPrecMMDag2}. The configurations underlying this demonstration have been obtained 
in a Monte-Carlo run with $y_b/y_t=0.024$ corresponding to the physical ratio of the bottom and top quark mass. 
Moreover, the condition numbers of the initial operator $\fermiMatDouble$ are significantly larger now than in the 
previously presented examples. Still the condition numbers of the operator 
$\precR[m]\newfermiMat\precR^2[m]\newfermiMat^\dagger\precR[m]$ are improved by 1-2 
orders of magnitude as compared to the case of the original operator $\fermiMatDouble$.

%\includeFigSingleSmall{MMdagCondNrComparisonlam0kap012220Nf3}
\includeFigSingleSmall{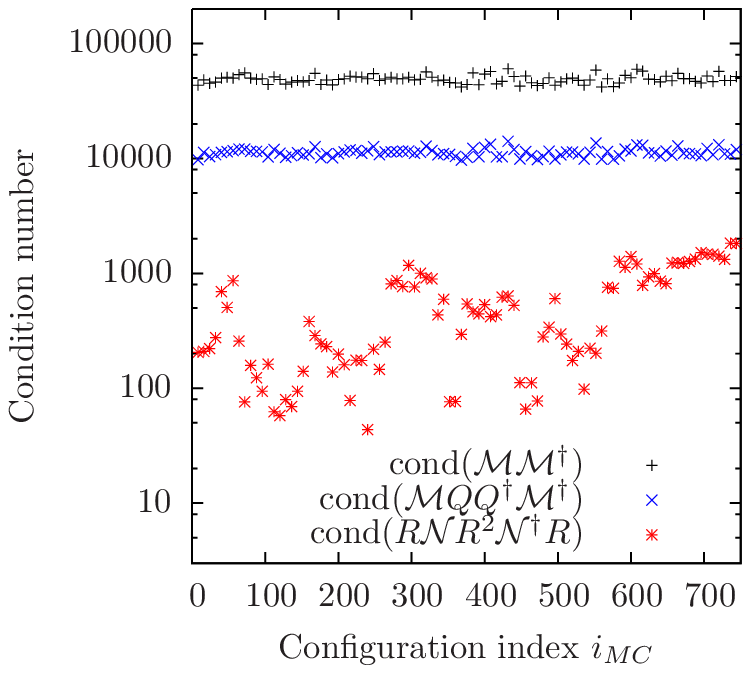}
{fig:ExampleOfPrecMMDag2}
{The condition numbers of the squared fermion matrix $\fermiMatDouble$ are compared to the corresponding
numbers of the preconditioned matrices $\fermiMat Q[m]Q^\dagger[m]\fermiMat^\dagger$ and
$R[m]\newfermiMat R^2[m]\newfermiMat^\dagger R[m]$, respectively. The underlying field configurations $\Phi$
have been generated in the non-generate Monte-Carlo run with $\kappa=0.12220$ specified in 
\tab{tab:Chap5EmployedRuns}. }
{Condition numbers of the squared fermion matrices $\fermiMatDouble$, $\fermiMat QQ^\dagger\fermiMat^\dagger$,
and $R\newfermiMat R^2\newfermiMat^\dagger R$ in the non-degenerate case.}

For the purpose of clarification let us shortly summarize the partition function we found to be
most suitable for the the numerical calculations so far. Up to some constant factor the partition
function in \eq{eq:PHMCPartitionFunction2} is equivalent to
\bea
\label{eq:PHMCPartitionFunction4}
Z_\Phi &=& \FuncIntAvg_{PR}\left[e^{iN_f\arg\det(\fermiMat\precP[m])} \cdot W[\Phi]     \right], \\
W[\Phi]&=& \det\left(\sqrt{\precR\newfermiMat\precR^2\newfermiMat^\dagger\precR}\cdot P(\precR\newfermiMat\precR^2\newfermiMat^\dagger\precR)  \right),
\eea
with the definition of the functional integral $\FuncIntAvg_{PR}$ being
\bea
\FuncIntAvg_{PR}\left[O[\Phi]  \right] &=& \int \intD{\Phi} \intD{\omega} \intD{\omega^\dagger}\,\, 
O[\Phi] \cdot e^{- \left(S_\Phi[\Phi]+ \frac{1}{2} \sum\limits_{i=1}^{N_f}
\omega^\dagger_i P(\precR\newfermiMat\precR^2\newfermiMat^\dagger\precR)\omega_i
\right)}, \quad\quad
\eea
where the preconditioning matrix $\precR\equiv\precR[m_\Phi]$ is held constant over the whole Monte-Carlo
run and $m_\Phi$ is intended to be fixed to $m_\Phi=\langle m \rangle$. This approach produces exactly
the same dynamics as \eq{eq:PHMCPartitionFunction2} does, but is more favorable from a numerical point 
of view due to the much smaller condition number of the operator $\precR\newfermiMat\precR^2\newfermiMat^\dagger\precR$ 
as compared to $\fermiMatDouble$. 
  
For completeness it is remarked that the actually relevant quantity determining the necessary size of the underlying 
approximation interval in a PHMC algorithm is not the observed maximal value of the condition number itself, but rather
the ratio of the largest and smallest eigenvalues observed in the whole Markov process given as
\bea
\label{eq:DefOfGammaQuant}
\gamma_{max} = \frac{\max\, \left| \nu \right|}{\min\, \left| \nu \right|},
\eea
where the respective maximizations and minimizations are performed over all generated field configurations $\Phi$ and
over all eigenvalues $\nu$ of the underlying fermion operator. This measure $\gamma_{max}$
generally differs from the observed maximal value of the condition number, since the largest eigenvalue $\nu_{max}$
associated to a given field configuration and a given fermion operator fluctuates during the Monte-Carlo process. 
The results on $\gamma_{max}$ for the
original and the preconditioned fermion operators as obtained in the same example Monte-Carlo runs already considered before 
are listed in \tab{tab:ExamplesOoofApproxPoly}. From this presentation one can also learn that some fraction of the 
gained improvement of the condition number $\cond(\precR[m]\newfermiMat\precR^2[m]\newfermiMat^\dagger\precR[m])$ 
as compared to $\cond(\fermiMatDouble)$ is lost by the requirement that the preconditioning parameter must be fixed.

However, the relevant quantity $\gamma_{max}$ of the operator 
$\precR[m_\Phi]\newfermiMat\precR^2[m_\Phi]\newfermiMat^\dagger\precR[m_\Phi]$ is still reduced by a typically factor 
of around $20-30$ in comparison to the corresponding finding of the original matrix $\fermiMatDouble$ as 
can be seen in \tab{tab:ExamplesOoofApproxPoly}.

\includeTabNoHLines{|c|c|r|c|c|c|}{
\cline{4-6}
\multicolumn{3}{c|}{} & \multicolumn{3}{c|}{Example setup} \\ \hline
                  & Operator                                                      &  $\gamma_{max}$   & $\lambda_P / \epsilon_P$ &   $N_P$    &     $\delta_P$ \\ \hline
$\kappa=0.12301$  & $\fermiMatDouble$                                             &  618.1            & $6.2\cdot 10^{2 }$       &   152      &  $8.6\cdot 10^{-6}$ \\
$\kappa=0.12301$  & $R[m]\newfermiMat R^2[m] \newfermiMat^\dagger R[m]$                 &  13.9             &       --                 &   --       &  --             \\ 
$\kappa=0.12301$  & $R[m_\Phi]\newfermiMat R^2[m_\Phi] \newfermiMat^\dagger R[m_\Phi]$  &  25.1             & $2.5\cdot 10^{1 }$       &   30       &  $5.7\cdot 10^{-6}$             \\ \hline 
$\kappa=0.29929$  & $\fermiMatDouble$                                             &  2561.8           & $2.6\cdot 10^{3 }$       &   310      &  $9.7\cdot 10^{-6}$             \\
$\kappa=0.29929$  & $R[m]\newfermiMat R^2[m] \newfermiMat^\dagger R[m]$                 &  67.0             &       --                 &   --       &  --             \\ 
$\kappa=0.29929$  & $R[m_\Phi]\newfermiMat R^2[m_\Phi] \newfermiMat^\dagger R[m_\Phi]$  &  121.4            & $1.2\cdot 10^{2 }$       &   66       &  $8.4\cdot 10^{-6}$             \\ \hline 
$\kappa=0.12220$  & $\fermiMatDouble$                                             &  619994.6         & $6.2\cdot 10^{5 }$       &   1520     &  $9.6\cdot 10^{-6}$             \\
$\kappa=0.12220$  & $R[m]\newfermiMat R^2[m] \newfermiMat^\dagger R[m]$                 &  1992.8           &       --                 &   --       &  --             \\ 
$\kappa=0.12220$  & $R[m_\Phi]\newfermiMat R^2[m_\Phi] \newfermiMat^\dagger R[m_\Phi]$  &  2268.4           & $2.3\cdot 10^{3 }$       &   292      &  $9.4\cdot 10^{-6}$             \\ \hline 
}
{tab:ExamplesOoofApproxPoly}
{The quantity $\gamma_{max}$ defined in \eq{eq:DefOfGammaQuant} is presented for the operators $\fermiMatDouble$,
$R[m]\newfermiMat R^2[m] \newfermiMat^\dagger R[m]$, and $R[m_\Phi]\newfermiMat R^2[m_\Phi] \newfermiMat^\dagger R[m_\Phi]$ 
as obtained in the Monte-Carlo runs specified in \tab{tab:Chap5EmployedRuns} for the given values of $\kappa$. 
The preconditioning parameter $m_\Phi$ was chosen at the start of the thermalization phase of each Markov process
with the intention to match the respectively resulting average magnetization $\langle m \rangle$. 
Moreover, example setups for typical approximation polynomials are given, covering the whole observed eigenvalue spectrum of
the respective operator. The chosen polynomial degree $N_P$ is the minimal even degree that leads to $\delta_P<10^{-5}$.
The upper bound $\lambda_P$ is one in all cases.}
{Maximal eigenvalue fluctuation $\gamma_{max}$ of the preconditioned and the unpreconditioned fermion matrices.}

To illustrate the actual impact of the achieved reduction of the fermion matrix condition numbers on the 
computational cost for the evaluation of the polynomial $P(\fermiMatDouble)$, or rather 
$P(\precR\newfermiMat\precR^2\newfermiMat^\dagger\precR)$ in the improved case, a couple of example
setups for the approximation polynomial $P(x)$ typically chosen for the observed value of $\gamma_{max}$
are presented in \tab{tab:ExamplesOoofApproxPoly}. In these example setups the ratio $\lambda_P/\epsilon_P$
of the approximation interval boundaries was chosen such that the whole observed eigenvalue spectrum is
covered. The presented value of the polynomial degree $N_P$ is given here as the minimal even degree fulfilling
the criterion $\delta_P<10^{-5}$ of the associated maximal relative deviation $\delta_P$, which will turn out 
to be a reasonable choice in \sect{sec:ExactReweighing} still leading to an exact algorithm.

One can observe in the presented listing that the polynomial degree $N_P$, specified in the above manner, decreases
quickly when the approximation interval is reduced while the requested maximal relative deviation $\delta_P$ is held constant.
From the presented numbers one finds that the achieved reduction of the quantity $\gamma_{max}$ typically translates 
into a decrease of the required polynomial degree $N_P$ by a factor of around $5$. Since the computation of the 
matrix-valued polynomial $P(\fermiMatDouble)$ consumes the major portion of the numerical resources, this achieved 
reduction of the polynomial degree turns, for the most part, into an overall improvement factor of the whole Monte-Carlo 
calculation. 

Concerning the applicability of the here considered preconditioning technique also in the case of QCD one faces the problem 
that a pendant of the here exploited correlation between the eigenvectors of the general fermion matrix $\fermiMat[\Phi]$ 
and those of $\fermiMat[\Phi']$ does apparently not exist in the case of QCD due to its local gauge invariance. The presented 
approach can therefore not directly be applied to the case of QCD, unless one succeeds to establish the aforementioned correlation, 
for instance, by means of gauge fixing procedures. While such an approach does not seem to be practical for the Markov process 
of generating the gauge field configurations itself, it might be interesting for the analysis of the latter field configurations 
with respect to the calculation of the fermion propagators, which requires a large number of fermion matrix inversions for each single 
gauge field configuration but only one gauge fixing procedure.

%-----------------------------------------------------------------------------------------------------
\section{Reducing auto-correlation times through Fourier acceleration}
\label{sec:FACC}

The auto-correlation time $\ACtime$ is a measure for the statistical correlation of 
the configurations generated in a Monte-Carlo process. It can roughly be interpreted as the 
number of subsequent configurations, which have to be considered as statistically dependent. 
More precisely, the auto-correlation time as defined in the following is not a direct property of the
field configurations alone but depends also on the choice of the considered observable $O[\Phi]$.
The true performance of a simulation algorithm in the sense of achieved accuracy of the numerical
result on $\langle O \rangle$ per computation time is thus not only determined by the run-time speed itself, 
\ie by the number of configurations generated in a given computation time, but also by the auto-correlation 
time $\ACtime$ of the considered observable.
  
This section intends to demonstrate how the method of Fourier acceleration\footnote{Special thanks go to
Prof. Julius Kuti and to Chris Schroeder for pointing out the Fourier acceleration technique to me.
Its applicability to a similar Higgs-Yukawa model has moreover been demonstrated in \Ref{Schroeder:2007zu}.
}~\cite{Batrouni:1985jn,Catterall:2001jg} 
can be adopted to the considered Higgs-Yukawa model in order to reduce the encountered auto-correlation times of the considered
observables, thus increasing the overall performance of the implemented simulation algorithm. For that purpose 
we start here with a more precise definition of the auto-correlation time $\ACtime$.
    
As discussed in \sect{sec:HMCAlgorithm} the general strategy for calculating the expectation value 
$\langle O \rangle$ of an observable $O$ in the Monte-Carlo approach is to estimate that quantity
by the ensemble average $\bar O_N$ obtained from a given finite set of $N$ field configurations according to 
\eqs{eq:DefOfExpValueInMarrrkovChain}{eq:DefOfSampleValueInMarrrkovChain}. 
In the case of statistically independent samples the variance $\sigma^2_{\bar O_N}$ of $\bar O_N$
would be given by $\sigma^2_{O}/N$, where $\sigma_{O}$ denotes the standard deviation of the single measurements $O_n$,
$n=1,\ldots, N$. In practice, however, the field configurations generated in the Monte-Carlo process, and thus the 
corresponding measurements $O_n$, are highly correlated, unless some method of direct sampling was used. For the error 
analysis it is therefore of great importance to consider the correlation among the obtained measurements $O_n$ to 
obtain a reliable error estimate for the ensemble average $\bar O_N$, and thus for the estimate of $\langle O \rangle$.

Starting from the definition of the variance $\sigma^2_{\bar O_N}$ of the ensemble average $\bar O_N$ being
\beq
\sigma_{\bar O_N}^2=  \left\langle  \left[ \bar O_N - \langle O \rangle \right]^2    \right\rangle
\eeq
where the assumed unbiasedness of the estimate of $\langle O \rangle$ through $\bar O_N$  has already been used, 
one directly arrives at the result that the variance $\sigma^2_{\bar O_N}$ still scales proportional to $1/N$ but 
with a proportionality constant $C(\infty)$ instead of $\sigma^2_{O}$ according to
\bea
\label{eq:DefGammaStrat1}
\sigma_{\bar O_N}^2&=&\frac{C(\infty)}{N}, \quad \quad  C(W) = \sum\limits_{t=-W}^W \Gamma(t), \quad \mbox{ with }\\
\label{eq:DefGammaStrat3}
\Gamma(t) &=& \frac{1}{N-|t|} \sum\limits_{{1\le n \le N} \atop {1\le n+t \le N}} \Big\langle \left[ O_n-\langle O\rangle  \right] \cdot 
\left[ O_{n+t}-\langle O\rangle  \right]\Big\rangle,
\eea
where terms of the order $O(\ACtime/N)$ have been neglected and the actually finite summation in \eq{eq:DefGammaStrat1}
has been extended to an infinite sum, which is well justified, since one usually has $N\gg\ACtime$. Moreover, the definition
in \eq{eq:DefGammaStrat3} only makes sense for $|t|< N$ which, however, is an irrelevant restriction in practice for the same
reason. It is further remarked that one finds $\Gamma(t)=0,\, \forall t\neq 0$ in the case of complete statistical independence 
and thus $C(\infty) = \Gamma(0) \equiv \sigma^2_O$ recovering the prior result. 

In a practical determination of $C(\infty)$, however, all expectation values in \eq{eq:DefGammaStrat3} 
have to be replaced by corresponding ensemble averages. Moreover, the window size $W$, in which the function $\Gamma(t)$ is to be summed up,
typically has to be set to a finite value $W\ll N$, since the determination of the correlation function $\Gamma(t)$ becomes
statistically unstable at large separations $t$. Choosing an appropriate value for $W$ is thus a trade-off between the latter
statistical uncertainty on the one hand and the systematic neglection of long-ranged correlations on the other hand. For a 
reliable determination of $C(\infty)$ one usually requests the observation of a plateau in the function $C(W)$ and takes its plateau 
value as an approximation for $C(\infty)$. This whole procedure is well known as the $\Gamma$-strategy~\cite{Wolff:2003sm}
in the literature.

The auto-correlation time is then defined through the decay rate of the expected exponential behaviour of
the correlation function $\Gamma(t)$ according to 
\beq
\Gamma(t) \propto \exp\left({-\frac{|t|}{\ACtime}}\right).
\eeq
It is thus directly connected to $C(\infty)$ via
\beq
\ACtime = \frac{1}{\log\left(1+2\frac{\Gamma(0)}{C(\infty)-\Gamma(0)}   \right)}
\eeq
which simplifies in the case of $C(\infty)/\Gamma(0)\gg 1$ to
\beq
\ACtime = \frac{C(\infty)}{2\Gamma(0)}.
\eeq

A typical example for an observable with very large auto-correlation time is presented in \fig{fig:ExampleForAutoCorrTime1}a.
Here, the magnetization $m$ has been selected as the considered observable. The corresponding auto-correlation time, expressed 
in terms of the function $C(W)$, is depicted in \fig{fig:ExampleForAutoCorrTime1}b. The presented data have been obtained in a PHMC 
calculation without applying any of the enhancement procedures which will be discussed in this section. For the given example
one finds the auto-correlation time of the underlying observable $m$ to be of order $O(100)$  which is a typical value for 
the chosen observable and the selected lattice volume. Such a large auto-correlation time obviously poses a 
severe difficulty for the reliable calculation of the expectation value of this observable. 

%\includeFigTriple{ObsMvsMCtimeLam0Kap012313Unimp}{ObsMAutoCorrCWLam0Kap012313Unimp}{PHMCForceLam0Kap012313}
\includeFigTriple{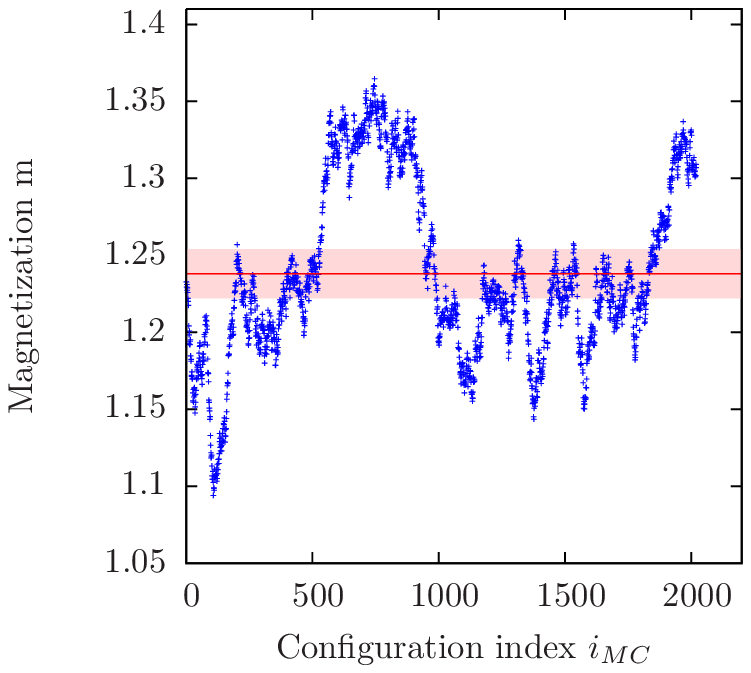}{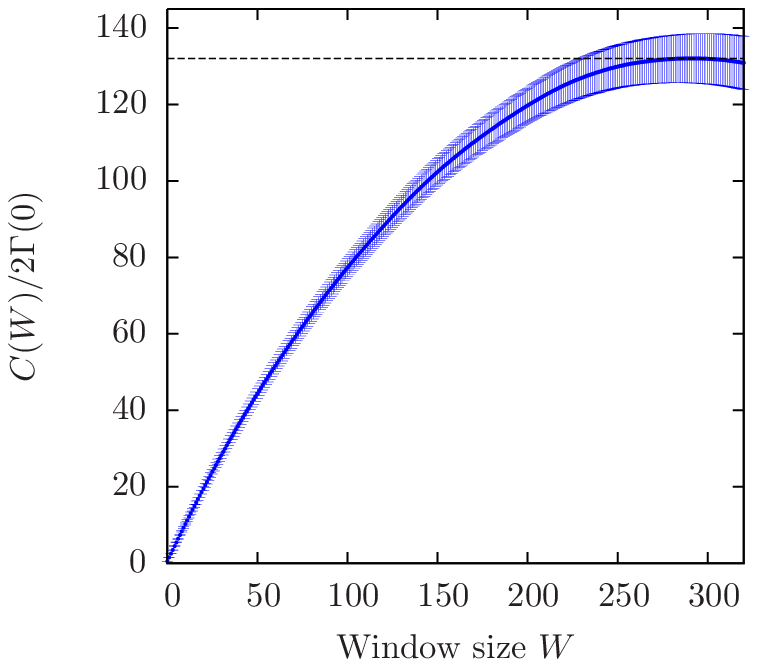}{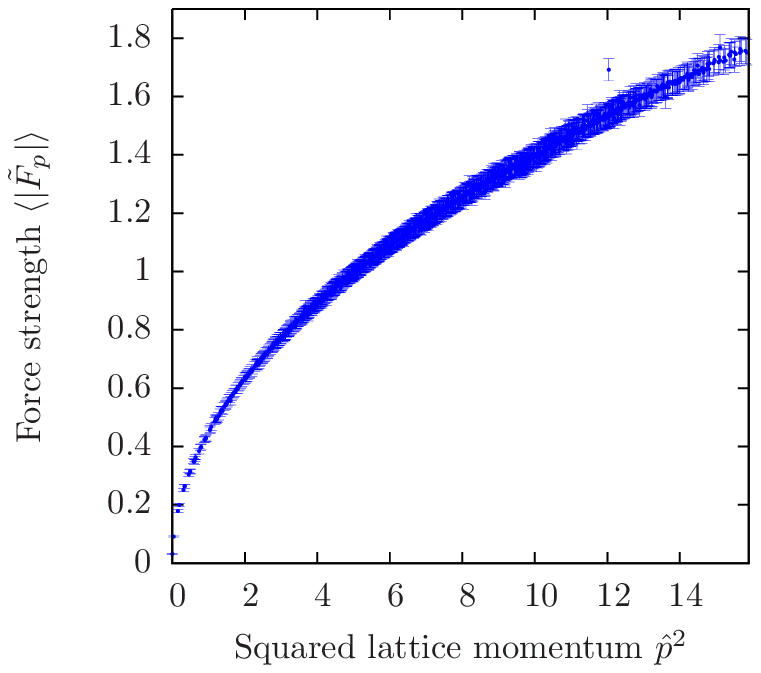}
{fig:ExampleForAutoCorrTime1}
{The numerical value of the magnetization $m$ is plotted in panel (a) versus the configuration index
$\confNR$ of the underlying field configuration $\Phi$. The horizontal line depicts the obtained
expectation value $\langle m \rangle$ and the highlighted band indicates its associated statistical uncertainty. 
Panel (b) shows the resulting function $C(W)/2\Gamma(0)$ versus the window size $W$, while the dashed line marks
its plateau value. The average force strength $\langle| \tilde F_p|\rangle$
in momentum space is plotted in panel (c) versus the squared lattice momentum $\hat p^2$. The presented
findings have been obtained in the Monte-Carlo run with $\kappa=0.12313$ specified in \tab{tab:Chap5EmployedRuns}
without employing the technique of Fourier acceleration.
}
{Example of the auto-correlation observed without employing Fourier acceleration.}

It is well known that observables dominated by low momentum modes generally suffer the most from large 
auto-correlation times, as can be formally shown at least for the free theory~\cite{Buendia:1989rj}. The 
auto-correlation time of the observable $m$, which purely reflects  
the amplitudes of the zero-momentum modes of the scalar field $\Phi$, can therefore be considered as a good indicator 
for the worst observable auto-correlation time in a given Monte-Carlo run. This is why we
will mainly consider the magnetization $m$ as the underlying observable for the analysis of the auto-correlation time
in the following. 

The origin of these extremely bad auto-correlation properties can be observed in \fig{fig:ExampleForAutoCorrTime1}c.
This plot shows the numerically calculated average molecular force $\langle |\tilde F_p[\Phi,\omega]| \rangle$ in momentum space defined as
\bea
\label{eq:DefOfForceMolDynF}
\tilde F_p[\Phi,\omega]  = \frac{1}{\sqrt{V}} \sum\limits_x e^{-ipx} \cdot F_x[\Phi,\omega]
\eea
versus the squared lattice momentum of the respective mode it is acting on. Interestingly, a very 
pronounced relation between $\hat p^2$ and the associated force is observed. The long auto-correlation times of the 
magnetization $m$ can directly be ascribed to the minimum of the measured average forces being located at $\hat p^2=0$. 
The much larger forces acting on the higher momentum modes, on the other hand, are the main sources for the discretization 
errors encountered in the numerical integration of the molecular dynamics equations of motion. The step size $\intStepSize$ 
of the numerical integrator has thus to be adjusted to sufficiently small values to allow for a stable integration of these
high-momentum modes, while the very same modes do not contribute to the considered observable $m$ at all. 
In fact, this is the general situation, which can, however, be more 
or less pronounced depending on the considered observable. The reasoning is that the low momentum modes determine
the long-range physics, which we are mainly interested in when performing the later lattice calculations. These low momentum modes, however, 
evolve the slowest in the unimproved Monte-Carlo algorithm considered so far.

A well-known cure to the described problem is the {\textit{Fourier-Acceleration}} technique~\cite{Batrouni:1985jn,Catterall:2001jg}. 
Its main idea is to assign different effective trajectory lengths to the momentum modes, such that
the low momentum modes see an enlengthened effective trajectory length during the integration of the equations
of motion. For that purpose a fictitious mass $m(p)$ is introduced for each momentum mode,
in such a way that those modes with smaller assigned masses move faster during the integration process. 
This can be achieved by replacing the standard setting for the function $f(\pi)=\pi^\dagger\pi/2$, which determines 
the sampling of the conjugate momenta $\pi$ according to \eq{eq:ProbDisDesityOfConjMom}, with some more appropriate choice 
incorporating the latter fictitious masses $m(p)$ according to
\beq
\label{eq:FourrierAcceleratedSamplingFunction1}
f^{(F)}(\pi) = \sum\limits_{p\in\ImpSpace} \frac{\tilde \pi_p^\dagger \tilde\pi_p}{2m(p)},
\eeq
where $\tilde\pi_p$ denotes the Fourier transform of the conjugate momenta $\pi_x$ given as
\beq
\tilde\pi_p = \frac{1}{\sqrt{V}}\sum\limits_x e^{-ipx} \cdot \pi_x.
\eeq
Obviously, the introduction of the aforementioned masses does not influence the physical results obtained
with this new choice of the function $f(\pi)$ as discussed in \sect{sec:HMCAlgorithm}. To understand the effect of 
these fictitious masses let us consider a solution $\xi^{(F)}(\MCtime)\equiv(\Phi^{(F)}(\MCtime), \pi^{(F)}(\MCtime), \omega^{(F)}(\MCtime))$ 
of the Fourier-accelerated equations of motion associated to the function $f^{(F)}(\pi)$ according to \eqs{eq:HMCEqOfMotion1}{eq:HMCEqOfMotion3}. 
This solution fulfills
\bea
\label{eq:FACCeomPhi}
\frac{d}{d\MCtime} \tilde\Phi^{(F)}_p(\MCtime) &=& \frac{\tilde\pi^{(F)}_p(\MCtime)}{m(p)}, \\
\label{eq:FACCeomMomenta}
\frac{d}{d\MCtime} \tilde\pi^{(F)}_p(\MCtime)  &=& - \tilde F_p[\Phi^{(F)}(\MCtime)],
\eea
where the here omitted pseudo fermion fields $\omega^{(F)}(\MCtime)$ are again assumed not to propagate during the integration step.
We now define the rescaled momenta $\tilde\pi^{(F,r)}_p\fhs{-2.0mm} =\fhs{-0.75mm} \tilde\pi^{(F)}_p/\sqrt{m(p)}$, such that
$\tilde\pi^{(F,r)}_p$ is Gauss-distributed with standard deviation one as it is the case for the original
momenta $\tilde\pi_p$ sampled according to the standard choice $f(\pi)=\pi^\dagger\pi/2$. The trajectory 
$\xi^{(F,r)}(\MCtime)\equiv(\Phi^{(F)}(\MCtime), \pi^{(F,r)}(\MCtime), \omega^{(F)}(\MCtime))$ then solves the differential
equations\footnote{For the sake of brevity the inner mappings $\Phi^{(F)}(\MCtime^p)\equiv \Phi^{(F)}(\MCtime(\MCtime^p))$ 
and $\pi^{(F,r)}(\MCtime^p) \equiv \pi^{(F,r)}(\MCtime(\MCtime^p))$ are implicit here.}
\bea
\label{eq:FACCeomPhiRes}
\frac{d}{d\MCtime^p} \tilde\Phi^{(F)}_p(\MCtime^p) &=& \tilde\pi^{(F,r)}_p(\MCtime^p), \\
\label{eq:FACCeomMomentaRes}
\frac{d}{d\MCtime^p} \tilde\pi^{(F,r)}_p(\MCtime^p)  &=& -\tilde F_p[\Phi^{(F)}(\MCtime^p)],
\eea
which coincide with the original equations of motion in the unimproved scenario given in \eqs{eq:HMCEqOfMotion1}{eq:HMCEqOfMotion3} 
differing only through the replacement $\MCtime \rightarrow \MCtime^p = \MCtime / \sqrt{m(p)}$. Integrating \eqs{eq:FACCeomPhi}{eq:FACCeomMomenta}
from $\MCtime=0$ to $\MCtime=\traLength$, which is actually done in the simulation algorithm, would thus formally correspond to an 
integration of the momentum modes in \eqs{eq:FACCeomPhiRes}{eq:FACCeomMomentaRes} from $\MCtime^p=0$ to $\MCtime^p=\traLength/\sqrt{m(p)}$, 
respectively, explaining the notion of the momentum dependent, effective trajectory length
\beq
\traLength^p = \traLength / \sqrt{m(p)}.
\eeq

The vital question is how these effective trajectory lengths and thus the fictitious masses should be chosen.
Here, it is intended to determine $m(p)$ such that the momenta $\tilde\pi_p$, resulting from the Gaussian 
sampling according to the distribution $f^{(F)}(\pi)$ in \eq{eq:FourrierAcceleratedSamplingFunction1}, are of 
the same scale as the corresponding forces $\tilde F_p$, which would be a 'natural' situation. We thus choose 
the fictitious masses as
\beq
\label{eq:SettingOfFicMasses}
m(p) = \frac{1}{4} \left\langle \left| \tilde F_p[\Phi,\omega]\right|  \right\rangle^2,
\eeq
where the specified factor $1/4$ is purely conventional, translating into an overall, momentum independent boost of the
effective trajectory lengths only. In a practical lattice calculation the average forces $\langle |\tilde F_p[\Phi,\omega]|\rangle$
can simply be measured in the early phase of the thermalization process of the
Monte-Carlo run before the Fourier acceleration is finally switched\footnote{Switching the Fourier acceleration on and off 
simply refers to setting the momentum masses to the setting in \eq{eq:SettingOfFicMasses} or to $m(p)=1$, recovering then
the usual, unimproved equations of motion.} on. For stability reasons one
should, however, define a maximal effective trajectory length that will not be exceeded. For the most performed lattice
calculations in this work the relative prolongation of the effective trajectory lengths with respect to the nominal value $\traLength$
has been restricted to $\traLength^p/\traLength \le 12$.

The effect of the presented Fourier acceleration technique shall now tested. For that purpose the Monte-Carlo run that produced 
the results presented in \fig{fig:ExampleForAutoCorrTime1} was repeated with the same parameter settings, but activated
Fourier acceleration technique, \ie with the choice of the function $f(\pi)$ according to \eq{eq:FourrierAcceleratedSamplingFunction1}
and employing the corresponding equations of motion listed in \eqs{eq:FACCeomPhi}{eq:FACCeomMomenta}. The obtained results for the
measured magnetizations are presented in \fig{fig:ExampleForAutoCorrTime2}a while panel (b) shows the associated auto-correlation time 
of that observable in terms of the function $C(W)$ resulting from this improved approach. The underlying choice of the momentum 
dependent, effective trajectory lengths $\traLength^p$ in relation to the nominal value $\traLength$ is shown in \fig{fig:ExampleForAutoCorrTime2}c.
One finds from these results that the benefit of the Fourier acceleration technique is excellent
for the given example. The auto-correlation time could be reduced by a factor of the order $O(100)$, which is a typical
finding for the given lattice volume in the considered regime of small Yukawa and quartic coupling constants.

It is remarked at this point that the Fourier acceleration technique is also applicable in the case of an infinite 
quartic self-coupling constant. In this scenario the field variables $\Phi_x$ are each constraint to a sphere. Again, 
effective trajectory lengths can be assigned to the different momentum modes by the introduction of 
fictitious masses $m(p)$ according to \eq{eq:FourrierAcceleratedSamplingFunction1} with the only difference that in this 
case the conjugate momenta $\pi$ possess only three components $\pi^i$, $i=1,2,3$.
A possible choice for the equations of motions leaving the action $S[\Phi,\pi,\omega]$ invariant is then given by
\bea
\label{eq:FACCLamInfinityPhiEqOfMotion}
\frac{d(\Phi_x\theta)}{d\MCtime} &=& \beta_x\theta \cdot \left(\Phi_x\theta\right), \quad \mbox{with}\quad 
\beta_x = \frac{1}{\sqrt{V}} \sum\limits_{p\in\ImpSpace} e^{ipx} \frac{\tilde\pi_p}{m(p)}, \quad \beta^0 \equiv 0, \\
\label{eq:FACCLamInfinityPEqOfMotion}
\frac{d(\pi_x\theta)}{d\MCtime} &=& -\left(F_x[\Phi,\omega]\theta\right) \cdot \left(\Phi_x\theta \right)^\dagger,
\eea
where the same notation as in \sect{sec:HMCAlgorithm} has been used. Again the movement in the would-be component $\pi_x^0$
is implicitly projected out in \eq{eq:FACCLamInfinityPEqOfMotion} and the given differential equation in \eq{eq:FACCLamInfinityPhiEqOfMotion} 
can be solved exactly for constant $\beta_x$ leading then to the finite step size result
\beq
\Phi_x(\MCtime+\intStepSize)\theta = e^{\intStepSize \beta_x\theta} \cdot (\Phi_x(\MCtime)\theta)
\eeq
which is applied in the actual numerical implementation to fulfill the constraint on the field $\Phi$
exactly. The construction of a suitable numerical integration scheme is then fully analogue to that presented in \sect{sec:HMCAlgorithm}.

%\includeFigTriple{ObsMvsMCtimeLam0Kap012313Imp}{ObsMAutoCorrCWLam0Kap012313Imp}{EffectiveTraLengthLam0Kap012313}
\includeFigTriple{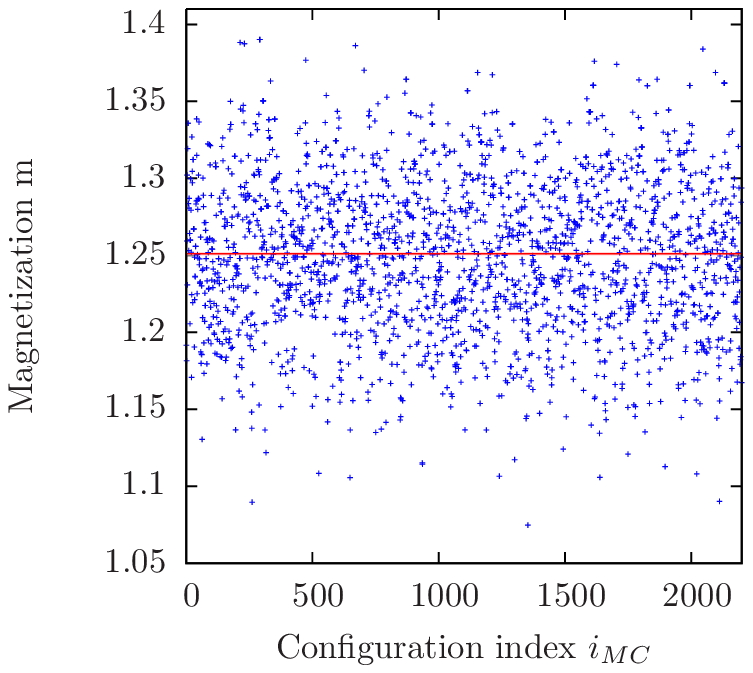}{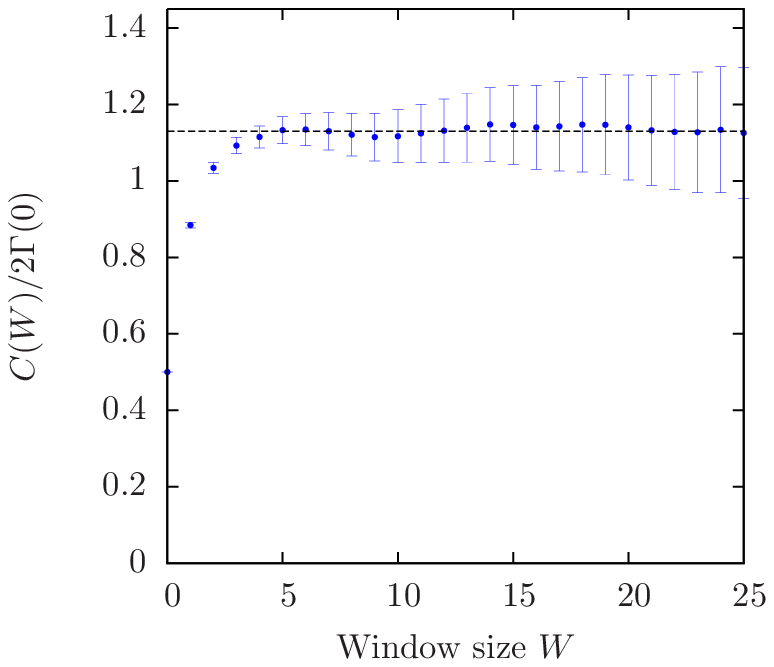}{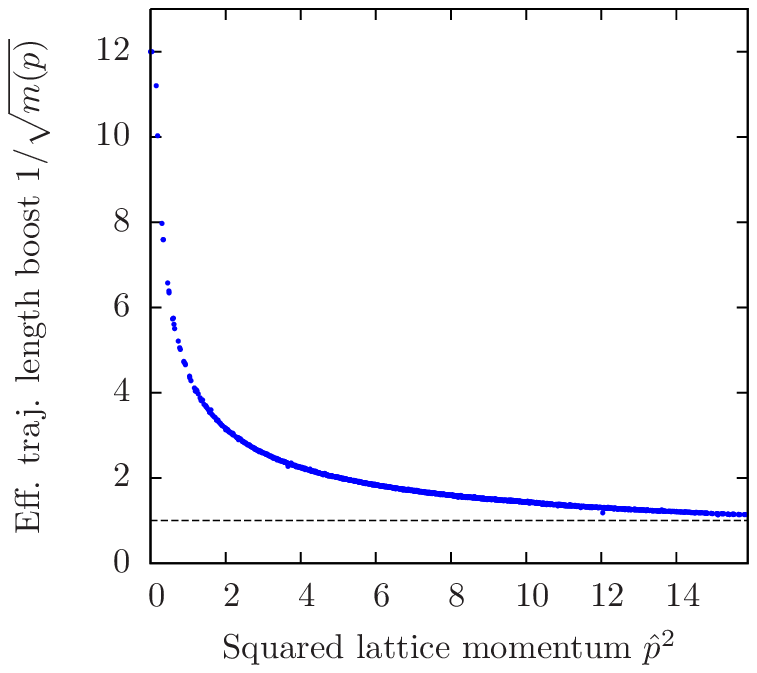}
{fig:ExampleForAutoCorrTime2}
{The numerical value of the magnetization $m$ is plotted in panel (a) versus the configuration index
$\confNR$ of the underlying field configuration $\Phi$. The  horizontal line depicts the obtained
expectation value $\langle m \rangle$ while the graphical presentation of its associated statistical uncertainty 
is hidden below that line and thus invisible in this plot.
Panel (b) shows the resulting function $C(W)/2\Gamma(0)$ versus the window size $W$, while the dashed line marks
its plateau value. The boost factor $1/\sqrt{m(p)}$ of the effective trajectory length $\traLength^p$ as compared
to its nominal value $\traLength$ is plotted in panel (c) versus the squared lattice momentum $\hat p^2$. 
The presented findings have been obtained in the Monte-Carlo run with $\kappa=0.12313$ specified in 
\tab{tab:Chap5EmployedRuns} employing the technique of Fourier acceleration.
}
{Example of the auto-correlation observed with employing Fourier acceleration.}

An example for the benefit of the Fourier acceleration technique at infinite quartic self-coupling constant is given in 
\fig{fig:ExamplePropagatorAutoCorrTimes}a, where the auto-correlation times of the magnetization $m$ obtained in two Monte-Carlo runs 
with the same parameter settings differing only through the usage or non-usage, respectively, of the presented improvement
technique are compared to each other. The auto-correlation time obtained for deactivated Fourier acceleration technique is much larger than 
that measured in the case of activated Fourier acceleration. The attainable improvement factor, 
being approximately $9.1$ in this example, is typically found to be smaller than in the prior case of small 
coupling constants. However, the achieved improvement is still crucial for the successful evaluation of the considered 
model in the regime of strong quartic coupling constants.

To provide an even clearer picture of what has been achieved the auto-correlation times of the Higgs propagator, \ie of the observable
$\tilde h_{p}\tilde h_{-p}$, are shown in \fig{fig:ExamplePropagatorAutoCorrTimes}b and \fig{fig:ExamplePropagatorAutoCorrTimes}c
versus the squared lattice momentum $\hat p^2$. These results are taken from the same Fourier-accelerated Monte-Carlo runs that have 
already served as examples before. One observes that the presented auto-correlation times are more or less independent of the momentum $p$.
Moreover, one sees that almost all auto-correlation times fluctuate around a value of $0.5$ in the given examples. Only the low momentum
modes exhibit a somewhat larger value of $\ACtime$, which is a remnant of the original situation where the zero momentum mode is clearly
the slowest mode of all. 

%\includeFigTriple{ObsMAutoCorrCWLamnanKap030400}{HiggsPropAutoCorrTimesLam0Kap012313Imp}{HiggsPropAutoCorrTimesLamnanKap030400Imp}
\includeFigTriple{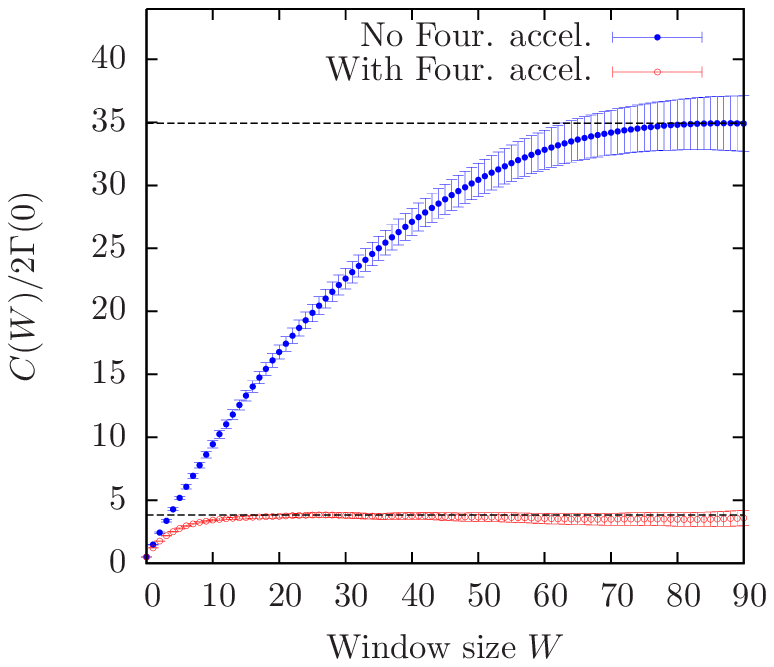}{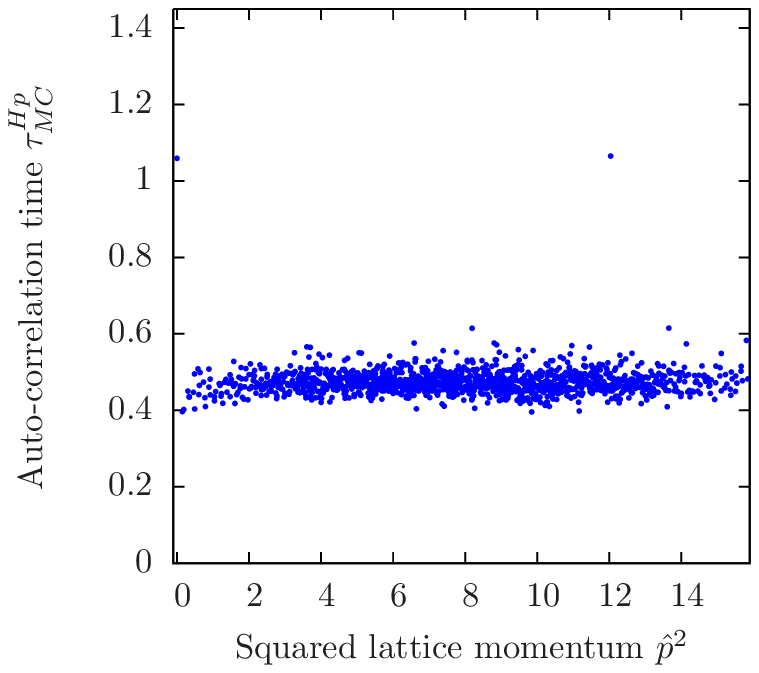}{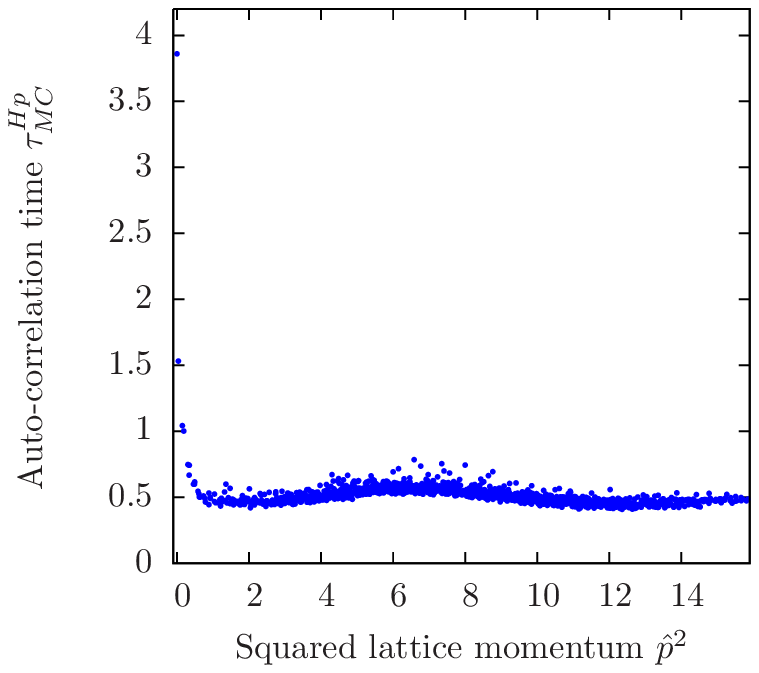}
{fig:ExamplePropagatorAutoCorrTimes}
{The functions $C(W)/2\Gamma(0)$ obtained in the Monte-Carlo runs with $\kappa=0.30400$ as specified in 
\tab{tab:Chap5EmployedRuns} with and without employing the technique of Fourier acceleration are compared
to each other in panel (a). The dashed lines indicate the corresponding plateau values. In panel (b) and
(c) the auto-correlation times $\ACtime^{Hp}$ of the Higgs propagator $\langle \tilde h_p \tilde h_{-p}\rangle$
are plotted versus the squared lattice momentum $\hat p^2$. The middle panel refers to the listed Monte-Carlo
run with $\kappa=0.12313$ and the right one to the setting $\kappa=0.30400$, both of them performed
using the technique of Fourier acceleration.
}
{Example of the effect of Fourier acceleration at infinite bare quartic coupling constants.}

For completeness it shall be remarked here that the case of finite but arbitrarily large quartic coupling constant $\hat\lambda\gg 1$
can also be treated very efficiently. In this scenario the field variables $\Phi_x$ are not exactly constraint to 
a sphere. The application of the equations of motion in \eqs{eq:FACCLamInfinityPhiEqOfMotion}{eq:FACCLamInfinityPEqOfMotion}
would thus not lead to an ergodic algorithm, while the application of \eqs{eq:FACCeomPhi}{eq:FACCeomMomenta}, on the other hand, would result 
in extremely small required step sizes $\intStepSize$ of the numerical integrator due to the almost-constraint induced by $\hat\lambda\gg 1$.
A possible solution to this problem would be to split up the dynamics of the field $\Phi$ into a tangential and a radial movement and to 
assign different fictitious masses to either of them. Here, however, an even simpler approach is implemented. For that purpose the conjugate 
momenta $\pi$ are extended to a seven-component field $\pi^0,\ldots,\pi^6$ and the Fourier accelerated version of the function $f(\pi)$ is given as
\beq
\label{eq:FourrierAcceleratedSamplingFunctionLargeLambda}
f^{(F)}(\pi) = \sum\limits_{i=0}^3 \sum\limits_{p\in\ImpSpace} \frac{ \left|\tilde \pi_p^i\right|^2}{2m_1(p)}
+\sum\limits_{i=4}^6 \sum\limits_{p\in\ImpSpace} \frac{\left|\tilde \pi_p^i\right|^2}{2m_2(p)}.
\eeq
Corresponding equations of motion can then easily be derived. The differential equations given as 
\bea
\label{eq:FACCLamInfinityPhiEqOfMotionLargeLambda}
\frac{d(\Phi_x\theta)}{d\MCtime} &=&  \alpha_x\theta  +  \beta_x\theta \cdot \left(\Phi_x\theta\right)  \\
\label{eq:FACCLamInfinityPEqOfMotionLargeLambda2}
\sum\limits_{i=0}^3 \frac{d(\pi^i_x\theta_i)}{d\MCtime} &=& -\left(F_x[\Phi,\omega]\theta\right) , \\
\label{eq:FACCLamInfinityPEqOfMotionLargeLambda3}
\sum\limits_{i=1}^3 \frac{d(\pi^{3+i}_x\theta_i)}{d\MCtime} &=& -\left(F_x[\Phi,\omega]\theta\right) \cdot \left(\Phi_x\theta \right)^\dagger, 
\eea
hold the total action $S[\Phi,\pi,\omega]$ constant and are thus chosen in this approach.
Here, the would-be $\theta_0$-movement in \eq{eq:FACCLamInfinityPEqOfMotionLargeLambda3} is again implicitly projected out and
the quantities $\alpha_x$, $\beta_x$ are defined as
\bea
\alpha^{j}_x &=& \frac{1}{\sqrt{V}} \sum\limits_{p\in\ImpSpace} e^{ipx} \frac{\tilde\pi^j_p}{m_1(p)}, \quad  j=0,\ldots, 3, \\ 
\beta^{j}_x  &=& \frac{1}{\sqrt{V}} \sum\limits_{p\in\ImpSpace} e^{ipx} \frac{\tilde\pi^{3+j}_p}{m_2(p)}, \quad  j=1,2, 3, \quad \beta^0 \equiv 0.
\eea

The momentum dependent fictitious masses $m_2(p)$ can then be determined such that the tangential movement of the field 
variables $\Phi_x$ is boosted to a desired speed, while the momenta $\pi^0,\ldots,\pi^3$ generate the missing radial 
movement, which can be adjusted to be arbitrarily slow via the masses $m_1(p)$. In the implemented approach, the tangential masses
$m_2(p)$ are chosen according to \eq{eq:SettingOfFicMasses} and the fictitious masses $m_1(p)$ are set to $m_1(p) = m_2(p)/\zeta^2$, where
$\zeta$ is a positive real number that determines the ratio of the effective trajectory lengths of the respective unconstrained movement
and the tangential movement.

The resulting algorithm is then ergodic, yields low 
auto-correlation times, and the numerical cost of the integration scheme does not diverge as the finite quartic coupling 
constant $\hat \lambda$ is sent to infinity, provided that the fictitious masses are properly adapted in that limit. It is 
this method that will later be used to evaluate the model at large quartic coupling constants, for instance at $\hat \lambda=100$, 
which would be rather problematic with the original approach.

As a technical remark it is added that one should project out the radial components of the force $F_x[\Phi,\omega]$ with respect to 
the radial direction given by $\Phi_x$ at each space-time point $x$, when determining the fictitious masses $m_2(p)$
according to \eq{eq:SettingOfFicMasses}. Otherwise these masses would be influenced by force components that actually do not act
on the conjugate momenta $\pi_x^4,\pi_x^5,\pi_x^6$. With the same rationale this remark also applies to the previously discussed 
scenario of an infinite quartic self-coupling constant. 

The alert reader might now wonder whether the reversibility of the integration scheme might be disturbed
due to positive Lyapunov exponents of the integrator~\cite{Edwards:1996vs,Liu:1997fs,Joo:2000dh} and the enlengthened
trajectory lengths. The effective trajectory lengths of the modes are, however, increased according 
to the force strength acting on that respective mode, \ie the most boosted trajectory lengths are assigned to
the originally slowest modes. Furthermore, the reversibility of the integration scheme was monitored during
the simulation runs but no worrisome changes of the reversibility properties have been observed. This monitoring 
was done by periodically performing test integrations during the simulation where an integration was followed by an 
integration with reversed momenta. This allows to interpret the difference between the final field configuration $\Phi^{fin}$ 
obtained after the two numerical integration processes and the original field configuration $\Phi^{orig}$ as a measure for 
the reversibility of the applied integration scheme. The exact definition of this reversibility measure is given as
\bea
\label{eq:DefOfReversMeas}
|\Delta \Phi| / \sqrt{V} &=&   \left(\frac{1}{{V}}  \sum\limits_x  \left|\Phi_x^{fin} - \Phi_x^{orig} \right|^2 \right)^{1/2}.
\eea
Typical examples of the latter quantity are given in \fig{fig:ExampleForReversibility} as obtained in the same Monte-Carlo 
runs that have already been considered before. This measure was evaluated for both, the Fourier accelerated as well as in the 
unimproved Monte-Carlo runs. From the presented findings one can conclude that the negative effect of the Fourier acceleration technique
on the reversibility properties of the underlying numerical integrator, though clearly detectable by means of the considered quantity, 
is still in an acceptable range from a practical point of view.

%\includeFigDouble{ReversibilityCheckLam0Kap012313}{ReversibilityCheckLamnanKap030400}
\includeFigDouble{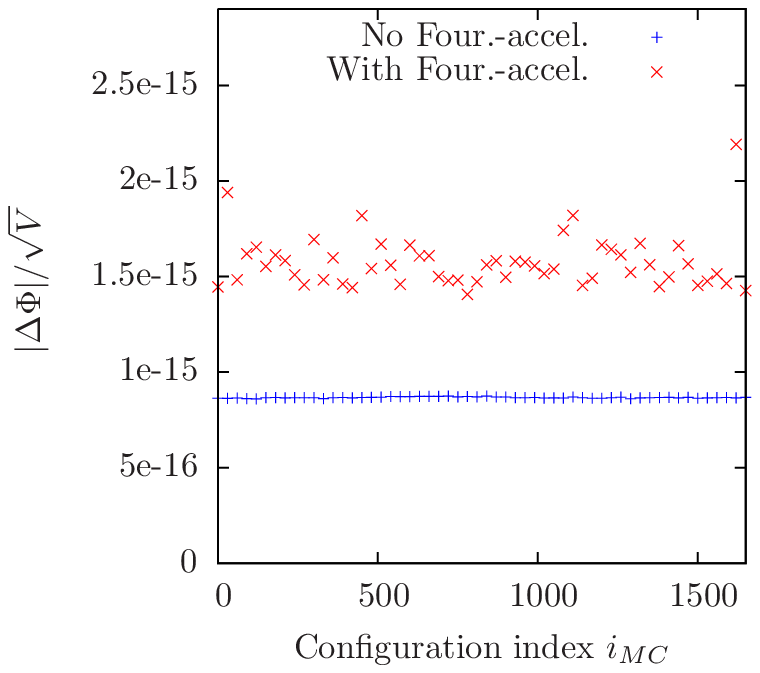}{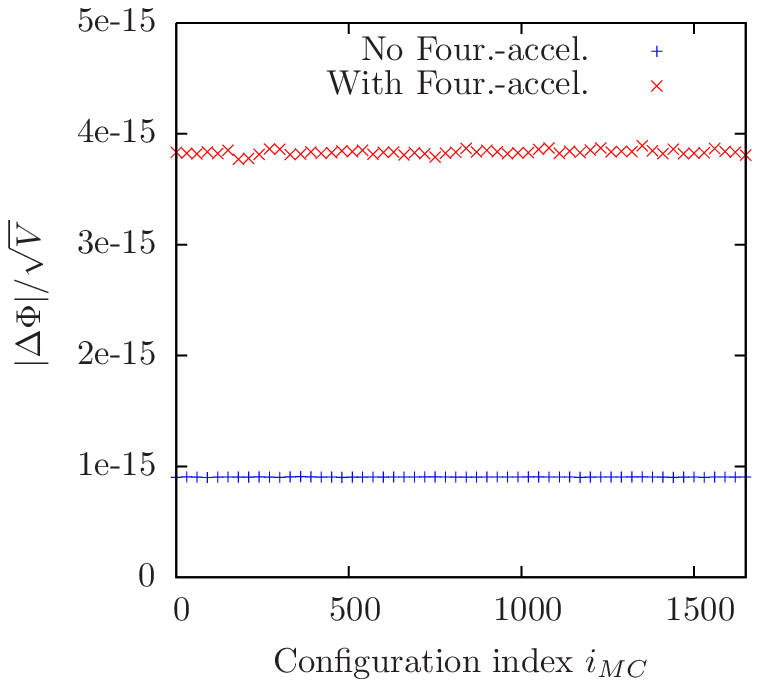}
{fig:ExampleForReversibility}
{The reversibility measure $|\Delta \Phi| / \sqrt{V}$ defined in \eq{eq:DefOfReversMeas} is plotted versus
the configuration index of the underlying field configuration, starting from which the measurement was performed
as described in the main text. The respective results in the cases of activated and deactivated Fourier
acceleration are compared to each other in each plot. The parameters of the underlying Monte-Carlo
runs are specified in \tab{tab:Chap5EmployedRuns}. Panel (a) refers to the listed parameter set with $\kappa=0.12313$
and panel (b) to the set with $\kappa=0.30400$.
}
{Example of the effect of Fourier acceleration on the reversibility properties of the algorithm.}

Furthermore, one may ask whether the achieved improvement in the auto-correlation time might be eaten up by
a reduced acceptance rate in the Metropolis accept/reject step caused by the enlengthened effective trajectory 
lengths $\traLength^p$. Indeed, the acceptance rate is affected and a smaller step size is required in the integration 
scheme to compensate for that. The benefit of the reduced auto-correlation, however, clearly overweighs
this drawback as demonstrated in \tab{tab:SummaryOfFacc}, showing the measure $\Upsilon$ for the numerical costs of the 
numerical integration as observed in a series of Monte-Carlo runs performed with and without the use of Fourier acceleration.
The aforementioned measure $\Upsilon$ is defined here as
\bea
\label{eq:DefOfNumCost1}
\Upsilon = \frac{C(\infty)}{\bar p_{acc}} \cdot N_{Force},
\eea
where $N_{Force}$ denotes the number of force evaluations required for the integration of one trajectory and the average
acceptance rate $\bar p_{acc}$ was tuned to be the smallest adjustable value above $80\%$ by varying the number of integration
steps $\intSteps$ for a given trajectory length $\traLength$.

In fact, one learns from the listing in \tab{tab:SummaryOfFacc} that the increase of the global trajectory length $\traLength$ 
is generally advantageous, also independently of the framework of Fourier acceleration, provided that the reversibility of the 
integrator is not lost, which has also been observed in the case of QCD~\cite{Meyer:2006ty}. This can be understood, at least for the here 
considered Higgs-Yukawa model, by considering the (P)HMC-update of the field $\Phi$ as a random walk in the 
high-dimensional configuration space. In this scenario the field configuration would have propagated the distance $D=\sqrt{N}\cdot d$ 
in configuration space after $N$ update steps with $d\propto \traLength$ denoting the average size of a single update step, \ie the update 
generated by a single integration of the equations of motion from $\MCtime=0$ to $\MCtime=\traLength$. If one interprets the 
auto-correlation time as the number of update steps necessary to propagate from one field configuration to the next {\textit{statistically
independent}} configuration, where statistical independence shall here be understood as a sufficiently large distance 
between the latter configurations in configuration space, then one would expect the auto-correlation time to scale proportional 
to $1/d^2$ and thus
\beq
\label{eq:ScalingOfACtime}
\ACtime \propto \frac{1}{\traLength^2}.
\eeq
The necessary decrease of the integrator step size $\intStepSize$ accompanying an enlengthening of the trajectory length, 
however, scales only with $\intStepSize\propto \traLength^{-1/r}$ where $r$ is the order of the selected integration scheme,
\ie $r=2$ for the leap-frog scheme. The pure cost of the integration scheme, neglecting the varying auto-correlation time,
thus scales proportional to $\traLength^{1+1/r}$, leading to the scaling behaviour $\Upsilon \propto \traLength^{1/r - 1}$
for the total numerical cost measure $\Upsilon$ including the effect on the auto-correlation time. This explains why it is 
expected to be advantageous to increase the global trajectory length under the aforementioned conditions, assuming that
$r>1$.

Examples for the actually observed scaling behaviour of the auto-correlation time $\ACtime$ and the numerical 
cost measure $\Upsilon$ with respect to the global trajectory length are presented in \tab{tab:SummaryOfFacc}. 
One finds that the auto-correlation time $\ACtime$ approximately scales in agreement with the estimate in \eq{eq:ScalingOfACtime}
for the unimproved Monte-Carlo runs. The numerical cost measures given in this listing also behave roughly as
expected.

In fact, the presented Fourier acceleration technique is based on these observations. In contrast to a naive increase 
of the global trajectory length $\traLength$, however, the Fourier acceleration technique influences the effective
trajectory lengths $\traLength^p$ individually, in dependence on the original speed of the respective modes, 
minimizing thus the associated negative effects on the numerical costs and the numerical stability. The Fourier 
acceleration technique is thus superior to the naive increase of the global trajectory length as can be seen 
in \tab{tab:SummaryOfFacc}. It is therefore the method of choice in the final implementation of the PHMC algorithm.

\includeTab{|c|c|c|c|c|c|c|}{
                   & FACC   & $\traLength$  & $\Nconf$   & $\langle m \rangle$  & $\ACtime$                        & $\Upsilon$ \\ \hline
$\kappa=0.12313$   & No     & 1.0           &  20120 &    $1.262\pm 0.010$              &    $481.3\pm 9.2$     &  $4839\pm 92$      \\   % 1 x 20015 / 20120   C=962.676 +-18.328
$\kappa=0.12313$   & No     & 2.0           &  2020  &    $1.238\pm 0.021$              &    $132.1\pm 6.4$     &  $2662\pm 129$   \\  % 2 x 2005/2020 
$\kappa=0.12313$   & No     & 4.0           &  2120  &    $1.257\pm 0.010$              &    $37.8 \pm 2.6$     &  $1198\pm 82 $   \\   %  3 x 2007/2120 
$\kappa=0.12313$   & Yes    & 2.0           &  21780 &    $1.251\pm 0.001$              &    $1.1 \pm 0.1$      &  $37  \pm 1 $    \\\hline %C=2.27 pm 0.07  3 x 20002 /21780 
$\kappa=0.30400$   & No     & 0.5           &  2020  &    $0.2000 \pm 0.0024$              &    $102.4\pm 9.0$  &  $1026 \pm 90 $     \\ % C = 204.7 +- 18.0  1 x 2015/ 2020
$\kappa=0.30400$   & No     & 1.0           &  2580  &    $0.2032 \pm 0.0015$              &    $34.9\pm 2.1$   &  $450  \pm 28 $     \\ % C= 69.9 +- 4.3    1 x 2002/2580 
$\kappa=0.30400$   & No     & 2.0           &  2100  &    $0.2017 \pm 0.0007$              &    $8.0\pm 0.5$    &  $251  \pm 17 $      \\ % C = 16.0 +- 1.1  3 x   2011/ 2100 
$\kappa=0.30400$   & Yes    & 1.0           &  22360 &    $0.2025 \pm 0.0001$              &    $3.8\pm 0.2$    &  $171  \pm 8 $       \\ % C=7.65 +- 0.37  4 x 20016/  22360 
}
{tab:SummaryOfFacc}
{The auto-correlation time $\ACtime$ and the numerical costs $\Upsilon$ as defined in \eq{eq:DefOfNumCost1} are presented
for two series of Monte-Carlo runs performed with varying nominal trajectory lengths $\traLength$ 
without employing the Fourier acceleration technique. These runs have been configured with the model
parameter sets listed in \tab{tab:Chap5EmployedRuns} specified by the given values of the hopping parameter
$\kappa$. These findings are compared to the respective results obtained in the case of activated Fourier
acceleration. The observed values of the average magnetization $\langle m \rangle$ and the number of 
underlying field configurations $N_{conf}$ are also given. All runs employed the same Omelyan integration scheme
introduced in \sect{sec:MultiPolynoms} with convergence order $r=4$.
}
{Comparison of the Fourier acceleration technique with the simple approach of an overall increase of the trajectory length.}

Concerning the applicability of the Fourier acceleration technique also in the case of QCD one faces the
problem that the exploited correlation between the force strength and the momentum modes does apparently not exit
in the case of QCD due to its local gauge invariance. The here presented technique can therefore not directly be 
applied to QCD.

%-----------------------------------------------------------------------------------------------------
\section{Krylov-space based exact reweighting}
\label{sec:ExactReweighing}

As already pointed out in \sect{sec:basicConceptsOfPHMC} the PHMC algorithm requires the consideration 
of a weight factor $W[\Phi]$ for each generated field configuration to guarantee its correctness. This is 
due to the finite degree, and thus finite accuracy of the chosen approximation polynomial\footnote{For simplicity 
$\fermiMatDouble$ will be used here and in the following 
as a placeholder standing for $\fermiMatDouble$ itself or its preconditioned version 
$\precR\newfermiMat\precR^2\newfermiMat^\dagger\precR$ introduced in \sect{sec:PreconOfMDouble} which
is actually used in the implementation of the PHMC algorithm.} $P(\fermiMatDouble)$.
A second source for the necessity of reweighting the generated configurations is the finite extent of the 
approximation interval $[\epsilon_P, \lambda_P]$, unless it is guaranteed by some specific boundary that the
eigenvalue spectrum of $\fermiMatDouble$ is constrained to the chosen interval. The reason is that there
is no boundary at all for the deviation between the polynomial and the approximated function outside the specified
approximation interval.

The definition of this weight factor has already been given in \eq{eq:DefOfWeightFactor1}. In the literature
one finds basically two different standard approaches to evaluate this reweighting factor, both of which can
be performed separately from the actual generation of the field configurations in the Monte-Carlo process.
The first is based on rewriting the determinant in \eq{eq:DefOfWeightFactor1} in terms of a Gaussian integration
over some complex vector $\hat\omega$ according to
\bea
\label{eq:AnsatzForDetOfWeightFac1}
W[\Phi]&=& \det\left(\fermiMat\fermiMat^\dagger\cdot P(\fermiMat\fermiMat^\dagger) \cdot P(\fermiMat\fermiMat^\dagger) \right)^\frac{N_f}{2} 
= \left(\frac{1}{C} \FuncIntAvg_W\left[ \Sigma(\Phi,\hat \omega) \right] \right)^{\frac{N_f}{2}},  \quad\quad\\
\label{eq:AnsatzForDetOfWeightFac3}
\Sigma(\Phi,\hat \omega) &=& \exp\left({-\frac{1}{2}\hat \omega^\dagger 
\left[\left(\fermiMat\fermiMat^\dagger \cdot P(\fermiMat\fermiMat^\dagger) \cdot P(\fermiMat\fermiMat^\dagger) \right)^{-1} -\ID\right] \hat \omega }\right),
\quad\quad \\
C &=& \FuncIntAvg_W\left[1\right]
\eea
where constant factors have been neglected, the usual setting $\alpha=1/2$ has been assumed, and the functional integral $\FuncIntAvg_W$ 
is defined as
\bea
\FuncIntAvg_W\left[O[\Phi, \hat \omega] \right]
&=&  \int \intD{\hat\omega} \intD{\hat\omega^\dagger}\,\, O[\Phi, \hat\omega]\cdot  e^{-\frac{1}{2}\hat\omega^\dagger \hat\omega}.
\eea
The weight factor $W[\Phi]$ can then be computed by means of a Monte-Carlo integration through randomly 
sampling a number of $N_s$ evaluation points where the integrand is to be evaluated. More precisely, the integral 
representation of the determinant in \eq{eq:AnsatzForDetOfWeightFac1} has been split up here into a Gaussian factor 
$\exp(-\frac{1}{2}\hat\omega^\dagger \hat\omega)$, according to which the $N_s$ evaluation points, \ie the complex vectors 
$\hat\omega$, are sampled, and the corresponding remaining factor $\Sigma(\Phi, \hat\omega)$, that has to be evaluated for 
the selected sampling points. In the given decomposition the numerical estimate of $W[\Phi]$ is then given 
as the resulting ensemble average. This approach is very well applicable in practice, since the expression $\Sigma(\Phi,\hat\omega)$ 
is close to one by construction.

The second method is based on an exact computation of all eigenvalues $\nu_i$, $i=1,\ldots,n$ of $\fermiMatDouble$ with
$\nu<\epsilon_P$. The idea here is to use an approximation polynomial $P(x)$ with sufficiently high degree such
that the approximation can be considered to be exact within the interval $[\epsilon_P, \lambda_P]$. 
This is a reasonable assumption, if the maximal relative deviation $\delta_P$ is sufficiently small, ideally of the order 
of the machine precision. In this scenario the weight factor is given as
\beq
\label{eq:AnsatzForDetOfWeightFac4}
W[\Phi]= \prod\limits_{i=1}^n \nu_i^{1/2} \cdot P(\nu_i).
\eeq

Both presented approaches have their specific advantages and disadvantages. The first ansatz has the very useful
benefit that it allows to perform
the actual Monte-Carlo calculation with a rather low degree $N_P$ of the approximation polynomial, speeding up
the actual generation of the field configurations. Its main and obvious drawback is
that it is not an exact method. The accuracy of the approximation for the true weight can be improved
to an arbitrary level of precision by increasing the number of sample points $N_s$ in the Monte-Carlo evaluation
of \eq{eq:AnsatzForDetOfWeightFac1}. This, however, is connected to a corresponding increase of the numerical
costs. In fact, each sample point requires the evaluation of the {\textit{inverse}} of $\fermiMat\fermiMat^\dagger \cdot P(\fermiMat\fermiMat^\dagger) \cdot P(\fermiMat\fermiMat^\dagger)$
applied to the vector $\omega$, which is usually computed by a CG-solver. Though the number of CG-iterations $N_{CG}$ is
small due to the small condition number of this composed operator, the total cost for one computation of 
$W[\Phi]$ sums up to $N_s\cdot N_{CG}\cdot (2N_P+1)$ applications of $\fermiMatDouble$. In addition to the 
inexactness of this approach it is also numerically rather expensive. 

The latter drawback also applies to the second presented method. This method can be configured to be exact
provided that a sufficiently high degree for the approximation polynomial was chosen. This, however, slows down
the whole Monte-Carlo calculation. For smaller polynomial degrees the Monte-Carlo process speeds up but the 
reweighting method becomes inexact. Moreover, the exact computation of the smallest eigenvalues can be costly
depending on the eigenvalue density and other factors. Still, this approach may be the method of choice in case
of an eigenvalue spectrum that consists of a bulk of eigenvalues of moderate norm and only 
very few, well separated eigenvalues with extremely small norm. 

Here, however, a third, new reweighting strategy shall be introduced that combines the ability of dealing with
low degrees of the approximation polynomial, which is the main feature of the first approach, with the exactness
of the second ansatz. Moreover, the new method will benefit from low numerical costs in comparison to
the presented alternatives. 

The new reweighting strategy is based on writing the partition function in \eq{eq:PHMCPartitionFunction} as 
\bea
\label{eq:PHMCPartitionFunctionNewReweighing}
Z_\Phi &=& \int \intD{\Phi} \intD{\omega} \intD{\omega^\dagger}\,\, 
W[\Phi,\omega] \cdot e^{iN_f\arg\det(\fermiMat)} \cdot
e^{- S_\Phi[\Phi]- \frac{1}{2} \sum\limits_{i=1}^{N_f}\omega^\dagger_i P(\fermiMat\fermiMat^\dagger)\omega_i
}
\eea
with the new reweighting factor
\bea
\label{eq:DefOfNewWeightFactor1}
W[\Phi,\omega]&=& e^{-\frac{1}{2} \sum\limits_{i=1}^{N_f}\omega^\dagger_i
\left[ \left(\fermiMatDouble\right)^{-\alpha}- P(\fermiMat\fermiMat^\dagger) \right]\omega_i},\quad \alpha=\frac{1}{2},
\eea
depending on the scalar field $\Phi$ and all $N_f$ pseudo-fermion fields $\omega_i,\, i=1,\ldots,N_f$. 
Since the expressions $\omega^\dagger_iP(\fermiMat\fermiMat^\dagger) \omega_i$ are computed
during the PHMC simulation anyhow, the only source for additional numerical costs is the
computation of $\omega^\dagger_i \left(\fermiMatDouble\right)^{-\alpha} \omega_i$. This term, however, has to be
calculated to the desired accuracy, \ie close to machine precision. 

This can be done by Krylov-space based methods. Here, a Lanczos-based approach
will be used, which was introduced in \Ref{Borici:1999ws} for the purpose of applying
the inverse square root of a positive, hermitian matrix on a given vector $\omega_i$.
Its main idea is to approximate $(\fermiMatDouble)^{-1/2}$ by the inverse square root of the projection of the
operator $\fermiMatDouble$ to the Krylov space\footnote{The definition of the Krylov-space and the details
of this algorithm are postponed to \appen{sec:DescriptionOfLancosApproach}.} $\krylovSpace_{N_L}(\fermiMatDouble, \omega_i)$ 
for some sufficiently large value of $N_L$, where this latter number specifies the dimension of the Krylov space.
One of the most important features of this 
approach is that the number of Lanczos iterations $N_L$ necessary for reaching the desired relative accuracy
$\delta_L$ will not exceed the number of needed iterations of a corresponding CG-algorithm to calculate
$\left(\fermiMatDouble\right)^{-1}\omega_i$ to the same relative precision, which has been proven in \Ref{vandenEshof:2002ms}.
This Lanczos-based algorithm is thus quite efficient. It will be described in more detail
and extended to the more general task of computing $\left(\fermiMatDouble\right)^{-\alpha}\omega_i$ 
for some arbitrary real number $\alpha\in\re$ in \sect{sec:DescriptionOfLancosApproach}.

For now, however, we take the announced algorithm for granted and continue with the discussion of the
new reweighting technique. As mentioned earlier the lastly presented method for calculating the reweighting 
factor does not only depend on the field configuration $\Phi$ but also on the pseudo-fermion fields $\omega_i$.
The weight $W[\Phi,\omega]$ is therefore calculated during the Monte-Carlo process whenever a new field configuration
$\Phi$ has been accepted, which is convenient and practicable due to relatively low additional numerical 
costs. It would, however, also be possible to calculate this weight factor separately from the actual Monte-Carlo
process. In that case the pseudo-fermion fields $\omega_i$ would need to be temporarily stored on disk
until the weight $W[\Phi,\omega]$ has been determined. 

The introduced method shall now be tested and compared to the Gaussian estimator for the weight factor $W[\Phi]$
according to \eq{eq:AnsatzForDetOfWeightFac1}. For that purpose the rescaled Gaussian weight factor $W[\Phi]/\langle W[\Phi] \rangle$,
specifying the relative weight of the respective configurations, has been determined
in an example Monte-Carlo run as presented in \fig{fig:ExampleForReweighingFac}a. The given
numbers are not exact by construction and are thus afflicted with statistical errors, which can be reduced by including 
a higher number $N_s$ of sampling points within the Gaussian estimation of the weight factor. Here, the presented numbers 
have been computed with $N_s=10$ sample points for each evaluation of $W[\Phi]$. In \fig{fig:ExampleForReweighingFac}c
the corresponding numerical cost for a {\textit{single}} sample point in terms of applications of the fermion operator 
$\fermiMatDouble$ is shown. It is remarked that the total numerical costs, that were invested here, is thus even larger 
by a factor of $N_s=10$. 

%\includeFigTriple{GaussWeightsLam0Kap012313}{KrylovWeightsLam0Kap012313}{KrylovGaussCostsLam0Kap012313}
\includeFigTriple{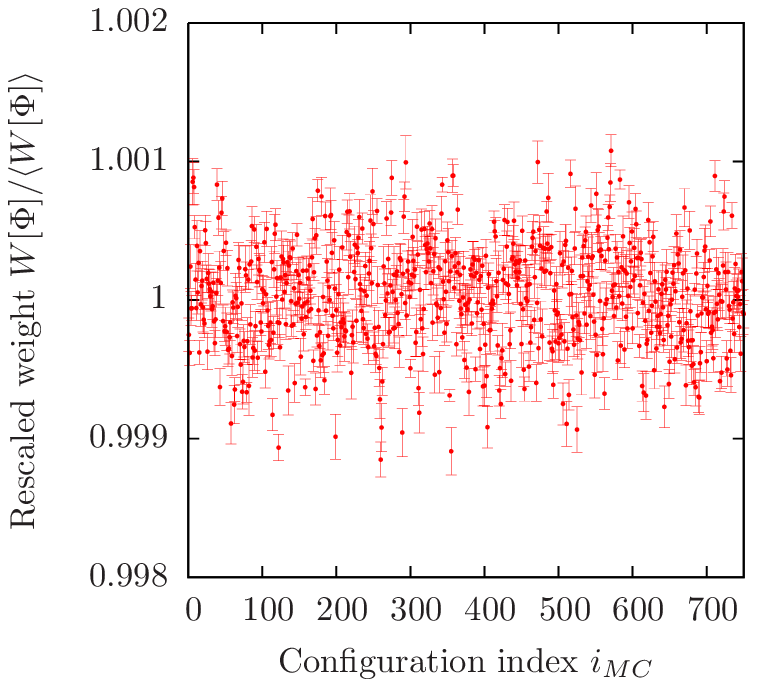}{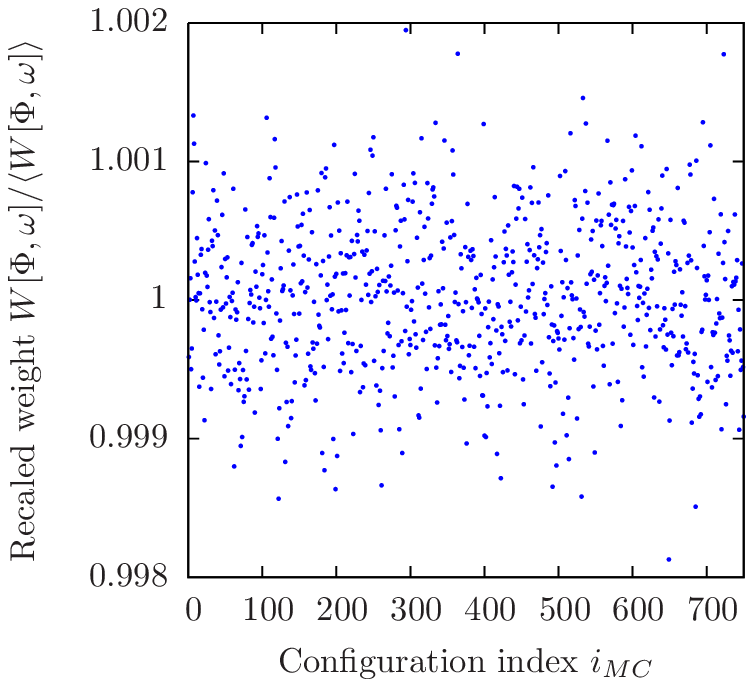}{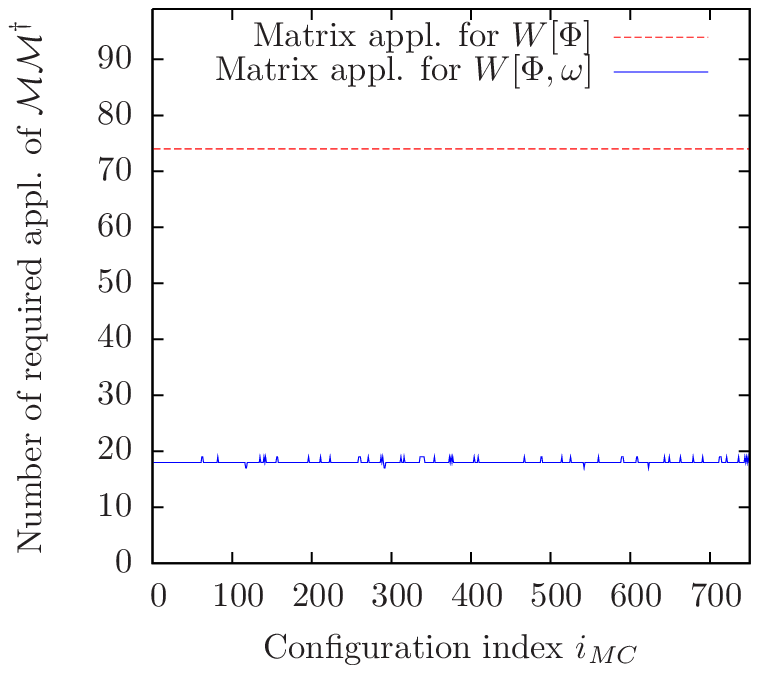}
{fig:ExampleForReweighingFac}
{The rescaled weight factors $W[\Phi]/\langle W[\Phi] \rangle$ and $W[\Phi,\omega]/\langle W[\Phi,\omega]\rangle$ defined
in \eq{eq:AnsatzForDetOfWeightFac1} and \eq{eq:DefOfNewWeightFactor1}, respectively, are presented in panels (a) and (b)
as obtained in the Monte-Carlo run specified in \tab{tab:Chap5EmployedRuns} with $\kappa=0.12313$. The associated numerical
cost in terms of applications of the fermion operator $\fermiMatDouble$ are shown in panel (c). Here the results from 
the Gaussian estimator have been obtained with 10 sample points per evaluation of the weight factor $W[\Phi]$, \ie
$N_s=10$. The numerical costs in panel (c), however, reflect only the cost of a single sample point.
}
{Comparison of weight factors resulting from the standard Gaussian approach and the presented Krylov-space 
based reweighting strategy.}

This is to be compared with the results obtained by the Krylov-space based reweighting strategy. In 
\fig{fig:ExampleForReweighingFac}b the corresponding scaled weight factors $W[\Phi,\omega]/\langle W[\Phi,\omega] \rangle$ 
calculated in the same Monte-Carlo run are presented. The latter weights are exact and do therefore not exhibit any statistical 
error. For clarification it is pointed out that the two presented weight factors are not supposed to coincide. 

The numerical cost associated to the new reweighting scheme are presented in \fig{fig:ExampleForReweighingFac}c and compared
to those of the Gaussian estimation for one single sampling point. One clearly observes that the Krylov-space based
method requires significantly less applications of the fermion operator $\fermiMatDouble$ than the Gaussian estimator.
Here, a factor of about $4\cdot N_s$ is observed. The lower numerical cost of the new reweighting technique and the fact
that the latter approach is exact thus constitute the advantages of the Krylov-space based method.

In the given example the fluctuation of $W[\Phi,\omega]$ has been adjusted to be of the order $0.1\%$. This can be done by 
an appropriate choice of the approximation interval and the degree of the approximation polynomial. 
A crude estimate of the expected fluctuation of $W[\Phi,\omega]$ can be given for the case where
all eigenvalues $\nu_j,\, j=1,\ldots,8V$ of $\fermiMatDouble$ are contained within the approximation 
interval $[\epsilon_P, \lambda_P]$. Let $w_j$ be the corresponding orthonormal eigenvectors. After the direct sampling
of the pseudo-fermion fields $\omega_i$, which will be described in detail in \sect{sec:DirectSampling},
one has
\beq
\omega_i = \sum\limits_{j=1}^{8V} c_{i,j} \cdot \frac{1}{\sqrt{P(\nu_j)}}\, w_j,
\eeq
where the real and imaginary parts of the complex coefficients $c_{i,j}$ are random variables sampled according to a 
Gauss distribution with standard deviation of 1, \ie $\langle |\RE(c_{i,j})|^2\rangle = \langle |\IM(c_{i,j})|^2\rangle=1$. 
The weight factor is then given as
\beq
W[\Phi,\omega] = \exp\left( -\frac{1}{2} \sum\limits_{i=1}^{N_f} z_i \right)
\eeq
with
\beq
z_i \equiv \omega_i^\dagger \left[\left(\fermiMatDouble\right)^{-\alpha} -P(\fermiMatDouble)   \right]\omega_i = 
\sum\limits_{j=1}^{8V} \left|c_{i,j}\right|^2 \cdot 
\frac{\nu_j^{-\alpha} - P(\nu_j)}{P(\nu_j)}.
\eeq
To roughly estimate the latter quantity $z_i$ we will assume here the eigenvalues $\nu_j$ to be independently 
and uniformly distributed within the interval $[\epsilon_P, \lambda_P]$. We further define the average relative
deviation $\bar \delta_P$ as
\bea
\bar \delta_P &=& \sqrt{\frac{R_P^2}{\lambda_P-\epsilon_P}},
\eea
where the expression $R_P^2$ has been given in \eq{eq:DefOfMeasureRP}. Under the aforementioned assumptions the variance of 
$z_i$, which is given as
\bea
\sigma^2_{z_i} &=& \langle z^2_i\rangle -\langle z_i\rangle^2
\eea
then becomes 
\bea
\sigma^2_{z_i} &=& \sum\limits_{j=1}^{8V}   \left\langle |c_{i,j}|^4 \cdot  \left[\frac{\nu_j-P(\nu_j)}{P(\nu_j)}  \right]^2 \right\rangle 
- \sum\limits_{j=1}^{8V} \left\langle |c_{i,j}|^2 \cdot  \frac{\nu_j-P(\nu_j)}{P(\nu_j)}   \right\rangle^2
\eea
which can roughly be estimated as
\beq
\sigma_{z_i} \approx \sigma^{est}_{z_i} = \sqrt{32 V} \cdot \bar \delta_P,
\eeq
when exploiting the relations $\langle |c_{ij}|^4\rangle = 8$ and $\langle |c_{ij}|^2\rangle = 2$ according to the above
assumptions and considering the eigenvalues $\nu_j$ to be constant over the averaging procedure. One would thus expect 
the standard deviation $\sigma_{lW}$ of the logarithm $\log(W[\Phi, \omega])$, which is defined as
\beq
\sigma^2_{lW} = \left\langle \left[\log(W[\Phi,\omega])\right]^2 \right\rangle  - \left\langle \log(W[\Phi,\omega])\right\rangle^2,
\eeq
to be approximated by
\beq
\sigma_{lW} \approx  \sigma_{lW}^{est} =\sqrt{8N_fV}  \cdot \bar \delta_P.
\eeq

Though this estimate is rather crude because all eigenvalues $\nu_j$ have been assumed to be
independently distributed, the given estimate $\sigma_{lW}^{est}$ nevertheless turns out to describe the 
actual behaviour of the quantity $\sigma_{lW}$ as observed in direct Monte-Carlo calculations fairly well as can 
be seen in \tab{tab:SummaryOfReweighing}. For this listing the example Monte-Carlo runs that have already been considered
several times before have been repeated with different setups of the underlying approximation polynomial. 

The crucial question, however, is whether the numerical results on the physically relevant observables are actually independent
of the choice of the approximation polynomial in a practical Monte-Carlo calculation. Moreover, it is important to ask how the 
resulting statistical uncertainty of the considered observables is affected by the reweighting strategy. For the purpose of
clarification the average magnetization $\langle m \rangle$ as well as the Higgs correlator mass $m_{Hc}$, the numerical
extraction method of which will be discussed in \sect{sec:HiggsTSCanalysisLowerBound}, have been computed in the aforementioned
series of Monte-Carlo calculations. The obtained results are presented in \tab{tab:SummaryOfReweighing} and one finds that 
the considered observables, serving here as typical representatives for the observables of physical interest in this study, are 
indeed independent of the underlying approximation polynomial apart from statistical deviations of the size of the given error 
estimates, as expected.

Concerning the effect on the statistical uncertainty one finds for both considered observables, though most clearly noticeable
for the average magnetization $\langle m\rangle $, that the latter uncertainty is not significantly affected by the the value of the 
relative deviations $\delta_P$ provided that $\delta_P$ is below a certain threshold value. Above that threshold, however, the statistical 
uncertainty of the considered observable rises quickly. The crossing of that threshold value occurs when the fluctuation of the weight 
factor becomes comparable to that of the considered, unweighted observable $O[\Phi]$ evaluated on the generated field configurations $\Phi$. 

Choosing an optimal fluctuation scale $\sigma_{lW}$ is thus a trade-off between pure simulation speed, which is proportional
to $1/N_P$, and the statistical uncertainty of the considered expectation value $\langle O \rangle$ increased by the fluctuations
of the weight factor $W[\Phi,\omega]$. In this work $\sigma_{lW}$ was typically chosen rather conservative, \ie not larger 
than the fluctuation of the most stable considered observable, which is the magnetization $m$ for the most cases. 
Depending on the observables that one is mostly interested in, one can, however, exploit the relative independence of the
underlying approximation polynomials to significantly increase the overall performance of the simulation algorithm in terms
of achieved accuracy per numerical cost.

Finally, one may ask why the Krylov-space based method for computing $[\fermiMatDouble]^{-1/2}\omega_i$ has not directly been employed 
for constructing a simulation algorithm in the spirit of an HMC-algorithm, having simply the CG-solver replaced by the latter 
method. This would indeed be possible, if one succeeds in providing an efficient ansatz for calculating the molecular dynamics 
forces in such an approach. 

It is further remarked that the here presented Krylov-space based, exact, and very efficient reweighting technique can immediately be 
applied also to the case of QCD.

\includeTab{|c|c|c|c|c|c|c|c|}{
\multicolumn{8}{|c|}{$\kappa =0.12313  $} \\ \hline
$N_P$ & $\delta_P$          & $\bar\delta_P$ \rule{0mm}{4mm}     & $\sigma_{lW}^{est}$ & $\sigma_{lW}$ & $N_{conf}$ & $\langle m \rangle$  & $m_{Hc}$ \\ \hline
8     & $4.0\cdot 10^{-3}$  & $5.7\cdot 10^{-4}$  & $5.8\cdot 10^{-1}$  & $6.8\cdot 10^{-1}$      &  2200      & 1.2497(239)  	& 0.1177(51)       \\ %$R=2.91289e-07$
10    & $1.1\cdot 10^{-3}$  & $1.4\cdot 10^{-4}$  & $1.4\cdot 10^{-1}$  & $1.3\cdot 10^{-1}$      &  2180      & 1.2515(49)         & 0.1144(37)       \\ %$R=1.75364e-08$
12    & $3.0\cdot 10^{-4}$  & $3.5\cdot 10^{-5}$  & $3.5\cdot 10^{-2}$  & $3.7\cdot 10^{-2}$      &  2200      & 1.2512(25)         & 0.1158(34)       \\ %$R=1.08837e-09$
14    & $8.0\cdot 10^{-5}$  & $8.8\cdot 10^{-6}$  & $9.0\cdot 10^{-3}$  & $8.6\cdot 10^{-3}$      &  2200      & 1.2541(15)         & 0.1127(32)       \\ %$R=6.9054e-11$ 
16    & $2.2\cdot 10^{-5}$  & $2.2\cdot 10^{-6}$  & $2.3\cdot 10^{-3}$  & $1.9\cdot 10^{-3}$      &  2180      & 1.2497(16)         & 0.1069(32)       \\ %$R=4.45516e-12$ 
18    & $5.9\cdot 10^{-6}$  & $5.7\cdot 10^{-7}$  & $5.8\cdot 10^{-4}$  & $5.3\cdot 10^{-4}$      & 21780      & 1.2506(5)	   	& 0.1116(12)   \\\hline%$R=2.91226e-13$
\multicolumn{8}{|c|}{$\kappa =0.30400  $}\\ \hline
8     & $6.7\cdot 10^{-3}$  & $9.0\cdot 10^{-4}$  & $9.2\cdot 10^{-1}$  & $5.8\cdot 10^{-1}$      & 2220       & 0.2019(31)         & 0.4060(94)       \\ %$R=7.42066e-07$
10    & $2.0\cdot 10^{-3}$  & $2.5\cdot 10^{-4}$  & $2.5\cdot 10^{-1}$  & $2.2\cdot 10^{-1}$      & 2280       & 0.2024(21)         & 0.4241(69)       \\ %$R=5.55406e-08$
12    & $6.1\cdot 10^{-4}$  & $6.8\cdot 10^{-5}$  & $7.0\cdot 10^{-2}$  & $5.4\cdot 10^{-2}$      & 2260       & 0.2021(3)          & 0.4230(67)       \\ %$R=4.2851e-09$
14    & $1.8\cdot 10^{-4}$  & $1.9\cdot 10^{-5}$  & $2.0\cdot 10^{-2}$  & $2.1\cdot 10^{-2}$      & 2280       & 0.2019(6)          & 0.4174(73)       \\ %$R=3.37968e-10$
16    & $5.6\cdot 10^{-5}$  & $5.4\cdot 10^{-6}$  & $5.6\cdot 10^{-3}$  & $5.0\cdot 10^{-3}$      & 22360      & 0.2025(1)          & 0.4243(23)        \\ %$R=2.71046e-11$
}
{tab:SummaryOfReweighing}
{The Monte-Carlo runs specified in \tab{tab:Chap5EmployedRuns} with $\kappa=0.12313$ and $\kappa=0.30400$ have been rerun
differing only in the setup of the underlying approximation polynomial and the number $\Nconf$ of generated field configurations.
While the approximation interval being 
$[0.1,1]$ for $\kappa=0.12313$ and $[0.085,1]$ for $\kappa=0.30400$ was unchanged, the polynomial degree $N_P$ was varied
resulting in the given values for the quantities $\delta_P$ and $\bar\delta_P$. The estimate $\sigma_{lW}^{est}$ is compared
to the actual fluctuation $\sigma_{lW}$ of $\log(W[\Phi,\omega])$ observed in the numerical simulation. Furthermore, the numerical
results on the average magnetization $\langle m \rangle$ and the Higgs correlator mass $m_{Hc}$ are shown. Note also the varying
statistics when comparing the presented statistical uncertainties.
}
{The independence of the lattice results from the chosen degree of the approximation polynomial.}

%-----------------------------------------------------------------------------------------------------
\section[Sampling of pseudo-fermion fields through polynomial approximation]{Direct sampling of pseudo-fermion fields through polynomial approximation}
\label{sec:DirectSampling}

The most efficient way for updating a field configuration in a Monte-Carlo process with respect to the auto-correlation 
would be its direct sampling according to the corresponding probability distribution, leading then to zero auto-correlation
time. While this type of sampling is not feasible for the scalar field $\Phi$ in the considered Higgs-Yukawa model except 
for the case of the free theory with $\lambda=y_t=y_b=0$, the pseudo-fermion fields $\omega_i$, more precisely the transformed 
pseudo-fermion fields $\eta_i = \sqrt{P(\fermiMatDouble)}\omega_i$ are Gaussian distributed according to \eq{eq:DefOfFermionActionSPF}
with the probability density $p(\eta_i) \propto exp(-\eta_i^\dagger\eta_i/2)$, where the index $i$ runs from $1$ to $N_f$.
The standard idea for updating the pseudo-fermion fields $\omega_i$ is thus to sample the transformed fields $\eta_i$ 
according to the latter Gaussian distribution $p(\eta_i)$ and to obtain $\omega_i$ from this sample through the 
inversion of the defining equation of $\eta_i$ according to
\beq
\label{eq:RelEtaOmega1}
\omega_i = \left[ P(\fermiMatDouble)\right]^{-\frac{1}{2}}\eta_i.
\eeq

There are several ways to solve \eq{eq:RelEtaOmega1} for a given vector $\eta_i$. A standard approach~\cite{Frezzotti:1998eu} 
applicable to QCD relies on the fact that the polynomial $P(Q^2)$ of the hermitian version of the QCD-Dirac operator 
$Q=\gamma_5\Dqcd$ can be written as the product
\beq
\label{eq:QCDsamplingOfOmegaDecomposition}
P(Q^2) =  \left[R(Q)\right]^\dagger \cdot R(Q) 
\eeq
where $R$ is an explicitly constructable~\cite{Frezzotti:1998eu} polynomial that can be determined on the basis of 
the square roots of the roots of the polynomial $P$. The pseudo-fermion fields are in this scenario not obtained from 
\eq{eq:RelEtaOmega1} but from
\bea
\label{eq:QCDSamplingOfOmega}
\omega_i = \left[R(Q)\right]^{-1}\eta_i = U^{-1} Q^2 P(Q^2) \left[ R(Q) \right]^\dagger \eta_i,\quad \mbox{with}\quad U=Q^2 P(Q^2)P(Q^2),
\eea
where the application of the inverse of the composed operator $U$ can be computed by means of a CG-algorithm. 

It is remarked here that the latter computation is very well feasible in practice despite the large number $4N_P+2$ of Dirac 
operator applications contained in the operator $U$. The reason is that the number $N_{CG}$ of necessary CG-iterations
is typically of order $O(1)$~\cite{Frezzotti:1998eu}, due to the operator $U$ exhibiting a very small condition number by construction,
which results from the polynomial $P(Q^2)$ being an approximation of $\left[Q^2\right]^{-1/2}$ in this example. For other functions 
underlying the latter approximation the definition of the operator $U$ has to be adapted analogously to still guarantee its condition number 
to be small. The total numerical costs of the approach in \eq{eq:QCDSamplingOfOmega} sum up to $N_{CG}\cdot(4N_P+2)+3N_P+2$ in terms 
of applications of $Q$. It should be noted that the condition number of $U$ depends on the accuracy of the approximation polynomial 
$P$. The number of required CG-steps $N_{CG}$ is thus expected to rise when one lowers the accuracy of the underlying polynomial. 

For the case of the considered Higgs-Yukawa model, however, the approach given in \eq{eq:QCDSamplingOfOmega} cannot directly 
be taken over. This is because there is no obvious way to decompose the polynomial $P(\fermiMatDouble)$ according to 
\eq{eq:QCDsamplingOfOmegaDecomposition} due to the lack of appropriate hermiticity properties of the fermion operator
$\fermiMat$ as discussed in \sect{sec:ComplexPhaseOfFermionDet}, which is in contrast to the above described situation in QCD.
Therefore, a different approach for the direct sampling of the pseudo-fermion fields will be used. 

Here, the idea is to compute \eq{eq:RelEtaOmega1} by means of a polynomial approximation $H(x)$ for the function $\left[P(x)\right]^{-1/2}$. 
The approximation interval for $H(x)$ is chosen to be $[0,\lambda_P]$, starting explicitly at zero. This is possible since the 
function $\left[P(x)\right]^{-1/2}$ has no singularity in the interval $[0,\lambda_P]$. The degree $N_H$ of this approximation 
polynomial $H(x)$ can be chosen sufficiently high such that its maximal relative deviation $\delta_H$, which is defined analogously 
to \eq{eq:DefOfDeltaPOfPolynom}, is of order of the machine precision. The pseudo-fermion fields are then directly obtainable through
\beq
\label{eq:NewDirectSamplingThroughPolApprox}
\omega_i = H(\fermiMatDouble) \eta_i.
\eeq

In general, there are many ways to construct such an approximation polynomial $H(x)$ for a given function $f(x)$ to be
approximated. The method described in \sect{sec:basicConceptsOfPHMC} could be adopted to the construction of an optimal 
approximation polynomial with respect to a minimal norm of the relative residual by defining an appropriate scalar product 
analogous to \eq{eq:DeterOfPolynomialPScalarProd}. Such an approach, however, would rely on the ability to calculate the 
aforementioned scalar product between the considered function $f(x)$ and a chosen set of basis polynomials to very high 
precision in order to determine the polynomial coefficients reliably as discussed in \sect{sec:basicConceptsOfPHMC}. This 
is straightforward for the case $f(x)=x^{-\alpha},\, \alpha \in \re$, since an explicit analytical expression for the considered 
scalar product can then trivially be derived. In the case considered here, however, the resulting integral expressions are more 
complicated hindering their obvious analytical solution. A numerical calculation of these expressions to the required numerical 
accuracy, on the other hand, would be extremely demanding with respect to the numerical costs. 

A method for the construction of the approximation polynomial $H(x)$, that evaluates the function $f(x)$ on a discrete, finite set of 
sample points only, is therefore favorable in this case. The standard Chebyshev approximation~\cite{Press:2007zu} is such a 
method. It determines the polynomial coefficients $c_i$ of $H(x)$ with respect to the 
Chebyshev polynomial basis ${T_i(x)}\,,i=0,\ldots,N_H$, \ie
\beq
\label{eq:RepOfApproxPolH}
H(x) = \sum\limits_{i=0}^{N_H} c_i \cdot T_i(x),
\eeq
by explicitly solving the set of linear equations
\bea
\label{eq:SetOfEqForStandardChebyApprox}
f(x_j) = \sum\limits_{i=0}^{N_H} c_i T_i(x_j), \quad j=0,\ldots,N_H
\eea
with respect to the coefficients $c_i$, where $\{x_j\}$ is a set of given sample points. In the standard Chebyshev approximation
these sample points are chosen as
\beq
\label{eq:SamplePointsForStandChebApprox}
x_j = \cos\left( \pi \frac{j+0.5}{N_H+1}  \right), \quad j=0,\ldots, N_H,
\eeq
rendering the solution of \eq{eq:SetOfEqForStandardChebyApprox} particularly simple by virtue of the orthogonal structure of
the vectors $t_i$ given as $(t_i)_j=T_i(x_j)$ with $i,j=0,\ldots,N_H$. It is well-known that the representation of $H(x)$ 
in terms of Chebyshev polynomials then allows for a very stable evaluation of \eq{eq:NewDirectSamplingThroughPolApprox}, 
since the numerically very stable recursion scheme
\bea
\label{eq:ChebyshevRecursion1}
T_0(x) &=& 1,\\
\label{eq:ChebyshevRecursion2}
T_1(x) &=& x,  \\
\label{eq:ChebyshevRecursion3}
T_{i+1}(x) &=& 2 x  T_i(x) - T_{i-1}(x)\, \quad \mbox{for } i\ge 1, 
\eea
can be used for the evaluation of the Chebyshev polynomials~\cite{Press:2007zu}. It should be noted that the given form of the 
sample points in \eq{eq:SamplePointsForStandChebApprox} as well as the recursion scheme
assumes the Chebyshev polynomials to be defined on the interval $[-1,1]$. Adequate rescalings to adopt the given formulas to
the here considered approximation interval $[0,\lambda_P]$ are implicitly assumed in the following.

The described standard Chebyshev approximation is easily implementable and yields reasonably accurate approximation polynomials.
Finding the optimal approximation polynomial for a given polynomial degree, however, requires a different algorithm, the
well known Remez-algorithm~\cite{Remez:321282,Press:2007zu} for instance. It is an iterative procedure that generates a sequence 
of approximation polynomials $H_n(x),\, n\in \N$. In each iteration step the polynomial $H_n(x)$ is constructed from a given set 
of $N_H+2$ sample points $\{x_j\}$ by solving a set of linear equations. The sample points are then tuned by relocating them to the 
positions of the extrema of the residual $f(x)-H_n(x)$. A common start constellation that typically yields good accuracy 
even without the aforementioned tuning is to choose the extrema of the Chebyshev polynomial $T_{N_H+1}$ as the start
constellation.

The tuned Remez algorithm is thus the method of choice to obtain the most accurate approximation polynomial constructable for
a given polynomial degree. Here, however, only the untuned version of the latter algorithm has been implemented, which already
works satisfactorily for our purpose. Examples of the required polynomial degree $N_H$ to approximate the function $[P(x)]^{-1/2}$ 
to a specified accuracy are given in \tab{tab:ExamplesForDirectSampling}. Here, the targeted maximal relative deviation of the approximant
$H(x)$ has been set to $\delta_H=10^{-12}$ and the presented parameters of the underlying polynomial $P(x)$ have been taken from some
actually performed lattice calculations as specified in \tab{tab:Chap5EmployedRuns} through the settings $\kappa=0.12313$ and
$\kappa=0.12220$. In this presentation the quantities $N_{H_C}$ and $N_{H_{Ru}}$ give the polynomial degrees required for the approximation
polynomial $H(x)$ as determined by the above described methods, \ie the standard Chebyshev and the untuned Remez algorithm, respectively, 
to reach the aspired value of $\delta_H$.

\includeTabNoHLines{|c|c|c|}{
\cline{2-3}
\multicolumn{1}{c|}{}                 &  $\kappa=0.12313$                    &  $\kappa=0.12220$                   \\ \hline
$\epsilon_P$                          &  0.1                                            & $5\cdot 10^{-4}$                              \\
$\lambda_P$                           &  1.0                                            & 1.0                                           \\
$N_P$                                 &  18                                             & 220                                           \\
$\delta_P$                            &  $5.9\cdot 10 ^{-6}$                            & $9.7\cdot 10^{-5}$                            \\
$\delta_H$                            &  $10^{-12}$                                     & $10^{-12}$                                    \\
$N_{H_C}$                             &  55                                             & 685                                           \\
$N_{H_{Ru}}$                          &  54                                             & 677                                           \\ \hline
$N_U(N_{CG}=1)$                              &  65                                             & 772                                           \\
$N_U(N_{CG}=2)$                              &  102                                            & 1213                                          \\
\hline
}
{tab:ExamplesForDirectSampling}
{The minimal polynomial degrees $N_{H_C}$ and $N_{H_{Ru}}$ of the approximation polynomial $H(x)$ as required by the standard Chebyshev 
and the untuned Remez algorithm, respectively, to undercut the specified value of $\delta_H$ are presented for two examples of the
underlying polynomial $P(x)$. The given parameters $\epsilon_P$, $\lambda_P$, $N_P$, and the associated value of $\delta_P$ are
taken from some actually performed lattice calculations as specified in \tab{tab:Chap5EmployedRuns} through the given values of 
the hopping parameter $\kappa$. These results are compared to the numerical would-be costs of the sampling approach in
\eq{eq:QCDSamplingOfOmega}, which are measured here in terms of applications of $\fermiMatDouble$ according to 
$N_U(N_{CG})=(N_{CG}\cdot(4N_P+2)+3N_P+2)/2$.
}
{Comparison between the standard approach and the polynomial based approach for sampling the pseudo fermion vector.}

The key observation is that the direct sampling of the pseudo-fermion fields can indeed be performed quite efficiently by means of
the considered polynomial based approach. One can infer from \tab{tab:ExamplesForDirectSampling} that the numerical costs for this 
kind of direct sampling with a given setting of $\delta_H=10^{-12}$ equals approximately three times the expenses of applying
$P(\fermiMatDouble)$ to a given pseudo fermion vector, at least in the considered examples.
This should be compared to the would-be costs of the -- here not applicable -- sampling method given in \eq{eq:QCDSamplingOfOmega},
which sum up to $N_U(N_{CG})=(N_{CG}\cdot(4N_P+2)+3N_P+2)/2$ in terms of applications of $\fermiMatDouble$. The resulting would-be 
costs for the considered examples are also listed in \tab{tab:ExamplesForDirectSampling} assuming the necessary number of CG-steps 
to be $N_{CG}=1$ and $N_{CG}=2$, respectively, which clearly is a non-pessimistic assumption. This is especially true in the
case where $\epsilon_P$ is not an absolute lower bound of the eigenvalue spectrum of the underlying fermion operator, since in that 
case the condition number of $U$ is no longer guaranteed to be of order $O(1)$. 

It is finally remarked that this method should be further improvable by applying the fully tuned Remez algorithm to find
the optimal approximant $H(x)$. Moreover, it is mentioned that the given approach for the direct sampling of the pseudo-fermion
fields is directly applicable also to the case of QCD, provided that a PHMC algorithm is employed for its numerical evaluation.

%-----------------------------------------------------------------------------------------------------
\section{Higher order multiple time-scale integrators}
\label{sec:MultiPolynoms}

The concept of multiple time-scale integration schemes~\cite{Sexton:1992nu} is based on the observation that it is often possible
to partition the total action $S[\Phi,\pi,\omega]$ into several contributions $S_i[\Phi,\omega],\,i=1,\ldots,n$ with 
\beq
S[\Phi,\pi,\omega]=\sum\limits_{i=1}^n S_i[\Phi,\omega] + f(\pi),
\eeq
such that the corresponding forces $F_i[\Phi,\omega]=dS_i/d\Phi$ acting on the conjugate momenta $\pi$ of the scalar field
$\Phi$ during the integration of the equation of motion differ strongly in strength and required computation time. 
Such a decomposition can then lead to a significant reduction of the overall numerical costs of performing the 
aforementioned molecular dynamics integration, provided that there is a correlation between the average strength 
of the forces $F_i[\Phi,\omega]$ and their numerical costs, such that the strongest forces require distinctly less computing 
resources for their calculation than the weaker ones. In this case the idea would be to construct a $n$-times nested 
integration scheme with different nested integration step sizes $\epsilon_i$, where longer step sizes would be assigned 
to the weaker, numerically more expensive forces while shorter step sizes would be used for the integration of the strong, 
rather inexpensive forces. 

Using the same notation as in \sect{sec:HMCAlgorithm} one can easily extend the leap-frog integration scheme given in 
\eqs{eq:LeapFrogScheme1}{eq:LeapFrogScheme4} to its multiple time-scale version. For that purpose we define
\bea
H[\Phi,\pi,\omega] &=& \sum\limits_{i=1}^n S_i[\Phi,\omega] + T[\pi] \quad \mbox{with}\quad T[\pi] = f(\pi).
\eea
The corresponding trajectory propagation operator for the given total trajectory length $\epsilon_0\equiv \traLength$ given as
\beq
e^{\traLength L(H)} = e^{\epsilon_0\left[\sum\limits_{i=1}^n L(S_i) + L(T)\right]},
\eeq
where the Liouville operator $L(H)$ has already been introduced in \eq{eq:DefOfLiouvilleOp},
can then again be approximated by using the Baker-Campbell-Hausdorff formula in \eq{eq:BCHformulaSimple} yielding
\bea
\label{eq:NestedLeapFrog1}
e^{\epsilon_{k-1}\left[\sum\limits_{i=k}^n L(S_i) + L(T)\right]} &=& \prod\limits_{j=1}^{N_k} e^{\epsilon_k\left[\sum\limits_{i=k}^n L(S_i) 
+ L(T)\right]}\quad
\mbox{with}\quad \epsilon_k = \epsilon_{k-1}/N_k \quad \mbox{and}\\
\label{eq:NestedLeapFrog2}
e^{\epsilon_k\left[\sum\limits_{i=k}^n L(S_i) + L(T)\right] + O\left(\epsilon_k^{3}\right)} &=&
e^{\frac{\epsilon_k}{2} L(S_{k})} \cdot
e^{\epsilon_k\left[\sum\limits_{i=k+1}^{n} L(S_i) + L(T)\right]} \cdot
e^{\frac{\epsilon_k}{2} L(S_{k})}, 
\eea
with $k=1,\ldots, n$ and $N_k$ denoting the number of integration steps for the nested integration on the $k$-th time-scale. 
The total number of integration steps with respect to the force $F_k[\Phi,\omega]$ is thus given as $N_1\cdot\ldots\cdot N_k$. 
It is remarked that the symplecticity and reversibility properties of this nested leap-frog scheme are directly inherited
from the single time-scale leap-frog scheme discussed in \sect{sec:HMCAlgorithm}.

The above presented nested leap-frog integration scheme can be further extended to a more general, higher order scheme by 
replacing \eq{eq:NestedLeapFrog2} with
\bea
\label{eq:GeneralNestedScheme1}
e^{\epsilon_k\left[\sum\limits_{i=k}^n L(S_i) + L(T)\right] + O\left(\epsilon_k^{r_k+1}\right)} &=&
\prod\limits_{j=1}^{m_k} 
e^{\epsilon_k c_{k,j} L(S_{k})} \cdot
e^{\epsilon_k d_{k,j}\left[\sum\limits_{i=k+1}^{n} L(S_i) + L(T)\right]} 
\eea
where the coefficients $c_{k,j},\, d_{k,j},\, j=1,\ldots,m_k$ specify the integration scheme applied on the $k$-th 
time scale and $r_k$ is the associated convergence order arising from this more general BCH-formula. 
For any desired order $r_k$ it is possible to select appropriate values for $m_k$ and $c_{k,j}$, $d_{k,j}$ with
$j=1,\ldots,m_k$ such that the desired convergence order in \eq{eq:GeneralNestedScheme1} is achieved. Furthermore, it can 
additionally be requested that $d_{k,m_k}=0$, $c_{k,j}=c_{k,m_k+1-j}$, and $d_{k,j}=d_{k,m_k-j}$,
which would then automatically guarantee the reversibility of the resulting integration scheme. The symplecticity 
follows again from the fact that all operators in \eq{eq:GeneralNestedScheme1} have a determinant of one, as discussed
in \sect{sec:HMCAlgorithm}. Examples for such sets of coefficients, taken from \Ref{Omelyan:2003zu}, are provided in 
\tab{tab:IntegratorParameters}.

\includeTab{|c|c|c|c|}{
                      &   LF                         & O2                            & O4                       \\ \hline
$r$                   &   2                          & 2                             & 4                        \\
$m$                   &   2                          & 3                             & 6                        \\
$c_1$                 &   0.5                        & 0.1931833275037836            & 0.08398315262876693      \\
$d_1$                 &   1                          & 0.5                           & 0.2539785108410595       \\
$c_2$                 &   0.5                        & $1-2 c_1$                     & 0.6822365335719091       \\
$d_2$                 &   0                          & 0.5                           & -0.03230286765269967     \\
$c_3$                 &   --                         & 0.1931833275037836            & $[1-2(c_1+c_2)]/2$       \\
$d_3$                 &   --                         & 0                             & $1-2(d_1+d_2)$           \\
$c_4$                 &   --                         & --                            & $[1-2(c_1+c_2)]/2$       \\
$d_4$                 &   --                         & --                            & -0.03230286765269967     \\
$c_5$                 &   --                         & --                            & 0.6822365335719091       \\
$d_5$                 &   --                         & --                            & 0.2539785108410595       \\
$c_6$                 &   --                         & --                            & 0.08398315262876693      \\
$d_6$                 &   --                         & --                            & 0                        \\
}
{tab:IntegratorParameters}
{The parameters $r$, $m$, $c_j$, and $d_j$ with $j$ running from 1 to $m$ are listed for the here considered integration schemes,
which have been labeled in the main text as the LF-, the O2-, and the O4-scheme. These numbers have been taken from 
\Ref{Omelyan:2003zu}.
}
{Parameters of the employed symplectic integrators.}

In the actual implementation of the PHMC algorithm two higher order integration schemes have been included
in addition to the standard leap-frog scheme. These are an order\footnote{The subscript $k$ is dropped
here and in the following when referring to the properties of the integrator in general, independent of
the respective time-scale.} $r=2$ and an order $r=4$ Omelyan integration scheme, the parameters of which are 
taken from \Ref{Omelyan:2003zu} and listed in \tab{tab:IntegratorParameters}.
These numerical schemes will be denoted as LF-, O2-, and O4-integrator, respectively, in the following. 
A particularity about the latter two Omelyan integrators implemented here is that the asymptotic number of required
force evaluations\footnote{'Asymptotic' means here averaged over an infinite number of subsequent integration steps
each of size $\epsilon$. The underlying rationale is that the first force in an integration step equals the
last force of the preceding integration step.} for a single integration step of size $\epsilon$, which is given by 
$m-1$, is larger than it would minimally have to be to reach the specified order. In fact, the leap-frog scheme needs 
one force evaluation less than the O2-integrator but both schemes have the same convergence order $r=2$. Furthermore,
there are also order $r=4$ integrators specified in \Ref{Omelyan:2003zu} which need less force evaluations than 
the considered O4-integrator. In fact, only three force evaluations are required to construct an order $r=4$
integrator. The arising additional freedom in the choice of the underlying parameters $c_{j}$, $d_{j}$ is used in 
the presented more complex schemes to minimize the first non-vanishing coefficient in the $\epsilon$-expansion 
of the residual $O(\epsilon^{r+1})$ appearing in \eq{eq:GeneralNestedScheme1}. It could be shown that this numerical 
investment leads to an overall performance advantage of the Omelyan integrators over their unimproved counterparts
with minimal nominal numbers of force evaluations by means of a higher acceptance rate~\cite{Omelyan:2003zu}.

The presented multiple time-scale integration scheme given in \eq{eq:GeneralNestedScheme1} shall now be applied
to the case of the considered Higgs-Yukawa model. The main question is then how the total action $S[\Phi,\pi,\omega]$ should 
be partitioned into the contributions $S_i[\Phi,\omega]$, such that the associated forces are either inexpensive or 
expensive but small compared to the other forces. An obvious idea is to separate the purely bosonic part of the 
action leading then to a two time-scale approach, \ie $n=2$, with the contributions
\bea
\label{eq:2TSPartitioning1}
S_1[\Phi,\omega] &=& S_{PF}[\Phi,\omega] \\
\label{eq:2TSPartitioning2}
S_2[\Phi,\omega] &=& S_\Phi[\Phi],
\eea
where $S_{PF}[\Phi,\omega]$ is the fermionic part of the total action as defined in \eq{eq:DefOfFermionActionSPF}. The outer integrator 
uses here a step size of $\epsilon_1\equiv \intStepSize=\traLength/N_1$ to integrate the fermionic contribution $S_1[\Phi,\omega]$, while 
the inner integrator applies the step size $\epsilon_2=\traLength/(N_1\cdot N_2)$. Though it is not clear how strong the inner, purely 
bosonic force $F_2[\Phi,\omega]$ actually is in comparison to the outer force, \ie the fermionic force, it is apparent that the numerical 
costs for the computation of $F_2[\Phi,\omega]$ are very low, making the presented approach thus a promising candidate for an overall 
reduction of the numerical expenses generated by the integration process. 

An explicit example for the relative sizes of the considered forces is presented in \fig{fig:HiggsVsYukawaForceStrength}.
Here, the averaged total force $\langle|\tilde F_p|\rangle$ in momentum space as given in \eq{eq:DefOfForceMolDynF} is presented 
versus the squared lattice momentum $\hat p^2$ together with the corresponding measurements of the partitioned forces 
$\langle |\tilde F_{1p}|\rangle$ and $\langle| \tilde F_{2p}|\rangle$, which are defined analogously. These results have 
been obtained in the Monte-Carlo run specified in \tab{tab:Chap5EmployedRuns} through the setting $\kappa=0.12313$.
One observes that the fermionic force is indeed significantly smaller than the bosonic contributions, except for the 
situation at small momenta. However, for the major fraction of momentum modes the aspired pronounced hierarchy of the 
considered forces is clearly observed, which is encouraging enough to investigate the actual benefit of the multiple 
time-scale integration scheme given in \eq{eq:GeneralNestedScheme1}.

%\includeFigDouble{ForceCompKap012313Overview}{ForceCompKap012313}
\includeFigDouble{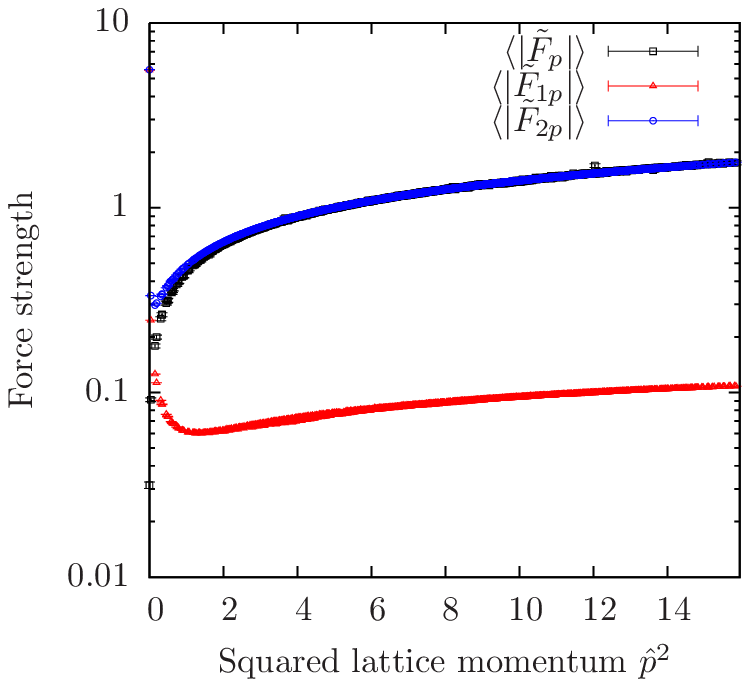}{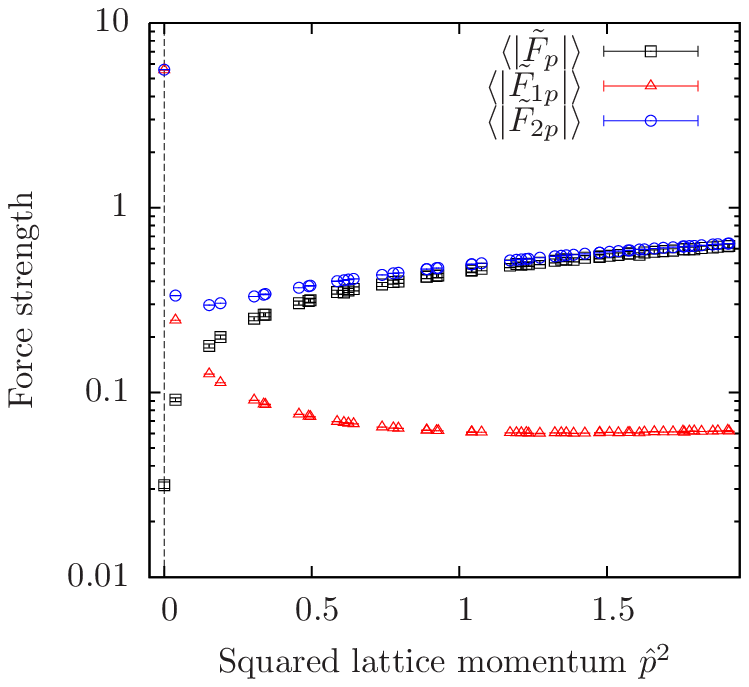}
{fig:HiggsVsYukawaForceStrength}
{The average force strengths $\langle| \tilde F_p |\rangle $, $\langle| \tilde F_{1p} |\rangle $, $\langle| \tilde F_{2p} |\rangle $
of the total force $F[\Phi,\omega]$ and the two partitions $F_1[\Phi,\omega]$, $F_2[\Phi,\omega]$ as defined through 
\eqs{eq:2TSPartitioning1}{eq:2TSPartitioning2} are presented versus the squared lattice momentum $\hat p^2$. These 
results have been obtained in the lattice calculation specified in \tab{tab:Chap5EmployedRuns} through the setting 
$\kappa=0.12313$. Both panels show the same data with panel (b), however, zooming in on the vicinity of the origin at $\hat p^2=0$.
The vertical dashed line is only meant to guide the eye.
}
{Comparison of the force strengths resulting from partitioning the total action in bosonic and fermionic contributions.}

For that purpose a measure for the numerical cost of the underlying integration scheme has to be specified. In analogy to
the cost measure given in \eq{eq:DefOfNumCost1} we define 
\beq
\label{eq:DefOfNumCost2}
\Upsilon = \frac{1}{\bar p_{acc}} \cdot N_{\fermiMatDouble},
\eeq
where $N_{\fermiMatDouble}$ is the number of applications of the fermion operator $\fermiMatDouble$ required for the 
numerical integration from $\MCtime=0$ to $\MCtime=\traLength$ and $\bar p_{acc}$ denotes the average acceptance rate.
The latter cost measure has then been evaluated based on the aforementioned Monte-Carlo data and the obtained results are
presented in \fig{fig:2TSimprovement}a.

%\includeFigDouble{IntegratorCompKap012313}{IntegratorCompKap030400}
\includeFigDouble{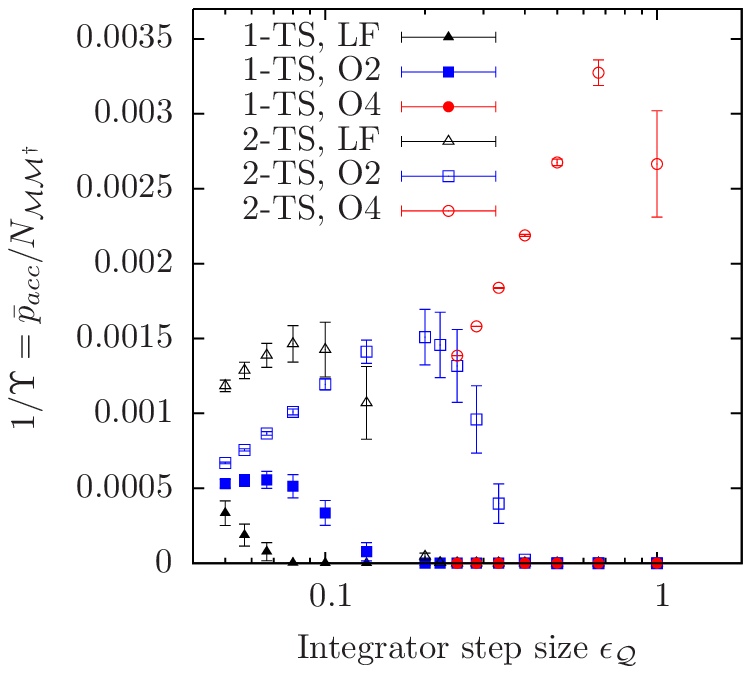}{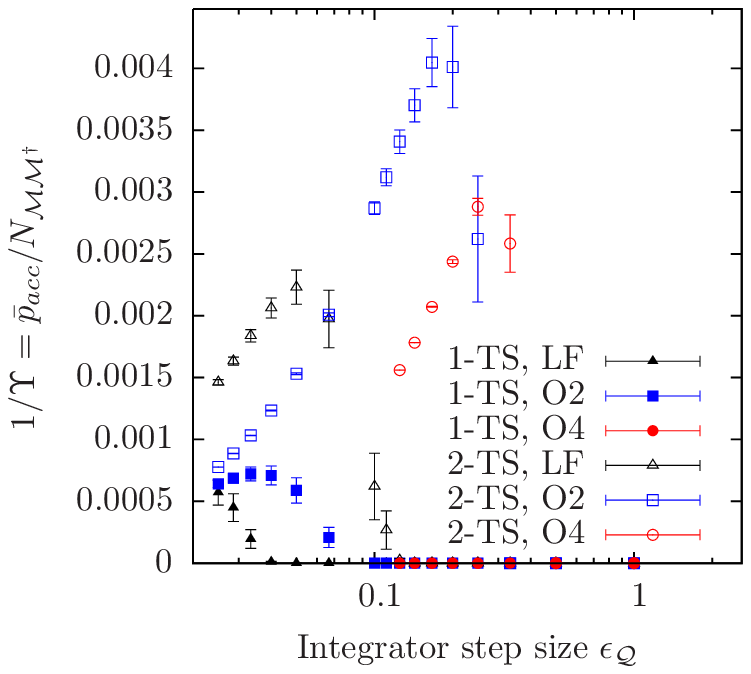}
{fig:2TSimprovement}
{The performance measure $1/\Upsilon=\bar p_{acc}/N_{\fermiMatDouble}$ is shown versus the step size $\intStepSize$
of the outer most integration scheme as observed for the six schemes listed in the plot legends. Here 1-TS refers to
the setting $n=1$ while 2-TS implies $n=2$ underlying time-scales. The labels LF, O2, and O4 specify the outer 
integration scheme. In the case $n=2$ the parameters of the inner scheme are set to $m_2=6$ and $N_2=5$. The presented
data in panels (a) and (b) have been obtained on the basis of the field configurations $\Phi$ generated in the lattice 
calculations specified in \tab{tab:Chap5EmployedRuns} through the settings $\kappa=0.12313$ and $\kappa=0.30400$,
respectively. The O4-scheme was only investigated for step sizes $\intStepSize\ge 0.25$ in panel (a) and $\intStepSize\ge 0.125$ 
in panel (b), which is why the curve of the 1-TS O4-integrator stays almost zero here.
}
{Dependence of the integrator performance $1/\Upsilon$ on the step size for different integrator schemes.}

Before continuing with the discussion of the given numbers, however, the details of the determination of the specified
cost measure $\Upsilon$ shall be briefly discussed. It has been evaluated for various integration schemes and for different 
outer step sizes $\intStepSize$. This has been done outside the actual Markov process by taking the field configurations 
after their completed generation in the respective Monte-Carlo run as the common starting points for all considered 
integration schemes. All considered schemes were then applied on the same basis of underlying start field configurations,
moreover, also with identical start values for the conjugate momenta $\pi$ as well as identical pseudo-fermion fields $\omega$,
which are both directly sampled as previously discussed. The observed differences in the acceptance rate are thus solely due to 
the properties of the integration scheme and are not induced by differences in the starting conditions. The value of the average 
acceptance rate was then determined by averaging over the observed variation $\Delta H$, given as the difference between the 
values of $H[\Phi,\pi,\omega]$ at the beginning and at the end of the integration process, according to 
\beq
\bar p_{acc} = \left\langle \min\left(1, \exp[-\Delta H]  \right) \right \rangle.
\eeq
For each considered integration scheme this experiment has been repeated for various settings of the steps size 
$\intStepSize\equiv \epsilon_1$ of the outer integrator. 

In \fig{fig:2TSimprovement}a the respective results of six different integration schemes are compared to each other. 
These are the LF-, the O2-, and the O4-scheme with only a single time-scale (1-TS), \ie $n=1$, and 
$S_1[\Phi,\omega] = S_{PF}[\Phi,\omega]+S_\Phi[\Phi]$,
as well as the nested LF-, O2-, and O4-schemes with $n=2$ (2-TS) and the partitioned forces chosen according
to \eqs{eq:2TSPartitioning1}{eq:2TSPartitioning2}. In the latter case the specified scheme, being LF, O2, or O4, 
refers to the outer integrator, while the inner integration of the bosonic forces is always performed based on the 
setting $m_2=6$, $N_2=5$ leading to a very accurate integration of the bosonic force contributions, such that the
fermionic force is the main source for the observed variation $\Delta H$.

With this background information at hand one can learn from \fig{fig:2TSimprovement}a that the separation of the bosonic
force onto a finer sampled time-scale indeed results in a substantial performance gain by a factor of approximately
6 in the considered example. One can also observe that the applied Omelyan O4-scheme is about 2 times more efficient
than the other $n=2$ time-scale schemes considered in the investigated setup.

The latter observation concerning the performance differences between the LF-, O2- and O4-schemes, all of
which applied with the same time-scale structure, depends, however, on the underlying model parameter
setup. This is illustrated by an additional example given in \fig{fig:2TSimprovement}b, showing the analogous results
evaluated for a different Monte-Carlo run with infinite quartic coupling constant, \ie $\lambda=\infty$. The detailed 
parameter setup underlying these results are specified in \tab{tab:Chap5EmployedRuns} through the setting $\kappa=0.30400$. 
In this plot one still observes the superiority of the $n=2$ time-scale integrators over their single time-scale
counterparts. The most efficient scheme is now, however, the O2-scheme exceeding the performance of the formerly
optimal scheme by around $\per{40}$ in this example. Testing different integration schemes for a given
set of model parameters can thus be worthwhile with respect to tuning the overall performance of the implemented 
algorithm.

Driven by the success of the considered multiple time-scale approach, the underlying technique for partitioning the 
forces shall be extended in the following, aiming at the further decomposition of the fermionic force $S_{PF}[\Phi,\omega]$.
Here, we shall have in mind the situation of non-degenerate Yukawa coupling constants set to its phenomenological
ratio. It has been discussed in \sect{sec:PreconOfMDouble} that the condition number of the fermion operator
$\fermiMatDouble$ becomes relatively large in that scenario leading then to the necessity of considering approximation
polynomials $P(\fermiMatDouble)$ of degree 200-300. It would thus be very desirable to have a means at hand with which 
also the fermionic force could be partitioned into numerically expensive, but weak, as well as strong, but inexpensive, 
forces. 

For that purpose the fermionic part of the action is rewritten here as a telescope sum according to
\bea
\label{eq:telescopeSumForFermionAction}
S_{PF}[\Phi,\omega] &=&  \frac{1}{2}\sum\limits_{j=1}^{N_F}
\omega_j^\dagger \left[P_1(\fermiMatDouble) -P_2(\fermiMatDouble)\right]\omega_j+\ldots \nonumber\\
&+& \omega_j^\dagger \left[P_{n-2}(\fermiMatDouble) -P_{n-1}(\fermiMatDouble)\right]\omega_j
+\omega_j^\dagger P_{n-1}(\fermiMatDouble)\omega_j,
\eea
where the polynomials $P_i(x),\,i=1,\ldots,{n-1}$ are approximation polynomials to the same function $f(x)=x^{-1/2}$
in our case. The polynomials differ only in their polynomial degree $N_{P_i}$, their maximal relative deviation $\delta_{P_i}$,
and the lower bound of their respective approximation intervals $[\epsilon_{P_i}, \lambda_P]$, while the upper 
interval bound $\lambda_P$ is identical for all polynomials. One can then decompose the total action into the $n\ge 2$
contributions $S_1,\ldots, S_n$ according to
\bea
\label{eq:SplittingOfAcrion1}
S_i[\Phi,\omega] &=& \frac{1}{2}\sum\limits_{j=1}^{N_F}
\omega_j^\dagger \left[P_i(\fermiMatDouble) -P_{i+1}(\fermiMatDouble)\right]\omega_j,\quad \mbox{for}\,\, i=1,\ldots,n-2,\,\,
\quad\quad\quad\\
\label{eq:SplittingOfAcrion2}
S_{n-1}[\Phi,\omega] &=& \frac{1}{2}\sum\limits_{j=1}^{N_F}
\omega_j^\dagger P_{n-1}(\fermiMatDouble)\omega_j,\quad \mbox{and } \\
\label{eq:SplittingOfAcrion3}
S_{n}[\Phi,\omega] &=& S_\Phi[\Phi].
\eea
It shall be remarked here, that no additional pseudo-fermion fields have been introduced in contrast to some
other multiple time-scale approaches~\cite{Urbach:2005ji,Hasenbusch:2002ai}. Moreover, all techniques for applying the 
aforementioned polynomials of 
$\fermiMatDouble$ to a given vector as well as the techniques for calculating the respective fermion forces are directly 
inherited from what has already been discussed in \sect{sec:basicConceptsOfPHMC}, since the difference of two polynomials 
is again a polynomial. The idea here is to use polynomials with very different degrees and relative accuracies, ordered 
according to their respective polynomial degree, such that $N_{P_1}\gg N_{P_2}\gg\ldots\gg N_{P_{n-1}}$. 
For an appropriate choice of these polynomials one can then expect to observe an ordering of the averaged force strengths
according to $\langle| \tilde F_{1p}| \rangle \ll \langle|\tilde F_{2p} |\rangle \ll \ldots \ll \langle|\tilde F_{(n-1)p}| \rangle$, 
as required for a profitable application of the nested integration scheme presented in \eqs{eq:NestedLeapFrog1}{eq:GeneralNestedScheme1}.

%\includeFigDouble{ForceCompKap012220}{IntegratorCompKap012220}
\includeFigDouble{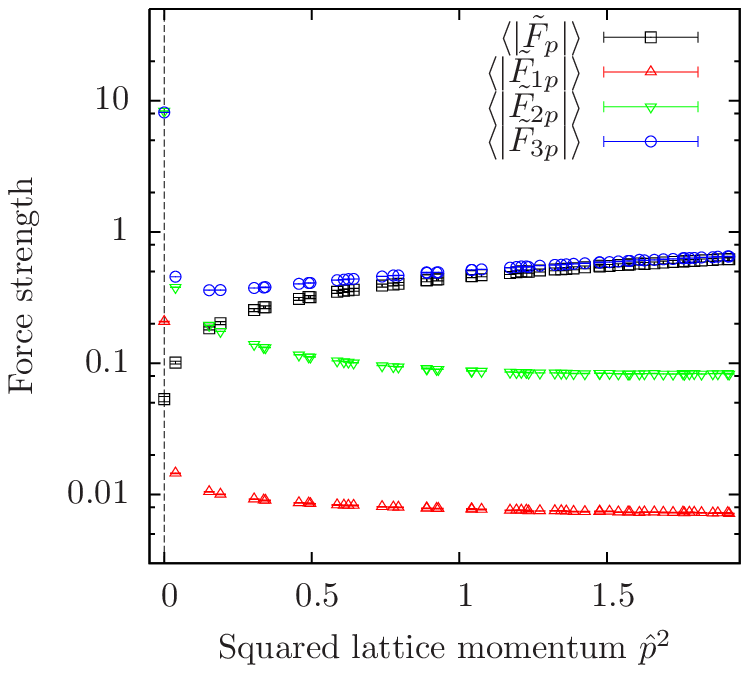}{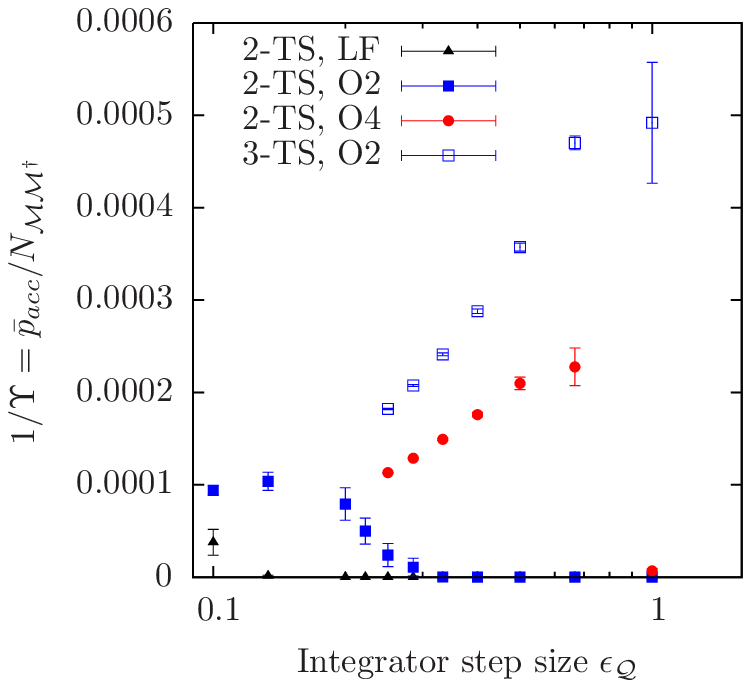}
{fig:3TSimprovement}
{The average force strengths $\langle| \tilde F_p |\rangle $, $\langle| \tilde F_{1p} |\rangle $, $\langle| \tilde F_{2p} |\rangle $, 
$\langle| \tilde F_{3p} |\rangle $ of the total force $F[\Phi,\omega]$ and the three partitions $F_1[\Phi,\omega]$, 
$F_2[\Phi,\omega]$, $F_3[\Phi,\omega]$ as defined through 
\eqs{eq:SplittingOfAcrion1}{eq:SplittingOfAcrion3} are presented in panel (a) versus the squared lattice momentum $\hat p^2$. 
The vertical dashed line is only meant to guide the eye.
In panel (b) the performance measure $1/\Upsilon=\bar p_{acc}/N_{\fermiMatDouble}$ is shown versus the step size $\intStepSize$
of the outer most integration scheme as observed for the four schemes listed in the plot legend. Here 2-TS refers to
the setting $n=2$ while 3-TS implies $n=3$ underlying time-scales. The labels LF, O2, and O4 specify the outer 
integration scheme. In the case $n=2$ the parameters of the inner scheme are set to $m_2=6$ and $N_2=5$, while we
have $m_2=6$, $N_2=1$, $m_3=6$, and $N_3=5$ for $n=3$. The presented data have been obtained on the basis of the lattice 
calculation specified in \tab{tab:Chap5EmployedRuns} through the setting $\kappa=0.12220$.
}
{Comparison of the force strengths resulting from partitioning the fermion action by means of the telescope
sum approach and the dependence of the integrator performance $1/\Upsilon$ on the step size for different 
integrator schemes in this setup.}

An example for the hierarchy of the force strengths $\langle|\tilde F_{ip}| \rangle$ is presented in 
\fig{fig:3TSimprovement}a. Here, the presented force strengths have been obtained in the Monte-Carlo run
specified in \tab{tab:Chap5EmployedRuns} through the setting $\kappa=0.12220$. 
In this example, the action $S_{PF}[\Phi,\omega]+S_\Phi[\Phi]$ was split up into three parts, \ie $n=3$, according to 
\eqs{eq:SplittingOfAcrion1}{eq:SplittingOfAcrion3}. The two underlying polynomials $P_1$ and $P_2$ have been
chosen to have the very distinct degrees $N_{P_1}=220$ and $N_{P_2}=24$. Also the values of the lower approximation
interval bound differ strongly. Here we have $\epsilon_{P_1}= 5\cdot 10^{-4}$ and $\epsilon_{P_2}= 6\cdot 10^{-3}$.
In the presented plot one clearly observes a very pronounced separation of the force strengths $\langle|\tilde F_{ip}| \rangle$
associated to the $n=3$ partitions $S_i[\Phi,\omega]$. Here, the strongest force is produced by the most inexpensive contribution, 
\ie the bosonic contribution $S_\Phi[\Phi]$, followed by the remaining two forces $F_2[\Phi,\omega]$ and $F_1[\Phi,\omega]$
in ascending order with respect to their numerical costs, as aspired. Again this observation only holds at sufficiently large 
momenta. For the major part of the momenta, however, the considered forces are clearly separated by a factor of approximately 
one order of magnitude, thus rendering the presented force partitioning a promising candidate for the intended performance 
improvement.

The actually obtained performance of this $n=3$ time-scale (3-TS) integration scheme is presented in \fig{fig:3TSimprovement}b.
Here, the parameters of the outer integrator have been chosen according to \tab{tab:IntegratorParameters} as specified through 
$m_1=3$, \ie the outer integrator is given here as the O2-scheme. The two inner integrators have been configured with $m_2=6$, 
$N_2=1$ and $m_3=6$, $N_3=5$. The resulting inverse costs $1/\Upsilon$ are then presented versus the varying value of the outer 
step size $\intStepSize\equiv\epsilon_1$. These results are compared to the $n=2$ time-scale LF-, O2-, and O4-schemes as already 
considered above. The latter are based, of course, on the finer polynomial $P_1$ with degree $N_{P_1}=220$. One clearly observes 
significant differences between the presented schemes with respect to the measure $1/\Upsilon$. In particular, one
sees that even the best considered setup for the $n=2$ time-scale approach, which was found to be far superior to the single time-scale 
schemes discussed above, can be beaten by the presented $n=3$ time-scale scheme improving the numerical performance by a factor 
of approximately 2.

It is finally remarked, that the presented approach of splitting up the fermion action in terms of a telescope sum 
according to \eq{eq:telescopeSumForFermionAction}, yielding then several fermion forces with potentially very different
force strengths, can directly be applied also to the case of QCD, provided that a PHMC algorithm is employed for its numerical
evaluation.

  \chapter{Results on the lower Higgs boson mass bound}
\label{chap:ResOnLowerBound}

The simulation algorithm detailed in the previous chapter shall now be applied to the actual determination of the lower
Higgs boson mass bounds. Before such mass bounds can be derived, however, a certain amount of preparation work 
has to be invested to build up the basis for the eventually intended calculations. First of all, the so far
unspecified details of the Higgs boson mass determination have to be clarified. This will be discussed in 
\sect{sec:ParticleMassDetDetailsLowerBound} together with the extraction method for the Goldstone renormalization
constant and the fermion mass computation. 

In a next step some analytical formulas for the dependence of the vacuum expectation value $v$, the Higgs boson self-energy, 
and the quark masses on the model parameters will be derived from lattice perturbation theory (LPT) in 
\sect{sec:PredFromPertTheory}. The observed good agreement between these analytical predictions and the corresponding
numerical results obtained in direct Monte-Carlo calculations can be understood as an indication for the correctness
of the implemented simulation algorithm. The analytical predictions on the quark masses will later also be helpful to 
understand the finally observed dependence of the numerically obtained quark masses on the cutoff $\Lambda$ and the 
lattice volume. 

In \sect{sec:EffPot}, the Higgs boson mass will alternatively also be estimated by calculating the effective potential 
introduced in \sect{sec:SmallYukawaCouplings}. Here, however, the already given result on the effective potential is extended 
by incorporating the missing $O(\lambda)$ contributions, allowing then to study the Higgs boson mass dependence on the model 
parameters $y_t$, $y_b$, and $\lambda$, in particular. From the obtained results it will be concluded that the lowest 
Higgs boson mass is indeed obtained at vanishing bare quartic coupling constant as one would have expected from the qualitative 
result in \eq{eq:perturbTheroyResult}.

With this preparation at hand the lower Higgs boson mass bounds can eventually be established by evaluating the Higgs boson masses
attainable at $\lambda=0$ with the Yukawa coupling constants adjusted such that the phenomenologically known values of the
quark masses are reproduced. This will be done in \sect{sec:SubLowerHiggsMassboundsDegenCase} for the degenerate scenario 
with equal top and bottom quark masses, which is numerically fairly well accessible and moreover conceptually well under control. 
This situation changes when proceeding to the lower Higgs boson mass bound determination in the physically more relevant
setup with $N_f=3$ and $y_b/y_t$ tuned to its phenomenological value, as discussed in \sect{sec:SubLowerHiggsMassboundsGenCase}.

Finally, the question for the universality of the obtained Higgs boson mass bounds will be addressed in \sect{sec:ModExtByHigherOrder},
where the model will be extended by some higher order self-interaction terms of the scalar field. Such an extension is not 
excluded by the usual renormalization arguments, since the Higgs-Yukawa sector can only be considered as a trivial theory 
with a non-removable cutoff by virtue of its triviality property. The effect of these additional coupling terms on the 
attainable Higgs boson masses is then investigated and it is found that the previously established lower mass bound can 
indeed significantly be altered.

%-------------------------------------------------------------------------------------------------------------------- 
\section{Details of the particle mass determination}
\label{sec:ParticleMassDetDetailsLowerBound}

The so far unspecified details concerning the computation of the physical observables defined in \sect{sec:SimStratAndObs} 
shall now be given in the subsequent sections. We will begin with the discussion of the Higgs boson mass determination based 
on the Higgs time-slice correlator $C_H(\Delta t)$ in \sect{sec:HiggsTSCanalysisLowerBound}, since this will be the method of choice in the weakly 
interacting regime of the considered Higgs-Yukawa model. The reasoning is that this approach does not rely on any analytical 
continuation of Euclidean results to Minkowski space-time and does therefore not suffer from the systematic uncertainties connected 
to such an analytical continuation in contrast to the mass determination based on the Higgs propagator analysis as proposed in \sect{sec:SimStratAndObs}.
This approach is justified as long as the Higgs boson can be treated as a stable particle, which seems acceptable in the weakly interacting 
regime of the considered model, as discussed in \sect{sec:UnstableSignature}, allowing then for a straightforward determination of 
the Higgs boson mass by studying the decay properties of the Higgs time-slice correlator $C_H(\Delta t)$.

The methods for the extraction of the Goldstone renormalization factor $Z_G$ and the Higgs boson masses $m_H$, $m_{Hp}$ based on the analysis 
of the respective propagators are then presented in \sects{sec:GoldPropAnalysisLowerBound}{sec:HiggsPropAnalysisLowerBound}. Finally, 
the fermion mass determination is discussed in \sect{sec:FermionTSCanalysisLowerBound}. 

The applicability of all presented computation schemes will be demonstrated for the selected Monte-Carlo 
runs specified in \tab{tab:Chap61EmployedRuns}. The listed lattice calculations, serving here as typical examples for the weakly 
interacting regime of the considered model, have been chosen to cover the full range of cutoffs $\Lambda$ that will be investigated
in the remainder of this chapter.

\includeTab{|cccccccc|}
{
\latVolExp{L_s}{L_t} & $N_f$ & $\kappa$  & $\hat \lambda$ & $\hat y_t$     & $\hat y_b/\hat y_t$ & $\langle m \rangle$ & $\Lambda$ \\
\hline
\latVol{32}{32}      & $1$   & $0.12301$ & $0$       & $0.35285$ & $1$       & $0.4197(13)$ & $\GEV{1163.9 \pm 3.6}$\\
\latVol{32}{32}      & $1$   & $0.12313$ & $0$       & $0.35302$ & $1$       & $1.2461(4)$  & $\GEV{393.5  \pm 1.3}$\\
}
{tab:Chap61EmployedRuns}
{The model parameters of the Monte-Carlo runs constituting the testbed for the computation schemes discussed in the subsequent
sections are presented together with the obtained values of the average magnetization $\langle m \rangle$ and the cutoff $\Lambda$
determined by \eq{eq:DefOfCutoffLambda}. The degenerate Yukawa coupling constants have been chosen here according to the tree-level
relation in \eq{eq:treeLevelTopMass} aiming at the reproduction of the phenomenologically known top quark mass.}
{Model parameters of the Monte-Carlo runs constituting the testbed for the considered computation schemes at small quartic
coupling constants.}

%-------------------------------------------------------------------------------------------------------------------- 
\subsection{Analysis of the Higgs time-slice correlator}
\label{sec:HiggsTSCanalysisLowerBound}

The Higgs correlator mass $m_{Hc}$ was defined in \eq{eq:DefOfHiggsMassFromTimeSliceCorr} via the exponential decay 
of the Higgs correlation function $C_H(\Delta t)$ at large time separations $1\ll\Delta t\le L_t/2$. 
Provided that the Higgs boson can be considered as a stable particle, meaning here that contributions 
to the correlation function $C_H(\Delta t)$ arising from lighter energy eigenstates of the underlying Hamiltonian are 
negligible, the determined correlator mass $m_{Hc}$ coincides with the physical pole mass $m_H$ given by the pole of 
the Higgs propagator, as discussed in \sect{sec:SimStratAndObs}.

In the context of the considered Higgs-Yukawa model, however, this condition is only approximately fulfilled
even in the weakly interacting region of the parameter space.
In fact, one would  expect the massless Goldstone modes to contaminate the correlation function $C_H(\Delta t)$. 
More precisely, it is not the single Goldstone modes that mix into $C_H(\Delta t)$
but the two Goldstone states, which have the same quantum numbers as the Higgs boson. The
single Goldstone modes are distinguishable from the Higgs modes by their isospin and
are thus projected out by the operator $O_H(t)$ underlying the Higgs time-slice
correlation function as defined in \eq{eq:HiggsFieldTSoperator}. The idea here is to respect the contribution arising from
the almost massless two Goldstone states by choosing an appropriate fit ansatz for the correlation function
$C_H(\Delta t)$. Though it is well-known that on a finite lattice volume the Goldstone particles are not 
exactly massless~\cite{Hasenfratz:1990fu}, the contamination induced by these states is here accounted for 
by incorporating an additional constant term $C$ into the fit ansatz according to
\beq
\label{eq:HiggsTimeCorrelatorFitAnsatz1}
f_{A,m_{Hc},C}(\Delta t) = A \cdot \cosh\Big[ m_{Hc} \cdot (\Delta t - L_t/2)\Big] + C
\eeq
where $A$, $C$, and $m_{Hc}$ denote the free fit parameters. Moreover, the fit is assumed to be performed at 
sufficiently large values of $\Delta t$ to avoid a contamination by excited states. This simple form is 
chosen here for the sake of the stability of the fit procedure. Additionally, the less flexible and thus 
even more stable fit approach
\beq
\label{eq:HiggsTimeCorrelatorFitAnsatz2}
f_{A,m_{Hc}}(\Delta t) = A \cdot \cosh\Big[ m_{Hc} \cdot (\Delta t - L_t/2)\Big]
\eeq
will also be applied to fit the Higgs correlation function $C_H(\Delta t)$. In this latter ansatz 
the constant term $C$ has been neglected making this ansatz insensitive to possible contaminations arising from 
lighter states on the one hand, but more stable on the other hand. The motivation for this second approach 
lies in the assumption that the contributions of the lighter two Goldstone states to the correlation function 
$C_H(\Delta t)$ might be negligible from a practical point of view in the weakly interacting regime of the considered 
model. In the following both approaches will therefore be applied and compared to each other.

The alert reader, however, may wonder whether the mass extraction based on the analysis of the long ranged decay 
properties of the correlation function $C_H(\Delta t)$ is justified at all, since this latter function is 
constructed from the operator $O_H(t)$, which is non-local in time. This non-locality arises from the rotation of the 
field $\varphi$ introduced in \sect{sec:modelDefinition} yielding the rotated field $\varphi^{rot}$ which actually
underlies the latter operator $O_H(t)$. By virtue of this rotation the 
usual arguments of the transfer matrix formalism can thus not directly be applied to the aforementioned operator $O_H(t)$ 
in the standard manner, as presented in \eq{eq:DefOfOpCorrFunctionInOpForm1}, 
to establish the connection between the long range behaviour of the correlator $C_H(\Delta t)$ and the particle
mass spectrum. The correlation function $C_H(\Delta t)$ associated to the rotated field $\varphi^{rot}$ has 
therefore explicitly been studied in the framework of the pure $\Phi^4$-theory~\cite{Gockeler:1991ty}. It 
was found that the operator $O_H(t)$, though non-local in time, still leads to an exponentially decaying
correlation function $C_H(\Delta t)$, allowing for the determination of the Higgs boson mass in the above described 
manner, with the downside, however, that the obtained Higgs boson mass is contaminated with additional finite 
volume contributions. 

For the purpose of testing these findings also in the broader context of the considered Higgs-Yukawa model
an alternative Higgs time-slice correlation function $C'_H(\Delta t)$ is defined here based on the local operator
\beq
O'_H(t) = \frac{1}{L_s^3}\sum\limits_{\vec x} \frac{\bar \varphi^\dagger_t}{|\bar\varphi_t|} \cdot \varphi_{t,\vec x}
\quad \mbox{with}\quad
\bar\varphi_t = \frac{1}{L_s^3}\sum\limits_{\vec x}  \varphi_{t,\vec x}
\eeq
according to
\beq
\label{eq:DefOfLocalCorrFunction}
C'_H(\Delta t) = \frac{1}{L_t} \sum\limits_{t=0}^{L_t-1} \Big\langle  O'_H(t+\Delta t)\cdot O'_H(t) \Big\rangle
-\Big\langle  O'_H(t+\Delta t) \Big\rangle\cdot\Big\langle O'_H(t) \Big\rangle,
\eeq 
where an adequate modulo operation is implicit to guarantee $0\le t+\Delta t<L_t$.

The given operator $O'_H(t)$ determines the direction $\bar\varphi_t/|\bar\varphi_t|$ of the Higgs mode in terms of the 
components of the field $\varphi$ on each time-slice separately and is thus local in time. The transfer matrix
formalism is therefore directly applicable to the corresponding correlation function $C'_H(\Delta t)$ in the 
standard manner. The above described procedure for extracting the Higgs boson mass from the
correlation function can thus be applied to $C'_H(\Delta t)$ without the aforementioned conceptual 
difficulties. The same fit formulas already given in \eq{eq:HiggsTimeCorrelatorFitAnsatz1} and \eq{eq:HiggsTimeCorrelatorFitAnsatz2} 
will therefore also be applied to $C'_H(\Delta t)$. 

Examples for the correlation functions $C_H(\Delta t)$ and $C'_H(\Delta t)$ obtained from the direct
Monte-Carlo calculations specified in \tab{tab:Chap61EmployedRuns} are presented in \fig{fig:HiggsCorrelatorExamplesAtWeakCoup}a,d.
These plots also show the respective fits resulting from the approaches given in \eq{eq:HiggsTimeCorrelatorFitAnsatz1} and 
\eq{eq:HiggsTimeCorrelatorFitAnsatz2}.
 
The other four plots in \fig{fig:HiggsCorrelatorExamplesAtWeakCoup} show the effective correlator masses 
$m_H^{eff}(\Delta t)$ obtained from the two different correlation functions $C_H(\Delta t)$ and $C'_H(\Delta t)$
combined with the two given fit functions. The effective masses are implicitly defined through
\bea
\label{eq:DefOfEffMassesForHiggs}
\frac{C_H(\Delta t +0.5) - C}{C_H(\Delta t -0.5) - C} = \frac{\cosh\Big[ m_{Hc}^{eff}(\Delta t) \cdot (\Delta t+0.5  - L_t/2)\Big]}
{\cosh\Big[ m_{Hc}^{eff}(\Delta t) \cdot (\Delta t-0.5  - L_t/2)\Big]},
\quad
\Delta t = \frac{1}{2},\ldots,L_t-\frac{1}{2} ,\quad
\eea
and can easily be computed by solving the given equation numerically.
Here the symbol $C_H$ stands either for the correlation function $C_H(\Delta t)$
or $C'_H(\Delta t)$ and $C$ is the fit parameter of \eq{eq:HiggsTimeCorrelatorFitAnsatz1}
or zero if the fit ansatz in \eq{eq:HiggsTimeCorrelatorFitAnsatz2} is used.

The effective masses are supposed to decrease monotonically with rising $\Delta t\le L_t/2$ and to 
converge to a plateau value for sufficiently large values of $\Delta t\le L_t/2$, which gives the 
actual correlator mass $m_{Hc}$ according to its definition in \eq{eq:DefOfHiggsMassFromTimeSliceCorr}. 
Since the statistical error of the effective masses generally grows with increasing $\Delta t\le L_t/2$, the 
choice of the value of $\Delta t$ where to read off the result for the Higgs boson mass is a trade-off 
between systematic and statistical uncertainties. To determine $m_{Hc}$ one would 
usually fit $m_{Hc}^{eff}(\Delta t)$ to a constant within a window of $\Delta t$ starting at a
value where the plateau building has already set in. 

Here, however, the statistical
errors of the effective masses are very large compared to the effect induced by the contamination
of the correlation functions through excited states. A reliable distinction between a phase 
of monotonic decrease of $m_{Hc}^{eff}(\Delta t)$ and the onset of a plateau building is thus
not very practical here. In the following the Higgs boson mass will therefore not be determined by
the effective masses themselves but through the overall cosh-fits given in \eq{eq:HiggsTimeCorrelatorFitAnsatz1}
and \eq{eq:HiggsTimeCorrelatorFitAnsatz2}. The Higgs boson masses resulting from these fits
are compared to the respective effective masses in \fig{fig:HiggsCorrelatorExamplesAtWeakCoup}. 
Although the presented correlator masses have even been obtained by fitting the Higgs correlator over the
full range of $\Delta t$, one finds that the obtained results on $m_{Hc}$ are perfectly consistent 
with the corresponding effective masses, thus justifying the chosen method of determining the correlator
masses.

%\includeFigTrippleDouble
%{HiggsCorrelatorKap012313L32}{HiggsCorrelatorEffectiveMassesKap012313L32}{HiggsCorrelatorEffectiveMassesWFCKap012313L32}
%{HiggsCorrelatorDOTSKap012313L32}{HiggsCorrelatorDOTSEffectiveMassesKap012313L32}{HiggsCorrelatorDOTSEffectiveMassesWFCKap012313L32}
\includeFigTrippleDouble
{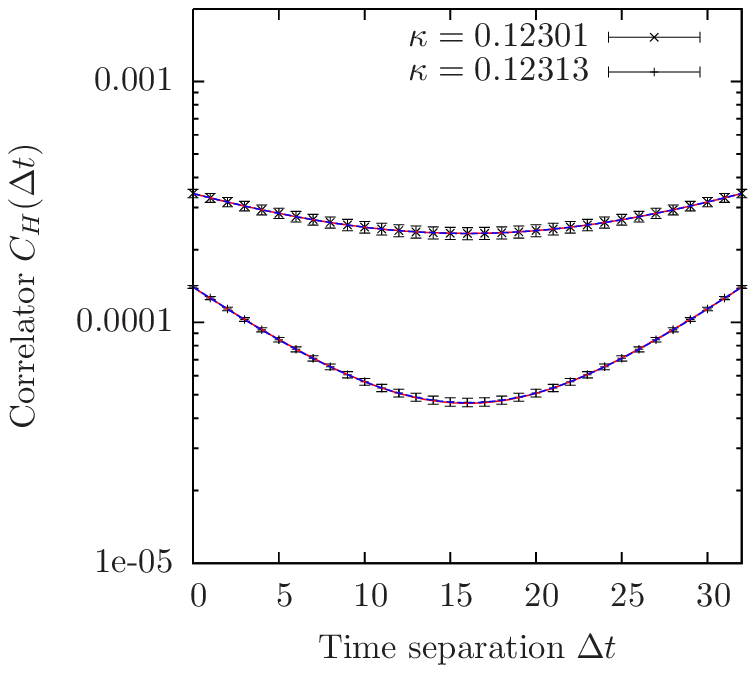}{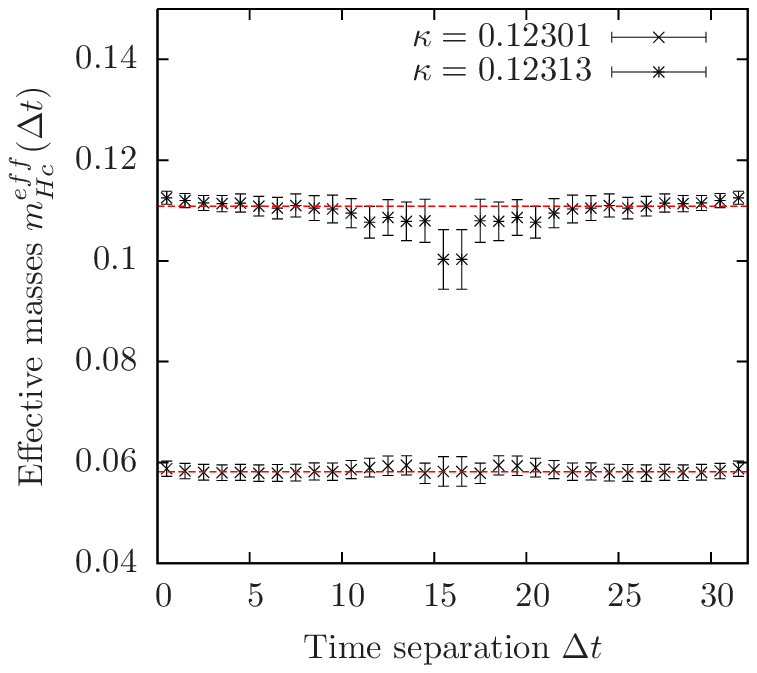}{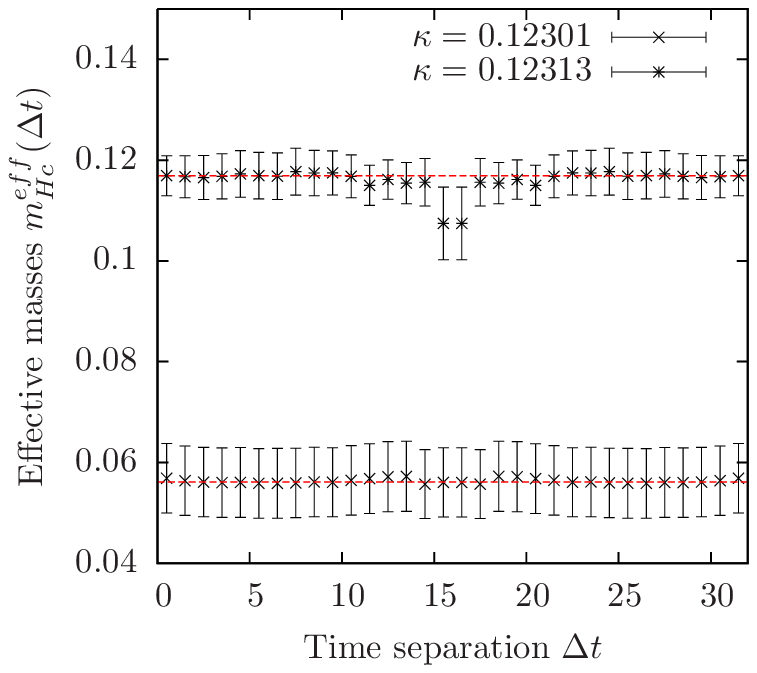}
{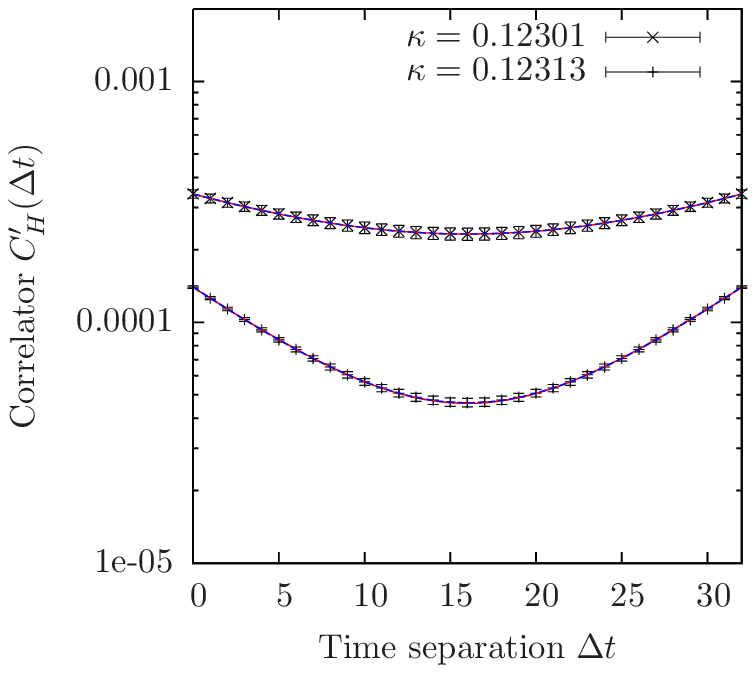}{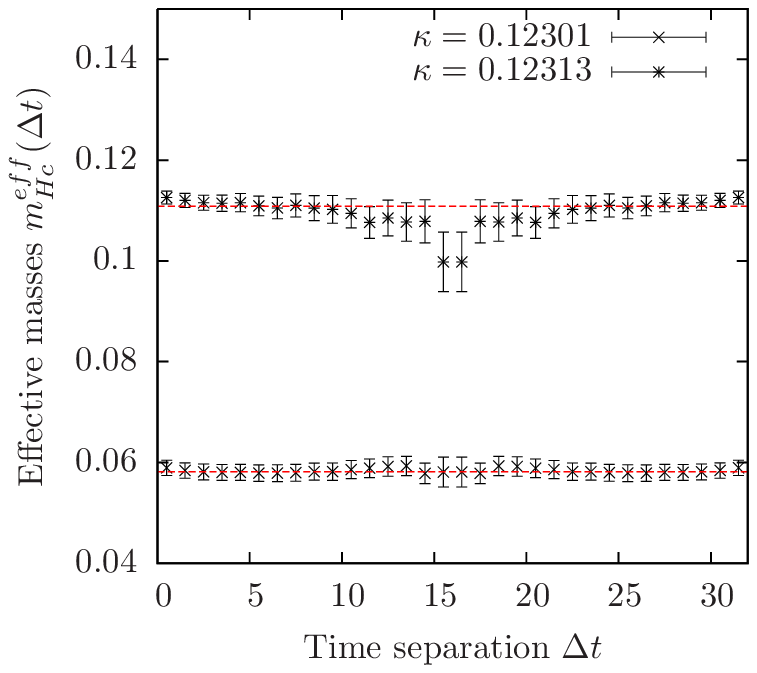}{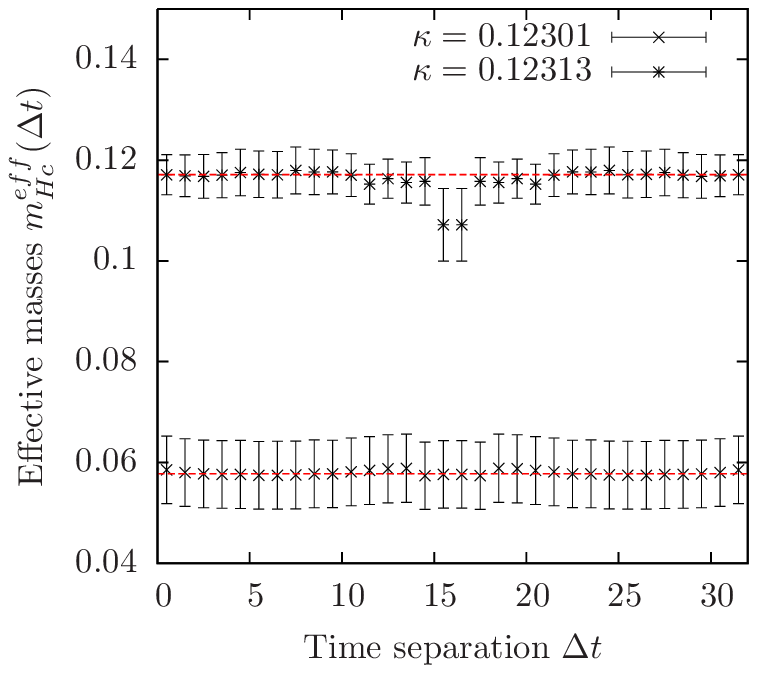}
{fig:HiggsCorrelatorExamplesAtWeakCoup}
{The Higgs time-slice correlation functions calculated in the Monte-Carlo runs specified in 
\tab{tab:Chap61EmployedRuns} are shown in panels (a) and (d) together with the respective fits 
obtained by the fit approaches in \eq{eq:HiggsTimeCorrelatorFitAnsatz1} and \eq{eq:HiggsTimeCorrelatorFitAnsatz2}
as depicted by the here almost coinciding dashed and solid curves, respectively. The corresponding effective masses are presented 
in the other four panels. The results depicted in the upper row belong to the correlation function $C_H(\Delta t)$, while the lower 
row refers to $C'_H(\Delta t)$. The effective masses in the middle column have been determined by \eq{eq:DefOfEffMassesForHiggs}
with $C=0$, while in panels (c) and (f) the constant $C$ was taken from
the fit ansatz $f_{A,m_{Hc},C}(\Delta t)$ to respect contaminations by lighter states.
The dashed horizontal lines depict the respective correlator masses $m_{Hc}$ obtained in the aforementioned fit procedures.
}
{Examples of Higgs time-slice correlation functions at small quartic coupling constants.}

The resulting Higgs correlator masses determined in the described manner are listed
in \tab{tab:SummaryOfHiggsPropCorrComp}. One learns from these findings that the mass
obtained from the non-local operator $O_H(t)$ and the local operator $O'_H(t)$ coincide
within the error bars here. From this observation it is concluded that it is indeed justified 
to extract the Higgs boson mass from the correlation function $C_H(\Delta t)$ associated to 
the non-local operator $O_H(t)$ as originally proposed in \sect{sec:SimStratAndObs}
at least in the weakly interacting regime of the considered model. 

\includeTab{|c|c|c|c|}{
                  &Correlator        &  $m_{Hc}$ from $f_{A,m_{Hc},C}(\Delta t)$ &  $m_{Hc}$ from $f_{A,m_{Hc}}(\Delta t)$\\ \hline
$\kappa=0.12301$  &   $C_H(\Delta t)$   &  0.056(7)                      &    0.058(2)                    \\
$\kappa=0.12301$  &   $C'_H(\Delta t)$  &  0.058(7)                      &    0.058(2)                    \\
$\kappa=0.12313$  &   $C_H(\Delta t)$   &  0.117(4)                      &    0.111(2)                     \\
$\kappa=0.12313$  &   $C'_H(\Delta t)$  &  0.117(4)                      &    0.111(2)                    \\
}
{tab:SummaryOfHiggsPropCorrComp}
{The Higgs correlator masses $m_{Hc}$ are listed as obtained in the fit procedures of the correlation
functions $C_H(\Delta t)$ and $C'_H(\Delta t)$ applying the fit approaches in \eq{eq:HiggsTimeCorrelatorFitAnsatz1}
and \eq{eq:HiggsTimeCorrelatorFitAnsatz2}, respectively. The underlying correlation functions have been
calculated numerically in the Monte-Carlo runs specified in \tab{tab:Chap61EmployedRuns}.
}
{Comparison of different extraction methods for the Higgs time-slice correlator mass at small quartic coupling constants.}

Furthermore, one sees in \tab{tab:SummaryOfHiggsPropCorrComp} that the masses obtained from 
the fit approach in \eq{eq:HiggsTimeCorrelatorFitAnsatz1}, which accounts for contributions
of massless states, seem to be somewhat larger than those determined by \eq{eq:HiggsTimeCorrelatorFitAnsatz2},
if at all distinguishable with respect to the given statistical uncertainties.
This is what one would have expected. However, the first approach is much more unstable than
the second ansatz which is manifest in the larger statistical uncertainty of the corresponding
results. Moreover, for the most cases the correlator masses determined with and without including
a constant term into the fit ansatz have turned out to be not clearly distinguishable in practice 
with respect to the given errors. 

One can therefore conjecture that it is favourable to use the less flexible
fitting procedure in \eq{eq:HiggsTimeCorrelatorFitAnsatz2} to extract the Higgs boson mass
from the correlation function due to the better stability properties of that ansatz. Furthermore,
we will use $C_H(\Delta t)$ as the underlying correlation function in the following. 
This is the approach that will be applied for the rest of this chapter.

%-------------------------------------------------------------------------------------------------------------------- 
\subsection{Analysis of the Goldstone propagator}
\label{sec:GoldPropAnalysisLowerBound}

The Goldstone renormalization constant $Z_G$ is required for determining the renormalized vacuum expectation
value of the scalar field $v_r$. It is thus needed for the fixation of the physical scale $a^{-1}$ of a given Monte-Carlo run
according to \eq{eq:FixationOfPhysScale}. This renormalization constant has been defined in \eq{eq:DefOfRenormalFactors} 
through a derivative of the inverse Goldstone propagator. As already pointed out in \sect{sec:SimStratAndObs} computing 
this derivative requires an analytical continuation $\tilde G_G^{c}(p_c)$ of the discrete lattice propagator, which was proposed to be 
obtained via a fit of the discrete lattice data. 

%\includeFigSingleMedium{LPTDiagramsForPurePhi4GoldstonePropagator}
\includeFigSingleMedium{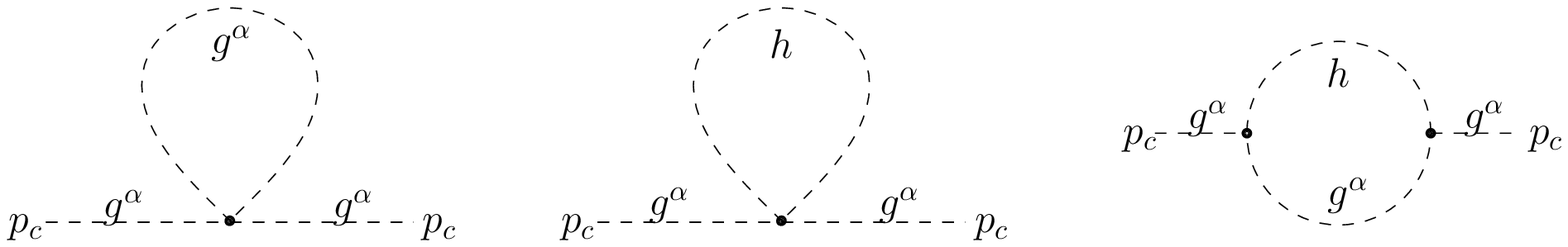}
{fig:GoldstonePropDiagrams}
{Illustration of the diagrams that contribute to the continuous space-time Goldstone propagator $\tilde G_G(p_c)$
in the Euclidean pure $\Phi^4$-theory at one-loop order.
}
{Diagrams contributing to the continuous space-time Goldstone propagator in the pure $\Phi^4$-theory at one-loop order.}

The idea here is to construct an appropriate fit function $f_G(p)$ based on a perturbative calculation of the Goldstone
propagator $\tilde G_G(p_c)$ in continuous Euclidean space-time. For simplicity the consideration is again restricted to the pure 
$\Phi^4$-theory. To one-loop order the only momentum dependent contribution to the Goldstone propagator is given
by the mixed Higgs-Goldstone loop illustrated on the right-hand side of \fig{fig:GoldstonePropDiagrams}. Adopting the 
same steps as detailed in \sect{sec:UnstableSignature} to the case of the considered Higgs-Goldstone loop, one can 
establish the result for the renormalized Goldstone propagator at one-loop order yielding 
\bea
\label{eq:GoldstonePropRenPT}
\tilde G^{-1}_G(p_c) &=& p_c^2 + m^2_{G} + 8\pi^{-2}\lambda_r^2 v_r^2 \cdot \left[
\OneGoldstoneLoopContEuc(p_c^2,m_H^2,m_G^2)  - \OneGoldstoneLoopContEuc(-m_{G}^2,m_H^2,m_G^2)\right] 
\eea
where the one-loop contribution $\OneGoldstoneLoopContEuc(p_c^2)$ is given as
\bea
2\OneGoldstoneLoopContEuc(p_c^2,m_H^2,m_G^2) &=& \frac{\sqrt{q}}{p_c^2} \cdot \arctanh\left( \frac{p_c^2+m^2_{G}-m^2_{H}}{\sqrt{q}} \right) 
+ \frac{m^2_{G} - m^2_{H}}{2p_c^2}\cdot \log\left( \frac{m^2_{H}}{m^2_{G}}\right)\quad \quad \\
&+&\frac{\sqrt{q}}{p_c^2} \cdot \arctanh\left(\frac{p^2+m^2_{H}-m^2_{G}}{\sqrt{q}}\right) \quad \mbox{with}\nonumber\\
q &=& \left(m^2_{G} - m^2_{H} + p_c^2  \right)^2 + 4m^2_{H} p_c^2.
\eea
The different structure of the one-loop contribution $\OneGoldstoneLoopContEuc(p_c^2,m_H^2,m_G^2)$ as compared to the corresponding 
result for the pure Higgs loop contribution $\OneBosonLoopContEuc(p_c^2,m_H^2)$ given in \eq{eq:DefOfContEuc1LoopBosContrib} is 
caused by the mixture of the different masses $m_H$ and $m_G$ appearing in the mixed Higgs-Goldstone loop illustrated in 
\fig{fig:GoldstonePropDiagrams}. For equal Higgs and Goldstone masses, \ie $m_G = m_H$, the given expression for the loop 
contribution $\OneGoldstoneLoopContEuc(p_c^2,m_H^2,m_G^2)$ recovers the result for the pure Higgs loop 
$\OneBosonLoopContEuc(p_c^2,m_H^2)$, as expected. For $m_G\neq 0$ the given formula can moreover be shown to be 
finite\footnote{More precisely, the given function converges to a finite constant in the limit $p_c\rightarrow 0$.} 
at $p_c=0$, as desired. 

In principle one can directly employ the expression in \eq{eq:GoldstonePropRenPT} as the sought-after fit function $f^{-1}_G(p)$. 
For clarification it is remarked at this point that instead of fitting the lattice propagator $\tilde G_G(p)$ itself with $f_G(p)$,
it is always the inverse propagator $\tilde G_G^{-1}(p)$ that is fitted with $f_G^{-1}(p)\equiv 1/f_G(p)$ in the following. However, 
for the actual fit procedure of the lattice data a modified version of \eq{eq:GoldstonePropRenPT} is used given as\footnote{According
to the symmetries of \eq{eq:GoldstoneFitAnsatz} one has $f_G(p)\equiv f_G(p^2)$, where appropriate mappings are implicitly assumed.
This type of shorthand notation will extensively be used in the following.}
\beq
\label{eq:GoldstoneFitAnsatz}
f^{-1}_G(p^2) = \frac{p^2 + \bar m^2_{G} +  A\cdot \left[ \OneGoldstoneLoopContEuc(p^2,\bar m_H^2, \bar m_G^2)  - 
\OneGoldstoneLoopContEuc(0,\bar m_H^2, \bar m_G^2)\right]}{Z_0},
\eeq
where $A$, $Z_0$, $\bar m_G$, and $\bar m_H$ are the free fit parameters. Two modifications have been applied here to the original result.
Firstly, the constant term $\OneGoldstoneLoopContEuc(-\bar m_{G}^2,\bar m_H^2,\bar m_G^2)$ in \eq{eq:GoldstonePropRenPT} has been replaced
by $\OneGoldstoneLoopContEuc(0,\bar m_H^2,\bar m_G^2)$ simply for convenience. These changes are 
nothing but a reparametrization of the fit parameters, which is the reason why the masses $\bar m_H^2$ and $\bar m_G^2$ are not denoted 
here as the Higgs and Goldstone boson masses $m^2_H$ and $m_G^2$, respectively. Strictly speaking, the different appearances of $\bar m_G^2$
in \eq{eq:GoldstoneFitAnsatz} now refer to slightly different numbers in this reparametrization which are, however, identified here with
each other for the sake of stability. Since the Goldstone mass is close to zero anyhow, this simplification is insignificant for a 
practical fit procedure. For clarification it is pointed out that in the here presented approach the Goldstone mass $m_G$ is actually not 
determined through the nominal value of the latter fit parameter $\bar m_G$ itself, but through the pole of the resulting analytical 
continuation $\tilde G_G^{c}(p_c)$ according to \eq{eq:DefOfHiggsAndGoldstoneMassByPole}.

More interestingly, however, a global factor $Z_0$ has been introduced in the denominator 
of \eq{eq:GoldstoneFitAnsatz} in the spirit of renormalization constant. This modification is purely heuristic and its sole 
purpose is to provide an effective description of the observed propagator, which is all we need at this point. It can, however, 
physically be motivated. For that purpose it is recalled that the given analytical result for the Goldstone propagator in 
\eq{eq:GoldstonePropRenPT} has been derived for the pure $\Phi^4$-theory only. In the following, however, the given fit 
ansatz will be applied to the full Higgs-Yukawa sector, including contributions from
fermions. These fermionic contributions will influence the overall slope of the propagator in addition to the 
effect of the already accounted Higgs-Goldstone loop. It will also change the Goldstone mass itself, but such a shift is 
already covered by the fit parameter $m_G$. The intention of the factor $Z_0$ is thus to make the given ansatz sufficiently flexible to 
allow for its application also in the presence of fermionic contributions to the Goldstone propagator. 

The alert reader may wonder, whether one should rather construct a fit ansatz from the renormalized result of the Goldstone
propagator derived in the full Higgs-Yukawa sector. This would indeed place the fit procedure on an even better conceptual footing.
However, it will turn out, that the given ansatz already works satisfactorily well for our purpose. 

More important seems to be the question what part of the lattice Goldstone propagator $\tilde G_G(p)$ should actually be included
into the fit procedure. It was already pointed out in \sect{sec:SimStratAndObs} that the consideration of the lattice propagator 
has to be restricted to small lattice momenta in order to suppress contaminations arising from discretization effects. For that purpose
the constant $\gamma$ was introduced specifying the set of momenta underlying the fit approach according to $\hat p^2 \le \gamma$.
In principle, one would want to choose $\gamma$ as small as possible. In practice, however, the fit procedure becomes increasingly
unstable when lowering the value of $\gamma$. In the following examples we will consider the settings $\gamma=1$, $\gamma=2$, 
and $\gamma=4$. To make the discretization effects associated to these not particularly small values of $\gamma$ less prominent in 
the intended fit procedure, the inverse lattice propagator $\tilde G^{-1}_G(p)$ is fitted with $f^{-1}_G(\hat p^2)$ instead of $f^{-1}_G(p^2)$, 
being a function of the squared lattice momentum $\hat p^2$, which is completely justified in the limit $\gamma\rightarrow 0$.

%\includeFigTrippleDouble
%{GoldstonePropagatorKap012301L32Pmax16}{GoldstonePropagatorKap012301L32Pmax1}{GoldstonePropagatorKap012301L32PmaxSmallest}
%{GoldstonePropagatorKap012313L32Pmax16}{GoldstonePropagatorKap012313L32Pmax1}{GoldstonePropagatorKap012313L32PmaxSmallest}
\includeFigTrippleDouble
{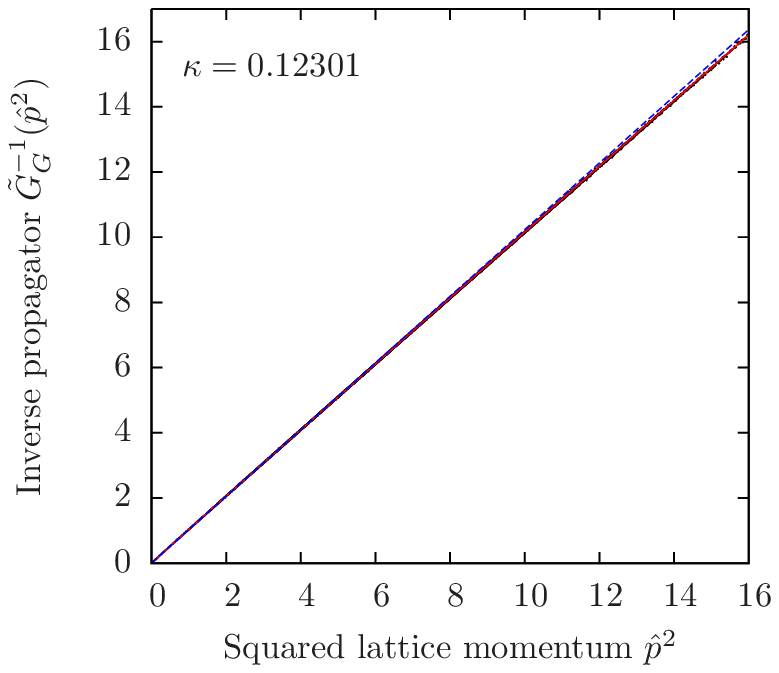}{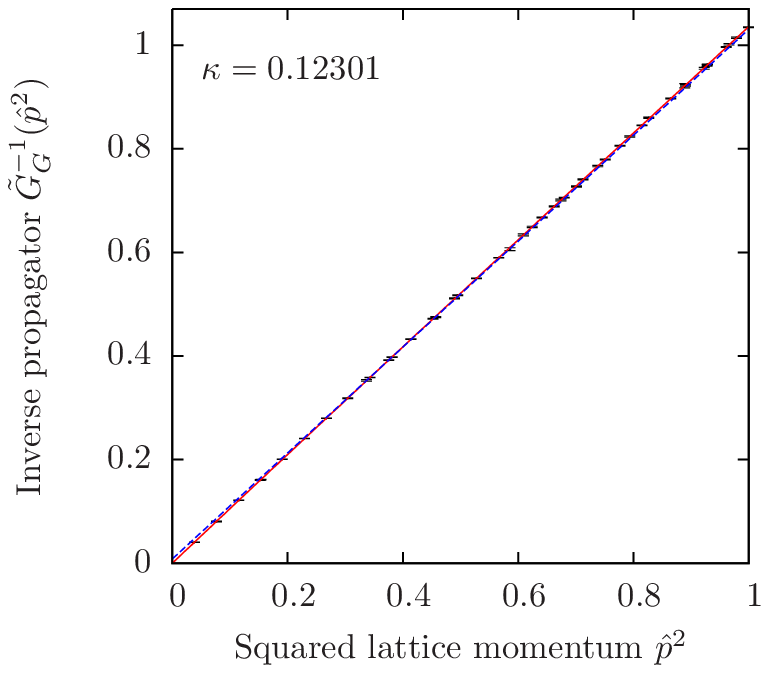}{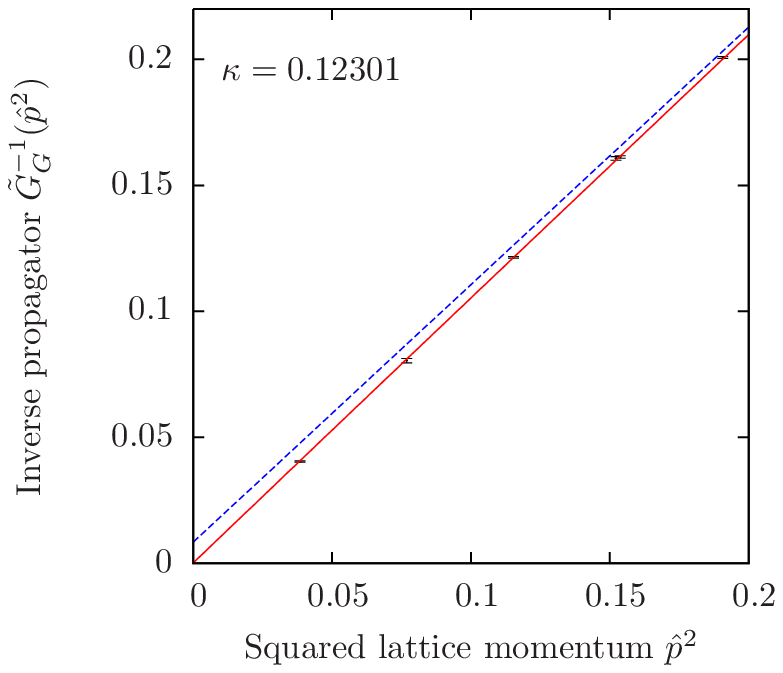}
{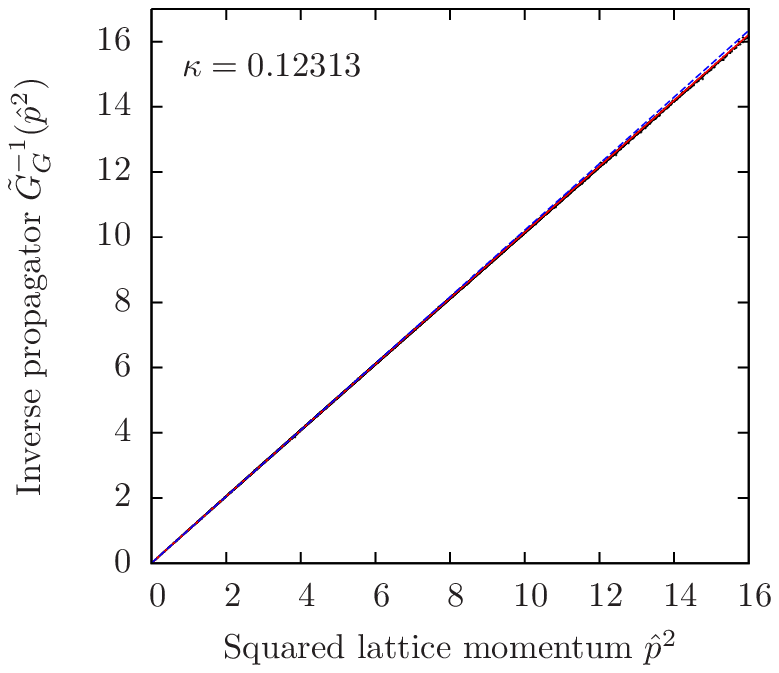}{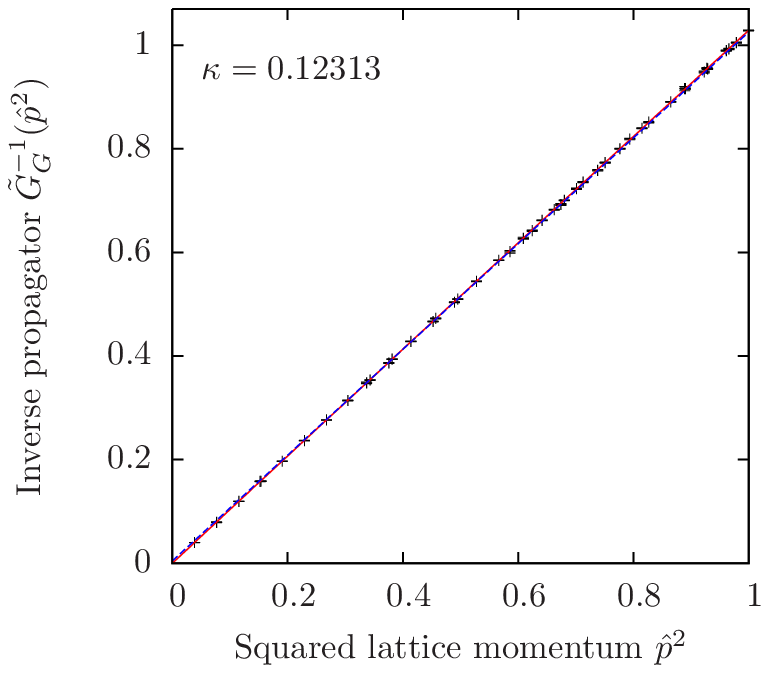}{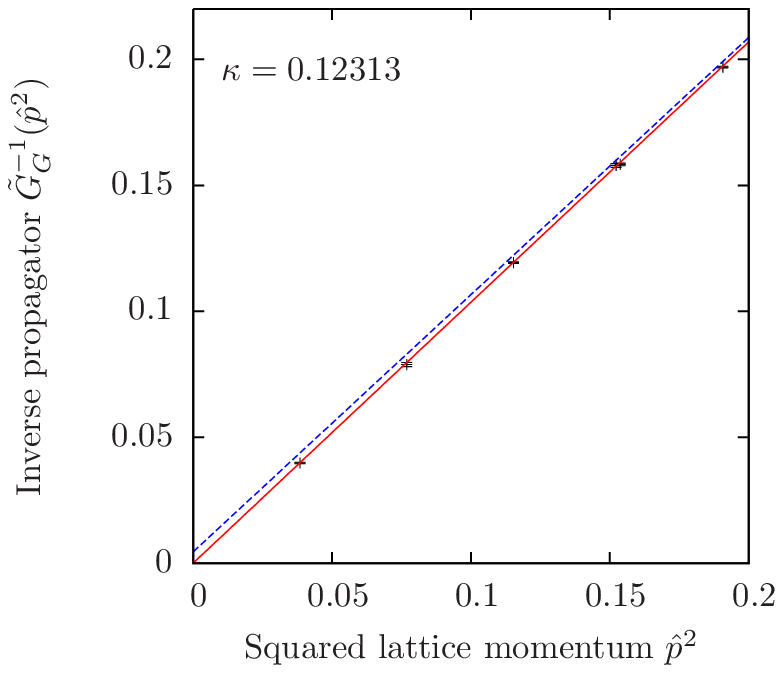}
{fig:GoldstonePropExample}
{The inverse lattice Goldstone propagators calculated in the Monte-Carlo runs specified in \tab{tab:Chap61EmployedRuns}
are presented versus the squared lattice momenta $\hat p^2$ together with the respective fits obtained from the 
fit approaches $f^{-1}_G(\hat p^2)$ (red solid line) and $l^{-1}_G(\hat p^2)$ (blue dashed line) for the setting $\gamma=4$. From 
left to right the three panel columns display the same data zooming in, however, on the vicinity of the origin at 
$\hat p^2 = 0$.}
{Examples of Goldstone propagators at small quartic coupling constants. }

Examples of the given fit procedure are presented in \fig{fig:GoldstonePropExample} where the inverse lattice Goldstone propagators as obtained
in the Monte-Carlo runs specified in \tab{tab:Chap61EmployedRuns} are shown together with their respective fits according to 
\eq{eq:GoldstoneFitAnsatz}. For a clearer demonstration of the quality of the applied fit ansatz $f_G(\hat p^2)$ the presented numerical data 
are also compared to the linear fit function
\beq
\label{eq:GoldstoneFitAnsatzLinear}
l_G^{-1}(\hat p^2) = \frac{\hat p^2 + m_G^2}{Z_G},
\eeq
where $m_G$ and $Z_G$ are the only free fit parameters. One learns already from this graphical comparison that the linear fit ansatz
is not suited for an adequate description of the Goldstone propagator at small momenta due to its pronounced curvature, being clearly 
observable thanks to the very small statistical errors of the lattice data. This becomes even more apparent when one considers the
corresponding values of the average squared residual per degree of freedom $\chi^2/dof$ listed in \tab{tab:GoldstonePropExampleResults}.
While this quantity is close to its expected value of one for the fit ansatz $f_G(\hat p^2)$, moreover approximately independent of the 
chosen value of $\gamma$ as desired, the corresponding result associated to the application of the linear fit approach $l_G(\hat p^2)$ is 
much larger and rises clearly with increasing values of $\gamma$.

\includeTabNoHLines{|c|c|c|c|c|c|c|c|}{
\cline{3-8}
\multicolumn{2}{c|}{}& \multicolumn{3}{c|}{fit ansatz $f_G(\hat p^2)$} &  \multicolumn{3}{c|}{linear fit ansatz $l_G(\hat p^2)$}\\ \hline
$\kappa$             & $\gamma$   & $Z_{G}$     & $m_{G}$    &  $\chi^2/dof$  & $Z_{G}$   & $m_{G}$   &  $\chi^2/dof$ \\ \hline
$0.12301$     & $1.0$        & 0.9641(37)  & 0.000(1)   &  0.71          & 0.9655(4) & 0.048(2)  &  2.94 \\
$0.12301$     & $2.0$        & 0.9654(45)  & 0.000(1)   &  0.92          & 0.9720(2) & 0.067(1)  &  5.22 \\
$0.12301$     & $4.0$        & 0.9699(17)  & 0.010(18)  &  0.95          & 0.9781(1) & 0.090(1)  &  6 94 \\\hline
$0.12313$     & $1.0$        & 0.9722(140) & 0.000(1)   &  0.90          & 0.9717(4) & 0.027(3)  &  1.25 \\
$0.12313$     & $2.0$        & 0.9775(44)  & 0.000(3)   &  0.91          & 0.9754(2) & 0.045(1)  &  2.44 \\
$0.12313$     & $4.0$        & 0.9784(65)  & 0.004(10)  &  1.09          & 0.9796(1) & 0.066(1)  &  4.01 \\\hline
}
{tab:GoldstonePropExampleResults}
{The results on the Goldstone renormalization factor $Z_G$ and the Goldstone mass $m_G$ obtained from
the fit approaches $f_G(\hat p^2)$ and $l_G(\hat p^2)$ as defined in \eq{eq:GoldstoneFitAnsatz} 
and \eq{eq:GoldstoneFitAnsatzLinear} are listed for several settings of the parameter 
$\gamma$ together with the corresponding average squared residual per degree of freedom $\chi^2/dof$
associated to the respective fits. The underlying Goldstone lattice propagators have been calculated
in the Monte-Carlo runs specified in \tab{tab:Chap61EmployedRuns}.
}
{Comparison of the Goldstone propagator properties obtained from different extraction schemes at small quartic coupling
constants.}

The Goldstone mass $m_G$ and the respective renormalization factor $Z_G$ can then be obtained according to \eq{eq:DefOfHiggsAndGoldstoneMassByPole}
and \eq{eq:DefOfRenormalFactors} from the analytical continuation of the lattice Goldstone propagator given by $\tilde G^{c}_G(p_c) = f_G(p^2_c)$
and $\tilde G^{c}_G(p_c) = l_G(p^2_c)$, respectively. The results corresponding to the various settings of $\gamma$ are listed
in \tab{tab:GoldstonePropExampleResults}. While the statistical errors associated to the numbers obtained from the
linear fit ansatz seem to be clearly favourable in comparison to the findings derived from the fit ansatz $f_G(\hat p^2)$,
the results arising from the linear ansatz $l_G(\hat p^2)$ are inconsistent for the different settings of $\gamma$. Especially the
obtained Goldstone mass $m_G$ depends significantly on the chosen fit interval specified by the value of $\gamma$. 
The linear fit ansatz $l_G(\hat p^2)$ can thus not be applied for a reliable determination of the aforementioned Goldstone 
propagator properties.

This is in contrast to the situation of the more elaborate fit ansatz $f_G(\hat p^2)$. All results derived by applying
this latter approach are consistent with respect to the given errors. This observed freedom in the
choice of $\gamma$ will be used to select the rather large setting $\gamma=4$ for the determination of the Goldstone
renormalization constant $Z_G$ in the later lattice calculations. The reason is that the given fit ansatz $f_G(\hat p^2)$ 
tends to become somewhat unstable on small lattices with low statistics and small values of $\gamma$. Choosing $\gamma=4$ 
then allows for an unique approach applicable to all considered lattice calculations.

%-------------------------------------------------------------------------------------------------------------------- 
\subsection{Analysis of the Higgs propagator}
\label{sec:HiggsPropAnalysisLowerBound}

We now turn to the analysis of the lattice Higgs propagator. In principle we will follow the same steps already
discussed in the preceding section. Again the idea is to determine an analytical continuation $\tilde G_H^{c}(p_c)$
of the lattice propagator $\tilde G_H(p)$ by fitting the latter to some adequate fit function $f_H(p)$. In \sect{sec:UnstableSignature}
we have already derived the perturbative one-loop result for the Higgs propagator of the pure $\Phi^4$-theory  
in continuous Euclidean space-time. Here, we will construct the sought-after fit function $f_H(p)$ based on that 
earlier result given in \eq{eq:StructureOfPropInNCompPhi4TheoryRenormalized}. 

However, with the same motivation already discussed for the case of the Goldstone propagator in the preceding section 
the original result in \eq{eq:StructureOfPropInNCompPhi4TheoryRenormalized} is slightly modified, leading then to 
the fit function\footnote{Again the inverse lattice propagator $\tilde G^{-1}_H(p)$ will be fitted with the inverse
expression $f_H^{-1}(p)$ instead of fitting $\tilde G_H(p)$ with $f_H(p)$.}
\bea
\label{eq:HiggsPropFitAnsatz}
f^{-1}_H(p^2) &=& \frac{p^2 + \bar m_H^2
+ A\cdot \left[ 36 \left( \OneBosonLoopContEuc(p^2, \bar m_H^2) - D_{H0}\right)  
+ 12 \left( \OneBosonLoopContEuc(p^2, \bar m_G^2) - D_{G0}\right) \right] }{Z_0},  \quad\quad
\eea
with the free fit parameters $\bar m_H^2$, $\bar m_G^2$, $A$, and $Z_0$. One of the aforementioned modifications
is that the constants $D_{H0}$ and $D_{G0}$ are given here as the loop contributions $\OneBosonLoopContEuc(p^2, \bar m_H^2)$
and $\OneBosonLoopContEuc(p^2, \bar m_G^2)$ evaluated at the origin $p^2=0$ instead of the location of the pole according to
\bea
D_{H0} &=& \OneBosonLoopContEuc(0, \bar m_H^2) = 1, \\
D_{G0} &=& \OneBosonLoopContEuc(0, \bar m_G^2) = 1,
\eea
which simplifies the practical fit procedure. Consequently, the explicit appearance of the imaginary contribution $-im_H\Gamma_H$
in \eq{eq:StructureOfPropInNCompPhi4TheoryRenormalized} 
is exactly canceled. These changes are nothing but a reparametrization of the fit parameters, which is the reason why the mass
$\bar m_H^2$ is not denoted here as the Higgs boson mass $m^2_H$. The only real discrepancy between the fit ansatz given in \eq{eq:HiggsPropFitAnsatz}
and the original perturbative result in \eq{eq:StructureOfPropInNCompPhi4TheoryRenormalized}, except for the introduction of
the overall factor $Z_0$, is that the two appearances of the parameter $\bar m_H^2$ in \eq{eq:HiggsPropFitAnsatz} would actually 
refer to two different parameters in this setup. For the sake of stability, however, these two parameters have been identified 
with each other here and are jointly referred to as $\bar m_H^2$. 

The other modification is the introduction of the additional fit parameter $Z_0$ which enters the given formula in \eq{eq:HiggsPropFitAnsatz}
in the spirit of a renormalization constant. Again this modification is purely heuristic with the intention to provide an effective
description of the observed propagator. The motivation for its introduction is the same as for the case of the previously discussed
Goldstone propagator.

Examples for the quality of the specified fit approach are shown in \fig{fig:HiggsPropExample}. Here the given fit function
was applied to the numerical lattice Higgs propagator calculated in the Monte-Carlo runs listed in \tab{tab:Chap61EmployedRuns}. 
Following the same rationale given in the preceding section the inverse lattice propagators are actually fitted with $f^{-1}_H(\hat p^2)$, 
being a function of the squared lattice momentum $\hat p^2$. In the presented plots the threshold value for
the fit interval was set to $\gamma=1$, which is a reasonable choice as discussed below, and one finds the performed fit 
to describe the numerical data very well. In addition, the obtained fit is compared to the naive ansatz of a linear
fit approach given as
\beq
\label{eq:HiggsPropFitAnsatzLinear}
l^{-1}_H(\hat p^2) = \frac{\hat p^2 + m_H^2}{Z_H},
\eeq
which is also presented in \fig{fig:HiggsPropExample}. In contrast to the preceding section, the linear fit also provides a
very good description of the numerical data, which is due to the threshold value $\gamma$ being chosen much smaller
here than for the analysis of the Goldstone propagator. However, when one considers the corresponding average squared residuals per 
degree of freedom $\chi/dof$ listed in \tab{tab:HiggsPropExampleResults} for a sequence of several different values of the 
parameter $\gamma$ one learns that the more elaborate fit ansatz in \eq{eq:HiggsPropFitAnsatz} yields significantly
better fit curves than $l_H(\hat p^2)$, as the threshold value $\gamma$ is increased.

%\includeFigTrippleDouble
%{HiggsPropagatorKap012301L32Pmax16}{HiggsPropagatorKap012301L32Pmax1}{HiggsPropagatorKap012301L32PmaxSmallest}
%{HiggsPropagatorKap012313L32Pmax16}{HiggsPropagatorKap012313L32Pmax1}{HiggsPropagatorKap012313L32PmaxSmallest}
\includeFigTrippleDouble
{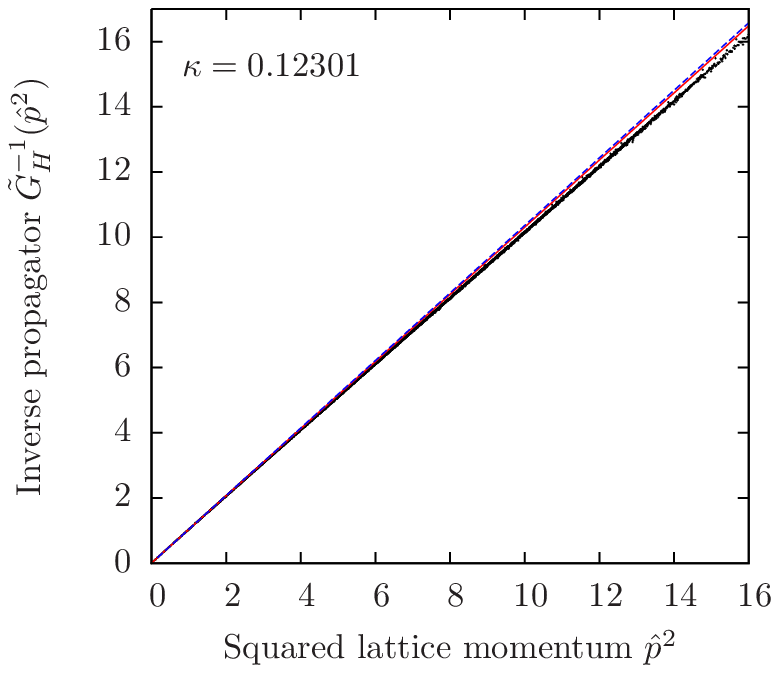}{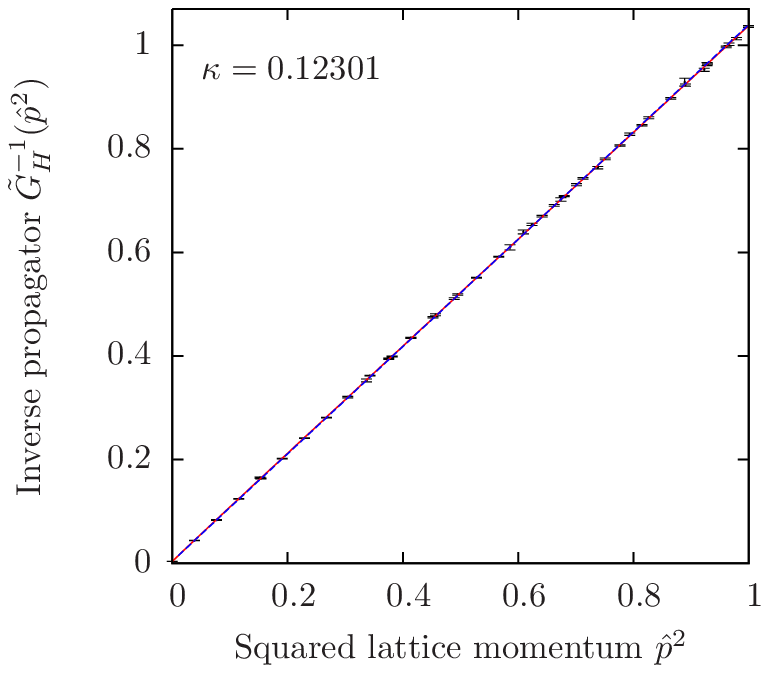}{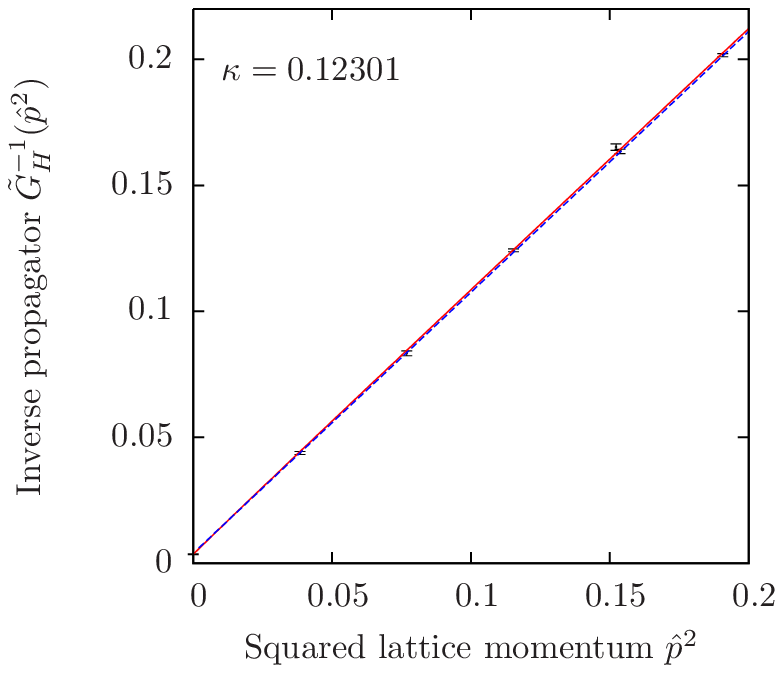}
{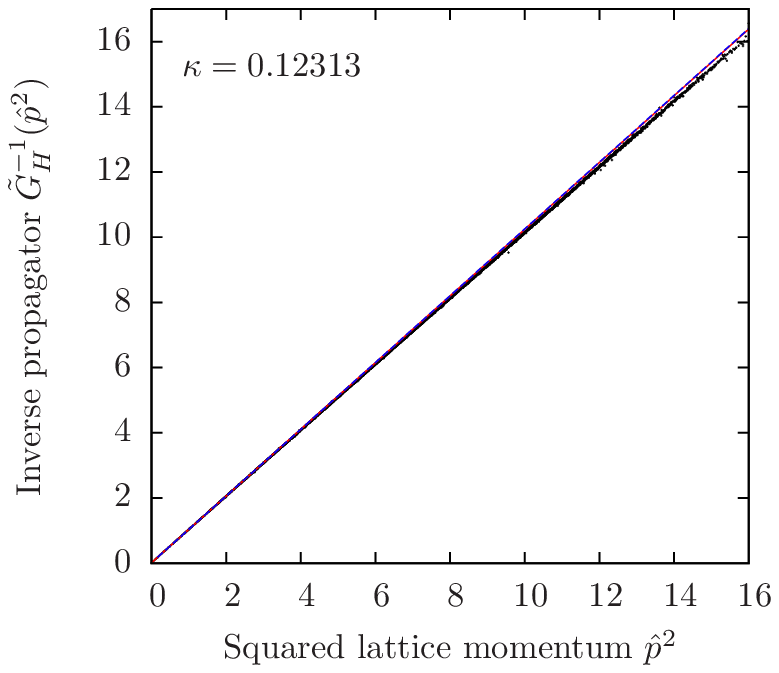}{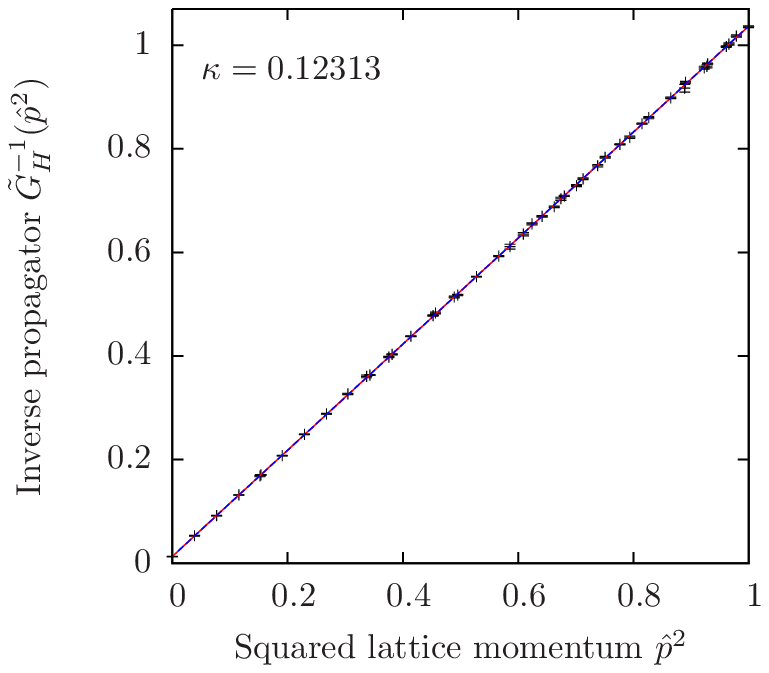}{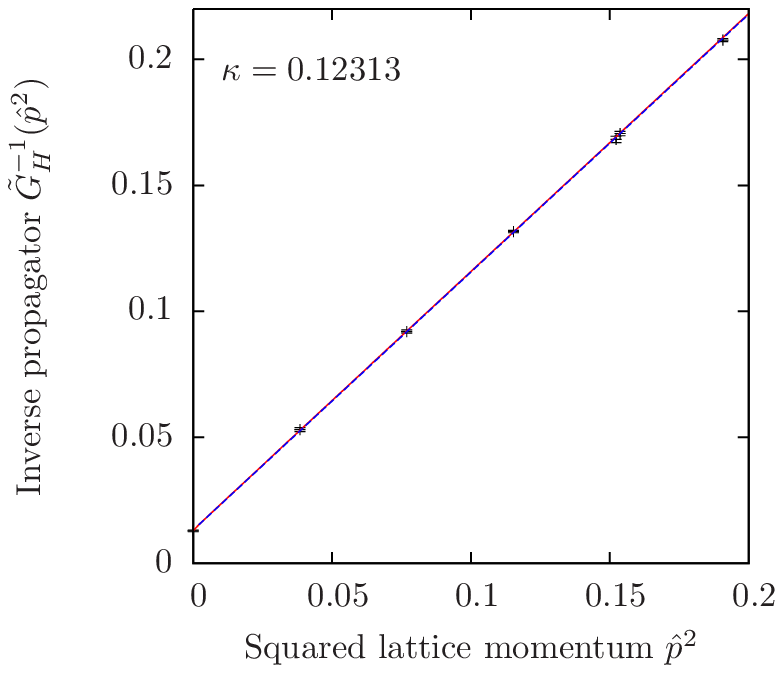}
{fig:HiggsPropExample}
{The inverse lattice Higgs propagators calculated in the Monte-Carlo runs specified in \tab{tab:Chap61EmployedRuns}
are presented versus the squared lattice momenta $\hat p^2$ together with the respective fits obtained from the 
fit approaches $f^{-1}_H(\hat p^2)$ (red solid line) and $l^{-1}_H(\hat p^2)$ (blue dashed line) for a threshold value of 
$\gamma=1$. From left to right the three panel columns display the same data zooming in, however, on the vicinity 
of the origin at $\hat p^2 = 0$.
}
{Examples of Higgs propagators at small quartic coupling constants.}

The Higgs propagator mass $m_{Hp}$ defined in \eq{eq:DefOfPropagatorMinkMass} and the Higgs pole mass $m_H$ together with its associated decay 
width $\Gamma_H$ given by the pole of the propagator on the second Riemann sheet according to \eq{eq:DefOfHiggsAndGoldstoneMassByPole} can then 
be obtained by defining the analytical continuation of the lattice propagator as $\tilde G_H^{c}(p_c) = f_H(p^2_c)$ and 
$\tilde G_H^{c}(p_c) = l_H(p^2_c)$, respectively. The results corresponding to the various performed fit procedures are listed in 
\tab{tab:HiggsPropExampleResults}. Since the linear function $l_H(p^2_c)$ can not exhibit a branch cut structure, however, the pole mass 
equals the propagator mass and the decay width is identical to zero for this linear fit approach. That is the reason why only the mass 
$m_H$ is specified in the latter scenario. 

\includeTabNoHLines{|c|c|c|c|c|c|c|c|c|}{
\cline{4-9}
\multicolumn{3}{c|}{} &  \multicolumn{4}{c|}{Fit ansatz $f_H(\hat p^2)$} &  \multicolumn{2}{c|}{Fit ansatz $l_H(\hat p^2)$}  \\ \hline
$\kappa$ & $\gamma$   & $m_{Hc}$   & $m_{Hp}$  &  $m_{H}$       &  $\Gamma_H$ &  $\chi^2/dof$   &  $m_{H}$  &  $\chi^2/dof$    \\ \hline
0.12301  &  1.0       & 0.058(2)   & 0.056(2)  & 0.056(2)       & 0.000(0)    & 1.26            &  0.062(2)  & 2.01\\
0.12301  &  2.0       & 0.058(2)   & 0.055(1)  & 0.055(1)       & 0.000(0)    & 1.39            &  0.068(2)  & 3.24\\
0.12301  &  4.0       & 0.058(2)   & 0.052(1)  & 0.052(1)       & 0.000(0)    & 1.56            &  0.080(3)  & 4.05\\ \hline
0.12313  &  1.0       & 0.111(2)   & 0.114(1)  & 0.114(1)       & 0.001(1)    & 1.01            &  0.113(1)  & 1.03\\
0.12313  &  2.0       & 0.111(2)   & 0.110(1)  & 0.110(1)       & 0.000(6)    & 1.07            &  0.116(1)  & 1.27 \\
0.12313  &  4.0       & 0.111(2)   & 0.106(1)  & 0.106(1)       & 0.000(0)    & 1.12            &  0.122(1)  & 1.68\\ \hline 
}
{tab:HiggsPropExampleResults}
{The results on the Higgs propagator mass $m_{Hp}$, the Higgs pole mass $m_H$, and the Higgs decay
width $\Gamma_H$ obtained from the fit approaches $f_H(\hat p^2)$ and $l_H(\hat p^2)$ are listed for several 
settings of the parameter $\gamma$ together with the corresponding average squared residual per degree 
of freedom $\chi^2/dof$ associated to the respective fit. For the linear fit ansatz $l_H(\hat p^2)$
only the pole mass is presented, since one finds $m_{Hp}\equiv m_H$ and $\Gamma_H\equiv 0$ when constructing
the analytical continuation $\tilde G_H^c(p_c)$ through $l_H(p_c^2)$. These results are compared to the
Higgs correlator mass $m_{Hc}$ as determined in \sect{sec:HiggsTSCanalysisLowerBound}. The underlying 
Higgs lattice propagators have been calculated in the Monte-Carlo runs specified in \tab{tab:Chap61EmployedRuns}.
}
{Comparison of the Higgs propagator properties obtained from different extraction schemes at small quartic
coupling constants.}

From the presented results one learns that the pole and propagator masses $m_H$ and $m_{Hp}$ obtained from the more sophisticated fit 
ansatz $f_H(\hat p^2)$ agree satisfactorily well with each other at $\gamma=1$ and $\gamma=2$, where they are also consistent with 
corresponding value of the Higgs boson correlator mass $m_{Hc}$ that has been discussed in \sect{sec:HiggsTSCanalysisLowerBound}. At 
the largest here presented threshold value, \ie$\gamma=4$, the latter results deviate significantly from the aforementioned numbers 
making obvious the fact, that the threshold value $\gamma$ has to be chosen sufficiently small to obtain reliable results for the 
sought-after propagator properties. Comparing the two different fit approaches, however, one finds that the more elaborate ansatz 
$f_H(\hat p^2)$ yields more stable fit results than the simple linear approach as the threshold value $\gamma$ is varied at least 
for the presented example with the smaller hopping parameter. It is remarked at this point that the observed differences between 
the latter two fit procedures are in fact rather mild in the here considered scenario of small bare coupling constants. They will, 
however, become much more pronounced in the strongly interacting regime of the model as discussed in \sect{sec:HiggsPropAnalysisUpperBound}.

Though the Higgs boson mass will actually not be determined by directly analyzing the Higgs propagator in the subsequent investigation
of the lower mass bounds, but rather by studying the time-slice correlation function as detailed before, it is nevertheless 
concluded from the numbers presented in \tab{tab:HiggsPropExampleResults} that selecting the threshold value to be $\gamma=1$ 
would be a reasonable choice for the final evaluation of the propagator properties, as already assumed in \fig{fig:HiggsPropExample}. 

It is remarked that the conceptual footing of the applied fit approach $f_H(\hat p^2)$ could be further improved by including 
also the fermion contributions. However, from the above findings one can conjecture that the applied fit ansatz, 
albeit derived from the pure $\Phi^4$-theory only, already provides a sufficient means for determining the Higgs propagator and pole 
masses in a consistent manner, at least in the here considered weakly interacting regime of the Higgs-Yukawa model.

In principle, the chosen fit approach $f_H(\hat p^2)$ also allows to access the decay width according to \eq{eq:DefOfHiggsAndGoldstoneMassByPole}
by numerically searching the pole of the analytically continued Higgs propagator $\tilde G_H^{c}(p_c)$ on the second Riemann sheet.
The resulting findings for $\Gamma_H$ are also listed in \tab{tab:HiggsPropExampleResults}. It is remarked at this point that the
latter numbers are very sensitive to small variations in the fit procedure due to the relative large number of free fit parameters
considered here. However, all obtained numbers are fully consistent with zero within the given errors, which is in reassuring
agreement with the assumption that the Higgs boson can be treated as a stable particle in the weakly interacting regime of the 
considered model.

%-------------------------------------------------------------------------------------------------------------------- 
\subsection{Analysis of the fermion time-slice correlator}
\label{sec:FermionTSCanalysisLowerBound}

The fermionic time-slice correlation function $C_f(\Delta t)$ has been defined in \eq{eq:DefOfFermionTimeSliceCorr} and
it has already been pointed out that the given correlation function would be identical to zero due to the symmetries of
the considered model if it was evaluated in the background of the unrotated scalar field $\varphi$. The fermionic correlator
is therefore computed after rotating the field $\varphi$ as described in \sect{sec:SimStratAndObs}, \ie in the background of $\varphi^{rot}$. 
As an example the top quark correlation functions $C_t(\Delta t)$, obtained in the lattice calculations specified in 
\tab{tab:Chap61EmployedRuns}, are presented in \fig{fig:TopCorrelatorExample}a. The corresponding bottom quark correlators 
are not shown, since they yield identical results up to statistical fluctuations in the here considered mass degenerate setup.
As already remarked in \sect{sec:SimStratAndObs} the full {\textit{all-to-all}} correlator can trivially be computed, 
explaining the low statistical uncertainties of the given results.

%\includeFigTriple{TopCorrelatorKap012313L32}{TopCorrelatorEffectiveMassesKap012301L32}{TopCorrelatorEffectiveMassesKap012313L32}
\includeFigTriple{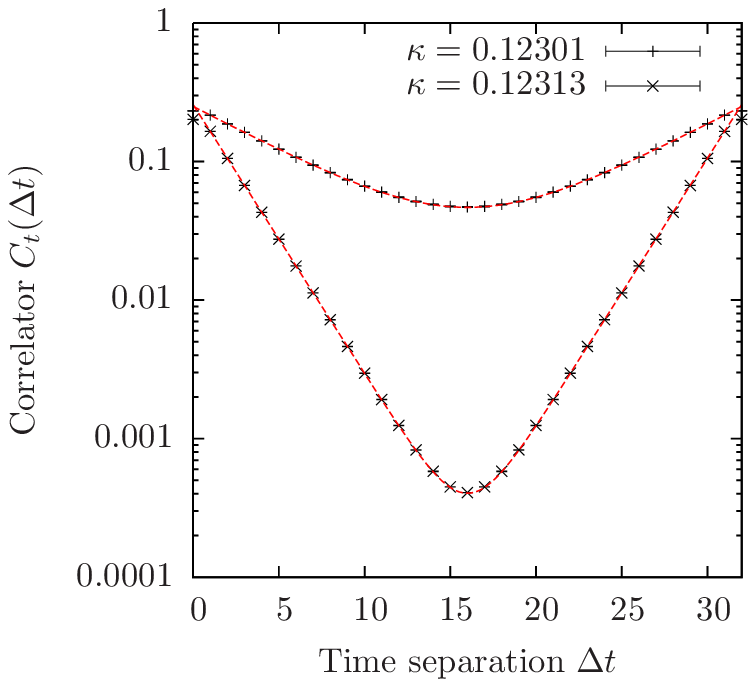}{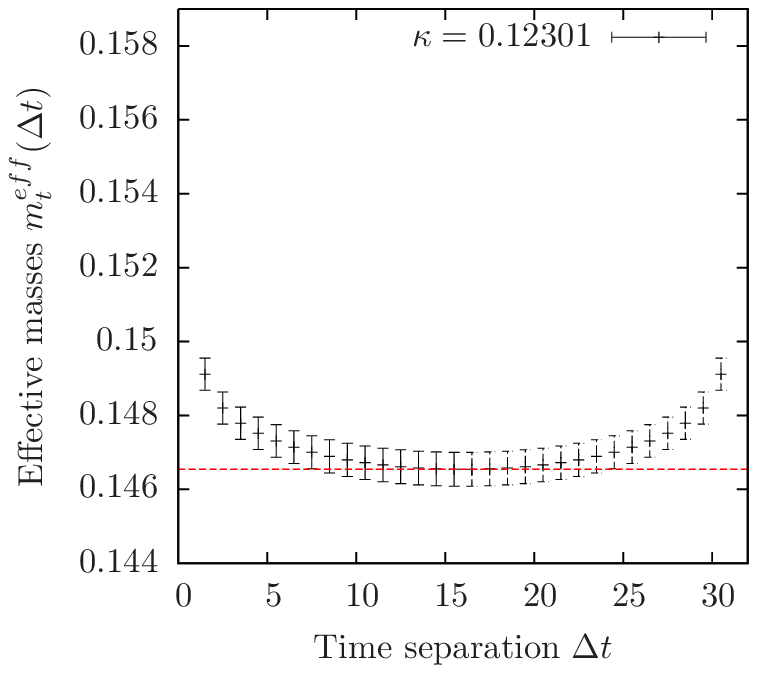}{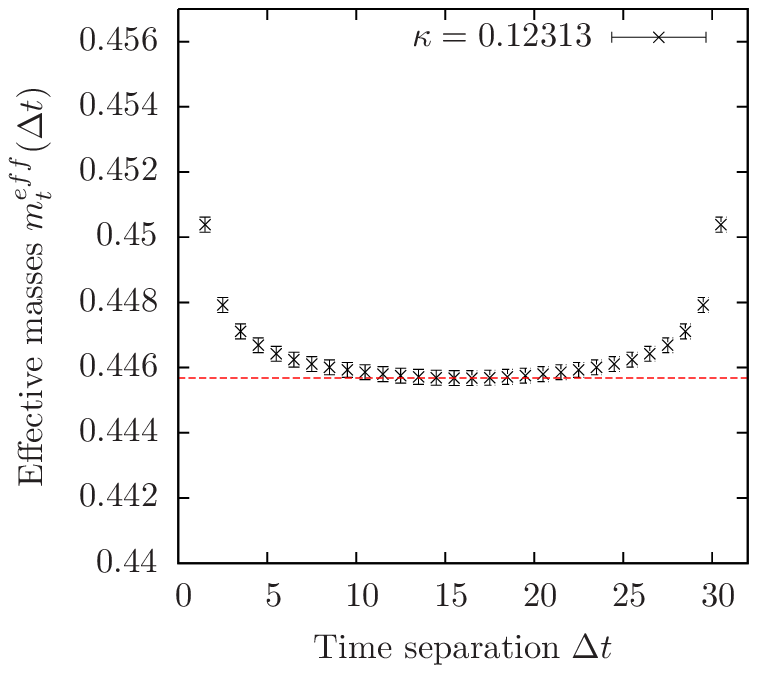}
{fig:TopCorrelatorExample}
{The top quark correlation functions $C_t(\Delta t)$ calculated in the Monte-Carlo runs specified in 
\tab{tab:Chap61EmployedRuns} are presented in panel (a) together with the respective fits obtained from
the fit approach in \eq{eq:FermionTimeCorrelatorFitAnsatz}. The corresponding effective masses 
$m_t^{eff}(\Delta t)$ are shown in panels (b) and (c). The horizontal dashed lines depict the result 
on the top quark mass determined by \eq{eq:FermionMassFromEffFermMass}.
}
{Examples of top quark time-slice correlation functions at small quartic coupling constants.}

\nin
The presented lattice correlation functions have been fitted with the fit approach
\beq
\label{eq:FermionTimeCorrelatorFitAnsatz}
f_{A,m_{f}}(\Delta t) = A \cdot \cosh\Big[ m_{f} \cdot (\Delta t - L_t/2)\Big]
\eeq  
which, however, was done here only for the purpose of illustration. The fermionic mass $m_f$, as specified in 
\eq{eq:DefOfFermionMassFromTimeSliceCorr}, is actually determined by means of the effective fermion 
masses according to
\beq
\label{eq:FermionMassFromEffFermMass}
m_f = m_{f}^{eff}\left(\frac{L_t-1}{2}\right)
\eeq
where the effective masses $m_{f}^{eff}(\Delta t)$ are again defined implicitly through
\bea
\label{eq:DefOfEffMassesForFermion}
\frac{C_f(\Delta t +0.5) }{C_f(\Delta t -0.5) } = \frac{\cosh\Big[ m_{f}^{eff}(\Delta t) \cdot (\Delta t+0.5  - L_t/2)\Big]}
{\cosh\Big[ m_{f}^{eff}(\Delta t) \cdot (\Delta t-0.5  - L_t/2)\Big]},
\quad
\Delta t = \frac{1}{2},\ldots,L_t-\frac{1}{2}. \quad
\eea
Here, no constant term has been included in contrast to the corresponding definition of the bosonic effective masses given in 
\eq{eq:DefOfEffMassesForHiggs}, since the considered fermions are stable particles in the framework of the considered Higgs-Yukawa
model, at least in the mass degenerate case with equal top and bottom Yukawa coupling constants.

The effective masses corresponding to the presented correlation functions $C_t(\Delta t)$ are shown in \fig{fig:TopCorrelatorExample}b,c.
Contrary to the case of the Higgs correlation function discussed in \sect{sec:HiggsTSCanalysisLowerBound} the obtained effective top 
quark masses $m_t^{eff}(\Delta t)$ are very accurate and allow for a clear observation of their expected monotonic decline
with rising values of $\Delta t <L_t/2$. The fermion masses can thus reliably be determined according to 
\eq{eq:FermionMassFromEffFermMass}. This is the approach that will be used for the rest of this work.

%-------------------------------------------------------------------------------------------------------------------- 
\section{Predictions from Lattice Perturbation Theory}
\label{sec:PredFromPertTheory}

In this section the basic concepts of lattice perturbation theory (LPT) in the context of the considered Higgs-Yukawa model 
will be briefly introduced. The resulting apparatus will then be applied in the following to derive analytical predictions 
for the symmetric-broken phase transition, the Higgs boson self energy, as well as the top and bottom quark masses.

It shall be remarked at this point that we will mostly use {\textit{bare}} perturbation theory throughout this section. 
For our purpose of testing the lattice results which are obtained from 
a regularized theory with accessible, bare quantities, namely the model parameters $\hat y_{b,t}$, $\kappa$, and $\hat\lambda$, 
this is the method of choice, provided that the corresponding perturbative series in the bare quantities converge sufficiently fast.
It will turn out that this is indeed mostly\footnote{A counterexample will be presented in \sect{sec:QuarkMassesFromPT}.}
the case with respect to our needs, at least in the weakly coupling regime of the model at the considered values of the 
cutoff $\Lambda$. 

The main concept of the here considered approach is to split up the total action 
$S[\varphi,\psi,\bar\psi]=S_\varphi[\varphi]+S_F[\varphi,\psi,\bar\psi]$ with its constituents defined in 
\eq{eq:BosonicLatticeHiggsActionContNot} and \eq{eq:DefYukawaCouplingTerm} into a Gaussian integrable part 
$S_0[\varphi,\psi,\bar\psi]$ and an interacting part $S_I[\varphi,\psi,\bar\psi]$. The idea is then to expand 
the exponential of the interacting contribution $\exp(-S_I)$ into a power series in terms of $S_I$, the summands 
of which can then be integrated with respect to the remaining weight $\exp(-S_0)$ according to
\beq
\label{eq:MainConceptOfPertTheory}
\int \intD{\varphi} \intD{\psi} \intD{\bar \psi}\,\, O[\varphi,\psi,\bar\psi]\cdot e^{-S} = 
\int \intD{\varphi} \intD{\psi} \intD{\bar \psi}\,\,  O[\varphi,\psi,\bar\psi] \cdot
\sum\limits_{n=0}^\infty \frac{1}{n!}\left[ -S_I  \right]^n  \cdot   e^{-S_0},
\eeq
where $O[\varphi,\psi,\bar\psi]$ is some considered observable.

The choice of how to split up the total action into a Gaussian part $S_0$ and the remainder $S_I$, however, is not unique. In principle,
this choice will not effect the final result of any calculation performed in such a perturbative expansion, provided that all terms of the 
summation in \eq{eq:MainConceptOfPertTheory} are respected. Especially, no apriori knowledge about spontaneous
symmetry breaking would then be needed and one could always expand the total action $S$ around 
the origin, \ie around $\varphi=0$. In practice, however, the infinite sum in \eq{eq:MainConceptOfPertTheory} is truncated after
a few considered orders. The corresponding result in such a practical calculation will thus strongly depend on the
chosen expansion point, \ie on the chosen division of the total action into $S_0$ and $S_I$. In case of a broken vacuum, 
for instance, the naive approach of expanding the total action $S$ around the origin, \ie $\varphi=0$, would lead to very bad 
convergence properties of the associated perturbative series. For this case, which we will restrict the consideration to,
it would be more appropriate to expand the total action around the non-vanishing vev $v$, which we - at this point - assume to be 
given. It will later be determined self-consistently.

Using the definitions of the Higgs and Goldstone modes $h_x, g^\alpha_x$ given in \eq{eq:DefOfHiggsAndGoldstoneModes} the total action 
can now be expanded around the non-vanishing vev $v$ by dividing $S$ into the Gaussian-integrable part $S_0$ and the interaction parts $S_I$
according to
\bea
\label{eq:SplitUpOfTotalActionIntoIntAndGaussPart1}
S_0 &=& \frac{1}{2}m_0^2V v^2 + \lambda Vv^4 + \bar\psi\fermiMatv\psi 
+ \sum\limits_{x}\frac{1}{2}\left[m_0^2+12\lambda v^2\right] h_xh_x + \sum\limits_{x,\mu}\frac{1}{2}\nabla_\mu^f h_x\nabla_\mu^f h_x\quad\quad \nonumber\\
&+& \sum\limits_x \sum\limits_{\alpha=1}^3 \frac{1}{2}\left[m_0^2+4\lambda v^2\right] g_x^\alpha g_x^\alpha  
+ \sum\limits_{x,\mu}\sum\limits_{\alpha=1}^3 \frac{1}{2} \nabla_\mu^f g^\alpha_x\nabla_\mu^f g^\alpha_x, \\
\label{eq:SplitUpOfTotalActionIntoIntAndGaussPart2}
S_I &=& \sum\limits_{x,y} \bar\psi_x \left[h_x \hat B_0 + \sum\limits_{\alpha=1}^3 g^\alpha_x \hat B_\alpha
\right]\GammaOp_{x,y}\psi_y \nonumber \\
&+&\lambda  \sum\limits_{x}\left[ h_x^4 +4 vh_x^3+ 2h_x^2\sum\limits_{\alpha=1}^3g^\alpha_x g^\alpha_x 
+ \sum\limits_{\alpha,\beta=1}^3g_x^\alpha g_x^\alpha g_x^\beta g_x^\beta\right] \nonumber \\
&+& \sum\limits_x\left[ m_0^2vh_x 
+4\lambda v^3h_x
+4\lambda vh_x\sum\limits_{\alpha=1}^3g_x^\alpha g_x^\alpha \right],
\eea
where the continuum notation of the lattice action in terms of the bare mass $m_0$, the coupling constants $y_{t,b}$ and $\lambda$
was used here for convenience, which are, however, trivially related to the lattice parameters $\kappa$, $\hat y_{t,b}$, and $\hat \lambda$ 
as discussed in \sect{sec:modelDefinition}. Moreover, it is remarked that the summation over the $N_f$ fermion generations is 
implicit in this notation and that the symmetry of the model has been exploited allowing to assume the vacuum expectation 
value of the scalar field $\Phi$ to point into the 0-direction. The so far unspecified expression $\fermiMatv$ and its momentum space 
representation $\fermiMatvMom(p)$ are furthermore defined as
\bea
\fermiMatv &=& \fermiMat[\Phi'], \quad \Phi'_x = \frac{v}{\sqrt{2\kappa}} \hat\Phi, \quad \hat\Phi_\mu =  \delta_{\mu,0}, \\
\fermiMatvMom(p) &=& \DMom(p) + \frac{v}{\sqrt{2\kappa}} \hat B_0 \GammaOpMom(p), \\
\DMom(p) &=& \Upsilon(0,p) \D(p) \Upsilon(p,0), \\
\GammaOpMom(p) &=& \Upsilon(0,p) \GammaOp(p) \Upsilon(p,0),
\eea
where the underlying expression for $\Upsilon(p,0)$, $\D(p)$, and $\GammaOp(p)$ have been given in \eq{eq:DefOfSpinorBasisTransMat}, 
\eq{eq:MatDiracSpinorRep}, and \eq{eq:DefOfVertexStructureMatSpinorRep}, respectively.

The expectation value of a considered observable $O[\varphi,\psi,\bar\psi]$ in the original theory can now be expressed as the expectation
value with respect to the Gaussian action $S_0$ reweighted by the exponential of the interacting contribution $S_I$
according to
\beq
\label{eq:ExpValueInPertTheo}
\langle O[\varphi,\psi,\bar\psi] \rangle = \frac{\langle O[\varphi,\psi,\bar\psi] \cdot e^{-S_I} \rangle_{S_0}}{\langle e^{-S_I} \rangle_{S_0}},
\eeq
where the expectation value with respect to $S_0$ is defined as
\beq
\langle O[\varphi,\psi,\bar\psi] \rangle_{S_0} = \frac{1}{Z_{S_0}} \FuncIntAvg_{S_0}[O[\varphi,\psi,\bar\psi]], \quad Z_{S_0}=\FuncIntAvg_{S_0}[1]
\eeq
with 
\beq
\FuncIntAvg_{S_0}[O[\varphi,\psi,\bar\psi]] = 
\int \intD{\varphi} \intD{\psi} \intD{\bar \psi}\,\,  O[\varphi,\psi,\bar\psi]\cdot e^{-S_0}.
\eeq

In this notation the weight factor $\exp(-S_I)$ gives rise to the perturbative expansion in terms 
of the coupling constants $y_b$, $y_t$, and $\lambda$ by virtue of its power series representation
in \eq{eq:MainConceptOfPertTheory}. Since the action $S_0$ is Gaussian, all field variables appearing
in the expansion of $\exp(-S_I)$ are statistically independent with respect to the statistical weight induced by $S_0$.
Moreover, the standard Wick theorem holds and all expectation values with respect to $S_0$
of odd powers of field variables vanish. The basic building blocks of this perturbative expansion are thus the 
bare, tree-level propagators generated by the Gaussian contribution $S_0$. In momentum space they read
\bea
\label{eq:HiggsPropagatorInMomSpace}
\langle \tilde h_{p} \tilde h_{p'} \rangle_{S_0} &=& \delta_{p+p',0} \cdot \frac{1}{\hat p^2 + m_0^2 + 12\lambda v^2}, \\
\label{eq:GoldstonePropagatorInMomSpace}
\langle \tilde g^\alpha_{p} \tilde g^\alpha_{p'} \rangle_{S_0} &=& \delta_{p+p',0} \cdot \frac{1}{\hat p^2 + m_0^2 + 4\lambda v^2},\quad \alpha=1,2,3, \\
\label{eq:FermionPropagatorInMomSpace}
\langle \tilde \psi_{p}\tilde {\bar\psi}_{p'}\rangle_{S_0} &=& \delta_{p,p'} \cdot \fermiMatvMom^{-1}(p),
\eea
where the Higgs-, Goldstone-, and quark fields in momentum space are given as
\bea
\label{eq:fieldsInMomSpace1}
\tilde h_p = \frac{1}{\sqrt{V}} \cdot \sum\limits_x\, e^{-ipx} h_x, && \tilde g^\alpha_p = \frac{1}{\sqrt{V}} \cdot \sum\limits_x\, e^{-ipx} g^\alpha_x, \\
\label{eq:fieldsInMomSpace2}
\tilde \psi_p = \frac{1}{\sqrt{V}} \cdot \sum\limits_x\, e^{-ipx} \psi_x, && \tilde {\bar\psi}_p = \frac{1}{\sqrt{V}} \cdot\sum\limits_x\, e^{+ipx} \bar\psi_x.
\eea

Furthermore, the structure of the interacting contribution $S_I$ in terms of the field variables in 
momentum space will be needed in the upcoming perturbative calculations. It is directly obtained from \eq{eq:SplitUpOfTotalActionIntoIntAndGaussPart2}
and \eqs{eq:fieldsInMomSpace1}{eq:fieldsInMomSpace2} yielding
\bea
\label{eq:InteractionContInMomSpace}
S_I &=& V^{-1/2}\fhs{-3mm}\sum\limits_{p_1\ldots p_3\in\ImpSpace}\fhs{-3mm}
\delta_{p_1,p_2+p_3} \cdot \tilde {\bar\psi}_{p_1}\left[\tilde h_{p_2} \hat B_0 + \sum\limits_{\alpha=1}^3 \tilde g^\alpha_{p_2} \hat B_\alpha  \right] 
\GammaOpMom(p_3)\tilde \psi_{p_3}\\
&+& \lambda V^{-1}\fhs{-3mm}\sum\limits_{p_1\ldots p_4\in\ImpSpace}\fhs{-3mm}
\delta_{p_1+p_2+p_3+p_4,0} \cdot \left[\tilde h_{p_1}\tilde h_{p_2}\tilde h_{p_3}\tilde h_{p_4}  
+2\tilde h_{p_1}\tilde h_{p_2}\sum\limits_{\alpha=1}^3\tilde g_{p_3}^\alpha \tilde g_{p_4}^\alpha    
+\fhs{-1mm}\sum\limits_{\alpha,\beta=1}^3\fhs{-1mm} \tilde g_{p_1}^\alpha \tilde g_{p_2}^\alpha \tilde g_{p_3}^\beta \tilde g_{p_4}^\beta        
\right] \nonumber\\
&+& V^{1/2} \left( m_0^2v + 4\lambda v^3 \right) \tilde h_0
+4\lambda vV^{-1/2} \fhs{-3mm}  \sum\limits_{p_1\ldots p_3\in\ImpSpace}\fhs{-3mm}
\delta_{p_1+p_2+p_3,0} \cdot  \tilde h_{p_1} \cdot \left[ \tilde h_{p_2}\tilde h_{p_3} 
+ \sum\limits_{\alpha=1}^3 \tilde g_{p_2}^\alpha \tilde g_{p_3}^\alpha\right].\nonumber
\eea

Finally, it is remarked here that \eqs{eq:HiggsPropagatorInMomSpace}{eq:GoldstonePropagatorInMomSpace} only hold if the expansion point,
which was chosen here to be the vacuum expectation value $v$, is chosen large enough to guarantee the positivity of the corresponding 
denominators. Also \eq{eq:FermionPropagatorInMomSpace} requires the expansion point $v$ to be non-zero. These requirements, however, are 
met when investigating the broken phase, as aspired.

%-------------------------------------------------------------------------------------------------------------------- 
\subsection{Phase diagram from one-loop LPT}
\label{sec:PhaseDiagramFromPT}

The apparatus introduced in the previous section shall now be applied to the calculation of the phase transition surface between the 
symmetric and the broken phase. More precisely, we will determine the dependence of the vev $v$ on the model parameters within 
the broken phase from one-loop lattice perturbation theory. The phase transition surface could then be obtained from these results
in the same manner already described in \sect{sec:SmallY}, if desired.

The main observation underlying the following calculation is that the total action $S$ could also have been expanded around the
expansion point $\breve v \neq v$ instead of the actual vacuum expectation value $v$ leading then to the same expressions already given in 
\eqs{eq:SplitUpOfTotalActionIntoIntAndGaussPart1}{eq:SplitUpOfTotalActionIntoIntAndGaussPart2} with the only difference that the field 
$h$ would now be given as the fluctuation of $\varphi$ around the 
expansion point $\breve v$. The scalar field $h$ would then no longer have the interpretation of being the Higgs field and 
its expectation value would not be zero. In fact, the expectation value $\langle \tilde h_0 \rangle$ of the zero momentum component 
vanishes if and only if the expansion point $\breve v$ coincides with the true vacuum expectation value $v$. More precisely, one
has 
\beq
\langle \tilde h_0\rangle = V^{1/2} \cdot (v - \breve v).
\eeq
The idea here is to calculate the expectation value $\langle \tilde h_0\rangle$ in lattice perturbation theory for a given 
expansion point $\breve v$. The true vacuum expectation value $v$ can then be determined in this perturbative approach by
identifying that value of the expansion point $\breve v$, where one finds $\langle \tilde h_0\rangle=0$.

In the following we will compute the expression $\langle \tilde h_0\rangle$ to one-loop order. The resulting equation 
$\langle \tilde h_0\rangle=0$ can then be solved numerically. Beginning with an arbitrary non-zero start value for 
the expansion point $\breve v_{0}$ one can, for instance, iteratively define the sequence $\breve v_{n}$, $n\in\N_0$ 
through the iterative relation
\beq
\label{eq:IterativeLPTVeVSeq}
\breve v_{n+1} = \breve v_{n} + V^{-1/2}\cdot \langle \tilde h_0\rangle(\breve v_{n}),
\eeq
where $\langle \tilde h_0\rangle(\breve v_{n})$ denotes the LPT result for $\langle \tilde h_0\rangle$ given the expansion point 
$\breve v$. The fix-point of \eq{eq:IterativeLPTVeVSeq} would then give a self-consistent determination of the vacuum expectation 
value $v$ based on lattice perturbation theory.
 
%\includeFigSingleMedium{LPTDiagramsForVEV}
\includeFigSingleMedium{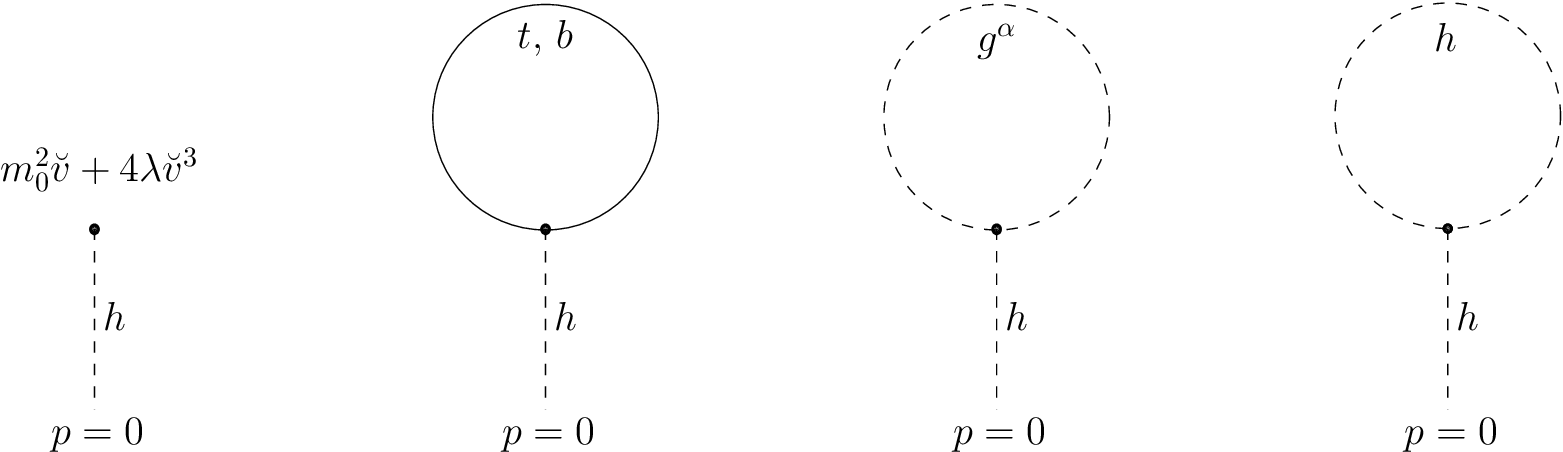}
{fig:FeynmanDiagramsForVev}
{Illustration of the diagrams contributing to the expectation value of $\langle \tilde h_0\rangle$ at one-loop order. 
The vertical dashed lines are Higgs propagators with zero momentum. 
}
{Diagrams contributing to the vacuum expectation value $\langle \tilde h_0\rangle$ at one-loop order.}

The missing step in the calculation of the LPT-prediction for the vev is thus to identify all diagrams\footnote{As usual
the vacuum bubbles and the denominator of \eq{eq:ExpValueInPertTheo} cancel each other at least up to finite volume
terms which, however, are neglected here.} that contribute to $\langle \tilde h_0\rangle$ at the considered 
loop order. From the explicit form of the interacting contribution $S_I$ in momentum space given in 
\eq{eq:InteractionContInMomSpace} one sees that four different Feynman diagrams contribute to this 
quantity at one-loop order as depicted in \fig{fig:FeynmanDiagramsForVev}. According to the form of 
the propagators explicitly given in \eqs{eq:HiggsPropagatorInMomSpace}{eq:FermionPropagatorInMomSpace} 
these diagrams translate into the analytical one-loop LPT result for the expectation value 
$\langle \tilde h_0\rangle(\breve v)$ reading
\bea
\label{eq:1loopLPTResultForExpOfH}
V^{-1/2}\cdot\langle \tilde h_0\rangle(\breve v) &\fhs{-1mm}=\fhs{-1mm}& \frac{-1}{m_0^2+12\lambda \breve v^2} 
\Bigg(m_0^2\breve v+4\lambda \breve v^3  - \frac{1}{V}\sum\limits_{p\in\ImpSpace} \Tr \left[
\hat B_0\GammaOpMom(p)\fermiMatvMom^{-1}(p)
\right] \nonumber\\
&\fhs{-1mm}+\fhs{-1mm}& 4\lambda \breve v \frac{1}{V} \sum\limits_{p\in\ImpSpace}\left[
\frac{3}{\hat p^2 + m_0^2 + 12\lambda \breve v^2}
+\frac{3}{\hat p^2 + m_0^2 + 4\lambda \breve v^2}
\right]
 \Bigg).
\eea

%\includeFigSingleMedium{LPTpredictionsVEV}
\includeFigSingleMedium{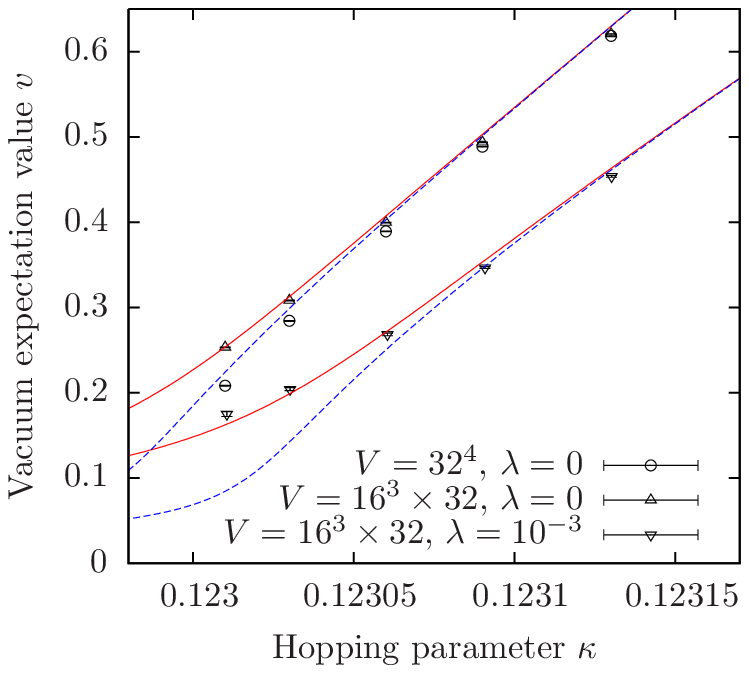}
{fig:LPTresultForVeV}
{The vacuum expectation value $v$ as obtained through the solution of the fix point equation given in \eq{eq:IterativeLPTVeVSeq} is 
presented by the solid ($V=16^3\times 32$) and dashed ($V=32^4$) lines versus the hopping parameter $\kappa$. These analytical results 
are compared to the corresponding lattice calculations of $v$ depicted by the triangular and circular symbols. Both, the analytical 
as well as the numerical computations, have been performed on two different finite lattice volumes and for two different values of the 
quartic coupling constant $\lambda$ as specified in the plot legend. The upper two curves refer to $\lambda=0$, while the lower
ones show the analytical $\lambda=10^{-3}$ curves. The numerical $\lambda=10^{-3}$ results, however, have not been computed 
on the \latticeX{32}{32}{,} since the generation of the underlying field configurations is expensive but not of further use for this 
study in contrast to the $\lambda=0$ data. The here degenerate Yukawa coupling constants have been chosen according to the tree-level 
relation in \eq{eq:treeLevelTopMass} aiming at the reproduction of the phenomenologically known value of the top quark mass.
}
{Dependence of the vacuum expectation value $v$ on $\kappa$ as obtained from one-loop lattice perturbation theory.}

In \fig{fig:LPTresultForVeV} the one-loop LPT predictions for the vacuum expectation value $v$ are plotted versus the hopping 
parameter $\kappa$ for different lattice volumes in the case of a vanishing and a non-vanishing value of the quartic coupling constant
$\lambda$. These perturbative results are compared to corresponding results obtained in direct Monte-Carlo simulations 
and good agreement is observed. 

From these findings one can conclude that the here presented approach of calculating the vev $v$ by means of 
lattice perturbation theory at one-loop order already provides a reasonably good description of the behaviour of the model 
in the broken phase within the weakly interacting regime.

%-------------------------------------------------------------------------------------------------------------------- 
\subsection{Higgs boson self-energy from one-loop LPT}
\label{sec:HiggsSelfEnergy}

In this section the Higgs propagator, and thus the Higgs boson self energy $\tilde \Sigma_H(p)$, will be computed to one-loop order
in lattice perturbation theory. For that purpose one usually decomposes the full Higgs propagator $\langle \tilde h_p \tilde h_{-p}\rangle$
into an infinite sum of products of bare propagators and one-particle-irreducible diagrams\footnote{Again, vacuum bubbles are canceled
with the denominator in \eq{eq:ExpValueInPertTheo} up to neglected finite volume terms.}, as illustrated in 
\fig{fig:FeynmanDiagramsForHiggsPropagator1PISum}. A one-particle-irreducible diagram, usually denoted as 1PI-diagram, 
is an amputated, connected diagram that cannot be disconnected by neglecting a single internal line only, where the 
property 'amputated' means that the external propagators have been divided out. In the given illustration, however, 
the notion 1PI has, more precisely, to be understood as the sum of all one-particle-irreducible diagrams with exactly two legs
(prior two their amputation), being Higgs-legs carrying the momentum $p$. For clarification it is remarked that the expansion point
around which the total action is split up into the Gaussian and the interacting contribution is assumed here to be the true
vacuum expectation value $v$.

%\includeFigSingleLarge{LPTHiggsPropagator1PISum}
\includeFigSingleLarge{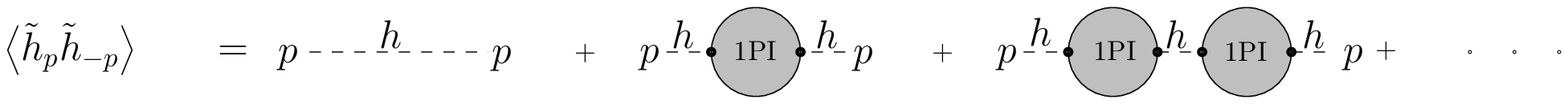}
{fig:FeynmanDiagramsForHiggsPropagator1PISum}
{Illustration of the decomposition of the full Higgs propagator $\langle \tilde h_p \tilde h_{-p}\rangle$
into an infinite sum of products of bare propagators $\langle \tilde h_p \tilde h_{-p}\rangle_{S_0}$, depicted
here as dashed lines, and one-particle-irreducible diagrams, sketched by the grey circles. 
The latter symbols have actually to be understood as the sum of all one-particle-irreducible diagrams. 
}
{Illustration of the propagator decomposition into one-particle-irreducible diagrams.}

By virtue of this decomposition the full Higgs propagator can then be written as an infinite geometric series according to 
\bea
\label{eq:HiggsPropSelfEnRel}
\langle \tilde h_p \tilde h_{-p}\rangle &=& \langle \tilde h_p \tilde h_{-p} \rangle_{S_0} \cdot 
\sum\limits_{n=0}^\infty \left[ \tilde \Sigma_H(p) \cdot \langle \tilde h_p \tilde h_{-p}\rangle_{S_0}
\right]^{n}
\eea
where the so-called self-energy $\tilde\Sigma_H(p)$ is the sum of the aforementioned one-particle-irreducible diagrams.
Summing this series is then straightforward yielding the result 
\bea
\label{eq:HiggsPropSelfEnRel2}
\langle \tilde h_p \tilde h_{-p}\rangle &=& \langle \tilde h_p \tilde h_{-p} \rangle_{S_0} \cdot 
\frac{1}{1-\tilde\Sigma_H(p)\langle \tilde h_p\tilde h_{-p} \rangle_{S_0}} \nonumber\\
&=& \frac{1}{\hat p^2 + m_0^2 + 12\lambda v^2 - \tilde\Sigma_H(p)}.
\eea

For the calculation of the self-energy $\tilde\Sigma_H(p)$ one thus needs to identify all one-particle-irreducible diagrams contributing 
to the Higgs propagator according to the aforementioned decomposition at a given loop order. For the one-loop order these 
diagrams are depicted in \fig{fig:FeynmanDiagramsForHiggsSelfEnery}. 

%\includeFigSingleMedium{LPTHiggsSelfEnergyDiagrams}
\includeFigSingleMedium{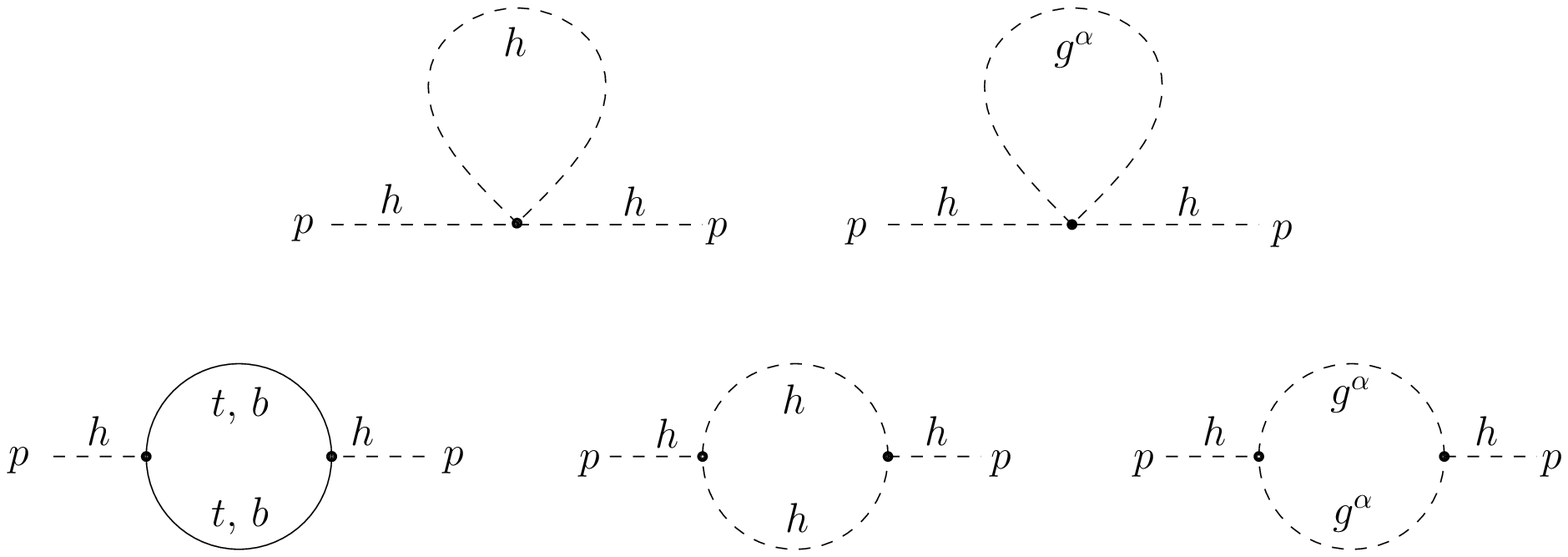}
{fig:FeynmanDiagramsForHiggsSelfEnery}
{Illustration of the one-particle-irreducible diagrams contributing to the Higgs propagator $\langle \tilde h_p \tilde h_{-p}\rangle$ 
at one-loop order.
}
{Diagrams contributing to the Higgs propagator in the full Higgs-Yukawa model at one-loop order.}

It is remarked here that the loop order does not coincide with the order in the quartic coupling constant $\lambda$ as one sees
in \fig{fig:FeynmanDiagramsForHiggsSelfEnery}. While the Higgs- and Goldstone-loop diagrams in the upper row give a contribution
proportional to $\lambda$, the Higgs- and Goldstone-loop diagrams in the second row are only proportional to $\lambda^2$. 
Here, the analytical expressions of all presented diagrams in \fig{fig:FeynmanDiagramsForHiggsSelfEnery} will be given, even though
the latter two diagrams do not add to the achieved order. This is because other diagrams of order $\lambda^2$, consisting of more than one 
loop, have been neglected. 

Using the results given in \sect{sec:PredFromPertTheory} the considered diagrams can be translated into an explicit analytical expression 
for the self-energy according to
\bea
\label{eq:LPTSelfEnergyResult0}
\tilde\Sigma_H(p) &=& \tilde\Sigma^{(1)}_H(p) + \tilde\Sigma^{(2)}_H(p), \\
\label{eq:LPTSelfEnergyResult1}
\tilde\Sigma^{(1)}_H(p) &=& -\frac{\lambda}{V} \sum\limits_{k\in\ImpSpace} 
\frac{12}{\hat k^2 + m_0^2 + 12\lambda v^2} + \frac{12}{\hat k^2 + m_0^2 + 4\lambda v^2} \nonumber\\
&-& \frac{1}{V} \sum\limits_{k\in\ImpSpace} \Tr \left[
\hat B_0\GammaOpMom(k)\fermiMatvMom^{-1}(k) \hat B_0 \GammaOpMom(k-p)\fermiMatvMom^{-1}(k-p)
\right],\\
\label{eq:LPTSelfEnergyResult2}
\tilde\Sigma^{(2)}_H(p)
&=&
16v^2\lambda^2 \frac{1}{V}\sum\limits_{k\in\ImpSpace} 
\frac{18}{\hat k^2+m_0^2+12\lambda v^2} \cdot \frac{1}{\hat t^2+m_0^2+12\lambda v^2}  \nonumber \\
&+&
16v^2\lambda^2 \frac{1}{V}\sum\limits_{k\in\ImpSpace} 
\frac{6}{\hat k^2+m_0^2+4\lambda v^2} \cdot \frac{1}{\hat t^2+m_0^2+4\lambda v^2},
\eea
where $\hat t^2$ denotes the squared lattice momentum of the relative momentum $t=p-k$ and the value of the vacuum expectation value $v$ 
is perturbatively fixed here as described in the preceding section. It is remarked that the given result $\tilde\Sigma^{(1)}_H(p)$ 
already contains all contributions up to the orders $y_t^2$, $y_b^2$, and $\lambda^1$. The aforementioned Higgs- and Goldstone-loop 
contributions of order $\lambda^2$ are here given separately in form of the expression $\tilde\Sigma^{(2)}_H(p)$ which, however, 
does not have a higher convergence order than $\tilde\Sigma^{(1)}_H(p)$, as discussed above.

Examples for the perturbative result of the self-energy $\tilde\Sigma_H(p)$ are presented in \fig{fig:LPTresultSelfEnergy}. These
analytical findings are compared to the corresponding numerical results obtained in direct Monte-Carlo calculations. 
The latter numerical results on the self-energy have been computed by means of \eq{eq:HiggsPropSelfEnRel2}, where
the underlying propagator $\langle\tilde h_p \tilde h_{-p}\rangle$ has directly been taken from the lattice calculation.
In the presented examples two cases, namely a vanishing and a non-vanishing quartic self-coupling constant 
$\lambda$, are considered. In both cases very good agreement is observed between the numerical and analytical results.

%\includeFigTriple{LPTpredictionsSelfEnergyLambdaZero}{LPTpredictionsSelfEnergyLambdaZeroZoom}{LPTpredictionsSelfEnergyLambdaNonZeroZoom}
\includeFigTriple{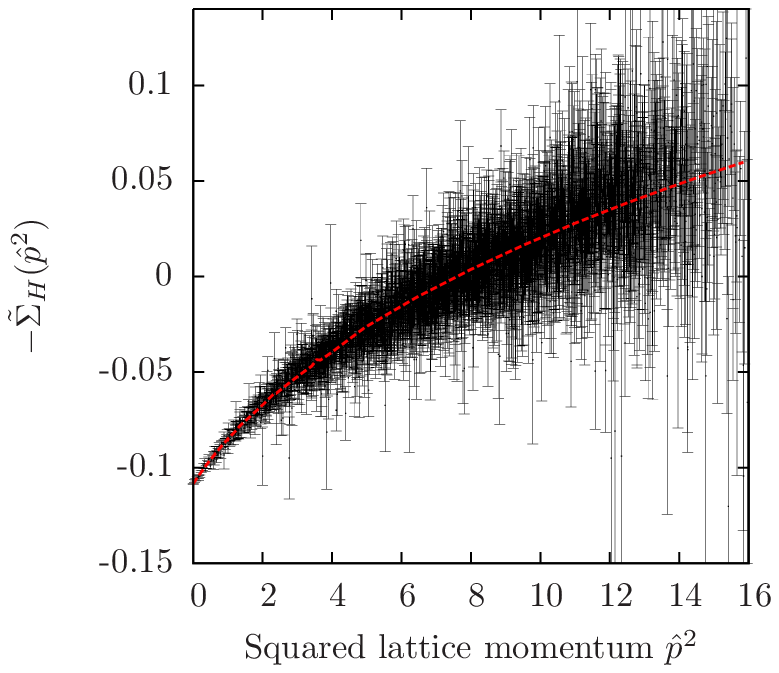}{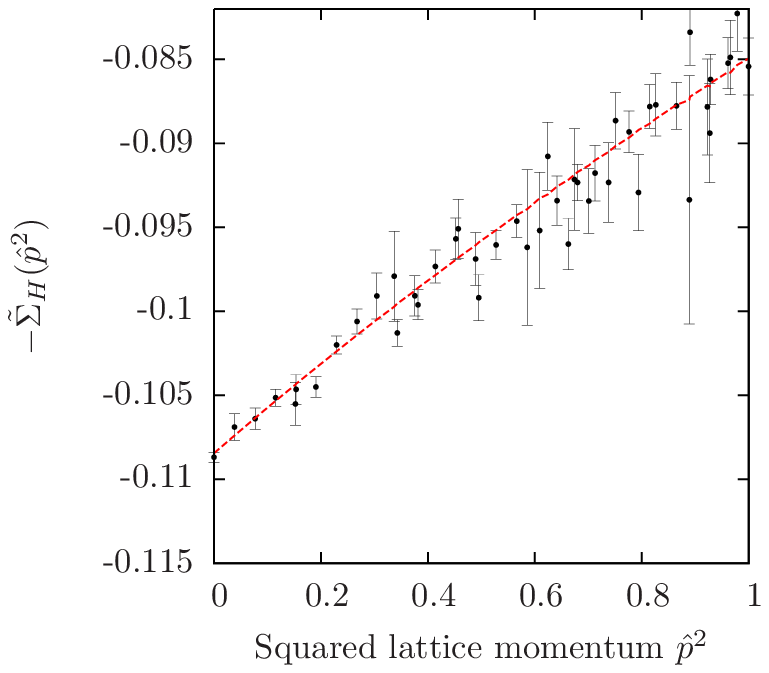}{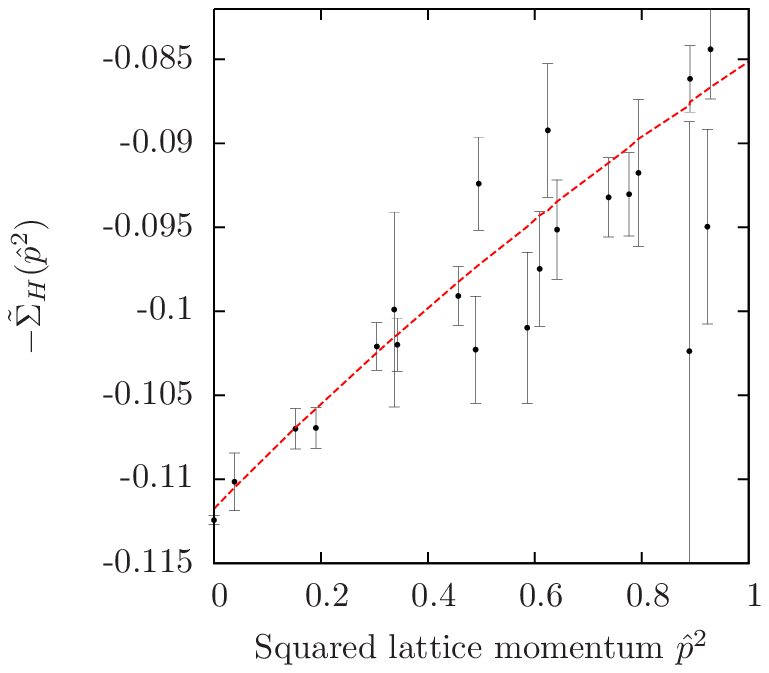}
{fig:LPTresultSelfEnergy}
{The negative self-energy $-\tilde\Sigma_H(\hat p^2)$ as obtained in two different Monte-Carlo runs is presented
versus the squared lattice momentum $\hat p^2$, where panel (a) and (b) actually show identical data with 
the latter, however, zooming in onto the vicinity of the origin at $\hat p^2=0$. The underlying lattice calculation 
for these two plots has been performed on a \lattice{32}{32}with $\lambda=0$ and $\kappa=0.12313$ leading then
to a cutoff of around $\GEV{400}$. The data shown in panel (c) have been generated on a \lattice{16}{32} with $\lambda=10^{-3}$
and $\kappa=0.12313$ resulting in a cutoff of approximately $\GEV{540}$. In both lattice calculations the degenerate Yukawa coupling
constant has been determined through the tree-level relation in \eq{eq:treeLevelTopMass} aiming at the reproduction
of the phenomenological value of the top quark mass. The presented numerical data are compared to the corresponding 
perturbative predictions given in \eq{eq:LPTSelfEnergyResult0} depicted by the red dashed lines. 
}
{The momentum dependence of the self-energy $\tilde\Sigma_H(\hat p^2)$ at quartic coupling constants.}

%-------------------------------------------------------------------------------------------------------------------- 
\subsection{Quark masses from one-loop LPT}
\label{sec:QuarkMassesFromPT}

The eventual aim of the evaluation of the considered Higgs-Yukawa model is to determine the range
of Higgs boson masses which are consistent with phenomenology. As a consequence the Yukawa
coupling constants need to be fine-tuned in order to reproduce the physical top and bottom
quark masses. In a first step one can use the tree-level relation given in \eq{eq:treeLevelTopMass} 
to choose the bare Yukawa coupling parameters underlying the actual
Monte-Carlo calculations as a first guess. To improve this tuning process of the Yukawa
coupling constants, however, it would be very useful to have an analytical,  next-to-leading order
calculation of the quark masses at hand. Here, the analytical result for the top and bottom quark
masses from one-loop lattice perturbation theory will be presented.

The first step is thus to calculate the fermion propagator $\langle \tilde \psi_{p}\tilde{\bar\psi}_{p}\rangle$. 
Following the same standard arguments summarized in \sect{sec:HiggsSelfEnergy}, the full propagator can again
be written in terms of the bare fermion propagator given in \eq{eq:FermionPropagatorInMomSpace} and 
the fermion self-energy $\tilde\Sigma_F(p)$ according to
\bea
\label{eq:DefOfFermionPropViaSelfEn}
\langle \tilde \psi_p \tilde{\bar\psi}_{p'}\rangle &=&
\delta_{p,p'} \cdot
\left(\begin{array}{*{2}{c}}
\langle \tilde t_p \tilde{\bar t}_p \rangle & 0 \\
0 & \langle \tilde b_p \tilde{\bar b}_p \rangle \\
\end{array}\right)
= \delta_{p,p'} \cdot \Big[\fermiMatvMom(p) - \tilde\Sigma_F(p)  \Big]^{-1},
\eea
where the fermion self-energy is given as the sum of all amputated, one-particle-irreducible 
diagrams with two external fermion legs (prior to their amputation) carrying momentum $p$. At one-loop 
order the only contributing diagrams are the fermion-Higgs and the fermion-Goldstone loops as depicted in 
\fig{fig:FeynmanDiagramsForFermionSelfEnery}. It is remarked, that no tad-pole diagrams have to be considered, 
if the expansion point, around which the total action is split up into a Gaussian and an interacting part, 
has been chosen as the vacuum expectation value $v$, which is assumed here. In that case all tadpole 
contributions mutually cancel, since otherwise one would have $\langle \tilde h_0 \rangle \neq 0$ as 
discussed in \sect{sec:PhaseDiagramFromPT}. 

%\includeFigSingleSmall{LPTFermionSelfEnergyDiagrams}
\includeFigSingleSmall{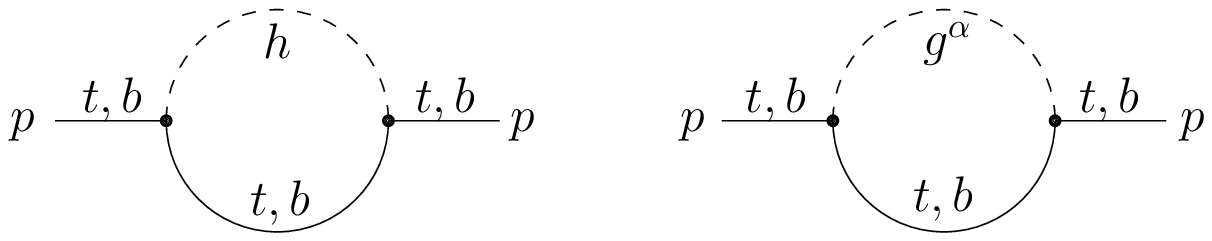}
{fig:FeynmanDiagramsForFermionSelfEnery}
{Illustration of the one-particle-irreducible diagrams contributing to the fermion propagators 
$\langle \tilde t_p \tilde{\bar t}_p \rangle$ and $\langle \tilde b_p \tilde{\bar b}_p \rangle$ 
at one-loop order.
}
{Diagrams contributing to the fermion propagator in the full Higgs-Yukawa model at one-loop order.}

Using the results given in \eqs{eq:HiggsPropagatorInMomSpace}{eq:FermionPropagatorInMomSpace} these diagrams directly translate 
into the analytical result for the fermion self-energy reading
\bea
\label{eq:ResultOnFermionSelfEn}
\tilde\Sigma_F(p) &=& \frac{1}{V}  \sum\limits_{k\in\ImpSpace} \frac{\hat B_0\GammaOpMom(k) \fermiMatvMom^{-1}(k) \hat B_0\GammaOpMom(p)}
{\hat t^2 + m_0^2 +12\lambda v^2} 
\nonumber\\
&+ &
\frac{1}{V}  \sum\limits_{k\in\ImpSpace}\sum\limits_{\alpha=1}^3 
\frac{\hat B_\alpha\GammaOpMom(k)\fermiMatvMom^{-1}(k) \hat B_\alpha\GammaOpMom(p)}
{\hat t^2 +m_0^2 + 4\lambda v^2},
\eea
where $\hat t^2$ denotes again the squared lattice momentum of the relative momentum $t=p-k$.

To extract the physical mass from this propagator, which is defined through the exponential decay
of the time-slice correlator $C_{t,b}(\Delta t)$ according to \eq{eq:DefOfFermionMassFromTimeSliceCorr}, 
one would have to calculate this time-slice correlator from the given one-loop result of the fermion 
propagator. This can be done by numerically performing a Fourier transformation of the propagator 
$\langle \tilde \psi_p \tilde{\bar\psi}_p\rangle$ at zero spatial momentum $\vec p =0$ along the 
time-direction according to 
\bea
C_{f}(\Delta t) &=& \frac{1}{V}\sum\limits_{p_t=0}^{L_t-1} e^{i\Delta t\cdot p_t} \cdot
\Big\langle 2\,\RE\,\Tr\,\left(\tilde f_{L,p_t, \vec 0}\cdot \tilde {\bar f}_{R,p_t,\vec 0}\right) \Big\rangle
\eea
with $f=t,b$. The latter left-handed fermion propagators underlying the above result are directly obtained from 
\eq{eq:DefOfFermionPropViaSelfEn} by applying appropriate projection operators according to \eq{eq:DefOfLeftHandedSpinors}
yielding then 
\bea
\label{eq:ResultOnLeftHandedFermionProp}
\left(\begin{array}{*{2}{c}}
\langle \tilde t_{L,p} \tilde{\bar t}_{R,p}\rangle  &  \\
 & \langle b_{L,p} \tilde{\bar b}_{R,p} \rangle\\
\end{array}\right)
&=& \hat P_- \left[\fermiMatvMom(p) - \tilde\Sigma_F(p)  \right]^{-1} P_-.
\eea
From this result it is straightforward to compute the effective fermion masses
according to \eq{eq:DefOfEffMassesForFermion}.

As an example the perturbatively obtained effective masses are compared to corresponding numerical results in \fig{fig:FermionCorrelatorExamplesAtWeakCoup},
which have been obtained in direct Monte-Carlo calculations. Here the value of the vacuum expectation value $v$ appearing
in \eq{eq:ResultOnFermionSelfEn} and \eq{eq:ResultOnLeftHandedFermionProp} has not been set to its perturbative prediction  
derived in \sect{sec:PhaseDiagramFromPT}, since that result would be to inaccurate for this purpose here. Instead, it has explicitly 
been set to its non-perturbatively obtained value as observed in the respective lattice calculations. However, even with
this setting the resulting perturbative predictions for the effective masses arising from \eq{eq:ResultOnFermionSelfEn}
are not particularly good, as can be observed in \fig{fig:FermionCorrelatorExamplesAtWeakCoup}. 

%\includeFigDouble{TopCorrelatorEffectiveMassesKap012301L32LPTcomp}{TopCorrelatorEffectiveMassesKap012313L32LPTcomp}
\includeFigDouble{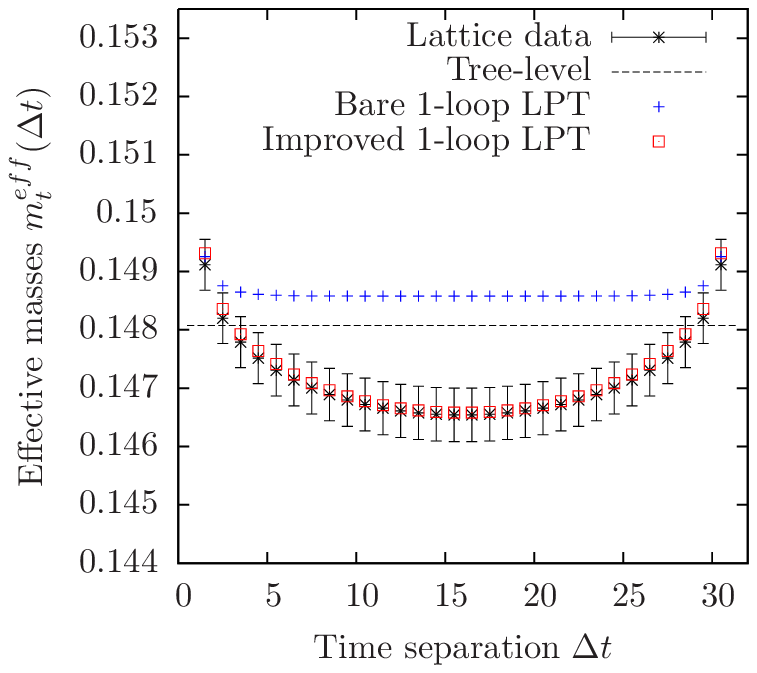}{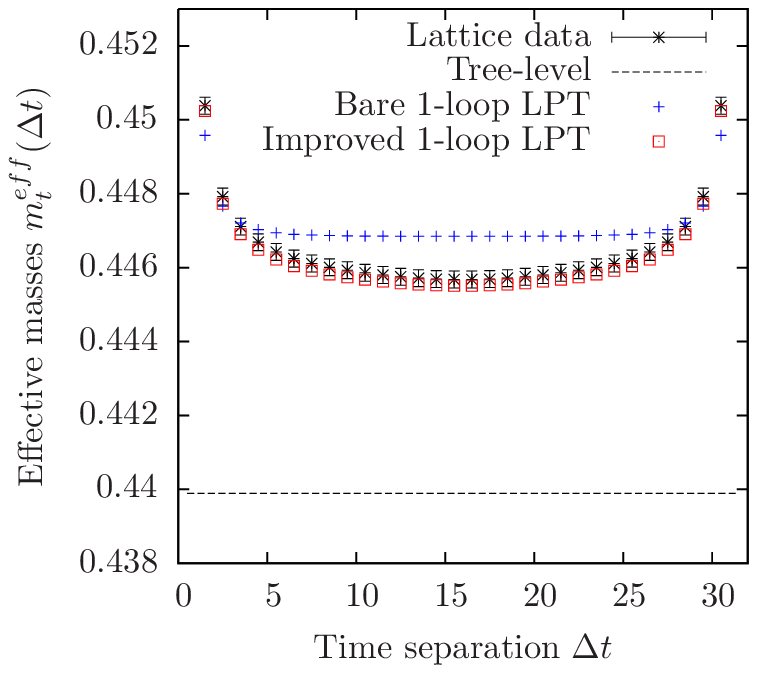}
{fig:FermionCorrelatorExamplesAtWeakCoup}
{The effective top quark masses $m_t^{eff}(\Delta t)$ are shown as calculated in two different Monte-Carlo
runs as specified in \tab{tab:Chap61EmployedRuns} through the value of the hopping parameter being $\kappa=0.12301$
on the left and $\kappa=0.12313$ one the right. The respective tree-level estimate is depicted by the black dashed line. 
These results are compared to the corresponding perturbative one-loop LPT predictions arising from \eq{eq:ResultOnFermionSelfEn}
and from the improved expression in \eq{eq:ResultOnFermionSelfEnImproved}. The explanation of the 'improved' results 
is given in the main text.
}
{Comparison of the numerically obtained effective top quark masses with corresponding predictions from lattice
perturbation theory.}
 
This is the announced example where the considered one-loop calculation in bare perturbation theory does not yield satisfactory 
results. However, the observed mismatch between the numerical and perturbative one-loop results can significantly be reduced already 
by an only partial renormalization procedure, where only the bosonic bare masses in \eq{eq:ResultOnFermionSelfEn} will be renormalized. 
For clarification it is remarked, that a full renormalization procedure is not desirable here, since the actual intention of the 
presented calculations still is the analytical prediction of the lattice results obtained for a given set of bare model parameters. 
This can be achieved by replacing the expressions $m_0^2+12\lambda v^2$ and $m_0^2+4\lambda v^2$ appearing in the denominators 
of \eq{eq:ResultOnFermionSelfEn} by the renormalized Higgs boson mass $m^2_H$ and the renormalized Goldstone mass $m_G^2$, which is 
assumed to be zero for simplicity. The value of the renormalized Higgs boson mass is taken here from the perturbative result of the 
self-energy derived in the preceding section according\footnote{This expression is not exactly equal to the actually introduced 
definition of the renormalized Higgs boson mass, but coincides with $m_H^2$ up to small perturbative corrections which, however, do
not change the achieved order of the subsequently given result on the self-energy.} to $m_H^2 \approx m_0^2 + 12\lambda v^2 - \tilde\Sigma_H(0)$. 
This replacement is perfectly justified in the sense that the achieved convergence order of the original result, which is $O(y^2_{t,b})$, 
is unaltered. Furthermore, the condition $k\neq p$ has to be imposed in the summation of the Goldstone loop,
due to the simplifying assumption $m_G=0$ which in fact only holds in infinite volume, in order to obtain a finite result of the 
discrete sum. This modification is justified in the infinite volume limit, since the then arising integral expression can be 
performed over the aforementioned singularity yielding a finite result due to the sufficiently low order of that singularity. 
The in this respect improved version of the self-energy then reads
\bea
\label{eq:ResultOnFermionSelfEnImproved}
\tilde\Sigma_F(p) &=& \frac{1}{V} \cdot  \sum\limits_{k\in\ImpSpace} \frac{\hat B_0\GammaOpMom(k) \fermiMatvMom^{-1}(k) \hat B_0\GammaOpMom(p)}
{\hat t^2 + m_H^2} 
\nonumber\\
&+ &
\frac{1}{V} \cdot  \sum\limits_{p\neq k\in\ImpSpace}\sum\limits_{\alpha=1}^3 
\frac{\hat B_\alpha\GammaOpMom(k)\fermiMatvMom^{-1}(k) \hat B_\alpha\GammaOpMom(p)}
{\hat t^2 + 0}.
\eea

The resulting perturbative predictions of the effective masses based on this improved one-loop LPT calculation are also 
depicted in \fig{fig:FermionCorrelatorExamplesAtWeakCoup} and one now finds excellent agreement between the numerical and 
analytical data, allowing thus for a very precise prediction of the fermion masses in the weakly interacting regime of 
the considered Higgs-Yukawa model.

%-------------------------------------------------------------------------------------------------------------------- 
\section{Predictions from the effective potential}
\label{sec:EffPot}

In \sect{sec:SmallYukawaCouplings} an expression for the effective potential in the large $N_f$-limit has been derived, which 
was used to predict the phase structure in the weakly interacting regime of the considered Higgs-Yukawa model. We will 
now apply these analytical formulas to predict also the Higgs boson mass at sufficiently small values of the coupling constants
$\lambda$ and $y_{t,b}$. In particular, we will investigate the dependence of the Higgs boson mass on the model 
parameters $y_b$, $y_t$, and $\lambda$. The knowledge about the latter dependence on the quartic coupling constant will later 
play a crucial role in the determination of the lowest attainable Higgs boson mass, \ie the lower Higgs boson mass bound.

%-------------------------------------------------------------------------------------------------------------------- 
\subsection{Dependence of the Higgs boson mass on \texorpdfstring{$y_b$ and $y_t$}{yb and yt}}
\label{sec:DepOfHiggsMassOnY}

From perturbation theory it is well known that the Higgs boson mass shift $\delta m_H^2$ decreases with 
rising Yukawa coupling constants $y_b$, $y_t$ according to the qualitative one-loop perturbation theory
result given in \eq{eq:perturbTheroyResult}. From this result one can, however, not conclude that the Higgs boson mass 
itself would also decrease with growing Yukawa coupling constants. In fact, the opposite is the case provided that the cutoff
$\Lambda$ and the quartic self-coupling constant $\lambda$ are held constant. This is because the phase 
transition line is also shifted when the Yukawa coupling parameters are varied, translating directly into an alteration 
of the bare mass parameter $m_0^2$ associated to the shifted phase transition point, as will be discussed in the following.

For the purpose of clarification we consider again the effective potential $\tilde U[\breve m_\Phi, \breve s_\Phi]$ given in 
\eq{eq:EffActionRewrittenForSmallYFinal}, which has been calculated in terms of the amplitudes $\breve m_\Phi$ and $\breve s_\Phi$ 
in the large $N_f$-limit with degenerate Yukawa coupling constants $\hat y_{t}= \hat y_b =  O(1/\sqrt{N_f})$ and $\hat \lambda = O(1/N_f)$. 
Since we are here only interested in considering the broken phase with $v\neq 0$ and $v_s = 0$, a less general expression is sufficient 
for this purpose. For convenience we therefore define $\breve U[\breve v] \equiv N_f\cdot \tilde U(\breve v / \sqrt{2 N_f\kappa}, 0)$, which 
then simplifies to 
\bea
\label{eq:EffPot}
\breve U[\breve v] &=&
\frac{1}{2} m_0^2 \breve v^2 + \lambda \breve v^4 + \breve U_{F}[\breve v]
\eea
up to constant terms with the fermionic contribution given as
\bea
\label{eq:DefOfFermionicContToEffPot}
\breve U_{F}[\breve v] &=&
\frac{-2N_f}{L_s^3\cdot L_t}\cdot \sum\limits_{p\in\ImpSpace} \log\left|\nu^+(p) + y_t \breve v \left(1-\frac{1}{2\rho}\nu^+(p)\right) 
\right|^2\nonumber\\
&+& \frac{-2N_f}{L_s^3\cdot L_t}\cdot \sum\limits_{p\in\ImpSpace}  \log\left|\nu^+(p) + y_b \breve v \left(1-\frac{1}{2\rho}\nu^+(p)\right)  \right|^2. 
\label{eq:FermionEffectivePot}
\eea
This result follows from \eq{eq:EffActionRewrittenForSmallYFinal} when setting $\breve s_\Phi$ to zero. Moreover, the original 
result has been extended here to the non-degenerate case with $y_t\neq y_b$.

For a given value of the vacuum expectation value $v$, and thus for a fixed location of the minimum of $\breve U[\breve v]$ according to
\beq
\label{eq:BareVevFromU}
\derive{\breve U[\breve v]}{\breve v}\Big|_{\breve v= v} = 0
\eeq
one can then trivially establish the relation
\beq
\label{eq:BareMassFromU}
m_0^2 = -4\lambda v^2 - \frac{1}{v} \frac{\mbox{d}}{\mbox{d}\breve v} \breve U_F[\breve v]\Big|_{\breve v = v}
\eeq
between the bare mass parameter $m_0^2$, the given vev $v$, and the coupling constants $y_{t,b}$ and $\lambda$.
It is remarked that the relations in \eqs{eq:BareVevFromU}{eq:BareMassFromU} hold, of course, only up to the considered 
order of the underlying calculation of the effective potential and moreover only for sufficiently large lattice volumes,
where fluctuations around the minimum of $\breve U[\breve v]$ can be neglected, which shall, however, be assumed
to be the case in the following. 

An estimate $m_{He}$ for the Higgs boson mass can then be obtained from the curvature of the effective potential at its
minimum yielding
\beq
\label{eq:PropMassFromU}
m^2_{He} = 8 \lambda v^2 - \frac{1}{v} \frac{\mbox{d}}{\mbox{d}\breve v} \breve U_F[\breve v] \Big|_{\breve v = v}
+ \frac{\mbox{d}^2}{\mbox{d}\breve v^2} \breve U_F[\breve v]\Big|_{\breve v = v}.
\eeq
In fact, the curvature of the effective potential gives rather the value of the inverse Higgs propagator $\tilde G^{-1}_H(p)$
at $p=0$ which is, however, directly related to the Higgs boson mass apart from the renormalization factor $Z_G$ and 
the fact that the actual Higgs boson mass is determined at a non-zero momentum configuration as discussed in \sect{sec:SimStratAndObs}.
For our purpose here it is sufficient to neglect these details and to consider $m_{He}$ as a reasonable estimate for the Higgs 
boson mass.
 
From the given finding in \eq{eq:PropMassFromU} one obtains an analytical prediction for the dependence of the Higgs
boson mass on the quartic coupling parameter $\lambda$ and the Yukawa coupling constants $y_{t,b}$ for a given value of
the vev $v$. This approach allows to discuss the parameter dependence of the Higgs boson mass for a fixed cutoff 
$\Lambda$, which is directly related to the vev $v$ according to \eqs{eq:FixationOfPhysScale}{eq:DefOfCutoffLambda} yielding 
$\Lambda\propto 1/v$ apart from the contribution of the renormalization constant $Z_G$ being, however,
close to one in the here considered model parameter setups.

In order to test the validity of the analytically predicted dependence of the Higgs boson mass on the 
degenerate Yukawa coupling constant $y_t=y_b$ at a fixed value of the vev $v$, \ie at an almost
fixed value of the cutoff $\Lambda$, the aforementioned analytical calculations are compared to corresponding results 
of direct Monte-Carlo calculations in \fig{fig:HiggsMassDepOnYukawaCoupling}. Both, the analytical as well as the numerical
calculations have been performed on a \lattice{16}{32} with $N_f=1$, $\lambda=0$, and degenerate Yukawa coupling constants.
The value of the vev $v$ has been adjusted here to correspond to an associated cutoff of approximately $\Lambda=\GEV{400}$,
which is a typical value for the later determination of the lower Higgs boson mass bounds.
 
In \fig{fig:HiggsMassDepOnYukawaCoupling}a the numerical and analytical results on the Higgs boson mass 
shift $\delta m_{Hc}^2=m_{Hc}^2-m_0^2$, defined according to the definition in \eq{eq:perturbTheroyResult}, are shown 
together with the corresponding bare mass parameters $m_0^2$, tuned here to yield the desired value of the vev $v$, 
versus the squared degenerate Yukawa coupling constant $y_t^2=y_b^2$ and very good agreement is observed. 
The same is true for the Higgs boson mass $m_{Hc}$ itself, which is presented in \fig{fig:HiggsMassDepOnYukawaCoupling}b.

%\includeFigDouble{YScanHiggsMassShiftFormEffPot}{YScanHiggsMassFormEffPot}
\includeFigDouble{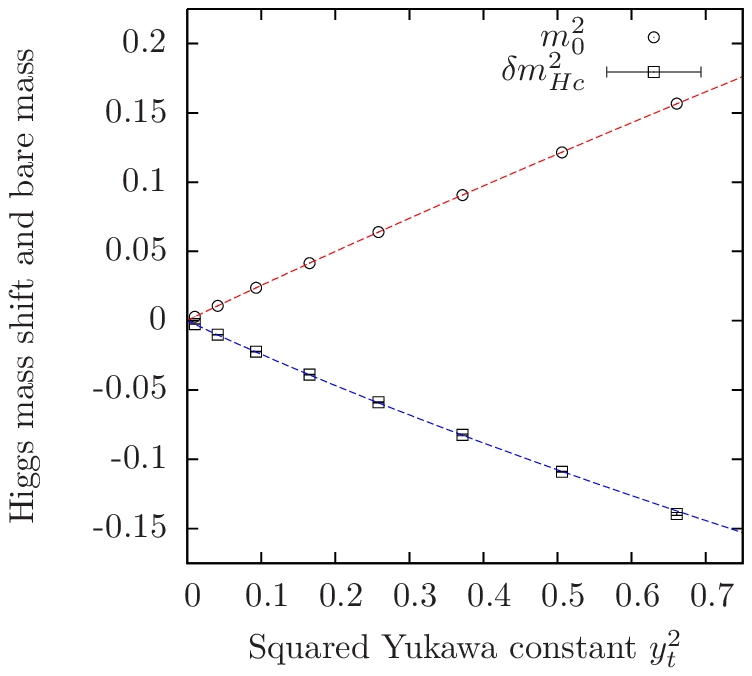}{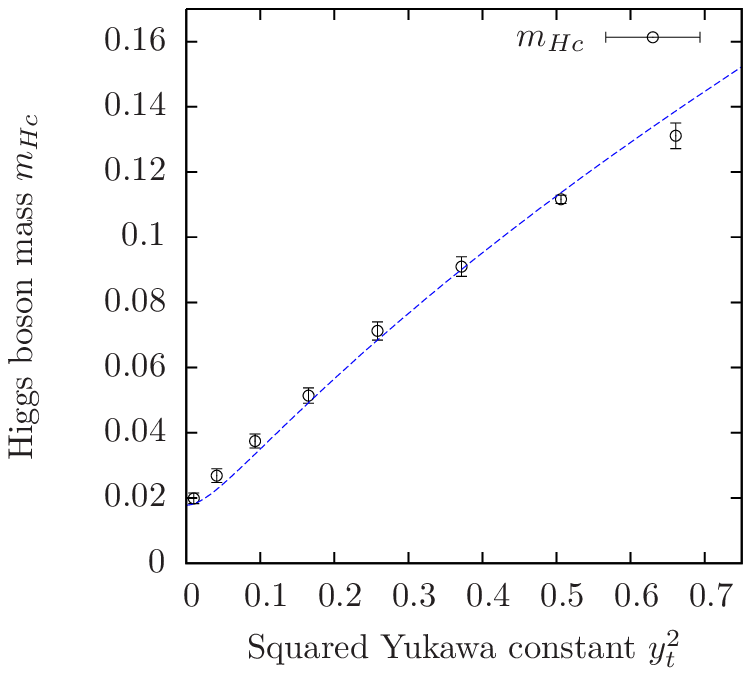}
{fig:HiggsMassDepOnYukawaCoupling}
{The dependence of the bare mass parameter $m_0^2$ and the mass shift $\delta m^2_{Hc}=m^2_{Hc}-m_0^2$
is shown versus the squared degenerate Yukawa coupling constant 
$y_t^2=y_b^2$ in panel (a). These numerical data have been obtained in direct Monte-Carlo calculations on a \lattice{16}{32}
with $\lambda=0$, $N_f=1$, and $\kappa$ tuned to sustain an approximately constant cutoff, where the actually obtained values of
$\Lambda$ fluctuate between $\GEV{370}$ and $\GEV{400}$. The corresponding values of the Higgs correlator mass $m_{Hc}$ 
are presented in panel (b). In both panels the given numerical data are compared to the 
respective predictions obtained from the effective potential formulas in \eq{eq:BareMassFromU} and \eq{eq:PropMassFromU}.
The vacuum expectation value $v$ underlying these analytical results has here been set to its numerically determined value 
computed in the highest statistics run, which is the one at $y_t^2\approx 0.5$. 
}
{Dependence of the Higgs correlator mass $m_{Hc}$ on the degenerate Yukawa coupling constant $y_t^2=y_b^2$ at small
quartic coupling constants.}

As expected from the qualitative result in \eq{eq:perturbTheroyResult} the mass shift $\delta m^2_{Hc}$ decreases quadratically 
with rising Yukawa coupling constants. The Higgs boson mass $m_{Hc}$ itself, however, grows with increasing $y_t^2=y_b^2$ as seen in 
\fig{fig:HiggsMassDepOnYukawaCoupling}b. As mentioned earlier, the reason is that the effect on the Higgs boson mass induced by the change 
of the mass shift $\delta m_{Hc}^2$ is overcompensated by the shift of the phase transition line when varying the Yukawa coupling 
constant while holding $\lambda$ and $\Lambda$ fixed. For constant cutoff and quartic self-coupling constant the resulting shift of 
the bare mass parameter $m_0^2$ is presented in \fig{fig:HiggsMassDepOnYukawaCoupling}a. 

So far, the presented results have been determined in the mass degenerate case, \ie for $y_t=y_b$, which is 
easier to access numerically. This brings up the question of how the results on the Higgs boson mass are 
influenced when pushing the top-bottom mass split to its physical value, \ie $m_b/m_t\approx 0.024$. From 
the qualitative result in \eq{eq:perturbTheroyResult} one expects the Higgs boson mass shift $\delta m_{Hc}^2$ to grow quadratically 
with decreasing $y_b$ and that is exactly what is observed in \fig{fig:HiggsMassDepOnYukawaCouplingSplitting}a. 
Here, the bare top quark Yukawa coupling constant $y_t$, the quartic coupling parameter $\lambda=0$, and the cutoff 
$\Lambda$, more precisely the vev $v$, are held constant, while $y_b$ is lowered to the physical ratio of $y_b/y_t$. 

The numerical data for the Higgs boson mass shift $\delta m_{Hc}^2$ are in good agreement with the predictions based on the
effective potential in \eq{eq:EffPot}, which are depicted by the dashed lines in \fig{fig:HiggsMassDepOnYukawaCouplingSplitting}a.
Again, the Higgs boson mass itself does not increase but {\textit{decrease}} with decreasing $y_b$ as shown in 
\fig{fig:HiggsMassDepOnYukawaCouplingSplitting}b due to the shift of the bare mass parameter $m_0^2$ as discussed earlier.

%\includeFigDouble{YbYtRatioScanHiggsMassShiftFormEffPot}{YbYtRatioScanHiggsMassFormEffPot}
\includeFigDouble{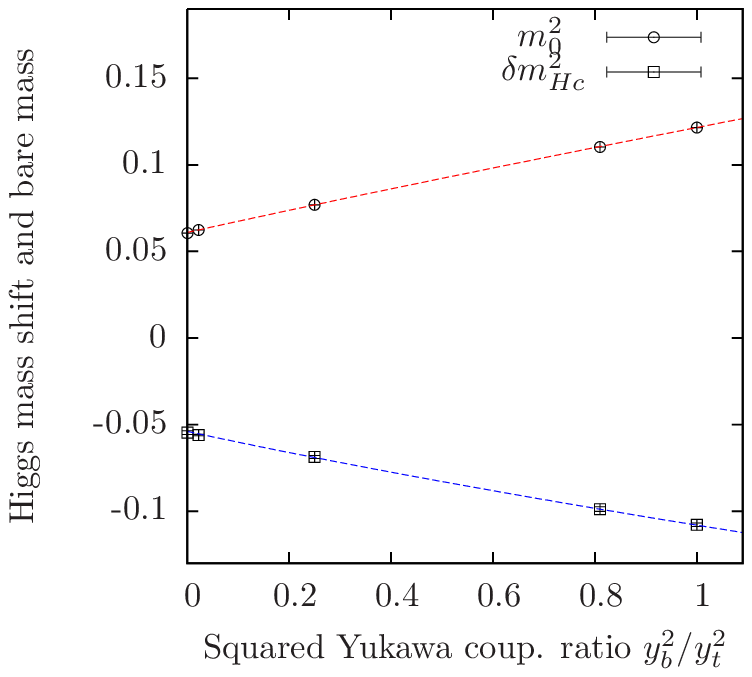}{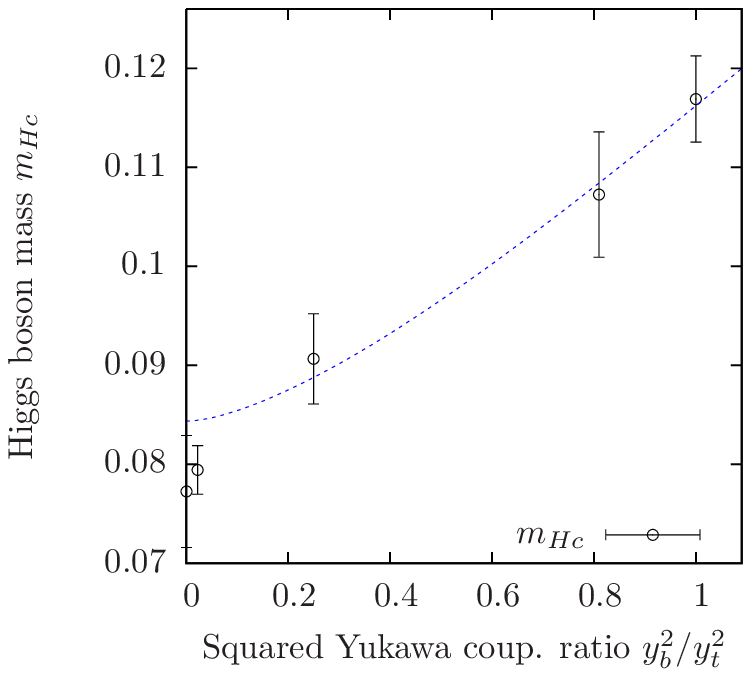}
{fig:HiggsMassDepOnYukawaCouplingSplitting}
{The dependence of the bare mass parameter $m_0^2$ and the mass shift $\delta m^2_{Hc}=m^2_{Hc}-m_0^2$
is shown versus the squared ratio of the Yukawa coupling constants
$y_b^2/y_t^2$ in panel (a). These numerical data have been obtained in direct Monte-Carlo calculations on a \lattice{12}{32}
with $\lambda=0$, $N_f=1$, and $\kappa$ tuned to sustain an approximately constant cutoff, where the actually obtained values of
$\Lambda$ fluctuate between $\GEV{385}$ and $\GEV{400}$. Here, the top quark Yukawa coupling constant has been fixed according to 
\eq{eq:treeLevelTopMass} aiming at reproducing the phenomenological value of the top quark mass.
The corresponding values of the Higgs correlator mass $m_{Hc}$ are presented in panel (b). In both panels the given
numerical data are compared to the respective predictions obtained from the effective potential formulas in \eq{eq:BareMassFromU} 
and \eq{eq:PropMassFromU}. The vacuum expectation value $v$ underlying the analytical results has here been set to its numerically 
determined value computed in the run at $y_b/y_t=1$. }
{Dependence of the Higgs correlator mass $m_{Hc}$ on the Yukawa coupling constant ratio $y_b^2/y_t^2$ at small
quartic coupling constants.}

From the very nice correspondence between the numerical and analytical results presented in \fig{fig:HiggsMassDepOnYukawaCoupling} 
and \fig{fig:HiggsMassDepOnYukawaCouplingSplitting} one can conclude that the effective potential $\breve U[\breve v]$ given in 
\eq{eq:EffPot} gives a good quantitative description of the considered model even at $N_f=1$, at least in the weakly interacting 
regime of the model at vanishing bare quartic self-coupling constant $\lambda=0$.

%-------------------------------------------------------------------------------------------------------------------- 
\subsection{Dependence of the Higgs boson mass on \texorpdfstring{$\lambda$}{the quartic coupling constant}}
\label{sec:DepOfHiggsMassOnLambda}

We now turn to the dependence of the Higgs boson mass on the quartic self-coupling parameter. From the qualitative result
in \eq{eq:perturbTheroyResult} one expects the mass shift $\delta m_{Hc}^2$ to grow linearly to lowest order with increasing 
values of $\lambda$. This expectation is again explicitly checked by direct Monte-Carlo simulations. In \fig{fig:HiggsMassDepOnQuarticCoupling}a
the numerical results on the Higgs boson mass shift $\delta m_{Hc}^2$ and the bare mass $m_0^2$ are presented versus the quartic coupling
parameter $\lambda$. These data were obtained on a \lattice{16}{32} with approximately constant cutoff, \ie for an approximately constant 
vev $v$, and degenerate Yukawa coupling constants fixed according to the tree-level relation in \eq{eq:treeLevelTopMass} aiming at the 
reproduction of the phenomenological value of the top quark mass. One clearly observes the expected linear increase of the mass shift 
$\delta m_{Hc}^2$ with rising quartic coupling constant, as expected.

%\includeFigDouble{LambdaScanHiggsMassShiftFormEffPot}{LambdaScanHiggsMassFormEffPot}
\includeFigDouble{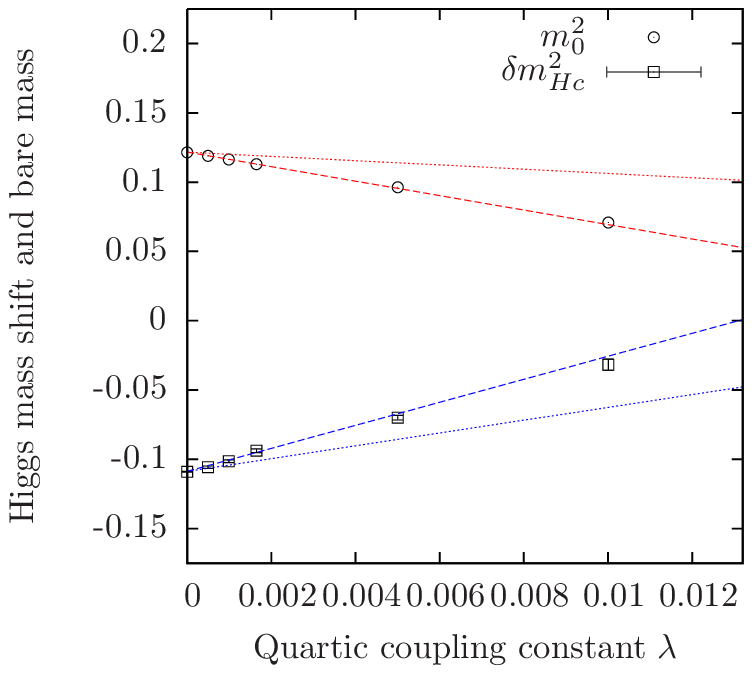}{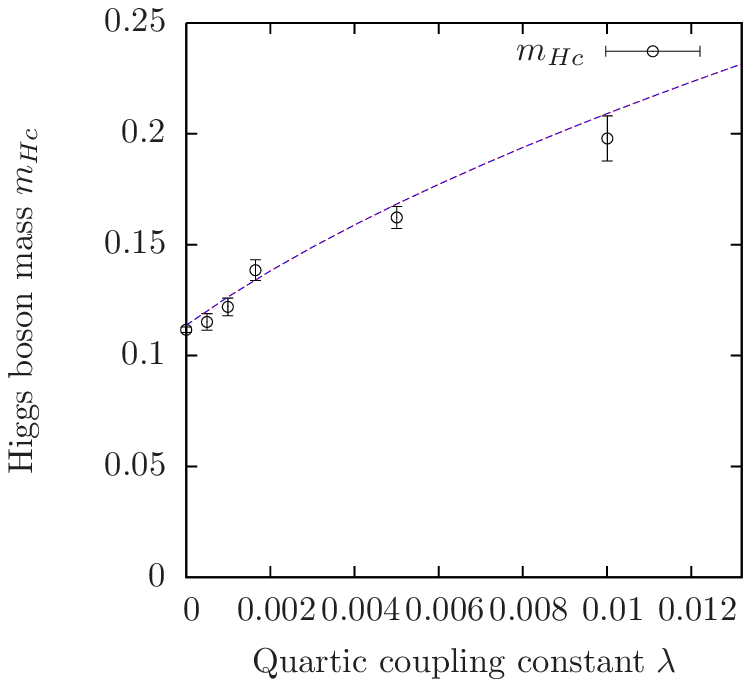}
{fig:HiggsMassDepOnQuarticCoupling}
{The dependence of the bare mass parameter $m_0^2$ and the mass shift $\delta m^2_{Hc}=m^2_{Hc}-m_0^2$ 
is shown versus the quartic coupling constant $\lambda$
in panel (a). These numerical data have been obtained in direct Monte-Carlo calculations on a \lattice{16}{32}
with $N_f=1$, degenerate Yukawa coupling constants fixed according to \eq{eq:treeLevelTopMass} aiming at 
reproducing the phenomenological value of the top quark mass, and $\kappa$ tuned to sustain an approximately constant 
cutoff, where the actually obtained values of $\Lambda$ fluctuate between $\GEV{385}$ and $\GEV{400}$. 
The corresponding values of the Higgs correlator mass $m_{Hc}$ are presented in panel (b). In both panels the given numerical 
data are compared to the respective predictions obtained from \eq{eq:BareMassFromU} and \eq{eq:PropMassFromU}. The results 
arising from the effective potential in \eq{eq:EffPot} are depicted by the dotted curves, while those obtained from the 
$O(\lambda)$-improved expression in \eq{eq:DesOfImprovedEffPotForLargeLam1Improved} are presented by the dashed lines. 
In panel (b) these curves coincide. The vacuum expectation value $v$ underlying these analytical calculations has here been 
set to its numerically determined value computed in the highest statistics run, which is the one at $\lambda=0$.}
{Dependence of the Higgs correlator mass $m_{Hc}$ on the quartic coupling constant $\lambda\ll 1$.}

As a first approach the presented numerical data are again compared to the analytical predictions from the effective potential
given in \eq{eq:EffPot}. The corresponding analytical expectations are depicted by the dotted curves in \fig{fig:HiggsMassDepOnQuarticCoupling}a.
Qualitatively, the analytical calculation correctly describes the behaviour of $\delta m_{Hc}^2$ and $m_0^2$. 
On a quantitative level, however, there is a huge discrepancy between the analytical predictions and the corresponding numerical
results concerning both, the mass shifts as well as the bare masses. 

The origin of this discrepancy is that the approximation of the bosonic contributions to $\breve U[\breve v]$ through $S_\Phi[\Phi']$,
which was justified in the limit $N_f\rightarrow \infty$ considered in \sect{sec:SmalYukCoup}, is just too crude at $N_f=1$ and $\lambda\neq 0$.
In fact, the given large $N_f$ result for the effective potential $\breve U[\breve v]$ does not even contain all contributions of order $O(\lambda)$. 
Here, the idea is thus to improve the analytical prediction to a practically sufficient level by determining all contributions
to the effective potential $\breve U[\breve v]$ being of order $O(\lambda)$. 

For that purpose we go back to the definition of the effective potential given in \eq{eq:DefOfEffPotInMomSpace}. 
Since we consider here only the dependence on the constant mode, the amplitude of which is specified by the parameter $\breve v$,
the latter definition simplifies to
\bea
\label{eq:DefOfEffPotLargeLam}
\breve U[\breve v] &\fhs{-2mm}=\fhs{-2mm}&  -\frac{1}{V}\log\left(\int \intD{\psi}\intD{\bar\psi}  \left[\prod\limits_{0\neq p\in\ImpSpace} 
\intd{\tilde h_p}\, \intd{\tilde g_p}\right]\,
e^{-S_I[\breve v, h, g,\psi,\bar\psi]} \cdot e^{-S_0[\breve v, h, g,\psi,\bar\psi]}
\Bigg|_{\tilde h_0=\breve v\cdot \sqrt{V},\,\, \tilde g_0^\alpha=0}\right)  \nonumber \\
&\fhs{-2mm}+\fhs{-2mm}& \frac{1}{2} m_0^2\breve v^2 + \lambda \breve v^4
\eea
with $\intd{\tilde g_p}\equiv \intd{\tilde g^1_p}\intd{\tilde g^2_p}\intd{\tilde g^3_p}$ and $\alpha=1,2,3$. The 
corresponding expressions for the Gaussian and the interacting contributions $S_0[\breve v, h, g,\psi,\bar\psi]$ 
and $S_I[\breve v, h, g,\psi,\bar\psi]$ are then given as
\bea
\label{eq:DefOfGaussActionEffPotLargeLam}
S_0[\breve v, h, g,\psi,\bar\psi] &=& 
\frac{1}{2} \sum\limits_{0\neq p \in \ImpSpace} \tilde h_{p} \left[ \hat p^2 + m_0^2\right] \tilde h_{-p}  
+ \frac{1}{2} \sum\limits_{\alpha =1 }^3\sum\limits_{0\neq p \in \ImpSpace} \tilde g^\alpha_{p} \left[ \hat p^2 + m_0^2\right] \tilde g^\alpha_{-p} \quad\quad \\
&+& \bar\psi \fermiMatvbreve  \psi \nonumber
\eea
and
\bea
\label{eq:DefOfInterActionEffPotLargeLam}
S_I[\breve v, h, g,\psi,\bar\psi] &\fhs{-4mm}=\fhs{-4mm}& 
\frac{\lambda}{V} \fhs{-1mm}\sum\limits_{p_1,\ldots,p_4}\fhs{-1mm} \delta_{p_1+p_2+p_3+p_4,0} 
\left(\tilde h_{p_1}\tilde h_{p_2} +  \sum\limits_{\alpha=1}^3 \tilde g^\alpha_{p_1}\tilde g^\alpha_{p_2} \right)
\left(\tilde h_{p_3}\tilde h_{p_4} +  \sum\limits_{\alpha=1}^3 \tilde g^\alpha_{p_3}\tilde g^\alpha_{p_4} \right) \quad \quad\nonumber \\
&\fhs{-4mm}-\fhs{-4mm}& \frac{\lambda}{V} \tilde h_0^4 + 
V^{-1/2}\fhs{-3mm}\sum\limits_{{p_1\ldots p_3\in\ImpSpace}\atop{p_2\neq 0}}\fhs{-3mm}
\delta_{p_1,p_2+p_3} \cdot \tilde {\bar\psi}_{p_1}\left[\tilde h_{p_2} \hat B_0 + \sum\limits_{\alpha=1}^3 \tilde g^\alpha_{p_2} \hat B_\alpha  \right] 
\GammaOpMom(p_3)\tilde \psi_{p_3} 
\eea
in the here considered notation. 

As already discussed in \sect{sec:SmallYukawaCouplings} a diagrammatic expansion of this effective potential can then be obtained by 
expanding $\exp(-S_I)$ into a power series. Here we are only interested in all those diagrams being of order $O(\lambda)$.
For the purpose of a brief overview they are sketched in \fig{fig:DiagramsForEffectivePotAt1OrderLambda}. Both depicted
diagrams are of first order in $\lambda$, the one on the right, however, does not depend on the parameter $\breve v$
in the chosen decomposition of the total action in \eq{eq:DefOfEffPotLargeLam}. It will therefore be neglected in the 
following. Calculating thus only the contributions of the remaining diagram then yields the result
\bea
\label{eq:DesOfImprovedEffPotForLargeLam1}
\breve U[\breve v] &=& \frac{1}{2} m_0^2 \breve v^2 + \lambda\breve v^4 + \breve U_F[\breve v] + \breve U_H[\breve v] + \breve U_G[\breve v], \\
\label{eq:DesOfImprovedEffPotForLargeLam2}
\breve U_H[\breve v] &=& \lambda\breve v^2V^{-1} \cdot \sum\limits_{0\neq p\in \ImpSpace}\frac{6}{\hat p^2 + m_0^2}, \\
\label{eq:DesOfImprovedEffPotForLargeLam3}
\breve U_G[\breve v] &=& \lambda\breve v^2V^{-1} \cdot \sum\limits_{0\neq p\in \ImpSpace}\frac{6}{\hat p^2 + m_0^2}, 
\eea
where the relation $1-\lambda x + O(\lambda^2)=\exp(-\lambda x + O(\lambda^2))$ was used to raise the contributions
from the considered diagrams to the exponent and thus into the effective action $\breve U[\breve v]$ without changing
the achieved order in $\lambda$. The given expression for $\breve U[\breve v]$ now contains all contributions of 
order $O(\lambda)$. For clarification it is pointed out that contributions of the form $y_{t,b}^n\lambda$ with $n>0$
are explicitly not meant to be covered by the notation $O(\lambda)$. 
It is further remarked that the numbers in the nominators of \eqs{eq:DesOfImprovedEffPotForLargeLam2}{eq:DesOfImprovedEffPotForLargeLam3} 
arise from the underlying combinatorics of the four point vertex. 

There is only one subtlety about the expressions in \eqs{eq:DesOfImprovedEffPotForLargeLam1}{eq:DesOfImprovedEffPotForLargeLam3} that
should be improved, in order to extend the applicability of the given analytical formulas. The problem is that
the considered expansion breaks down as soon as the bare mass $m_0^2$ becomes negative. The reason is that
the Gaussian integration in \eq{eq:DefOfEffPotLargeLam} induced by the contribution $S_0$ becomes divergent in 
that case. Though this scenario is not encountered in \fig{fig:HiggsMassDepOnQuarticCoupling} due to the presented 
bare quartic coupling constants $\lambda$ being chosen sufficiently small, it will later play a role in 
\sect{sec:ModExtByHigherOrder}, where negative bare mass parameters $m_0^2<0$ will be encountered. We therefore 
shall discuss this issue already at this point.

%\includeFigSingleSmall{LPTDiagramsForEffectivePotentialAtFirstOrderLambda}
\includeFigSingleSmall{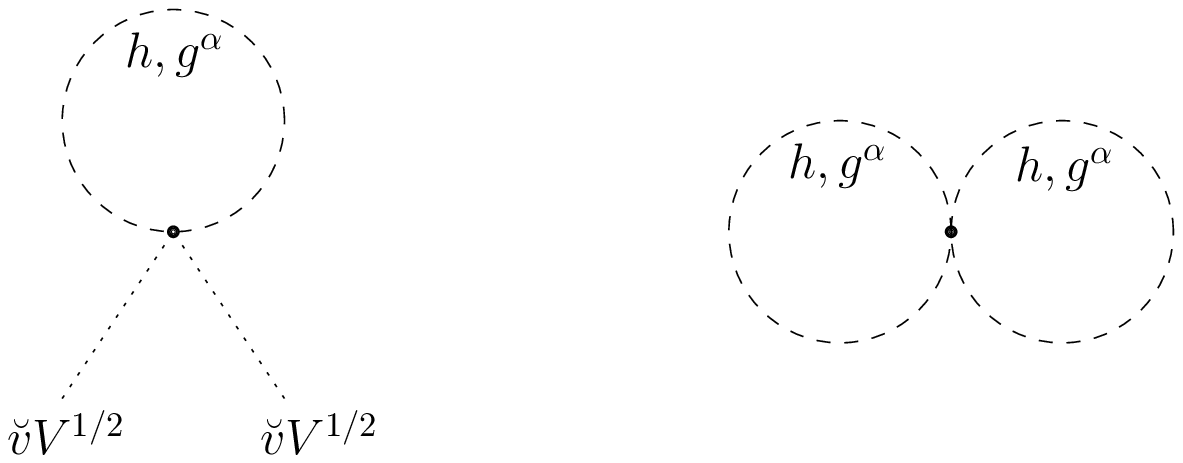}
{fig:DiagramsForEffectivePotAt1OrderLambda}
{Illustration of the purely bosonic diagrams arising from the expansion of $\exp(-S_I)$ and contributing to the effective potential $\breve U[\breve v]$ 
at order $O(\lambda)$. The closed propagator loops are summed over all momenta $0\neq k\in\ImpSpace$ excluding the constant mode due to the 
definition of the constraint effective potential in \eq{eq:DefOfEffPotLargeLam}. The dotted lines actually do not indicate propagators but 
only a multiplication with the outer amplitude $\tilde h_{0}=V^{1/2} \breve v$. The diagram on the right does not depend on the outer amplitude 
and will therefore be neglected in the following.}
{Purely bosonic diagrams contributing to the effective potential at order $O(\lambda)$.}

The aforementioned problem can be fixed by a partial renormalization procedure, where one simply replaces the bare mass $m_0$ 
in the denominators of \eqs{eq:DesOfImprovedEffPotForLargeLam2}{eq:DesOfImprovedEffPotForLargeLam3} with the renormalized 
Higgs boson mass $m_H$ and the renormalized Goldstone mass $m_G$, respectively. More precisely, we assume here a vanishing mass 
for the case of the Goldstone contribution $\breve U_G[\breve v]$ and take the Higgs boson mass estimate $m_{He}$ in 
\eq{eq:PropMassFromU} as the value for the renormalized Higgs boson mass leading then to the improved expression 
for the effective potential
\bea
\label{eq:DesOfImprovedEffPotForLargeLam1Improved}
\breve U[\breve v] &=& \frac{1}{2} m_0^2 \breve v^2 + \lambda\breve v^4 + \breve U_F[\breve v] + \breve U_H[\breve v] + \breve U_G[\breve v], \\
\label{eq:DesOfImprovedEffPotForLargeLam2Improved}
\breve U_H[\breve v] &=& \lambda\breve v^2V^{-1} \cdot \sum\limits_{0\neq p\in \ImpSpace}\frac{6}{\hat p^2 + m_{He}^2}, \\
\label{eq:DesOfImprovedEffPotForLargeLam3Improved}
\breve U_G[\breve v] &=& \lambda\breve v^2V^{-1} \cdot \sum\limits_{0\neq p\in \ImpSpace}\frac{6}{\hat p^2 + 0}.
\eea
For clarification it is pointed out that the considered replacements are justified, since they do not
change the achieved order $O(\lambda)$ of the final result. This is because the bare mass $m_0^2$, 
which is fixed according to \eq{eq:BareMassFromU} for a given value of the vev $v$, coincides with zero
and $m^2_{He}$ up to corrections of order $O(\lambda^ny^{2m}:\, n,m\in\N_0,\, n+m\ge 1)$. 

The Higgs boson mass estimate $m_{He}$ can then be determined in a self-consistent manner by iteratively calculating the
effective potential $\breve U[\breve v]$ from \eq{eq:DesOfImprovedEffPotForLargeLam1Improved} based on the
value for $m_{He}$ that has been obtained from \eq{eq:PropMassFromU} in the preceding iteration.

The results on the mass shifts and bare masses obtained by the improved effective potential in 
\eq{eq:DesOfImprovedEffPotForLargeLam1Improved} are shown in 
\fig{fig:HiggsMassDepOnQuarticCoupling}a depicted by the dashed lines. The improved results are in very 
good agreement with the numerical data. Of course, this correspondence only holds for sufficiently small
values of the quartic self-coupling constant according to the neglection of higher order contributions.
It is also interesting to note that the $\lambda$-dependence of the Higgs boson mass itself is already well 
described by the large $N_f$ potential in \eq{eq:EffPot} as seen in \fig{fig:HiggsMassDepOnQuarticCoupling}b. 
This is because the contribution to $m_{H_e}$ arising from the corrections to $m_0^2$ and $\delta m_H^2$, as 
induced by the consideration of $\breve U_H[\breve v]$ and $\breve U_G[\breve v]$, exactly cancels. From the 
above observations one can thus conclude that the improved effective potential in \eq{eq:DesOfImprovedEffPotForLargeLam1Improved} 
gives a good quantitative description of the Higgs boson mass in the weakly interacting regime of the considered 
Higgs-Yukawa model.

The main result of the present section is, that the lightest Higgs boson masses are indeed observed at vanishing bare 
quartic coupling constant, \ie $\lambda=0$. This can be inferred from the analytical result in \eq{eq:DesOfImprovedEffPotForLargeLam1Improved}.
Moreover, it can directly be observed in \fig{fig:HiggsMassDepOnQuarticCoupling}b, where the dependence of the 
Higgs boson mass on the bare parameter $\lambda$ is explicitly presented. That this dependence remains monotonically
rising with $\lambda$ also at much larger values of the bare quartic coupling constant will explicitly be demonstrated
in \sect{sec:DepOfHiggsMassonLargeLam}. Anticipating the latter observation at this point, the search for the lower 
Higgs boson mass bound can therefore safely be restricted to the scenario of vanishing bare quartic coupling constant,
\ie $\lambda=0$.

Finally, it is remarked that the decomposition of the total action into a Gaussian and an interacting contribution 
as given in \eq{eq:DefOfEffPotLargeLam} is not unique as already discussed in \sect{sec:SmallYukawaCouplings}. 
For instance, terms of the form $\lambda\breve v^2\tilde h_p\tilde h_{-p}$ and $\lambda\breve v^2\tilde g^\alpha_p\tilde g^\alpha_{-p}$ 
could also have been assigned to the Gaussian part $S_0$, which would then have led to a different result 
for $\breve U[\breve v]$ matching the here provided expression only up to the considered order $O(\lambda)$. It has, 
however, turned out in practice that the effective potential given in \eq{eq:DesOfImprovedEffPotForLargeLam1Improved} 
yields better agreement with the here considered direct Monte-Carlo calculations than this alternative approach.

%-------------------------------------------------------------------------------------------------------------------- 
\section{Lower Higgs boson mass bounds}
\label{sec:SubLowerHiggsMassbounds}

Given the knowledge about the $\lambda$-dependence of the renormalized Higgs boson mass, which was investigated in 
the preceding section coming to the concluding result that the lightest Higgs boson masses are indeed obtained
at vanishing bare quartic self-coupling constant as expected, we can now safely determine the cutoff-dependence of the lower 
Higgs boson mass bound $m_H^{low}(\Lambda)$ by evaluating the Higgs boson mass at $\lambda=0$ for several values 
of the cutoff $\Lambda$. 

This will be done for the mass degenerate case with $y_b=y_t$ and $N_f=1$ in \sect{sec:SubLowerHiggsMassboundsDegenCase}
as well as for the actual physical situation with $y_b/y_t=0.024$ and $N_f=3$ in \sect{sec:SubLowerHiggsMassboundsGenCase}.
However, a couple of restrictions limit the range of accessible energy scales. On the one hand all particle 
masses have to be small compared to $\Lambda$ to avoid unacceptably large cutoff effects, on the other hand all 
masses have to be large compared to the inverse lattice side lengths to bring the finite volume effects to a tolerable level.
As a minimal requirement we demand here that all particle masses $\hat m$ in lattice units fulfill 
\bea
\label{eq:RequirementsForLatMass}
\hat m < 0.5& \quad \mbox{and} \quad & \hat m\cdot L_{s,t}>2,
\eea
which already is a rather loose condition in comparison to the common situation in QCD, where one usually demands
at least $\hat m \cdot L_{s,t}>3$.

It will, however, be seen later that even these less restrictive requirements cannot be fulfilled in the actual physical 
setup $y_b/y_t=0.024$, $N_f=3$ with our available resources in contrast to the mass degenerate case with equal top 
and bottom quark masses. This is due to the immense lattice size that would be required to accommodate the
bottom quark with a mass of $m_b/a=\GEV{4.2}$ and the top quark with a mass of $m_t/a=\GEV{175}$ simultaneously
on the same lattice. 

Besides this technical difficulty there is also the conceptual problem in the non-degenerate case 
that the complex phase of the fermion determinant is not bound to real values. Though the fluctuation of
the complex phase seems to be strongly suppressed in actual Monte-Carlo simulations of physical interest as discussed 
in \sect{eq:ComplexPhaseOfFermionDetGenCase}, the true impact of this complex phase on the considered observables 
remains unclear in the context of the large scale computations that are to be performed here. 

We therefore perform the major part of the lower Higgs boson mass bound analysis in the mass degenerate
case with $y_b=y_t$ and $N_f=1$.

%-------------------------------------------------------------------------------------------------------------------- 
\subsection{Degenerate case with \texorpdfstring{$y_b=y_t$ and $N_f=1$}{yb=yt and Nf=1}}
\label{sec:SubLowerHiggsMassboundsDegenCase}

As a consequence of the chosen requirements given in \eq{eq:RequirementsForLatMass} concerning the cutoff $\Lambda$,
the lattice side lengths $L_{s,t}$, and the lightest and heaviest masses in the particle spectrum 
we will be able to investigate the lower Higgs boson mass bound $m_H^{low}(\Lambda)$ at 
energy scales $\Lambda$ ranging approximately from $\GEV{350}$ to $\GEV{1100}$. 
This follows when assuming a lattice with side lengths $L_s=L_t=32$, a degenerate top/bottom quark mass of 
$175\, \mbox{GeV}$, and a Higgs boson mass in the range of $\GEV{40}$ to $\GEV{70}$. These numbers
are justified aposteriori by the final results of the obtained Higgs boson mass bounds.
According to \eqs{eq:FixationOfPhysScale}{eq:DefOfCutoffLambda} the aforementioned range of energy scales 
is directly related to an interval of eligible values of the inverse vev $1/v$. Assuming\footnote{This
estimate is provided here only for the purpose of a first orientation. In the actual evaluation of the lattice 
calculations the respective value of $Z_G$ is, of course, explicitly determined.} a Goldstone renormalization 
constant of $Z_G=1$ the allowed values of the vacuum expectation value $v$ are given here as $1/v \in [1.4, 4.5]$. 

In the following we will consider a series of Monte-Carlo runs performed at five different values of the hopping
parameter $\kappa$ to cover the whole specified interval of energy scales. These lattice calculations will be run on
various lattice volumes to study the finite size effects and to allow ultimately for an infinite volume extrapolation
of the obtained lattice results. The model parameters underlying these Monte-Carlo calculations are listed in 
\tab{tab:SummaryOfParametersForLowerHiggsMassBoundRuns}. In this scenario the degenerate Yukawa coupling constants
have been fixed according to the tree-level relation in \eq{eq:treeLevelTopMass} aiming at the reproduction of the 
phenomenologically known top quark mass and the quartic coupling constant has been set to $\lambda=0$ for the aforementioned
reasons.

\includeTab{|ccccccccc|}
{
$\kappa$ & $L_s$                    & $L_t$ & $N_f$ &  $\hat \lambda$ & $\hat y_t$     & $\hat y_b/\hat y_t$ & 1/v & $\Lambda$ \\
\hline
0.12301  & 10,12,14,16,18,20,24,32  &   32  &  1    &  0              & 0.35285        & 1                 & $\approx 4.8$   & $\GEV{\approx 1160}$\\
0.12303  & 10,12,14,16,18,20,24,32  &   32  &  1    &  0              & 0.35288        & 1                 & $\approx 3.5$   & $\GEV{\approx 850}$\\
0.12306  & 10,12,14,16,18,20,24,32  &   32  &  1    &  0              & 0.35292        & 1                 & $\approx 2.6$   & $\GEV{\approx 630}$\\
0.12309  & 10,12,14,16,18,20,24,32  &   32  &  1    &  0              & 0.35296        & 1                 & $\approx 2.0$   & $\GEV{\approx 500}$\\
0.12313  & 10,12,14,16,18,20,24,32  &   32  &  1    &  0              & 0.35302        & 1                 & $\approx 1.6$   & $\GEV{\approx 400}$\\
}
{tab:SummaryOfParametersForLowerHiggsMassBoundRuns}
{The model parameters of the Monte-Carlo runs underlying the subsequent lattice calculation of the lower Higgs boson mass bound
are presented. In total, a number of 40 Monte-Carlo runs have been performed for that purpose. The available statistics of 
generated field configurations $\Nconf$ varies depending on the respective lattice volume. In detail we have $\Nconf\approx 20,000$ 
for $10\le L_s\le 16$, $\Nconf\approx10,000$ for $18\le L_s\le 20$, and $\Nconf\approx5,000$ for $24\le L_s\le 32$.
The numerically determined values of $1/v$ and $\Lambda$ are also approximately given. These numbers vary, of course, depending
on the respective lattice volumes and serve here only for the purpose of a rough orientation. The degenerate Yukawa coupling 
constants have been chosen according to the tree-level relation in \eq{eq:treeLevelTopMass} aiming at the reproduction of the 
phenomenologically known top quark mass.}
{Model parameters of the Monte-Carlo runs underlying the lattice calculation of the lower Higgs boson mass bound.}

The numerically obtained Higgs correlator masses $m_{Hc}$ resulting in these lattice calculations 
are presented in \fig{fig:LowerHiggsCorrelatorBoundFiniteVol}.
To illustrate the influence of the finite lattice volume those results, belonging to the same parameter
sets, differing only in the underlying lattice size, are connected by dotted lines to guide the eye. From 
these findings one learns that the model indeed exhibits strong finite volume effects when approaching the upper limit 
of the defined interval of reachable cutoffs, as expected, \ie when the size of the correlation length of the Higgs 
particle becomes comparable to the lattice side extension.

In \fig{fig:LowerHiggsCorrelatorBoundFiniteVol} the Higgs boson masses $m_{Hc}$ have actually been given in units of the
measured vacuum expectation value $v$, and the obtained data were plotted versus $1/v$. This presentation is
equivalent to showing the Higgs boson masses in physical units versus the cutoff $\Lambda$ up to a global scale
factor and up to corrections induced by a non-constant renormalization factor $Z_G\neq 1$. The chosen illustration,
however, allows for direct comparability between the given numerical results and the analytical predictions derived 
from the effective potential in the preceding section which are depicted by the dashed curves
in \fig{fig:LowerHiggsCorrelatorBoundFiniteVol}.
By comparing the analytical and numerical results to each other one observes that the analytical calculations 
describe the actually measured cutoff dependence of the Higgs boson masses very well. The analytical calculations
also correctly predict the observed finite volume effects of the Higgs boson mass $m_{Hc}$. From this finding
it can be concluded that the behaviour of the model is very well understood by means of the effective potential
calculations discussed in \sect{sec:EffPot}.

%\includeFigSingleMedium{HiggsMassVsCutoffAtZeroCouplingLATUNITS}
\includeFigSingleMedium{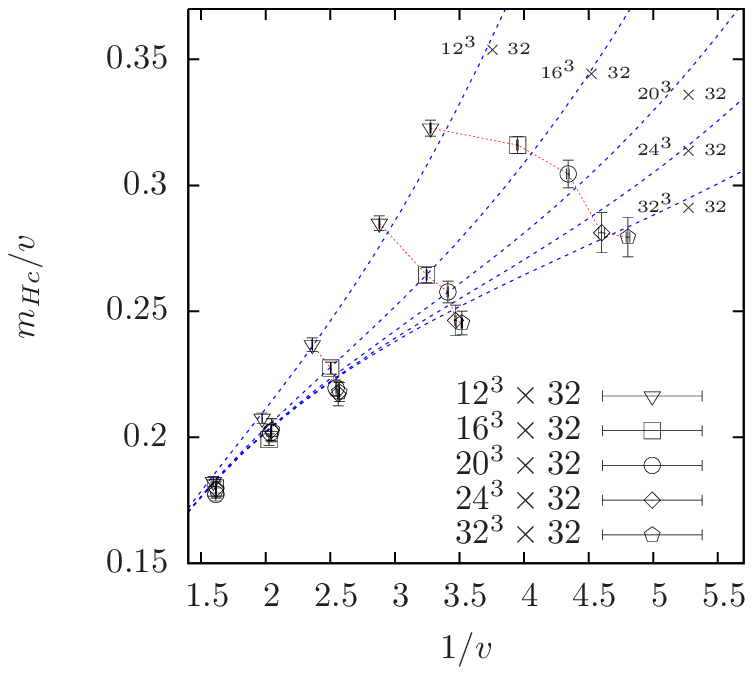}
{fig:LowerHiggsCorrelatorBoundFiniteVol}
{The Higgs correlator mass $m_{Hc}$ is presented in units of the vacuum expectation value $v$ 
versus $1/v$. These results have been determined in the direct Monte-Carlo calculations specified in
\tab{tab:SummaryOfParametersForLowerHiggsMassBoundRuns}. Those runs with identical parameter sets differing 
only in the underlying lattice volume are connected via dotted lines to illustrate the effects of the 
finite volume. Here, however, only a subset of the specified runs is presented for the sake of readability. 
The analytical predictions for the respective lattice sizes derived from the effective potential discussed 
in \sect{sec:EffPot} are depicted by the dashed curves.}
{Dependence of the Higgs correlator mass on $1/v$ at small quartic coupling constants.}

To obtain the desired lower Higgs boson mass bounds $m_H^{low}(\Lambda)$ these finite volume results on the Higgs
boson mass have to be extrapolated to the infinite volume limit and the renormalization factor $Z_G$ has to be
properly considered. For that purpose the finite volume dependence
of the Monte-Carlo results on the renormalized vev $v_r=v/\sqrt{Z_G}$ and the Higgs boson mass $m_{Hc}$ is explicitly shown in 
\fig{fig:FiniteVolumeEffectsOfLowerHiggsMassBound}a and \fig{fig:FiniteVolumeEffectsOfLowerHiggsMassBound}b, 
respectively. One sees in these plots that the finite volume effects are rather mild at $\kappa=0.12313$ 
corresponding to the lowest presented cutoff in \fig{fig:LowerHiggsCorrelatorBoundFiniteVol}
with $m_{Hc}\cdot L_{s,t}>3.2$ on the \lattice{32}{32}while the renormalized vev, and thus the associated 
cutoff $\Lambda$, as well as the Higgs boson mass itself vary strongly with increasing lattice size $L_s$ at the 
smaller presented hopping parameters.

It is well known from lattice investigations of the pure $\Phi^4$-theory~\cite{Hasenfratz:1989ux,Hasenfratz:1990fu,Gockeler:1991ty}
that the vev as well as the mass receive strong contributions from the Goldstone modes, inducing finite volume effects 
of algebraic form starting at order $O(L_s^{-2})$. The next non-trivial finite volume contribution was shown to 
be of order $O(L_s^{-4})$. In \fig{fig:FiniteVolumeEffectsOfLowerHiggsMassBound} the obtained data are 
therefore plotted versus $1/L_s^2$. Moreover, the aforementioned observation justifies to apply the linear fit ansatz 
\beq
\label{eq:LinFit}
f^{(l)}_{v,m}(L_s^{-2}) = A^{(l)}_{v,m} + B^{(l)}_{v,m}\cdot L_s^{-2}
\eeq
to extrapolate these data to the infinite volume limit, where the free fitting parameters $A^{(l)}_{v,m}$ and 
$B^{(l)}_{v,m}$ with the subscripts $v$ and $m$ refer to the renormalized vev $v_r$ and the Higgs boson mass 
$m_{Hc}$, respectively. 

%\includeFigDouble{HiggsMassVsCutoffAtZeroCouplingFiniteVolumeEffectsVeV}{HiggsMassVsCutoffAtZeroCouplingFiniteVolumeEffectsHiggsMass}
\includeFigDouble{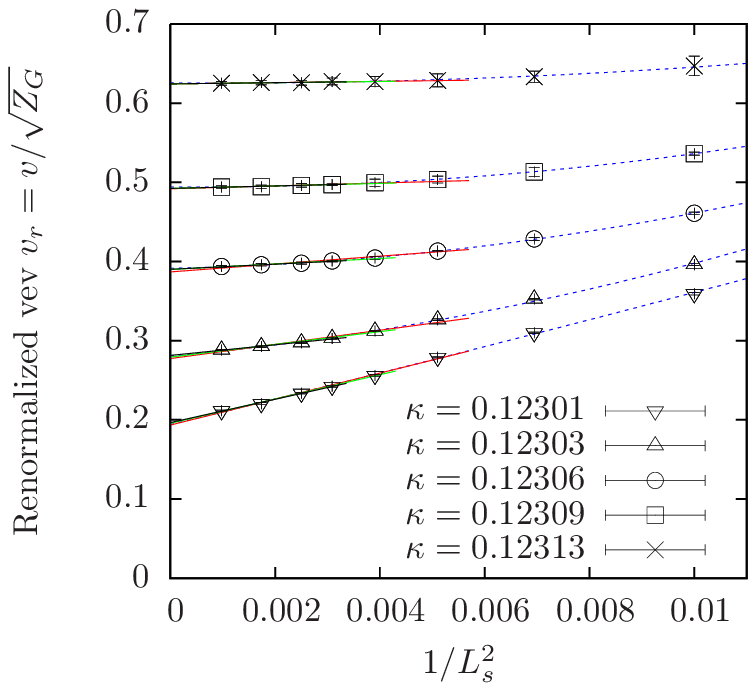}{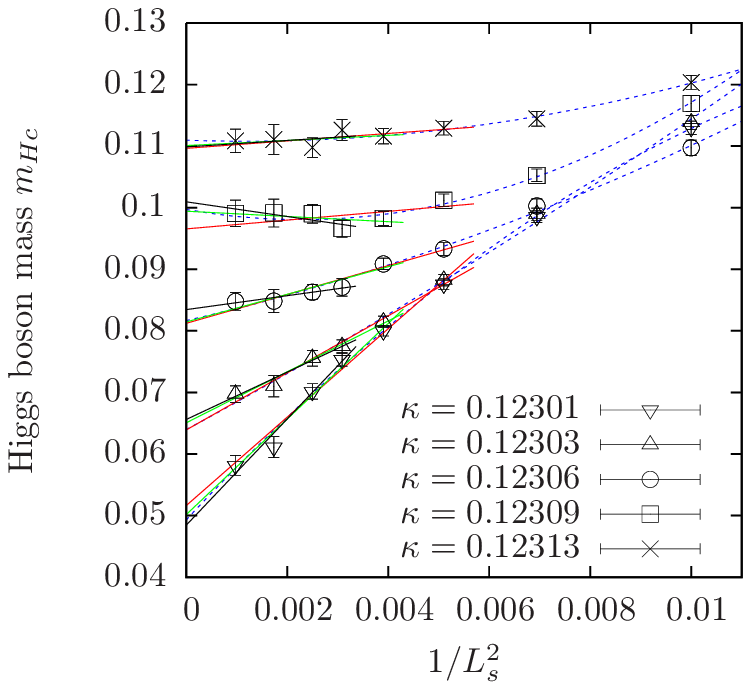}
{fig:FiniteVolumeEffectsOfLowerHiggsMassBound}
{The dependence of the renormalized vev $v_r = v/\sqrt{Z_G}$ on the squared inverse lattice side length $1/L^2_s$ 
is presented in panel (a) as calculated in the direct Monte-Carlo calculations specified in \tab{tab:SummaryOfParametersForLowerHiggsMassBoundRuns}.
In panel (b) the respective dependence of the Higgs correlator mass $m_{Hc}$ on $1/L^2_s$ is shown as obtained in the 
same Monte-Carlo runs. In both plots the dashed curves display the parabolic fits according to the fit ansatz in 
\eq{eq:ParaFit}, while the solid lines depict the linear fits resulting from \eq{eq:LinFit} for the three threshold 
values $L_s'=14$ (red), $L_s'=16$ (green), and $L_s'=18$ (black).
}
{Dependence of the renormalized vev $v_r = v/\sqrt{Z_G}$ and the Higgs correlator mass $m_{Hc}$ on the squared 
inverse lattice side length $1/L^2_s$ at small quartic coupling constants.}

To respect the presence of higher order terms in $1/L_{s}^{2}$ only the largest lattice sizes are included into 
this linear fit. Here, we select all lattice volumes with $L_s\ge L'_s$. As a consistency check, testing the dependence 
of the resulting infinite volume extrapolations on the choice of the fit procedure, the threshold value $L'_s$ is 
varied. The respective results are listed in \tab{tab:ResultOfLowerHiggsMassFiniteVolExtrapolation}.
Moreover, the parabolic fit ansatz 
\beq
\label{eq:ParaFit}
f^{(p)}_{v,m}(L_s^{-2}) = A^{(p)}_{v,m} + B^{(p)}_{v,m}\cdot L_s^{-2} + C^{(p)}_{v,m}\cdot L_s^{-4}
\eeq
is additionally considered. It is applied to the whole range of available lattice sizes. The deviations
between the various fitting procedures with respect to the resulting infinite volume extrapolations of the
considered observables can then be considered as an additional, systematic uncertainty of the obtained values.
The respective fit curves are displayed in \fig{fig:FiniteVolumeEffectsOfLowerHiggsMassBound}a,b and the 
corresponding infinite volume extrapolations of the renormalized vev and the Higgs boson mass, which 
have here been obtained as the average over the three specified linear fits, are listed in 
\tab{tab:ResultOfLowerHiggsMassFiniteVolExtrapolation}.

\includeTab{|c|c|c|c|c|c|}{
\multicolumn{6}{|c|}{Vacuum expectation value $v$} \\ \hline
$\kappa$  	 & $A^{(l)}_v$, $L'_s=14$  & $A^{(l)}_v$, $L'_s=16$  & $A^{(l)}_v$, $L'_s=18$         & $A^{(p)}_v$         & $v_r$               \\ \hline
$\,0.12301\,$  	 & $\, 0.1935(18)\, $   & $\, 0.1956(12)\, $   & $\, 0.1966(13)\, $  & $\, 0.1940(20)\, $  & $\, 0.1952(15)(16)$  \\ 
$\,0.12303\,$  	 & $\, 0.2774(22)\, $   & $\, 0.2791(19)\, $   & $\, 0.2813(11)\, $  & $\, 0.2805(20)\, $  & $\, 0.2793(18)(20)$  \\ 
$\,0.12306\,$  	 & $\, 0.3868(17)\, $   & $\, 0.3898(7) \, $   & $\, 0.3902(6) \, $  & $\, 0.3914(6) \, $  & $\, 0.3889(11)(23)$  \\ 
$\,0.12309\,$  	 & $\, 0.4919(7) \, $   & $\, 0.4923(4) \, $   & $\, 0.4924(4) \, $  & $\, 0.4941(2) \, $  & $\, 0.4922(5)(11)$  \\ 
$\,0.12313\,$  	 & $\, 0.6244(3) \, $   & $\, 0.6244(4) \, $   & $\, 0.6244(6) \, $  & $\, 0.6256(6) \, $  & $\, 0.6244(5)(6)$  \\ \hline
\multicolumn{6}{|c|}{Higgs correlator mass $m_{Hc}$} \\ \hline
$\kappa$  	 & $A^{(l)}_m$, $L'_s=14$  & $A^{(l)}_m$, $L'_s=16$  & $A^{(l)}_m$, $L'_s=18$         & $A^{(p)}_m$         & $m_{Hc}$               \\ \hline
$\,0.12301\,$  	 & $\, 0.0516(17)\, $   & $\, 0.0501(20)\, $   & $\, 0.0484(24)\, $  & $\, 0.0493(15)\, $  & $\, 0.0500(21)(16)$  \\ 
$\,0.12303\,$  	 & $\, 0.0639(10)\, $   & $\, 0.0651(9) \, $   & $\, 0.0656(11)\, $  & $\, 0.0640(14)\, $  & $\, 0.0649(10)(9)$  \\ 
$\,0.12306\,$  	 & $\, 0.0812(11)\, $   & $\, 0.0815(15)\, $   & $\, 0.0834(5) \, $  & $\, 0.0817(12)\, $  & $\, 0.0820(11)(11)$  \\ 
$\,0.12309\,$  	 & $\, 0.0966(20)\, $   & $\, 0.0994(16)\, $   & $\, 0.1010(18)\, $  & $\, 0.0997(15)\, $  & $\, 0.0990(18)(22)$  \\ 
$\,0.12313\,$  	 & $\, 0.1096(10)\, $   & $\, 0.1101(14)\, $   & $\, 0.1099(22)\, $  & $\, 0.1110(11)\, $  & $\, 0.1099(16)(7)$  \\ 
}
{tab:ResultOfLowerHiggsMassFiniteVolExtrapolation}
{The results of the infinite volume extrapolations of the Monte-Carlo data for the renormalized vev $v_r$ and the Higgs 
boson mass $m_{Hc}$ are presented as obtained from the parabolic ansatz in \eq{eq:ParaFit} and the linear approach in 
\eq{eq:LinFit} for the considered threshold values $L_s'=14$, $L_s'=16$, and $L_s'=18$. 
The final results on $v_r$ and $m_{Hc}$, displayed in the very right column, are determined here by averaging over
the three linear fit approaches. An additional, systematic uncertainty of these final results is specified
in the second pair of brackets taken from the largest observed deviation among all respective fit results, including
the parabolic fit.}
{Infinite volume extrapolation of the Monte-Carlo data for the renormalized vev $v_r$ and the Higgs 
boson mass $m_{Hc}$ at small quartic coupling constants.}

However, before the obtained results on the Higgs boson mass $m_{Hc}$ can actually be considered as the sought-after 
lower Higgs boson mass bound $\lowBound$, it has to be checked, that the physical top quark mass is actually 
reproduced in the chosen model parameter setups specified in \tab{tab:SummaryOfParametersForLowerHiggsMassBoundRuns}.
This is not obvious, since the tree-level relation given in \eq{eq:treeLevelTopMass} has been used so far to select the value
of the bare Yukawa coupling constants underlying the performed lattice calculations.

For that purpose the corresponding finite volume top quark masses\footnote{To save numerical resources, only every eighth
field configuration has been evaluated in the quark mass analysis.} are presented in \fig{fig:TopCorrelatorMassesFiniteVol}a.  
Those results belonging to the same parameter sets, differing only in the underlying lattice size,
are again connected by dotted lines to guide the eye and the solid horizontal line depicts the tree-level 
expectation. One clearly observes significant deviations between the latter tree-level expectation and the 
actual numerical results, in particular on the smaller lattice volumes and for large values of the cutoff. 

Here, the presented top quark masses have again been given in units of $v$, while the obtained data were plotted versus
$1/v$, allowing thus for a direct comparison with the analytical predictions derived in \sect{sec:QuarkMassesFromPT}. 
These analytical expectations are depicted by the dashed lines in \fig{fig:TopCorrelatorMassesFiniteVol}a and 
excellent agreement is observed. From this presentation one can deduce that the top quark mass, and 
thus the renormalized Yukawa coupling constant, decreases slightly with growing cutoff when holding the bare 
parameters $y_t$, $y_b$, and $\lambda$ fixed, which is the behaviour that one would have expected in a trivial theory.

%\includeFigDouble{TopMassVsCutoffAtZeroCouplingLATUNITS}{HiggsMassVsCutoffAtZeroCouplingFiniteVolumeEffectsTopMass}
\includeFigDouble{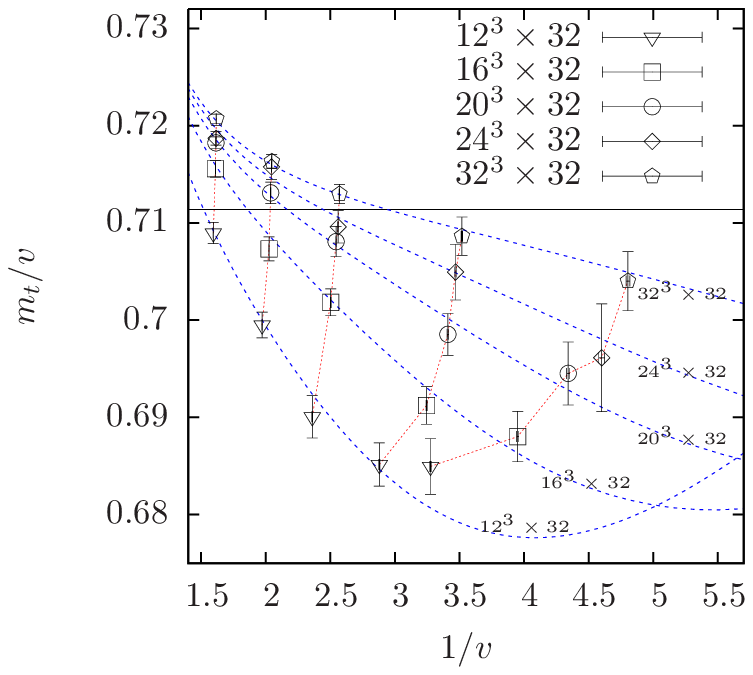}{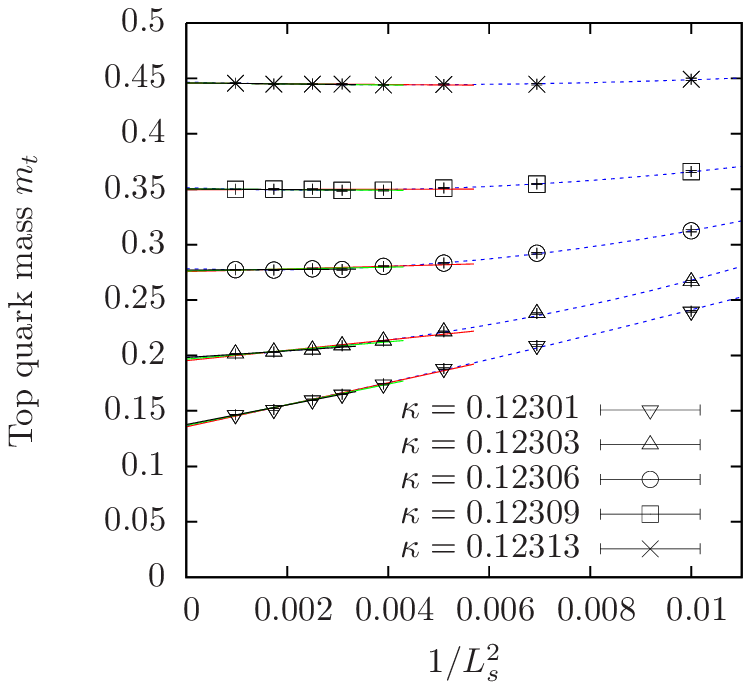}
{fig:TopCorrelatorMassesFiniteVol}
{The top quark mass $m_{t}$ is presented in units of the vacuum expectation value $v$ in panel (a) 
versus $1/v$. These results have been determined in the direct Monte-Carlo calculations specified in
\tab{tab:SummaryOfParametersForLowerHiggsMassBoundRuns}. Those runs with identical parameter sets differing 
only in the underlying lattice volume are connected via dotted lines to illustrate the effects of the 
finite volume. Here, however, only a subset of the specified runs is presented for the sake of readability. 
The analytical predictions for the respective lattice sizes derived from the perturbative calculation discussed 
in \sect{sec:QuarkMassesFromPT} are depicted by the dashed curves. The solid horizontal line depicts the tree-level
expectation. The corresponding top quark masses are plotted in lattice units versus $1/L_s^2$ in panel (b). 
The dashed curves display the parabolic fits according to the fit ansatz in 
\eq{eq:ParaFit}, while the solid lines depict the linear fits resulting from \eq{eq:LinFit} for the three threshold 
values $L_s'=14$ (red), $L_s'=16$ (green), and $L_s'=18$ (black).
}
{Dependence of the top quark mass on $1/v$ at small quartic coupling constants.}

One can in principle exploit the observed very good agreement between the analytical predictions and the actual
numerical results on the top quark mass to establish an efficient tuning of the bare Yukawa coupling constants 
beyond a trial-and-error approach, allowing ultimately to hold the fermion masses constant for each considered
cutoff\footnote{Apart from the top quark mass being constrained by the triviality property of the Higgs-Yukawa sector.} 
and lattice volume in some follow-up Monte-Carlo calculations. The aforementioned cutoff dependence of the actually 
measured top quark mass, however, becomes less prominent as the lattice volume increases and we will 
shortly see that in the infinite volume extrapolation the targeted value of the top quark mass is actually  
reproduced within the given errors. For that reason such an improvement has not been investigated in this study.

\includeTab{|c|c|c|c|c|c|}{
\multicolumn{6}{|c|}{Top mass} \\ \hline
$\kappa$  	 & $A^{(l)}_t$, $L'_s=14$  & $A^{(l)}_t$, $L'_s=16$  & $A^{(l)}_t$, $L'_s=18$         & $A^{(p)}_t$         & $v_r$               \\ \hline
$\,0.12301\,$  	 & $\, 0.1357(13)\, $   & $\, 0.1369(10)\, $   & $\, 0.1377(10)\, $  & $\, 0.1363(15)\, $  & $\, 0.1368(11)(10)$  \\ 
$\,0.12303\,$  	 & $\, 0.1953(17)\, $   & $\, 0.1968(14)\, $   & $\, 0.1981(14)\, $  & $\, 0.1983(13)\, $  & $\, 0.1967(15)(15)$  \\ 
$\,0.12306\,$  	 & $\, 0.2758(8) \, $   & $\, 0.2764(7) \, $   & $\, 0.2772(5) \, $  & $\, 0.2784(4) \, $  & $\, 0.2765(7)(13)$  \\ 
$\,0.12309\,$  	 & $\, 0.3497(5) \, $   & $\, 0.3503(3) \, $   & $\, 0.3503(4) \, $  & $\, 0.3514(4) \, $  & $\, 0.3502(4)(9)$  \\ 
$\,0.12313\,$  	 & $\, 0.4457(3) \, $   & $\, 0.4460(2) \, $   & $\, 0.4460(4) \, $  & $\, 0.4466(3) \, $  & $\, 0.4459(3)(5)$  \\ 
}
{tab:ResultOfLowerTopMassFiniteVolExtrapolation}
{The results of the infinite volume extrapolations of the Monte-Carlo data for the top quark mass $m_t$ are presented
as obtained from the parabolic ansatz in \eq{eq:ParaFit} and the linear approach in \eq{eq:LinFit} for the considered 
threshold values $L_s'=14$, $L_s'=16$, and $L_s'=18$. The final results on $m_{t}$, displayed in the very right column, 
are determined here by averaging over the three linear fit approaches. An additional, systematic uncertainty of these 
final results is specified in the second pair of brackets taken from the largest observed deviation among all respective 
fit results, including the parabolic fit.
}
{Infinite volume extrapolation of the Monte-Carlo data on the top quark mass $m_t$ at small quartic coupling constants.}

The already announced infinite volume extrapolation of the top quark mass is here performed in exactly the same manner
as described above. The numerically obtained finite volume top quark masses are plotted versus $1/L_s^2$ in 
\fig{fig:TopCorrelatorMassesFiniteVol}b together with the parabolic and linear fits arising from the fit approaches 
in \eq{eq:LinFit} and \eq{eq:ParaFit}. The corresponding results of the appropriately relabeled fit parameters are listed 
in \tab{tab:ResultOfLowerTopMassFiniteVolExtrapolation} and the final infinite volume extrapolations, also shown in that 
table, are again determined as the average over the three linear fit results as specified by the three listed values of 
the threshold parameter $L_s'$.

The obtained infinite volume extrapolations of the top quark mass are presented in \fig{fig:LowerHiggsMassBoundPredictionInifinteVol}a
depicted by the square symbols.
From these results one learns that the numerically computed top quark masses, extrapolated to the infinite
volume limit, are indeed consistent with the targeted value of the top quark mass within the given errors as already
stated above, thus finally justifying the underlying fixation of the bare Yukawa coupling constant according to the 
tree-level relation in \eq{eq:treeLevelTopMass}, at least for the here achieved accuracy.

The main result of this section, however, is presented in \fig{fig:LowerHiggsMassBoundPredictionInifinteVol}b, showing
the cutoff dependence of the lower Higgs boson mass bound as obtained from the infinite volume extrapolations summarized in 
\tab{tab:ResultOfLowerHiggsMassFiniteVolExtrapolation}. These numerical results, depicted by the square symbols here,
are compared to the analytical infinite volume predictions obtained from the effective potential given in 
\eqs{eq:EffPot}{eq:DefOfFermionicContToEffPot}, where the discrete sums have been replaced by corresponding integrals to 
obtain the analytical infinite volume predictions.

According to the relatively good agreement between the presented analytical and numerical findings it is furthermore reasonable 
to show also some analytical infinite volume predictions for a couple of other physical setups, which will not
be backed up by equally precise Monte-Carlo calculations here, with the only exception of the, however far less accurate,
data depicted by the circular symbols which will be discussed below. In \fig{fig:LowerHiggsMassBoundPredictionInifinteVol}b
three additional setups with varying values of $N_f$ and $y_b/y_t$ are additionally presented to study the influence of 
these parameters on the lower Higgs boson mass bound. One can infer from this presentation that the number of fermion 
generations (or here equivalently the number of colours) as well as the bottom-top mass splitting have a strong impact 
on the lower mass bound. 

%\includeFigDouble{TopMassFromInfiniteVolumeResultsVersusCutoff}{HiggsMassFromLargeNfForInfiniteVolume}
\includeFigDouble{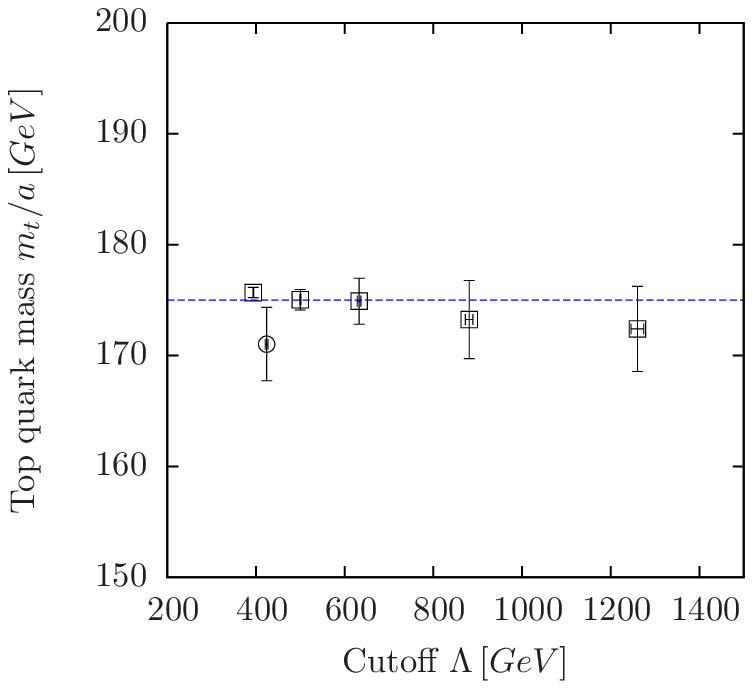}{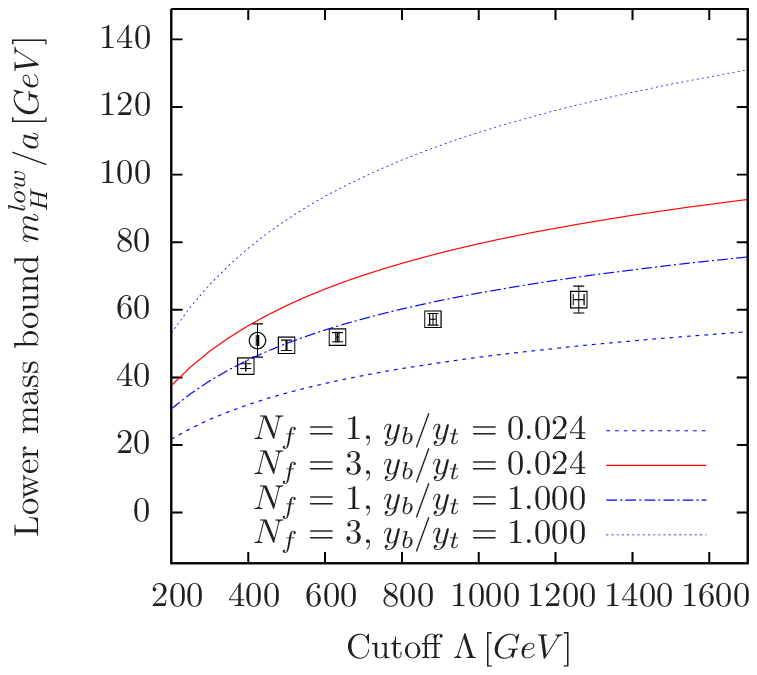}
{fig:LowerHiggsMassBoundPredictionInifinteVol}
{The cutoff dependence of the top quark masses extrapolated to the infinite volume limit is presented in panel (a). 
The dashed horizontal line depicts the tree-level expectation. In panel (b) the various curves show the cutoff 
dependence of the lower Higgs boson mass bound in the infinite volume limit
derived from the effective potential discussed in \sect{sec:EffPot} for different physical setups ($N_f, y_b/y_t$). 
The dash-dotted curve corresponds to the scenario ($N_f=1, y_b=y_t$) considered in this section, while
the solid curve represents the setup ($N_f=3, y_b/y_t=0.024$) coming closest to the actual physical situation 
in the Standard Model.
These analytical findings are compared to the corresponding infinite volume extrapolations of the obtained lattice 
results. In both panels the square symbols represent the results from \tab{tab:ResultOfLowerHiggsMassFiniteVolExtrapolation}
and \tab{tab:ResultOfLowerTopMassFiniteVolExtrapolation}, while the circles depict the top quark and Higgs boson masses 
obtained in the scenario ($N_f=3, y_b/y_t=0.024$), which is discussed in \sect{sec:SubLowerHiggsMassboundsGenCase}.
}
{Cutoff dependence of the top quark mass and the lower Higgs boson mass bound extrapolated to the infinite volume 
limit.}

The solid curve in \fig{fig:LowerHiggsMassBoundPredictionInifinteVol}b with $N_f=3$ and $y_b/y_t=0.024$ 
comes closest to the actual physical situation in the Standard Model and one observes that the lower Higgs
boson mass bound is significantly increased in this physically more relevant scenario as compared to the so far
discussed degenerate case with $N_f=1$. The aforementioned circular symbols in 
\fig{fig:LowerHiggsMassBoundPredictionInifinteVol} actually represent some infinite volume extrapolated 
results of corresponding lattice data computed in this latter physical setup, as will be
discussed in the subsequent section. Though this latter numerical result on the lower mass bound, available here
only at one selected cutoff parameter $\Lambda$, indeed tends to be larger than the previously presented numbers, 
as expected from the analytical predictions, the associated uncertainty of this result is too large to clearly 
distinguish between the considered degenerate and non-degenerate scenarios.

The concluding remark of this section is thus that the cutoff dependence of the infinite volume lower Higgs boson 
mass bounds $m_H^{low}(\Lambda)$ can indeed be calculated by means of the applied lattice techniques at least
in the mass degenerate case for an interval of cutoffs $\Lambda$ ranging approximately from $\GEV{350}$ to 
$\GEV{1100}$. From the obtained results it can be concluded that the lower Higgs boson mass bound is a manifest 
property of the pure Higgs-Yukawa sector that evolves directly from the Higgs-fermion interaction for a given 
Yukawa coupling parameter. For growing cutoff this lower mass constraint rises monotonically with flattening slope 
as expected from perturbation theory. Moreover, the quantitative size of the (analytically obtained) lower bound 
for $N_f=3$ and $y_b/y_t=0.024$ depicted by the solid line in \fig{fig:LowerHiggsMassBoundPredictionInifinteVol}
is comparable to the magnitude of the perturbative results based on vacuum stability considerations~\cite{Hagiwara:2002fs} 
which were discussed and presented in \sect{chap:Introduction}. A direct quantitative comparison is, however, non-trivial 
due to the different regularization schemes, blocking a direct matching of the underlying cutoff parameters $\Lambda$.

%-------------------------------------------------------------------------------------------------------------------- 
\subsection{Physical setup with \texorpdfstring{$y_b/y_t=0.024$ and $N_f=3$}{yb/yt=0.024 and Nf=3}}
\label{sec:SubLowerHiggsMassboundsGenCase}

In the preceding section the lower Higgs boson mass bound $\lowBound$ has been derived for the degenerate case
with equal top and bottom quark masses and $N_f=1$. Now, we want to extend these previous considerations in two
respects. First of all, we intend to reproduce the phenomenologically known values for both, the top quark and the bottom quark
masses, by setting the ratio of the Yukawa coupling constants to its phenomenological value,
which will be done here again with the help of the tree-level relation in \eq{eq:treeLevelTopMass}. The other modification is
that we will now properly respect the so far neglected colour index of the fermions by setting the fermion generation number
to three, \ie $N_f=3$, in the subsequent lattice calculations. For clarification it is pointed out, that there are
still no gauge fields included within the model. It is only the multiplicity induced by the fermion colour index
that will now appropriately be accounted for.

However, corresponding lattice calculations in this physical setup are much more demanding than the computations performed in the
mass degenerate case. From a technical perspective this is mainly due to the much larger condition numbers in the non-degenerate
scenario as observed in \sect{sec:PreconOfMDouble}. Moreover, there is the conceptual uncertainty arising from the fluctuating complex 
phase of the fermion determinant as pointed out in \sect{eq:ComplexPhaseOfFermionDetGenCase}. Finally, the huge discrepancy between the 
top and bottom masses would require enormous lattice volumes to accommodate both particles simultaneously on the same lattice, at least 
when retaining the minimal conditions in \eq{eq:RequirementsForLatMass}. 

\includeTab{|cccccccc|}
{
$\kappa$ & $L_s$                    & $L_t$ & $N_f$ &  $\Nconf$  & $\hat \lambda$ & $\hat y_t$     & $\hat y_b/\hat y_t$ \\
\hline
0.122204   & 10  &   32  &  3    &  10620   & 0              & 0.35169        & 0.024                 \\
0.122204   & 12  &   32  &  3    &  10860   & 0              & 0.35169        & 0.024                 \\
0.122204   & 14  &   32  &  3    &  6680   & 0              & 0.35169        & 0.024                 \\
0.122204   & 16  &   32  &  3    &  4600   & 0              & 0.35169        & 0.024                 \\
0.122204   & 18  &   32  &  3    &  2200   & 0              & 0.35169        & 0.024                 \\
0.122204   & 20  &   32  &  3    &  1240   & 0              & 0.35169        & 0.024                 \\
}
{tab:SummaryOfParametersForLowerHiggsMassBoundRunsNf3}
{The model parameters of the Monte-Carlo runs underlying the subsequent lattice calculation of the lower Higgs boson mass bound
in the non-degenerate case are presented. The Yukawa coupling constants have been chosen according to the tree-level relation 
in \eq{eq:treeLevelTopMass} aiming at the reproduction of the phenomenologically known top and bottom quark masses. The specified
hopping parameter $\kappa$ was selected with the intention of finally reaching a cutoff $\Lambda$ of approximately $\GEV{400}$.
}
{Model parameters of the Monte-Carlo runs underlying the lattice calculation of the lower Higgs boson mass bound
in the non-degenerate case.}

Subsequently, it will nevertheless be tried to determine a lower Higgs boson mass bound also in this physically more significant
setup. For that purpose a series of lattice calculations has been performed on different lattice volumes to allow later for an
infinite volume extrapolation. The model parameters underlying these computations are listed in
\tab{tab:SummaryOfParametersForLowerHiggsMassBoundRunsNf3}. Here, only one value for the hopping parameter $\kappa$, and thus
for the associated cutoff $\Lambda$, has been investigated due to the aforementioned difficulties. The given model parameters
were chosen such that the resulting cutoff becomes approximately $\GEV{400}$, since this setting clearly obeys the requirement
$\hat m<0.5$ in \eq{eq:RequirementsForLatMass}, while generating the least prominent finite volume effects of all parameter 
setups considered in the preceding section. 

%\includeFigDoubleDoubleHere{HiggsMassVsCutoffAtZeroCouplingAtNf3NonDegFiniteVolumeEffectsVeV}{HiggsMassVsCutoffAtZeroCouplingAtNf3NonDegFiniteVolumeEffectsHiggsMass}
%{HiggsMassVsCutoffAtZeroCouplingAtNf3NonDegFiniteVolumeEffectsTopMass}{HiggsMassVsCutoffAtZeroCouplingAtNf3NonDegFiniteVolumeEffectsBottomMass}
\includeFigDoubleDoubleHere{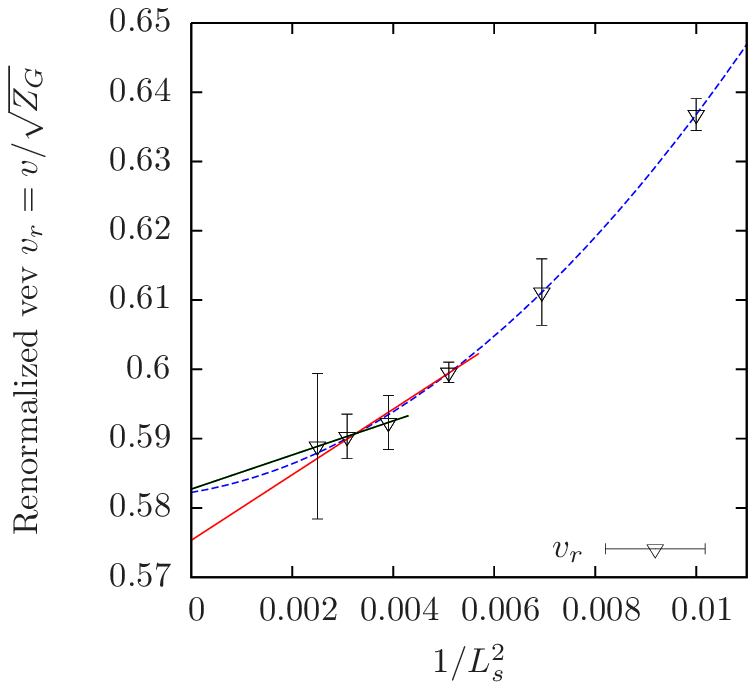}{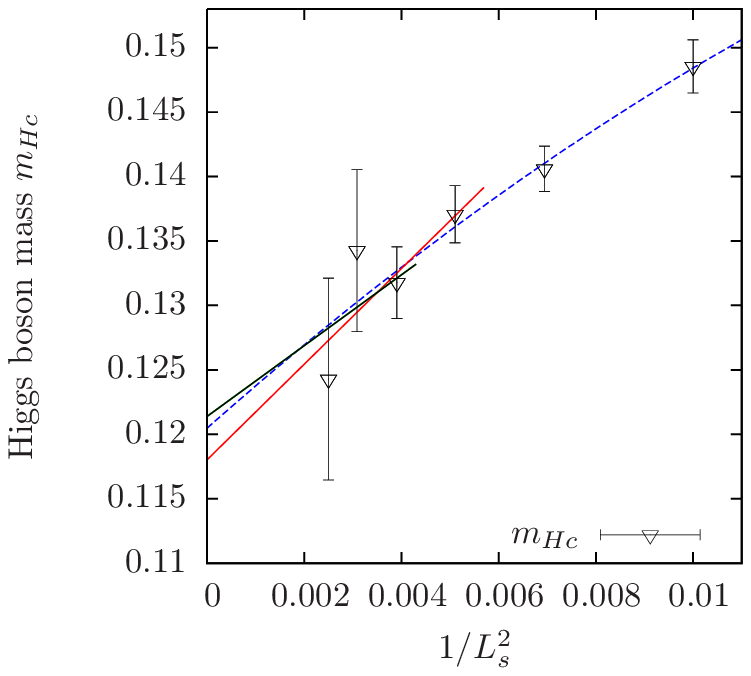}
{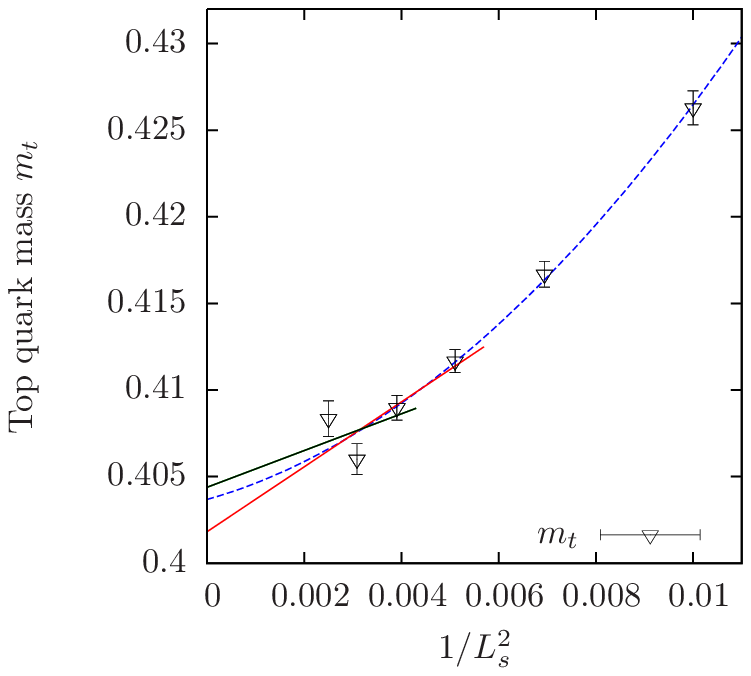}{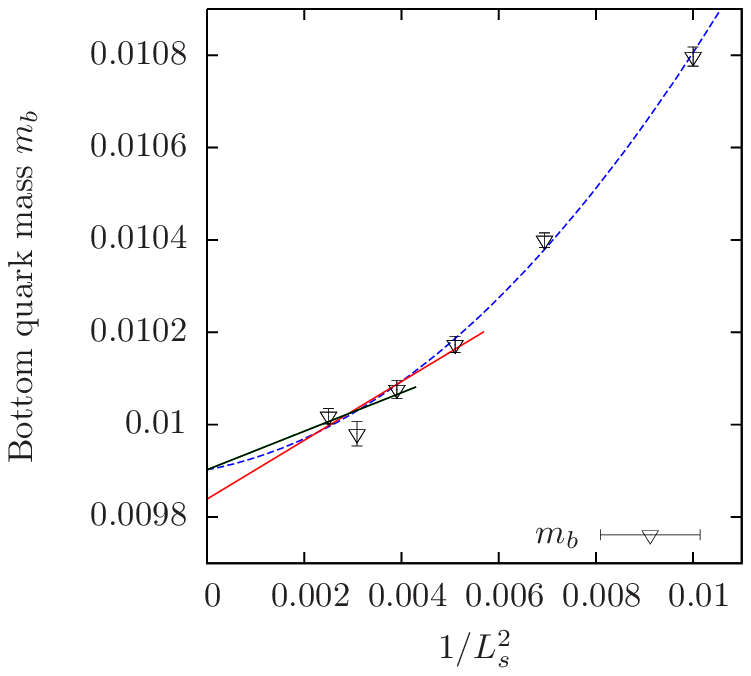}
{fig:InfiniteVolumeExtrapolationAtNf3NonDeg}
{The dependence of the renormalized vev $v_r = v/\sqrt{Z_G}$, the Higgs correlator mass $m_{Hc}$, the top quark mass
$m_t$, and the bottom quark mass $m_b$ on the squared inverse lattice side length $1/L^2_s$ 
is presented. These results have been obtained in the direct Monte-Carlo calculations specified in 
\tab{tab:SummaryOfParametersForLowerHiggsMassBoundRunsNf3}.
In all panels the dashed curves display the parabolic fits according to the fit ansatz in 
\eq{eq:ParaFit}, while the solid lines depict the linear fits resulting from \eq{eq:LinFit} for the two threshold 
values $L_s'=14$ (red), and $L_s'=16$ (black).
}
{Dependence of the renormalized vev $v_r = v/\sqrt{Z_G}$, the Higgs correlator mass $m_{Hc}$, the top quark mass
$m_t$, and the bottom quark mass $m_b$ on the squared inverse lattice side length $1/L^2_s$ at zero bare quartic
coupling constants.}

The corresponding numerical results of the lattice calculations specified in \tab{tab:SummaryOfParametersForLowerHiggsMassBoundRunsNf3}
are presented in \fig{fig:InfiniteVolumeExtrapolationAtNf3NonDeg}, where the renormalized vev $v_r=v/\sqrt{Z_G}$, the Higgs correlator
mass $m_{Hc}$, as well as the top and bottom quark masses are plotted versus $1/L_s^2$. One directly observes that the obtained results are 
distinctly less accurate than the corresponding results in the mass degenerate case, which is due to the much lower statistics obtained
in this scenario. At the same time the finite volume effects are clearly stronger, as expected, thus rendering the infinite
volume extrapolation of the given numerical results more problematic. 

For the latter purpose we apply again the linear and parabolic fit approaches given in \eq{eq:LinFit} and \eq{eq:ParaFit}. 
The fit curves resulting for the respective observables are also depicted in \fig{fig:InfiniteVolumeExtrapolationAtNf3NonDeg} and
the corresponding results of the extrapolation to the infinite volume limit are listed in
\tab{tab:ResultOfLowerHiggsMassFiniteVolExtrapolationNf3NonDeg}. In principle, it is tried here to apply the same extrapolation strategy
already discussed in the preceding section. Due to the smaller number of available lattice results, however, the linear
fit procedure has been performed with only two values of the threshold parameter $L_s'$ as presented in the aforementioned table. 
The final extrapolation results have then been obtained by averaging over the one parabolic and the two linear fits. The translation
of these numbers into physical units is also provided in \tab{tab:ResultOfLowerHiggsMassFiniteVolExtrapolationNf3NonDeg}.

\includeTabNoHLines{|c|c|c|c|c|}{
\hline
observable	 & $A^{(l)}_{v,m,t,b}$, $L'_s=14$  & $A^{(l)}_{v,m,t,b}$, $L'_s=16$  & $A^{(p)}_{v,m,t,b}$ & Final result of ext.             \\ \hline
$\,v_r\,$  	 & $\, 0.5753(23)\, $   & $\, 0.5827(10)\,   $  & $\, 0.5822(20)\, $  & $\, 0.5801(19)(37)$   \\ 
$\,m_{Hc}\,$  	 & $\, 0.1180(59)\, $   & $\, 0.1214(164)\,  $  & $\, 0.1205(59)\, $  & $\, 0.1200(106)(17)$   \\ 
$\,m_t\,$  	 & $\, 0.4018(31)\, $   & $\, 0.4044(68)\,   $  & $\, 0.4037(32)\, $  & $\, 0.4033(47)(13)$   \\ 
$\,m_b\,$  	 & $\, 0.00984(6)\, $   & $\, 0.00990(12)\,  $  & $\, 0.00990(7)\, $  & $\, 0.00988(9)(3)$   \\ \hline
\multicolumn{1}{c}{} & \multicolumn{4}{|c|}{Final results of extrapolation in physical units}    \\ \cline{2-5}
\multicolumn{1}{c|}{} &   $\Lambda=\GEV{246}/v_r$  &  $m_{Hc}/a$     &   $m_t/a$        &  $m_b/a$   \\ \cline{2-5}
\multicolumn{1}{c|}{} &   $\GEV{424.1 \pm 3.0}$   &  $\GEV{50.9 \pm 4.6}$  &   $\GEV{171.0 \pm 2.4}$  &  $\GEV{4.19 \pm 0.05}$          \\ \cline{2-5}
}
{tab:ResultOfLowerHiggsMassFiniteVolExtrapolationNf3NonDeg}
{The results of the infinite volume extrapolations of the Monte-Carlo data for the renormalized vev $v_r$, the Higgs correlator 
mass $m_{Hc}$, the top quark mass $m_t$, and the bottom quark mass $m_b$ are presented as obtained from the parabolic ansatz 
in \eq{eq:ParaFit} and the linear approach in \eq{eq:LinFit} for the considered threshold values $L_s'=14$ and $L_s'=16$. 
The final extrapolation results on the aforementioned observables, displayed in the very right column, are determined here by averaging over
all performed fit approaches, including the parabolic fit. An additional, systematic uncertainty of these final results 
is specified in the second pair of brackets taken from the largest observed deviation among all respective fit results.
These final extrapolation results are furthermore presented in physical units in the bottom line of the table, where the 
previously separated statistical and systematic uncertainties have been combined into a total error. These are the numbers
underlying the presentation in \fig{fig:LowerHiggsMassBoundPredictionInifinteVol}.
}
{Infinite volume extrapolation of the Monte-Carlo data for the renormalized vev $v_r$, the Higgs correlator 
mass $m_{Hc}$, the top quark mass $m_t$, and the bottom quark mass $m_b$ st zero bare quartic coupling constants.}

As intended, the phenomenological values of the top and bottom quark masses are approximately reproduced, at least clearly for the
bottom quark mass, while the top quark mass may require some more tuning of the underlying bare Yukawa coupling constant. One can also
observe that the obtained lower Higgs boson mass bound tends to be somewhat larger than the corresponding result in the 
mass degenerate case. This has been anticipated by the analytical predictions depicted in \fig{fig:LowerHiggsMassBoundPredictionInifinteVol}, 
where the here obtained final results have already been presented in comparison with the lower mass bounds arising in the degenerate case. 
The associated uncertainty of the presented result, however, is too large to clearly distinguish between the analytical curves
with $N_f=1, y_b/y_t=1$ and $N_f=3, y_b/y_t=0.024$ illustrated in \fig{fig:LowerHiggsMassBoundPredictionInifinteVol}.

%-------------------------------------------------------------------------------------------------------------------- 
\section{Model extension by higher order bosonic self-interaction terms}
\label{sec:ModExtByHigherOrder}

So far, the given results on the lower Higgs boson mass bound have been derived in the Higgs-Yukawa model as 
originally specified in \sect{sec:modelDefinition}. In particular, the self-interaction of the scalar field $\varphi$ 
is completely determined by the $\lambda|\varphi|^4$-coupling term. Since the Higgs-Yukawa sector can, however,
only be considered as an effective field theory with a non-removable cutoff $\Lambda$ due to its triviality property
as discussed in \sect{chap:Introduction}, the usual renormalization arguments do not exclude the consideration of 
higher dimensional operators in the framework of this effective theory. This observation then opens up the question 
of how the obtained lower Higgs boson mass bounds would be altered when extending the underlying model by
incorporating such higher dimensional operators into the theory. 

To give at least a partial answer in this matter we will study here the effect of the additional $\lambda_6|\varphi|^6$ 
and $\lambda_8|\varphi|^8$ coupling terms. They are included into the original form of the purely bosonic part of the 
lattice action given in \eq{eq:BosonicLatticeHiggsActionContNot} yielding then the extended lattice action 
\bea
\label{eq:BosonicLatticeHiggsActionContNotWithExtension}
S^{ext}_{\varphi}[\varphi] &=& \sum\limits_{x,\mu} 
\frac{1}{2} \nabla^f_\mu \varphi^\dagger_x
\nabla^f_\mu \varphi_x
+ \sum\limits_{x} \frac{1}{2} m_0^2 \varphi^\dagger_x\varphi_x
+\sum\limits_{x} \lambda\left(\varphi_x^\dagger \varphi_x \right)^2   \\
&+&\sum\limits_{x} \lambda_6\left(\varphi_x^\dagger \varphi_x \right)^3   
+\sum\limits_{x} \lambda_8\left(\varphi_x^\dagger \varphi_x \right)^4. \nonumber
\eea

Here, we will begin with deriving some analytical predictions of the effect of these new coupling terms. For that purpose
the effective potential calculation performed in \sect{sec:DepOfHiggsMassOnLambda} is revisited. The idea is to calculate all 
contributions to $\breve U[\breve v]$ as defined in \eq{eq:DefOfEffPotLargeLam}, that are of order $O(\lambda)$, $O(\lambda_6)$, 
or $O(\lambda_8)$. For that purpose the total action is again split up into a purely Gaussian contribution $S_0$ as already given in 
\eq{eq:DefOfGaussActionEffPotLargeLam}, the extended tree-level part $m_0^2\breve v^2/2 + \lambda\breve v^4+\lambda_6\breve v^6+\lambda_8\breve v^8$, 
and the remaining interacting contribution $S_I$, where the latter now becomes
\bea
\label{eq:DefOfInterActionEffPotLargeLamModExt}
S_I &\fhs{-4mm}=\fhs{-3mm}& 
\frac{\lambda}{V} \fhs{-1mm}\sum\limits_{k_1,\ldots,k_4}\fhs{-1mm} \delta_{k_1+\ldots+k_4,0} 
\left(\tilde h_{k_1}\tilde h_{k_2} +  \sum\limits_{\alpha=1}^3 \tilde g^\alpha_{k_1}\tilde g^\alpha_{k_2} \right) \cdot
\left(\tilde h_{k_3}\tilde h_{k_4} +  \sum\limits_{\alpha=1}^3 \tilde g^\alpha_{k_3}\tilde g^\alpha_{k_4} \right) \\
&\fhs{-4mm}+\fhs{-3mm}&
\frac{\lambda_6}{V^2} \fhs{-1mm}\sum\limits_{k_1,\ldots,k_6}\fhs{-1mm} \delta_{k_1+\ldots+k_6,0} 
\left(\tilde h_{k_1}\tilde h_{k_2} +  \sum\limits_{\alpha=1}^3 \tilde g^\alpha_{k_1}\tilde g^\alpha_{k_2} \right) \cdot \ldots \cdot
\left(\tilde h_{k_5}\tilde h_{k_6} +  \sum\limits_{\alpha=1}^3 \tilde g^\alpha_{k_5}\tilde g^\alpha_{k_6} \right) \nonumber\\
&\fhs{-4mm}+\fhs{-3mm}&
\frac{\lambda_8}{V^3} \fhs{-1mm}\sum\limits_{k_1,\ldots,k_8}\fhs{-1mm} \delta_{k_1+\ldots+k_8,0} 
\left(\tilde h_{k_1}\tilde h_{k_2} +  \sum\limits_{\alpha=1}^3 \tilde g^\alpha_{k_1}\tilde g^\alpha_{k_2} \right) \cdot \ldots \cdot
\left(\tilde h_{k_7}\tilde h_{k_8} +  \sum\limits_{\alpha=1}^3 \tilde g^\alpha_{k_7}\tilde g^\alpha_{k_8} \right) \nonumber\\
&\fhs{-4mm}-\fhs{-3mm}& \frac{\lambda}{V} \tilde h_0^4  - \frac{\lambda_6}{V^2} \tilde h_0^6  - \frac{\lambda_8}{V^3} \tilde h_0^8 
+V^{-1/2}\fhs{-3mm}\sum\limits_{{p_1\ldots p_3\in\ImpSpace}\atop{p_2\neq 0}}\fhs{-3mm}
\delta_{p_1,p_2+p_3} \cdot \tilde {\bar\psi}_{p_1}\left[\tilde h_{p_2} \hat B_0 + \sum\limits_{\alpha=1}^3 \tilde g^\alpha_{p_2} \hat B_\alpha  \right] 
\GammaOpMom(p_3)\tilde \psi_{p_3}.\quad \nonumber
\eea

The next step is then the identification of all diagrams arising from the expansion of $\exp(-S_I)$ and contributing
to $\breve U[\breve v]$ at the considered orders. For the
case of the $|\varphi|^6$-coupling term these diagrams are sketched in \fig{fig:DiagramsForEffectivePotAt1OrderLambda6ModelExt}, while
the corresponding diagrams generated by the four point vertex have already been illustrated in \fig{fig:DiagramsForEffectivePotAt1OrderLambda}. 
The $|\varphi|^8$-diagrams are not explicitly listed here, since they are fully analogous to the already presented ones. It is further 
remarked that the contribution on the very left in \fig{fig:DiagramsForEffectivePotAt1OrderLambda6ModelExt} does actually not generate
any dependence on $\breve v$ and could thus also be neglected for our purpose here. It is nevertheless computed just for the sake
of uniformity.

%\includeFigSingleMedium{LPTDiagramsForEffectivePotentialLambda6Extension}
\includeFigSingleMedium{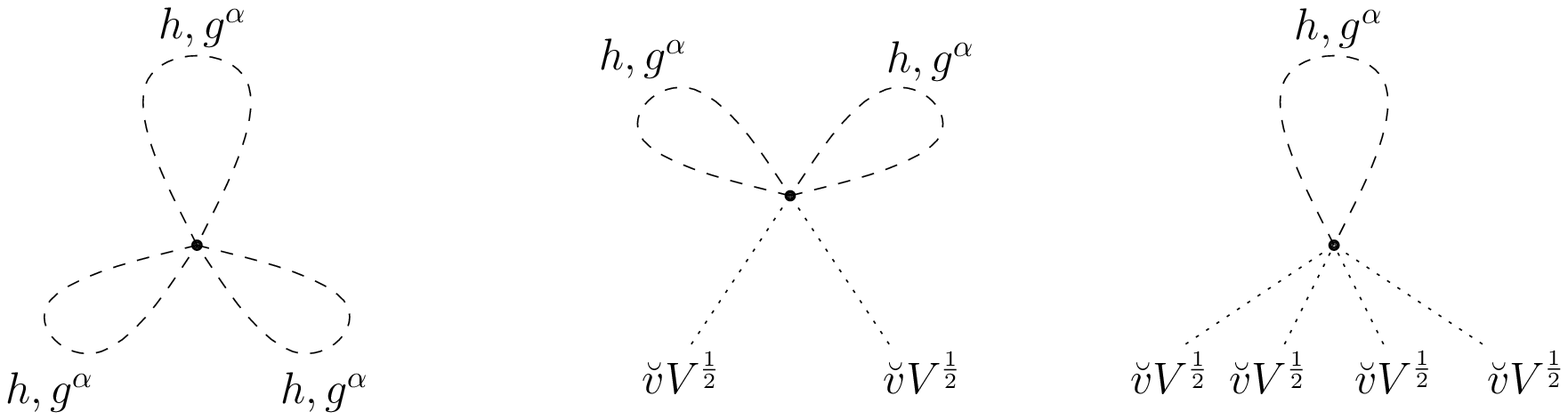}
{fig:DiagramsForEffectivePotAt1OrderLambda6ModelExt}
{Illustration of the purely bosonic diagrams arising from the expansion of $\exp(-S_I)$ with $S_I$ defined in \eq{eq:DefOfInterActionEffPotLargeLamModExt}
and contributing to the effective potential $\breve U[\breve v]$ at order $O(\lambda_6)$. The closed propagator loops are summed over
all momenta $0\neq k\in\ImpSpace$ excluding the constant mode due to the definition of the constraint effective potential in \eq{eq:DefOfEffPotLargeLam}. 
The dotted lines actually do not indicate propagators but only a multiplication with the outer amplitude $\tilde h_{0}=V^{1/2} \breve v$. The 
diagram on the left does not depend on the outer amplitude and is listed here only for the sake of completeness.
}
{Purely bosonic diagrams contributing to the effective potential at order $O(\lambda_6)$.}

Following the same steps already discussed in \sect{sec:DepOfHiggsMassOnLambda} one then directly arrives at the final result 
\bea
\label{eq:EffPotForModExt}
\breve U[\breve v] &\fhs{-2mm}=\fhs{-2mm}& 
\frac{1}{2}m_0^2\breve v^2 + \lambda\breve v^4+\lambda_6\breve v^6+\lambda_8\breve v^8
+ \breve U_F[\breve v] 
+\lambda\cdot \sum\limits_{\alpha=1}^2\sum\limits_{\beta=0}^\alpha\, C_{4,\alpha,\beta} \cdot \breve v^{4-2\alpha}  P_H^{\alpha-\beta}  P_G^{\beta}  
\quad\quad\\
&\fhs{-2mm}+\fhs{-2mm}& \lambda_6\cdot \sum\limits_{\alpha=1}^3\sum\limits_{\beta=0}^\alpha\, C_{6,\alpha,\beta} \cdot \breve v^{6-2\alpha}  P_H^{\alpha-\beta}  P_G^{\beta}  
+ \lambda_8\cdot \sum\limits_{\alpha=1}^4\sum\limits_{\beta=0}^\alpha\, C_{8,\alpha,\beta} \cdot \breve v^{8-2\alpha} P_H^{\alpha-\beta}  P_G^{\beta} \nonumber
\eea
for the effective potential $\breve U[\breve v]$ including all contributions up to the orders $O(\lambda)$, $O(\lambda_6)$, and $O(\lambda_8)$, where
the Higgs and Goldstone propagator sums $P_H$ and $P_G$ are given as
\bea
\label{eq:DefOfHiggsAndGoldstonePropSums}
P_H = \frac{1}{V}\sum\limits_{0\neq p \in \ImpSpace} \frac{1}{\hat p^2 + m_{He}^2} &\mbox{ and }&
P_G = \frac{1}{V}\sum\limits_{0\neq p \in \ImpSpace} \frac{1}{\hat p^2 + 0}.
\eea
Here, the bare mass $m_0^2$ that one would originally have obtained in the denominators of \eq{eq:DefOfHiggsAndGoldstonePropSums} has
again been replaced by the estimates of the renormalized Higgs and Goldstone masses for the same reasons already discussed in 
\sect{sec:DepOfHiggsMassOnLambda}. The so far unspecified coefficients $C_{4,\alpha,\beta}$, $C_{6,\alpha,\beta}$, and 
$C_{8,\alpha,\beta}$ are then given by the multiplicities of the associated diagrams, which have been determined here with the 
help of some trivial, self-written computer program counting the respective numbers of Wick contractions. The resulting coefficients are 
listed in \tab{tab:CoeffForEffPotLambda8ModExt}, where one also finds the numbers $C_{4,1,0}$ and $C_{4,1,1}$ that recover 
the expression for the effective potential previously presented in \eq{eq:DesOfImprovedEffPotForLargeLam1Improved} for the case 
of vanishing model extension parameters, \ie $\lambda_6=\lambda_8=0$.

\includeTabNoHLines{|c|c|c|c|c|c|}{
\cline{1-4}
\multicolumn{4}{|c|}{$C_{4,\alpha,\beta}$}  & \multicolumn{2}{c}{}  \\\cline{1-4}
              &  $\,\,\,\beta=0\,\,\,$   &  $\,\,\,\beta=1\,\,\,$   &  $\,\,\,\beta=2\,\,\,$   &    \multicolumn{2}{c}{}   \\ \cline{1-4}
$\alpha=1$    &    6         &    6         &   --         &   \multicolumn{2}{c}{} \\       
$\alpha=2$    &    3         &    6         &   15         &   \multicolumn{2}{c}{}  \\       
\cline{1-5}
\multicolumn{5}{|c|}{$C_{6,\alpha,\beta}$}  & \multicolumn{1}{c}{}  \\\cline{1-5}
              &  $\,\,\,\beta=0\,\,\,$   &  $\,\,\,\beta=1\,\,\,$   &  $\,\,\,\beta=2\,\,\,$   &  $\,\,\,\beta=3\,\,\,$   &  \multicolumn{1}{c}{}   \\ \cline{1-5}
$\alpha=1$    &  15          &    9         &   --         &   --         &    \multicolumn{1}{c}{} \\       
$\alpha=2$    &  45          &    54        &   45         &   --         &    \multicolumn{1}{c}{}  \\       
$\alpha=3$    &  15          &    27        &   45         &   105        &    \multicolumn{1}{c}{}\\       
\cline{1-6}
\multicolumn{6}{|c|}{$C_{8,\alpha,\beta}$} \\\hline
              &  $\,\,\,\beta=0\,\,\,$   &  $\,\,\,\beta=1\,\,\,$   &  $\,\,\,\beta=2\,\,\,$   &  $\,\,\,\beta=3\,\,\,$   &  $\,\,\,\beta=4\,\,\,$   \\ \hline
$\alpha=1$    &  28          &    12        &   --         &   --         &    --        \\       
$\alpha=2$    &  210         &    180       &   90         &   --         &    --        \\       
$\alpha=3$    &  420         &    540       &   540        &   420        &    --        \\       
$\alpha=4$    &  105         &    180       &   270        &   420        &    945       \\       \hline
}
{tab:CoeffForEffPotLambda8ModExt}
{Listing of the coefficients $C_{4,\alpha,\beta}$, $C_{6,\alpha,\beta}$, and $C_{8,\alpha,\beta}$ appearing
in \eq{eq:EffPotForModExt}. These numbers have been determined by evaluating the multiplicities of the bosonic
diagrams arising from the expansion of $\exp(-S_I)$ and being of order $O(\lambda)$, $O(\lambda_6)$, or 
$O(\lambda_8)$. This combinatoric work was done with the help of some trivial, self-written computer program counting the 
respective number of Wick contractions.
}
{The coefficients $C_{4,\alpha,\beta}$, $C_{6,\alpha,\beta}$, and $C_{8,\alpha,\beta}$ underlying the perturbative
expansion of the effective potential including higher order bosonic self-interactions.}

To check the validity of the given analytical formula in \eq{eq:EffPotForModExt} the arising predictions for the Higgs boson 
mass dependence on the new model parameters $\lambda_6$ and $\lambda_8$ are explicitly compared to corresponding results 
obtained in direct Monte-Carlo calculations as presented in \fig{fig:HiggsMassDepOnLam6Lam8Coupling}. Here, the parameters $\lambda_6$
and $\lambda_8$ have been varied, while the quartic coupling constant and the targeted quark masses have been
kept constant and the hopping parameter has been tuned with the intention to sustain a fixed value of the cutoff $\Lambda$.
One clearly observes a very good agreement between the numerical and the analytical data at least for sufficiently
small values of the respective coupling constants. For the $|\varphi|^8$-coupling term the onset of the expected deviation between 
the numerical and analytical results at larger values of the coupling parameter can also be observed. From these findings one can 
conclude that the behaviour of the extended model is well understood for not too large values of the coupling parameters.
  
The derived analytical result for the effective potential given in \eq{eq:EffPotForModExt} can now be used to identify those
bare model parameters that would generate a given target value of the vev $v$ and a given target value of the Higgs boson mass 
$m_{He}$ by solving the set of equations
\bea
\label{eq:DefEqForModExtParaSetups1}
0 &=& \derive{}{\breve v} \breve U[\breve v] \Bigg|_{\breve v = v} \quad \mbox{and}\\
\label{eq:DefEqForModExtParaSetups2}
m^2_{He} &=& \derive{^2}{\breve v^2} \breve U[\breve v] \Bigg|_{\breve v = v}
\eea
with respect to the bare model parameters. The solution to these equations is underdetermined. We can therefore assign
specific values to some of the parameters. Here, we use this freedom to choose the degenerate Yukawa coupling constants still 
according to the tree-level relation in \eq{eq:treeLevelTopMass} aiming at the reproduction of the phenomenological value of 
the top quark mass and the eight-point coupling parameter is explicitly chosen to zero, \ie $\lambda_8=0$. Furthermore, the 
constant $\lambda_6$ can freely be chosen from a so far unspecified interval of eligible values. 

%\includeFigDoubleDoubleHere{Lambda6ScanHiggsMassFormEffPot}{Lambda6ScanHiggsMassShiftFormEffPot}
%{Lambda8ScanHiggsMassFormEffPot}{Lambda8ScanHiggsMassShiftFormEffPot}
\includeFigDoubleDoubleHere{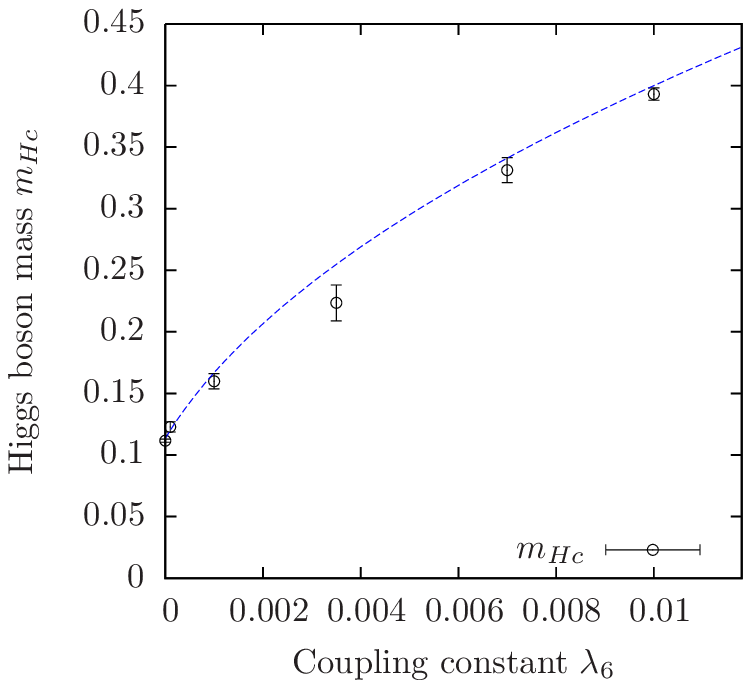}{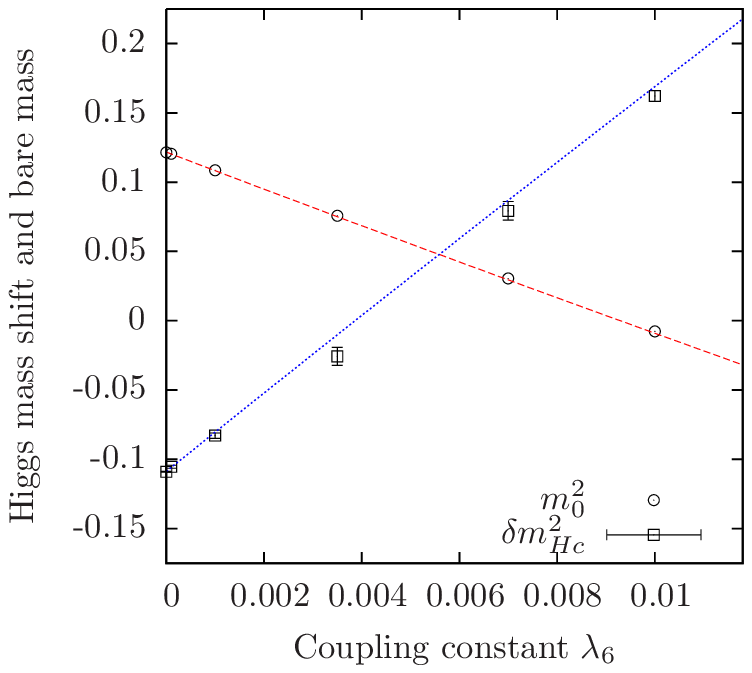}
{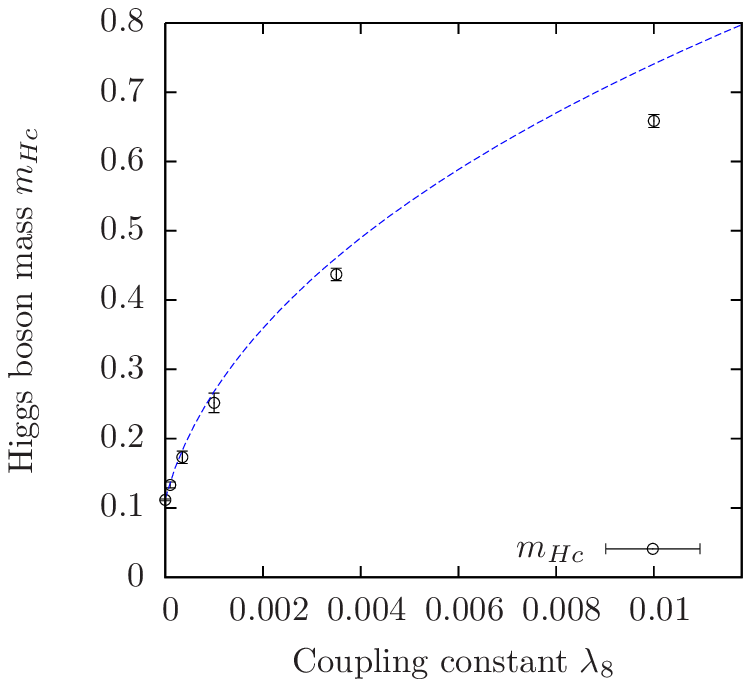}{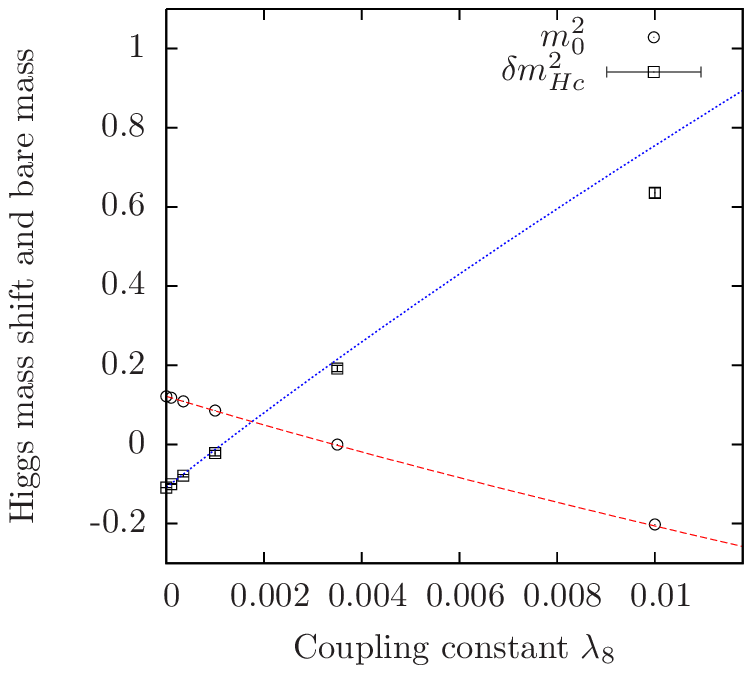}
{fig:HiggsMassDepOnLam6Lam8Coupling}
{The dependence of the Higgs correlator mass $m_{Hc}$ is shown versus the six-point coupling constant $\lambda_6$
in panel (a) and versus the eight-point coupling constant $\lambda_8$ in panel (c). These numerical data have been obtained 
in direct Monte-Carlo calculations on a \lattice{16}{32} with $N_f=1$, $\lambda=0$, degenerate Yukawa coupling constants 
fixed according to \eq{eq:treeLevelTopMass} aiming at reproducing the phenomenological value of the top quark mass, 
and $\kappa$ tuned to sustain an approximately constant cutoff, where the actually obtained values of $\Lambda$ fluctuate 
between $\GEV{380}$ and $\GEV{400}$ in the first mentioned setup and between $\GEV{360}$ and $\GEV{400}$ for the $\lambda_8$-scan. 
The corresponding values of the bare mass parameter $m_0^2$ and the mass shift $\delta m^2_{Hc}=m^2_{Hc}-m_0^2$ are presented in 
panel (b) and (d), respectively. In all panels the given numerical data are compared to the corresponding predictions obtained 
from the effective potential given in \eq{eq:EffPotForModExt} in the same manner discussed in \sect{sec:EffPot}. In both 
scenarios the vacuum expectation value $v$ underlying these analytical calculations has been set to its numerically determined 
value computed in the highest statistics run, which is the one at $\lambda_6=\lambda_8=0$.
}
{Dependence of the Higgs correlator mass $m_{Hc}$ on $\lambda_6$ and $\lambda_8$.}

Here, the idea is to find out whether the lower Higgs boson mass bound established in the preceding
section can be undercut in the framework of the considered model extension. For that purpose a couple 
of parameter setups are presented in \tab{tab:ModExtTargetHiggsMassParameters} solving 
\eqs{eq:DefEqForModExtParaSetups1}{eq:DefEqForModExtParaSetups2} for different choices of $\lambda_6$ and varying
targeted Higgs boson masses $m_{He}$, while the choice of the targeted vev is intended to generate a constant cutoff value of 
approximately $\Lambda^*\equiv\GEV{246}/v = \GEV{400}$. Here, the star symbol in the superscript indicates that the 
renormalization factor $Z_G$ was assumed to be one for the conversion between the vev $v$ and the cutoff. In the actual
lattice computation the renormalization constant is, of course, fully respected.

\includeTab{|c|c|c|c|c|c|c|c|c|}{
\latVolExp{L_s}{L_t} &  $y_t=y_b$   &  $N_f$  &   $m_0^2$   &    $\lambda$              &     $\lambda_6$      &     $\lambda_8$     &   $\Lambda^*$      &    $m_{He}\cdot \Lambda^*$ \\ \hline
\latVol{16}{32}      &  0.71138   &  1      &   0.14262       & $-6.5428 \cdot 10^{-3}$     &     $10^{-3}$        &     0               &   $\GEV{400}$      &    $\GEV{35}$              \\
\latVol{16}{32}      &  0.71138   &  1      &   0.13188       & $-2.1853 \cdot 10^{-3}$     &     $10^{-4}$        &     0               &   $\GEV{400}$      &    $\GEV{35}$              \\
\latVol{16}{32}      &  0.71138   &  1      &   0.14912       & $-7.7850 \cdot 10^{-3}$     &     $10^{-3}$        &     0               &   $\GEV{400}$      &    $\GEV{25}$              \\
\latVol{16}{32}      &  0.71138   &  1      &  0.13836       & $-3.4250 \cdot 10^{-3}$     &     $10^{-4}$        &     0               &   $\GEV{400}$      &    $\GEV{25}$              \\
\latVol{16}{32}      &  0.71138   &  1      &   0.15345       & $-8.6132 \cdot 10^{-3}$     &     $10^{-3}$        &     0               &   $\GEV{400}$      &    $\GEV{15}$              \\
\latVol{16}{32}      &  0.71138   &  1      &   0.14268       & $-4.2515 \cdot 10^{-3}$     &     $10^{-4}$        &     0               &   $\GEV{400}$      &    $\GEV{15}$              \\
\latVol{16}{32}      &  0.71138   &  1      &   0.15561       & $-9.0274 \cdot 10^{-3}$     &     $10^{-3}$        &     0               &   $\GEV{400}$      &    $\GEV{5}$             \\
}
{tab:ModExtTargetHiggsMassParameters}
{The model parameters of the Monte-Carlo runs underlying the subsequent lattice calculation of the Higgs boson mass in the 
extended model are presented. The degenerate Yukawa coupling constants have been chosen according to the tree-level relation 
in \eq{eq:treeLevelTopMass} aiming at the reproduction of the phenomenologically known top quark mass. 
Furthermore, $\lambda_8$ was set to zero and for the six-point coupling constant the two values $\lambda_6=10^{-3}$
and $\lambda_6=10^{-4}$ have been considered. The values of $\lambda$ and $m_0^2$ were then adjusted to solve 
\eqs{eq:DefEqForModExtParaSetups1}{eq:DefEqForModExtParaSetups2} for the respectively given target values 
$\Lambda^*\equiv \GEV{246/v}$ and $m_{He}$. 
}
{Model parameters of the Monte-Carlo runs underlying the performed lattice calculation of the Higgs boson mass in the 
extended model including higher order bosonic self-interactions.}

The proposed model parameter setups in \tab{tab:ModExtTargetHiggsMassParameters} have then been used to configure corresponding direct
lattice calculations. The actually obtained numerical results on the cutoff $\Lambda$ and the Higgs boson mass $m_{Hc}$ are finally 
presented in \fig{fig:TrueHiggsMassVersusTargetdMass} versus the respective targeted values of the Higgs boson mass. One clearly observes
that the lower Higgs boson mass bound can indeed significantly be shifted towards smaller values when extending the model by the considered
higher order self-interaction terms. In the presented examples the obtained Higgs boson masses are up to approximately two times smaller 
than the previously derived lower mass bound discussed in \sect{sec:SubLowerHiggsMassboundsDegenCase}. This situation occurs here at 
negative values of the quartic coupling constant $\lambda$. For clarification it is pointed out that a negative value of $\lambda$ does not 
render the model unstable in the here considered cases, since the higher order coefficient $\lambda_6$ was chosen to be positive.

One also learns from the given results that the trial of adjusting arbitrarily small Higgs boson masses by means of the here presented 
method has failed. This may be due to the existence of a lower Higgs boson mass bound also in the framework of the extended model.
It may, however, simply be the analytical determination procedure of the underlying bare model parameters that needs to be further
improved. Since the intended retention of the cutoff value $\Lambda\approx\GEV{400}$ has also failed, the latter explanation seems 
more likely. A direct investigation of this question based on a trial-and-error approach is, however, not very practical, since the 
generation of the observed low Higgs boson masses requires a precise tuning of the underlying bare model parameters. The latter question
thus remains open here.

As a concluding remark it is summarized that the established lower Higgs boson mass bound in the preceding section can clearly
be undercut in the here considered model extension. The latter mass bounds can thus not be considered to be universal, in the sense 
that they are not independent of the specific form of the underlying interactions.

%\includeFigDouble{TargetedMassScan}{TargetedCutoffScan}
\includeFigDouble{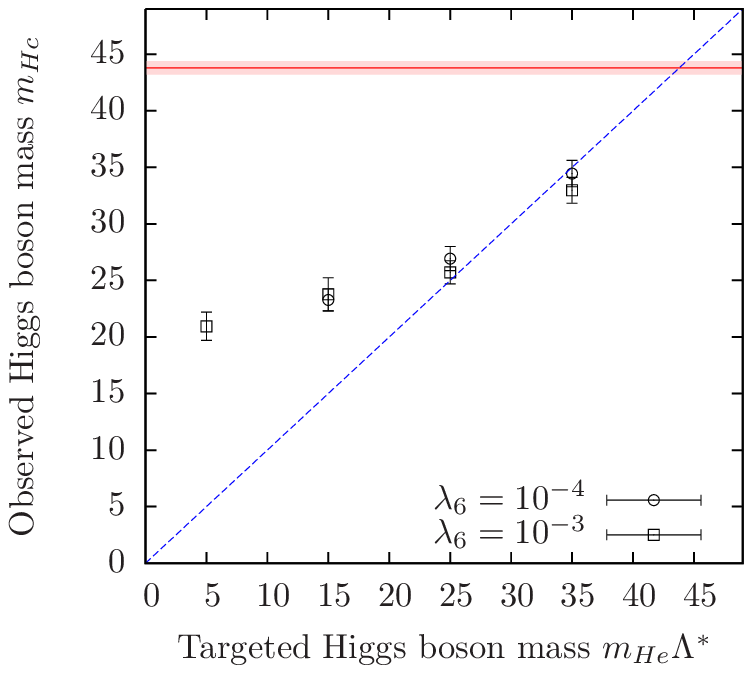}{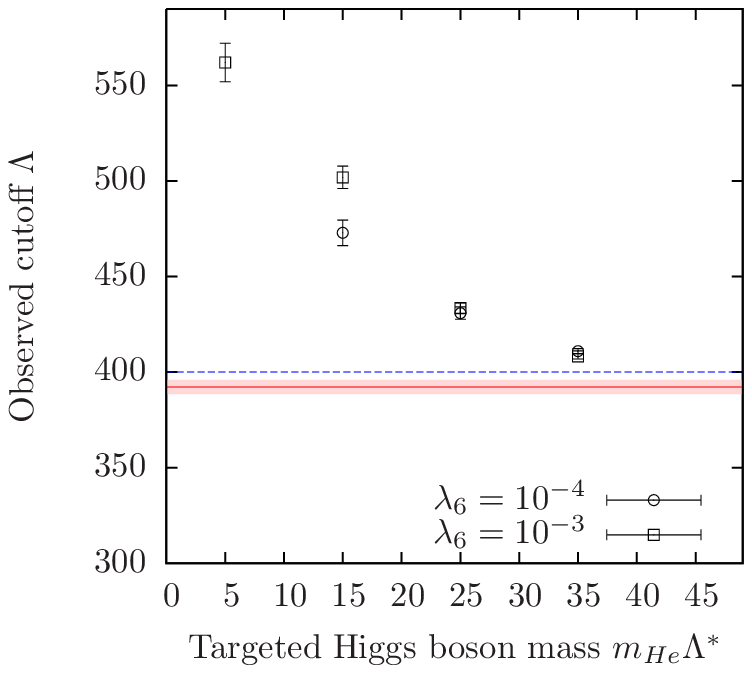}
{fig:TrueHiggsMassVersusTargetdMass}
{The Higgs correlator mass $m_{Hc}$ as obtained in the direct Monte-Carlo calculations specified in 
\tab{tab:ModExtTargetHiggsMassParameters} is shown in panel (a) versus the targeted Higgs boson mass $m_{He}\Lambda^*$ also given in
the aforementioned table. The diagonal dashed line is meant to guide the eye. In panel (b) the corresponding results on the
measured value of $\Lambda$ are presented. The dashed horizontal line depicts the targeted value $\Lambda^*$ taken from 
\tab{tab:ModExtTargetHiggsMassParameters}. In both panels the solid horizontal line indicates the respective result obtained 
in the Monte-Carlo run with $\lambda=0$, $V=16^3\times 32$, and $\kappa=0.12313$ as specified in \tab{tab:Chap61EmployedRuns},
which served as an estimator for the lower Higgs boson mass bound in the original model as discussed 
in \sect{sec:SubLowerHiggsMassboundsDegenCase}. The highlighted band represents the corresponding size of the 
associated error of the latter quantities.
}
{The Higgs correlator mass $m_{Hc}$ obtained in direct Monte-Carlo calculations versus the targeted Higgs boson mass.}

  \chapter{Results on the upper Higgs boson mass bound}
\label{chap:ResOnUpperBound}

The sought-after upper Higgs boson mass bound $m_H^{up}(\Lambda)$ and in particular its dependence
on the cutoff parameter $\Lambda$ shall finally be investigated in this chapter. The aforementioned 
bound will be established here by means of direct lattice calculations performed in the strongly
interacting regime of the considered Higgs-Yukawa model as outlined in \sect{sec:SimStratAndObs}. Before 
we actually begin with the discussion of the obtained lattice data, it is, however, worthwhile
to briefly recall some of the major results concerning the upper Higgs boson mass bound that have
already been established in previous lattice investigations. 

For the pure $\Phi^4$-theory, thus neglecting the coupling to the fermions, the latter mass bound has 
been studied in great detail. Numerous lattice studies, \seeegs{Hasenfratz:1987eh,Hasenfratz:1988kr,Bhanot:1988ua}
and the references therein, examined in particular the case of an infinite bare quartic self-coupling constant
$\lambda=\infty$, where the Higgs boson mass is expected to be maximal. By virtue of the cluster 
algorithm~\cite{Swendsen:1987ce,Wolff:1988uh} the achieved numerical performance of these investigations 
could later be enhanced, which finally led to very precise results on the vacuum expectation value 
$v$ and the Higgs boson mass itself~\cite{Heller:1993yv,Gockeler:1994rx}. For the orientation of the 
reader a single number shall be cited here taken from a rather early study~\cite{Kuti:1987nr}, which 
estimated the upper mass bound to be approximately $m_H^{up}(\Lambda)\approx\GEV{640}$ at a cutoff of 
$\Lambda\approx\TEV{1.3}$ in the here considered notation.

Accompanied by these numerical investigations the pure lattice $\Phi^4$-theory has been solved 
analytically~\cite{Luscher:1987ay,Luscher:1987ek,Luscher:1988uq} in the sense that most low energy 
observables can be calculated analytically 
for any set of bare model parameters, especially for arbitrary values of the bare quartic coupling 
constant $\lambda$, except for model parameters in the broken phase far away from the critical line. 
These latter points in the parameter space are, however, not of physical interest with respect to the 
here considered lattice approach, since the cutoff scale associated to these parameter sets would be low
and in particular not large compared to the observable particle masses.

The aforementioned analytical results have been obtained by combining a high temperature expansion with 
perturbative calculations performed in the framework of renormalized perturbation theory. The latter high 
temperature calculation expands all quantities in terms of the hopping parameter $\kappa$ instead of the 
quartic coupling constant $\lambda$ by cutting off the infinite sum in the expression
\beq
\label{eq:HighTempExpansion}
Z_\Phi = \int \prod\limits_{x} \Bigg[ \intd{\Phi_x} e^{-\Phi^\dagger_x \Phi_x - \hat\lambda\left(\Phi^\dagger_x\Phi_x - 1\right)^2} \Bigg]\,\,
\sum\limits_{n=0}^{\infty} \frac{1}{n!} \left(\hat\kappa\sum\limits_{x,\mu} \Phi^\dagger_x\left[\Phi_{x+\mu}+\Phi_{x-\mu}\right]  \right)^n
\eeq
after a finite number of summands, where $Z_\Phi$ denotes the partition function of the pure $\Phi^4$-theory
here. The high temperature expansion can thus be applied to any choice of the bare quartic self-coupling 
constant $\lambda\ge 0$, in particular to $\lambda=\infty$. Its convergence properties are 
best for small values of the hopping parameter $\kappa$, \ie in the symmetric phase. In fact, the respective
expansion coefficients for the renormalized mass, the field renormalization factor, and the renormalized
quartic coupling constant have explicitly been worked out up to 14-th order~\cite{Luscher:1988gk} and excellent 
results have been obtained for $\kappa\in[0,0.95\cdot \kapCrit]$, where $\kapCrit$ is the critical hopping 
parameter of the pure $\Phi^4$-theory for the respective value of $\lambda$. For even larger hopping parameters, 
however, this approach finally breaks down.

The main finding of this high temperature expansion is that the renormalized quartic coupling constant 
$\lambda_r$ is always in the perturbative regime at $\kappa=0.95\cdot \kapCrit$ independent of the 
choice of the bare coupling parameter $\lambda$. The results determined from the high temperature expansion 
at this value of the hopping parameter can therefore be evolved to larger values of $\kappa$, and thus into
the here physically relevant broken phase, by integrating the Callan-Symanzik equations, where the coefficients 
of the associated $\beta$-functions can be obtained from renormalized perturbation theory as long as $\lambda_r$
remains within the perturbative regime.

By performing the latter integration of the Callan-Symanzik equations it could be shown that the renormalized 
quartic coupling constant monotonically decreases with growing hopping parameter $\kappa<\kapCrit$, \ie 
in the symmetric phase, until it finally vanishes completely at the critical line in full agreement with
the triviality picture of the pure $\Phi^4$-theory. The model thus remains in the perturbative regime for all hopping 
parameters in the interval $[0.95\cdot\kapCrit, \kapCrit]$. For $\kappa>\kapCrit$, \ie in the broken phase, 
the renormalized quartic coupling constant increases with growing hopping parameter and the perturbative regime 
is eventually left at some value $\kappa=\kappa_{np}>\kapCrit$.

With this combined approach of high temperature expansion and renormalized perturbation theory it is 
therefore possible to calculate the Higgs boson mass and the renormalized quartic coupling constant for all 
values of the bare coupling constant $\lambda\ge 0$ and for all hoping parameters $\kappa<\kappa_{np}$. 
These calculations have been performed in great detail in \Refs{Luscher:1987ay,Luscher:1987ek,Luscher:1988uq} 
where one can also find extensive listings of the obtained analytical results on the Higgs boson 
mass\footnote{The underlying definitions of the Higgs boson mass and the renormalized quartic coupling
constant considered in \Ref{Luscher:1988uq} are the quantities $m_{Hp}$ and $\lambda_r$ given in
\eq{eq:DefOfPropagatorMinkMass} and \eq{eq:DefOfRenQuartCoupling}.} $m_{Hp}$ and the renormalized quartic coupling 
constant $\lambda_r$ at various values of $\kappa$ and $\lambda$. Moreover, the logarithmically corrected scaling 
laws of $m_{Hp}$, $\lambda_r$, and the vev $v$ have also been derived by considering the aforementioned Callan-Symanzik 
equations in the vicinity of the critical line~\cite{Luscher:1987ay,Luscher:1987ek,Luscher:1988uq}. For the broken 
phase with the underlying model being the four-component $\Phi^4$-theory in infinite volume they read
\bea
\label{eq:ScalingBehOfVeV}
v &\propto& \left(\kappa-\kapCrit\right)^{1/2} \cdot \left|\log\left(\kappa-\kapCrit  \right)   \right|^{1/4}, \\
\label{eq:ScalingBehOfMass}
m_{Hp} &\propto& \left(\kappa-\kapCrit\right)^{1/2} \cdot \left|\log\left(\kappa-\kapCrit  \right)   \right|^{-1/4},\\
\label{eq:ScalingBehOfLamRen}
\lambda_r &\propto& \left|\log\left(\kappa-\kapCrit  \right)   \right|^{-1}.
\eea

Combining these scaling laws with the additional finding~\cite{Luscher:1988uq} $Z_G = C_G(\hat \lambda) + O(\lambda_r)$, 
where the constant $C_G(\hat \lambda)$ is also provided in \Ref{Luscher:1988uq}, one immediately arrives at the analytically 
expected functional form of the dependence of the Higgs boson mass and the renormalized quartic coupling constant $\lambda_r$ 
on the cutoff. In the limit $\Lambda\rightarrow\infty$ and for fixed bare quartic coupling constant $\hat \lambda$ it is 
given as 
\bea
\label{eq:StrongCouplingLambdaScalingBeaviourMass}
\frac{m_{Hp}}{a} &=& A_m \cdot \left[\log(\Lambda^2/\mu^2) + B_m \right]^{-1/2}, \\
\label{eq:StrongCouplingLambdaScalingBeaviourLamCoupling}
\lambda_r &=& A_\lambda \cdot \left[\log(\Lambda^2/\mu^2) + B_\lambda \right]^{-1},
\eea
where double-logarithmic terms have been neglected, $\mu$ denotes some
unspecified scale, and $A_{m,\lambda}\equiv A_{m,\lambda}(\mu)$, $B_{m,\lambda}\equiv B_{m,\lambda}(\mu)$ 
are constants.

So far, all summarized results refer to the pure $\Phi^4$-theory. It is thus worthwhile to ask 
whether these scaling laws still hold in the considered Higgs-Yukawa model including the
coupling to the fermions. In that respect it is remarked that the same functional dependence has 
also been observed in an analytical study~\cite{Fodor:2007fn} of a Higgs-Yukawa model in continuous 
Euclidean space-time based, however, on an one-component Higgs field. In that study the running of 
the renormalized coupling constants with varying cutoff has been investigated by means of renormalized 
perturbation theory in the large $N_f$-limit. Furthermore, the scaling behaviour of the renormalized 
Yukawa coupling constant has also been derived. It was found to be
\bea
\label{eq:StrongCouplingLambdaScalingBeaviourYCoupling}
y_r &=& A_y \cdot \left[\log(\Lambda^2/\mu^2) + B_y \right]^{-1/2},
\eea
where $A_y\equiv A_y(\mu)$ and $B_y\equiv B_y(\mu)$ are again so far unspecified constants and $y_r$ stands 
here for the renormalized top and bottom Yukawa coupling constants $y_{t,r}$ and $y_{b,r}$, respectively, 
as defined in \eq{eq:DefOfRenYukawaConst}.

In the following sections these analytically derived scaling laws will explicitly be compared to the 
results of direct lattice calculations performed in the considered Higgs-Yukawa model. In particular,
the question of how much the here respected fermion dynamics influences the previously observed results
in the pure $\Phi^4$-theory will be addressed. For that purpose we begin with the details of the
Higgs boson mass determination in \sect{sec:CorrVsPropMass}, which are different from the earlier discussed approach 
applied to the weakly interacting regime of the model. The crucial question, at what values of the bare
coupling parameter $\lambda$ one actually observes the largest Higgs boson masses, is then considered
in \sect{sec:DepOfHiggsMassonLargeLam}. Given this knowledge about the $\lambda$-dependence of the Higgs boson mass, 
the sought-after upper mass bound is then finally established in \sect{sec:UpperMassBounds}.

%--------------------------------------------------------------------------------------------------------------------------------------------------
\section{Details of the particle mass determination}
\label{sec:CorrVsPropMass}

In analogy to the discussion in \sect{sec:ParticleMassDetDetailsLowerBound} the details of the particle mass determination
shall now be discussed for the case of the strongly interacting regime of the considered Higgs-Yukawa sector, with the 
latter notion referring here to the setting $\lambda\gg 1$. It is basically tried to follow the same approaches already 
introduced for the weakly coupling scenario. However, some minor modifications in the aforementioned computation schemes 
have been conducted to make them successfully applicable also in the here considered case of large quartic coupling constants.

In particular, the Higgs boson mass determination will eventually be based on the analysis of the Higgs propagator as proposed
in \sect{sec:SimStratAndObs}. The reasoning is twofold. Firstly, the latter method yields significantly more stable results
on the Higgs boson mass $m_{Hp}$ than the alternative extraction scheme based on the Higgs time-slice correlator. Apart from
this rather technical downside there is the additional
argument that the time-slice correlator is expected to receive significant contributions from lighter states due
to the Higgs boson becoming increasingly unstable with an eventually no longer negligible decay width $\Gamma_H$
when entering the strongly interacting regime as discussed in \sect{sec:SimStratAndObs}, thus complicating the 
approach of determining the Higgs boson mass from the latter time-slice correlator also from a conceptual point of 
view.

Again, the considered computation schemes will be tested in the framework of direct Monte-Carlo calculations, the model parameters
of which are listed in \tab{tab:Chap71EmployedRuns}. These Monte-Carlo runs have been selected to cover the full range of cutoffs
that will be investigated for the actual determination of the upper Higgs boson mass bound in the considered Higgs-Yukawa model.

\includeTab{|cccccccc|}
{
\latVolExp{L_s}{L_t} & $N_f$ & $\kappa$  & $\hat \lambda$ & $\hat y_t$     & $\hat y_b/\hat y_t$ & $\langle m \rangle$ & $\Lambda$ \\
\hline
\latVol{32}{32}      & $1$   & $0.30039$ & $\infty$       & $0.55139$ & $1$       & $ 0.1301(3)$ & $\GEV{ 2373.0\pm 6.4}$\\
\latVol{32}{32}      & $1$   & $0.30400$ & $\infty$       & $0.55038$ & $1$       & $ 0.1984(1)$ & $\GEV{ 1548.1\pm 1.8}$\\
}
{tab:Chap71EmployedRuns}
{The model parameters of the Monte-Carlo runs constituting the testbed for the subsequently discussed computation schemes 
are presented together with the obtained values of the average magnetization $\langle m \rangle$ and the cutoff $\Lambda$
determined by \eq{eq:DefOfCutoffLambda}. The degenerate Yukawa coupling constants have been chosen here according to the tree-level
relation in \eq{eq:treeLevelTopMass} aiming at the reproduction of the phenomenologically known top quark mass.
}
{Model parameters of the Monte-Carlo runs constituting the testbed for the considered computation schemes at large quartic
coupling constants.}

%-------------------------------------------------------------------------------------------------------------------- 
\subsection{Analysis of the Goldstone propagator}
\label{sec:GoldPropAnalysisUpperBound}

With respect to the determination of the renormalization constant $Z_G$ and the Goldstone mass $m_G$ we follow exactly the
same approach introduced in \sect{sec:GoldPropAnalysisLowerBound}. As an example the Goldstone propagators obtained in the
lattice calculations specified in \tab{tab:Chap71EmployedRuns} are presented in \fig{fig:GoldstonePropExampleArStrongCoup}.
These numerical data are again fitted\footnote{Again, it is actually the inverse propagator $\tilde G_G^{-1}(p)$ that is 
fitted with $f_G^{-1}(\hat p^2)$.} with the fit formula $f_G(\hat p^2)$ given in \eq{eq:GoldstoneFitAnsatz}, which is based on an 
heuristically motivated modification of the renormalized one-loop result on the Goldstone propagator of the pure $\Phi^4$-theory 
in continuous Euclidean space-time, as detailed in \sect{sec:GoldPropAnalysisLowerBound}. 
One can then already observe from the graphical presentation in \fig{fig:GoldstonePropExampleArStrongCoup} that the considered
fit ansatz $f_G(\hat p^2)$ describes the numerical data significantly better than the simple linear fit formula $l_G(\hat p^2)$ 
given in \eq{eq:GoldstoneFitAnsatzLinear}. 

Furthermore, it has already been discussed in the aforementioned section that the propagators should only be considered at small 
momenta to avoid unacceptably large discretization effects. For that purpose the constant $\gamma$ has been introduced,
such that the applied fit procedures are restricted to the momenta with $\hat p^2\le \gamma$. In principle, the constant
$\gamma$ should be chosen as small as possible. For too small values of $\gamma$, however, the fit procedure becomes unstable. 
To find an optimal setting for the latter threshold value, the dependence of the fit results on the parameter $\gamma$ is listed
in \tab{tab:GoldstonePropExampleResultsAtStrongCoup}, where the presented Goldstone mass $m_G$ and the respective renormalization 
factor $Z_G$ have been obtained according to \eq{eq:DefOfHiggsAndGoldstoneMassByPole} and \eq{eq:DefOfRenormalFactors} from the 
analytical continuation of the lattice Goldstone propagator given by $\tilde G^{c}_G(p_c) = f_G(p^2_c)$ and $\tilde G^{c}_G(p_c) = l_G(p^2_c)$, 
respectively. 
 
At first glance one notices that the linear ansatz $l_G(\hat p^2)$ yields more stable results than $f_G(\hat p^2)$. These
results are, however, inconsistent with themselves when varying the parameter $\gamma$. One can also observe
in \tab{tab:GoldstonePropExampleResultsAtStrongCoup} that the associated average squared residual per degree of freedom 
$\chi^2/dof$ significantly differs from one at the selected values of $\gamma$, making apparent that the simple linear
fit ansatz is not suited for the reliable determination of the Goldstone propagator properties.

%\includeFigTrippleDouble
%{GoldstonePropagatorKap030039L32Pmax16}{GoldstonePropagatorKap030039L32Pmax1}{GoldstonePropagatorKap030039L32PmaxSmallest}
%{GoldstonePropagatorKap030400L32Pmax16}{GoldstonePropagatorKap030400L32Pmax1}{GoldstonePropagatorKap030400L32PmaxSmallest}
\includeFigTrippleDouble
{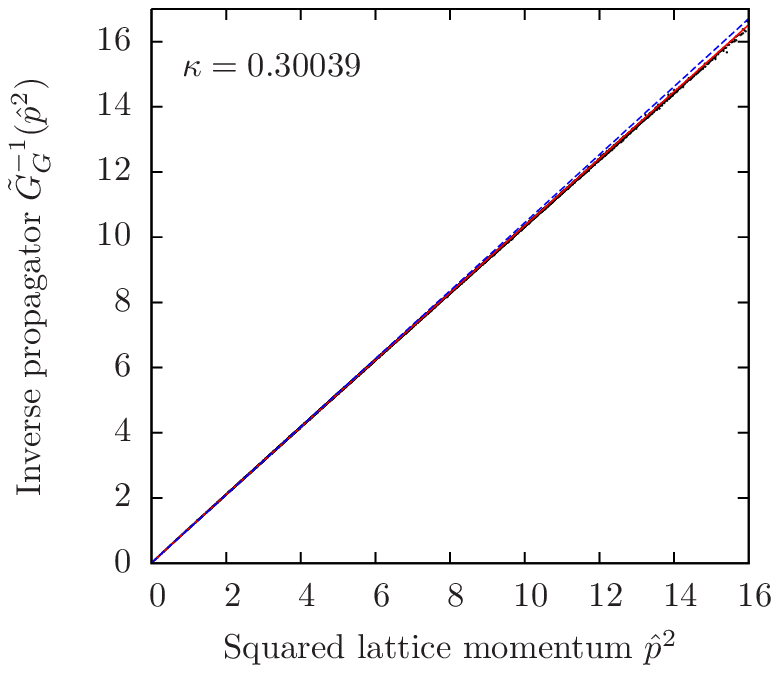}{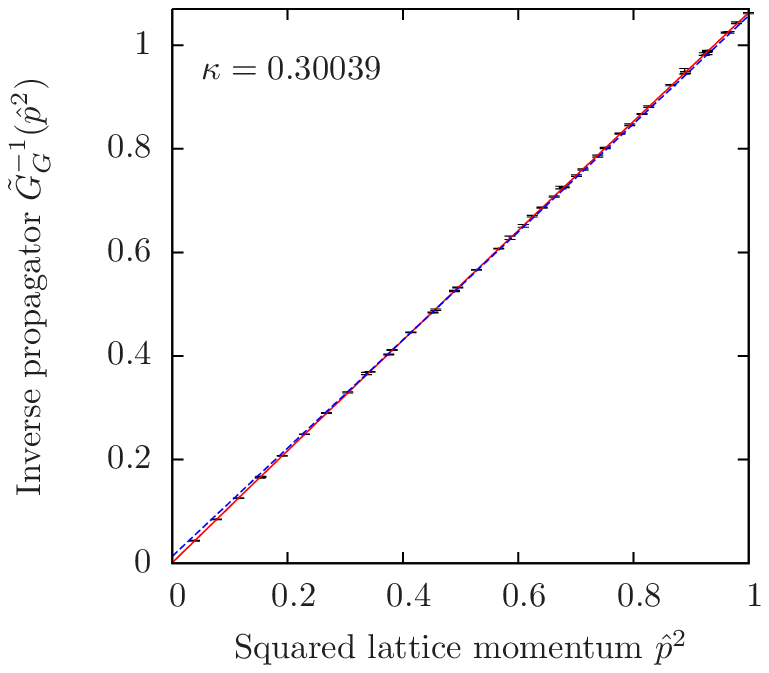}{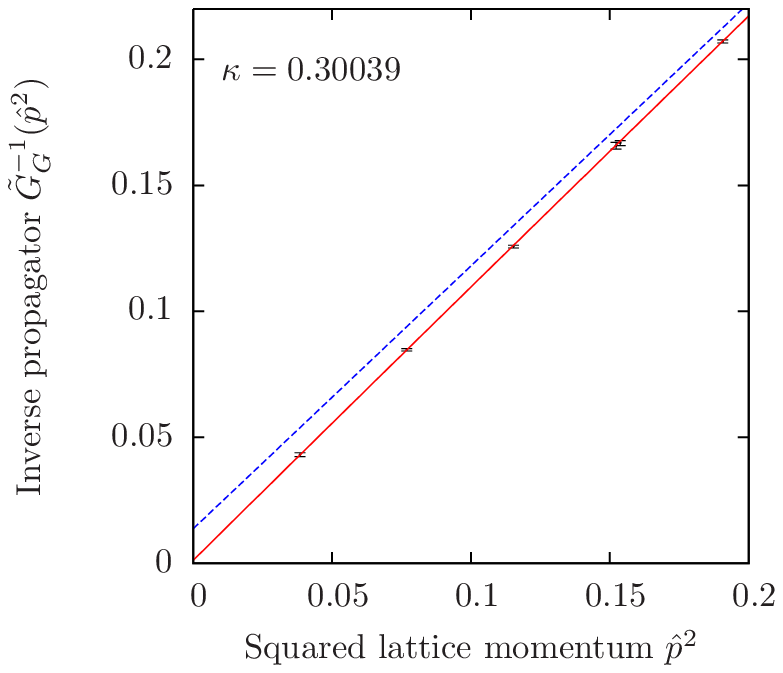}
{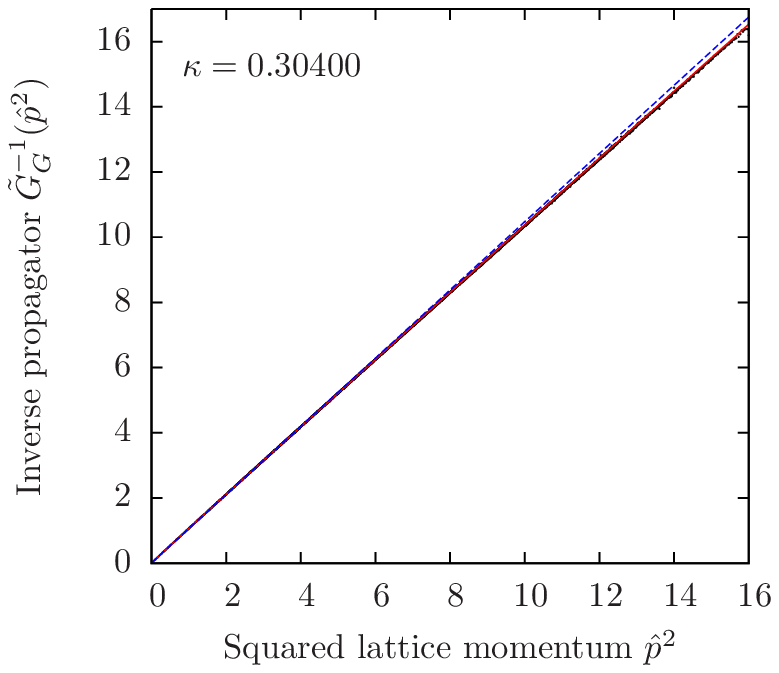}{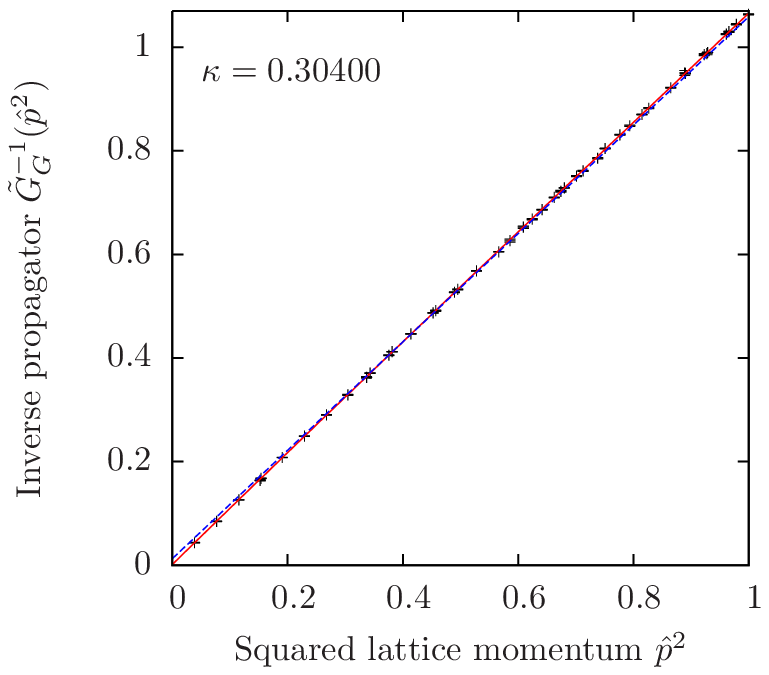}{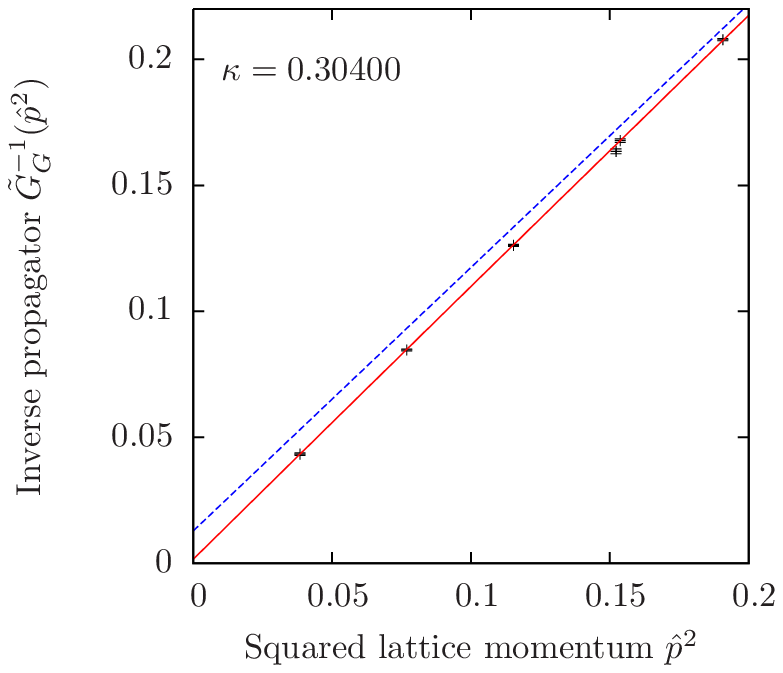}
{fig:GoldstonePropExampleArStrongCoup}
{The inverse lattice Goldstone propagators calculated in the Monte-Carlo runs specified in \tab{tab:Chap71EmployedRuns}
are presented versus the squared lattice momenta $\hat p^2$ together with the respective fits obtained from the 
fit approaches $f^{-1}_G(\hat p^2)$ in \eq{eq:GoldstoneFitAnsatz} (red solid line) and $l^{-1}_G(\hat p^2)$ 
in \eq{eq:GoldstoneFitAnsatzLinear} (blue dashed line) with $\gamma=4.0$. From left to right 
the three panel columns display the same data zooming in, however, on the vicinity of the origin at $\hat p^2 = 0$.
}
{Examples of Goldstone propagators at large quartic coupling constants.}
\includeTabNoHLines{|c|c|c|c|c|c|c|c|}{
\cline{3-8}
\multicolumn{2}{c|}{}& \multicolumn{3}{c|}{fit ansatz $f_G(\hat p^2)$} &  \multicolumn{3}{c|}{linear fit ansatz $l_G(\hat p^2)$}\\ \hline
$\kappa$             & $\gamma$   & $Z_{G}$     & $m_{G}$    &  $\chi^2/dof$  & $Z_{G}$   & $m_{G}$   &  $\chi^2/dof$ \\ \hline
$0.30039$     & $1.0$        & 0.9380(107) & 0.027(15)  &  1.00          & 0.9422(5) & 0.067(2)  & 2.61  \\
$0.30039$     & $2.0$        & 0.9431(52)  & 0.028(11)  &  0.81          & 0.9507(3) & 0.089(2)  & 4.79  \\
$0.30039$     & $4.0$        & 0.9457(27)  & 0.033(8)   &  0.94          & 0.9585(2) & 0.114(2)  & 6.19  \\\hline
$0.30400$     & $1.0$        & 0.9400(90)  & 0.029(10)  &  1.41          & 0.9403(4) & 0.066(1)  & 4.40  \\
$0.30400$     & $2.0$        & 0.9426(36)  & 0.032(7)   &  1.07          & 0.9476(2) & 0.084(1)  & 6.53  \\
$0.30400$     & $4.0$        & 0.9478(18)  & 0.038(4)   &  1.06          & 0.9559(1) & 0.111(1)  & 9.67  \\\hline
}
{tab:GoldstonePropExampleResultsAtStrongCoup}
{The results on the Goldstone renormalization factor $Z_G$ and the Goldstone mass $m_G$, obtained from
the fit approaches $f_G(\hat p^2)$ and $l_G(\hat p^2)$ as defined in \eq{eq:GoldstoneFitAnsatz} and 
\eq{eq:GoldstoneFitAnsatzLinear}, are listed for several settings of the parameter 
$\gamma$ together with the corresponding average squared residual per degree of freedom $\chi^2/dof$
associated to the respective fit. The underlying Goldstone lattice propagators have been calculated
in the Monte-Carlo runs specified in \tab{tab:Chap71EmployedRuns}.
}
{Comparison of the Goldstone propagator properties obtained from different extraction schemes at large quartic
coupling constants.}

In contrast to that the more elaborate fit ansatz $f_G(\hat p^2)$ yields much better values of $\chi^2/dof$ being close
to the expected value of one as can be seen in \tab{tab:GoldstonePropExampleResultsAtStrongCoup}. Moreover, the results on
the renormalization constant $Z_G$ and the Goldstone mass $m_G$ obtained from this ansatz remain consistent with respect to 
the specified errors when varying the constant $\gamma$. 
For the rest of this chapter the aforementioned quantities $Z_G$ and $m_G$ will therefore always be determined by means of the
here presented method based on the fit ansatz $f_G(\hat p^2)$ with a threshold value of $\gamma=4$, since this setting yields 
the most stable results, while still being consistent with the findings obtained at smaller values of $\gamma$.

%-------------------------------------------------------------------------------------------------------------------- 
\subsection{Analysis of the Higgs propagator}
\label{sec:HiggsPropAnalysisUpperBound}

Concerning the analysis of the Higgs propagator we will again maintain the same strategy that was already introduced in 
\sect{sec:HiggsPropAnalysisLowerBound}. As an example the lattice Higgs propagators as obtained in the Monte-Carlo runs
specified in \tab{tab:Chap71EmployedRuns} are presented in \fig{fig:HiggsPropExampleAtStrongCoup}. These numerical data
are again fitted\footnote{Again, it is actually the inverse propagator $\tilde G_H^{-1}(p)$ that is fitted with $f_H^{-1}(\hat p^2)$.} 
with the fit ansatz $f_H(\hat p^2)$ given in \eq{eq:HiggsPropFitAnsatz}. The only modification that has been conducted 
is that the Goldstone boson mass parameter $\bar m_G$ appearing in the aforementioned fit formula is not treated 
as a free parameter here in contrast to the approach originally discussed in \sect{sec:HiggsPropAnalysisLowerBound}. Instead, 
it is fixed to the value of $m_G$ resulting from the analysis of the Goldstone propagator by the method described in the 
previous section. The purpose of this modification is to achieve a higher stability of the considered fit procedure, 
which otherwise would yield only unsatisfactory results with respect to the associated statistical uncertainties, 
in contrast to the situation of the weakly interacting regime discussed in \sect{sec:HiggsPropAnalysisLowerBound}.

Again, one can observe, however less clearly as compared to the previously discussed examples for the case of the Goldstone 
propagator, that the more elaborate fit ansatz $f_H(\hat p^2)$ describes the lattice data more accurately than the simple 
linear fit approach $l_H(\hat p^2)$ given in \eq{eq:HiggsPropFitAnsatzLinear}. This is better observable in the lower row 
of \fig{fig:HiggsPropExampleAtStrongCoup} than in the upper row, where the differences tend to be rather negligible. The 
reason why the observed differences between the two fit approaches are less pronounced here, as compared to the situation
in the preceding section, simply is, that the threshold value $\gamma$ was chosen here to be $\gamma=1$ which will be motivated 
below. This setting is much smaller than the value underlying the previously discussed examples for the Goldstone propagators
and causes the linear fit to come closer to the more elaborate ansatz $f_H(\hat p^2)$.

The Higgs propagator mass $m_{Hp}$ defined in \eq{eq:DefOfPropagatorMinkMass} and the Higgs pole mass $m_H$ together with its 
associated decay width $\Gamma_H$ given by the pole of the propagator on the second Riemann sheet according to 
\eq{eq:DefOfHiggsAndGoldstoneMassByPole} can then again be obtained by defining the analytical continuation of the lattice 
propagator as $\tilde G_H^{c}(p_c) = f_H(p^2_c)$ and $\tilde G_H^{c}(p_c) = l_H(p^2_c)$, respectively. The results arising from 
the considered fit procedures are listed in \tab{tab:HiggsPropExampleResultsAtStrongCoup} for several values of the threshold
value $\gamma$. Since the linear function $l_H(p^2_c)$ can not exhibit a branch cut structure, however, the pole mass equals the
propagator mass and the decay width is identical to zero when applying the linear fit approach. That is the reason why only 
the Higgs boson mass $m_H$ is presented in the latter scenario. 

One observes in \tab{tab:HiggsPropExampleResultsAtStrongCoup} that the Higgs boson masses obtained from the linear fit ansatz
$l_H(\hat p^2)$ are again inconsistent with the respective values obtained at varying values of the threshold parameter $\gamma$,
thus rendering this latter approach unsuitable for the description of the Higgs propagator. This becomes also manifest through
the presented values of the average squared residual per degree of freedom $\chi^2/dof$ associated to the linear ansatz, which 
are clearly off the expected value of one, apart from one presented exception\footnote{In the first line of 
\tab{tab:HiggsPropExampleResultsAtStrongCoup} one finds $\chi^2/dof=1.13$ for $l_H(\hat p^2)$ being smaller than the 
corresponding value $\chi^2/dof=1.17$ for $f_H(\hat p^2)$. This is possible although the set of functions parametrizable 
by $l_H(\hat p^2)$ is a subset of the corresponding set associated to $f_H(\hat p^2)$, since the number of degrees of 
freedom differs for the latter two fit approaches according to $dof=N_{data}-N_{para}$, where $N_{data}$ is the number 
of data points and $N_{para}$ is the number of free fit parameters.} at $\gamma=0.5$. 

%\includeFigTrippleDouble
%{HiggsPropagatorKap030039L32Pmax16}{HiggsPropagatorKap030039L32Pmax1}{HiggsPropagatorKap030039L32PmaxSmallest}
%{HiggsPropagatorKap030400L32Pmax16}{HiggsPropagatorKap030400L32Pmax1}{HiggsPropagatorKap030400L32PmaxSmallest}
\includeFigTrippleDouble
{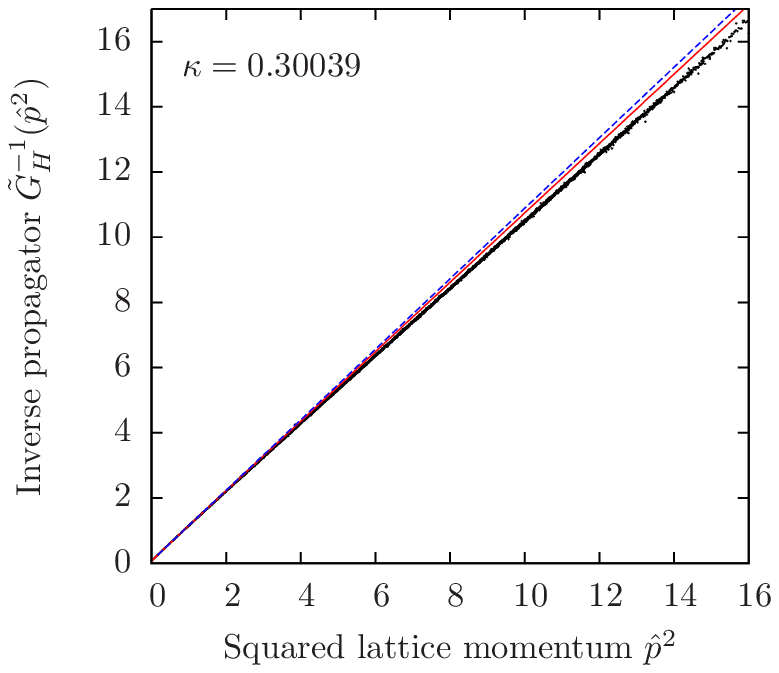}{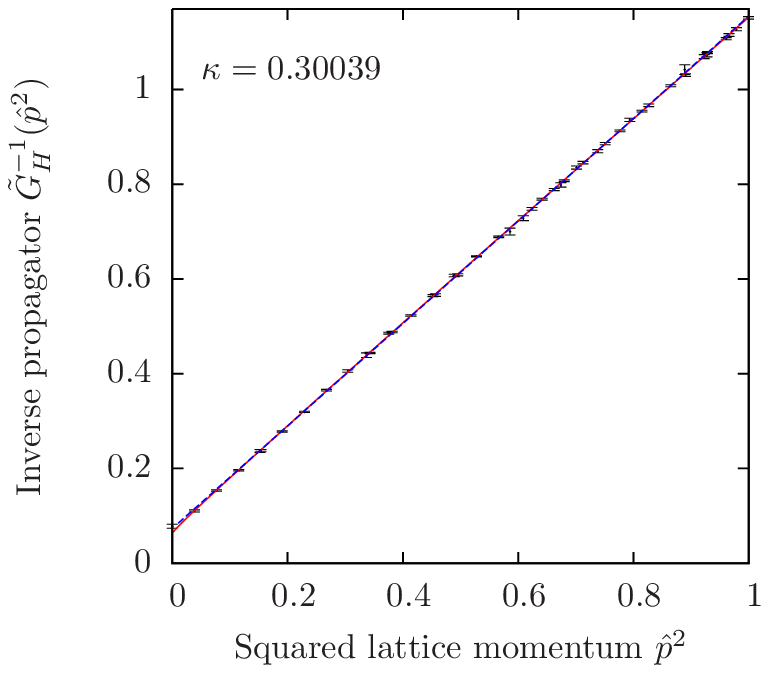}{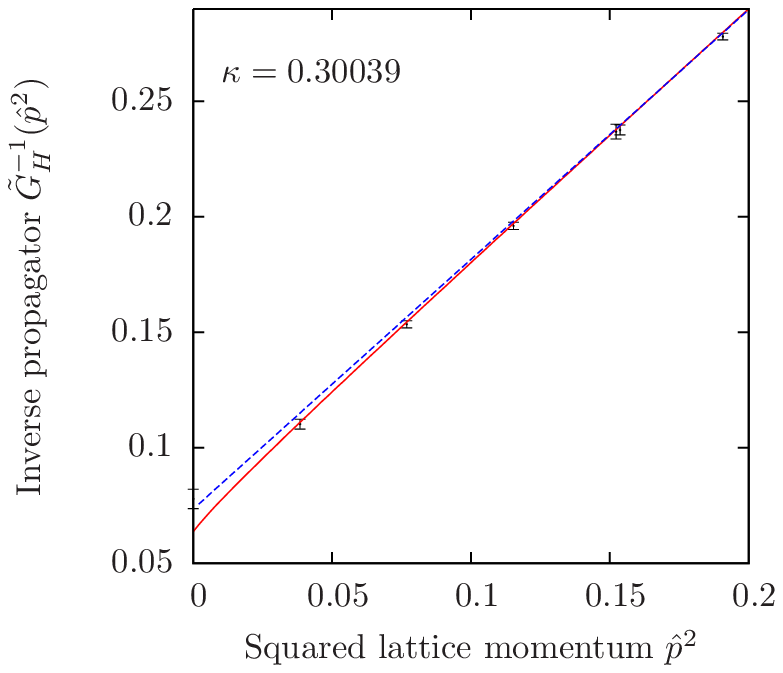}
{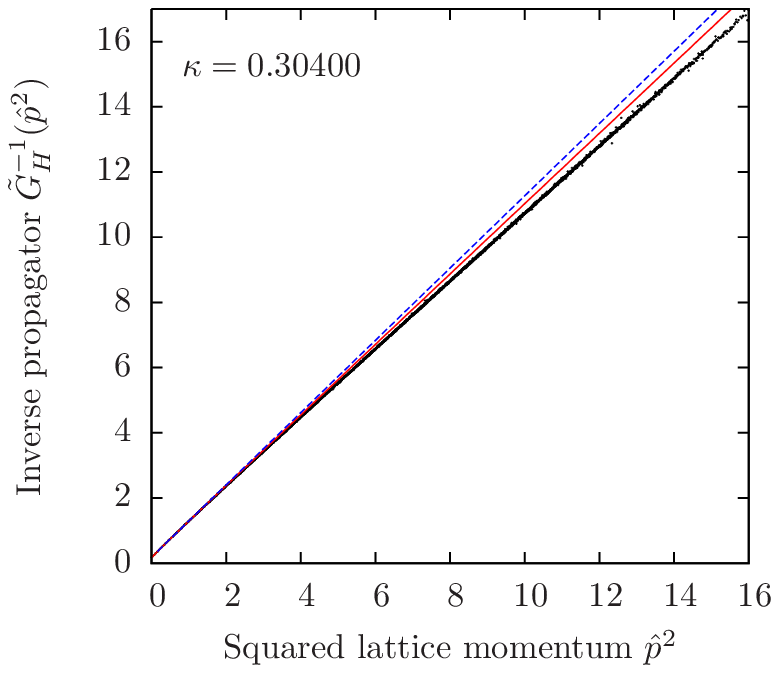}{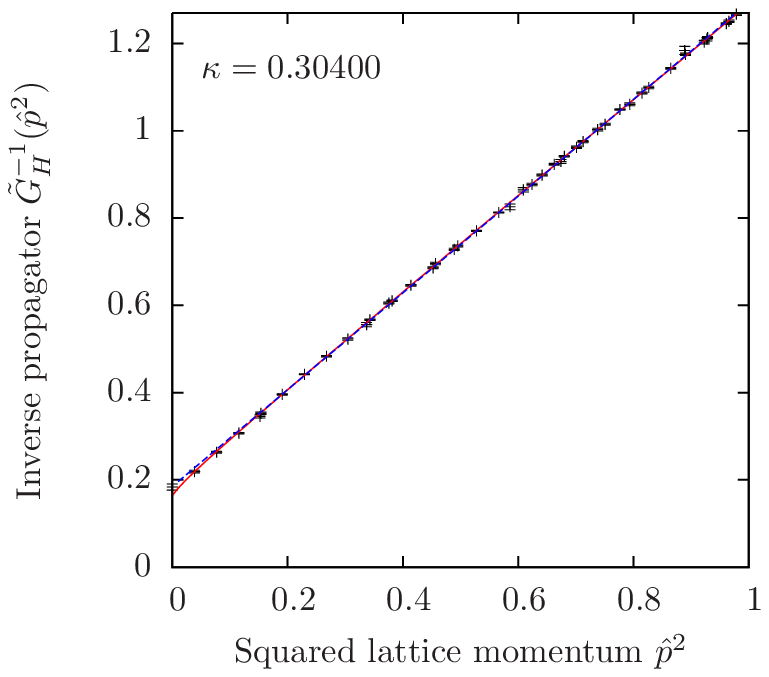}{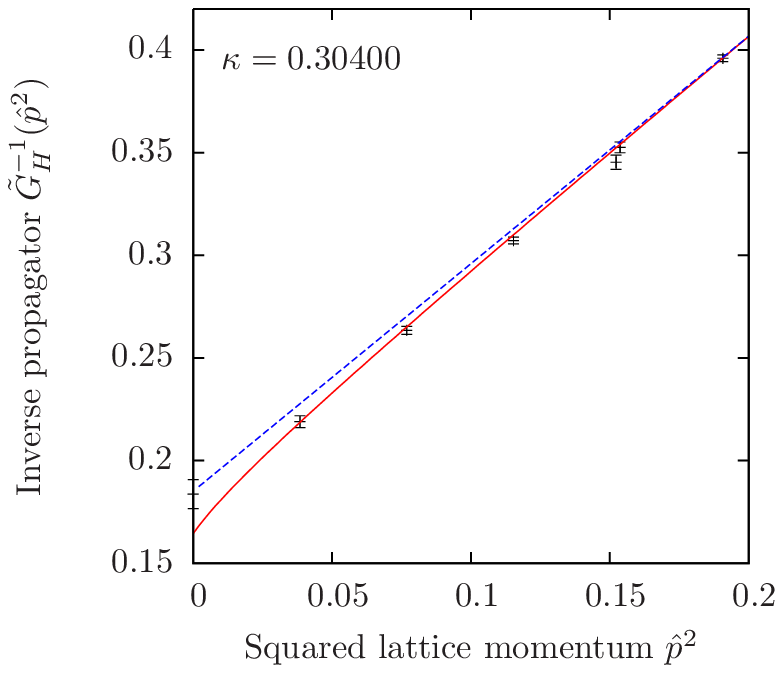}
{fig:HiggsPropExampleAtStrongCoup}
{The inverse lattice Higgs propagators calculated in the Monte-Carlo runs specified in \tab{tab:Chap71EmployedRuns}
are presented versus the squared lattice momenta $\hat p^2$ together with the respective fits obtained from the 
fit approaches $f^{-1}_H(\hat p^2)$ in \eq{eq:HiggsPropFitAnsatz} (red solid line) and $l^{-1}_H(\hat p^2)$ 
in \eq{eq:HiggsPropFitAnsatzLinear} (blue dashed line) with $\gamma=1.0$. From left to right the three 
panel columns display the same data zooming in, however, on the vicinity of the origin at $\hat p^2 = 0$.
}
{Examples of Higgs propagators at large quartic coupling constants.}

\includeTabNoHLines{|c|c|c|c|c|c|c|c|c|}{
\cline{4-9}
\multicolumn{3}{c|}{} &  \multicolumn{4}{c|}{Fit ansatz $f_H(\hat p^2)$} &  \multicolumn{2}{c|}{Fit ansatz $l_H(\hat p^2)$}  \\ \hline
$\kappa$ & $\gamma$   & $m_{Hc}$   & $m_{Hp}$  &  $m_{H}$       &  $\Gamma_H$ &  $\chi^2/dof$   &  $m_{H}$  &  $\chi^2/dof$    \\ \hline
0.30039  &  0.5       & 0.249(6)   & 0.253(2)  & 0.296(83)      & 0.000(0)    &  1.17           &  0.253(2)  & 1.13\\
0.30039  &  1.0       & 0.249(6)   & 0.252(2)  & 0.253(2)       & 0.035(8)    &  1.20           &  0.261(2)  & 1.62\\
0.30039  &  2.0       & 0.249(6)   & 0.246(2)  & 0.249(2)       & 0.062(4)    &  1.09           &  0.273(1)  & 2.58\\ \hline
0.30400  &  0.5       & 0.395(5)   & 0.405(3)  & 0.406(3)       & 0.039(13)   &  1.43           &  0.399(2)  & 1.75\\
0.30400  &  1.0       & 0.395(5)   & 0.409(1)  & 0.410(1)       & 0.054(7)    &  1.16           &  0.409(1)  & 2.23\\
0.30400  &  2.0       & 0.395(5)   & 0.409(1)  & 0.412(1)       & 0.087(4)    &  1.27           &  0.423(1)  & 4.63 \\ \hline
}
{tab:HiggsPropExampleResultsAtStrongCoup}
{The results on the Higgs propagator mass $m_{Hp}$, the Higgs pole mass $m_H$, and the Higgs decay
width $\Gamma_H$ obtained from the fit approaches $f_H(\hat p^2)$ in \eq{eq:HiggsPropFitAnsatz}
and $l_H(\hat p^2)$ in \eq{eq:HiggsPropFitAnsatzLinear} are listed for several 
settings of the parameter $\gamma$ together with the corresponding average squared residual per degree 
of freedom $\chi^2/dof$ associated to the respective fit. For the linear fit ansatz $l_H(\hat p^2)$
only the pole mass is presented, since one finds $m_{Hp}\equiv m_H$ and $\Gamma_H\equiv 0$ when constructing
the analytical continuation $\tilde G_H^c(p_c)$ through $l_H(p^2_c)$. These results are compared to the
Higgs correlator mass $m_{Hc}$ as determined in \sect{sec:HiggsTSCanalysisUpperBound} from the correlator $C_H(\Delta t)$
analyzed by means of \eq{eq:HiggsTimeCorrelatorFitAnsatz1}. The underlying 
Higgs lattice propagators have been calculated in the Monte-Carlo runs specified in \tab{tab:Chap71EmployedRuns}.
}
{Comparison of the Higgs propagator properties obtained from different extraction schemes at large quartic coupling
constants.}

Again the situation is very different in case of the more elaborate fit ansatz $f_H(\hat p^2)$ yielding significantly smaller values 
of $\chi^2/dof$. The presented results on the propagator mass $m_{Hp}$ as well as the pole mass $m_H$ are also in much better agreement
with the corresponding values obtained at varied threshold parameter $\gamma$. Moreover, the values of $m_{Hp}$ and $m_H$ are consistent 
with each other with respect to the given errors. These numbers are also compared to the correlator mass $m_{Hc}$, taken here from the 
forthcoming discussion in the subsequent section. In the latter comparison significant deviations are observed in some of the presented 
examples, which is, however, not too surprising due to the large systematic uncertainties of the correlator mass $m_{Hc}$,
as discussed in \sect{sec:HiggsTSCanalysisUpperBound}. It is also pointed out that the statistical errors of the propagator masses 
$m_{Hp}$ are much smaller than for the correlator mass $m_{Hc}$, which is one of the already mentioned arguments to determine the 
Higgs boson mass by means of the propagator analysis in the strongly interacting regime. 

From the findings presented in \tab{tab:HiggsPropExampleResultsAtStrongCoup} one can conclude that selecting the threshold value
to be $\gamma=1$ for the analysis of the Higgs propagator is a very reasonable choice, which leads to consistent and satisfactory 
results. This is the setting that will be used for the subsequent investigation of the upper Higgs boson mass bounds to determine 
the propagator mass $m_{Hp}$ as well as the pole mass $m_H$. It is further remarked that the here chosen value of $\gamma$ is much 
smaller than the value $\gamma=4$ selected in the preceding section for the analysis of the Goldstone propagator. While this large 
setting worked well in the latter scenario, it leads to inconsistent results in the here considered case and has therefore been excluded 
from the given presentation.

For clarification it is pointed out that the applied fit formula $f_H(\hat p^2)$, which is based on a perturbative calculation performed 
in continuous Euclidean space-time, is expected to become inconsistent with the lattice data if $\gamma$ is chosen too large due to 
the effect of discretization errors as discussed in \sect{sec:HiggsPropAnalysisLowerBound}. However, apart from that the perturbative 
one-loop calculation underlying the fit ansatz $f_H(\hat p^2)$ is not expected to describe the Higgs propagator very well in the here
considered strongly interacting regime of the model, since the one-loop calculation does not even contain all bosonic (momentum-dependent) 
next-to-leading order contributions, which are the order $O(\lambda^2)$ contributions in this case\footnote{The $O(\lambda)$ contributions
are momentum-independent as discussed in \sect{sec:HiggsSelfEnergy}.}. While the main purpose of the Higgs propagator
analysis, which is the Higgs boson mass determination, could satisfactorily be achieved with the here presented approach, it would 
still be worthwhile to extend the analysis at this point by considering a fit ansatz based on a higher order perturbative calculation 
of the Higgs propagator, since this would further improve the conceptual footing of the applied strategy. In such an extended fit ansatz
one should then also include the fermionic contributions. 

Finally, the obtained results on the decay width shall be addressed. Unfortunately, one finds in \tab{tab:HiggsPropExampleResultsAtStrongCoup}
that the extracted decay constants $\Gamma_H$ are not consistent when varying the parameter $\gamma$, which would be a minimal
requirement to trust in the presented numbers. The problem causing the observed inconsistency is as follows. For a too small threshold value
$\gamma$ the curvature of the Higgs propagator, which is eventually connected to the decay width $\Gamma_H$ through the analytical
continuation of the lattice Higgs propagator as discussed in \sect{sec:SimStratAndObs}, can no longer reliably be resolved due to
the statistical uncertainties of the lattice data, leading then ultimately to zero decay width. For larger values of $\gamma$, on the 
other hand, the effects described in the previous paragraph become the main source for the observed inconsistencies. 

One therefore has to conclude that the here presented attempt to determine the decay width $\Gamma_H$ of the Higgs boson through the 
analysis of the lattice Higgs propagator by means of the fit ansatz $f_H(\hat p^2)$, which is based on an one-loop perturbative 
calculation in the pure $\Phi^4$-theory, was not successful. Two further steps could be performed to improve on this unsatisfactory 
finding. On the one hand higher statistics would be helpful to reduce the statistical errors of the lattice data. On the other hand, 
and more importantly, an enhanced fit ansatz based on a higher order perturbative calculation and respecting also the fermionic contributions
is required to finally obtain reliable results also in this matter. 

This approach for determining $\Gamma_H$, however, will not further be pursued in this study. Instead, L\"uscher's well-known 
method~\cite{Luscher:1990ux,Luscher:1991cf} for determining the decay properties will be applied to the case of the considered 
Higgs-Yukawa model in \chap{chap:ResOnDecayWidth}, where the question of the Higgs boson decay width will finally be revisited.

%-------------------------------------------------------------------------------------------------------------------- 
\subsection{Analysis of the Higgs time-slice correlator}
\label{sec:HiggsTSCanalysisUpperBound}

We now turn to the discussion of the Higgs correlator mass $m_{Hc}$ in the strongly interacting regime of the considered
Higgs-Yukawa model, referring here to the setting $\lambda\gg 1$. For that purpose we follow the same steps that have been presented
for the scenario of weak interactions in \sect{sec:HiggsTSCanalysisLowerBound}. At that point it has already been discussed that 
the time-slice correlation function $C_H(\Delta t)$ underlying the definition of the Higgs correlator mass $m_{Hc}$ in 
\eq{eq:DefOfHiggsMassFromTimeSliceCorr} is actually lacking an obvious connection to the transfer matrix formalism, complicating 
thus the correct physical interpretation of the considered correlator. For that reason an alternative definition of a Higgs time-slice 
correlation function manifestly invalidating the aforementioned concern has additionally been investigated as defined in
\eq{eq:DefOfLocalCorrFunction}. Moreover, two different fit
approaches given in \eq{eq:HiggsTimeCorrelatorFitAnsatz1} and \eq{eq:HiggsTimeCorrelatorFitAnsatz2} have been applied to extract the 
correlator mass $m_{Hc}$ from the respective correlation function, where the first included also a $\Delta t$-independent term 
in the fit ansatz to account for the contributions of massless states, while the second only contained a single $\cosh$-function.
In all considered cases, however, the respective Higgs boson mass $m_{Hc}$ was found to be consistent with the other obtained results 
on that quantity, which allowed for a reliable determination of the Higgs boson mass from the time-slice correlation function in the 
case of weak interactions.

Here, we want to repeat this previous analysis for the case of the strongly interacting regime of the model. For that purpose
the aforementioned time-slice correlation functions $C_H(\Delta t)$ and $C'_H(\Delta t)$, as observed in the lattice calculations
specified in \tab{tab:Chap71EmployedRuns}, are presented in \fig{fig:HiggsCorrelatorExamplesAtStrongCoup}a,d together with the 
respective fits according to \eq{eq:HiggsTimeCorrelatorFitAnsatz1} and \eq{eq:HiggsTimeCorrelatorFitAnsatz2}. One observes in these 
plots that the measured correlation functions are clearly less accurately determined than in the weakly interacting regime discussed in 
\sect{sec:HiggsTSCanalysisLowerBound}. One reason explaining this observation is that the Higgs boson mass is much larger here than in the 
formerly considered scenario, leading thus to a steeper decay of the underlying correlation functions, which renders their reliable 
determination more demanding with respect to the required number of statistically independent field configurations generated in the 
lattice calculations. 

Moreover, one observes that the fit ansatz in \eq{eq:HiggsTimeCorrelatorFitAnsatz1}, incorporating the constant $C$ to respect contributions 
from massless states, seems to be better suited to describe the actually measured form of the correlation functions. However, the latter 
constant, as determined in the fit procedure, just turns out to have the wrong sign in the presented examples, implying 
that one has to be careful with the interpretation of that observation. In fact, the statistical errors of the presented data are just 
too large to draw a clear conclusion in this matter.

%\includeFigTrippleDouble
%{HiggsCorrelatorKap030400L32}{HiggsCorrelatorEffectiveMassesKap030400L32}{HiggsCorrelatorEffectiveMassesWFCKap030400L32}
%{HiggsCorrelatorDOTSKap030400L32}{HiggsCorrelatorDOTSEffectiveMassesKap030400L32}{HiggsCorrelatorDOTSEffectiveMassesWFCKap030400L32}
\includeFigTrippleDouble
{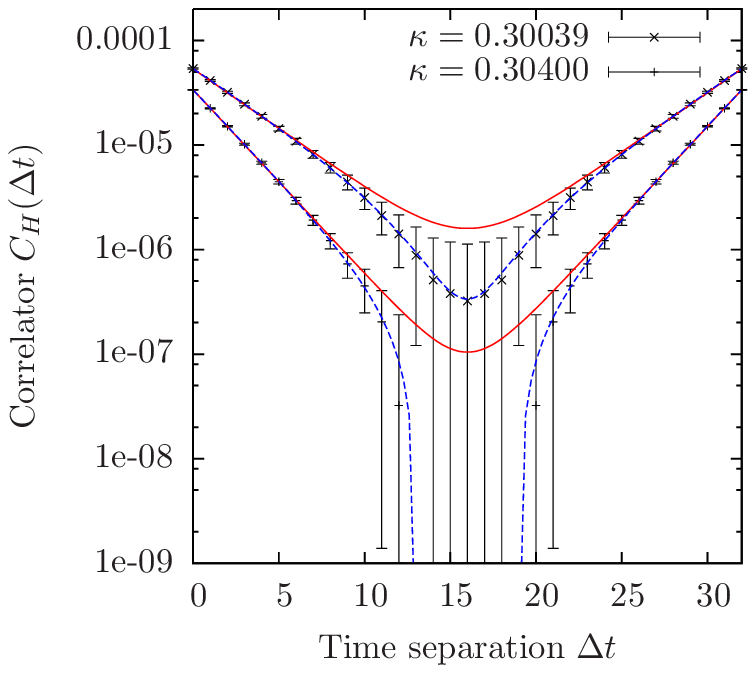}{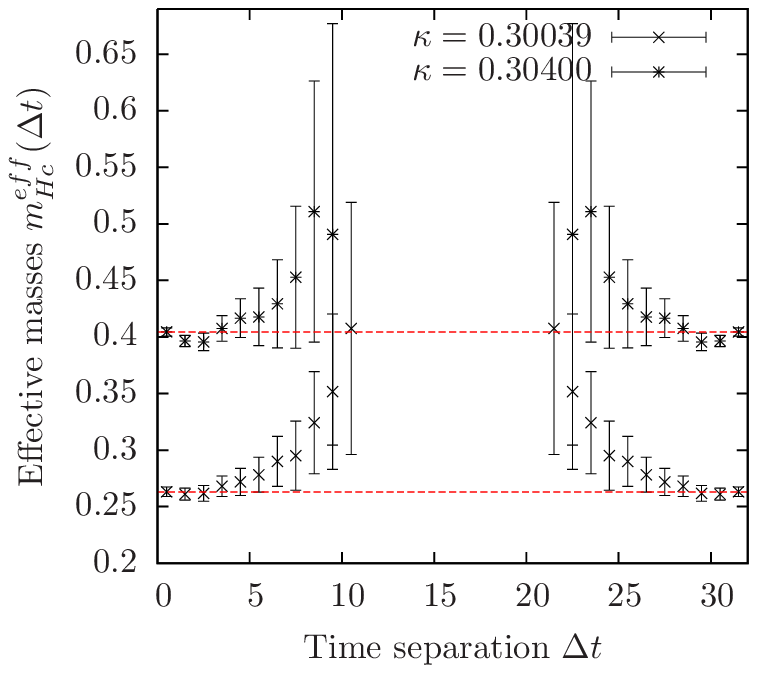}{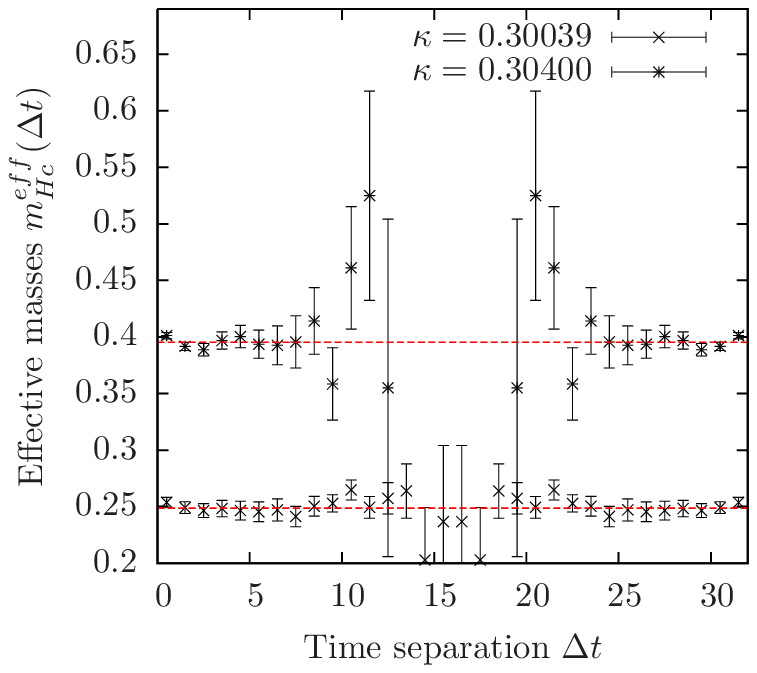}
{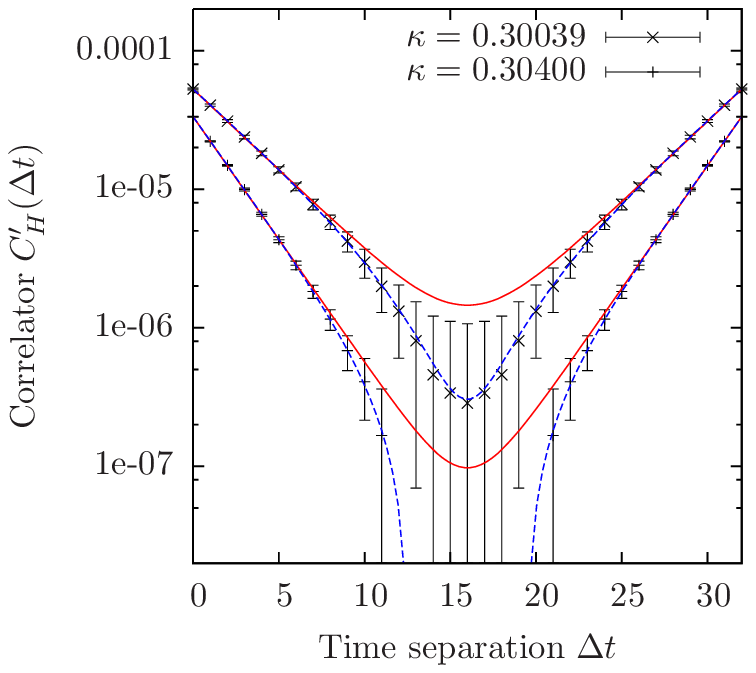}{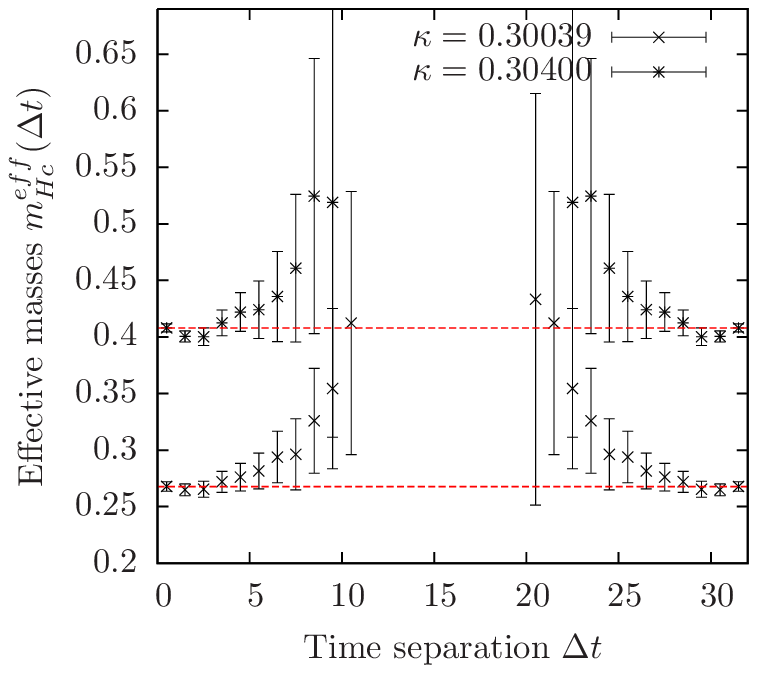}{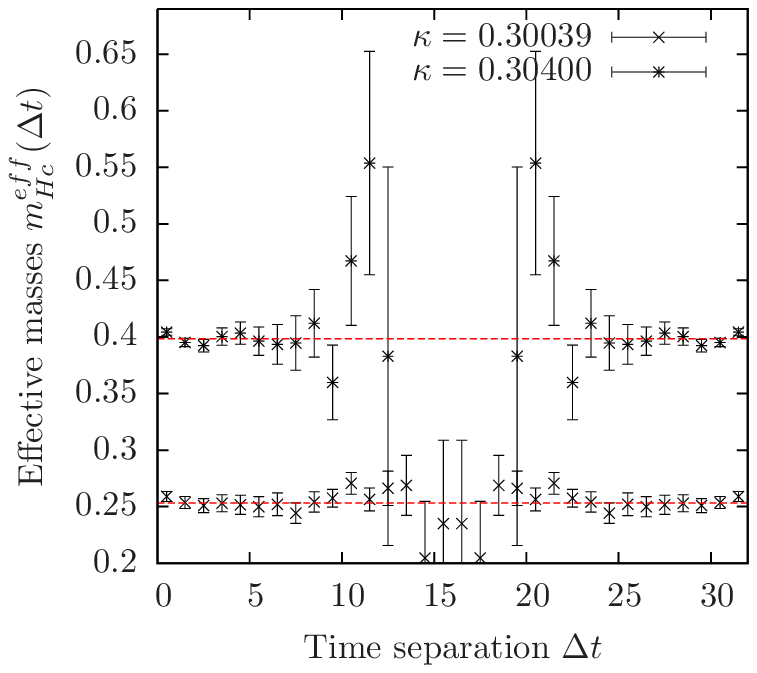}
{fig:HiggsCorrelatorExamplesAtStrongCoup}
{The Higgs time-slice correlation functions calculated in the Monte-Carlo runs specified in 
\tab{tab:Chap71EmployedRuns} are shown in panels (a) and (d) together with the respective fits 
obtained by the fit approaches in \eq{eq:HiggsTimeCorrelatorFitAnsatz1} and \eq{eq:HiggsTimeCorrelatorFitAnsatz2}
depicted by the dashed and solid curves, respectively. The corresponding effective masses are presented in the other 
four panels. For the sake of a better readability data points with too large statistical uncertainties have been 
removed from the plots. The results depicted in the upper row belong to the correlation function $C_H(\Delta t)$, 
while the lower row refers to $C'_H(\Delta t)$. The effective masses in the middle column have been determined 
by \eq{eq:DefOfEffMassesForHiggs} with $C=0$, while in panels (c) and (f) the constant $C$ was taken from
the fit ansatz $f_{A,m_{Hc},C}(\Delta t)$ to respect contaminations by lighter states.
The dashed horizontal lines depict the correlator masses $m_{Hc}$ obtained in the aforementioned fit procedures.
}
{Examples of the Higgs time-slice correlation functions at large quartic coupling constants.}

The associated effective masses as defined in \eq{eq:DefOfEffMassesForHiggs} are shown in the other four panels in 
\fig{fig:HiggsCorrelatorExamplesAtStrongCoup}. One sees that the statistical uncertainties associated to the presented effective
masses rise quickly with increasing values of $\Delta t$, rendering a reliable determination of the correlator mass $m_{Hc}$
from these data very problematic. The sought-after correlator masses are therefore again determined here by the overall $\cosh$-fits
in \eq{eq:HiggsTimeCorrelatorFitAnsatz1} and \eq{eq:HiggsTimeCorrelatorFitAnsatz2} applied to the whole range of time separations 
$\Delta t$. The then resulting values of $m_{Hc}$ are compared to the corresponding effective masses in \fig{tab:Chap71EmployedRuns} 
and satisfactory agreement is observed with respect to the specified statistical errors.

The obtained results on the correlator masses are summarized in \tab{tab:SummaryOfHiggsPropCorrCompAtStrongCoup}. From this
listing one learns that the masses $m_{Hc}$ derived from the correlators $C_{H}(\Delta t)$ and $C'_H(\Delta t)$ with the same
fit approach are consistent with each other in the here considered examples. It is, however, remarked that this situation 
changes dramatically on smaller lattice volumes, where the aforementioned masses differ significantly. An example of the volume 
dependence of these discrepancies observed in the strongly interacting regime is presented in 
\fig{fig:VolumeDepOfVariousMassDefAtLargeCoup}. One sees in the given presentation that these discrepancies vanish in the 
infinite volume limit, as already discussed in \sect{sec:HiggsTSCanalysisLowerBound}. The eventual convergence of the considered 
correlator masses with increasing lattice volume can also be understood from the qualitative argument that the magnetizations 
defined on neighbouring time-slices should become increasingly correlated when the lattice volume is enlarged. In the infinite 
volume limit the definitions of $C_H(\Delta t)$ and $C'_{H}(\Delta t)$ should therefore finally become identical to each other. 
On too small lattices, however, the two considered correlators yield significantly different results on the Higgs boson mass
$m_{Hc}$, thus blocking its reliable determination on the basis of the here presented approach. For clarification it is remarked
that the here discussed situation is in contrast to the scenario of the weakly interacting regime, where the aforementioned correlator 
masses in the investigated lattice calculations, though not explicitly demonstrated in this study, were found to be consistent 
with each other independently of the considered lattice volume. 

\includeTab{|c|c|c|c|}{
                  &Correlator        &  $m_{Hc}$ from $f_{A,m_{Hc},C}(\Delta t)$ &  $m_{Hc}$ from $f_{A,m_{Hc}}(\Delta t)$\\ \hline
$\kappa=0.30039$  &   $C_H(\Delta t)$   & 0.249(6)                       & 0.263(4)                       \\
$\kappa=0.30039$  &   $C'_H(\Delta t)$  & 0.253(6)                       & 0.268(4)                       \\
$\kappa=0.30400$  &   $C_H(\Delta t)$   & 0.395(5)                       & 0.404(4)                        \\
$\kappa=0.30400$  &   $C'_H(\Delta t)$  & 0.399(5)                       & 0.408(4)                       \\
}
{tab:SummaryOfHiggsPropCorrCompAtStrongCoup}
{The Higgs correlator masses $m_{Hc}$ are listed as obtained in the fit procedures of the correlation
functions $C_H(\Delta t)$ and $C'_H(\Delta t)$ applying the fit approaches in \eq{eq:HiggsTimeCorrelatorFitAnsatz1}
and \eq{eq:HiggsTimeCorrelatorFitAnsatz2}, respectively. The underlying correlation functions have been
calculated numerically in the Monte-Carlo runs specified in \tab{tab:Chap71EmployedRuns}.
}
{Comparison of different extraction schemes for the Higgs time-slice correlator mass at large quartic coupling
constants.}

Comparing also the obtained correlator masses $m_{Hc}$ arising from analyzing the same correlation functions, however, with 
the two different fit approaches in \eq{eq:HiggsTimeCorrelatorFitAnsatz1} and \eq{eq:HiggsTimeCorrelatorFitAnsatz2} one finds in 
\tab{tab:SummaryOfHiggsPropCorrCompAtStrongCoup} that the observed deviation is insignificant with respect to the given errors 
for the case $\kappa=0.30400$ and at most weakly significant in the other case. A similar picture is inferred from 
\fig{fig:VolumeDepOfVariousMassDefAtLargeCoup}. As pointed out before one would have expected the resulting correlator 
mass $m_{Hc}$ obtained by including a constant term into the fit ansatz to be larger than in the case, where only a single $\cosh$-function 
is considered. Since this is not observed here, one has to be careful with the interpretation of the numbers listed in 
\tab{tab:SummaryOfHiggsPropCorrCompAtStrongCoup}. However, the statistical uncertainty of the presented data is just 
too poor to draw reasonable conclusions in this matter.

Finally, it is concluded that the analysis of the correlation functions $C_H(\Delta t)$ and $C'_H(\Delta t)$ in the here presented
manner does not yield a reliable determination of the Higgs boson mass on (small) finite lattice volumes in the here considered
strongly interacting regime of the model due to conceptual uncertainties. It also remains unclear how much the obtained
correlator masses are systematically influenced by the contributions of lighter states due to the unstable nature of the Higgs boson.
Moreover, the considered Higgs correlator masses suffer from much larger statistical uncertainties than the previously 
discussed propagator masses. In the subsequent investigation of the upper Higgs boson mass bound the Higgs boson mass will therefore 
always be determined by the method based on the analysis of the Higgs propagator as discussed in \sect{sec:HiggsPropAnalysisUpperBound}. 

%\includeFigDouble{VolumeDepOfVariousMassesKap030039}{VolumeDepOfVariousMassesKap030400}
\includeFigDouble{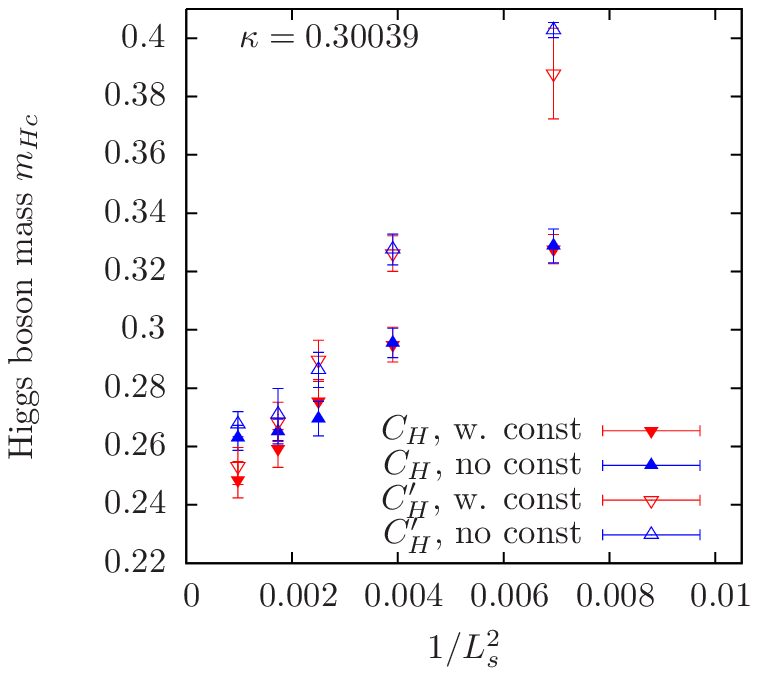}{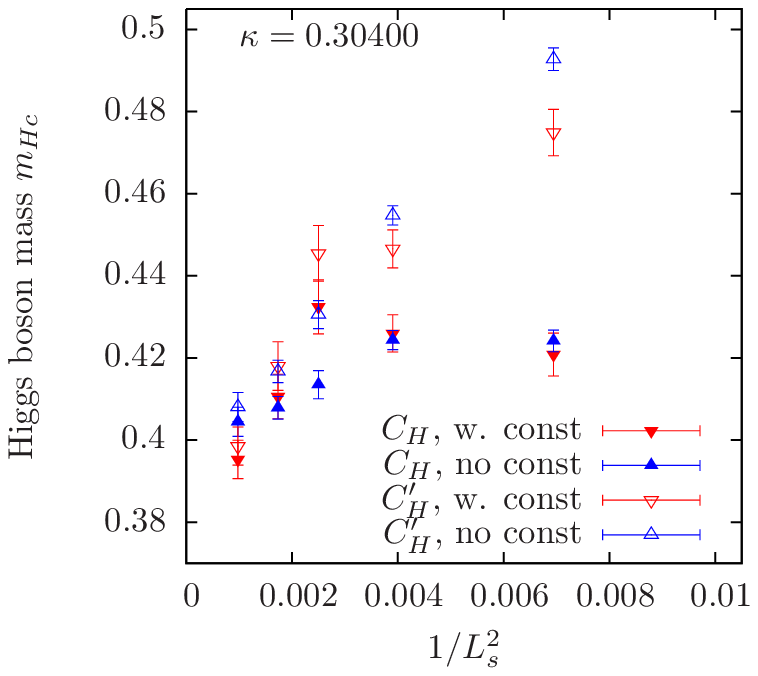}
{fig:VolumeDepOfVariousMassDefAtLargeCoup}
{The Higgs correlator masses $m_{Hc}$ as obtained from the correlation functions $C_H(\Delta t)$ and $C'_H(\Delta t)$ fitted
with and without including a constant into the fit ansatz according to \eq{eq:HiggsTimeCorrelatorFitAnsatz1} and 
\eq{eq:HiggsTimeCorrelatorFitAnsatz2} are presented versus $1/L_s^2$ to display the encountered finite volume effects.
These results have been obtained in direct Monte-Carlo calculations as specified in \tab{tab:Chap71EmployedRuns}, however,
with varying lattice volumes. 
}
{Volume dependence of the different extraction schemes for the Higgs time-slice correlator mass at large 
quartic coupling constants.}

%--------------------------------------------------------------------------------------------------------------------------------------------------
\section{Dependence of the Higgs boson mass on the bare \texorpdfstring{coupling constant $\lambda$}{quartic coupling constant}}
\label{sec:DepOfHiggsMassonLargeLam}

We now turn to the question whether the largest Higgs boson mass is indeed obtained at infinite bare quartic coupling constant 
for a given set of quark masses and a given cutoff $\Lambda$ as one would expect from the results of the effective potential 
calculation performed in \sect{sec:DepOfHiggsMassOnLambda}. Since the underlying formulas of that consideration, however, were 
derived in the weakly coupling regime, the actual dependence of the Higgs boson mass on the bare quartic coupling constant $\lambda$ 
in the scenario of strong interactions will explicitly be checked here by means of direct Monte-Carlo calculations. The final answer 
of what bare coupling constant produces the largest Higgs boson mass will then be taken as input for the investigation of the upper 
mass bound in the subsequent section.

Numerical results on the Higgs propagator mass $m_{Hp}$ are therefore plotted versus the bare quartic coupling constant $\lambda$ in 
\fig{fig:StrongCoulingHiggsMassDepOnQuartCoup}a. The presented data have been obtained for a cutoff that was intended to be kept
constant at approximately $\Lambda\approx \GEV{1540}$ by an appropriate tuning of the hopping parameter, while the degenerate Yukawa 
coupling constants were fixed according to the tree-level relation in \eq{eq:treeLevelTopMass} aiming at the reproduction of the top 
quark mass. One clearly observes that the Higgs boson mass monotonically rises with increasing values of the bare coupling constant
$\lambda$ until it finally converges to the $\lambda=\infty$ result, which is depicted by the horizontal line in the presented plot.
From this result one can conclude that the largest Higgs boson mass is indeed obtained at infinite bare quartic coupling constant,
as expected. The forthcoming search for the upper mass bound will therefore be restricted to the scenario of $\lambda=\infty$.

%\includeFigDouble{LambdaScanHiggsMassAtLambdaLarge}{LambdaScanRenLamCoupAtLambdaLarge}
\includeFigDouble{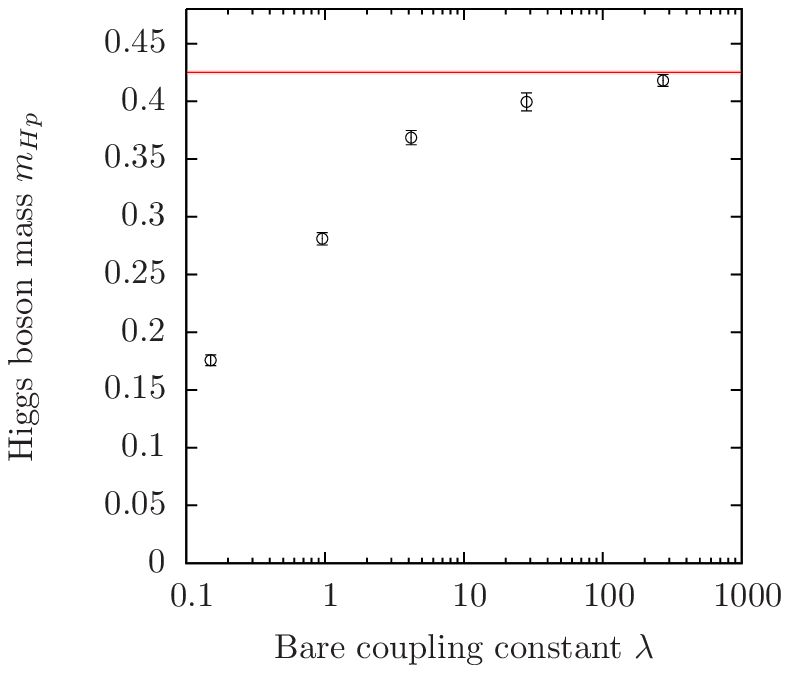}{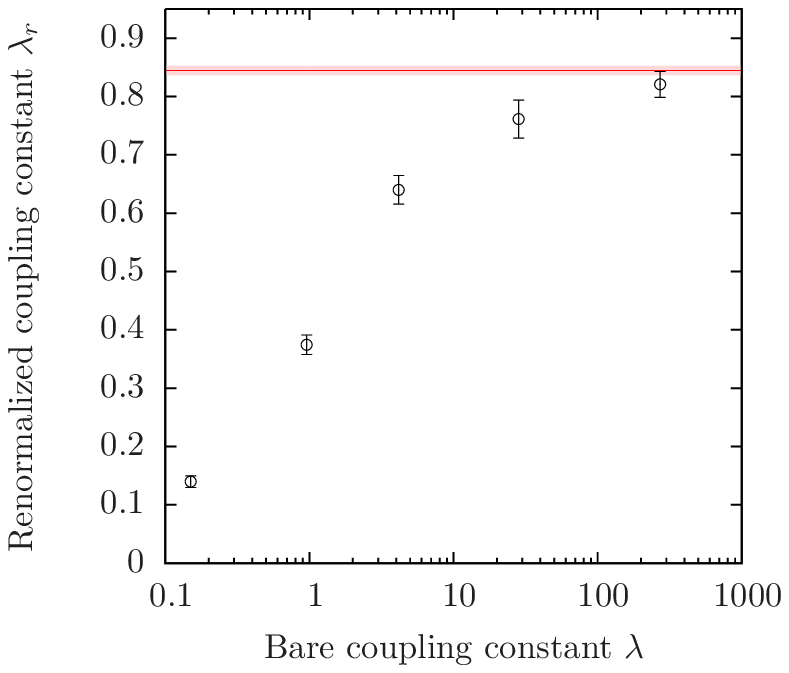}
{fig:StrongCoulingHiggsMassDepOnQuartCoup}
{The Higgs boson mass $m_{Hp}$ and the renormalized quartic coupling constant $\lambda_r$ are shown versus the bare coupling
constant $\lambda$ in panels (a) and (b), respectively. These results have been obtained in direct Monte-Carlo calculations 
on a \lattice{16}{32} with the degenerate Yukawa coupling constants fixed according to the tree-level relation in 
\eq{eq:treeLevelTopMass} aiming at the reproduction of the top quark mass. The hopping parameter was tuned with the intention
to hold the cutoff constant, while the actually obtained values of $\Lambda$ fluctuate here between $\GEV{1504}$ and
$\GEV{1549}$. The horizontal lines depict the corresponding results at infinite bare coupling constant $\lambda=\infty$, and
the highlighted bands mark the associated statistical uncertainties.
}
{Dependence of the Higgs boson mass $m_{Hp}$ and the renormalized quartic coupling constant $\lambda_r$ on the 
bare quartic coupling constant $\lambda$.}

Furthermore, the corresponding behaviour of the renormalized quartic coupling constant $\lambda_r$ as defined in \eq{eq:DefOfRenQuartCoupling} 
is presented in \fig{fig:StrongCoulingHiggsMassDepOnQuartCoup}b. As expected one observes a monotonically rising dependence
of $\lambda_r$ on the bare coupling constant $\lambda$, eventually converging to the $\lambda=\infty$ result depicted by the 
horizontal line. From this finding one can also conclude that the model should remain treatable by means of renormalized
perturbation theory also in the regime of large bare quartic coupling constants $\lambda\gg 1$.

Finally, it is remarked that the numerical computations would have become increasingly difficult with rising values of 
$\lambda$ due to the ever increasing number of required integration steps of the underlying molecular dynamics integration as 
discussed in \sect{sec:FACC}, if one had tried to integrate the equations of motion in \eqs{eq:FACCeomPhi}{eq:FACCeomMomenta}
directly. To circumvent this problem of increasing stiffness of the latter differential equations the alternative approach
in \eqs{eq:FACCLamInfinityPhiEqOfMotionLargeLambda}{eq:FACCLamInfinityPEqOfMotionLargeLambda3} has been used here. With its 
help different integration step sizes can be assigned to the tangential and radial update steps of the field variables $\Phi_x$, 
making the integration possible for any finite value of $\lambda$.

\vs{20mm}

%--------------------------------------------------------------------------------------------------------------------------------------------------
\section{Upper Higgs boson mass bounds}
\label{sec:UpperMassBounds}

In this section the cutoff dependent upper Higgs boson mass bound $m_H^{up}(\Lambda)$ shall finally be determined 
by means of direct Monte-Carlo calculations performed in the considered Higgs-Yukawa model. Given the knowledge 
about the dependence of the Higgs boson mass on the bare quartic self-coupling constant $\lambda$, which was studied 
in the preceding section, the search for the desired upper mass bound can safely be restricted to the scenario of an 
infinite bare quartic coupling constant, \ie $\lambda=\infty$. Moreover, we will restrict the investigation here to the
mass degenerate case with equal top and bottom quark masses and $N_f=1$ due to the much larger numerical costs of 
the physically more relevant setup with $y_b/y_t=0.024$ and $N_f=3$, which would clearly exceed those of the 
corresponding scenario in the weakly coupling regime discussed in \sect{sec:SubLowerHiggsMassboundsGenCase}. 
In the considered setup the Yukawa coupling constants will then again be fixed according to the tree-level relation 
in \eq{eq:treeLevelTopMass} aiming at the reproduction the physical top quark mass.

For the same reasons already discussed in \sect{sec:SubLowerHiggsMassbounds} the accessible energy scales $\Lambda$ 
are again restricted by \eq{eq:RequirementsForLatMass}. Employing a top mass of $\GEV{175}$ and a Higgs boson 
mass of below $\GEV{700}$, which will turn out to be justified after the upper mass bound has eventually been established, 
it should be possible to reach energy scales between $\GEV{1400}$ and $\GEV{2800}$ on a \latticeX{32}{32}{.}

\includeTab{|ccccccccc|}
{
$\kappa$ & $L_s$                    & $L_t$ & $N_f$ &  $\hat \lambda$ & $\hat y_t$     & $\hat y_b/\hat y_t$ & 1/v & $\Lambda$ \\
\hline
0.30039  & 12,16,20,24,32  &   32  &  1    &  $\infty$         & 0.55139        & 1          & $\approx 7.7$   & $\GEV{\approx 2370 }$\\
0.30148  & 12,16,20,24,32  &   32  &  1    &  $\infty$         & 0.55239        & 1          & $\approx 6.5$   & $\GEV{\approx 1990 }$\\
0.30274  & 12,16,20,24,32  &   32  &  1    &  $\infty$         & 0.55354        & 1          & $\approx 5.6$   & $\GEV{\approx 1730 }$\\
0.30400  & 12,16,20,24,32  &   32  &  1    &  $\infty$         & 0.55470        & 1          & $\approx 5.0$   & $\GEV{\approx 1550 }$\\ \hline
0.30570  & 12,16,20,24,32  &   32  &  1    &  $\infty$         & 0              & --         & $\approx 9.0$   & $\GEV{\approx 2810 }$\\
0.30680  & 12,16,20,24,32  &   32  &  1    &  $\infty$         & 0              & --         & $\approx 7.1$   & $\GEV{\approx 2220 }$\\
0.30780  & 12,16,20,24,32  &   32  &  1    &  $\infty$         & 0              & --         & $\approx 6.2$   & $\GEV{\approx 1910 }$\\
0.30890  & 12,16,20,24,32  &   32  &  1    &  $\infty$         & 0              & --         & $\approx 5.5$   & $\GEV{\approx 1700 }$\\
0.31040  & 12,16,20,24,32  &   32  &  1    &  $\infty$         & 0              & --         & $\approx 4.9$   & $\GEV{\approx 1500 }$\\
}
{tab:SummaryOfParametersForUpperHiggsMassBoundRuns}
{The model parameters of the Monte-Carlo runs underlying the subsequent lattice calculation of the upper Higgs boson mass bound
are presented. In total, a number of 45 Monte-Carlo runs have been performed for that purpose. The available statistics of 
generated field configurations $\Nconf$ varies depending on the respective lattice volume. In detail we have $\Nconf\approx 20,000$ 
for $12\le L_s\le 16$, $\Nconf\approx10,000-15,000$ for $L_s = 20$, $\Nconf\approx8,000-16,000$ for $L_s=24$, 
$\Nconf\approx 3,000-5,000$ for $L_s=32$.
The numerically determined values of $1/v$ and $\Lambda$ are also approximately given. These numbers vary, of course, depending
on the respective lattice volumes and serve here only for the purpose of a rough orientation. The degenerate Yukawa coupling 
constants in the upper four rows have been chosen according to the tree-level relation in \eq{eq:treeLevelTopMass} aiming at the 
reproduction of the phenomenologically known top quark mass. In the other cases it is exactly set to zero recovering the pure 
$\Phi^4$-theory.}
{Model parameters of the Monte-Carlo runs underlying the lattice calculation of the upper Higgs boson mass bound.}

For the purpose of investigating the cutoff dependence of the upper mass bound a series of direct Monte-Carlo
calculations has been performed with varying hopping parameters $\kappa$ associated to cutoffs covering approximately
the given range of reachable energy scales. At each value of $\kappa$ the Monte-Carlo computation has been
rerun on several lattice sizes to examine the respective strength of the finite volume effects, ultimately allowing for
the infinite volume extrapolation of the obtained lattice results. In addition, a corresponding series of Monte-Carlo
calculations has been performed in the pure $\Phi^4$-theory, which will finally allow to address the question for the
fermionic contributions to the upper Higgs boson mass bound. The model parameters underlying these two series of lattice 
calculations are presented in \tab{tab:SummaryOfParametersForUpperHiggsMassBoundRuns}.

%\includeFigDouble{HiggsMassVsCutoffAtInfiniteCouplingLATUNITS}{HiggsMassVsCutoffAtInfiniteCouplingLATUNITSPurePhi4}
\includeFigDouble{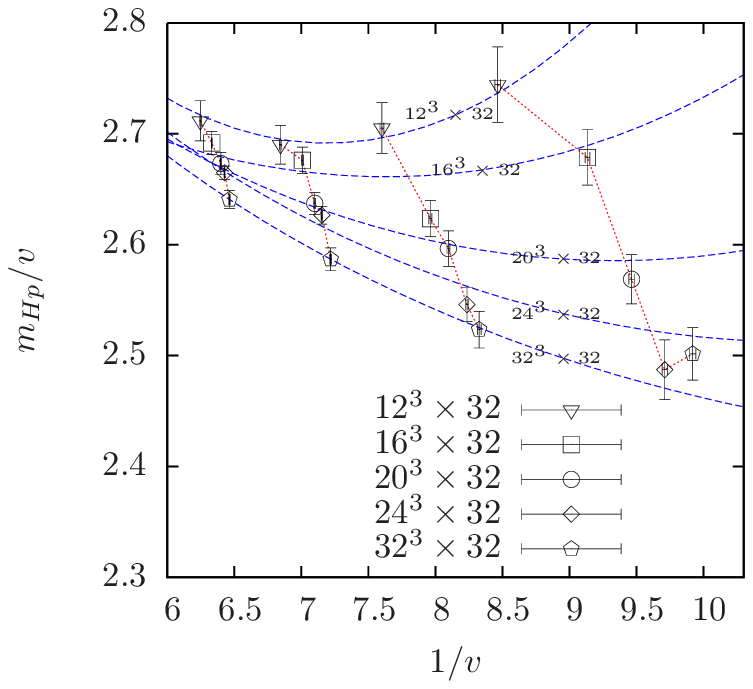}{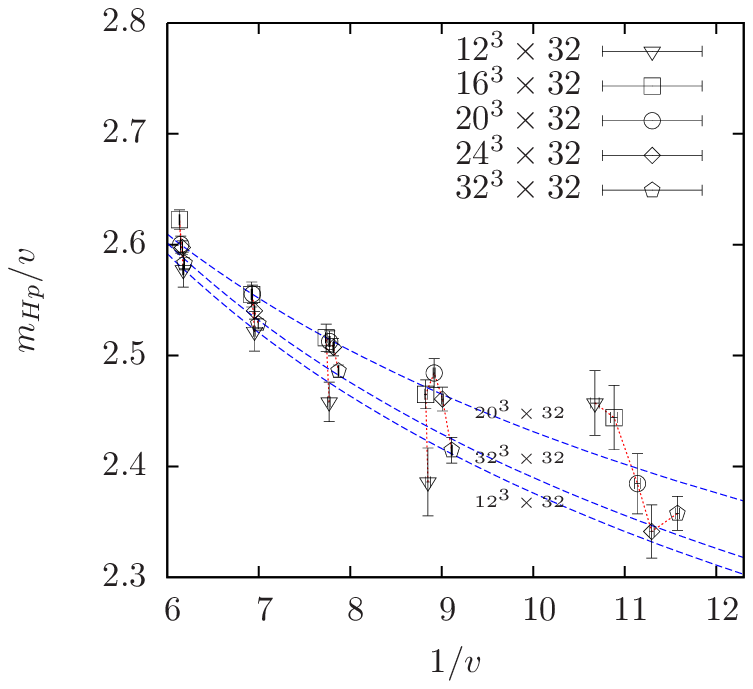}
{fig:UpperHiggsCorrelatorBoundFiniteVol}
{The Higgs propagator mass $m_{Hp}$ is presented in units of the vacuum expectation value $v$ 
versus $1/v$. These results have been determined in the direct Monte-Carlo calculations specified in
\tab{tab:SummaryOfParametersForUpperHiggsMassBoundRuns}. Those runs with identical parameter sets differing 
only in the underlying lattice volume are connected via dotted lines to illustrate the effects of the 
finite volume. The dashed curves depict the fits of the lattice results according to the finite size expectation
in \eq{eq:RenormHiggsMassFitFormula} as explained in the main text. Panel (a) refers to the full Higgs-Yukawa model,
while panel (b) shows the corresponding results of the pure $\Phi^4$-theory.
}
{Dependence of the Higgs propagator mass on $1/v$ at infinite bare quartic coupling constant.}

The numerically obtained Higgs boson masses $m_{Hp}$ resulting in these lattice calculations are presented in 
\fig{fig:UpperHiggsCorrelatorBoundFiniteVol}, where panel (a) refers to the full Higgs-Yukawa model while panel (b)
displays the corresponding results in the pure $\Phi^4$-theory. To illustrate the influence of the finite lattice volume 
those results, belonging to the same parameter sets, differing only in the underlying lattice size, are connected by dotted 
lines to guide the eye. From these findings one learns that the model indeed exhibits strong finite volume effects when 
approaching the upper limit of the defined interval of reachable cutoffs, as expected. 

In \fig{fig:UpperHiggsCorrelatorBoundFiniteVol}a one sees that the Higgs boson mass seems to increase with
the cutoff $\Lambda$ on the smaller lattice sizes. This, however, is only a finite volume effect. On the larger 
lattices the Higgs boson mass decreases with growing $\Lambda$ as expected from the triviality property of the 
Higgs sector. In comparison to the results obtained in the pure $\Phi^4$-theory shown in \fig{fig:UpperHiggsCorrelatorBoundFiniteVol}b
the aforementioned finite size effects are much stronger and can thus be ascribed to the influence of the coupling to the
fermions. 

Since the top quark is the lightest physical particle in the here considered scenario it is obvious that the coupling of the Higgs
boson to the fermions will generate the major part of the observed finite size effects apart from the Goldstone contributions. The 
investigation of the setup with the actual physical splitting of the top and bottom quark mass would therefore be even more demanding 
in the strongly coupling regime than it has been in the previously discussed situation in \sect{sec:SubLowerHiggsMassboundsGenCase}, 
which is one of the reasons why this analysis will not be pursued in the following. 

At this point it is worthwhile to ask whether the observed finite volume effects can also be understood by some quantitative consideration.
For the weakly interacting regime this could be achieved by computing the effective potential for a given finite volume
as discussed in \sect{sec:EffPot}. In contrast to that scenario the calculation of the effective potential in terms of the
bare model parameters is not directly useful in the present situation, since the underlying perturbative expansion 
would break down due to the bare quartic coupling constant being infinite here. This problem can be cured by extending the already
started partial renormalization procedure in \sect{sec:DepOfHiggsMassOnLambda}, such that the bare quartic coupling constant also 
becomes replaced by its renormalized counterpart. Starting from \eq{eq:DesOfImprovedEffPotForLargeLam1Improved}
and assuming a definition of the renormalized quartic coupling constant $\lambda_r$ fulfilling the relation 
$\lambda = \lambda_r + O(\lambda_r^2)$ such as the definition given in \eq{eq:DefOfRenQuartCoupling} one can directly derive an 
estimate for the Higgs boson mass in terms of $\lambda_r$ 
according to
\beq
m_{He}^2 = 8\lambda_r v^2 
-\frac{1}{v} \derive{}{\breve v} \breve U_F[\breve v]\Bigg|_{\breve v = v} + \derive{^2}{\breve v^2} 
\breve U_F[\breve v]\Bigg|_{\breve v = v}
\eeq
which respects all contributions of order $O(\lambda_r)$. It is remarked that the contributions $\breve U_H[\breve v]$
and $\breve U_G[\breve v]$ in \eq{eq:DesOfImprovedEffPotForLargeLam1Improved} do not contribute to $m_{He}$, as discussed
in \sect{sec:DepOfHiggsMassOnLambda}. Combining the above result with the expected scaling laws given in 
\eqs{eq:ScalingBehOfVeV}{eq:ScalingBehOfLamRen} a crude estimate for the observed behaviour of the Higgs boson mass
presented in \fig{fig:UpperHiggsCorrelatorBoundFiniteVol} can be established according to
\bea
\label{eq:RenormHiggsMassFitFormula}
m_{He}^2 &=& \frac{8v^2 A'_\lambda}{\log(v^{-2}) + B'_\lambda}  -\frac{1}{v} \derive{}{\breve v} \breve U_F[\breve v]\Bigg|_{\breve v = v} + \derive{^2}{\breve v^2} 
\breve U_F[\breve v]\Bigg|_{\breve v = v}, 
\eea
where double-logarithmic terms have been neglected and $A'_\lambda$, $B'_\lambda$ are so far unspecified parameters.

Since the value of the renormalized quartic coupling constant $\lambda_r$ is not known apriori, the idea is here to use the result in 
\eq{eq:RenormHiggsMassFitFormula} as a fit ansatz with the free fit parameters $A'_\lambda$ and $B'_\lambda$ to fit the observed 
finite volume behaviour of the Higgs boson mass presented in \fig{fig:UpperHiggsCorrelatorBoundFiniteVol}. These lattice data
have been given in units of the vacuum expectation value $v$, plotted versus $1/v$, to allow for the intended
direct comparison with the analytically derived finite volume expectation in \eq{eq:RenormHiggsMassFitFormula}. The resulting 
fits are depicted by the dashed curves in \fig{fig:UpperHiggsCorrelatorBoundFiniteVol}, where the free parameters $A'_\lambda$ and $B'_\lambda$
have independently been adjusted for every presented series of constant lattice volume in the full Higgs-Yukawa model and the pure 
$\Phi^4$-theory, respectively. Applying the above fit ansatz simultaneously to all available data does not lead to satisfactory results, 
since the renormalized quartic coupling constant itself also depends significantly on the underlying lattice volume, as will be 
seen later in this section. 

One can then observe in \fig{fig:UpperHiggsCorrelatorBoundFiniteVol} that the established crude estimate describes the actual finite
volume cutoff-dependence of the presented Higgs boson mass satisfactorily well, at least for the case of the full Higgs-Yukawa model. 
This observation also holds for the results obtained in the pure $\Phi^4$-theory, except for the smallest investigated lattice volumes 
at the smallest considered values of the vacuum expectation value $v$, which deviate significantly from the expected behaviour in 
\eq{eq:RenormHiggsMassFitFormula}. This, however, is not too surprising for the following reason. In the infinite volume limit one would 
suppose the assumed scaling laws in \eqs{eq:ScalingBehOfVeV}{eq:ScalingBehOfLamRen} to describe the behaviour of the considered model the better, 
the closer one approaches the critical line. On a finite lattice volume, however, these scaling laws eventually break down in the latter limit 
$\kappa\rightarrow \kapCrit$. This can easily be understood from the fact that the vacuum expectation value $v$ as defined in \eq{eq:DefOfVEV}, 
for instance, can never vanish exactly on a finite lattice volume in contrast to the expected scaling in \eq{eq:ScalingBehOfVeV}.

Apart from this latter remark one can thus conclude that the observed finite volume behaviour of the Higgs boson mass presented in 
\fig{fig:UpperHiggsCorrelatorBoundFiniteVol} can be well understood by means of the analytical expectation in 
\eq{eq:RenormHiggsMassFitFormula}.

%\includeFigDoubleDoubleHere{HiggsMassVsCutoffAtInfiniteCouplingFiniteVolumeEffectsVeV}{HiggsMassVsCutoffAtInfiniteCouplingFiniteVolumeEffectsVeVPurePhi4}
%{HiggsMassVsCutoffAtInfiniteCouplingFiniteVolumeEffectsHiggsMass}{HiggsMassVsCutoffAtInfiniteCouplingFiniteVolumeEffectsHiggsMassPurePhi4}
\includeFigDoubleDoubleHere{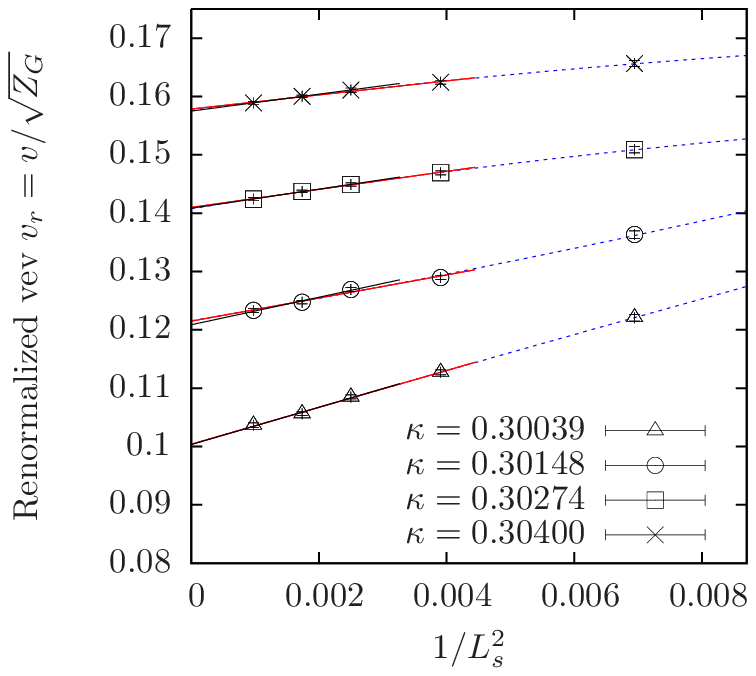}{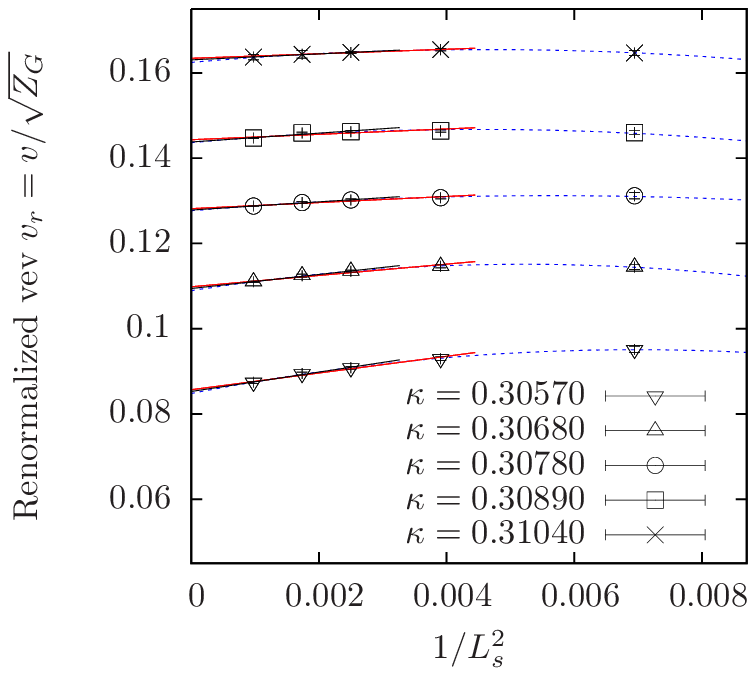}
{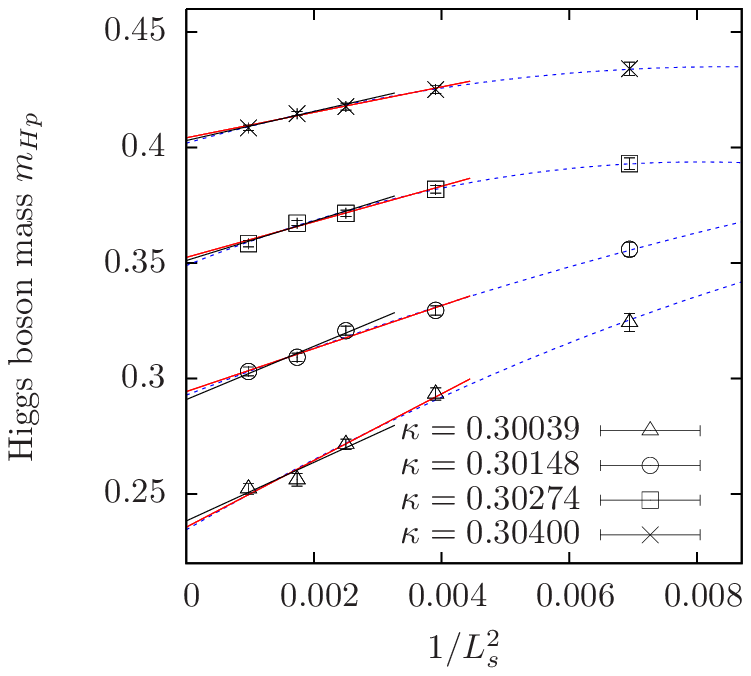}{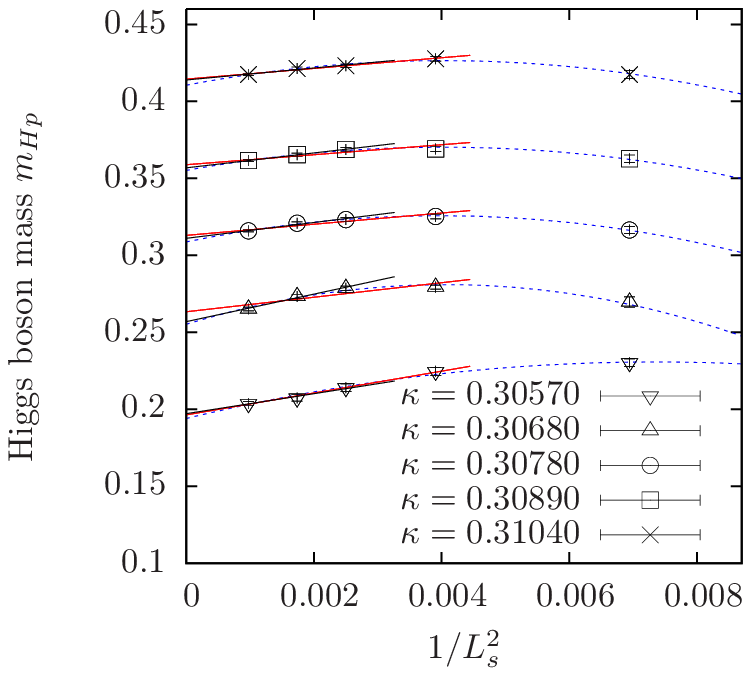}
{fig:FiniteVolumeEffectsOfUpperHiggsMassBoundVEVandMH}
{The dependence of the renormalized vev $v_r = v/\sqrt{Z_G}$ and the Higgs propagator mass $m_{Hp}$
on the squared inverse lattice side length $1/L^2_s$ is presented in the upper and the lower panel rows, respectively, as
determined in the direct Monte-Carlo calculations specified in \tab{tab:SummaryOfParametersForUpperHiggsMassBoundRuns}.
Panels (a) and (c) show the results for the full Higgs-Yukawa model, while panels (b) and (d) refer to the pure $\Phi^4$-theory.
In all plots the dashed curves display the parabolic fits according to the fit ansatz in 
\eq{eq:ParaFit}, while the solid lines depict the linear fits resulting from \eq{eq:LinFit} for the two threshold 
values $L_s'=16$ (red), and $L_s'=20$ (black).
}
{Dependence of the renormalized vev $v_r = v/\sqrt{Z_G}$ and the Higgs propagator mass $m_{Hp}$
on the squared inverse lattice side length $1/L^2_s$ at infinite bare quartic coupling constant.}

\includeTab{|c|c|c|c|c|}{
\multicolumn{5}{|c|}{Vacuum expectation value $v$} \\ \hline
$\kappa$  	 & $A^{(l)}_v$, $L'_s=16$  & $A^{(l)}_v$, $L'_s=20$         & $A^{(p)}_v$         & $v_r$               \\ \hline
$\,0.30039\,$  	 & $\, 0.1004(3) \, $   & $\, 0.1003(6) \, $  & $\, 0.1004(5) \, $  & $\, 0.1004(5)(1)$  \\ 
$\,0.30148\,$  	 & $\, 0.1215(5) \, $   & $\, 0.1209(6) \, $  & $\, 0.1216(8) \, $  & $\, 0.1213(6)(4)$  \\ 
$\,0.30274\,$  	 & $\, 0.1410(1) \, $   & $\, 0.1408(1) \, $  & $\, 0.1408(1) \, $  & $\, 0.1409(1)(1)$  \\ 
$\,0.30400\,$  	 & $\, 0.1579(2) \, $   & $\, 0.1575(1) \, $  & $\, 0.1576(2) \, $  & $\, 0.1577(2)(2)$  \\ 
$\,0.30570\,$  	 & $\, 0.0857(4) \, $   & $\, 0.0852(4) \, $  & $\, 0.0848(2) \, $  & $\, 0.0852(3)(5)$  \\ 
$\,0.30680\,$  	 & $\, 0.1099(4) \, $   & $\, 0.1094(2) \, $  & $\, 0.1089(1) \, $  & $\, 0.1097(3)(5)$  \\ 
$\,0.30780\,$  	 & $\, 0.1282(3) \, $   & $\, 0.1278(1) \, $  & $\, 0.1277(2) \, $  & $\, 0.1279(2)(3)$  \\ 
$\,0.30890\,$  	 & $\, 0.1443(5) \, $   & $\, 0.1438(5) \, $  & $\, 0.1436(4) \, $  & $\, 0.1439(5)(4)$  \\ 
$\,0.31040\,$  	 & $\, 0.1634(2) \, $   & $\, 0.1630(1) \, $  & $\, 0.1625(2) \, $  & $\, 0.1630(2)(5)$  \\  \hline
\multicolumn{5}{|c|}{Higgs propagator mass $m_{Hp}$} \\ \hline
$\kappa$  	 & $A^{(l)}_m$, $L'_s=16$  & $A^{(l)}_m$, $L'_s=20$         & $A^{(p)}_m$         & $m_{Hp}$               \\ \hline
$\,0.30039\,$  	 & $\, 0.2356(41)\, $   & $\, 0.2382(70)\, $  & $\, 0.2344(67)\, $  & $\, 0.2361(61)(19)$  \\ 
$\,0.30148\,$  	 & $\, 0.2943(29)\, $   & $\, 0.2908(39)\, $  & $\, 0.2928(40)\, $  & $\, 0.2926(36)(18)$  \\ 
$\,0.30274\,$  	 & $\, 0.3524(20)\, $   & $\, 0.3510(38)\, $  & $\, 0.3489(23)\, $  & $\, 0.3508(28)(18)$  \\ 
$\,0.30400\,$  	 & $\, 0.4042(14)\, $   & $\, 0.4030(25)\, $  & $\, 0.4018(15)\, $  & $\, 0.4030(19)(12)$  \\ 
$\,0.30570\,$  	 & $\, 0.1964(10)\, $   & $\, 0.1971(16)\, $  & $\, 0.1940(25)\, $  & $\, 0.1958(18)(16)$  \\ 
$\,0.30680\,$  	 & $\, 0.2633(42)\, $   & $\, 0.2568(20)\, $  & $\, 0.2552(30)\, $  & $\, 0.2584(32)(41)$  \\ 
$\,0.30780\,$  	 & $\, 0.3130(17)\, $   & $\, 0.3110(14)\, $  & $\, 0.3087(7) \, $  & $\, 0.3109(13)(22)$  \\ 
$\,0.30890\,$  	 & $\, 0.3589(17)\, $   & $\, 0.3568(3) \, $  & $\, 0.3552(10)\, $  & $\, 0.3570(12)(19)$  \\ 
$\,0.31040\,$  	 & $\, 0.4145(8) \, $   & $\, 0.4139(14)\, $  & $\, 0.4105(15)\, $  & $\, 0.4130(13)(20)$  \\ 
}
{tab:ResultOfUpperHiggsMassFiniteVolExtrapolation1}
{The results of the infinite volume extrapolations of the Monte-Carlo data for the renormalized vev $v_r$ and the Higgs 
boson mass $m_{Hp}$ are presented as obtained from the parabolic ansatz in \eq{eq:ParaFit} and the linear approach in 
\eq{eq:LinFit} for the considered threshold values $L_s'=16$ and $L_s'=20$. 
The final results on $v_r$ and $m_{Hp}$, displayed in the very right column, are determined here by averaging over
the parabolic and the two linear fit approaches. An additional, systematic uncertainty of these final results is specified
in the second pair of brackets taken from the largest observed deviation among all respective fit results.}
{Infinite volume extrapolation of the Monte-Carlo data for the renormalized vev $v_r$ and the Higgs 
boson mass $m_{Hp}$ at infinite bare quartic coupling constant.}

To obtain the desired upper Higgs boson mass bounds $m_H^{up}(\Lambda)$ these finite volume results have to be extrapolated 
to the infinite volume limit and the renormalization factor $Z_G$ has to be properly considered. For that purpose the finite 
volume dependence of the Monte-Carlo results on the renormalized vev $v_r=v/\sqrt{Z_G}$ and the Higgs boson mass $m_{Hp}$, 
as obtained for the two scenarios of the full Higgs-Yukawa model and the pure $\Phi^4$-theory, is explicitly shown in 
\fig{fig:FiniteVolumeEffectsOfUpperHiggsMassBoundVEVandMH}. One sees in these plots that the finite volume effects are 
rather mild at the largest investigated hopping parameters $\kappa$ corresponding to the lowest considered values of 
the cutoff $\Lambda$, while the renormalized vev as well as the Higgs boson mass itself vary strongly with increasing 
lattice size $L_s$ at the smaller presented hopping parameters, as expected.

The aforementioned infinite volume extrapolation is then performed by applying the same strategy as in 
\sect{sec:SubLowerHiggsMassboundsDegenCase}. For that purpose the finite volume lattice data are fitted with the linear 
and the parabolic fit approaches given in \eq{eq:LinFit} and \eq{eq:ParaFit}, respectively. The resulting fits 
are displayed in \fig{fig:FiniteVolumeEffectsOfUpperHiggsMassBoundVEVandMH}. According to the smaller number 
of investigated lattice volumes as compared to the situation in \sect{sec:SubLowerHiggsMassboundsDegenCase}, only
two different threshold values $L'_s$ have been considered here. The respective fit parameters are listed in 
\tab{tab:ResultOfUpperHiggsMassFiniteVolExtrapolation1}, where the presented final infinite volume results have been 
obtained by averaging over the parabolic fit and the two linear fits. Again, an additional systematic uncertainty is
specified, which has been derived from the deviation between the latter fit results.

%\includeFigDouble{InfiniteVolumeExtrapolationUpperBoundMH}{InfiniteVolumeExtrapolationUpperBoundMHHighLambdaExtrapolation}
\includeFigDouble{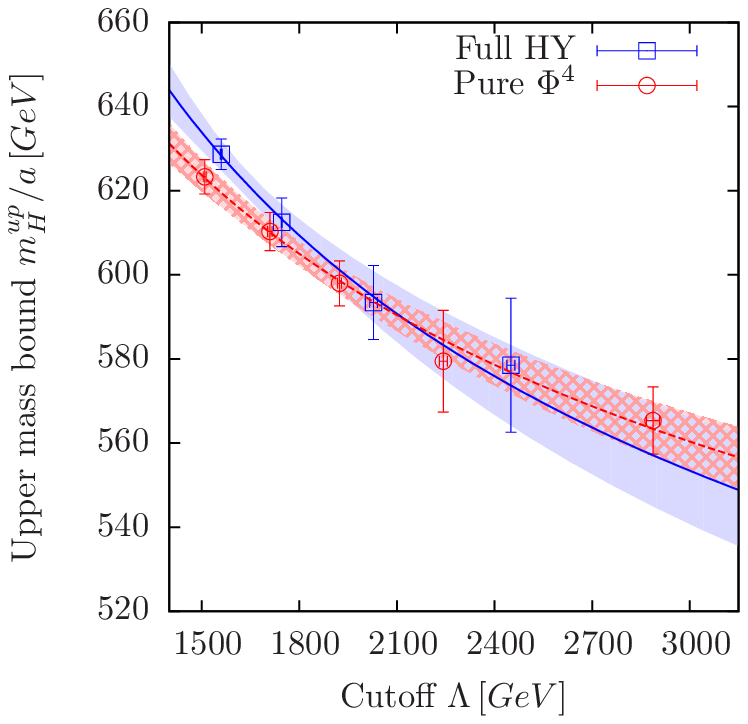}{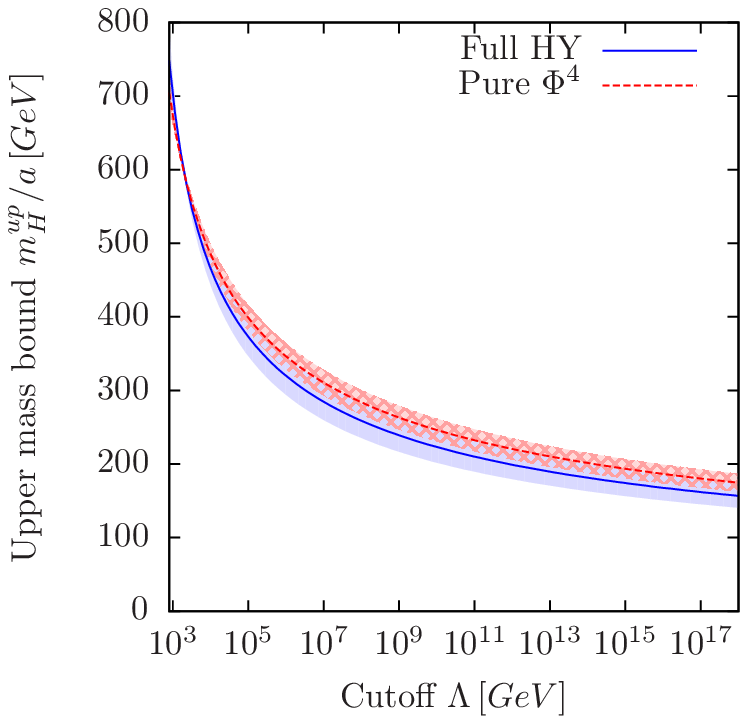}
{fig:UpperMassBoundFinalResult}
{The cutoff dependence of the upper Higgs boson mass bound is presented in panel (a) as obtained from the infinite volume extrapolation results
in \tab{tab:ResultOfUpperHiggsMassFiniteVolExtrapolation1}. The dashed and solid curves are fits of the data arising from the full Higgs-Yukawa model
and the pure $\Phi^4$-theory, respectively, with the analytically expected cutoff dependence in \eq{eq:StrongCouplingLambdaScalingBeaviourMass}.
Panel (b) shows the aforementioned fit curves extrapolated to larger values of the cutoff $\Lambda$. In both panels the highlighted bands 
reflect the uncertainty of the respective fit curves.
}
{Cutoff dependence of the upper Higgs boson mass bound.}

The sought-after cutoff dependent upper Higgs boson mass bound, and thus the main result of the current chapter, can then directly
be obtained from the latter infinite volume extrapolation. The bounds arising in the full Higgs-Yukawa model and the pure
$\Phi^4$-theory are jointly presented in \fig{fig:UpperMassBoundFinalResult}a. In both cases one clearly observes 
the expected decrease of the upper Higgs boson mass bound with rising cutoff $\Lambda$. Moreover, the obtained results
can very well be fitted with the expected cutoff dependence given in \eq{eq:StrongCouplingLambdaScalingBeaviourMass}, as 
depicted by the dashed and solid curves in \fig{fig:UpperMassBoundFinalResult}a, where $A_m$, $B_m$ are the respective free 
fit parameters.

Concerning the effect of the fermionic dynamics on the upper Higgs boson mass bound one finds in \fig{fig:UpperMassBoundFinalResult}
that the individual results on the Higgs boson mass in the full Higgs-Yukawa model and the pure $\Phi^4$-theory at single cutoff 
values $\Lambda$ are not clearly distinguishable from each other with respect to the associated uncertainties. Respecting all 
presented data simultaneously by considering the aforementioned fit curves also does not lead to a much clearer picture, as can be 
observed in \fig{fig:UpperMassBoundFinalResult}, where the uncertainties of the respective fit curves are indicated by the highlighted 
bands. At most, one can infer a mild suggestion from the presented results, being that the inclusion of the fermion dynamics causes 
a somewhat steeper descent of the upper Higgs boson mass bound with increasing cutoff $\Lambda$. A definite answer regarding the 
latter effect, however, remains missing here due to the large statistical uncertainties. The clarification of this issue would require 
the consideration of higher statistics as well as the evaluation of more different lattice volumes to improve the reliability of the 
above infinite volume extrapolations.

On the basis of the latter fit results one can extrapolate the presented fit curves to very large values of the 
cutoff $\Lambda$ as illustrated in \fig{fig:UpperMassBoundFinalResult}b. It is intriguing to compare these large cutoff 
extrapolations to the results arising from the consideration of the Landau pole presented in \fig{fig:PerturbativeHiggsMassBounds}. 
One observes good agreement with that perturbatively obtained upper mass bound even though the here presented data have been calculated 
in the mass degenerate case and for $N_f=1$. This, however, is not too surprising according to the observed relatively mild dependence 
of the upper mass bound on the fermion dynamics. For clarification it is remarked that a direct comparison between the 
aforementioned perturbative and numerical results is non-trivial due to the different underlying regularization schemes as already 
discussed in \sect{sec:SubLowerHiggsMassboundsDegenCase}. With growing values of $\Lambda$, however, the cutoff dependence becomes 
less prominent, thus rendering such a direct comparison increasingly reasonable in that limit. 

%\includeFigTriple{HiggsMassVsCutoffAtInfiniteCouplingFiniteVolumeEffectsRenLambdaCoup}{HiggsMassVsCutoffAtInfiniteCouplingFiniteVolumeEffectsRenLambdaCoupPurePhi4}{HiggsMassVsCutoffAtInfiniteCouplingFiniteVolumeEffectsTopMass}
\includeFigTriple{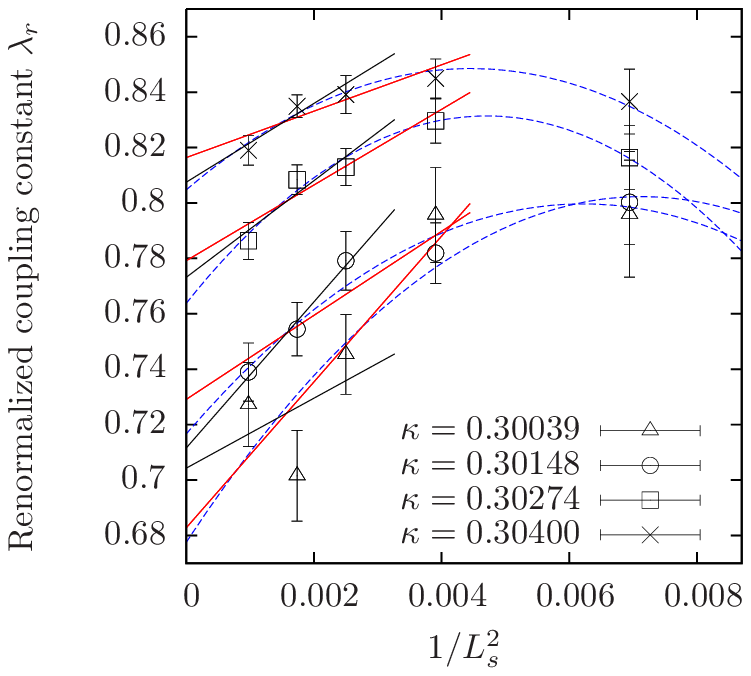}{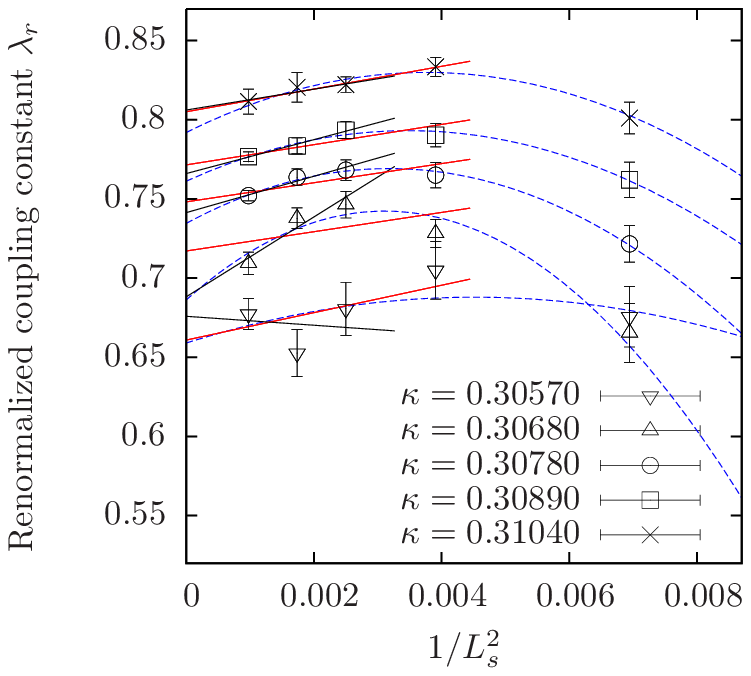}{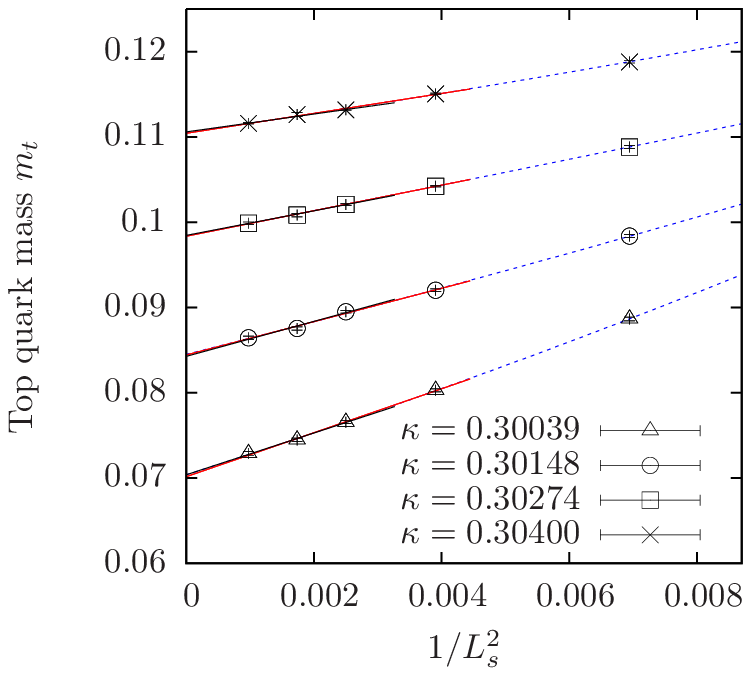}
{fig:FiniteVolumeEffectsOfUpperHiggsMassBoundLamR}
{The dependence of the renormalized quartic coupling constant $\lambda_r$ as well as the top quark mass $m_t$
on the squared inverse lattice side length $1/L^2_s$ 
is presented as calculated in the direct Monte-Carlo calculations specified in \tab{tab:SummaryOfParametersForUpperHiggsMassBoundRuns}.
Panels (a) and (c) show the results for the full Higgs-Yukawa model, while panel (b) refers to the pure $\Phi^4$-theory.
In all plots the dashed curves display the parabolic fits according to the fit ansatz in 
\eq{eq:ParaFit}, while the solid lines depict the linear fits resulting from \eq{eq:LinFit} for the two threshold 
values $L_s'=16$ (red), and $L_s'=20$ (black).
}
{Dependence of the renormalized quartic coupling constant $\lambda_r$ and the top quark mass $m_t$
on the squared inverse lattice side length $1/L^2_s$ at infinite bare quartic coupling constant.}

It is further remarked that the numerical determination of the presented cutoff dependence of the Higgs boson mass $m_{Hp}/a$
in physical units is rather demanding from a numerical point of view, since the vev $v$ as well as the mass $m_{Hp}$ both scale 
proportional to $\sqrt{\kappa-\kappa_c}$ to lowest order in $\kappa$ according to \eq{eq:ScalingBehOfVeV} and \eq{eq:ScalingBehOfMass}. 
This leading contribution thus cancels exactly when calculating the Higgs boson mass in physical units, \ie $m_{Hp}/a$, which 
would lead to a cutoff-independent Higgs boson mass, if only leading terms would be resolvable. It is only the subleading 
logarithmic contribution to the vev $v$ and the mass $m_{Hp}$ that yield the scaling behaviour presented in 
\eq{eq:StrongCouplingLambdaScalingBeaviourMass}. The numerical data thus have to be precise enough to resolve the logarithmic 
contributions in order to observe the correct scaling behaviour. This is a demanding task from a numerical point of view, which 
could, however, satisfactorily be solved with relative errors being below the $1\%$-level in the best cases thanks to the algorithmic 
improvements presented in \chap{chap:SimAlgo}. 

Finally, the question for the cutoff dependence of the renormalized quartic coupling constant $\lambda_r$ and -- in the case
of the full Higgs-Yukawa model -- the top quark mass with its associated value of the renormalized Yukawa coupling constant $y_{t,r}$
shall be addressed. For that purpose we follow exactly the same steps as above. The underlying finite volume lattice results on 
the renormalized quartic coupling constant and the top quark mass are fitted again with the parabolic and the linear fit 
approaches in \eq{eq:LinFit} and \eq{eq:ParaFit} as presented in \fig{fig:FiniteVolumeEffectsOfUpperHiggsMassBoundLamR}.

\includeTabHERE{|c|c|c|c|c|}{
\multicolumn{5}{|c|}{Renormalized quartic coupling constant $\lambda_r$} \\ \hline
$\kappa$  	 & $A^{(l)}_\lambda$, $L'_s=16$  & $A^{(l)}_\lambda$, $L'_s=20$         & $A^{(p)}_\lambda$         & $\lambda_r$               \\ \hline
$\,0.30039\,$  	 & $\,0.6827(280)\, $   & $\,0.7043(460)\, $  & $\,0.6775(452)\, $  & $\, 0.6882(406)(134)$  \\ 
$\,0.30148\,$  	 & $\,0.7291(118)\, $   & $\,0.7116(66) \, $  & $\,0.7166(134)\, $  & $\, 0.7191(110)(88)$  \\ 
$\,0.30274\,$  	 & $\,0.7791(79) \, $   & $\,0.7731(139)\, $  & $\,0.7638(81) \, $  & $\, 0.7720(103)(77)$  \\ 
$\,0.30400\,$  	 & $\,0.8164(71) \, $   & $\,0.8074(97) \, $  & $\,0.8047(67) \, $  & $\, 0.8095(79)(59)$  \\ 
$\,0.30570\,$  	 & $\,0.6609(182)\, $   & $\,0.6760(288)\, $  & $\,0.6590(288)\, $  & $\, 0.6653(258)(85)$  \\ 
$\,0.30680\,$  	 & $\,0.7171(201)\, $   & $\,0.6882(149)\, $  & $\,0.6862(182)\, $  & $\, 0.6972(179)(155)$  \\ 
$\,0.30780\,$  	 & $\,0.7482(56) \, $   & $\,0.7414(37) \, $  & $\,0.7346(24) \, $  & $\, 0.7414(41)(68)$  \\ 
$\,0.30890\,$  	 & $\,0.7716(47) \, $   & $\,0.7660(17) \, $  & $\,0.7612(34) \, $  & $\, 0.7663(35)(52)$  \\ 
$\,0.31040\,$  	 & $\,0.8051(23) \, $   & $\,0.8061(45) \, $  & $\,0.7919(88) \, $  & $\, 0.8010(59)(71)$  \\  \hline
\multicolumn{5}{|c|}{Top quark mass $m_{t}$} \\ \hline
$\kappa$  	 & $A^{(l)}_t$, $L'_s=16$  & $A^{(l)}_t$, $L'_s=20$         & $A^{(p)}_t$         & $m_{t}$               \\ \hline
$\,0.30039\,$  	 & $\, 0.0701(2) \, $   & $\, 0.0704(4) \, $  & $\, 0.0704(3) \, $  & $\, 0.0703(3)(2)$  \\ 
$\,0.30148\,$  	 & $\, 0.0844(3) \, $   & $\, 0.0843(6) \, $  & $\, 0.0845(4) \, $  & $\, 0.0844(5)(1)$  \\ 
$\,0.30274\,$  	 & $\, 0.0983(1) \, $   & $\, 0.0984(2) \, $  & $\, 0.0984(1) \, $  & $\, 0.0984(1)(1)$  \\ 
$\,0.30400\,$  	 & $\, 0.1104(1) \, $   & $\, 0.1106(1) \, $  & $\, 0.1105(2) \, $  & $\, 0.1105(1)(1)$  \\ 
}
{tab:ResultOfUpperHiggsMassFiniteVolExtrapolation2}
{The results of the infinite volume extrapolations of the Monte-Carlo data for the renormalized quartic coupling
constant $\lambda_r$ and the top quark mass $m_{t}$ are presented as obtained from the parabolic ansatz in 
\eq{eq:ParaFit} and the linear approach in \eq{eq:LinFit} for the considered threshold values $L_s'=16$ and $L_s'=20$. 
The final results on $\lambda_r$ and $m_{t}$, displayed in the very right column, are determined here by averaging over
the parabolic and the two linear fit approaches. An additional, systematic uncertainty of these final results is specified
in the second pair of brackets taken from the largest observed deviation among all respective fit results.}
{Infinite volume extrapolation of the Monte-Carlo data for the renormalized quartic coupling
constant $\lambda_r$ and the top quark mass $m_{t}$ at infinite quartic coupling constant.}

The corresponding infinite volume extrapolations are listed in \tab{tab:ResultOfUpperHiggsMassFiniteVolExtrapolation2},
where the final extrapolation result is obtained by averaging over all performed fit approaches, and an additional systematic
error is again estimated from the deviations between the various fit procedures.

The sought-after cutoff dependence of the aforementioned renormalized coupling constants can then directly be obtained
from the latter infinite volume extrapolations. The respective results are presented in \fig{fig:FinalResultsOnRenCoupAtStrongCoup} and
one observes the renormalized coupling parameters to decrease with growing cutoff $\Lambda$ as expected in a trivial theory.
Again, the obtained numerical results are fitted with the analytically expected scaling behaviour given in 
\eq{eq:StrongCouplingLambdaScalingBeaviourLamCoupling} and \eq{eq:StrongCouplingLambdaScalingBeaviourYCoupling}. As already 
discussed for the case of the Higgs boson mass determination, the individual measurements of $\lambda_r$ in the two considered 
models at single cutoff values $\Lambda$ are not clearly distinguishable. Respecting the available data simultaneously by means of 
the aforementioned fit procedures also leads at most to the mild suggestion that the inclusion of the fermion dynamics results in a 
somewhat steeper descent of the renormalized quartic coupling constant with rising cutoff $\Lambda$ as compared to the pure $\Phi^4$-theory. 
A definite conclusion in this matter, however, cannot be drawn at this point due to the large statistical uncertainties 
encountered in \fig{fig:FinalResultsOnRenCoupAtStrongCoup}.

%\includeFigDouble{InfiniteVolumeExtrapolationUpperBoundLamRen}{InfiniteVolumeExtrapolationUpperBoundRenY}
\includeFigDouble{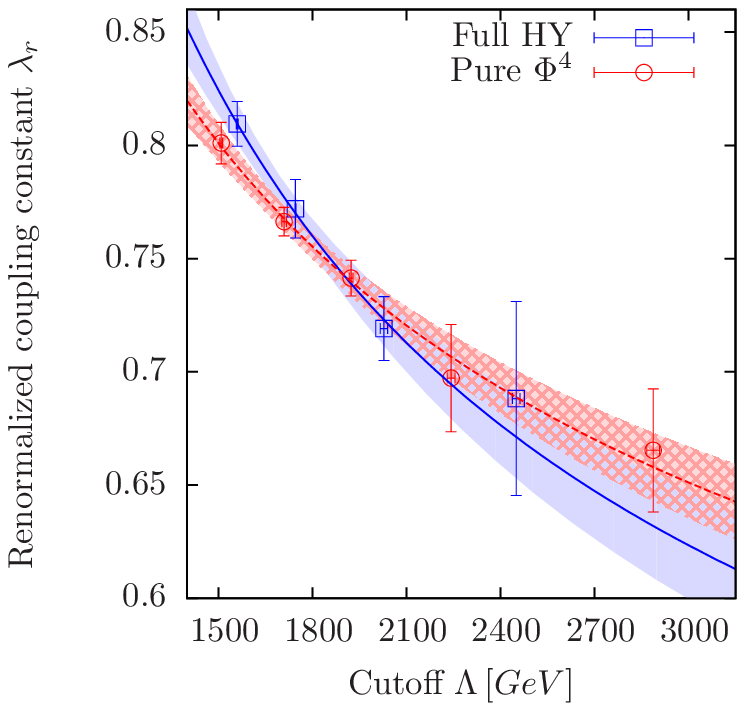}{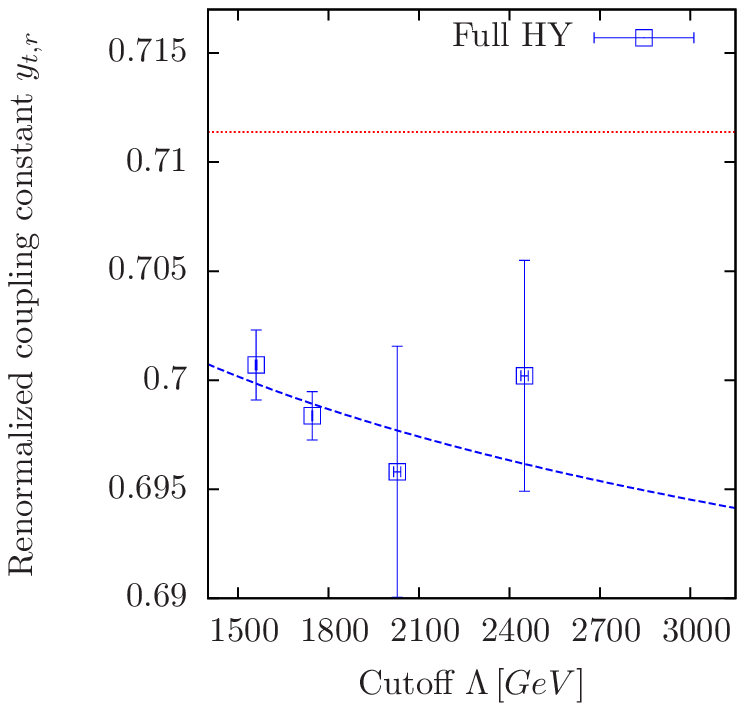}
{fig:FinalResultsOnRenCoupAtStrongCoup}
{The cutoff dependence of the renormalized quartic and Yukawa coupling constants is presented in panels (a) and (b), respectively,
as obtained from the infinite volume extrapolation results in \tab{tab:ResultOfUpperHiggsMassFiniteVolExtrapolation2}. The dashed 
and solid curves are fits with the respective analytically expected cutoff dependence in \eq{eq:StrongCouplingLambdaScalingBeaviourLamCoupling} 
and \eq{eq:StrongCouplingLambdaScalingBeaviourYCoupling}. The horizontal line in panel (b) indicates the bare degenerate Yukawa coupling 
constant underlying the performed lattice calculations. 
}
{Cutoff dependence of the renormalized quartic constant and the renormalized Yukawa coupling constant at infinite bare 
quartic coupling constant.}

Furthermore, the renormalized Yukawa coupling constant is compared to its bare counterpart depicted by the horizontal
line in \fig{fig:FinalResultsOnRenCoupAtStrongCoup}b. Since the latter bare quantity was chosen according to the tree-level
relation in \eq{eq:treeLevelTopMass} aiming at the reproduction of the physical top quark mass, one can directly infer from this
presentation how much the actually measured top quark mass differs from its targeted value of $\GEV{175}$. Here, one observes
a significant discrepancy of up to $\proz{2}$, which can in principle be fixed in follow-up lattice calculations, if
desired. According to the observed rather weak dependence of the upper Higgs boson mass bound on the Yukawa coupling constants,
however, such an adjustment would most likely not even be resolvable with the here achieved accuracy. The same rationale also
constitutes an additional reason why it was not tried to proceed to the investigation of the physically more relevant 
scenario of non-degenerate Yukawa coupling constants. 

As a concluding remark it is summarized that the upper Higgs boson mass bound in the full Higgs-Yukawa model is very close
to the one that arises in the pure $\Phi^4$-theory. With the here available statistics a difference between the latter two mass 
bounds could not clearly be resolved. It is only mildly suggested by the performed fit procedures that the descent of the upper 
Higgs boson mass bound with increasing cutoff $\Lambda$ may be somewhat steeper in the case of the full Higgs-Yukawa model than 
in the pure $\Phi^4$-theory. For a definite answer in this matter, however, higher statistics would be required. 

  \chapter{First results on the decay width of the Higgs boson}
\label{chap:ResOnDecayWidth}

As a last consideration in the present work, only meant here as a brief and preliminary outlook towards some
possible future developments that could be targeted in a further investigation of the underlying Higgs-Yukawa 
model, we will revisit the question for the decay properties of the Higgs boson in this chapter, having in mind
that the determination of the decay width through the analysis of the Higgs propagator as presented in 
\sect{sec:HiggsPropAnalysisUpperBound} did not yield satisfactory results. For that purpose L\"uscher's well-known 
method~\cite{Luscher:1990ux,Luscher:1991cf} for extracting the scattering phase shifts, and thus the decay width, 
from some finite volume lattice calculations shall be applied to the considered Higgs-Yukawa model.
Subsequently, we will begin with a brief summary of the main concepts underlying the latter method, assuming the reader 
to be familiar with the basic theoretical background of scattering processes\footnote{An introduction to 
this subject can be found, for instance, in \Ref{Newton:2002zu,Taylor:2006zu}. For the proofs and further details of 
L\"uscher's method the reader is referred to \Ref{Luscher:1990ux,Luscher:1991cf}.} in general. 

To discuss the main ideas of L\"uscher's approach we first introduce an abstract system of two identical bosons 
with mass $m_\pi$ in a cubic box of continuous three-dimensional space with finite volume $L^3$. L\"uscher's crucial 
observation has then been that the energy spectrum\footnote{Here and in the following we will only consider the
center-of-mass frame, such that 'energy' always means 'center-of-mass energy' in this section.}
in the considered {\textit{finite}} volume $L^3$ is connected to the scattering 
phase shifts $\delta_l$ of the two-boson system in {\textit{infinite}} volume, where $l\in \N_0$ denotes the angular 
momentum quantum number. To elaborate more on the latter statement we introduce the center-of-mass
energy $E$ of the two-boson system in the finite box $L^3$. Moreover, the squared relative momentum
$k^2$ in the center-of-mass frame is defined according to 
\bea
\label{eq:RelOfKandW}
E &=& \sqrt{(2m_\pi)^2 + (2k)^2}.
\eea
It is remarked that this quantity is not restrained to the discrete set of finite volume momenta. In fact, $k$ is 
the relative momentum\footnote{With some abuse of notation the actual vector character of $k$ is omitted here.}
in the center-of-mass frame that one would observe for a given center-of-mass energy $E$ in 
the case of an infinitely large volume. This definition of the relative momentum $k$ has been established here, 
since the later given relation between $E$ and the scattering phase shifts $\delta_l$ actually connects $E$ with 
$\delta_l(k)$, \ie the infinite volume scattering phase shifts at the relative momentum $k$.

Furthermore, it is already emphasized at this point that the relations given in the following are only valid in the elastic 
scattering regime, that is for
\beq
\label{eq:ElasticRegime}
2m_\pi <E < 4m_\pi.
\eeq
 
The next crucial observation is that the announced relation between the scattering phase shifts $\delta_l$ and the 
observable energies $E$ is greatly restrained by symmetry arguments. For clarification we 
consider the eigenstate $\Psi$ associated to a given energy eigenvalue $E$. If $\Psi$ is invariant
under an adequately chosen representation of a symmetry group, one can directly exclude all those phase shifts $\delta_l$ from
the announced relation to $E$, the associated wave function of which, given in terms of the respective Legendre
polynomials, is not invariant under the latter representation\footnote{For the here omitted details, the reader is 
referred to \Ref{Luscher:1990ux,Luscher:1991cf}.}.

The symmetry group in the here considered scenario is the cubic group $O(3,\Z)$ according to the continuous rotational
symmetry being broken by the finite volume $L^3$. The trivial representation of this group, which is denoted as $A_1^+$,
simply performs a discrete cubic rotation on a singlet wave function. As a consequence of the Legendre polynomials,
more precisely the associated wave functions, being only invariant under $A_1^+$ for $l=0,4,6,\ldots$ the phase shifts 
$\delta_1(k)$, $\delta_2(k)$, $\delta_3(k)$, $\delta_5(k)$, $\ldots$ can not contribute to the energy $E$, 
provided that the underlying state $\Psi$ is invariant under $A_1^+$. Similar statements can also be established for other 
representations of the considered symmetry group $O(3,\Z)$. 

Moreover, it is a general result in scattering theory, known as threshold behaviour~\cite{Sakurai:1993zu}, that the phase 
shifts $\delta_l(k)$ are suppressed at low momentum $k$ according to $\delta_l(k)\propto k^{2l+1}$. Combining this general 
observation with the aforementioned exact symmetry constraints, one finds that the announced relation between $E$ and $\delta_l(k)$ 
actually reduces to a relation between $E$ and $\delta_0(k)$ with the next to leading order contribution given by 
$\delta_4(k)\propto k^{9}$ at sufficiently small values of $k$, provided that the underlying state $\Psi$ is invariant under $A_1^+$. 
Analogously, it would reduce to a relation between $E$ and $\delta_1(k)+O(k^{7})$, if $\Psi$ was invariant under the vector 
representation $T_1^-$, thus allowing to access also this scattering phase shift in a practical calculation, provided one 
can study the center-of-mass energies $E$ of states obeying that symmetry~\cite{Gockeler:1994rx}. 

Here, however, we will only be interested in the scattering phase shift $\delta_0(k)$ at zero angular momentum, \ie $l=0$.
The corresponding relation to the energy $E$ has been worked out in detail in \Ref{Luscher:1990ux}. For this 
scenario and up to the aforementioned corrections it is given as 
\bea
\label{eq:LuschersMainRelation}
\delta_0(k) &=& -\omega\left(q\right) \mbox{ modulo } \pi, \\
q &=& \frac{kL}{2\pi},
\eea
where $k$ is obtained for a given finite volume center-of-mass energy $E$ according to \eq{eq:RelOfKandW} and the function
$\omega(q)$ is defined\footnote{This relation defines $\omega(q)$ only up to an integer multiple of $\pi$. This ambiguity, 
however, is canceled in \eq{eq:LuschersMainRelation} by the modulo operation and is therefore ignored here.} as
\bea
\tan(-\omega(q)) &=& \frac{q\pi^{3/2}}{\zetaF(1,q^2)},
\eea
with $\zetaF(1,q^2)$ denoting the zeta function according to
\bea
\label{eq:DefOfZetaFunction}
\zetaF(s,q^2) &=& \frac{1}{\sqrt{4\pi}}  \sum\limits_{\vec n\in \Z^3} \left(\vec n^2 - q^2  \right)^{-s}.
\eea

The physical interpretation of \eq{eq:LuschersMainRelation} is that the center-of-mass energies $E$ which are observable
in a finite volume $L^3$ are constrained to a discrete set of energy values, determined by the condition that the respectively
associated relative momentum $k$ solves the presented relation in \eq{eq:LuschersMainRelation} for a given infinite volume
phase shift $\delta_0(k)$. Conversely, the given relation can be used to extract the aforementioned infinite volume scattering 
phase shift, once the energy levels $E$ observable in the finite volume $L^3$ are known, for instance by means of direct 
lattice calculations. It is pointed out, that the finite volume underlying such lattice calculations is not a disadvantage here that
has to be overcome by some sort of infinite volume extrapolation. Instead, it is exactly the finiteness of the considered volume
$L^3$ that is in fact exploited in \eq{eq:LuschersMainRelation} to derive the infinite volume scattering phase shifts in this
approach. One has, however, to make sure that discretization effects are negligible in the intended lattice calculations, since
the presented relations have been derived in the continuum.

It is further remarked that the definition of the zeta-function in \eq{eq:DefOfZetaFunction} is only meaningful for sufficiently 
large real parts of $s$, \ie $\RE(s)>3/2$. At smaller real parts the zeta-function is instead defined through an analytic 
continuation of \eq{eq:DefOfZetaFunction}. For the calculation of its value at $s=1$ an integral representation has been
devised in \Ref{Luscher:1990ux} according to 
\bea
\sqrt{4 \pi} \zetaF(1, q^2) &=& (2 \pi)^3 \int\limits_0^1 \intd{t} \Big [ (4 \pi t)^{-\frac{3}{2}} (e^{t q^2} - 1) +
(4 \pi t)^{-\frac{3}{2}} e^{t q^2} \sum\limits_{\vec{n} \neq \vec{0}} e^{-\frac{\pi^2}{t} \vec{n}^2}  \Big]  \\
&+&\sum_{\vec{n} \in \Z^3} \frac{e^{q^2 - \vec{n}^2}}{\vec{n}^2 - q^2} - 2 \pi^{\frac{3}{2}}, \nonumber
\eea
which is well suited for a numerical evaluation. With the help of this expression
it is straightforward to determine the phase shift $\delta_0(k)$ according to \eq{eq:LuschersMainRelation} at the
momenta $k$ determined through \eq{eq:RelOfKandW}, once some center-of-mass energies $E$ have been computed for some
finite volumes $L^3$, such that the underlying states $\Psi$ are invariant under the symmetry group representation $A_1^+$.
This is the strategy, mainly taken over from \Ref{Gockeler:1994rx}, that will be applied in the following.

Before we continue, however, with the determination of the aforementioned energy levels $E$, the question for
the applicability of the discussed approach in the context of the here investigated Higgs-Yukawa model has to be 
addressed, since the above considerations were performed within a system of two identical bosons only, not respecting
the existence of other particles such as the top and bottom quarks, for instance. However, the fermion dynamics can completely 
be integrated out, leading then to a solely bosonic system with a more complicated, effective coupling structure. Since 
no particular form of the underlying interaction has been assumed in \Ref{Luscher:1990ux} for the derivation of 
\eq{eq:LuschersMainRelation}, it thus follows that the so far given relations concerning the connection between the
infinite volume scattering phase shifts and the bosonic, finite volume energy spectrum also hold in the considered Higgs-Yukawa model, 
respecting then also the fermionic contributions through the aforementioned effective coupling structure. In this picture 
the Goldstone bosons in the latter model can then be identified as the above introduced abstract bosonic particles with 
mass $m_\pi$, while the Higgs boson would be considered as a resonance in that system~\cite{Gockeler:1994rx}. 
 
It is further remarked that the so far stated relations have been obtained for the case of massive bosons, \ie for $m_\pi>0$.
The Goldstone bosons in the considered Higgs-Yukawa model, however, would be massless. To overcome this problem, an
external, space-time independent current $J\neq 0$ is explicitly introduced into the Higgs-Yukawa model in the spirit 
discussed in \sect{sec:FuncFormInEucTime}. The extended form of the bosonic action is then given as
\bea
S_{\Phi,J}[\Phi] &=& S_\Phi[\Phi] + J^0\sum\limits_x \Phi_x^0,
\eea
where the latter current $J$ is assumed here and in the following to point into the 0-direction of the four component 
field $\Phi$. As a consequence the Goldstone bosons acquire a mass $m_G\equiv m_\pi>0$, thus finally establishing the 
applicability of the above discussed relations also for the case of the considered Higgs-Yukawa model. 

For the purpose of determining the sought-after scattering phase shifts $\delta_0(k)$, and thus eventually also the  
width of the Higgs resonance, a series of lattice calculations has been performed with varying physical lattice volume 
as listed in \tab{tab:SummaryOfParametersForDecayWidth}. In these runs all model parameters were kept constant, while 
only the nominal number of lattice sites was varied, leading to an approximately constant physical scale $a^{-1}$ and 
thus to a variation of the actual physical volume. The external current $J$ has been chosen such that 
the expected Higgs resonance mass lies in the elastic scattering region of the two Goldstone bosons according to 
\eq{eq:ElasticRegime}, while the degenerate Yukawa coupling constants were again fixed by the tree-level relation 
in \eq{eq:treeLevelTopMass} aiming at the reproduction of the top quark mass.
 
\includeTab{|ccccccccc|}
{
$\kappa$ & $L_s$                    & $L_t$ & $N_f$ &  $\Nconf$  & $\hat \lambda$ & $\hat y_t$     & $\hat y_b/\hat y_t$ & $|J|$ \\
\hline
0.13000    & 12  &   12  &  1    & 21440    & 0.01           & 0.36274        & 1.0    & 0.005               \\
0.13000    & 16  &   16  &  1    & 20760    & 0.01           & 0.36274        & 1.0    & 0.005               \\
0.13000    & 20  &   20  &  1    & 22100    & 0.01           & 0.36274        & 1.0    & 0.005               \\
0.13000    & 24  &   24  &  1    & 10880    & 0.01           & 0.36274        & 1.0    & 0.005               \\
0.13000    & 32  &   32  &  1    &  8780    & 0.01           & 0.36274        & 1.0    & 0.005               \\
0.13000    & 36  &   36  &  1    &  4440    & 0.01           & 0.36274        & 1.0    & 0.005               \\
0.13000    & 40  &   40  &  1    &  2360    & 0.01           & 0.36274        & 1.0    & 0.005               \\
}
{tab:SummaryOfParametersForDecayWidth}
{The model parameters of the Monte-Carlo runs underlying the subsequent lattice calculation of the Higgs resonance
are presented. The Yukawa coupling constants have been chosen according to the tree-level relation 
in \eq{eq:treeLevelTopMass} aiming at the reproduction of the phenomenologically known top quark mass. 
The given value of the bare quartic coupling constant $\hat\lambda$ corresponds to a rather small renormalized quartic coupling constant of 
$\lambda_r\approx 0.1$, chosen to allow for a direct comparison with perturbation theory in a regime where it is
clearly applicable. The specified parameter $\kappa$ and the external current $J$ were selected such that the resulting
Higgs resonance lies within the elastic scattering region according to \eq{eq:ElasticRegime}.
}
{Model parameters of the Monte-Carlo runs underlying the presented lattice calculation of the Higgs resonance.}

For clarification it is remarked that we start here with a rather small value of the bare quartic coupling constant $\hat\lambda$
in this exploratory study, which will later be seen to correspond to a renormalized quartic coupling constant of approximately
$\lambda_r \approx 0.1$, being around one order of magnitude smaller than for the case of the upper mass bound considered in 
\sect{sec:UpperMassBounds}. The rationale behind this setting is that it allows for a direct comparison with renormalized 
perturbation theory in a regime where it can perfectly be trusted. Eventually, one is, however, rather interested in the 
case of the maximally possible renormalized quartic coupling constant, given at $\hat \lambda=\infty$. The latter scenario 
would therefore be the prime target of some follow-up investigations.

The remaining step for the determination of the infinite volume scattering phase shift $\delta_0(k)$ is thus the computation
of the center-of-mass energies $E$ observable in the aforementioned finite volume calculations. Since we are here only interested
in the zero angular momentum scatterings phase shift $\delta_0(k)$, we restrict the consideration to states $\Psi$ which
are invariant under the representation $A_1^+$ as discussed above. We therefore consider only correlation functions based on lattice 
observables $O_i$, $i=1,\ldots,N_O$ that obey the latter symmetry such as  
\bea
\label{eq:ScatteringOperator1}
O_1(t) &=& \frac{1}{L_s^3} \sum\limits_{\vec x} \Phi^0_{t,\vec x},   \\
\label{eq:ScatteringOperator2}
O_2(t) &=& \frac{1}{L_s^6} \sum\limits_{\vec x, \vec y} \Phi^0_{t,\vec x}  \Phi^0_{t,\vec y},   \\
\label{eq:ScatteringOperator3}
O_3(t) &=& \frac{1}{3L_s^6} \sum\limits_{\vec x, \vec y} \sum\limits_{\alpha=1}^3 \Phi^\alpha_{t,\vec x}  \Phi^\alpha_{t,\vec y},   \\
\label{eq:ScatteringOperator4}
O_4(t) &=& \frac{1}{|K_1|\cdot L_s^6} \sum\limits_{\vec k'\in K_1}\sum\limits_{\vec x, \vec y} e^{-i\vec k'(\vec x-\vec y)}\cdot \Phi^0_{t,\vec x}  \Phi^0_{t,\vec y},   \\
\label{eq:ScatteringOperator5}
O_5(t) &=& \frac{1}{|K_1|\cdot 3L_s^6} \sum\limits_{\vec k'\in K_1}\sum\limits_{\vec x, \vec y} e^{-i\vec k'(\vec x-\vec y)}\cdot\sum\limits_{\alpha=1}^3 \Phi^\alpha_{t,\vec x}  \Phi^\alpha_{t,\vec y},
\eea
where the set $K_1$ of lattice momenta is defined as
\bea
K_1 &=& \left\{\epsilon\frac{2\pi}{L_s} \hat e_\mu : \quad \epsilon = \pm 1,\, \mu=1,2,3   \right\}
\eea
with $\hat e_\mu$ denoting the unit vector pointing in direction $\mu$ and $|K_1|$ denotes here the number of elements of $K_1$, 
\ie $|K_1|=6$. The operators in \eqs{eq:ScatteringOperator1}{eq:ScatteringOperator5} would then generate states\footnote{When
considering the given expressions in the operator formalism, their application to the vacuum generates a state $\Psi$, which is
invariant under $A_1^+$. As lattice observables they are invariant themselves under an appropriate representation of the symmetry
group corresponding to $A_1^+$.} $\Psi$, that are invariant under $A_1^+$. It is remarked here that the introduction of the external
current $J$ explicitly breaks the original $O(4)$ symmetry of the model. The 0-component of the field $\Phi$ is now directly 
associated to the Higgs mode, while the other three components refer to the Goldstone modes. The given operators $O_2(t),\ldots, O_5(t)$ 
thus generate two particle states with isospin zero and relative lattice momentum $k'=0$ as well as $k' = 4\pi/L_s$, respectively. 
It is further remarked that the operator $O_1(t)$ only consists of a single field variable unlike the other operators. It
is thus not directly a two particle operator. In fact, it directly corresponds to the expression given in \eq{eq:HiggsFieldTSoperator}, 
which was used to determine the Higgs boson mass in \sect{sec:HiggsTSCanalysisLowerBound}. However, the states generated through 
$O_1(t)$ have exactly the same quantum numbers as those resulting from the other introduced operators, such that $O_1(t)$ 
can equally well be considered for our purpose.

We now define the $n\times n$ time-slice correlation matrix $C(\Delta t)$ according to
\bea
\label{eq:DefOfOrigCorrMatrix}
C_{i,j}(\Delta t) &=& \frac{1}{L_t}\sum\limits_{t=0}^{L_t-1} \langle O_i(t+\Delta t) O_j(t)\rangle
-\langle O_i(t+\Delta t)\rangle  \langle O_j(t)\rangle,
\eea
where adequate modulo operations are again implicit to guarantee $0\le t+\Delta t<L_t$. In the operator formalism one
would then have 
\bea
\label{eq:DefOfCorrMatrix}
C_{i,j}(\Delta t) &=& \sum\limits_{n\neq 0} W_{n,i}^* e^{-E_n\Delta t}  W_{n,j}
\eea
with 
\bea
W_{n,i} &=& \langle n | \hat O_i(0) | \Omega\rangle,
\eea
where $E_0=0$ has been assumed without loss of generality, $\Omega\equiv |0\rangle$ is the, in this case non-degenerate, 
vacuum state, the index $n$ runs over all energy eigenstates $|n\rangle$ of the underlying Hamiltonian except the vacuum state, 
$E_n$ denotes the associated energy eigenvalues, and $\hat O_i$ is the counterpart of the lattice observable $O_i$ in the 
operator formalism. If one cuts off the infinite sum in \eq{eq:DefOfCorrMatrix} after the first $N_O$ summands, where 
the energy levels $E_n$ are assumed to be labeled in ascending order and only states $|n\rangle$ with $W_{n,i}\neq 0$ for
at least one operator $\hat O_i$ are counted, one obtains
\bea
C(\Delta t) &=& W^\dagger \diag\left(e^{-E_n\Delta t}  \right) W + O\left( e^{-E_{N_O+1}\Delta t} \right),
\eea
with $W$ and the given diagonal matrix being $N_O\times N_O$-matrices here. This approximation is reasonable for sufficiently large
values of $N_O$ and $\Delta t$, which is assumed to be the case in the following. Subsequently, the given error term
is therefore neglected.

If $W$ was unitary, the eigenvalues $\lambda_\nu(\Delta t)$, $\nu=1,\ldots,N_O$ of the hermitian correlation matrix $C(\Delta t)$ would 
then directly be given as $\exp(-E_1\Delta t),\ldots, \exp(-E_{N_O}\Delta t)$. Conversely, one could determine the energy eigenvalues 
$E_1,\ldots, E_{N_O}$ by computing the eigenvalues of $C(\Delta t)$ without  picking up contaminations by lighter states. For a sufficiently 
large number $N_O$ of independent operators the low lying energy eigenvalues moreover become shielded against contaminations from excited 
states. The correlation matrix analysis would thus allow to extract these low lying energy eigenvalues as well as excited levels with
very well controllable systematic effects. 

The matrix $W$, however, is not unitary in general, such that the eigenvalues of the hermitian correlation matrix $C(\Delta t)$
are in fact some linear combinations of the exponentials listed above. The eigenvalues of $C(\Delta t)$ can nevertheless be
shown~\cite{Luscher:1990ck} to converge to the aforementioned exponential expressions for sufficiently large time separations $\Delta t$,
thus allowing to apply the above approach for the determination of the sought-after energy levels also in the general case, provided
that sufficiently large time separations $\Delta t$ are considered. 

Alternatively, the latter mixture can be resolved by studying the generalized correlation matrix $C'(\Delta t)$ instead 
of $C(\Delta t)$ given as
\bea
\label{eq:DefOfGenCorMatrix}
C'(\Delta t) &=& \left[C(\Delta t_0) \right]^{-1} C(\Delta t)\\
\label{eq:SpectralDecompOfGenC}
&=& W^{-1} \diag\left(e^{-E_n(\Delta t-\Delta t_0)}  \right) W
 + O\left( e^{-E_{N_O+1}\Delta t} \right)
 + O\left( e^{-E_{N_O+1}\Delta t_0} \right). \quad \quad \quad
\eea
Provided that both error terms in \eq{eq:SpectralDecompOfGenC} can be neglected, meaning that $\Delta t$ as well as 
$\Delta t_0$ are sufficiently large, the eigenvalues $\lambda'_\nu(\Delta t)$, $\nu=1,\ldots,N_O$ of the generalized correlation
matrix $C'(\Delta t)$ are given as $\exp(-E_1(\Delta t-\Delta t_0)),\ldots, \exp(-E_{N_O}(\Delta t-\Delta t_0))$, where $\Delta t_0$ 
is some reference time separation, thus directly allowing to read off the sought-after energy levels from the eigenvalues 
$\lambda'_\nu(\Delta t)$. This method is known in the literature as the generalized eigenvalue approach to the analysis 
of the correlation matrix~\cite{Luscher:1990ck}. 

Both approaches have been applied to the available data sets and it was found that the theoretically superior analysis based
on the generalized correlation matrix in \eq{eq:DefOfGenCorMatrix} leads here to larger statistic uncertainties of the extracted
energy levels, rendering the determination of the sought-after decay properties of the Higgs boson less stable. For that reason,
only the first analysis approach on the basis of the initial correlation matrix $C(\Delta t)$ in \eq{eq:DefOfOrigCorrMatrix} 
will be considered in the following.

Examples for the dependence of the eigenvalues $\lambda_\nu(\Delta t)$ on the time separation $\Delta t$ are presented 
in \fig{fig:CorrelationMatrixExamples}. The associated energy eigenvalues $E_1,\ldots, E_{N_O}$ are here extracted by 
fitting the lattice data on each eigenvalue $\lambda_\nu(\Delta t)$ to the single $\cosh$-ansatz given in \eq{eq:HiggsTimeCorrelatorFitAnsatz2}. 
For that purpose one has to assign the measured eigenvalues properly to each other, which can, for instance, be done 
by matching the respective eigenvectors.

%\includeFigTriple{CorrelationMatrixAnalysisNoGenEWL32}{CorrelationMatrixAnalysisNoGenEWL36}{CorrelationMatrixAnalysisNoGenEWL40}
\includeFigTriple{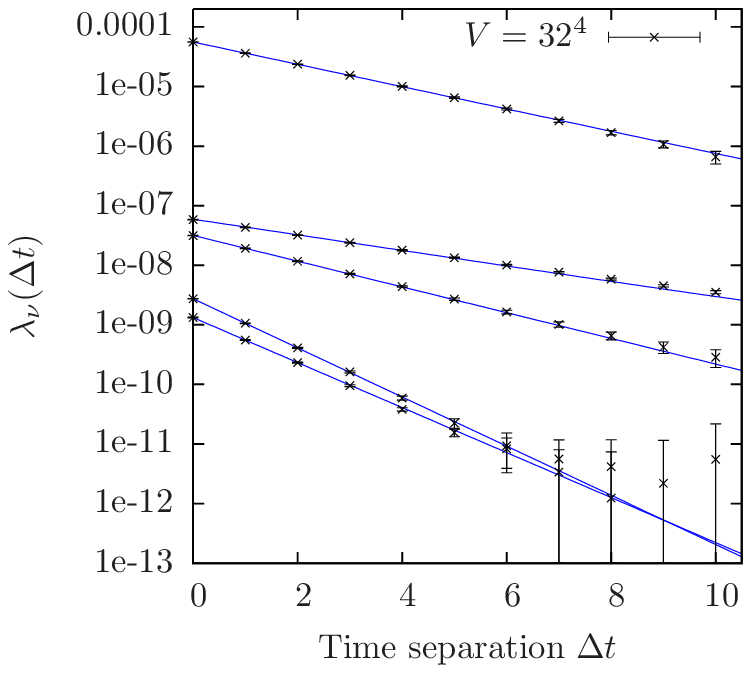}{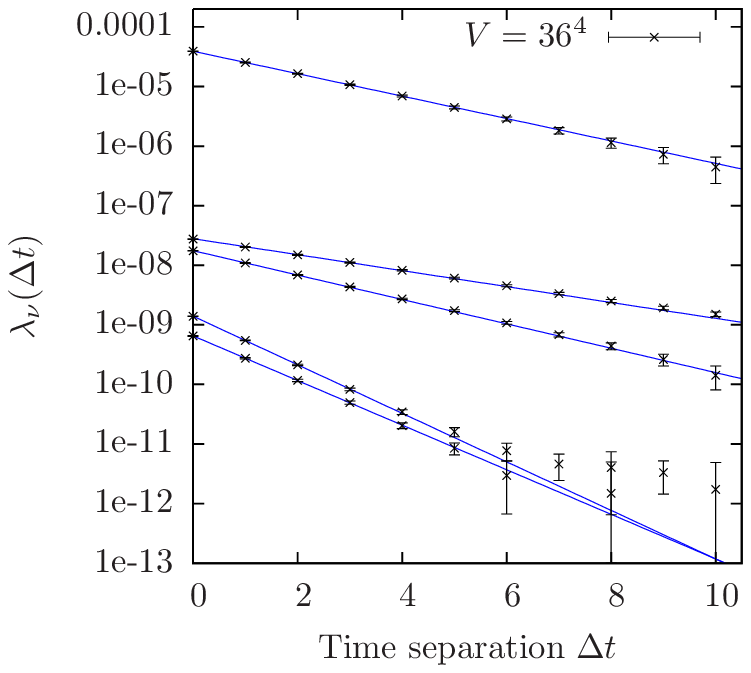}{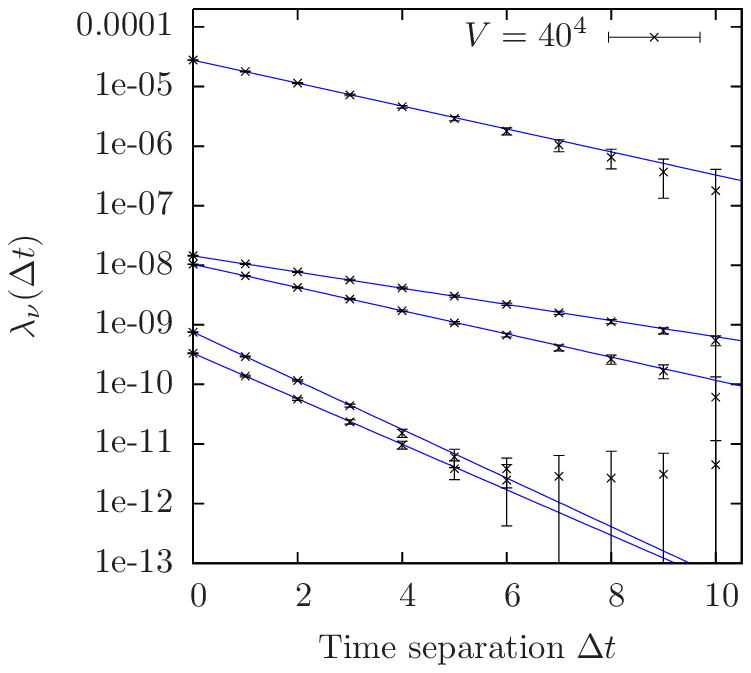}
{fig:CorrelationMatrixExamples}
{As an example, the eigenvalues $\lambda_\nu(\Delta t)$ of the correlation matrix $C(\Delta t)$ as defined in 
\eq{eq:DefOfOrigCorrMatrix} are presented versus the time separation $\Delta t$ in an interval, where the statistical 
uncertainties remain at a reasonable level. The observables underlying the correlation matrix were given in 
\eqs{eq:ScatteringOperator1}{eq:ScatteringOperator5} and the three panels show the respective data for the three largest considered 
lattice volumes $V=32^4$, $V=36^4$, and $V=40^4$. The solid lines depict the fits of the measured eigenvalues resulting from
the single $\cosh$-ansatz in \eq{eq:HiggsTimeCorrelatorFitAnsatz2}.
}
{Examples for the dependence of the eigenvalues $\lambda_\nu(\Delta t)$ of the correlation matrix $C(\Delta t)$ on the 
time separation $\Delta$.}

Analogously, the non-zero Goldstone mass $m_\pi\equiv m_G$ is determined here by applying the above correlation
matrix analysis also to the three Goldstone mode observables
\bea
O_{G_\alpha}(t) &=& \frac{1}{L_s^3} \sum\limits_{\vec x} \Phi^\alpha_{t,\vec x},  \quad \alpha=1,2,3, 
\eea
where again the initial correlation matrix in \eq{eq:DefOfOrigCorrMatrix} is considered due to the degeneracy of 
the Goldstone mass spectrum. The final numerical result on $m_G$ is then determined as the average 
over the resulting three degenerate Goldstone masses.

With the Goldstone mass\footnote{It is remarked that the Goldstone mass was found here to be independent of the lattice
volume with respect to the achieved accuracy.} and the energy eigenvalues $E_n$ at hand as observed in the 
performed finite volume lattice calculations, one can thus finally determine the scattering phase shifts $\delta_0(k)$ at the
associated momenta $k$ according to \eq{eq:RelOfKandW} and \eq{eq:LuschersMainRelation}. For that purpose the consideration
is restricted to only that part of the observed energy spectrum that lies within the elastic region specified
by \eq{eq:ElasticRegime}. The resulting scattering phase shifts are presented in \fig{fig:FirstResultScatteringPhases}.
Since the phase shifts are only specified up to an integer multiple of $\pi$, this freedom has explicitly
been used to present the measured phase shifts $\delta_0(k)$ in a monotonically increasing form.

In order to extract the Higgs resonance width $\Gamma_{Hr}$ from these numerical results on the scattering
phase shifts, we will fit the latter data, more precisely the associated cross sections $\sigma(k)$, with a Breit-Wigner 
function, which is a common approach for the description of resonance phenomena in general.
Such an ansatz is, of course, not expected to describe the actual phase shifts exactly, which is why more elaborate 
fit approaches are used in other studies. Here, however, the achieved accuracy of the lattice data is 
not high enough to clearly resolve such differences. Moreover, the application of a Breit-Wigner fit
provides a model-independent, general approach for the extraction of the decay width, not relying on any additional
information, such as a perturbatively derived fit formula, for instance.
\vs{-1mm}

%\includeFigSingleMedium{ScatteringPhasesAtLam001}
\includeFigSingleMedium{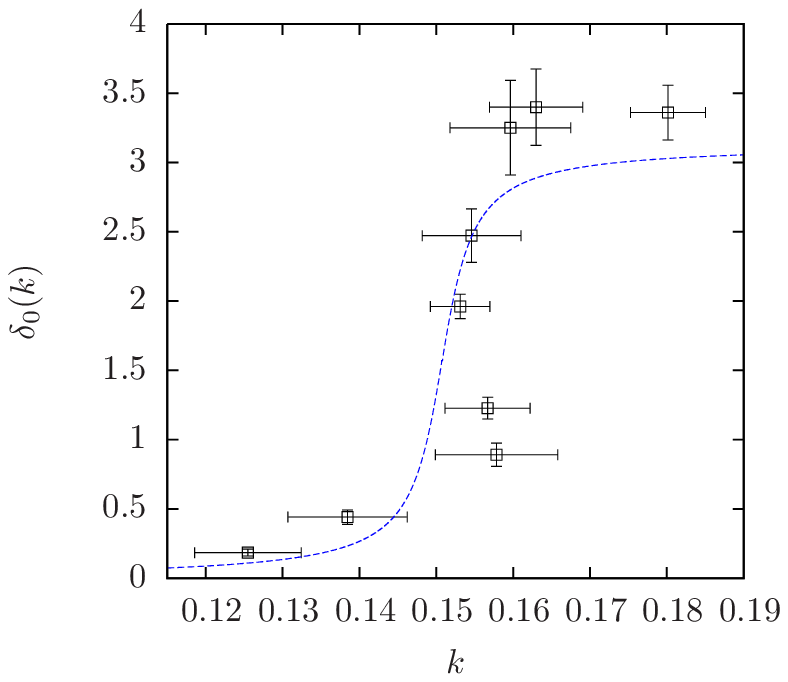}
{fig:FirstResultScatteringPhases}
{The scattering phase shifts $\delta_0(k)$ at zero angular momentum as obtained from \eq{eq:LuschersMainRelation}
are presented versus the momentum $k$ determined through \eq{eq:RelOfKandW} on the basis of the finite volume
energy spectrum observed in the performed lattice calculations specified in \tab{tab:SummaryOfParametersForDecayWidth}.
The dashed curve is a fit according to the ansatz in \eq{eq:FitAnsatzForScatteringPhases}. The indefiniteness of
the phase shifts with respect to integer multiples of $\pi$ has been used to present the data in a monotonically 
ascending form.}
{Scattering phase shifts $\delta_0(k)$ at zero angular momentum at a small quartic coupling constant.}

For that purpose the connection between the scattering phase shifts $\delta_l(k)$ and the cross sections $\sigma(k)$
needs to be known. By virtue of the optical theorem~\cite{Sakurai:1993zu} one has
\bea
\sigma(k) &=& \frac{4\pi}{k^2} \sum\limits_{l\in\N_0} (2l+1) \sin^2(\delta_l(k)) \\
&\approx& \frac{4\pi}{k^2}  \sin^2(\delta_0(k)),
\eea
where all higher angular momentum contributions have been assumed to be negligible, which can again be justified 
according to the aforementioned threshold behaviour at sufficiently small momentum $k$. As motivated above the
resulting cross sections will then be fitted by a relativistic Breit-Wigner function given as
\bea
\label{eq:FitAnsatzForScatteringPhases}
f(E) &=& \frac{A(m_{Hr},m_\pi, \Gamma_{Hr})}{(E^2-m_{Hr}^2)^2 + m_{Hr}^2\Gamma_{Hr}^2},
\eea
where the energy $E\fhs{-0.75mm}\equiv\fhs{-0.75mm} E(k)$ is related to the momentum $k$ by \eq{eq:RelOfKandW} and 
$A(m_{Hr}, m_\pi, \Gamma_{Hr})$ $=$ $16\pi m^2_{Hr}\Gamma_{Hr}^2/(m^2_{Hr}-4m_\pi^2)$ is not a free
fit parameter but a function of $m_{Hr}$, $m_\pi$, and $\Gamma_{Hr}$, which can easily be worked out such that
$f(E)$ and $\sigma(k)$ coincide at the resonance $E=m_{Hr}$, where one has $\sin^2(\delta_0(k))=1$. 
The two free fit parameters in this approach are thus the Higgs resonance mass $m_{Hr}$ and its
width $\Gamma_{Hr}$. The fit curve obtained from this ansatz is presented in \fig{fig:FirstResultScatteringPhases},
while the underlying fit parameters, and thus the results on the resonance mass and width, are listed
in \tab{tab:ResultOfScatteringAnalysis}.

\includeTabNoHLines{|c|c|c|}{
\cline{2-3}
\multicolumn{1}{c|}{} &  Resonance analysis                & Reference value  \\ \hline
Mass       	 & $\,m_{Hr}      = 0.4281(31)\, $   &  $\,m_{Hp}      = 0.4328(32)\, $ \\ 
Decay width   	 & $\,\Gamma_{Hr} = 0.0086(33)\, $   &  $\,\Gamma_{H}  = 0.0076(2)\, $ \\  \hline
}
{tab:ResultOfScatteringAnalysis}
{The Higgs resonance mass $m_{Hr}$ and its width $\Gamma_{Hr}$ are presented as obtained from fitting the
scattering phase shifts $\delta_0(k)$ with the fit ansatz in \eq{eq:FitAnsatzForScatteringPhases}.
Moreover, the Higgs propagator mass $m_{Hp}$ and its width $\Gamma_H$ are listed as reference values to
compare with. The source of the latter numbers is \tab{tab:ResultOfScatteringExampleInfiniteVolExtrapolation},
where the decay width has been computed by means of \eq{eq:FinalFormulaForGamma}.
}
{Comparison of the extracted resonance parameters with corresponding reference values obtained from
perturbation theory in combination with direct lattice calculations.}

\includeTabNoHLines{|c|c|c|c|c|}{
\hline
observable	 & $A^{(l)}_{v,m,\lambda}$, $L'_s=16$  & $A^{(l)}_{v,m,\lambda}$, $L'_s=20$  & $A^{(p)}_{v,m,\lambda}$ & Final ext. result   \\ \hline
$\,v_r\,$  	 & $\, 0.4147(5) \, $   & $\, 0.4147(6) \,   $  & $\, 0.4148(7) \, $  & $\, 0.4147(6)(1)$   \\ 
$\,m_{Hp}\,$  	 & $\, 0.4335(25)\, $   & $\, 0.4313(28)\,   $  & $\, 0.4336(36)\, $  & $\, 0.4328(30)(12)$   \\ 
$\,m_{Gp}\,$  	 & $\, 0.1534(12)\, $   & $\, 0.1540(16)\,   $  & $\, 0.1543(18)\, $  & $\, 0.1539(16)(5)$   \\ 
$\,\lambda_r\,$  & $\, 0.1197(16)\, $   & $\, 0.1181(17) \,  $  & $\, 0.1195(22)\, $  & $\, 0.1191(19)(8)$   \\ \hline
\multicolumn{3}{|c|}{Final results of extrapolation in physical units}   & \multicolumn{2}{c}{}   \\ \cline{1-3}
$\Lambda=\GEV{246}/v_r$  &  $m_{Hp}/a$     &   $m_{Gp}/a$      & \multicolumn{2}{c}{}    \\ \cline{1-3}
$\GEV{593.2 \pm 0.9}$   &  $\GEV{256.7  \pm 2.0}$  &   $\GEV{ 91.3 \pm 1.0}$     & \multicolumn{2}{c}{} \\ \cline{1-3}
}
{tab:ResultOfScatteringExampleInfiniteVolExtrapolation}
{The results of the infinite volume extrapolations of the Monte-Carlo data for the renormalized vev $v_r$, the Higgs and 
Goldstone propagator masses $m_{Hp}$ and $m_{Gp}$, as well as the renormalized quartic coupling constant $\lambda_r$ are 
presented as obtained from the parabolic ansatz in \eq{eq:ParaFit} and the linear approach in \eq{eq:LinFit} for the considered 
threshold values $L_s'=16$ and $L_s'=20$. The final results on the aforementioned observables, displayed in the very right column, 
are determined here by averaging over all performed fit approaches, including the parabolic fit. An additional, systematic 
uncertainty is specified in the second pair of brackets taken from the largest observed deviation 
among all respective fit approaches. These final extrapolation results are furthermore presented in physical units in the bottom 
line of the table, where the previously separated statistical and systematic uncertainties have been combined into a total error. 
}
{Infinite volume extrapolation of the obtained Monte-Carlo data for the renormalized vev $v_r$, the Higgs and 
Goldstone propagator masses $m_{Hp}$ and $m_{Gp}$, as well as the renormalized quartic coupling constant $\lambda_r$.}

For clarification it is remarked that the resonance width $\Gamma_{Hr}$ as obtained through the above approach
is given here as the full width of the Breit-Wigner function at half of its maximal height, which is commonly
referred to as the FWHM-value in the literature. This definition of $\Gamma_{Hr}$ is somewhat arbitrary and does not exactly
correspond~\cite{Liu:2008hy} to the original definition of the Higgs decay width $\Gamma_H$ given through the 
location of the pole of the Higgs propagator continued onto the second Riemann sheet as specified in \sect{sec:UnstableSignature}.
It can, however, be shown in the framework of some exactly solvable model~\cite{Liu:2008hy} that these kinds of
different definitions become equivalent in the limit of infinitesimally small decay widths. In the following we will 
therefore be ignorant towards a potential systematic discrepancy between the definitions of $\Gamma_{Hr}$ and $\Gamma_H$.

At this point it would be interesting to compare these findings on the resonance width to some analytical expectations.
Thanks to the relatively small value of the renormalized quartic self-coupling constant $\lambda_r$ arising from
the chosen setting $\hat\lambda=0.01$, a reliable relation between the decay width $\Gamma_H$, the renormalized
coupling constant $\lambda_r$, the renormalized vacuum expectation value $v_r$, and the boson masses $m_H$ and 
$m_G$ can be established by means of renormalized perturbation theory. For that purpose we reconsider the 
renormalized result on the Higgs propagator in \eq{eq:StructureOfPropInNCompPhi4TheoryRenormalized}, which has 
been established through a perturbative one-loop calculation in the pure $\Phi^4$-theory. 

It has already been pointed out in \sect{sec:UnstableSignature} that the given expression in 
\eq{eq:StructureOfPropInNCompPhi4TheoryRenormalized} has to obey the minimal condition of being 
real for entirely real momenta $p$, which directly constitutes the sought-after relation
according to
\bea
\label{eq:ConditionForRealityOfHiggsProp}
0 &=& -im_H\Gamma_H - 36i\pi^{-2}\lambda_r^2v_r^2\IM(C_{H0}) - 4i\pi^{-2}(n-1)\lambda_r^2v_r^2\IM(C_{G0}),
\eea
where the definitions of $C_{H0}$ and $C_{G0}$ have been given in \eqs{eq:DefOfRenHiggsPropConstantCH0}{eq:DefOfRenHiggsPropConstantCG0}. 
This constraint has been solved numerically for the calculation of the spectral functions in \fig{fig:SpectralFunctionsFromPT}. 
Here, however, an analytical approximation shall be presented. For that purpose we note that
\bea
\label{eq:ImValueOfCH0}
\IM(C_{H0}) &=&  0 + O(\lambda_r), \\
\label{eq:ImValueOfCG0}
\IM(C_{G0}) &=&  -\frac{\pi}{2}  \sqrt{\frac{4m_G^2-m_H^2}{-m_H^2}} + O(\lambda_r),
\eea
where the value $\pi/2$ in \eq{eq:ImValueOfCG0} arises from the branch cut of the $\arctanh$-function in \eq{eq:DefOfContEuc1LoopBosContrib}
and the minus sign originates from the continuation onto the second Riemann sheet. Combining the relations
in \eqs{eq:ConditionForRealityOfHiggsProp}{eq:ImValueOfCG0} one directly finds the perturbative one-loop
result for the sought-after relation in the pure $\Phi^4$-theory to be
\bea
\Gamma_H &=& 2\pi^{-1} \lambda_r^2v_r^2 (n-1) \frac{\sqrt{m_H^2 - 4m_G^2}}{m_H^2},
\eea
where terms of order $O(\lambda_r^3)$ have been neglected and $n$ denotes the number of components of the scalar
field $\Phi$, \ie $n=4$ in our case. According to the here considered definition of the renormalized quartic 
coupling constant $\lambda_r$ in \eq{eq:DefOfRenQuartCoupling}, the latter expression 
becomes\footnote{The given result is identical to the formula applied in \Ref{Gockeler:1994rx}.}
\bea
\label{eq:FinalFormulaForGamma}
\Gamma_H &=& \frac{n-1}{4\pi} \lambda_r  \frac{m_H^2 - m_G^2}{m_H^2}   \sqrt{m_H^2 - 4m_G^2},
\eea
where $m_H$ and $m_{Hp}$ have been identified with each other, which is justified here, since it only induces negligible 
corrections to $\Gamma_H$ being of order $O(\lambda_r^3)$ according to \eq{eq:ConnectionMhMhp}. The same substitution 
will also be exploited in the following.

%\includeFigDoubleDoubleHere{ScatteringAnalysisFiniteSizeEffectsVeV}{ScatteringAnalysisFiniteSizeEffectsHiggsMass}
%{ScatteringAnalysisFiniteSizeEffectsGoldstoneMass}{ScatteringAnalysisFiniteSizeEffectsLambdaRen}
\includeFigDoubleDoubleHere{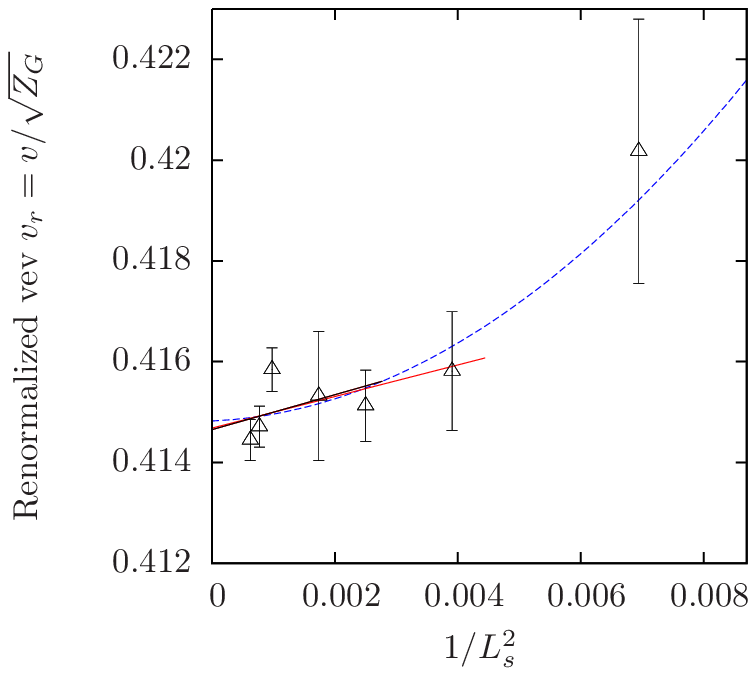}{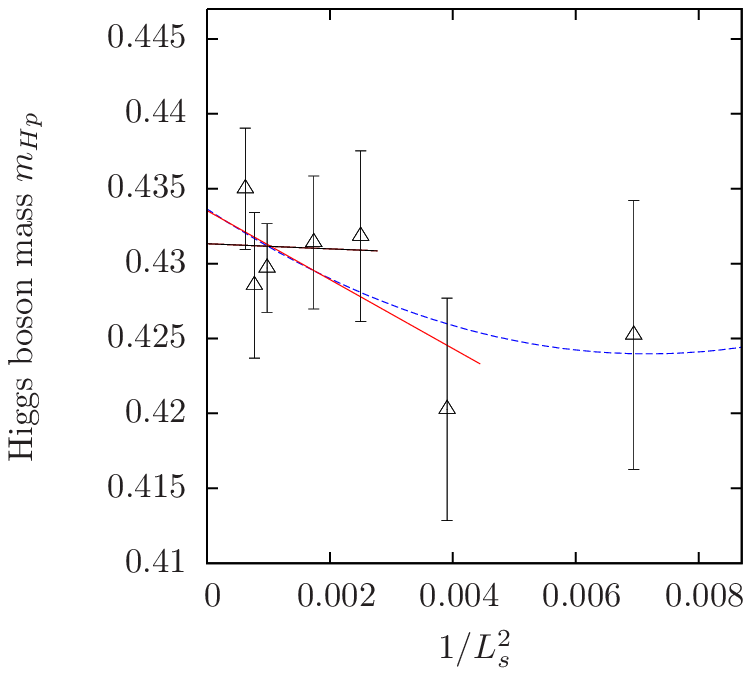}
{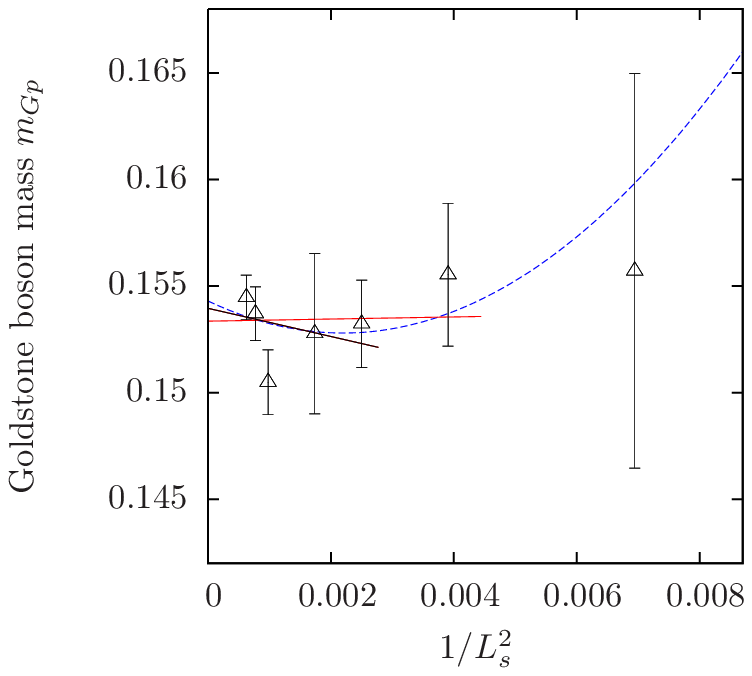}{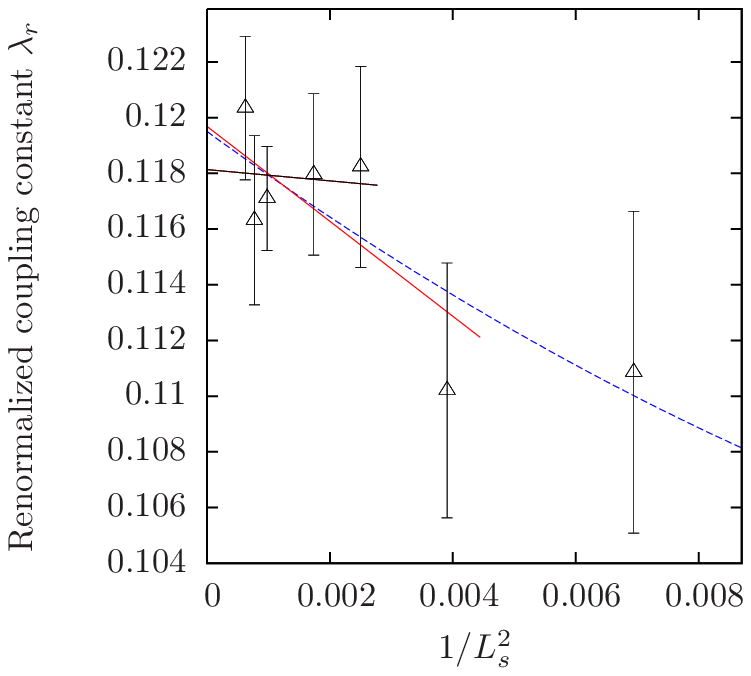}
{fig:InfiniteVolExtrapolForCompWithScatteringPhases}
{The dependence of the renormalized vev $v_r = v/\sqrt{Z_G}$, the Higgs and Goldstone propagator masses $m_{Hp}\approx m_H$ 
and $m_{Gp}\equiv m_G$, as well as the renormalized quartic coupling constant $\lambda_r$ on the squared inverse lattice 
side length $1/L^2_s$ is presented. These results have been obtained in the direct Monte-Carlo calculations specified in 
\tab{tab:SummaryOfParametersForDecayWidth}.
In all panels the dashed curves display the parabolic fits according to the fit ansatz in 
\eq{eq:ParaFit}, while the solid lines depict the linear fits resulting from \eq{eq:LinFit} for the two threshold 
values $L_s'=16$ (red) and $L_s'=20$ (black).
}
{Dependence of the renormalized vev $v_r = v/\sqrt{Z_G}$, the Higgs and Goldstone propagator masses $m_{Hp}\approx m_H$ 
and $m_{Gp}\equiv m_G$, as well as the renormalized quartic coupling constant $\lambda_r$ on the squared inverse lattice 
side length $1/L^2_s$ at a small quartic coupling constant.}

For the purpose of performing the aspired comparison with the perturbative calculation the infinite volume results 
on the renormalized quantities $\lambda_r$, $m_H\approx m_{Hp}$, and $m_G\equiv m_{Gp}$ need to be determined.
This is done here in exactly the same manner as already detailed in \sect{sec:UpperMassBounds}. 
The finite volume lattice data for the aforementioned quantities as well as for the renormalized vev $v_r$ are therefore 
presented in \fig{fig:InfiniteVolExtrapolForCompWithScatteringPhases}. These data are again plotted versus $1/L_s^2$ and
fitted with the linear and parabolic fit approaches in \eq{eq:LinFit} and \eq{eq:ParaFit}. The resulting fit curves 
are also displayed in \fig{fig:InfiniteVolExtrapolForCompWithScatteringPhases} and the underlying fit parameters
are listed in \tab{tab:ResultOfScatteringExampleInfiniteVolExtrapolation}, where the final infinite volume
extrapolations are again determined as the respective average over all performed fit procedures.

With these data at hand the perturbative result on the decay width $\Gamma_H$ can be obtained from \eq{eq:FinalFormulaForGamma}
as listed in \tab{tab:ResultOfScatteringAnalysis}. Comparing the results on $\Gamma_H$ and the Higgs propagator mass $m_{Hp}$
with the previously discussed resonance parameters $\Gamma_{Hr}$ and $m_{Hr}$ one observes satisfactory agreement 
with respect to the specified uncertainties. 

As a concluding remark it is therefore summarized that the decay properties of the Higgs boson can indeed be determined in 
the considered Higgs-Yukawa model by virtue of L\"uscher's method~\cite{Luscher:1990ux,Luscher:1991cf}, even though the 
achieved accuracy and thus the underlying statistics of generated field configurations certainly has to be increased in 
follow-up investigations. At this point, however, only a preliminary and 
brief outlook has been given on what could be done next. In particular, the quartic coupling constant has here been fixed to a 
rather small value to allow for a reliable comparison with perturbation theory. In a follow-up investigation one could be interested 
in increasing the underlying renormalized quartic coupling constant $\lambda_r$ up to its maximal value by sending the bare parameter 
$\lambda$ to infinity, which would eventually allow to study the maximally possible decay width of the Higgs boson in the here
considered Higgs-Yukawa model. 

  \chapter{Summary and conclusions}
\label{chap:Conclusions}

The aim of the present work has been the non-perturbative determination of the upper and lower mass bounds of the 
Standard Model Higgs boson based on first principle computations, in particular not relying on additional information such as 
the triviality property of the Higgs-Yukawa sector or indirect arguments like vacuum stability considerations.
The motivation for the consideration of the aforementioned mass bounds finally lies in the ability to draw conclusions
on the energy scale $\Lambda$ at which a new, so far unspecified theory of elementary particles definitely has to 
substitute the Standard Model, once the Higgs boson and its mass $m_H$ will have been discovered experimentally.
In that case the latter scale $\Lambda$ can be deduced by requiring consistency between the observed mass $m_H$
and the mass bounds $\lowBound$ and $\upBound$ intrinsically arising from the Standard Model under the assumption 
of being valid up to the cutoff scale $\Lambda$. 

The Higgs boson might, however, very well not exist at all, especially since the Higgs sector, can only be considered 
as an effective theory of some so far undiscovered, extended theory, due to its triviality property. In such a scenario, 
a conclusion about the validity of the Standard Model can nevertheless be drawn, since the non-observation of the Higgs 
boson at the LHC would eventually exclude its existence at energies below, lets say, $\TEV{1}$ thanks to the large accessible 
energy scales at the LHC. An even heavier Higgs boson is, however, definitely excluded without the Standard Model becoming 
inconsistent with itself according to the results in \chap{chap:ResOnUpperBound} and the requirement that the cutoff $\Lambda$ be 
clearly larger than the mass spectrum described by that theory. In the case of non-observing the Higgs boson at the LHC,
one can thus conclude on the basis of the latter results, that new physics must set in already at the TeV-scale.
For the rest of this section, however, we assume the discovery of the Higgs boson with mass $m_H$ allowing
then to draw conclusions on $\Lambda$ in the previously described manner.

For the purpose of establishing the aforementioned mass bounds, the lattice approach has been employed to allow for a 
non-perturbative investigation of a Higgs-Yukawa model serving here as a reasonable simplification of the full Standard Model, 
containing only those fields and interactions which are most essential for the Higgs boson mass determination. 
The model therefore consists of the scalar field $\varphi$, composed out of the Higgs and the Goldstone modes, as well as 
the top-bottom doublet and their mutual interactions, which together are responsible for the emergence of the lower Higgs 
boson mass bound. Moreover, the considered Higgs-Yukawa model also describes the self-interaction of the scalar field 
$\varphi$, which is expected to yield the largest contribution to the upper Higgs boson mass bound as the bare quartic 
coupling constant is sent to infinity, thus motivating the non-perturbative approach.

According to the chiral nature of the electroweak interaction in the full Standard Model, left- and right-handed
particles couple differently among each other. To maintain this chiral gauge coupling structure also on the lattice
a consistent formulation of chiral gauge theories on the lattice would be required. Due to the neglection 
of any gauge fields in the here considered Higgs-Yukawa model, however, the latter local gauge symmetry reduces to a 
global symmetry, which can then be realized on the lattice by virtue of L\"uscher's proposals in \Ref{Luscher:1998pq}.

In the present work the aforementioned proposal for the construction of chirally invariant lattice Higgs-Yukawa models 
has been applied to the situation of the actual Standard Model Higgs-fermion coupling structure, 
\ie for $\varphi$ being a complex doublet equivalent to one Higgs and three Goldstone modes. The resulting chirally
invariant lattice Higgs-Yukawa model, constructed here on the basis of the Neuberger overlap operator, then obeys a global 
$\SUtwoTimesUoneY$ symmetry, as desired. 

The fundamental strategy underlying the determination of the cutoff-dependent upper and lower Higgs boson mass bounds
has then been the numerical evaluation of the maximal interval of Higgs boson masses attainable within the considered Higgs-Yukawa 
model in consistency with phenomenology. The latter condition refers here to the requirement of reproducing the phenomenologically
known values of the top and bottom quark masses as well as the renormalized vacuum expectation value $v_r$ of the scalar field, where 
the latter condition was used here to fix the physical scale of the performed lattice calculations. Applying this strategy thus 
requires the evaluation of the model to be performed in the broken phase, however close to a second order phase transition to
a symmetric phase, in order to allow for the adjustment of arbitrarily large cutoff scales, at least from a conceptual point of view. 

As a preparatory consideration the phase structure of the underlying Higgs-Yukawa model has therefore been investigated, which
has been done here in the limit of an infinite number $N_f$ of degenerate fermion generations. Though the actual physical
situation in the Standard Model corresponds to the finite and even small setting $N_f=1$, or rather $N_f=3$ to respect also the 
colour index despite the general absence of the gauge fields themselves, it was still possible to derive the qualitative phase 
structure of the Higgs-Yukawa model by means of this approach. The considered order parameters have been chosen to be the vacuum 
expectation value $v$ as well as its staggered counterpart $v_s$ which finally allowed to locate the symmetric ($v=0$, $v_s=0$) 
and the broken ($v\neq 0$, $v_s=0$) phases of the model in bare parameter space. In fact, a rich phase structure could be observed 
including also phases with $v=0$ and $v_s\neq 0$ as well as phases with $v\neq 0$ and $v_s\neq 0$, where the vacuum expectation 
values $v$ and $v_s$ are simultaneously non-zero. 

This investigation of the phase structure in the large $N_f$-limit has primarily targeted the scenario of small Yukawa coupling 
constants, since this is the relevant setup for the considered purpose. However, the model has also been studied in the limit of 
large Yukawa coupling constants. Interestingly, a symmetric phase also emerges in this scenario, the existence of which, however, is 
obscured by strong finite volume effects, hindering its observation on too small lattices. Though this part of the phase structure
is not of direct interest for the present study, it seems to be valuable for other questions, such as the study of 
mass generation in this strong coupling symmetric phase~\cite{Giedt:2007qg} or the investigation of nucleon bound
states induced by pion exchange, which is assumed to be describable by means of a Higgs-Yukawa model with strong Yukawa coupling
constants~\cite{deSoto:2005jj}, however with a very different interpretation of the underlying field content. In both cases, 
\ie for small as well as for large Yukawa coupling constants, the phase transitions between the symmetric and the 
broken phase were found to be of second order, making the model thus suitable for direct lattice calculations. To confirm 
the validity of the performed analytical calculations, the above results have explicitly been compared to direct Monte-Carlo 
computations at finite values of $N_f$ and good agreement has been observed.

With this information about the second order phase transition line between the symmetric and broken phases at hand, the next step
towards the aim of the present study has been the preparation of a suitable simulation algorithm, capable of
evaluating the considered model for arbitrary, especially for odd values of $N_f$. For that purpose a PHMC 
algorithm~\cite{Frezzotti:1997ym} has been implemented. This underlying concept, however, has been extended by a number of algorithmic
improvements applicable at least in the case of the here considered Higgs-Yukawa model, which finally led to a substantial
and indispensable performance gain. Among them are a preconditioning technique, reducing the condition number of the underlying 
fermion operator by 1-2 orders of magnitude, the well-known Fourier acceleration approach finally resulting in drastically smaller
auto-correlation times, as well as an exact, Krylov-space based reweighting strategy allowing to further reduce the degree
of the employed approximation polynomials while keeping the algorithm exact. Some of these techniques, for instance the
latter reweighting strategy, can directly be applied also to the case of QCD, while others are not directly adaptable,
as discussed in \chap{chap:SimAlgo}. In total, the presented enhancements induced a very substantial performance gain which was
crucial for the success of the performed lattice calculations.

Equipped with these technical and conceptual preparations the question for the lower Higgs boson mass bound could finally be
addressed. For that purpose it has explicitly been confirmed by direct lattice calculations that the lightest Higgs boson masses 
are indeed generated at vanishing quartic self-coupling constant, as suggested by analytical calculations performed in the framework 
of the constraint effective potential, thus allowing to restrict the search for the lower Higgs boson mass bound to the scenario of 
$\lambda=0$. For the mass degenerate case with equal top and bottom quark masses and $N_f=1$ the finite volume lattice results on 
the Higgs boson mass obtained at $\lambda=0$ for a series of cutoffs and lattice sizes have then been presented and compared to 
their respective analytical predictions based on the aforementioned effective potential calculations and very good agreement 
has been observed. These numerical finite size results were then extrapolated to the infinite volume limit to actually establish 
the desired cutoff-dependent lower Higgs boson mass bound $\lowBound$, as summarized in \fig{fig:FinalSummaryPlot}. 

The main effort concerning the investigation of the lower bound has been spent on the above mentioned degenerate scenario with
$y_t=y_b$ and $N_f=1$, since it is reliably accessible from a numerical as well as from a conceptual point of view. This situation
changes when one proceeds to the physically more relevant case of $N_f=3$ and non-degenerate Yukawa coupling constants fixed according
to the phenomenologically known values of the top and bottom quark masses. In that scenario one 
becomes confronted with the conceptual problem of a fluctuating complex phase of the fermion determinant leading to unknown
systematic uncertainties on the numerical results arising from the employed simulation algorithm. In addition, the numerical
requirements themselves are also significantly increased in this parameter setup according to larger condition numbers of the 
fermion matrix as well as stronger finite volume effects induced by the large ratio of the bottom and top quark masses.
The numerical investigation of the lower Higgs boson mass bound has nevertheless been tried also in this scenario and the obtained
results are in agreement with the analytical prediction arising from the aforementioned effective potential calculation,
though a clear resolution of the differences between the considered degenerate $N_f=1$ results and the non-degenerate $N_f=3$ results
could not be achieved. However, assuming the latter effective potential calculations to remain trustworthy also in the non-degenerate 
scenario, which is in agreement with the observed lattice results, one would then find a cutoff-dependent lower Higgs boson mass bound of 
approximately $\lowBound = \GEV{80}$ at a cutoff scale of $\Lambda=\GEV{1000}$, as summarized in \fig{fig:FinalSummaryPlot}.
  
The main result of the presented findings is that a lower Higgs boson mass bound is a manifest property of the pure
Higgs-Yukawa sector that evolves directly from the Higgs-fermion interaction for a given set of Yukawa coupling constants.
With growing cutoff this lower mass constraint was found to rise monotonically with flattening slope as expected
from perturbation theory. Moreover, the quantitative size of the obtained lower bound is comparable to the magnitude of the
perturbative results based on vacuum stability considerations~\cite{Hagiwara:2002fs} as presented in \fig{fig:PerturbativeHiggsMassBounds}. 
A direct quantitative comparison between the latter approaches is, however, non-trivial due to the different underlying 
regularization schemes in combination with the strong cutoff-dependence of the considered constraints.

The question for the universality of the derived lower mass bound has then briefly been addressed by studying its dependence on the
extension of the underlying Lagrangian by some higher order self-interaction terms of the scalar field, \ie $\lambda_6|\varphi|^6$
and $\lambda_8|\varphi|^8$ in this case. Such a consideration is not excluded by the usual renormalization arguments, since the 
Higgs-Yukawa sector can only be considered as an effective theory due to its triviality property. With the help of adequate 
analytical calculations in the framework of the constraint effective potential, several bare parameter setups have been 
estimated for this extended model, which supposedly lead to Higgs boson masses undercutting the lower mass bounds previously 
established in the pure Higgs-Yukawa model. By means of direct lattice Monte-Carlo computations it could then be shown, that the 
selected parameter setups indeed generate Higgs boson masses which are significantly pushed below the aforementioned lower bound. 
The interpretation of this finding is that the lower Higgs boson mass bound arising in the pure Higgs-Yukawa model cannot be 
considered as being universal, \ie it is not independent of the specific form of the underlying interactions. However, the Higgs 
boson mass could not be made arbitrarily small, though the aforementioned effective potential calculations would have suggested that. 
This might be an indication for the existence of a true lower bound also in this extended scenario. It might, however, also be 
that only the applied method for finding adequate bare coupling parameters in this extended model, which would potentially lead to even 
smaller Higgs boson masses, has to be further improved. Moreover, one may ask whether smaller masses can be attained by including 
even more coupling terms within the Lagrangian. This issue poses an interesting problem for future investigations.

The question for the upper Higgs boson mass bound has then been addressed by applying the same conceptual strategy also in the
regime of large quartic self-coupling constants $\lambda$. In a first step it has explicitly been confirmed by direct lattice
calculations that the largest attainable Higgs boson masses are indeed observed in the case of an infinite bare quartic coupling constant,
as suggested by the aforementioned effective potential calculations. Consequently, the search for the upper Higgs boson mass bound
has subsequently been constrained to the bare parameter setting $\lambda=\infty$. The resulting finite volume lattice data on the Higgs
boson mass turned out to be sufficiently precise to allow for their reliable infinite volume extrapolation, yielding then a cutoff-dependent
upper bound of approximately $\upBound=\GEV{630}$ at a cutoff of $\Lambda=\GEV{1500}$. These results were moreover precise enough
to actually resolve their cutoff-dependence as summarized in \fig{fig:FinalSummaryPlot}, which is in very good agreement with 
the analytically expected logarithmic decline, and thus with the triviality picture of the Higgs-Yukawa sector. It is pointed out again, 
that this achievement has been numerically demanding, since the latter logarithmic decline of the upper bound $\upBound$ is 
actually only induced by subleading logarithmic contributions to the scaling behaviour of the considered model close to its phase 
transition, which had to be resolved with sufficient accuracy. By virtue of the analytically expected 
functional form of the cutoff-dependent upper mass bound, which was used here to fit the obtained numerical data, an extrapolation 
of the latter results to much higher energy scales could also be established, being again in good agreement with the corresponding 
perturbatively obtained bounds~\cite{Hagiwara:2002fs} presented in \fig{fig:PerturbativeHiggsMassBounds}. A direct comparison has, 
however, again been avoided here for the previously discussed reasons. 

%\includeFigSingleMedium{FinalSummaryPlot}
\includeFigSingleMedium{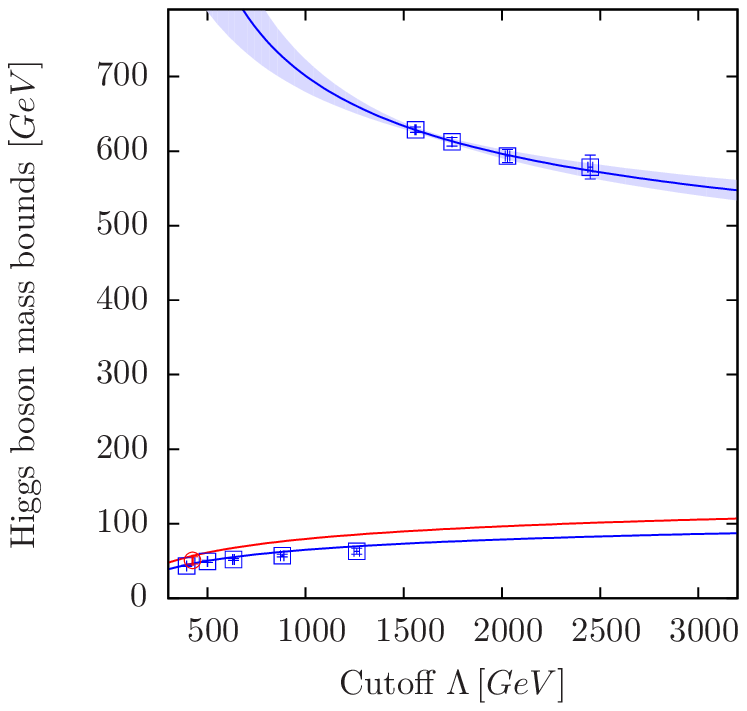}
{fig:FinalSummaryPlot}
{The upper and lower Higgs boson mass bounds taken from \fig{fig:UpperMassBoundFinalResult} and 
\fig{fig:LowerHiggsMassBoundPredictionInifinteVol}, respectively, are jointly presented
versus the cutoff $\Lambda$. The squared symbols depict the numerical results obtained in the mass degenerate scenario
with equal, non-zero top and bottom Yukawa coupling constants and $N_f=1$ as discussed in 
\sects{sec:SubLowerHiggsMassboundsDegenCase}{sec:UpperMassBounds}, while the circular symbol
refers to the non-degenerate setup with $N_f=3$ considered in \sect{sec:SubLowerHiggsMassboundsGenCase}. The upper 
curve and the highlighted band display the analytical fit of the numerical data on the upper mass bound and its associated 
uncertainty. The lower two curves represent the analytical predictions on the lower bound as derived from the effective 
potential. The lower of the latter two curves refers to the degenerate scenario with $N_f=1$, while the other one shows 
the analytical mass bound in the physically more relevant, non-degenerate setup with $N_f=3$. 
}
{Cutoff-dependence of the upper and lower Higgs boson mass bounds.}

The interesting question for the fermionic contribution to the observed upper Higgs boson mass bound has then been addressed
by explicitly comparing the latter findings to the corresponding results arising in the pure $\Phi^4$-theory. For the considered
energy scales this potential effect, however, turned out to be not very well resolvable with the available accuracy of the lattice 
data. The performed fits with the expected analytical form of the cutoff-dependence only mildly suggest the upper mass bound in the 
full Higgs-Yukawa model to decline somewhat steeper with growing cutoffs than the corresponding results in the pure $\Phi^4$-theory. 
To obtain a clearer picture in this respect, a higher accuracy of the numerical data and thus a higher statistics of the underlying 
field configurations would be needed.

It has moreover also been tried to derive some first information on the decay properties of the Higgs boson directly from the analysis
of the curvature of the Higgs propagator based on its renormalized perturbative one-loop expression, which was analytically
derived in continuous Euclidean space-time. This approach, however, turned out to be not successful in the here considered 
scenarios of large quartic coupling constants, since it was not possible to stabilize the corresponding fit result on the decay
width $\Gamma_H$, meaning here to retrieve results on $\Gamma_H$ independently of the momentum interval in which the propagator 
has been analyzed. In that respect it might be interesting to ask, whether this problem can be overcome by considering perturbative
calculations of the Higgs propagator up to higher orders in the renormalized quartic coupling constant and including also the
fermionic contributions, while simultaneously increasing the underlying statistics of field configurations.

In the present work, however, this ansatz has not further been pursued. Instead, a first account on L\"uscher's 
method~\cite{Luscher:1990ux,Luscher:1991cf} for determining the Higgs boson decay width by studying the volume dependence of 
the finite size lattice mass spectrum  has been given. This brief and preliminary outlook was only meant here as a conceptual 
study, demonstrating the applicability of the latter method in the framework of the considered Higgs-Yukawa model. For that purpose 
a model parameter setup associated to a rather small renormalized quartic self-coupling constant has been chosen, since 
this allowed to derive a reliable reference value for the Higgs boson decay width from perturbation theory. In the subsequent 
analysis it was then found that the Higgs boson decay width $\Gamma_H$ can indeed be determined with the help of L\"uscher's 
approach in the here considered Higgs-Yukawa model. The observed numerical value for $\Gamma_H$ was moreover found to be in 
agreement with the corresponding result derived from a perturbative one-loop calculation in the pure $\Phi^4$-theory, 
thus encouraging to repeat similar analyses also in the physically more interesting scenario of maximal values for the renormalized 
quartic self-coupling constant, \ie at $\lambda=\infty$. This will then eventually allow to establish an upper bound also for 
the decay width of the Standard Model Higgs boson on the basis of the here considered Higgs-Yukawa model.

Such future extensions of the here presented investigation can directly be pursued with the already implemented software 
code. In this respect one might, however, want to improve the applied analysis by considering the so-called moving-frame, 
referring here to the frame, relative to which the involved momenta and energies are determined.
Concerning the relatively large observed statistical uncertainties in \chap{chap:ResOnDecayWidth} one would also
want to significantly increase the underlying statistics of field configurations. By virtue of the algorithmic improvements
presented in \chap{chap:SimAlgo} and according to the applied parallelization technique discussed in \appen{sec:xFFT}
this can directly be achieved by simultaneously rerunning multiple replicas of the lattice calculations considered in 
\chap{chap:ResOnDecayWidth}. With such an ansatz it is apparently possible to increase the field configuration
statistics, at least for some small number of selected runs, by clearly more than one order of magnitude\footnote{This is one
of the projects initiated and pursued by Jim Kallarackal.}. This should then allow to study the Higgs boson decay properties 
with higher precision. It should moreover allow to clearly identify the fermionic contributions to the upper Higgs boson mass bound.

In addition to that it might be worthwhile to repeat the presented Higgs propagator analyses presented in 
\sects{sec:HiggsPropAnalysisLowerBound}{sec:HiggsPropAnalysisUpperBound} on the basis of some improved analytical 
fit formula, including also higher order and in particular fermionic contributions, which might then, together with
the aforementioned significantly increased statistics, allow to extract some decay properties also from this approach.

% usw.

% Anhang %%%%%%%%%%%%%%%%%%%%%%%%%%%%%%%%%%%%%%%%%%%%%%%%%%%%%%%%%%%%%%%%%%%%%%%

  \appendix

% Kapitel des Anhangs %%%%%%%%%%%%%%%%%%%%%%%%%%%%%%%%%%%%%%%%%%%%%%%%%%%%%%%%%%

 \chapter{The Lanczos method for \texorpdfstring{computing $f(\fermiMatDouble)\omega,\, f(x)=x^{-\alpha},\, \alpha\in\re$}
{applying the inverse square root of MM}}
\label{sec:DescriptionOfLancosApproach}
    
In \Ref{Borici:1999ws} a Lanczos-based, iterative method was presented, capable of computing 
$\LanczosMatrix^{-1/2}\omega$ for a given positive, hermitian matrix $\LanczosMatrix$ and a given vector $\omega$.
Its main idea is to approximate $\LanczosMatrix^{-1/2}$ by the inverse square root of the projection of the 
operator $\LanczosMatrix$ onto the Krylov space 
\bea
\krylovSpace_{N_L}(\LanczosMatrix, \omega) &=& \mbox{span}\left\{\LanczosMatrix^{i}\omega: \quad i=0,\ldots,N_L-1  \right\}
\eea
for some sufficiently large value of $N_L$. In this section the latter method shall be extended to the more general
case of computing $f(\LanczosMatrix)\omega$, where $f:\re\rightarrow \re$ is a given real function\footnote{The function 
$f:\re\rightarrow \re$ is a real function acting on real numbers. With some abuse of notation, however, it 
will also be applied to hermitian matrices in the sense that it actually acts on their respective eigenvalues 
in the spectral representation.} with $f(x)=x^{-\alpha}$, $\alpha\in\re$. For clarification it is remarked that the subsequent 
discussion, which is based here on the latter general operator $\LanczosMatrix$, directly includes also
the special case of $\LanczosMatrix=\fermiMatDouble$, which is actually of interest in the present work.

Here, we begin with the discussion of the original version of the aforementioned Lanczos-based algorithm as specified in 
\Ref{Borici:1999ws}. For a given prescribed relative accuracy $\LanczosRelAcc$ it iteratively generates a sequence of orthonormal 
vectors $q_i,\, i=0,\ldots,N_L$ by the successive application of the operator $\LanczosMatrix$, starting from 
$q_0 = \omega / |\omega|$. The number of Lanczos iterations $N_L$ specifies the total of performed matrix applications
of the considered operator $\LanczosMatrix$. The vectors $\{ q_i:\,\, i=0,\ldots,N_L-1\}$ then form an orthonormal basis 
of $\krylovSpace_{N_L}(\LanczosMatrix, \omega)$. Moreover, the Lanczos algorithm generates a tridiagonal $N_L\times N_L$-matrix 
$T_{N_L}$, which is by construction the representation of the operator $\LanczosMatrix$ projected onto the Krylov space 
$\krylovSpace_{N_L}(\LanczosMatrix, \omega)$ in the basis $\{q_i:\,\, i=0,\ldots,N_L-1\}$. In detail this algorithm 
reads~\cite{Borici:1999ws}:
\bea
\label{eq:LanczosAlgorithm}
&&\beta_{-1} := 0,\,\rho_{-1} := 0, \,\rho_0 := 1 / |\omega|, \,q_{-1} := 0, \,q_0 := \omega/|\omega| \nonumber \\
&&\mbox{for} \,i := 0, \ldots \nonumber\\
&&\quad v := \LanczosMatrix q_i, \quad \alpha_i := q_i^{\dag} v \nonumber\\
&&\quad v := v - q_i \alpha_i - q_{i-1} \beta_{i-1} \nonumber\\
&&\quad \beta_i := |v|, \quad q_{i+1} := v / \beta_i \nonumber\\
&&\quad \rho_{i+1} := - \frac{\rho_i \alpha_i + \rho_{i-1} \beta_{i-1}}{\beta_i} \nonumber\\
\label{eq:StoppingCriterionFoLanczos}
&&\quad \mbox{if}\, \frac{\rho_0}{|\rho_{i+1}|} < \LanczosRelAcc, \, \mbox{then}\, N_L := i+1,\, \mbox{STOP}\\
&&\mbox{end for}\nonumber
\eea
The aforementioned tridiagonal matrix $T_{N_L}$ is then defined on the basis of the computed values $\alpha_i$ and 
$\beta_i$ according to
\bea
(T_{N_L})_{i,j} &=& 
\left\{
\begin{array}{*{3}{lcl}}
\alpha_i   & : &  \mbox{for}\,\,\, i=j \\
\beta_i    & : &  \mbox{for}\,\,\, i=j-1 \\
\beta_j    & : &  \mbox{for}\,\,\, j=i-1 \\
0          & : &  \mbox{else}\\
\end{array} \right\},
\eea
where the above indices $i,j$ of the $N_L\times N_L$ matrix $T_{N_L}$ take the values $i,j=0,\ldots, N_L-1$.
The approximation for $\LanczosMatrix^{-1/2}\omega$ is then given as
\beq
\label{eq:LanczosApproach1}
Q T_{N_L}^{-\frac{1}{2}}  |\omega| \hat e_0
\stackrel{N_L\rightarrow N_V}{\longrightarrow}
\LanczosMatrix^{-\frac{1}{2}}\omega, \quad Q=(q_0,\ldots,q_{N_L-1}),
\eeq
where $N_V$ is the problem size, \ie the size of the vector $\omega$, and $\hat e_0$ denotes the unit vector with length $N_L$
in 0-direction, according to $[\hat e_0]_j = \delta_{j,0}$, $j=0,\ldots,N_L-1$. The inverse square root of the $N_L\times N_L$-matrix 
$T_{N_L}$, which is well defined since the matrix $T_{N_L}$ is symmetric and positive by construction, can then be computed 
by a spectral decomposition of $T_{N_L}$ in terms of its eigenvalues and eigenvectors, which are, for instance, obtainable
with the help of some standard linear algebra software package~\cite{LAPACK:1999zu} due to the typically small size of the 
problem and the tridiagonal form of $T_{N_L}$. 

In principle, the above approach can be applied to any function $f:\re \rightarrow \re$ by replacing the inverse
square root in \eq{eq:LanczosApproach1} with that general function. The vital
questions are, whether the expression in \eq{eq:LanczosApproach1} then still converges and - from a more pragmatic point
of view - whether one has a reliable and practical stopping criterion at hand indicating the actually achieved
relative accuracy, that has been reached after a certain number of Lanczos iterations, to be below the 
prescribed value $\LanczosRelAcc$. Here, the notion 'practical' refers to the numerical costs of that criterion, 
since a direct test would always be possible, at least in principle. 

For the case of the inverse square root, \ie for $f(x) = x^{-\alpha}$ with $\alpha=1/2$ such a stopping criterion has been 
devised in \Ref{Borici:1999ws} as presented in \eq{eq:StoppingCriterionFoLanczos}. In fact, the given condition is just  
the standard CG-stopping criterion\footnote{Up to finite machine precision rounding errors which are, however, ignored 
here and in the following.}~\cite{Borici:1999ws,vandenEshof:2002ms}, which guarantees the residual
\beq
\label{eq:DefOfResidualGamma0}
\Delta_{N_L}(0) = \omega - \LanczosMatrix Q T^{-1}_{N_L} |\omega|\hat e_0
\eeq
of a corresponding CG-process~\cite{Press:2007zu} for applying the inverse of $\LanczosMatrix$ to $\omega$
after $N_L$ iterations to be bounded by
\beq
\left|\Delta_{N_L}(0) \right| \le  \LanczosRelAcc\cdot |\omega|.
\eeq
The main statement of \Ref{Borici:1999ws}, which was proven in \Ref{vandenEshof:2002ms}, is that the latter CG-stopping 
criterion is inherited also to the case of the inverse square root. 
 
Here, one further step shall be taken. It will be shown in the following that the same condition given in 
\eq{eq:StoppingCriterionFoLanczos} can also be exploited to monitor and proof the convergence of
\beq
\label{eq:LanczosApproach2}
Q f(T_{N_L}) |\omega| \hat e_0
\stackrel{N_L\rightarrow N_V}{\longrightarrow}
f(\LanczosMatrix)\omega,
\eeq
provided that the real, non-zero function $f:\re \rightarrow \re$ can be written in the form
\bea
\label{eq:FormOfF}
f(x) &\fhs{-3mm}=\fhs{-3mm}&  \int\limits_0^\infty \mbox{d}t \frac{a(t)}{x + b(t)}, \quad \mbox{with}\quad \forall t,t'\ge 0: a(t),b(t) \in \re,\,\,
b(t)\ge 0,\,\, a(t)\cdot a(t')\ge 0. \quad \quad\quad
\eea
Examples of functions lying in this class will be given later. For now we continue with the proof, which is just
an extension of the one originally given in \Ref{vandenEshof:2002ms}. For that purpose we define the generalized 
residuals\footnote{The definition of the generalized residuals in \eq{eq:DefOfResidual} becomes identical to \eq{eq:DefOfResidualGamma0}
for $\gamma=0$.}
\beq
\label{eq:DefOfResidual}
\Delta_{N_L}(\gamma) = \omega - \left(\LanczosMatrix + \gamma\ID \right) Q \left[T_{N_L}+\gamma\ID_{N_L}\right]^{-1} |\omega|\hat e_0,
\eeq
where $0\le\gamma\in\re$ is some non-negative real number, $\ID$ is the identity acting on the same space as $\LanczosMatrix$, 
and $\ID_{N_L}$ is the $N_L\times N_L$ identity matrix. In the following we will need the explicit relation~\cite{vandenEshof:2002ms} 
between $\Delta_{N_L}(\gamma)$ and $\Delta_{N_L}(0)$ given as
\beq
\label{eq:ExactRelBetResiduals}
\Delta_{N_L}(\gamma) = \zeta_{N_L}(\gamma) \cdot \Delta_{N_L}(0) \quad \mbox{with} \quad
\zeta_{N_L}(\gamma) = \prod\limits_{k=0}^{N_L-1} \frac{\beta_{N_L,k}}{\beta_{N_L,k}+\gamma},
\eeq
where $\beta_{N_L,k}$ denotes the $k-th$ eigenvalue of $T_{N_L}$. The above relation directly follows from
the representation~\cite{Paige:1995} of the generalized residuals $\Delta_{N_L}(\gamma)$ according to
\bea
\Delta_{N_L}(\gamma) &=& \frac{\pi_{N_L,\gamma}(\LanczosMatrix+\gamma\ID)\omega}{\pi_{N_L,\gamma}(0)},
\eea
which is given here in terms of the so-called Lanczos polynomials~\cite{Paige:1995} defined as
\bea
\pi_{N_L,\gamma}(x) &=& \det\left(x\ID_{N_L} - \left[ T_{N_L}+ \gamma\ID_{N_L}  \right]   \right),
\eea
such that $\pi_{N_L,\gamma}(x)$ is a polynomial of degree $N_L$. With this preparation at hand one finds in particular that 
$0 < \zeta_{N_L}(\gamma) \le 1$, since $T_{N_L}$ is positive and $\gamma\ge 0$. 

The actually relevant residual, which will be considered here to prove the convergence of \eq{eq:LanczosApproach2}, 
is defined as 
\bea
\Delta_{f,N_L} &=&  \omega - \left[f(\LanczosMatrix)\right]^{-1} Q f(T_{N_L}) |\omega| \hat e_0. 
\eea
The sought-after proof can then be established through
\bea
\left| \Delta_{f,N_L} \right |
&\fhs{-3mm}=\fhs{-3mm}&\Bigg| \left[f(\LanczosMatrix)\right]^{-1}\fhs{-2mm} \int\limits_0^\infty \mbox{d}t\,
a(t)\cdot \left[\LanczosMatrix+b(t)\right]^{-1} 
\Big(\omega -   \left[\LanczosMatrix+b(t)\right] Q\left[T_{N_L} + b(t) \right]^{-1} |\omega| \hat e_0  \Big) \Bigg| \quad \quad\quad\\
&\fhs{-3mm}=\fhs{-3mm}& \Big| X\, \Delta_{N_L}(0)  \Big| \\
\label{eq:ProofOfLanczos3}
&\fhs{-3mm}\le\fhs{-3mm}& |\Delta_{N_L}(0)|    \le \LanczosRelAcc\cdot |\omega|,
\eea
where the operator $X$ is defined as
\bea
X &=& \left[f(\LanczosMatrix)\right]^{-1}  \int\limits_0^\infty \mbox{d}t\,
a(t)\cdot \zeta_{N_L}(b(t))\cdot \left[\LanczosMatrix+b(t)\right]^{-1}.
\eea
The missing link for arriving at \eq{eq:ProofOfLanczos3} is to show that the operator norm 
$|X|\equiv\sup_v |Xv|/|v|$ is bounded from above by one, \ie $|X|\le 1$, where the latter supremum is 
performed over all vectors $v$. For that purpose let $\nu$ be an arbitrary eigenvalue of $\LanczosMatrix$ 
with associated eigenvector $\xi$. Exploiting the requirements $b(t)\ge 0$ and $a(t)\cdot a(t')\ge 0$
specified in \eq{eq:FormOfF} as well as the aforementioned relation $0 <\zeta_{N_L}(b(t))\le 1$, one 
directly finds 
\bea
\left| X\xi\right| &=& \left| \left[f(\nu)\right]^{-1}   \int\limits_0^\infty \mbox{d}t\,
\zeta_{N_L}(b(t))\, \frac{a(t)}{\nu+b(t)}  \xi  \right| \\
&\le& \left| \left[f(\nu)\right]^{-1}  \int\limits_0^\infty \mbox{d}t\,
 \frac{a(t)}{\nu+b(t)} \right| \cdot \left|\xi  \right| = \left|\xi  \right|,
\eea
which proofs $|X|\le 1$ and thus \eq{eq:ProofOfLanczos3}. 
Under the assumption that the standard CG-algorithm converges for the given matrix $A$ and the vector $\omega$
we have now shown that the approximation in \eq{eq:LanczosApproach2} also converges and that the standard CG-stopping
criterion given in \eq{eq:StoppingCriterionFoLanczos} guarantees the relative error $|\Delta_{f,N_L}|/|\omega|$ to be smaller
than $\LanczosRelAcc$, provided that the considered function $f(x)$ can be written in the form of \eq{eq:FormOfF}.
This is indeed the case for the inverse square root according to the integral representation
\bdm
x^{-\frac{1}{2}} = 
\frac{2}{\pi} \int\limits_0^\infty \mbox{d}t \left(x +t^2\right)^{-1},
\edm
which was used in \Refs{Borici:1999ws,vandenEshof:2002ms}. Besides this particular case, the required
preconditions are also fulfilled for arbitrary inverse roots, \ie $f(x) = x^{-\alpha},\,0<\alpha<1$, 
according to the more general integral identity
\bea
\label{eq:MoreGeneralIntIdentity}
x^{-\alpha} &=&
\frac{\sin\left((1-\alpha)\pi  \right)} {(1-\alpha)\pi} 
\int\limits_0^\infty \mbox{dt} \left(x + t^{\frac{1}{1-\alpha}}\right)^{-1}, \quad 0<\alpha<1,
\eea
which was obtained here with the help of some computer algebra system~\cite{Mathematica:2007yu}. Combining this Lanczos-based
method with direct applications of the operator $\LanczosMatrix$ and its inverse $\LanczosMatrix^{-1}$, 
computable by the standard CG-algorithm, one can thus calculate $f(\LanczosMatrix)\omega$ for any function of the form 
$f(x)=x^{-\alpha},\, \alpha\in \re$, as announced.

Finally, an example for the convergence of this approach shall be given. For that purpose we consider a test setup 
consisting of three different functions $f_i(x)$, $i=1,2,3$ given as
\bea
\label{eq:DefOfKrylovTestFunc1}
f_1(x) &=& x^{-1}, \\
\label{eq:DefOfKrylovTestFunc2}
f_2(x) &=& x^{-1/2}, \\
\label{eq:DefOfKrylovTestFunc3}
f_3(x) &=& x^{-1/3},
\eea
where the first case corresponds to the standard case of a matrix inversion computable by the usual CG-algorithm,
the second example refers to the calculation of the inverse square root as proposed in \Ref{Borici:1999ws}, and 
the third case belongs to the extended scenario discussed above. For the purpose of determining the achieved relative 
accuracy the following measure
\bea
\label{eq:DefOfRelAccForKrylovApproach}
\delta_i &\equiv& \frac{|\Delta_{f_i,N_L}|}{|\omega|} \fhs{2mm}=\fhs{2mm}  \frac{1}{|\omega|} \cdot 
\left| \omega - A\left( \prod\limits_{j=1}^{i-1} f_i(\LanczosMatrix)\right) 
Qf_i\left(T_{N_L}\right) |\omega| \hat e_0 \right|, \quad i=1,2,3\quad\quad
\eea
is introduced. In \fig{fig:ExampleForLanczosApproach}a the dependence of the latter relative accuracy on the number $N_L$
of performed Lanczos iterations is presented for the three considered test cases. These results have been obtained
for the particular case $\LanczosMatrix=\fermiMatDouble$, which is actually relevant for this study. Here, however, the
unpreconditioned version of the fermion matrix $\fermiMatDouble$ according to its original definition in 
\eq{eq:DefOfFermiMatM} has been applied. The underlying lattice volume was chosen to be $V=8^3\times 16$, the degenerate 
Yukawa coupling constants were set to $\hat y_t=\hat y_b=1$, and the scalar field $\Phi$ appearing within the operator $\fermiMat$ 
as well as the pseudo fermion vector $\omega$ were both randomly sampled according to a standard Gauss distribution. The 
presented data then display the results obtained in a single inversion process starting with identical field configurations $\Phi$
and $\omega$ in each considered test case.
 
For the scenario $f_1(x)$ at sufficiently large values of $N_L$ it is well known that the relative deviation $\delta_1$ 
will finally vanish exponentially with the number of performed Lanczos iterations $N_L$. Before this asymptotic behaviour
sets in, however, the relative deviation $\delta_1$ is not expected to decrease monotonically. This is clearly
seen in \fig{fig:ExampleForLanczosApproach}a. Moreover, one also observes in that presentation that the 
relative deviations $\delta_2$, $\delta_3$ for the other two test cases are bounded by $\delta_1$, as demonstrated above.

In \fig{fig:ExampleForLanczosApproach}b the achieved relative accuracies $\delta_i$ are presented versus the prescribed 
relative accuracy specified through $\LanczosRelAcc$. For the case $f_1(x)$ this approximately yields a straight line 
with slope of one, since the estimate $|\Delta_{f_1,N_L}|=1/\rho_{N_L}$ for the norm of the residual in 
\eq{eq:StoppingCriterionFoLanczos} is exact in this scenario and $1/\rho_{N_L}\approx \LanczosRelAcc |\omega|$. In the 
two other cases the actually obtained relative accuracy is a more complicated function of $\LanczosRelAcc$ with the very 
essential condition, however, that one always has $\delta_i\le\LanczosRelAcc$, as proven above.

%\includeFigDouble{KrylovExampleAchievedPrecVsIterations}{KrylovExampleAchievedPrecVsPrescribedPrec}
\includeFigDouble{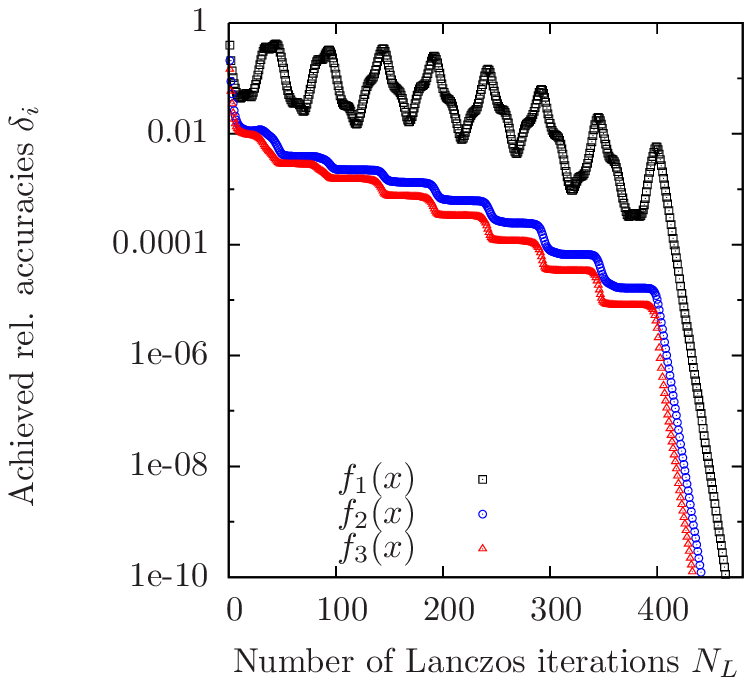}{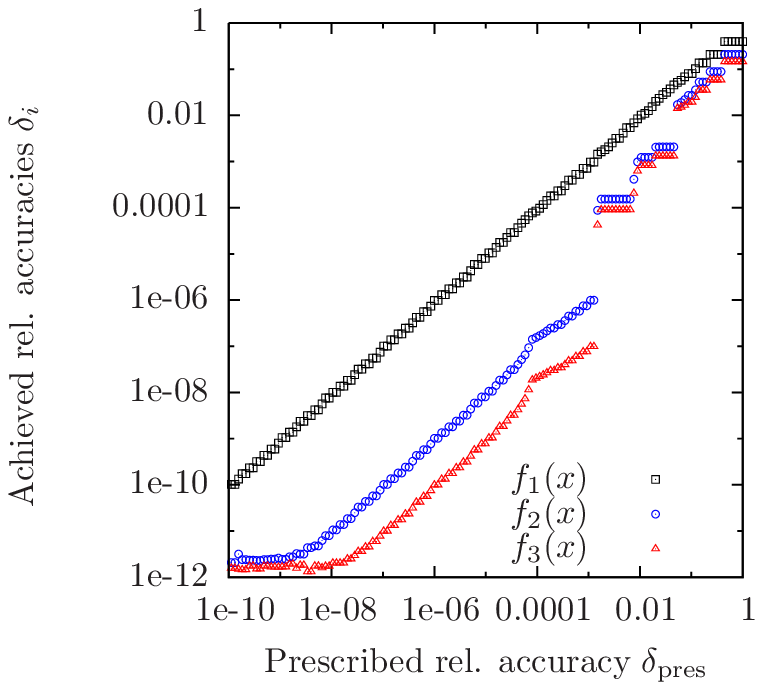}
{fig:ExampleForLanczosApproach}
{The relative accuracies $\delta_i$ as defined in \eq{eq:DefOfRelAccForKrylovApproach} are presented in panel (a) versus
the number of performed Lanczos iterations $N_L$ for the three test cases $f_1(x)$, $f_2(x)$, and $f_3(x)$
specified in \eqs{eq:DefOfKrylovTestFunc1}{eq:DefOfKrylovTestFunc3}. The same data on the relative accuracies
are also shown versus the prescribed relative accuracy $\LanczosRelAcc$ in panel (b). The underlying Lanczos iterations were performed
for the operator $\LanczosMatrix=\fermiMatDouble$ with a lattice volume of $V=8^3\times 16$, and degenerate Yukawa coupling constants 
chosen to be $\hat y_t=\hat y_b=1$. All three inversion problems started with the identical field configurations $\Phi$ and $\omega$,
which were randomly sampled according to a standard Gauss distribution. 
}
{Dependence of the relative accuracies $\delta_i$ of the considered Krylov-space based methods on the number of performed 
Lanczos iterations $N_L$ and on the prescribed relative accuracy $\LanczosRelAcc$.}

 %-----------------------------------------------------------------------------------------------------
\chapter{Fast implementations of FFT in four dimensions}
\label{sec:xFFT}

According to the absence of gauge fields in the here considered Higgs-Yukawa model
the Dirac operator $\D$ can very efficiently be constructed in momentum space as
discussed in \sect{sec:HMCAlgorithm}. Since the operator $B$ in \eq{eq:DefOfMatB}, on the other hand, 
is local in position space, the aforementioned strategy requires the pseudo fermion
vector $\omega$ to be transformed from position to momentum space or vice versa
each time one of the operators $\D$ and $B$, respectively, is applied to $\omega$.
It is well known that this can very efficiently be done by means of a Fast Fourier
Transform\footnote{The notion 'Fourier transform' will always refer to a discrete
Fourier transform of a vector or a multi-dimensional array of complex numbers in this
section.} (FFT), the numerical complexity of which being of order $O(V\cdot\log(V))$.
Since the complexity class of applying the latter operators $\D$ and $B$ is only of 
order $O(V)$ according to their respective diagonal structure, the aforementioned
Fourier transformations thus constitute the limiting factor of a practical implementation of
the considered computation strategy with respect to the overall numerical performance
in the limit of large lattice volumes.

The availability of an efficient software code for performing Fast Fourier Transformations
on an actually given computer hardware is therefore of significant importance for the success
of the above computation strategy in practice. In this section some remarks on the technical
details concerning such software implementations shall therefore be given, including also
the here applied form of their parallelization.

As a first step we begin with the consideration of a one-dimensional Fourier transform
of the complex vector $\nu_x$ with the index $x$ running from $0$ to $L-1$. Neglecting any normalization
constants the Fourier transform $\tilde \nu_p$ of $\nu_x$ is here defined as
\bea
\label{eq:DefOfFFTProblem}
\tilde \nu_p &=& \sum\limits_{x=0}^{L-1} \nu_x \cdot e^{-2\pi ipx/L}, \quad p=0,\ldots,L-1.
\eea
Implementing this formula exactly as written would then lead to an algorithm lying in 
the complexity class $O(L^2)$, thus rendering its application in a practical approach
unfeasible. The well known cure to this problem is the Fast Fourier Transform. In fact,
the notion 'Fast Fourier Transform' does not refer to one specific algorithm but rather
to a whole class of algorithms that can compute \eq{eq:DefOfFFTProblem} with a numerical
complexity of order $O(L\cdot\log(L))$. The probably most widely known approach
is the (radix-2) Cooley-Tukey algorithm~\cite{Cooley:1965zu}. In case of $L$ being a 
power of $2$, which shall always be assumed in the following, it computes $\tilde \nu_p$ 
according to\footnote{$\FT(o', s', L', \nu', p')$ is defined through \eq{eq:DefOfFFTrecExp}, 
while \eq{eq:FFTrecExpEquiv1} only holds for $L'\ge 2$.}
\bea
\tilde \nu_p &=& \FT(0, 1, L, \nu, p), \\
\label{eq:DefOfFFTrecExp}
\FT(o', s', L', \nu', p') &=& \sum\limits_{j=0}^{L'-1} e^{-2\pi ip'j/L'} \cdot \nu'_{js'+o'} \\
\label{eq:FFTrecExpEquiv1}
&=& \sum\limits_{j=0}^{L'/2-1} e^{-2\pi ip'2j/L'} \cdot \nu'_{2js'+o'} \\
&+& e^{-2\pi ip'/L'} \cdot\sum\limits_{j=0}^{L'/2-1} e^{-2\pi ip'2j/L'} 
\cdot \nu'_{2js'+o'+s'} \nonumber
\eea
through the recursive formula
\bea
\FT(o', s', L', \nu', p') &=& \FT(o', 2s', L'/2, \nu', p') \\ \nonumber
&+& e^{-2\pi ip'/L'} \cdot \FT(o'+s', 2s', L'/2, \nu', p') 
\eea
for $L'>1$ and else
\bea
\FT(o', s', 1, \nu', p') &=& \nu'_{o'}.
\eea
Computing this recursive algorithm literally as described above would again lead to
the numerical expenses being of order $O(L^2)$. The crucial observation, however, is
that 
\bea
\FT(o', s', L', \nu', p') &=& \FT(o', s', L', \nu', p'+L')
\eea
such that the quantity $\FT(o', s', L', \nu', p')$ only needs to be computed for
$p'=0,\ldots,L'-1$. If one further exploits $\exp(-2\pi ip'/L')=-\exp(-2\pi ip''/L')$
for $p''=p'+L'/2$ and $L'\ge 2$ one finds that a total number of
\bea
N_{CT}^{ADD} &=& 3 L\cdot \log_2(L) 
\eea
real additions and 
\bea
N_{CT}^{MUL} &=& 2 L\cdot \log_2(L) 
\eea
real multiplications are required for the here presented Cooley-Tukey algorithm.
For clarification it is remarked that a single complex multiplication translates into
four real multiplications plus two real additions. Moreover, trivial multiplications,
\ie multiplications with $\pm1$ or $\pm i$, have been counted here as full multiplications.
However, the given recursive formulation, though very helpful for the mathematical analysis 
of the underlying numerical operations, is usually not directly employed for the 
practical implementation of the algorithm. For that purpose some more effort
has to be invested into an appropriate bookkeeping of the computed interim results
$\FT(o', s', L', \nu', p')$. A very convenient approach is based on the reordering of the
start vector $\nu\rightarrow \hat \nu$ according to $\hat \nu_{\hat x} = \nu_{x}$ such that
$\hat x$ is the bit-reversal of the binary representation of the number $x$ in the binary
number system. These details, however, shall not be discussed here\footnote{A basic 
introduction to the implementation of the Cooley-Tukey algorithm can, for instance, be 
found in \Ref{Press:2007zu}. An advanced approach is presented in \Ref{Burrus:2009zu}.}.

As already pointed out above, the (radix-2) Cooley-Tukey approach is only one variant of a
FFT algorithm. In fact, there are other methods, for instance split-radix algorithms~\cite{Yavne:1968zu,duhamel:1984zu}
and even more efficient variations thereof~\cite{Frigo:2007}, requiring at least $\proz{20}$ less 
floating point operations, while still belonging to the same complexity class $O(L\cdot \log(L))$. 
Since the aforementioned number of floating point operations counts all performed real 
multiplications plus the additions, one would thus expect these more elaborate 
algorithms to be clearly superior to the Cooley-Tukey approach in terms of required runtime, 
\ie the time needed to perform a specific FFT.

This, however, is not necessarily the case in practice. The reason is that the true 
runtime speed of a certain software code on a given hardware depends on many more
variables than just the total number of required elementary operations and the clock-speed
of the processor in terms of operations per time. Among these factors are, for instance,
the achieved level of cache coherence and data locality, as discussed later in this
section. 

\includeFigSingleMedium{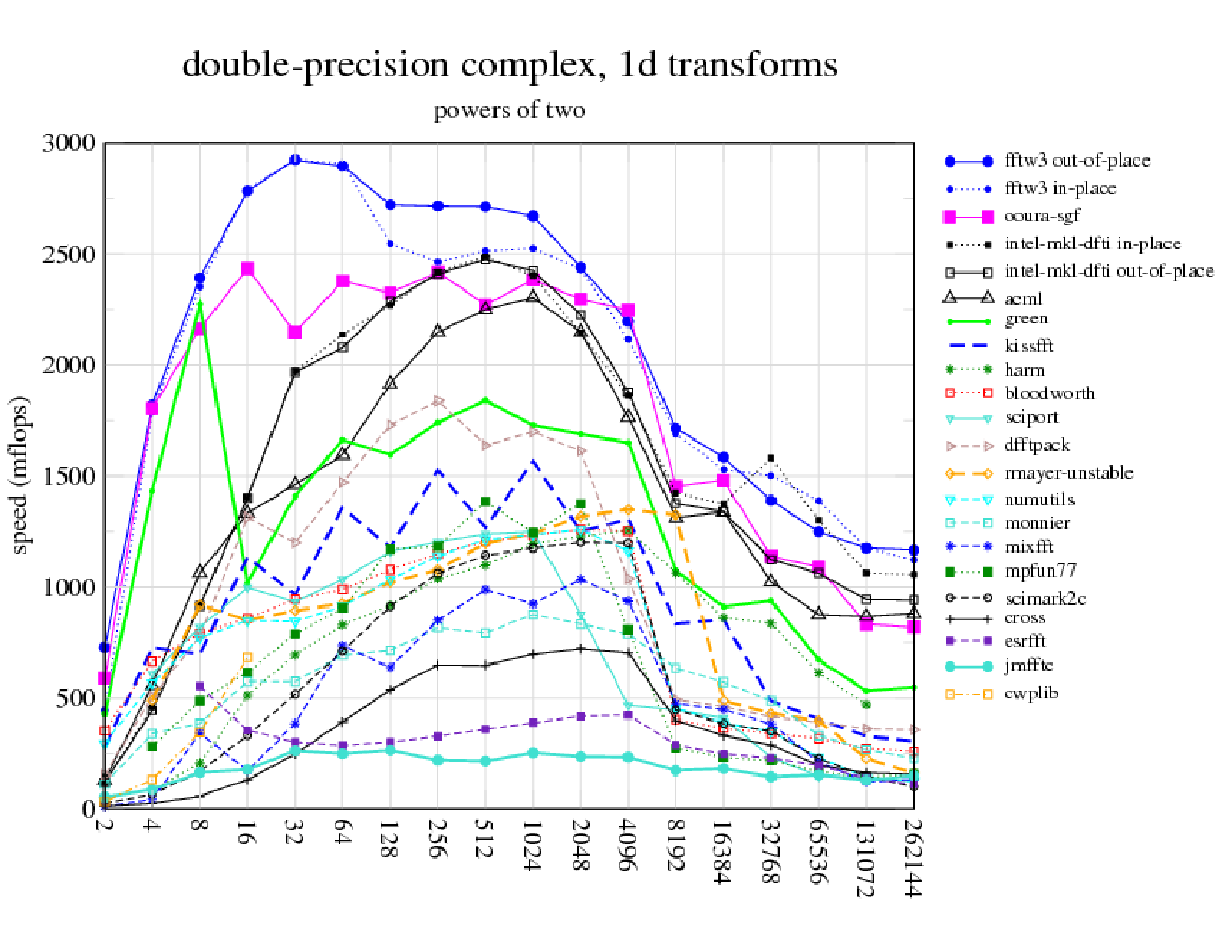}
{fig:FFTWcomparison1DdoubleComplex}
{The speeds of various FFT implementations is presented versus the vector length $L$
in terms of MFLOPS as defined in \eq{eq:DefOfMFlops1D}. The underlying complex
vector $\nu_x$ has been processed in double-precision. Details about the various
competitors of the FFTW implementation~\cite{FFTW05} can be found in \Ref{FFTWHomepage},
which is also the source of this plot. These measurements have been made on a
{\textit{Dual Core AMD Opteron(tm) Processor 275, 2.2 GHz}}. The solid lines are only meant 
to guide the eye.
}
{Speed of various FFT implementations in MFLOPS.} 

According to the large variety of hardware platforms it is therefore very difficult
to devise a particular algorithm that is optimal for all underlying hardware 
architectures. A tremendous collection of various FFT algorithms and corresponding software
realizations thereof has therefore been implemented and assembled into a single software 
library known as the 'Fasted Fourier Transform in the West' (FFTW)~\cite{FFTW05}, such that 
the library automatically picks that software code that runs most efficiently on the respective
hardware architecture. This selection procedure will be referred to in the following
as the internal FFTW tuning process.
 
In \fig{fig:FFTWcomparison1DdoubleComplex}, which was taken from the 
FFTW homepage~\cite{FFTWHomepage}, the performance of the FFTW library is
compared to the corresponding runtime speeds of other publicly available FFT
implementations, as detailed in \Ref{FFTWHomepage}. The here specified 
MFLOPS-numbers (million floating point operations per second) have been 
determined according to
\bea
\label{eq:DefOfMFlops1D}
\mbox{MFLOPS} &=& 5L\cdot \log_2(L) / t_{\mu s},
\eea
where $t_{\mu s}$ denotes here the number of micro seconds that the respective implementation
needed to perform the Fast Fourier Transform of the one-dimensional complex vector $\nu_x$ with
vector length $L$. For clarification it is remarked that this quantity does not give
the correct number of floating point operations that the respective algorithm actually
performed. In fact, the factor $5L\cdot \log_2(L)$ is just a reasonable normalization
of the measured time $t_{\mu s}$, which is, however, motivated by the number of floating
point operations $N_{CT}^{ADD}+N_{CT}^{MUL}$ that the above presented Cooley-Tukey approach
would have required\footnote{Apart from the aforementioned unnecessary counting of trivial multiplications.}.

From this presentation one can infer that the FFTW library indeed provides the fastest one-dimensional,
complex FFT implementation among all here considered competing software codes. 
However, one also learns from \fig{fig:FFTWcomparison1DdoubleComplex} that the achieved
numerical performance significantly decreases with growing vector length $L$ due to a reduction
of the underlying cache coherency and data locality, as will later be discussed in this section.

Concerning the technical details of this comparison the underlying hardware architecture has been a 
{\textit{Dual Core AMD Opteron(tm) Processor 275, 2.2 GHz, 64 bit mode}} as further specified in \Ref{FFTWHomepage} and the
transformed complex vector $\nu_x$ has been represented in double precision, \ie with 64 bit per
real number. This setup has been chosen among the various speed monitorings available in \Ref{FFTWHomepage},
since it comes closest to the scenario of the lattice calculations considered in the context of this work,
which have mainly been performed on the {\textit{HP XC4000 System}}~\cite{XC4000} at the {\textit{Scientific Supercomputing Center Karlsruhe}},
built on the basis of {\textit{AMD Opteron(tm) Processors}}, however with a higher clock-speed of $2.6\, GHz$. Though the hardware 
architecture actually underlying the {\textit{HP XC4000 System}} is not exactly identical to the one tested above, it is 
nevertheless reasonable to assume the FFTW library to provide the most efficient one-dimensional FFT implementation 
also on that machine.

For the purpose of this study, however, we are mainly interested in four-dimensional Fourier transforms. More precisely,
it is the complex pseudo fermion field $\omega_{x,j}\in \Comp$ that needs to be transformed with $x\equiv(x_0,x_1,x_2,x_3)$
denoting the four-dimensional lattice site index and $j=0,\ldots,N_I-1$, $N_I=8$ being the inner spinor index associated to the 
fermion doublets. This multi-dimensional Fourier transform can, however, trivially be decomposed into a series of one-dimensional
Fourier transforms computed along each respective space-time axis according to
\bea
\tilde \omega_{(p_0,p_1,p_2,p_3),j} &=& \sum\limits_{x_0=0}^{L_0-1} e^{-2\pi i x_0p_0/L_0} \sum\limits_{x_1=0}^{L_1-1} \ldots
\sum\limits_{x_2=0}^{L_2-1} \ldots \sum\limits_{x_3=0}^{L_3-1} e^{-2\pi i x_3p_3/L_3} \cdot \omega_{(x_0,x_1,x_2,x_3),j} \quad\quad
\eea
where $L_\mu$ denotes the lattice size in direction $\mu$ and all global normalization factors have again be discarded
for simplicity. The FFTW library, however, directly comes along with routines automatically exploiting and optimizing this 
decomposition scheme internally. The resulting MFLOPS-rates as observed for a typical test case with $N_I=8$ and the lattice side
lengths being a power of two are presented in \fig{fig:FFTComparisonMlops}a, where the presented MFLOPS-rates have here been 
determined according to
\bea
\label{eq:DefOfMFlops4D}
\mbox{MFLOPS} &=& N_I\cdot 5V\cdot \log_2(V) / t_{\mu s},
\eea
with $V=L_0L_1L_2L_3$ being the lattice volume and $t_{\mu s}$ denoting the execution time in micro seconds. Again, this quantity 
does not give the exact number of floating point operations actually performed, but it is motivated by the operation count of the 
multi-dimensional Cooley-Tukey algorithm in the previously described sense. 
It is further remarked that the presented MFLOPS-rates have directly been obtained on the aforementioned 
{\textit{HP XC4000 System}} with version 3.2.2 of the FFTW library, since no four-dimensional performance tests were 
available on the FFTW homepage~\cite{FFTWHomepage} at the time this work has been compiled.

As expected one observes in \fig{fig:FFTComparisonMlops}a that the numerical performance of the FFTW implementation significantly 
decreases with growing lattice volume. On the largest presented volumes one moreover finds it to be of similar magnitude, even somewhat
larger than the results in \fig{fig:FFTWcomparison1DdoubleComplex}, when mapping the respective one-dimensional and four-dimensional 
volumes adequately to each other, while also respecting the higher clock-speed of $2.6\, GHz$ as compared to the $2.2\, GHz$
underlying the measurements in \fig{fig:FFTWcomparison1DdoubleComplex}.

It is remarked that the notions 'fully tuned' and 'partly tuned' appearing in the legend of \fig{fig:FFTComparisonMlops} refer to the set
of FFT implementations that are actually considered inside the FFTW library for the internal optimization procedure. In the fully tuned
case, associated to the optimization mode {\textit{FFTW\_MEASURE}} | {\textit{FFTW\_EXHAUSTIVE}} in \Ref{FFTW05}, the system explicitly 
and individually measures the true wall clock runtime for all available FFT implementations contained in the FFTW library and 
selects then the actually fastest code. In the other case, referred to as 'partly tuned', which is specified through the flag 
setting {\textit{FFTW\_MEASURE}}, this optimization is restricted to a certain subset of these implementations, resulting then in
a reduced amount of time required by the optimization process. From the comparison in \fig{fig:FFTComparisonMlops}a one infers that
the performance of the partly tuned approach is almost identical to that of the fully tuned FFTW code in the example presented here. 
In the following we will nevertheless always consider the fully tuned approach with the intention to obtain the optimal performance 
out of the FFTW routines.

However, it turns out that the achieved performance of the already very efficient FFTW implementation of the four-dimensional complex 
Fourier transform can be pushed to an even higher level at least in the here considered setup which, however, is typical for the needs 
of this study. For that purpose a new implementation of a complex FFT in four dimensions has been built up, which will be referred to 
in the following as the 'xFFT' implementation. Since this development has only been pursued in the spirit of some auxiliary work with 
the main motivation of speeding up the lattice calculations actually performed in the context of the considered Higgs-Yukawa model, the
xFFT implementation has only reached a rather low level of sophistication in the sense that it can only be applied to four-dimensional 
lattices with equal side extensions given by powers of two. Moreover, it is especially tuned for {\textit{AMD Opteron(tm) Processors}} and
the case of large lattice volumes. On small lattice sizes it thus runs rather inefficient. However, on the largest here presented
volume the performance of the xFFT code on the considered hardware architecture is approximately $\proz{35}$ higher than that of the 
fully tuned FFTW implementation in terms of MFLOPS as can be inferred from \fig{fig:FFTComparisonMlops}a.

%\includeFigDouble{FFTcomparison1CoreInMFlops}{FFTcomparison2CoreInMFlops}
\includeFigDouble{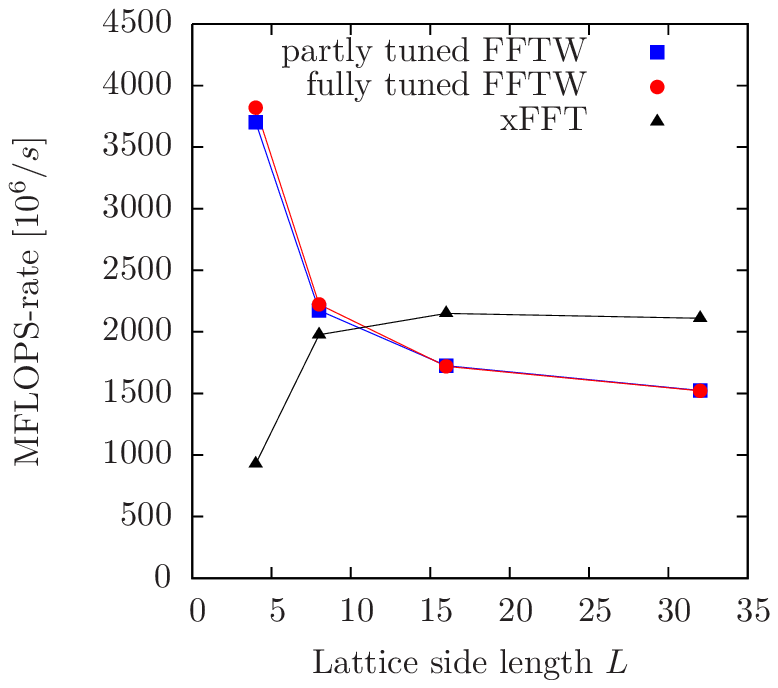}{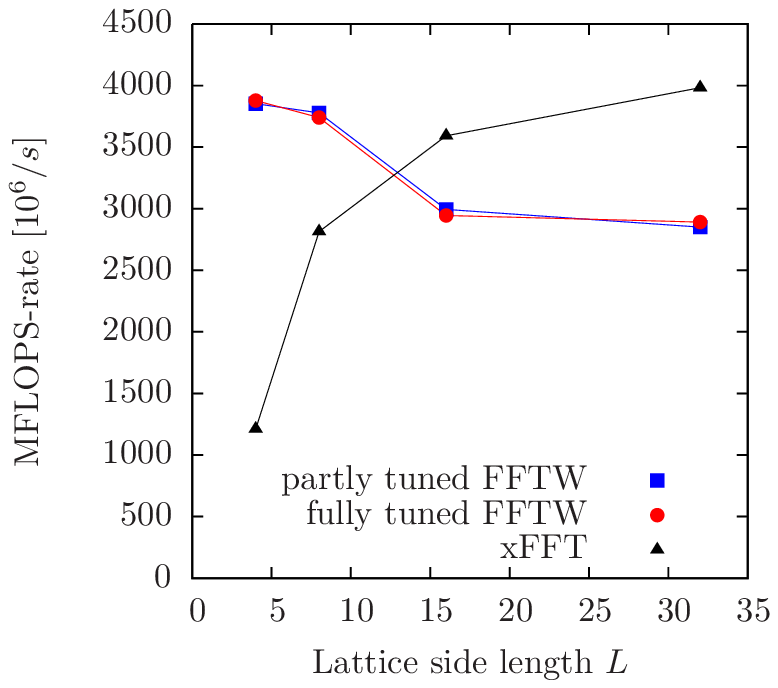}
{fig:FFTComparisonMlops}
{The speed of the FFTW implementation (version 3.2.2) and the xFFT implementation are compared to each other in terms of MFLOPS as defined
in \eq{eq:DefOfMFlops4D}. The underlying numerical task is the computation of a four-dimensional FFT of the complex vector
$\omega_{x,j}$ with $x$ being the lattice site index and $j=0,\ldots,N_I-1$ denoting the inner spinor index. This comparison has been 
performed with $N_I=8$ on varying lattice volumes $L^4$, the side length of which being restricted to powers of two as specified
on the horizontal axis. All numerical calculations were performed with double-precision. The notions of 'partly tuned FFTW'
and 'fully tuned FFTW' refer to the sets of FFT implementations that are used in the internal FFTW optimization procedure as discussed in
the main text. Panel (a) presents the situation of a single CPU core computing the whole transform, while panel (b) shows the 
achieved speed, when employing two CPU cores for the same calculations. These runs were performed on the {\textit{HP XC4000 System}} 
based on AMD machines as discussed in the main text.
}
{Comparison of the speed of the FFTW implementation and the xFFT implementation in MFLOPS.}

For clarification it is explicitly pointed out that the xFFT implementation is not based on a superior algorithm in the sense
that it would require a lower number of floating point operations. In fact, rather the opposite is the case, since the algorithmic 
foundation of the latter implementation is the multi-dimensional Cooley-Tukey approach which is clearly not the most efficient
algorithm in terms of floating point operations as discussed above. The improved performance is thus completely due to a more
efficient software realization of the underlying algorithm. 

To elaborate more on that the reader is assumed to be familiar with the basic concepts of the hardware architecture underlying modern
computer systems\footnote{A general introduction to modern computer hardware architecture can be found in \Ref{Hennessy:2006zu}. An 
introduction to corresponding software optimization strategies for the particular case of {\textit{AMD}} systems can be found in 
\Ref{AMDOpti:2009zu}.}. As a first step towards an understanding of how this improved efficiency level can be achieved, we will 
estimate the performance of the competing implementations FFTW and xFFT in terms of the ratio between the observed MFLOPS-rate as 
specified in \eq{eq:DefOfMFlops4D} and the theoretically possible rate $\Ppeak$, which is denoted as peak performance. The 
considered AMD processor core possesses a separate addition and a separate multiplication SSE-unit that can independently of each other 
perform two double precision operations simultaneously which, however, asymptotically takes 2 CPU-cycles in pipelining mode to 
complete. In the considered setup the total number of double precision floating point operations that can theoretically be performed 
in each second thus sums up to $\Ppeak= 0.5\cdot 2\cdot 2\cdot 2.6\cdot 10^9/s=5.2\cdot 10^9/s$. If one compares this number with the 
MFLOPS-rates presented in \fig{fig:FFTComparisonMlops}a for the largest presented volume, one finds that the performance of the FFTW 
implementation is around $\proz{29}$ while the one of the xFFT code is around $\proz{40}$. 

From this comparison one can infer that it is not so much the number of floating point operations that limit the speed of the considered
FFT implementations but rather other, administrative processes, such as the transportation of data from the main memory to the processor
and vice versa. To minimize the required data traffic the xFFT implementation is explicitly based on two-dimensional Fourier transforms
that are, for instance, first performed along the dimensions (0,1) and then along the dimensions (2,3). This approach then guarantees that 
the overall data amount of the transformed vector $\omega$, being in total $16\cdot N_I V$ bytes, is only transported twice from main memory 
to the processor and back for one four-dimensional FFT, provided that the data of each individual two-dimensional Fourier transform, which 
will be referred to as a two-dimensional data slice in the following, can simultaneously be stored in the data cache. Contrary to that one 
would need four such data transfers, if one would first perform all one-dimensional transforms along the first space-time dimension, then 
all one-dimensional transforms along the second dimension and so on. This observation is well-known in the literature. The crucial question 
is, whether the aforementioned data slices can indeed simultaneously be stored in the data cache.

To address the latter issue, let us now consider the specific example of the $V=32^4$ volume. In that scenario the latter data slices have the size 
$16\cdot L^2N_I$ bytes, \ie $128\,KB$ here. From the perspective of the total data amount, these data slices should perfectly fit at least 
into the Level 2 cache. There is, however, a crucial obstacle that can actually block the effective caching of the considered data slices. 
This problem arises from the method how the cache cell address $\Acache$ that is supposed to hold a particular datum, read from the
main memory address $\Amem$, is determined. In practice this is, at least in part, done by a modulo operation according to
\bea
\Acache = \Amem \mbox{ modulo } \Ncache,
\eea
where $\Ncache$ is the size of the cache, \ie usually a power of two. This is called an associative cache mapping strategy in the
literature. The problem now arises, for instance, when a one-dimensional Fourier transform is performed along the outer most 
index\footnote{To map the multi-dimensional vector index $(x_0,x_1,x_2,x_3),j$ of $\omega_{x,j}$ into the main memory, a 
single sequential address has to be computed. This is done here according to $\Amem = 16(j+x_3N_I+x_2L_3N_I+x_1L_2L_3N_I+x_0 L_1L_2L_3 N_I)$. 
In that sense $x_0$ would be the outer most index here.} of the vector $\omega$. In that case the memory addresses involved in that particular 
transformation would all differ by multiples of $d=16\cdot N_IL_1L_2L_3$. In the here considered case one would typically have $d>\Ncache$, 
while $d$ is also a power of two like $\Ncache$, which finally results in all data involved in the aforementioned Fourier transform to 
be mapped to the same cache address, thus kicking each other out of the cache. The result is a so-called low cache coherency which leads to 
the necessity of reloading the same data multiple times from the main memory. 

This problem is manifestly circumvent in the xFFT implementation by using internal buffer structures that explicitly break the
latter translation invariance of the cache addresses associated to the buffer addresses storing the aforementioned two-dimensional
data slices. It is remarked that these cache coherency considerations have a severe impact on the overall performance.

Another crucial point is the efficient prefetching of the required data as well as the associated question of data locality. 
The notion of 'prefetching' refers here to the possibility that the processor can notify the memory controller that a certain 
memory address will be requested in the future, thus allowing the memory controller to begin already with the data loading
process, while the processor can simultaneously continue with some other operations. In an optimal scenario this can
prevent the processor from having to wait for data that have not yet been delivered to the processor. An efficient prefetching 
strategy is thus vital for the success of many software implementations. Due to the above described bit-reversed reordering of 
the transformed vector $\omega$ this is especially true, and unfortunately also more complicated, in the case of a FFT 
implementation. 

In such a prefetching strategy it is very beneficial to request the data in sequential order, which is known as the problem 
of data locality. The rationale is that the memory latency times for delivering the datum at a certain memory address depend significantly 
on the previously requested memory address. This is because the main memory is built in form of a $N\times N$ matrix consisting of
$N$ rows and $N$ columns, where each matrix cell can store information. When a certain memory address is requested, the matrix
row containing that address has to be activated, unless it has already been activated in the previous memory access. This row
activation takes some time which is known as the row address strobe (RAS-latency). In a subsequent step the column of the matrix
containing the requested memory address has to be activated, which again takes some additional time known as column address strobe
(CAS-latency). If one addresses memory in some incoherent order, such as in a bit-reversed order for instance, one is confronted 
with the penalty of having to wait for the RAS- and CAS-latency times in each memory access. If one, on the other hand, accesses
the same data in a coherent order, one profits from the memory matrix row staying activated as long as possible, thus reducing
the amount of RAS-latency penalties encountered. 

In the implemented xFFT code special effort has been invested to improve the order in which the data is requested from the main 
memory to optimize data locality. In total the above described techniques sum up to the performance advantage on the considered
AMD architecture already presented in \fig{fig:FFTComparisonMlops}a. It is further remarked that the xFFT implementation, which 
has originally been designed for the latter architecture, is less advantageous on Intel machines. On an Intel Core 2 CPU it was, 
however, still found to be faster than the FFTW implementation in the here considered scenario by around $\proz{15}$.

Finally, some remarks shall be given concerning the parallelization of the Fourier transform. It is well known that the computation
of the underlying lower dimensional Fourier transforms can efficiently be assigned to different, remote computer nodes, \seeeg{FFTWHomepage}.
With this approach the Fourier transform can be split up into a large number of smaller tasks running on remote nodes, which finally
leads to a significantly smaller wall clock runtime. The efficiency of this approach in terms of achieved versus theoretical operation counts, 
however, suffers from the additional requirement of network communication and is thus usually smaller than that of a single node computation. 
In the context of the here considered Higgs-Yukawa model it was therefore found to be more suitable, at least for the studied lattice sizes,
to use a shared memory parallelization on a single computer mode. Though the scalability of such an approach is restricted by the number 
of available CPU cores on that single node, the advantage of this method lies in the higher achieved efficiency, thus ultimately allowing to 
generate more field configurations in total with a given amount of numerical resources than it would have been possible with a 
remote node parallelization ansatz. The applicability of the shared memory approach, of course, depends on the question, whether a single 
node provides enough performance to process a single lattice calculation, which, however, is the case thanks to the algorithmic improvements
in \chap{chap:SimAlgo}.

%\includeFigSingleMedium{FFTScaling}
\includeFigSingleMedium{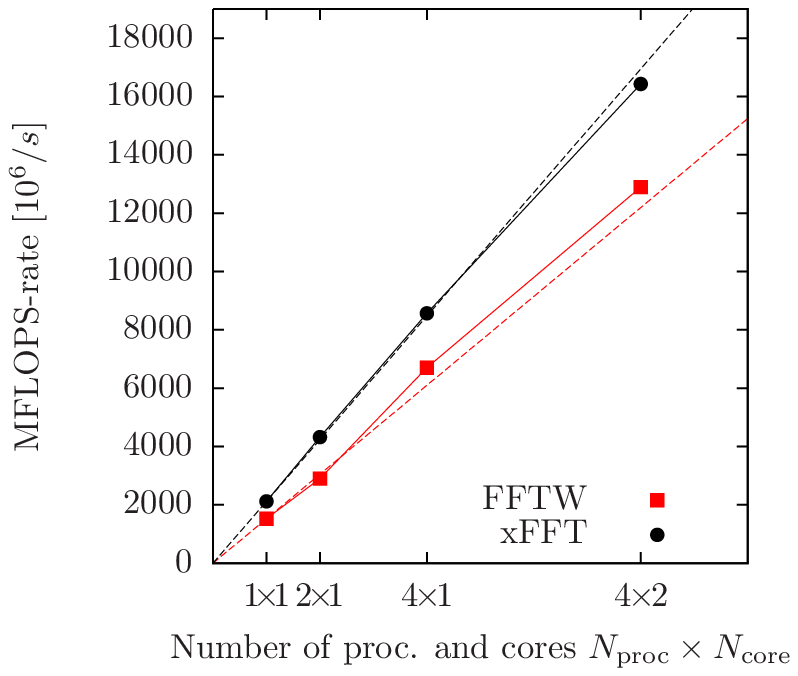}
{fig:FFTScaling}
{The speeds of the FFTW implementation (version 3.2.2) and the xFFT implementation are compared to each other in terms of MFLOPS as defined
in \eq{eq:DefOfMFlops4D}. The underlying numerical task is the computation of a four-dimensional FFT of the complex vector
$\omega_{x,j}$ with $x$ being the lattice site index and $j=0,\ldots,N_I-1$ denoting the inner spinor index. This comparison 
has been performed with $N_I=8$ on a \latticeX{32}{32}{.}The numerical task is here distributed to $\Nproc$ processors with 
$\Ncore$ cores each as specified on the horizontal axis, while the data $\omega_{x,j}$ are distributed to the $\Nproc$ memory segments
associated to the $\Nproc$ processors according to the underlying NUMA-architecture as explained in the main text. The dashed lines
are straight lines laid through the origin and the data point at $1 \times 1$ of the FFTW and the xFFT measurement, respectively,
to demonstrate the would-be behaviour of exactly linear scaling. The solid lines only guide the eye. All numerical calculations were 
performed with double-precision. The FFTW results were obtained after the full internal tuning procedure of the FFTW implementation has 
been applied. These runs were performed on the {\textit{HP XC4000 System}} based on AMD machines as discussed in the main text.
}
{Speed scaling behaviour of the FFTW implementation and the xFFT implementation with rising number of involved CPU cores.} 

An example of the MFLOPS-rates observed for the FFT computation when employing two CPU cores on a single node is presented in 
\fig{fig:FFTComparisonMlops}b. In both cases, \ie for the FFTW as well as the xFFT implementation, one observes a good scaling
behaviour in comparison with \fig{fig:FFTComparisonMlops}a at least on the largest presented lattice volume. If one naively employs
an increasing number of CPU cores for the FFT computation on a single node, however, the achieved MFLOPS-rates will finally 
run into a plateau. This is because the memory bus bandwidth will eventually become the limiting factor. 

A decent way to overcome this limitation is to exploit the Non-Uniform Memory Architecture (NUMA)~\cite{AMDNUMA:2006zu} underlying
the AMD systems installed in the aforementioned {\textit{HP XC4000 System}}. In this setup each computer node possesses $\Nproc$ processors
each consisting of two CPU-cores. Attached to each processor is a separate memory bus connecting the respective processor to
its individually associated main memory. The result is that each processor can access its own main memory independently from the
work load of the other processors and their respective memory buses. Moreover, the processors are also connected to each other 
by some additional bus, denoted as Hyper-Transport bus, which is dedicated to the inter-processor communication. 

The idea here is to split up the pseudo fermion vector $\omega_{x,j}$ according to its inner index $j$ into $\Nproc$ pieces and to
distribute these pieces to the $\Nproc$ different main memories. This allows then each of the $\Nproc$ processors to perform
the Fourier transform on its respective sub vector without interfering with the computations on the other processors. On each processor
the respective work load can then efficiently be shared among the two available cores as demonstrated in \fig{fig:FFTComparisonMlops}b.
The scaling of the total MFLOPS-rate resulting from this approach is presented in \fig{fig:FFTScaling} for the case of a node with
$\Nproc=4$ having $\Ncore=2$ CPU cores each and very good scaling behaviour is observed in both cases, \ie for the FFTW as well as the 
xFFT implementation, while again the xFFT approach yields significantly better performance. 

It is further remarked that this distributed storage of the vector $\omega$ is maintained during the whole lattice
computation, since most other operations on the vector $\omega$, such as vector additions, scalar products and even the application 
of the operator $B$ can easily be split up into corresponding non-interfering tasks, provided that the inner index is adequately distributed. 
In such a scenario it is only the application of the momentum space representation of the Dirac operator $\D$ being non-diagonal only with 
respect to the inner index $j$ that then finally mixes vector components assigned to different processor memories. In this approach the 
application of the operator $\D$ is thus the only operation that does not scale with the number $\Nproc$. It can, however, still be efficiently 
implemented thanks to the fast Hyper-Transport buses connecting the $\Nproc$ processors with each other.
  
As a concluding remark it is summarized that high-performance implementations of Fast Fourier Transforms in four dimensions are 
indeed available, which thus allows for the efficient construction of the overlap Dirac operator as described in \sect{sec:HMCAlgorithm}. 
Though not explicitly discussed here and not implemented within the xFFT approach, this includes also the situation of lattice 
side lengths being non-powers of two decomposable, however, into small prime factors as discussed, for instance, in \Ref{FFTW05}. 

% usw.

% Abkuerzungen* %%%%%%%%%%%%%%%%%%%%%%%%%%%%%%%%%%%%%%%%%%%%%%%%%%%%%%%%%%%%%%%%

% \include{abbreviations}

% Danksagung* %%%%%%%%%%%%%%%%%%%%%%%%%%%%%%%%%%%%%%%%%%%%%%%%%%%%%%%%%%%%%%%%%%

  \chapter*{Acknowledgments}
\label{chap:Acknowledgments}

Over the last years I received a lot of warm and inspiring support from many 
individuals and organizations, my supervisors, colleagues, friends, relatives
and other supporters, I feel deeply indebted to. Here, I would like to take
the opportunity to express my gratitude to these persons and to name them
individually.

First of all, I would like to cordially thank my direct supervisor Dr. habil. Karl Jansen
for guiding and mentoring this work in a very close, friendly, inspiring, and 
supportive manner, for sharing his ideas and great experience on
Higgs-Yukawa models in particular and lattice theory in general, and especially
for the endless hours he spent with me on discussing the open questions of this research
project. His experience was extremely helpful for identifying promising investigation 
strategies and safely avoiding conceptual impasses. Moreover, I am  
thankful for him providing me the opportunity to participate in a series of international 
conferences and schools, which greatly promoted the success of this work.

Of equal importance has been the continuous and ongoing support of my supervising 
professor Prof. Dr. Michael M\"uller-Preussker that has always been constituting the robust and 
reliable fundament, which is required for the successful completion of a PhD project.
It has been his scientific advice and shared experience, from which I could
tremendously profit over the last years. Moreover, the friendly and warm atmosphere
in his working group has driven the time spent on the project into a very pleasant experience.
Especially, I would like to thank him for his personal engagement 
that allowed me to benefit from a PhD scholarship provided by the Deutsche Telekom Stiftung 
and to participate in the annual Nobel laureate meeting in Lindau 2008. For all that I
feel indebted to him.

Moreover, I would like to express my gratitude to the Deutsche Telekom Stiftung for supporting
me by providing a PhD scholarship and for granting additional funding for traveling and attending
conferences. The generous financial aid coming along with that scholarship enabled me in particular to spend
a considerable fraction of my PhD research project abroad. I am therefore very thankful to the
numerous people enabling the work of this organization. In particular, I would like to express 
my special gratitude to Prof. Dr. Dr. h.c. Manfred Erhardt for continuously supporting and mentoring me 
over the last couple of years in a personal manner and from a non-physical perspective. 

During my half-year stay in the research group of Prof. Julius Kuti at the University of California, San Diego,
I received a lot of warm and friendly support from the local group members and collaborators, among them Prof. 
Kieran Holland, Dr. Daniel Nogradi, Chris Schroeder, Pamela Smilo, and in particular from Prof. Julius Kuti, 
who originally invited me to San Diego, and from whom I could learn a lot about Fourier acceleration techniques 
and many other details of Higgs-Yukawa systems. I am very thankful for the helpful and inspiring discussions
and moreover for the great hospitality that I experienced during that stay.

Special thanks go to Jim Kallarackal, who helped me a lot in broadening my understanding of and knowledge on 
general quantum field theory through uncountable discussions, for conversations and collaborations on specific 
issues of the considered Higgs-Yukawa model, for instance contributing certain routines of the software code 
analyzing the generated field configurations, and for his helpful comments on this manuscript. 

Also, I would like to thank Prof. Jaume Carbonell, Dr. Jean-Christian Angl\`es d'Auriac, and Dr. Feliciano de Soto
for inviting me to the University of Grenoble, France, for a research visit. In particular, I am grateful
for discussions about the details of alternative physical applications of the studied Higgs-Yukawa model in the
framework of nuclear theory as well as for their great hospitality.
 
Moreover, I thank Prof. Dr. Andreas Wipf, Prof. Dr. Holger Gies, Dr. Tobias K\"astner, and Christian Wozar for 
inspiring discussions about related scalar-fermion systems and for their hospitality.

I am also grateful to Dr. Dru Renner, Dr. Georg von Hippel, and Dr. Rainer Sommer for many helpful hints and 
comments on general lattice techniques and quantum field theory. 

During my daily work time I always felt surrounded by friendly and cooperative colleagues. In 
particular, I would like to name my current and former office workmates Xu Feng, Jenifer Gonzalez Lopez, 
Isaac Hailperin, Petra Kovacikova, Marina Marinkovic, and Andreas Nube, who made working here a very pleasant
experience. I would like to thank all of them for many motivating and instructive conversations -- and moreover --
simply for the great time over the last years. 

Additional funding for my PhD position before and after the above scholarship and for attending conferences 
has been provided by the DFG through the DFG-project {\textit{Mu932/4-1}}, the Humboldt-University Berlin, 
the Wilhelm und Else Heraeus-Stiftung, as well as DESY in Zeuthen, which I am very grateful for.

The major fraction of the computational workload has been performed on the {\textit{HP XC4000 System}}
at the {\textit{Scientific Supercomputing Center}} of the University Karlsruhe, Germany. The work of the
local staff maintaining and running that machine has always been characterized by a high degree of reliability
and professionalism which, together with the generous allocation of computing resources, greatly promoted 
the success of this study. I also would like to acknowledge the support of the {\textit{SGI system HLRN-II}} at 
the {\textit{HLRN Supercomputing Service}} Berlin-Hannover, Germany. 

Last but not least I would like to thank my family, in particular my parents Brigitte and Dankward Gerhold as well as
my sister Rosemarie Gerhold, and my girlfriend Cornelia Kipping, who have always been there for me. Without their great moral 
support this work would not have been possible. Thank you!

%Own Publications
%\nociteownPublications{ownGerhold:2009ub}
%\nociteownPublications{ownGerhold:2008mb}
%\nociteownPublications{ownGerhold:2007pj}
%\nociteownPublications{ownGerhold:2007gx}
%\nociteownPublications{ownGerhold:2007yb}
%\nociteownPublications{ownGerhold:2006kw}
%\nociteownPublications{ownGerhold:2006rc}
%\nociteownPublications{ownGerhold:2006bh}
%\nociteownPublications{ownGerhold:2006sk}
%\nociteownPublications{ownIlgenfritz:2005ga}

% Lebenslauf* %%%%%%%%%%%%%%%%%%%%%%%%%%%%%%%%%%%%%%%%%%%%%%%%%%%%%%%%%%%%%%%%%%

% \include{cv} % Achtung! Ab sofort keine Veroeffentlichtung erlaubt.

% Literaturverzeichnis %%%%%%%%%%%%%%%%%%%%%%%%%%%%%%%%%%%%%%%%%%%%%%%%%%%%%%%%%

% Verwendung von bibtex ist pflichtig.
  \bibliography{bibliography}  
  \bibliographystyle{unsrtnat} %bzw. andere auf unserer Website aufgelisteten Stile
%  \bibliographystyle{alphadidiEN}  %  plainnat, abbrvnat, unsrtnat, alphadidiEN

%\bibliographyownPublications{ownPublications}  
%\bibliographystyleownPublications{unsrtnat} %bzw. andere auf unserer Website aufgelisteten Stile

\renewcommand{\bibname}{Own Publications}

% Verzeichnisse %%%%%%%%%%%%%%%%%%%%%%%%%%%%%%%%%%%%%%%%%%%%%%%%%%%%%%%%%%%%%%%%
 
  \listoffigures
  \newpage

  \listoftables
  \newpage  

% Selbststaendigkeiterklaerung %%%%%%%%%%%%%%%%%%%%%%%%%%%%%%%%%%%%%%%%%%%%%%%%%

  \selectlanguage{ngerman}
\chapter*{Selbst\"andigkeitserkl\"arung}

\vs{20mm}

\noindent Hiermit erkl\"are ich, die vorliegende Arbeit selbst\"andig 
ohne fremde Hilfe verfa{\ss}t und nur die angegebene Literatur 
und Hilfsmittel verwendet zu haben.\\
\vspace{5cm}

\noindent\dcauthorname\ \dcauthorsurname \\  %%\hspace{-.6cm}
Berlin, den \dcdatesubmitted \\

  \end{document}